\documentclass{aa}  
\relax
\usepackage{graphicx}
\usepackage{txfonts}
\usepackage{pdflscape}
\usepackage{color}

\def\colort{\textcolor{blue}}

\usepackage{amstext}
\usepackage[colorlinks=true,linkcolor=black,citecolor=blue]{hyperref}%

\begin{document} 
\title{Torus and polar dust dependence on AGN properties}

   \author{I. Garc\'ia-Bernete\inst{1}\fnmsep\thanks{E-mail: igbernete@gmail.com}, O. Gonz\'alez-Mart\'in\inst{2}, C. Ramos Almeida\inst{3,4}, A. Alonso-Herrero\inst{5}, M. Mart\'inez-Paredes\inst{6}, M.\,J. Ward\inst{7}, P.F. Roche\inst{1}, J.\,A. Acosta-Pulido\inst{3,4}, E. L\'opez-Rodr\'iguez\inst{8}, D. Rigopoulou\inst{1} and D. Esparza-Arredondo\inst{2}\\}
 
   \institute{$^1$Department of Physics, University of Oxford, Keble Road, Oxford OX1 3RH, UK \\
   $^2$Instituto de Radioastronom\'ia and Astrof\'sica (IRyA-UNAM), 3-72 (Xangari), 8701, Morelia, Mexico\\ 
   $^3$Instituto de Astrof\'isica de Canarias, Calle V\'ia L\'actea, s/n, E-38205 La Laguna, Tenerife, Spain\\
   $^4$Departamento de Astrof\'sica, Universidad de La Laguna, E-38206 La Laguna, Tenerife, Spain\\
   $^5$Centro de Astrobiolog\'ia (CAB), CSIC-INTA, Camino Bajo del Castillo s/n, E-28692, Villanueva de la Ca\~nada, Madrid, Spain\\
   $^6$Korea Astronomy and Space Science Institute 776, Daedeokdae-ro, Yuseong-gu, Daejeon, Republic of Korea (34055)\\
   $^7$Centre for Extragalactic Astronomy, Durham University, South Road, Durham DH1 3LE, UK\\
   $^8$Kavli Institute for Particle Astrophysics and Cosmology (KIPAC), Stanford University, Stanford, CA 94305, USA\\}

\titlerunning{Torus and polar dust dependence on AGN properties}
\authorrunning{Garc\'ia-Bernete et al.}

   \date{}

  \abstract
   {We present a statistical analysis of the properties of the obscuring material around active galactic nuclei (AGN). This study represents the first of its kind for an ultra-hard X-ray (14–195~keV; Swift/BAT) volume-limited (D$_{\rm L}<$40~Mpc) sample of 24 Seyfert (Sy) galaxies (BCS$_{40}$ sample) using high angular resolution infrared data and various torus models: smooth, clumpy and two-phase torus models and clumpy disc$+$wind models. We find that the smooth, clumpy and two-phase torus models (i.e. without including the polar dusty wind component) and disc$+$wind models provide best fits for a comparable number of galaxies, 8/24 (33.3\%) and 9/24 (37.5\%), respectively. We find that the best-fitted models depend on the hydrogen column density (N$_{\rm H}^{\rm X-ray}$), which is related to the X-ray (unobscured/obscured) and/or optical (Sy1/Sy2) classification. In particular, smooth, clumpy and two-phase torus models best reproduce the infrared (IR) emission of AGN with relatively high hydrogen column density (median value of log (N$_{\rm H}^{\rm X-ray}$\,cm$^{-2}$)=23.5$\pm$0.8; i.e. Sy2). However, clumpy disc$+$wind models provide the best fits to the nuclear IR spectral energy distributions (SEDs) of Sy1/1.8/1.9 (median value of log (N$_{\rm H}^{\rm X-ray}$\,cm$^{-2}$)=21.0$\pm$1.0), specifically in the near-IR (NIR) range. The success of the disc$+$wind models in fitting the NIR emission of Sy1 galaxies is due to the combination of adding large graphite grains to the dust composition and self-obscuration effects caused by the wind at intermediate inclinations. In general, we find that the Seyfert galaxies having unfavourable (favourable) conditions, i.e. nuclear hydrogen column density and Eddington ratio, for launching IR dusty polar outflows are best-fitted with smooth, clumpy and two-phase torus (disk$+$wind) models confirming the predictions from simulations. Therefore, our results indicate that the nature of the inner dusty structure in AGN depend on the intrinsic AGN properties.}

      
   \maketitle

%________________________________________________________________

\section{Introduction}
Active galactic nuclei (AGN) are powered by accretion of material onto supermassive black holes (SMBHs), which releases energy in the form of radiation and/or mechanical outflows to the interstellar medium (ISM) of the host galaxy. The impact of the energy released by AGN in its surrounding environment has been proposed as a key mechanism responsible for regulating star formation in galaxies \citep{HopkinsQuataert10}. Although they comprise a relatively small fraction of the galaxies in the local universe ($\sim$10\%), AGN are now considered to be a short phase ($<$100~Myr; e.g. \citealt{Hopkins05}) that might take place in all galaxies (e.g. \citealt{Hickox14}).

Nearby Seyfert (Sy) galaxies are intermediate luminosity AGN which are close enough ($\sim$tens of Mpc) to study their nuclear emission and characterize the properties of the nuclear obscurer on $\sim$100~pc scales (at the average angular resolution of 8-10~m-class ground-based telescopes $\sim$0.3\arcsec ~at 10\,$\rm{\mu}$m). The dusty torus/disc\footnote{Hereafter we will use the terms dusty torus and disc interchangeably. This term does not necessarily refer to a geometrically thick torus. Note that the majority of torus models use a flared disc geometry (i.e. a disc whose thickness increases with the distance from the centre).} is the key piece of the AGN unified model \citep{Antonucci93}. Depending on its orientation, it obscures the central engines of type 2 AGN, and provides a direct view of the central engine in the case of type 1 AGN. This nuclear dust absorbs a significant part of the AGN radiation and, then, reprocesses it to emerge in the infrared (IR; e.g.  \citealt{Pier92}).

Early works using direct imaging and interferometric data have found a relatively compact torus ($\sim$0.1-10~pc) in the mid-IR (MIR; $\sim$5--30~$\mu$m; e.g. \citealt{Jaffe04,Packham05,Tristram07,Tristram09,Tristram14,Radomski2008,Burtscher09,Burtscher13,Raban09,Lopez-Gonzaga16,Leftley18}). Recently, Atacama Large Millimeter/submillimeter Array (ALMA) observations of
the cold dust in Seyfert galaxies have spatially resolved the sub-mm counterpart of the torus (e.g. \citealt{Gallimore16,Garcia-Burillo16,Garcia-Burillo21,Imanishi18,Impellizzeri19}). \citet{Garcia-Burillo21} found that the bulk of the nuclear cold dust emission in Sy galaxies is equatorial with a median diameter of $\sim$42\,pc. These dusty molecular tori have been also detected in molecular gas observations of Seyfert galaxies and low-luminosity AGNs (e.g. \citealt{Herrero18,Herrero19,Herrero21,Combes19,Garcia-Burillo21}). These results suggest the multi-phase nature of the torus structure.

Due to the small angular size of the dusty and molecular tori, specially in the IR, 8-10~m-class ground-based telescopes cannot resolve it. Thus, comparing torus models to the observed nuclear IR spectral energy distributions (SEDs) is a powerful tool to constrain the properties of the nuclear dusty structure. Torus models can be broadly grouped in two categories: dynamical (i.e. radiation hydrodynamical and magneto-hydrodynamical simulations; e.g. \citealt{Wada02,Schartmann08,Wada12,Dorodnitsyn17,Kudoh20,Takasao22}) and static (i.e. radiative transfer models; e.g. \citealt{Pier92,Efstathiou95,Fritz06,Nenkova08A,Nenkova08B,Hoenig10B,Hoenig17,Stalevski2012,Stalevski16,Siebenmorgen2015}). The dynamical models include processes such as supernovae and AGN feedback. However, they require large computational times and thus it is more difficult to compare them with observations. On the other hand, static torus models can be easily compared with the observations, assuming various geometries and compositions of the dust (see \citealt{Ramos17,Honig19} for reviews).

For the sake of simplicity, the first geometrical torus models assumed a uniform distribution of the dust (e.g. \citealt{Pier92,Fritz06}). However, pioneering works showed that a clumpy distribution of the dust was necessary to prevent the destruction of grains \citep{Krolik88}. Therefore, a clumpy formalism has been employed in the majority of torus models (e.g. \citealt{Nenkova08A,Nenkova08B,Hoenig10B,Hoenig17}). Moreover, several hydrodynamical simulations predict that the torus is a multiphase structure (e.g. \citealt{Wada02,Schartmann14}) with a combination of smooth and clumpy dust distributions (i.e. two-phase torus models; e.g. \citealt{Stalevski2012,Stalevski16,Siebenmorgen2015}).

Since the first torus models were developed, our view of the dusty
torus has changed considerably. For instance, recent observations using IR interferometry have motivated a more complex scenario to explain the IR nuclear emission of Seyfert galaxies. \citet{Hoenig13} suggested that a significant fraction of the MIR emission is produced by dust located in the polar direction, whereas the near-infrared (NIR) flux is produced by a clumpy and compact disk (i.e. the dusty torus). Thus, some of the geometrical torus models also include a polar dust component (e.g. \citealt{Hoenig17}; hereafter clumpy disc$+$wind models). This polar emission has been detected at small scale (few pc) so far in six sources of the 23 observed using IR interferometry \citep{Lopez-Gonzaga16, Leftley18, Rosas22, Isbell22}. In addition, previous works also showed a large scale polar dust component (up to few hundred parsec; e.g. \citealt{Bock00,Radomski03,Packham05,Asmus14,Asmus19,Herrero21}).

Although the nuclear dust properties of nearby Seyfert galaxies have been extensively studied in the literature, only few works compared SED fits with different torus models to investigate which of them better reproduces the SED of Seyfert galaxies (e.g. \citealt{Gonzalez-Martin19A,Gonzalez-Martin19B}; hereafter GM19A,19B; \citealt{Esparza-Arredondo19,Esparza-Arredondo21}) and type-1 QSOs \citep{Martinez-Paredes21}. Given the different assumptions used to build the various available torus models, it is crucial to compare how they fit the observational data. Using \emph{Spitzer}/IRS spectra ($\sim$5-35\,$\mu$m), GM19B found that the clumpy disc$+$wind models \citep{Hoenig17} reproduce well the MIR emission of Sy1, whereas Sy2 are almost equally fitted by clumpy torus models (\citealt{Nenkova08A,Nenkova08B}; $\sim$43\% of the Sy2s) or clumpy disc$+$wind models ($\sim$40\% of the Sy2s). However, this study was limited by the spatial resolution ($\sim$4$\arcsec$) and spectral coverage (5-30\,$\mu$m) of Spitzer/IRS. Furthermore, \citet{Ramos14} reported that the minimum combination of subarcsecond angular resolution data needed to constrain torus model parameters is N-band spectroscopy (8--13\,$\mu$m) and NIR photometry (at least two data-points) when using the clumpy torus models by \citet{Nenkova08A,Nenkova08B}. However, there is a lack of detailed studies comparing different torus models to high angular resolution NIR-to-MIR data of Sy galaxies.

In this work, we investigate for the first time how various geometrical torus models (i.e. smooth, clumpy, two-phases) and disk$+$wind models fit the nuclear IR ($\sim$1-30~$\mu$m) SED of the ultra-hard X-ray volume-limited sample of Sy galaxies (BCS$_{40}$ sample) presented in \citet{Bernete16}. This will allow us to better understand the geometry, chemical composition, grain sizes and distribution of the nuclear dust. In addition, this will help to test the validity of the various torus models. 

The paper is organized as follows. Section \ref{sample} describes the sample selection. Section \ref{sec:models} gives a summary of the models used throughout this paper. The nuclear IR SED modelling is presented in Section \ref{results}. The main results are included in Section \ref{comparison} and discussed in Section \ref{discussion}.  Finally, in Section \ref{conclusions} we summarize the main conclusions of this work.  

\begin{table*}
\footnotesize
\caption{Main properties of the BCS$_{40}$ sample.}
\centering
\begin{tabular}{lccccccccc}
\hline
Name &	R.A.&	Dec.&D$_{L}$ &Spatial&Sy &log (N$_{\rm H}$)&log (L$_{\rm int}^{\rm X-ray}$)&log (M$_{\rm BH}$/M$_{\sun}$)&log ($\lambda_{\rm Edd}$)\\
 	&	(J2000)&	(J2000)&(Mpc) &scale& type& (cm$^{-2}$)&(erg s$^{-1}$)&&\\
 	&				&& &(pc/arcsec)& &&&&\\
 (1)&(2)&(3)&(4)&(5)&(6)&(7)&(8)&(9)&(10)\\	
\hline
NGC\,1365		&	03h33m36.4s&	-36d08m25s&21.5&103&1.8&22.21&42.32&7.92 (a)&-2.44		\\
NGC\,2110		&	05h52m11.4s&	-07d27m22s&32.4&155&2.0&22.94&42.69&9.25 (b)&-3.40		\\
ESO\,005-G004		&	06h05m41.6s&	-86d37m55s&24.1&116&2.0&	24.34 &42.78 & 6.98 (c) &-1.04\\
NGC\,2992		&	09h45m42.0s&	-14d19m35s&34.4&164&1.9	&21.72&42.00&5.42 (b)&-0.26	\\
MCG-05-23-016		&	09h47m40.1s&	-30h56m55s&35.8&171&2.0& 22.18&43.20&7.98 (a)&-1.62	\\
NGC\,3081		&	09h59m29.5s&	-22d49m35s&34.5&164&2.0 &23.91&42.72&8.41 (b)&-2.53		\\
NGC\,3227		&	10h23m30.6s&	+19d51m54s&20.4&98&1.5&20.95&42.10&6.62 (b)&-1.36		\\
NGC\,3783		&	11h39m01.7s&	-37d44m19s&36.4&173&1.2	& 20.49&43.43&7.14 (a)&-0.55	\\
UGC\,6728		&	11h45m16.0s&	+79d40m53s&32.1&153&1.2	& 20.00&41.80&5.32 (b)& -0.36	\\
NGC\,4051		&	12h03m09.6s&	+44d31m53s&12.9&62&1.2&20.00&41.33&5.60 (b)& -1.11		\\
NGC\,4138		&	12h09m29.8s&	+43d41m07s&17.7&85&1.9&22.89&41.23&7.30 (b)&-2.91		\\
NGC\,4151		&	12h10m32.6s&	+39d24m21s&20.0&96&1.5& 22.71&42.31&7.43 (a)&-1.96		\\
NGC\,4388*		&	12h25m46.7s&	+12d39m44s&17.0&82&2.0&23.52&43.05&6.99 (b)&-0.78		\\
NGC\,4395		&	12h25m48.8s&	+33d32m49s&3.84&19&1.8&21.04&40.50&4.88 (a)&-1.22		\\
NGC\,4945		&	13h05m27.5s&	-49d28m06s&4.36&21&2.0&24.80&42.69&7.78 (a)&-1.93		\\
NGC\,5128/CenA	&	13h25m27.6s&	-43d01m09s&4.28&21&2.0&23.02&42.39&7.94 (a)&-2.39	\\
MCG-06-30-015		&	13h35m53.7s&	-34d17m44s&26.8&128&1.2&20.85&42.74&7.42 (a)&-1.52	\\
NGC\,5506		&	14h13m14.9s&	-03d12m27s&30.1&144&1.9&22.44&42.99&8.29 (a)&-2.14		\\
NGC\,6300		&	17h16m59.5s&	-62d49m14s&14.0&68&2.0&23.31&41.84&7.01 (a)&-2.01		\\
NGC\,6814		&	19h42n40.6s&	-10d19m25s&25.8&123&1.5	&20.97&42.31&6.46 (b)&-0.99	\\
NGC\,7172		&	22h02m01.9s&	-31d52m11s&37.9&180&2.0	&22.91&42.76&8.45 (b)&-2.53	\\
NGC\,7213		&	22h09m16.3s&	-47d10m00s&25.1&120&1.5& 20.00&41.95&7.37 (c)&-2.26		\\
NGC\,7314		&	22h35m46.2s&	-26d03m02s&20.9&100&1.9&21.60&42.33&7.24 (b)&-1.75		\\
NGC\,7582		&	23h18m23.5s&	-42d22m14s&22.1&106&2.0&24.33&42.86&7.52 (a)& -1.50		\\
\hline
\end{tabular}						 
\tablefoot{ Right ascension (R.A.), declination (Dec.) and Seyfert type. *This galaxy is part of the Virgo Cluster \citep{Binggeli85}. We assumed a cosmology with H$_0$=70 km~s$^{-1}$~Mpc$^{-1}$, $\Omega_m$=0.3, and $\Omega_{\Lambda}$=0.7, and a velocity-field corrected using the \citet{Mould00} model, which includes the influence of the Virgo cluster, the Great Attractor, and the Shapley supercluster. The X-ray hydrogen column density and intrinsic 2--10~keV X-ray luminosity were taken from \citet{Ricci17}. References for the BH masses: (a) GB19; (b) \citet{Koss17}; (c) \citet{Vasudevan10}. Note that the Eddington ratio is derived following the methodology as in \citet{Ricci17c}.} 
\label{tab1}
\end{table*}
\section{Sample selection}
\label{sample}
Our sample consists of 24 Seyfert galaxies selected from the nine-month catalog \citep{Tueller2008} observed with {\textit{Swift/BAT}}. This sample was previously presented in \citet{Bernete16} (hereafter BAT Complete Seyfert sample at D$_L<$40\,Mpc, BCS$_{40}$ sample). The ultra-hard 14-195 keV band used to select the parent sample is far less sensitive to the effects of obscuration than optical or softer X-ray wavelengths, making this AGN selection one of the least biased for N$\rm{_H}$ $\rm{<}$10$\rm{^{24}}$~cm$\rm{^{-2}}$ to date (see e.g. \citealt{Winter2009,Winter2010,Weaver2010,Ichikawa2012,Ricci15,Ueda15}). 
We selected all the Seyfert galaxies in the nine-month catalog with luminosity distances D$_L<$40\,Mpc. We used this distance limit to ensure a resolution element of $\leqslant$50\,pc in the MIR, considering the average angular resolution of 8-10~m-class ground-based telescopes ($\sim$0.3\arcsec ~at 10\,$\rm{\mu}$m). The sample contains 8~Sy1 (Sy1, Sy1.2 and Sy1.5), 6~Sy1.8/1.9 and 10 Sy2 galaxies. This sample covers an AGN luminosity range log(L${_{\textrm{bol}}}$\,erg~s$^{-1}$)$\sim$41.75-44.75\footnote{Throughout this paper we adopt the standard notation log(x)$\equiv$ log$_{10}$(x).} and X-ray hydrogen column densities of N$_{\rm H}^{\rm X-ray}\sim$1$\times$10$^{20}$-6$\times$10$^{24}$~cm$^{-2}$. Note that we use bolometric luminosities derived by using the commonly employed bolometric correction of 20 (L${_{\textrm{bol}}}$=20~$\times$~L${_{2-10~\textrm{keV}}}$; e.g. \citealt{Vasudevan09}) at the 2-10~keV luminosities listed in Table \ref{tab1}. The main properties of the BCS$_{40}$ sample are shown in Table\,\ref{tab1}.

\section{Torus models} \label{sec:models}

\begin{figure*}
\centering
\par{
\includegraphics[width=14.5cm]{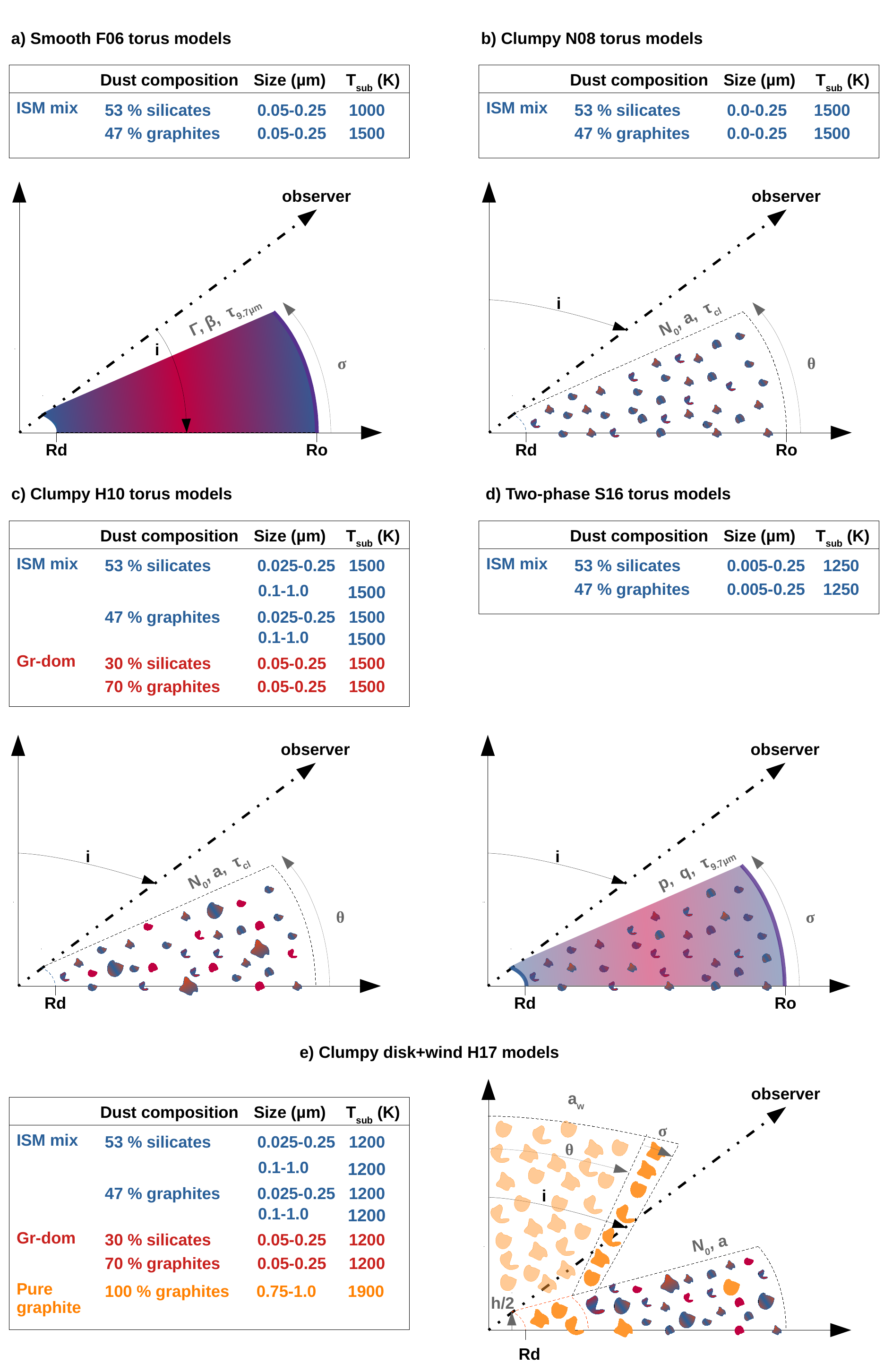}	
\par}
\caption{Scheme showing the different dust geometries and compositions of the various torus models used in this work. See Appendix\,\ref{nuclear_fits} for further details on the individual model parameters.
Note that the clumpy disc H17D models are not represented in this figure. However, the latter models consist of clumpy disc component of the clumpy disc$+$wind H17 models.}
\label{torus_geo}
\end{figure*}

We have chosen six models comprising different dust compositions, distributions and geometries (see also \citealt{Gonzalez-Martin19A} and references therein). In Figure\,\ref{torus_geo} we summarize the dust geometries and compositions, sublimation temperatures, and main parameters of each model used in this work. We also present a brief description of the models below:\\

$\rm{\bullet}$ \underline{Smooth torus model} by \cite{Fritz06} (hereafter {\textit{smooth F06 torus models}}): The model used a simple toroidal geometry, consisting of a flared disc represented as two concentric spheres having the polar cones removed. These two spheres delimit the inner and the outer torus radius, respectively. For the composition of dust the model considered a standard Galactic mix of 53\% silicates and 47\% graphite. The silicate and graphite grains have radii of $\rm{a_{g}}$ = 0.025 - 0.25 and $\rm{a_{g}}$ = 0.005 - 0.25 $\rm{\mu m}$, respectively. The parameters of the model are the viewing angle toward the torus, $i$, the half opening angle of the torus, $\sigma$, the exponent of the logarithmic azimuthal and radial density distribution, $\gamma$ and $\beta$, respectively, the ratio between external and internal radii, $\rm{Y = R_{\rm o}/R_{\rm d}}$, and the edge-on optical depth at 9.7 $\mu m$, $\tau_{9.7 \mu m}$ (see Table \ref{fritz_tab_parameters} of Appendix \ref{nuclear_fits}).

$\rm{\bullet}$ \underline{Clumpy torus model} by \cite{Nenkova08A,Nenkova08B} (hereafter {\textit{clumpy N08 torus models}}): The model used a formalism that accounts for the concentration of dust in clouds, forming a torus-like structure. They assumed spherical dust grains and a standard Galactic mix of 53\% silicates and 47\% graphite. The parameters of the model are: the view angle toward the torus, $i$, the number of clouds, $N_{0}$, the half opening angle of the torus, $\sigma$, the ratio between external and internal radii, $\rm{Y = R_{\rm o}/R_{\rm d}}$, the slope of the radial density distribution, $q$, and the optical depth of the individual clouds, $\tau_{\nu}$ (see Table\,\ref{nenkova_tab_parameters} of Appendix\,\ref{nuclear_fits}).

$\bullet$ \underline{Clumpy toroidal model} by \citet{Hoenig10B} \citep[see also][]{Hoenig06,Hoenig10A} (hereafter {\textit{clumpy H10 torus models}}): These are radiative transfer models of three-dimensional clumpy dust tori using optically thick dust clouds and a low torus volume filling factor. The majority of the models use a standard ISM dust mixture of graphite (47\%) and silicate (53\%) dust grains with a classical MRN size distribution (\citealt{Mathis77}) and a maximum size of 0.25 $\mu$m. By contrast, clumpy H10 torus models also include ISM-like large grains with sizes between 0.1 and 1.0 $\mu$m (i.e. using the same graphite/silicate mixture) and a population of graphite dominated grains (high refractory material), with 70\% fraction of graphites (30\% silicates) and maximum sizes of 0.25\,$\mu$m. The parameters of this library of SEDs are: the viewing angle $i$, the number of clouds along an equatorial line-of-sight $\rm{N_0}$, the half-opening angle of the distribution of clouds $\rm{\theta}$, the radial dust-cloud distribution power law index $a$, and the opacity of the clouds $\rm{\tau_{cl}}$. The outer torus radius $\rm{R_{\rm o}}$ is fixed to the inner radius as $\rm{R_{\rm o}=150\,R_{d}}$ (see Table \ref{hoenig10_tab_parameters} of Appendix \ref{nuclear_fits}).

$\rm{\bullet}$ \underline{Two-phase torus model} by \cite{Stalevski16} (hereafter {\textit{two-phase S16 torus models}}): The model used a torus geometry with a two-phase dusty medium, consisting of high-density clumps embedded in a smooth dusty component of low density. The dust chemical composition is set to a mixture of silicate and graphite grains. Model parameters are: the viewing angle toward the observer, $i$, the ratio between the outer and the inner radius of the torus, $Y = \rm{R_{\rm o}/R_{\rm d}}$, the half opening angle of the torus, $\sigma$, the indices that set dust density gradient with the radial $p$ and polar $q$ distribution of dust, and the $9.7 \mu m$ average edge-on optical depth, $\tau_{9.7\mu m}$ (see Table \ref{stalev_tab_parameters} of Appendix \ref{nuclear_fits}).

$\rm{\bullet}$ \underline{Clumpy disc and outflow model} by \cite{Hoenig17} (hereafter {\textit{clumpy disc$+$wind H17 models}}): The model consists of a clumpy disc plus a polar outflow. The authors used the same dust composition as in the clumpy H10 torus models, but they also included a second population of large pure-graphite grains (0.75-1.0 $\mu$m) which are more resilient than small silicates in hard environments (see e.g. \citealt{Waxman00,Perna03,Schartmann08,Lu16,Almeyda17,Garcia-Gonzalez17,Hoenig17}). The parameters for this model are the viewing angle, $i$, and the number of clouds in the equatorial plane, $N_{0}$, the exponent of the radial distribution of clouds in the disc, $a$, the optical depth of individual clouds in the disc, $\tau_{cl}$ (fixed to 50), the index of the dust cloud distribution power-law along the wind, $a_{w}$, the half-opening angle of the wind, $\theta$, the angular width of the hollow wind cone, $\sigma$, and the wind-to-disc ratio, $f_{wd}$ (defines the ratio between the number of clouds along the cone and $N_{0}$) (see Table \ref{hoenig17_tab_parameters} of Appendix\,\ref{nuclear_fits}). Note that this model assumes a fixed clouds radius (R$_{\rm cl}=$0.035$\times$r$_{\rm sub}$).

$\rm{\bullet}$ \underline{Clumpy disc} by \cite{Hoenig17} (hereafter {\textit{clumpy disc H17D models}}): The model consists of clumpy disc component of the previously described clumpy disc$+$wind H17 models (i.e. removing the wind component). Thus, the authors used the same dust grain composition and dust sublimation formalism as in the clumpy disc$+$wind H17 models (see text above and Table \ref{hoenig17_tab_parameters} of Appendix \ref{nuclear_fits}). Note that we include this model to further investigate the impact of the pure-graphite polar dust component on the fits (see Section \ref{discussion_dust_compo}).

It is worth noting that the main differences between the various models employed in this study are: a) nuclear dust geometry (i.e. torus, disc+wind), b) dust distribution (i.e. smooth, clumpy or two-phase) and c) dust composition and the treatment of the sublimation temperature of the dust grains. In particular, clumpy disc$+$wind H17 models are significantly different from the other torus models described above, both in the dust geometry and composition (see Section \ref{sec:models}). \citet{Hoenig17} proposed that a polar dusty outflow is launched near the dust sublimation zone and thus the polar dust composition should be similar to the dust in the inner regions of the disc (see \citealt{Hoenig17,Isbell21}). Therefore, they only included a population of large pure-graphite grains in the polar dust component assuming that it is swept-up dust from the inner wall of the torus/disc where silicate grains are destroyed by the intense emission from the AGN. In contrast, they included both silicate and graphite grains in the torus/disc component. To account for the different dust compositions, \citet{Hoenig17} used a physically motivated dust sublimation model considering that larger grains are heated less efficiently than smaller grains. This leads to various grain radial layers (species and sizes), where large graphite grains are hotter and closer to the AGN (e.g. \citealt{Schartmann08}). Note that this sublimation temperature treatment is not taken into account in the other torus models considered here (e.g. \citealt{Nenkova08A,Nenkova08B,Hoenig10B,Stalevski16}), although the smooth F06 torus models use different sublimation temperatures (T$_{\rm sub}^{\rm silicates}$=1000\,K and T$_{\rm sub}^{\rm graphites}$=1500\,K). Note that for simplicity throughout
this work we will use the term {\textit{torus models}} to refer to the smooth, clumpy and two-phases torus models (i.e. those models that do not include the dusty polar component).

\section{SED fitting with torus models}
\label{results}

\subsection{Accretion disc contribution}
\label{accretion_disk}
The subarcsecond resolution NIR fluxes of type 2 AGN are dominated by emission from hot AGN-heated dust with very little or no contribution from the accretion disc. However, another contribution to the NIR emission can be stellar emission from the host galaxy. To separate, as much as possible, the nuclear NIR emission from the stellar emission the highest possible spatial resolution is required (see e.g. \citealt{herrero98,herrero03}). Then, we assume that the flux contained in the scaled PSF (i.e. scaled PSF-star to the peak of the galaxy emission at different percentages; e.g. \citealt{Bernete2015,Bernete16}; GB19 and references therein) corresponds to the unresolved component and it is practically uncontaminated by star formation. 

In the case of type 1 AGNs, the NIR emission is mainly produced by very hot dust and the direct emission from the accretion disc of the AGN (see e.g. \citealt{caballero16}, GB19, \citealt{Landt19} and references therein). To quantify the contribution from the accretion disc to the nuclear NIR emission, we followed the same procedure as described in \citet{caballero16} using a semi-empirical model consisting of a template for the accretion disc and two blackbodies to fit the optical and NIR emission of each galaxy individually (see GB19). In GB19 we found that the accretion disc contribution to the nuclear IR SEDs ($\sim$0.4 arcsec) of Sy1s was, on average, 46$\pm$28, 23$\pm$13, and 11$\pm$5\% in the J, H, and K bands, respectively. Therefore, we subtracted the accretion disc component in the NIR range of each source prior to fitting the nuclear IR SEDs with the various torus models (see GB19 for further details). 

\begin{table*}
\begin{center}
\caption{Summary of models producing the best fit for each galaxy.}
\begin{tabular}{lll}
\hline \hline
Object          & Best model--HR & Best model--LR\\ 
\hline
\multicolumn{3}{c}{Sy1 galaxies}\\

MCG-06-30-015   &  H17 & H17\\
NGC3227         &  H17 (H17D) & H17 (H17D)\\
NGC3783         &  H17 & H17\\
NGC4051         &  \dots & \dots\\
NGC4151         &  H17   & H17\\
NGC6814         &  H17 & H17\\
NGC7213         &  H17D & H17D\\
UGC6728         &  H17/H17D (H10) & H17/H17D\\ 
\vspace{0.01cm}\\
\multicolumn{3}{c}{Sy1.8/1.9 galaxies}\\
NGC1365         &  H17 & H17\\
NGC2992         &  H17 & H17\\
NGC4138         &  H17/H17D (H10) & H17/H17D\\
NGC4395         &  \dots  & \dots\\
NGC5506         &  H17  & H17\\
NGC7314         &  H17    & H17\\ 
\vspace{0.01cm}\\
\multicolumn{3}{c}{Sy2 galaxies}\\
ESO005-G004     &  \dots  & \dots  \\
MCG-05-23-016   &  H17D (S16/H10/H17/N08) & H17D/H10 (H17) \\
NGC2110         &  H17D (H17/H10/N08) & H17D (H17/H10)\\
NGC3081         &  S16  (H17D/F06/H10/H17/N08) & H17D/H10/S16/N08 (H17)\\
NGC4388         &  H10 & H10\\
NGC4945         &  N08 (F06/H10/S16)  & N08/S16 (H10/F06)\\
NGC5128         &  H10 (F06/H17/S16/H17D)  & H10/H17D/H17/S16 (F06)\\
NGC6300         &  \dots  & \dots  \\
NGC7172         &  \dots  & \dots  \\
NGC7582         &  F06 (H17/S16/H17D)  & F06/S16 (H17D/H17/H10)\\
\hline \hline
\end{tabular}
\tablefoot{The best fit for each galaxy is selected according to the Akaike information criterion ($\epsilon$<0.01; see Section \ref{comparison}). Comparably good fits ($\rm{(\chi^2_{red}-\chi^2_{red,min})<0.5}$) are shown in within parenthesis. Objects without an assigned model cannot be reproduced by any of the models.} 
\label{tab:summaryfit}
\end{center}
\end{table*}
\subsection{SED fitting procedure}
\label{sedfittingsect}

Using the torus models described in Section \ref{sec:models} and XSPEC \citep{Arnaud96}, which is a command-driven and interactive spectral-fitting program within the HEASOFT\footnote{https://heasarc.gsfc.nasa.gov} software, we fitted all the nuclear NIR-to-MIR SEDs of our sample of Seyfert galaxies. XSPEC provides an easy way to incorporate new models using additive tables\footnote{\citet{Gonzalez-Martin19A} showed how to create an XSPEC additive table for each of the models employed.} together with a wide range of tools to perform spectral fittings to the data. 

\begin{figure}
\centering
\includegraphics[width=1.\columnwidth]{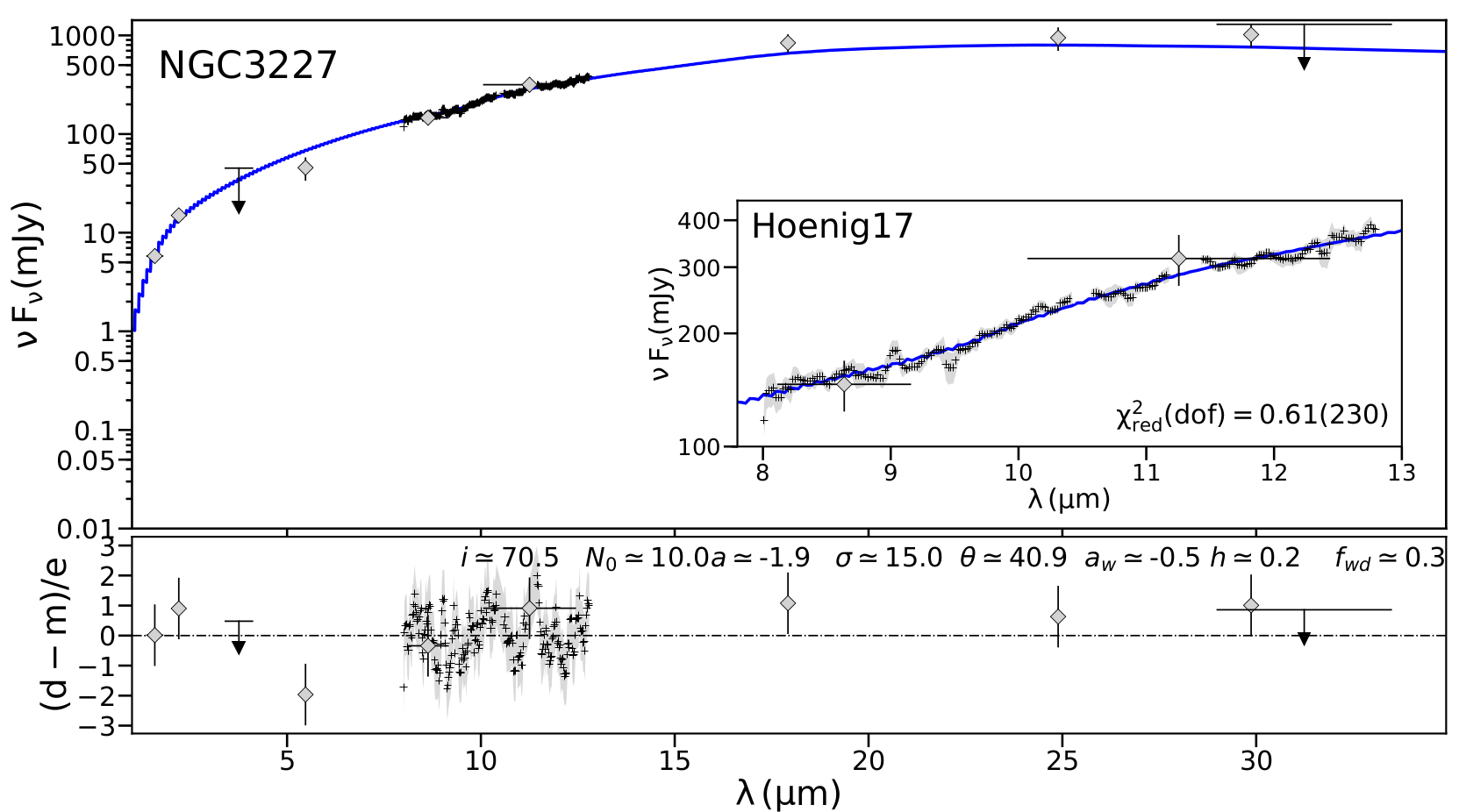}
\caption{Example of the nuclear IR SED of NGC\,3227 fitted with the clumpy disc$+$wind H17 models (top panel) and its residuals (bottom panel). The grey diamonds correspond to the high angular resolution photometric points. The black arrows represent low angular resolution data, which are treated as upper limits. Black crosses correspond to the high angular resolution N-band spectrum. The solid blue line is the best-fitted model.}
\label{example_fit}
\end{figure}

To construct high angular resolution NIR-to-MIR SEDs for the whole sample we compiled the highest angular resolution IR ($\sim$1-30~$\mu$m) nuclear fluxes available from the literature. The published MIR photometry and N-band spectroscopy (7.5--13~$\mu$m) used in this work was obtained with 8-10 m-class ground-based telescopes and different instruments (e.g. Gran Telescopio CANARIAS/CanariCam, Very Large Telescope/VISIR, Gemini/T-ReCS and MICHELLE). The nuclear NIR fluxes are from both ground- and space-based (i.e. Hubble Space Telescope; \emph{HST}) data (see Table 2 of GB19). In this work, we used the nuclear IR SEDs as in GB19 (see e.g. Fig. \ref{example_fit} and Appendix \ref{nuclear_fits}). We converted the N-band spectra and IR photometric data into XSPEC format using the {\sc flx2xsp} task within HEASOFT.

We masked  regions containing narrow spectral features, the 11.3 $\rm{\mu m}$ feature attributed to polycyclic aromatic hydrocarbon molecules (PAHs), and [S\,IV]$\lambda$10.5$\rm{\mu m}$ and [Ne\,II]$\lambda$12.8$\rm{\mu m}$ emission lines, to reveal the IR continuum. Note that other weak emission lines are not masked, since they do not affect the fit and that the other PAH emission bands are relatively weak in this sample. Previous studies also showed the importance of including an IR extinction law for fitting the IR SED of Sy galaxies (e.g. \citealt{Ramos11b}; hereafter RA11; \citealt{Herrero11}; hereafter AH11; \citealt{Ramos14} \& \citealt{Bernete19}; hereafter GB19). This is especially important for sources with very deep silicate features which generally show prominent dust lanes and/or are hosted in highly inclined galaxies (e.g. AH11, \citealt{gonzalez-martin13}). To do so, we use the IR extinction curve of \citet{Pei92}, which is already included as a multiplicative component within XSPEC. 

High spatial resolution N-band spectroscopy provides information on the silicate feature around 9.7$\rm{\mu m}$ which is important for the restriction of the model parameters (see e.g. \citealt{Martinez-Paredes20}). However, including spectral and photometric data in the fit is not a straightforward task from the statistical point of view. The $\chi^2$ statistic method takes every point into account equally so the best fit would tend to match the N-band spectral region over the photometric points. To avoid this, we perform the spectral fitting into two steps. We first fit the photometric data and low spectral resolution N-band spectrum (low-resolution, LR, fit). The LR spectra are computed to match the average bandpass of the photometric data. Then, we compute the 3$\rm{\sigma}$ errors for each parameter, which we use as priors for the SED fitting using the full resolution N-band spectra (high resolution, HR, fit). Note that the same methodology has been used in \citet{Martinez-Paredes21} for a sample of QSOs.

We compute the $\chi^2$ statistics for both the LR and HR fits. We consider the fit to be acceptable if the reduced $\rm{\chi^2}$ (for both HR and LR; see e.g. GM19A,19B) is $\rm{\chi^2_{red}<2}$. Among them, the best fit is the one providing minimum $\rm{\chi^2_{red}}$ and we consider two fits equally good if $\rm{(\chi^2_{red} - \chi^2_{red,min})<0.5}$. In Appendix \ref{nuclear_fits}, we present the results of the nuclear IR SED fitting process with the various torus models (see Section \ref{sec:models}). Tables\,\ref{appendixfitSy1}, \ref{appendixfitSyInt}, and \ref{appendixfitSy2} of Appendix \ref{nuclear_fits} show the results for Sy1, Sy1.8/1.9, and Sy2 galaxies. 

To further evaluate the goodness of the fits we use two methods. First, we use a qualitative method for performing a visual inspection of the residuals. To do so, we construct the average residuals using the entire N-band spectra and IR photometry, but excluding the upper limits (i.e. lower angular resolution data). For consistency, we use the same wavelength grid for all photometry (1.6, 2.2, 5.5, 18.0, 25.0 and 30\,$\mu$m) employing a quadratic interpolation of nearby values for each galaxy. Secondly, we use the Akaike information criterion (AIC). This method allows an evaluation of the best fit by comparing the minimum $\rm{\chi^2_{red}}$ with as comparably good
fits those ($\rm{(\chi^2_{red}-\chi^2_{red,min})<0.5}$). To determine which one is the best fit to the data, we compare the Akaike weights  \citep{Emmanoulopoulos16} of the models providing acceptable fits ($\epsilon$=W$_{\rm model1}$/W$_{\rm model2}$). We consider the best fit when $\epsilon$<0.01, which means that a given fit is, at least, 10 times better than other good fits (see e.g. \citealt{Martinez-Paredes21}). Finally, to investigate whether the goodness of the fit depends on the Seyfert type and other AGN properties, we use Fisher's exact test\footnote{The Fisher's exact test is valid for all sample sizes, but it is commonly employed when sample sizes are small.} that is commonly employed for testing for the independence (see Section \ref{best_models}). In Table \ref{tab:summaryfit} we present the best (and comparably good) fits to the LR and HR SEDs, which are practically the same. Therefore in the following, we only discuss the HR results.

\section{Comparison of the various torus models}
\label{comparison}

\begin{figure*}
\centering
\par{
\includegraphics[width=17.5cm]{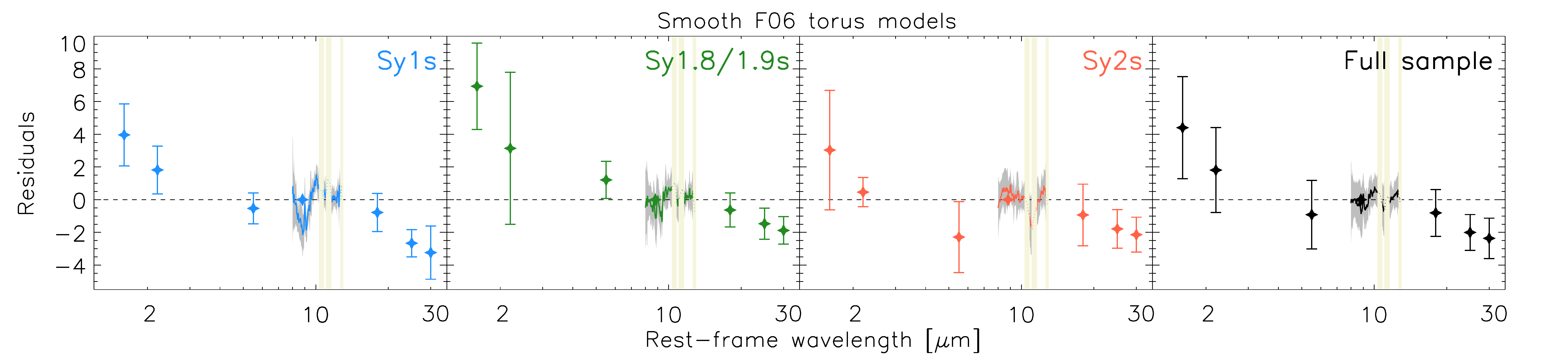}	
\includegraphics[width=17.5cm]{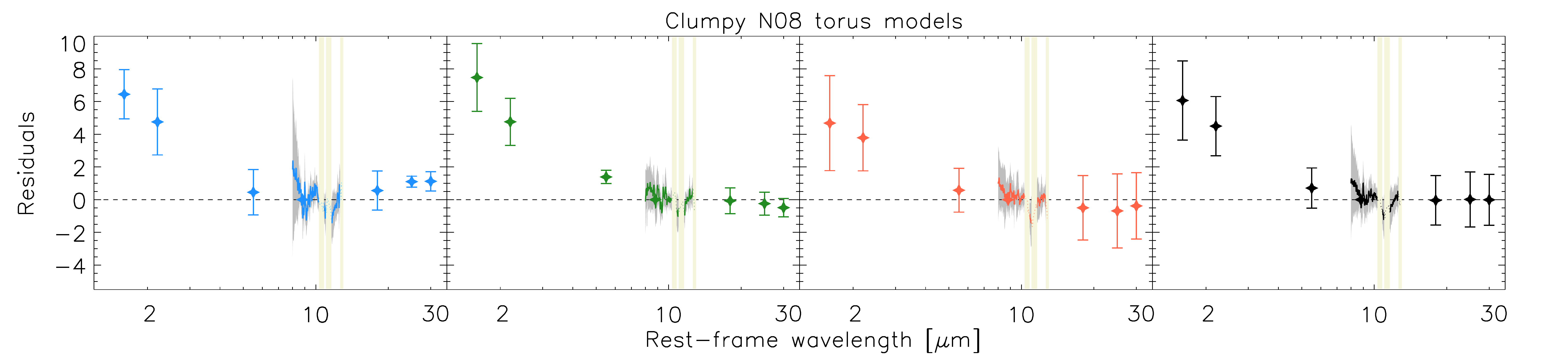}	
\includegraphics[width=17.5cm]{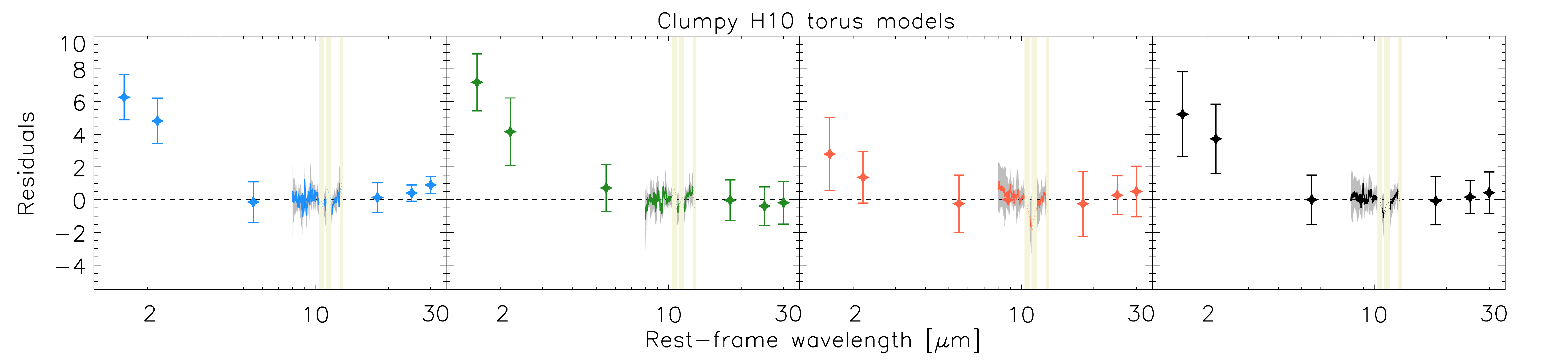}	
\includegraphics[width=17.5cm]{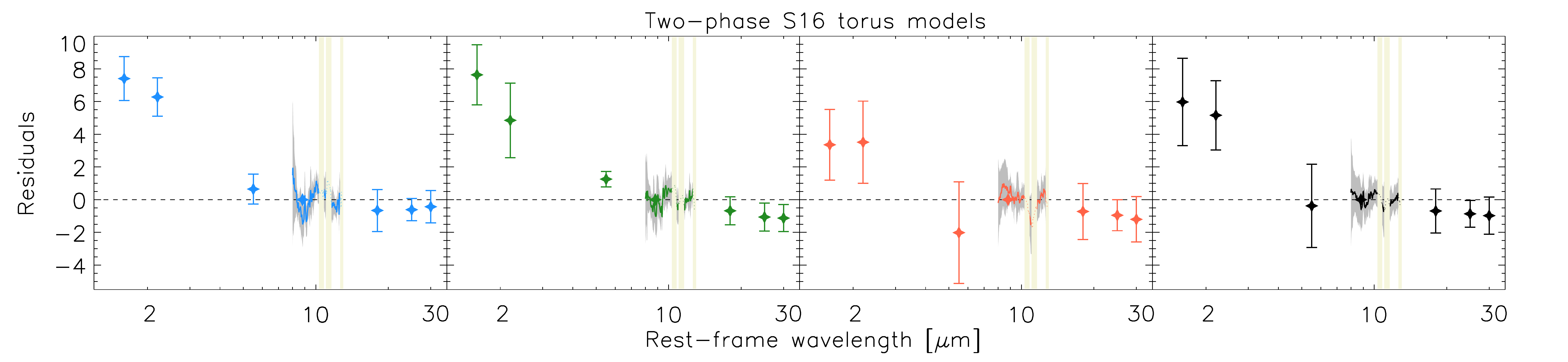}	
\includegraphics[width=17.5cm]{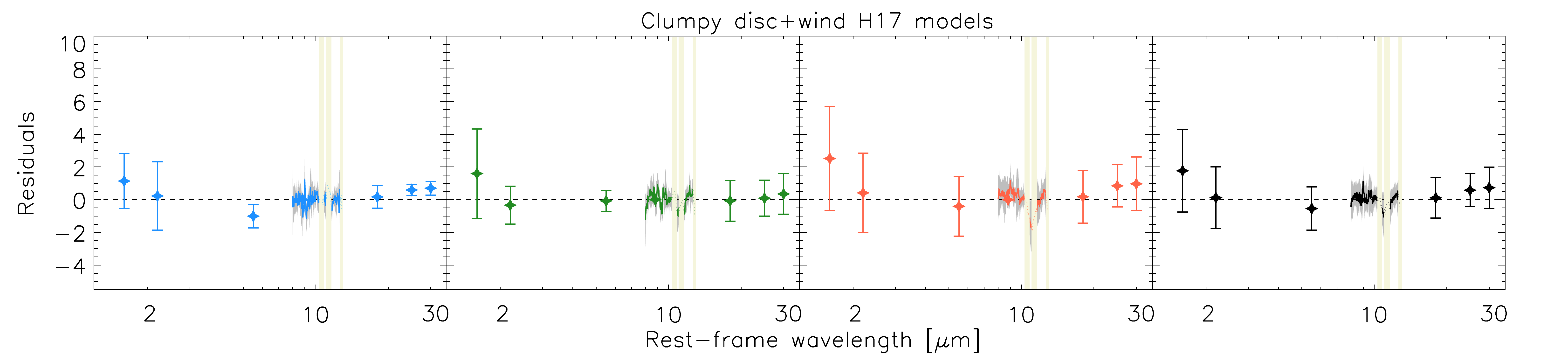}	
\includegraphics[width=17.5cm]{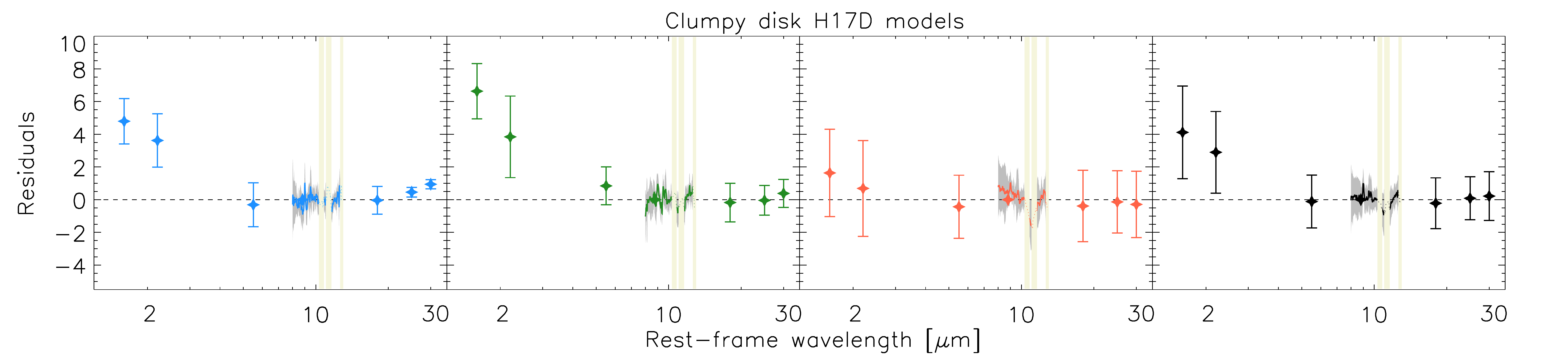}	
\par}
\caption{Average residuals (units as in Fig. \ref{example_fit}) of the spectral fitting for each torus model used in this work. Blue, green, red and black stars (and solid lines) correspond to Sy1, Sy1.8/1.9, Sy2 and the full sample, respectively. The regions masked in the fitting process are highlighted in beige vertical lines.}
\label{Residuals_histogram}
\end{figure*}

\subsection{Best model fits}
\label{best_models}

 \begin{figure*}
\centering
\par{
\includegraphics[width=5.8cm]{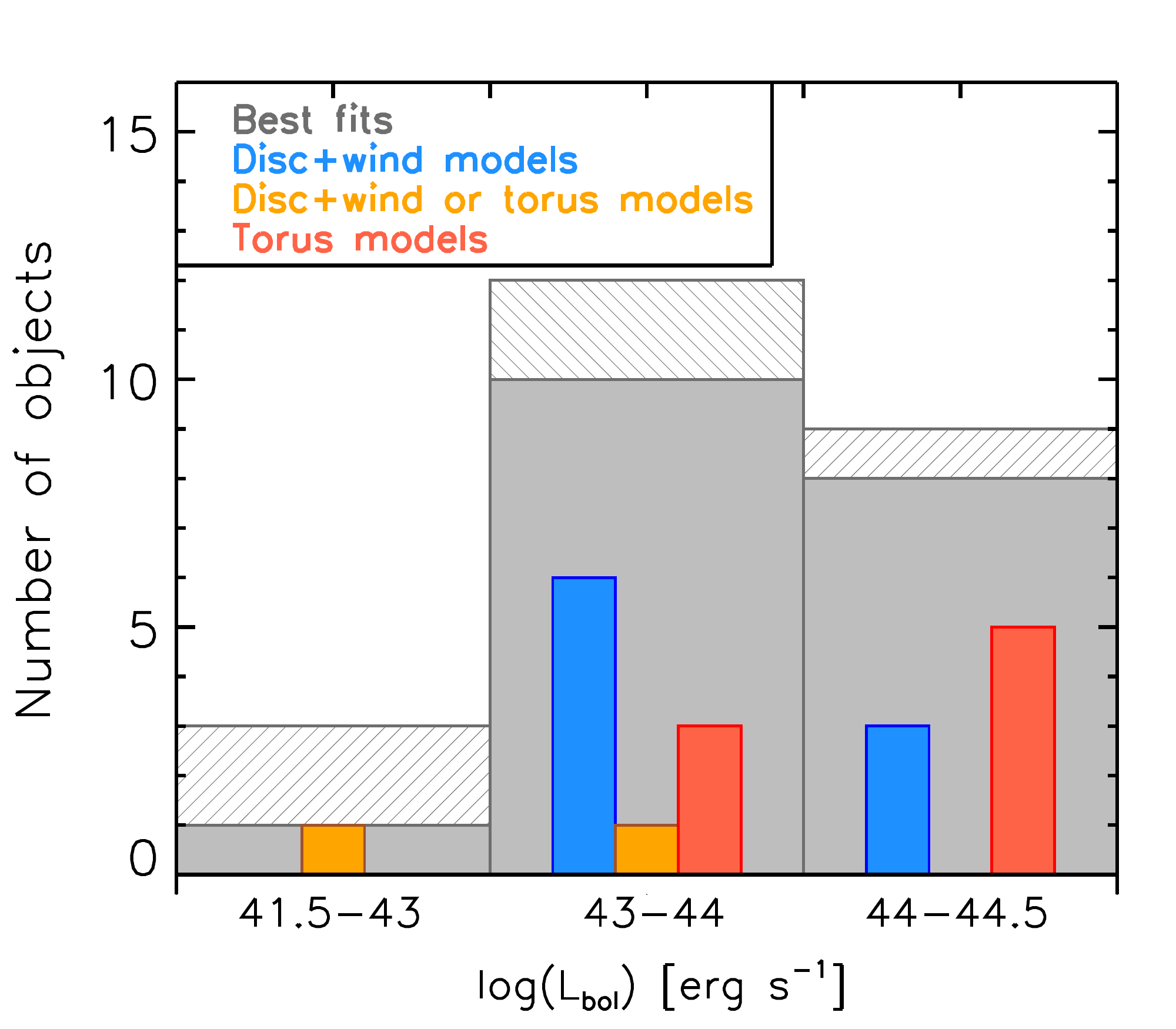}	
\includegraphics[width=5.8cm]{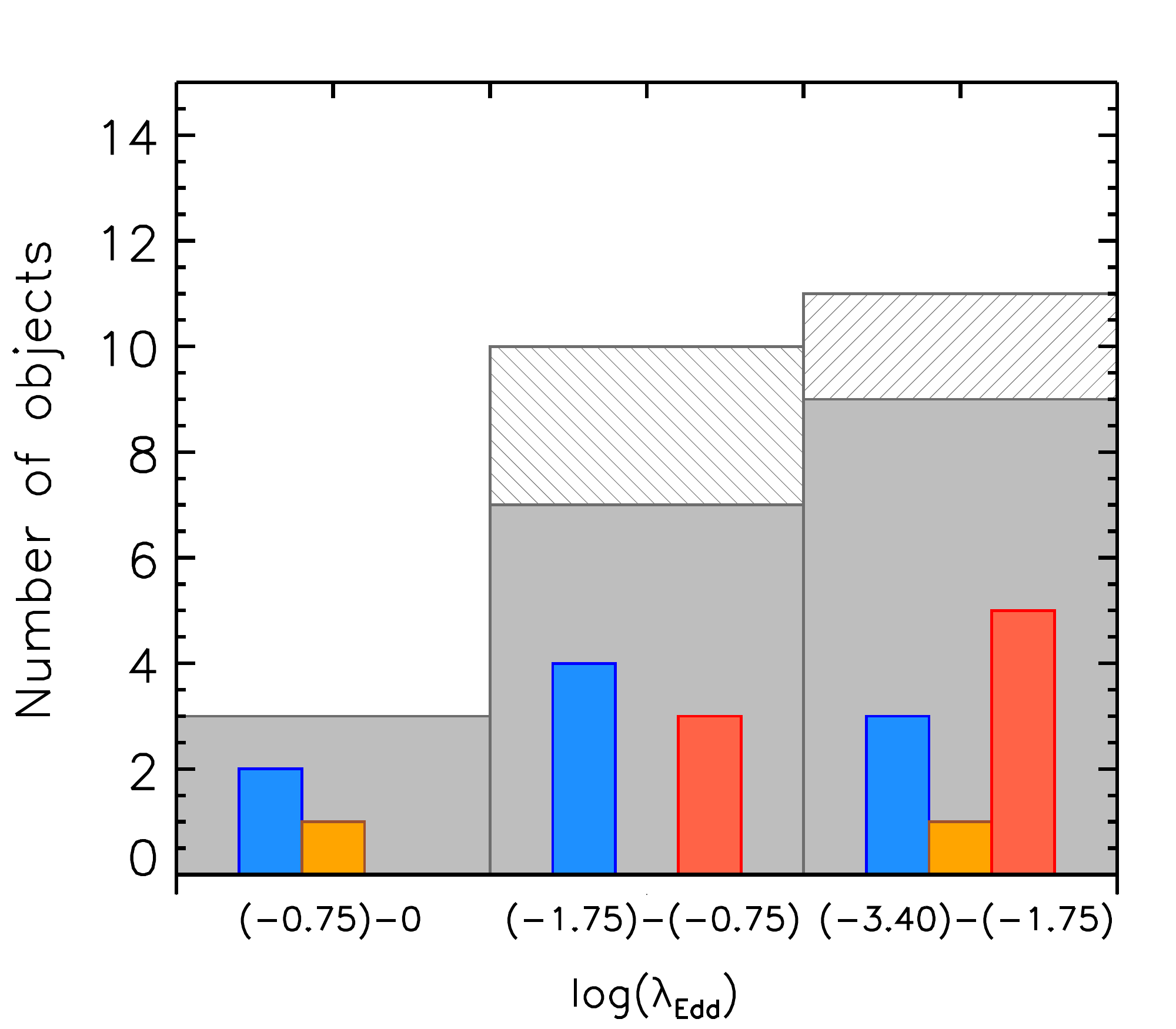}	
\includegraphics[width=5.8cm]{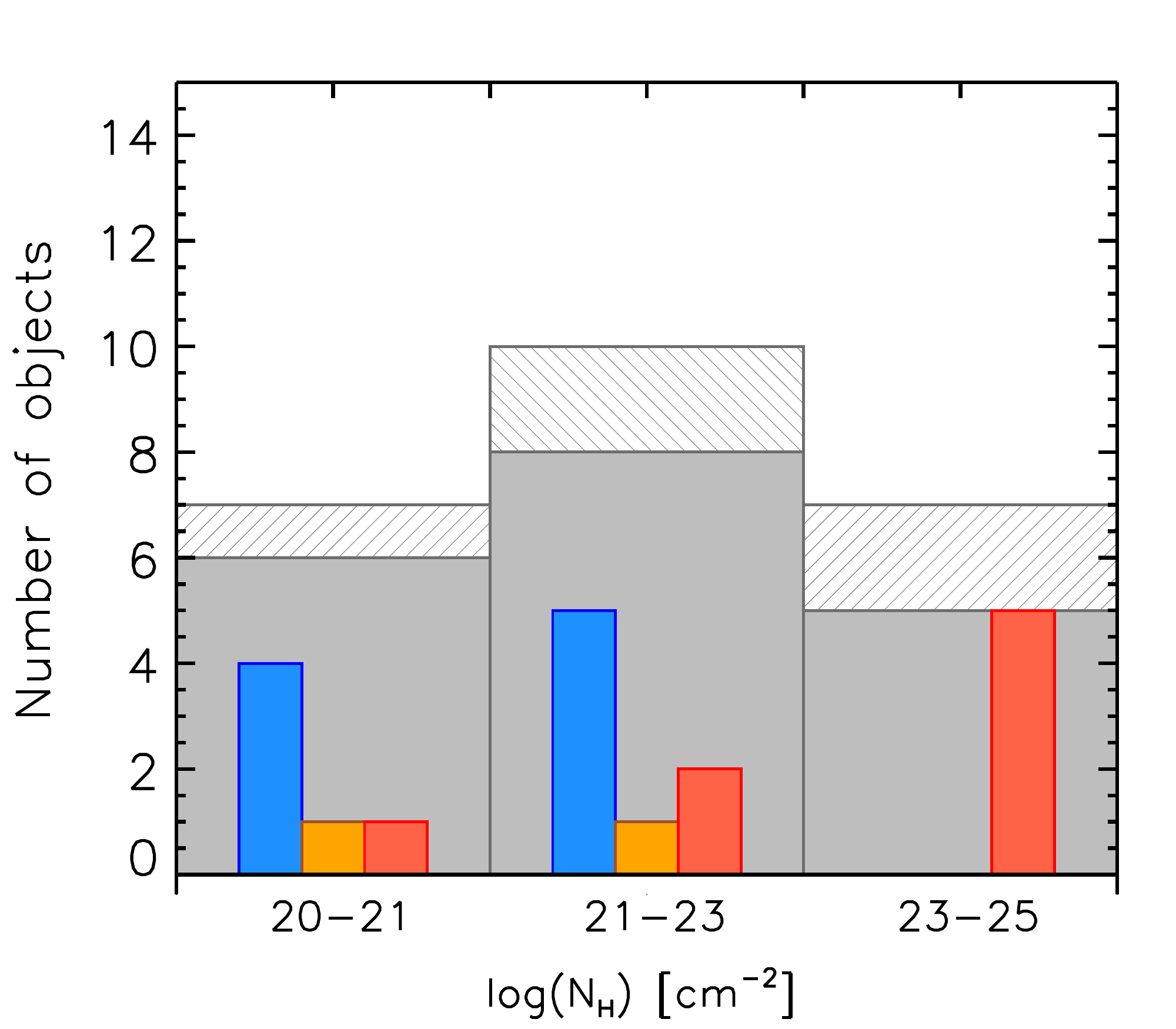}	
\par}
\caption{Best fit distributions of the BCS$_{40}$ sample per bolometric luminosity (left panel), Eddington ratio (central panel) and hydrogen column density (right panel) bin. The grey hatched and filled histograms are the distribution of unfitted and fitted sources. The blue and red filled histograms correspond to sources best fitted by clumpy disc$+$wind H17 models and torus models respectively. The orange filled histograms are sources equally fitted by clumpy disc$+$wind H17 models and torus models.} 
\label{fit_histogram}
\end{figure*} 

\subsubsection{Average fitting residuals}
\label{average_residual}

To determine which are the best suited models for reproducing the entire nuclear NIR-to-MIR SED of the various Sy groups, we first use a qualitative analysis of the average residuals of the spectral fitting. Fig.\,\ref{Residuals_histogram} presents these average residuals of our sample for each of the models considered in this work (see Section\,\ref{sec:models}). The average residuals are computed by grouping the various Seyfert types: from left to right panels of Fig.\,\ref{Residuals_histogram} are Sy1, Sy1.8/1.9, Sy2 and the full sample. From visual inspection of Fig.\,\ref{Residuals_histogram}, the average residuals of Sy1/1.8/1.9 indicate a clear excess at NIR emission for smooth, clumpy and two-phase torus models (i.e. torus models). This NIR excess was first reported by \citet{Neugebauer79} using a sample of quasars, and confirmed by \citet{Edelson86} in Seyfert galaxies. The clumpy disc H17D models generally produce slightly smaller residuals in the IR emission of Sy1 galaxies than other torus models used in this work. However, the models including the polar dust component produce the flattest residuals in the NIR for the entire sample (see Fig.\,\ref{Residuals_histogram}). 

The N-band spectra are equally well-fitted with most of the models, except for the clumpy N08 torus models and two-phase S16 torus models in the 8--10\,$\mu$m range. However, this can be in part due to contamination from the 7.7\,$\mu$m PAH band. On the other hand, the clumpy H10 torus models, clumpy disc$+$wind H17 models and clumpy disc H17D models show flatter average fitting residuals for the N-band spectra than the other models. Furthermore, the 18--30\,$\mu$m range is generally well reproduced by the various torus models, with the only exception of the smooth F06 torus models, which slightly over-predicts the emission above 20\,$\mu$m. Therefore, the clumpy disc$+$wind H17 models produce the best fits in the entire NIR-to-MIR range of the Sy1 galaxies in our sample. 

\begin{figure*}
\centering
    \renewcommand{\arraystretch}{0.01}
    \begin{tabular}{c c c}
    \textbf{{\hspace{0.9cm} smooth F06 torus models}} & \textbf{{\hspace{0.9cm} clumpy N08 torus models}} & \textbf{{ \hspace{0.9cm} clumpy H10 torus models}}\\
          \vspace{0.5cm}
      \includegraphics[width=5.8cm]{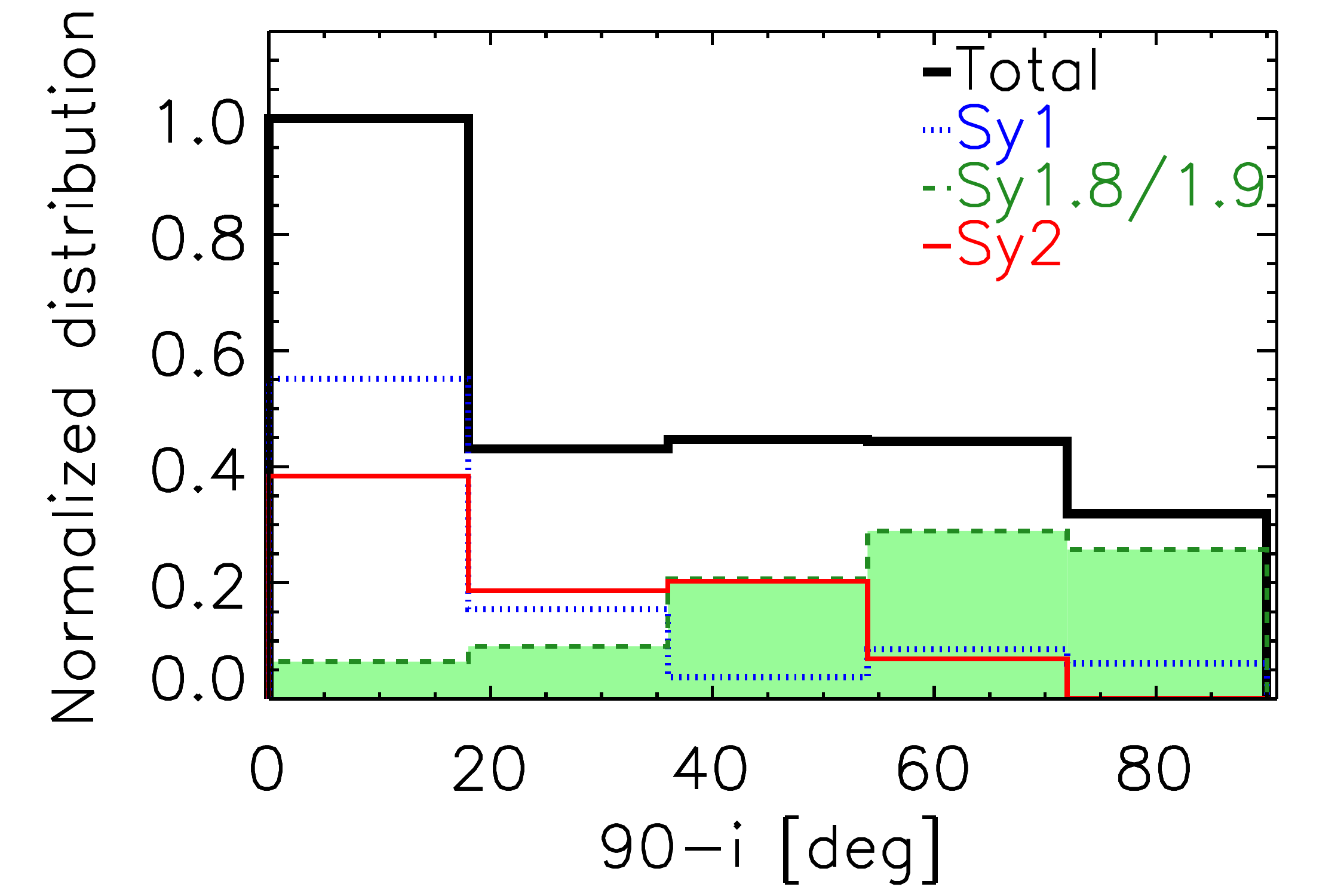}		 & \includegraphics[width=5.8cm]{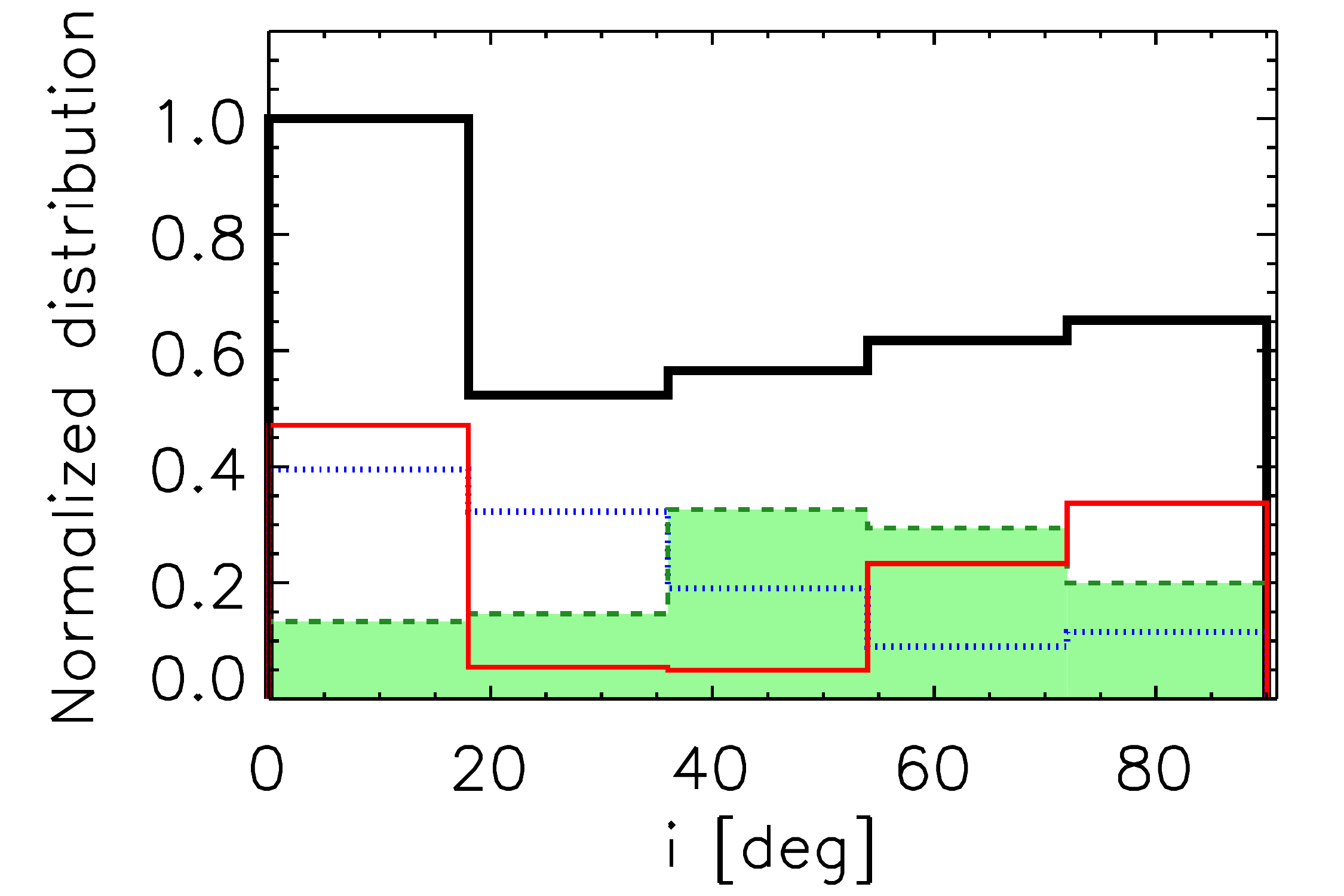}    &
      \includegraphics[width=5.8cm]{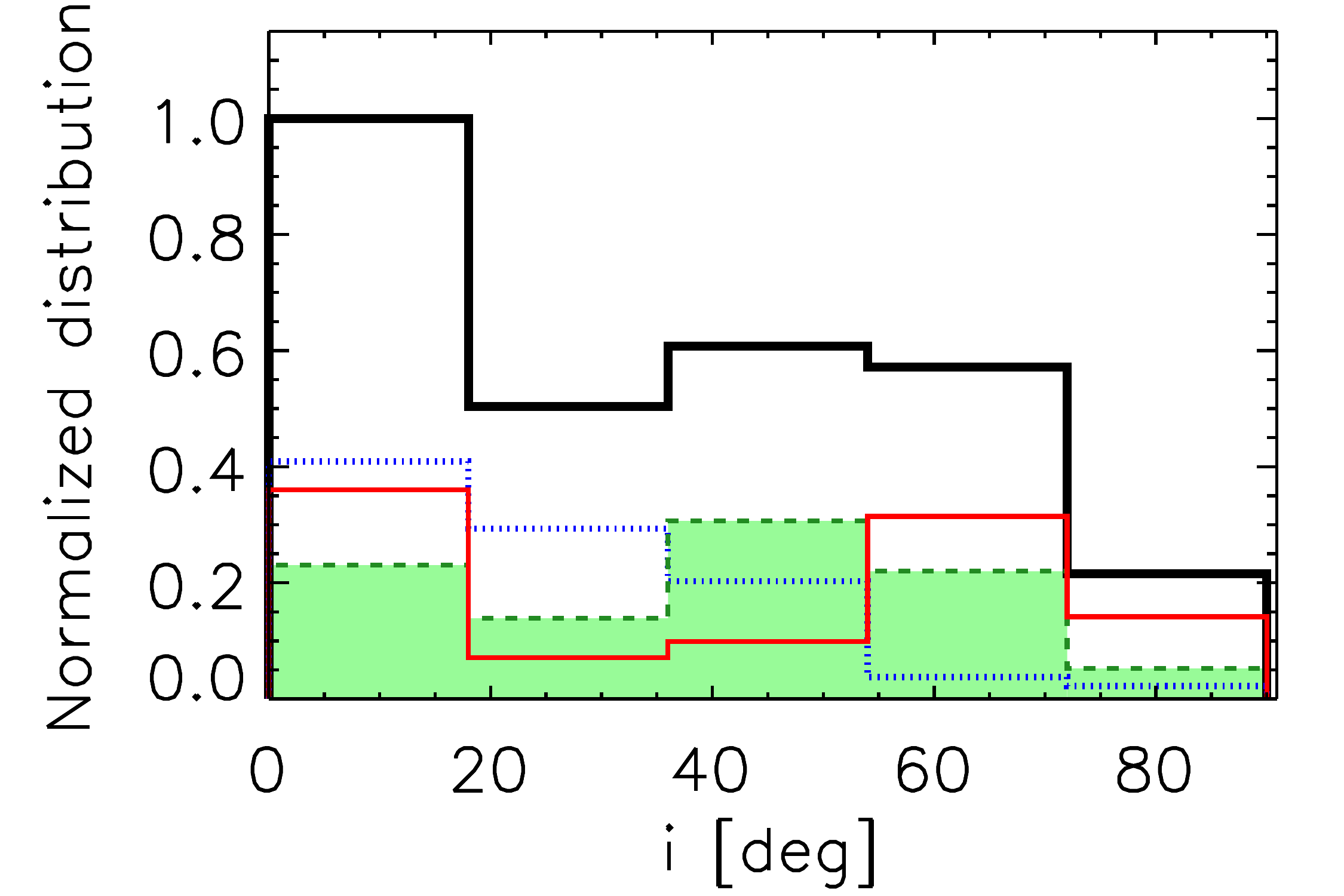}\\	
         \textbf{{\hspace{0.8cm} two-phase S16 torus models}} & \textbf{{\hspace{0.8cm} clumpy disc$+$wind H17 models}}  \\
      		 \includegraphics[width=5.8cm]{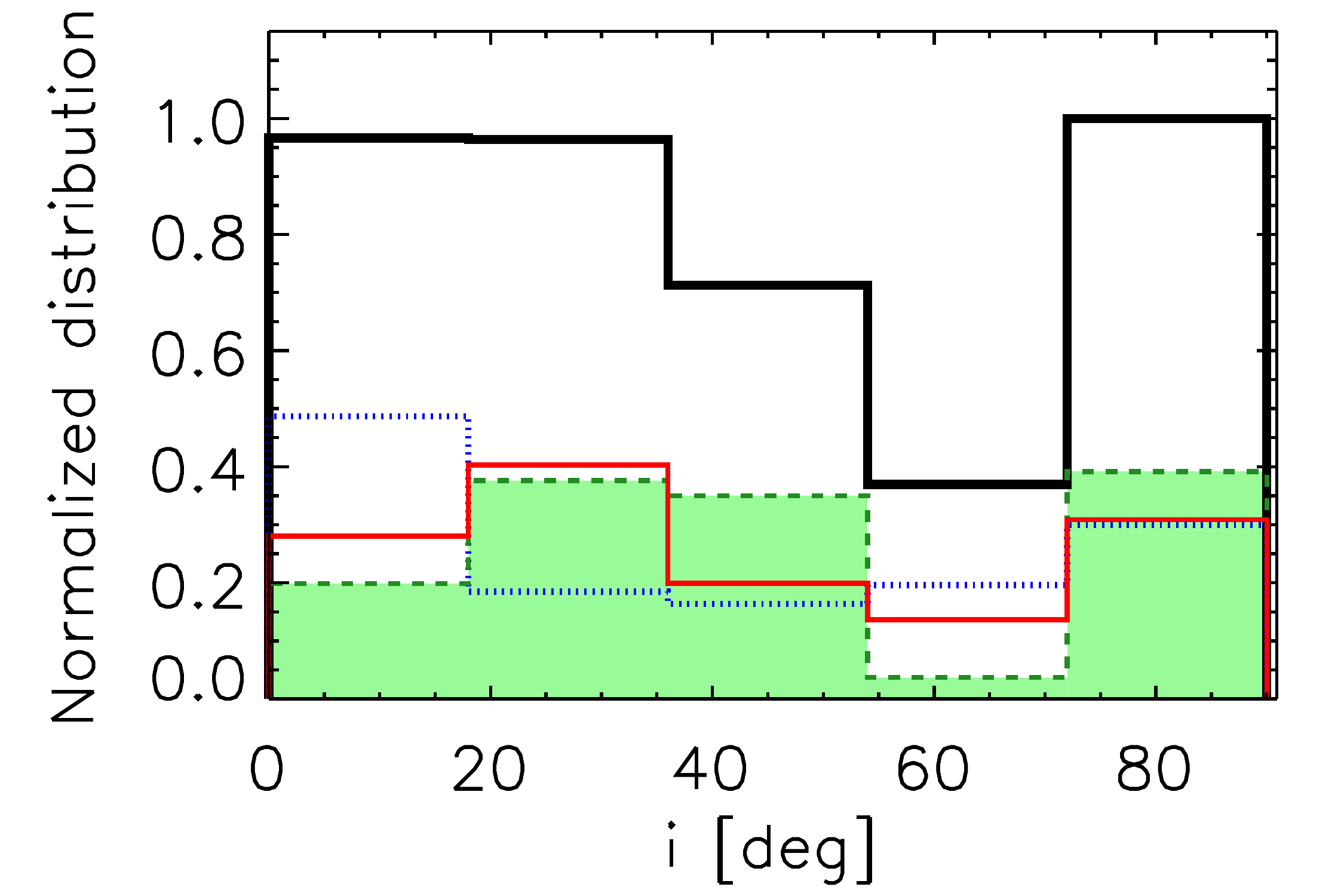} & \includegraphics[width=5.8cm]{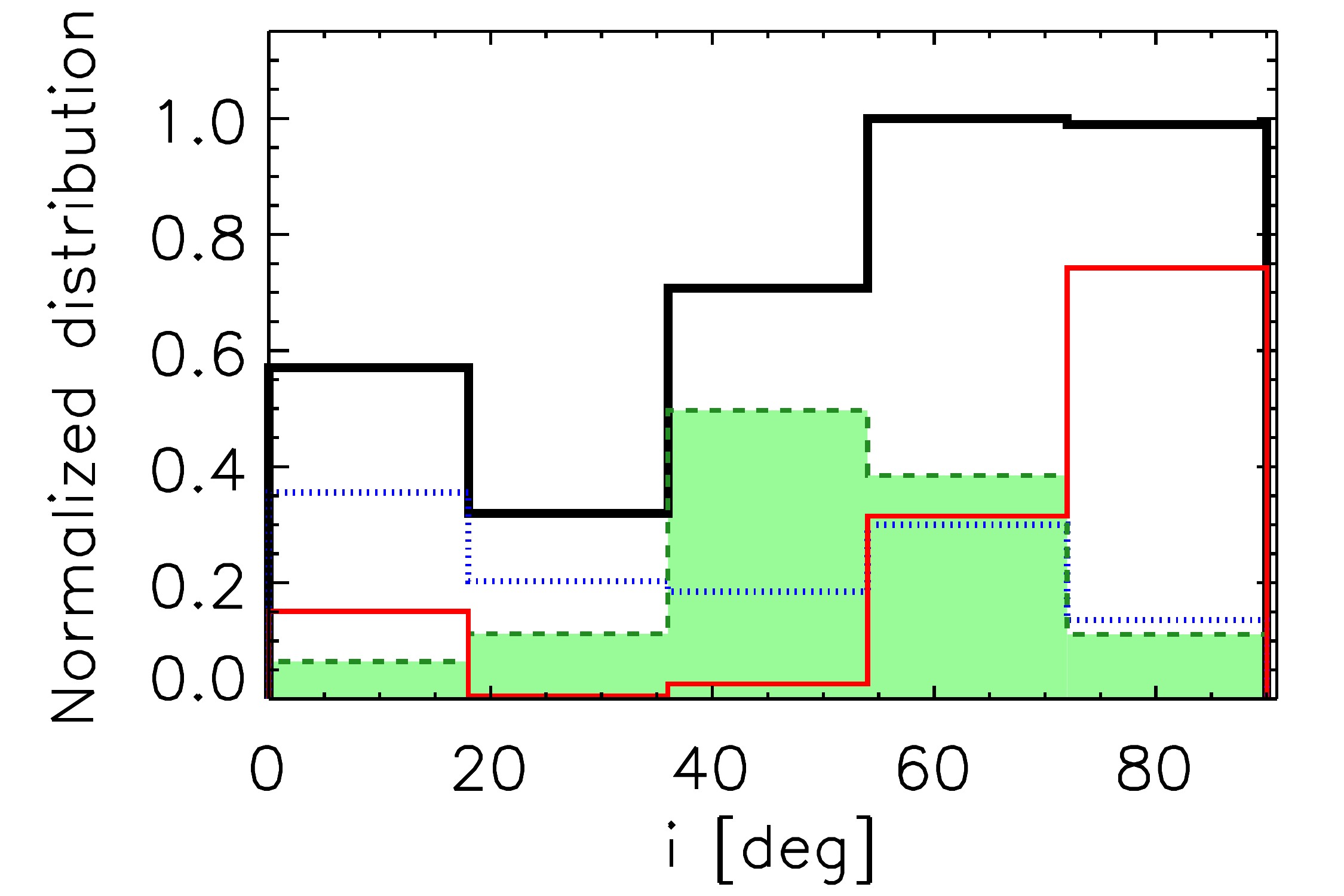}		\\
    \end{tabular}
\caption{Comparison between the torus/disc inclination combined probability distributions for different models considered here. Blue dotted, green dashed, red solid and black solid lines represent the parameter distributions of Sy1, Sy1.8/1.9, Sy2 and the entire sample, respectively. Note that 90-i$_{\rm F06}$=i$_{\rm N08}$=i$_{\rm H10}$=i$_{\rm S16}$=i$_{\rm H17}$.}
\label{i_distribution}
\end{figure*}

\begin{figure*}
\centering
    \renewcommand{\arraystretch}{0.01}
    \begin{tabular}{c c c}
    \textbf{{\hspace{0.9cm} smooth F06 torus models}} & \textbf{{\hspace{0.9cm} clumpy N08 torus models}} & \textbf{{ \hspace{0.9cm} clumpy H10 torus models}}\\
          \vspace{0.5cm}
      \includegraphics[width=5.8cm]{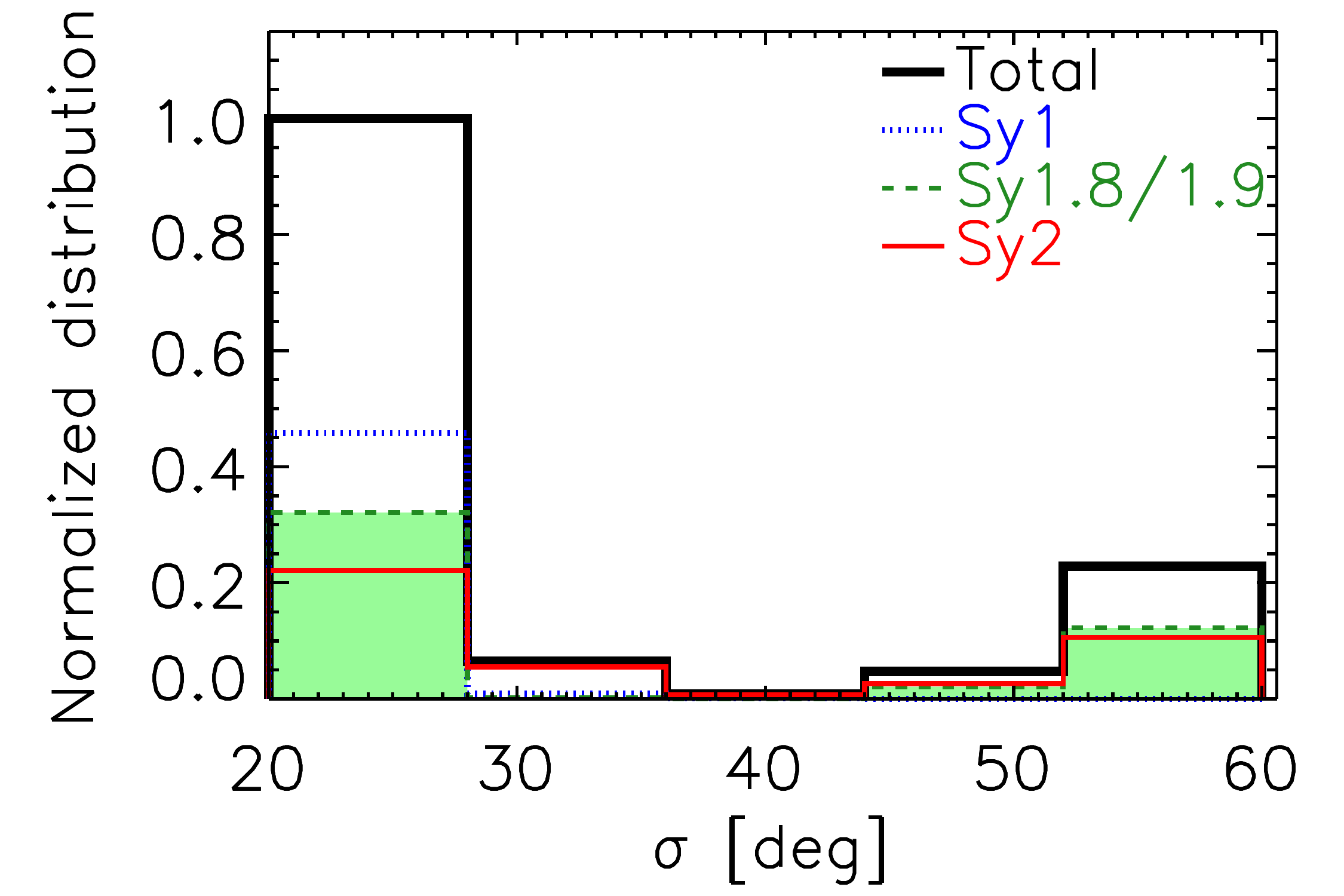}		 & \includegraphics[width=5.8cm]{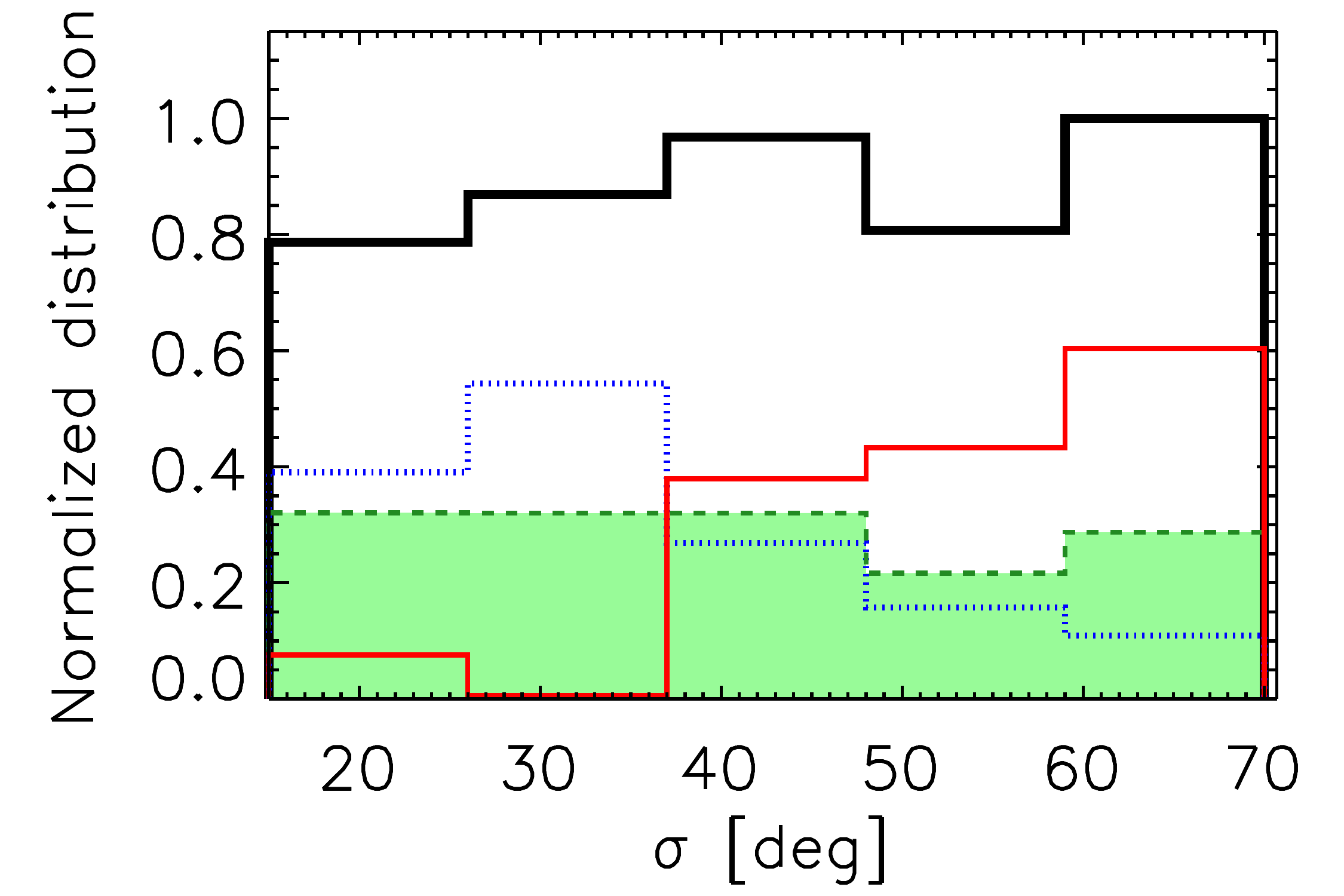}    &
      \includegraphics[width=5.8cm]{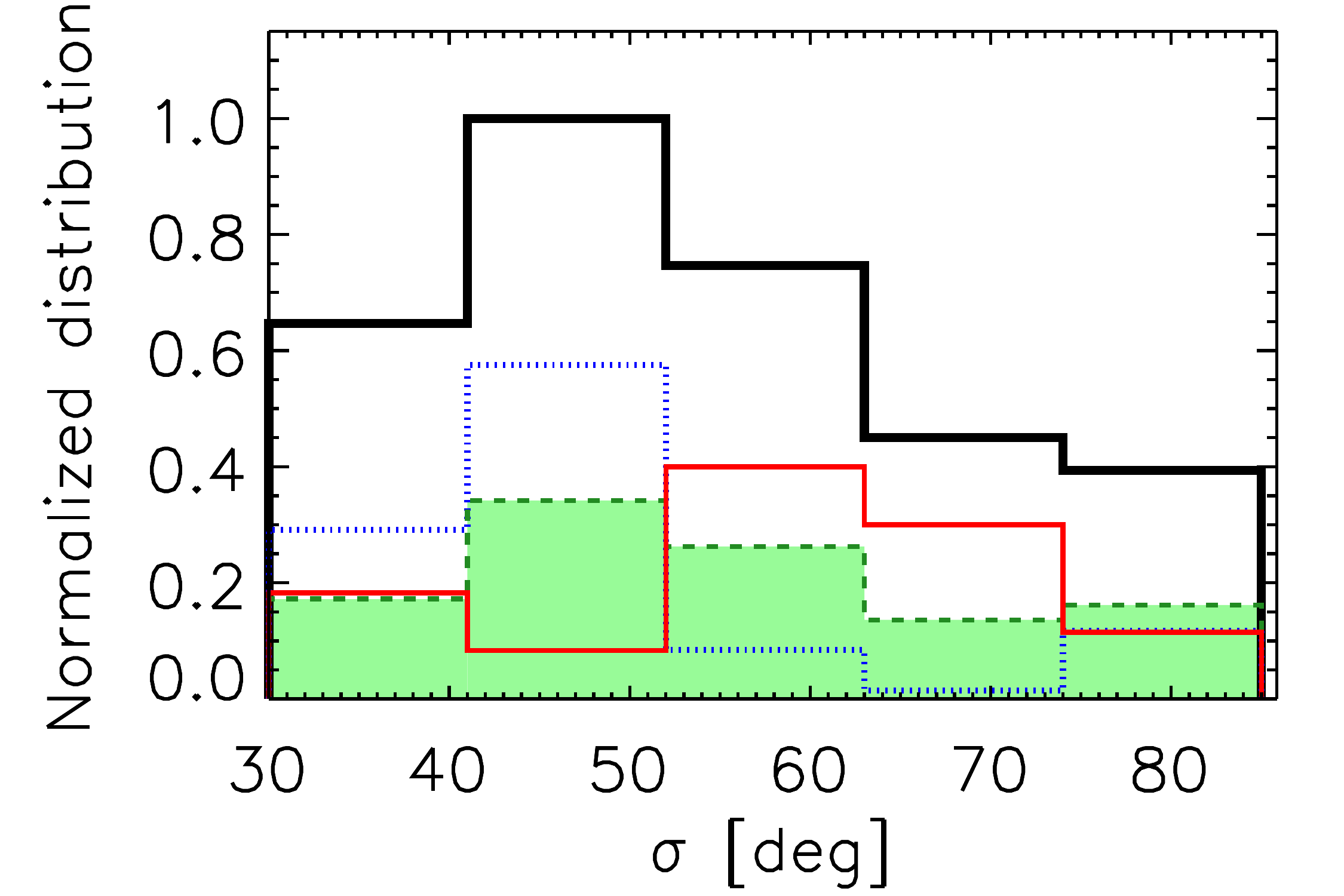}\\	
         \textbf{{\hspace{0.8cm} two-phase S16 torus models}} & \textbf{{\hspace{0.8cm} clumpy disc$+$wind H17 models}}  \\
      		 \includegraphics[width=5.8cm]{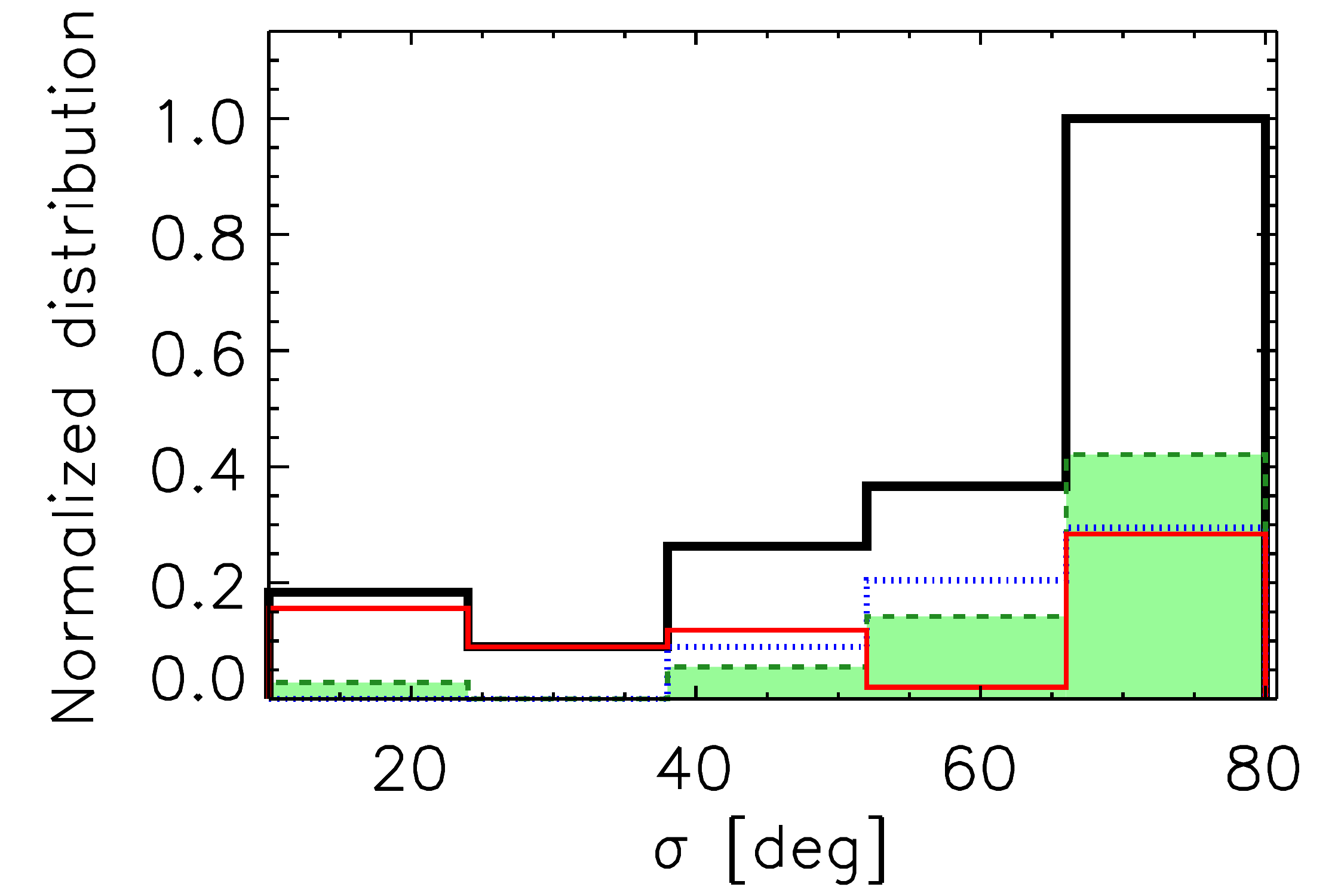} & \includegraphics[width=5.8cm]{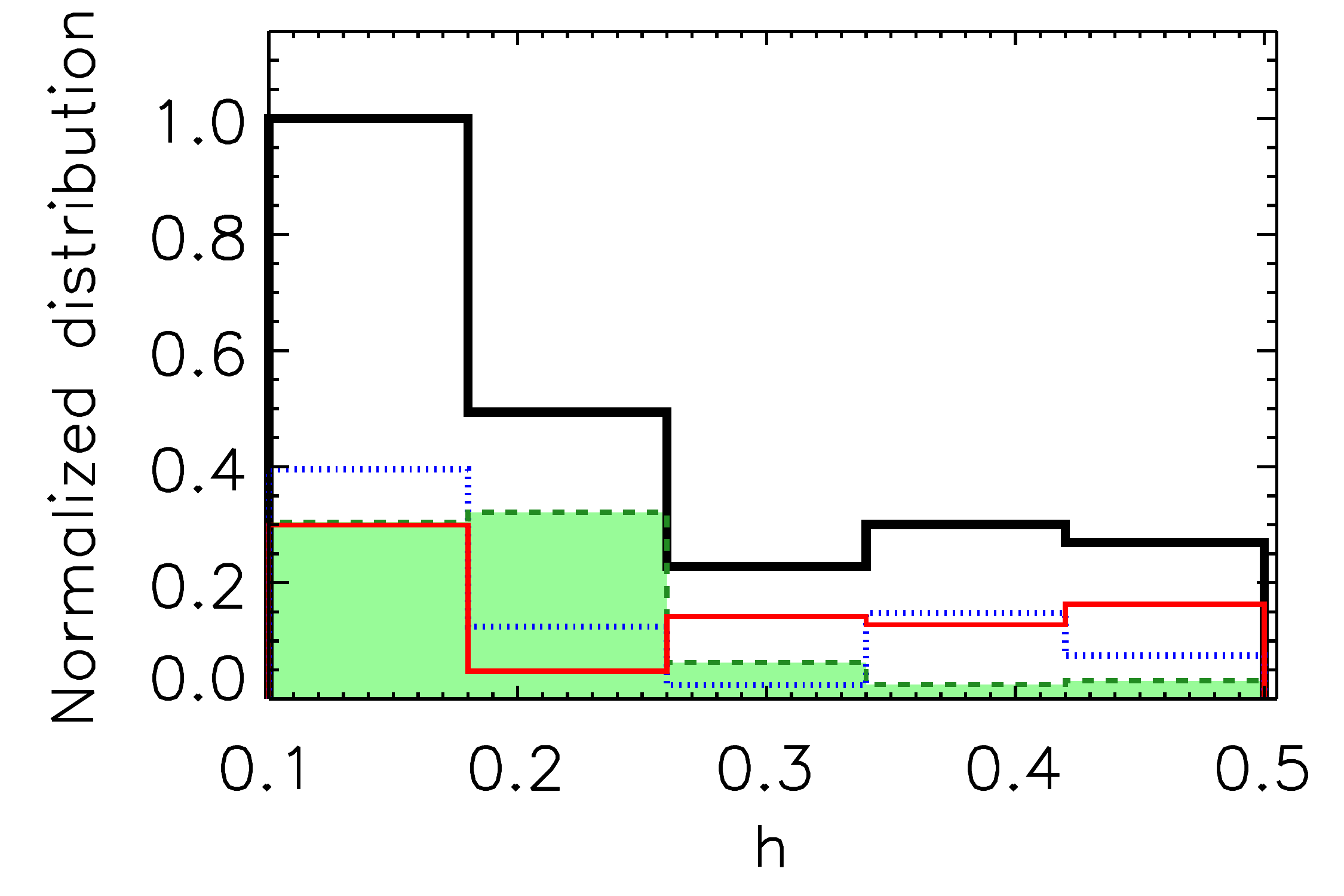}		\\
    \end{tabular}
\caption{Same as Fig. \ref{i_distribution} but for the torus/disc angular width. Note that $\sigma_{\rm H10}$=90-$\theta_{\rm H10}$. In the case of the clumpy disc$+$wind H17 models, the h parameter is the scale height of the dusty disc.}
\label{sigma_distribution}
\end{figure*}

\subsubsection{Quantitative methods}
\label{best_models_agn_prop}
In general, all models provide acceptable fits ($\rm{\chi^2_{red}<2}$) to the majority (19/24) of the nuclear IR SEDs (see Appendix\,\ref{nuclear_fits}). Using the AIC method (see Section\,\ref{sedfittingsect}), we find that the fraction of best fits provided by clumpy disc$+$wind and torus models is similar, 37.5 and 33.3\% (i.e. 9/24 and 8/24 sources), respectively. Furthermore, 2/24 galaxies (8.3\%) are equally fitted by clumpy disc$+$wind or torus models and 5/24 are not well fitted by any of the models used in this work (see Table\,\ref{tab:summaryfit}). According to the best fits, the IR SEDs of Sy1 (and Sy1.8/1.9) galaxies are best reproduced by clumpy disc$+$wind H17 models, whereas torus models are best suited to Sy2 galaxies (see Table\,\ref{statistics}). Using Fisher's exact test we find that these differences are statistically significant.

\begin{table}
\begin{center}
\caption{Summary of the Fisher's exact test results.}
\begin{tabular}{lcc}
\hline \hline
Test          & Samples &  p-value\\ 
(1)&(2)&(3)\\
\hline
{\bf{Disc$+$Wind best fits}}	& {\bf{Sy1 vs. Sy2}} & {\bf{\colort{$<$0.05}}}\\
{\bf{Disc$+$Wind best fits}}	& {\bf{Sy1/1.8/1.9 vs. Sy2}} & {\bf{\colort{$<$0.05}}}\\
Disc$+$Wind acceptable fits	& Sy1 vs. Sy2 & 0.37\\
Disc$+$Wind acceptable fits	& Sy1/1.8/1.9 vs. Sy2 & 0.31\\
{\bf{Torus best fits}}	& {\bf{Sy1 vs. Sy2}} & {\bf{\colort{$<$0.05}}}\\
{\bf{Torus best fits}}	& {\bf{Sy1/1.8/1.9 vs. Sy2}} & {\bf{\colort{$<$0.05}}}\\
Torus acceptable fits	& Sy1 vs. Sy2 & 0.34\\
Torus acceptable fits	& Sy1/1.8/1.9 vs. Sy2 & 0.12\\
Unfitted sources	& Sy1 vs. Sy2 & 0.59\\
Unfitted sources	& Sy1/1.8/1.9 vs. Sy2 & 1.00\\
\hline
\end{tabular}
\tablefoot{In bold we indicate distributions that can be considered statistically different (i.e. p-value<0.05).} 
\label{statistics}
\end{center}
\end{table}

The difference in the results for Sy1 and Sy2 galaxies confirms the trend first reported by GM19B using lower spatial resolution \emph{Spitzer}/IRS MIR spectra of a sample of AGN. However, in this work we confirm them using, for the first time, an ultra-hard X-ray selected sample of Seyferts and high-spatial resolution NIR-to-MIR data that allow us to better isolate the nuclear emission. Moreover, we do not find a clear trend between the models producing the best fits and AGN luminosity or Eddington ratio (see left and central panels of Fig.\,\ref{fit_histogram}). However, the right panel of Fig.\,\ref{fit_histogram} shows that it depends on the line-of-sight hydrogen column density.

In particular, clumpy disc$+$wind H17 models better reproduce the IR emission of AGN with relatively low hydrogen column densities (median value of log (N$_{\rm H}^{\rm X-ray}$\,cm$^{-2}$)=21.0$\pm$1.0; i.e. Sy1 and Sy1.8/1.9 galaxies) than torus models. On the other hand, torus models better reproduce the SEDs of AGN with high X-ray hydrogen column densities (median value of log (N$_{\rm H}^{\rm X-ray}$\,cm$^{-2}$)=23.5$\pm$0.8; i.e. Sy2s). This is in good agreement with theoretical predictions reported by \citet{Venanzi20}, where the authors found that for nuclear column densities of log(N$_{\rm H}$\,cm$^{-2}$)$<$23 the IR
radiation pressure becomes effective and polar outflows start to emerge (see also AH21). 

\subsection{Torus model parameters}
\label{torus_parameter}
In this section we investigate the main differences between the derived torus model parameters for the BCS\,40 sample using, for the first time, high angular resolution data and various models (see Section \ref{sec:models}). Note that we find similar fits using clumpy disc H17D and clumpy H10 torus models. Therefore, in the following, we will not discuss the individual parameters of clumpy disc H17D models.

A general trend is that, even for acceptable fits ($\rm{\chi^2_{red}<2}$), the model parameters are not well constrained (see Tables \ref{appendixfitSy1}, \ref{appendixfitSyInt}, and \ref{appendixfitSy2}). This result is independent of the torus model, Seyfert type, X-ray absorption along the line of sight or AGN luminosity. Nevertheless, rather than looking at the individual fits (see Appendix\,\ref{nuclear_fits}), we focus on the global statistics of the torus model parameters. For this purpose, we derived the combined probability distributions by concatenating together the individual arrays of the parameter probability distributions for all objects in each subgroup (see e.g. GB19). To quantify the differences between the combined probability distributions, we use the Kullback-Leibler divergence (KLD; \citealt{Kullback51}). This approach takes into account the overall shape of the combined distribution which always has a positive value. The larger the value the greater the difference of the distribution. This value is equal to zero for the case of two identical distributions. RA11 suggested that for values larger than 1 (boldface in Table \ref{fritz_tab_kdl}, \ref{nenkova_tab_kdl}, \ref{hoenig10_tab_kdl}, \ref{hoenig17_tab_kdl}, \ref{stalev_tab_kdl} in Appendix \ref{Combined_distribution}), two combined distributions may be considered to be significantly different.

Only a few model parameters can be directly compared between the various torus models, for example, the torus/disc inclination angle and its width. According to the KLD test, the differences in the torus/disc inclination angle between Sy subgroups are significant for the smooth F06 torus models, the clumpy H10 torus models and the clumpy disc$+$wind H17 models (see Fig. \ref{i_distribution}). In general, more edge-on values of the torus/disc inclination are needed for Sy2s than Sy1s. In particular, clumpy disc$+$wind H17 model results show the following trend for the disc inclination: i$_{\rm Sy1}$ $<$i$_{\rm Sy1.8/1.9}$ $<$i$_{\rm Sy2}$. The differences in the angular width of the torus/disc between Sy subgroups are also significant for the various models (see Fig.\,\ref{sigma_distribution} and Appendix \ref{Combined_distribution}). In general, the angular widths of the torus of Sy2 galaxies are larger than those of Sy1s (see Fig. \ref{sigma_distribution}). The only exception is found for the two-phase S16 torus models, which require a larger angular width of the torus for Sy1/1.8/1.9 than for Sy2 galaxies. Besides, for the clumpy disc$+$wind H17 models, there are no statistically significant differences between the angular width of Sy1 and Sy2 discs. This is likely related with the fact that clumpy disc$+$wind H17 models have relatively ``thin'' discs and, thus, it would be difficult to find differences between Sy1 and Sy2 discs. Summarizing, our results indicate that generally Sy1 galaxies have tori with smaller angular width and more face-on values of the torus inclination than those of type 2 Seyferts.

\begin{figure*}
\centering
\par{
\includegraphics[width=8.8cm]{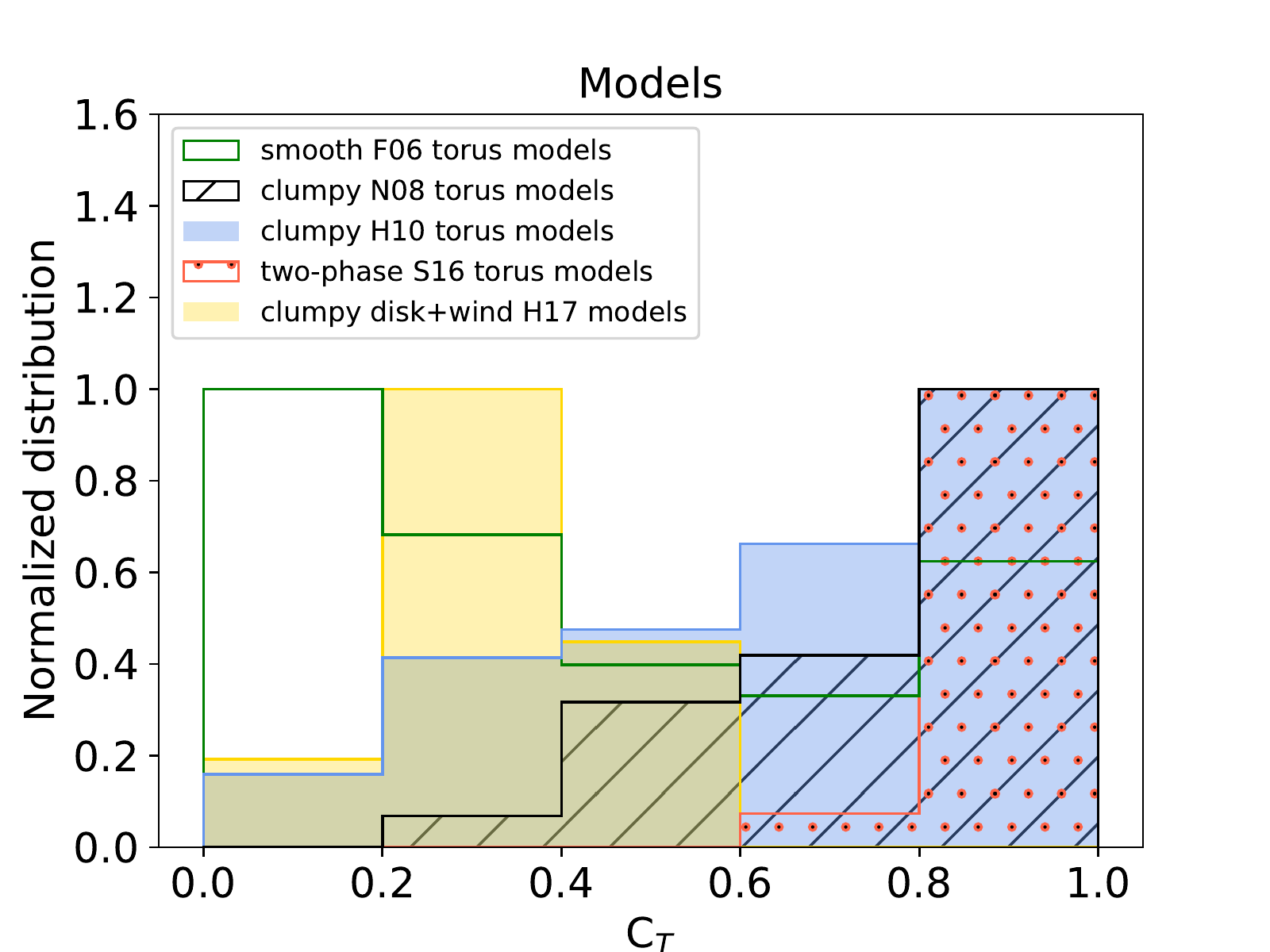}
\includegraphics[width=8.8cm]{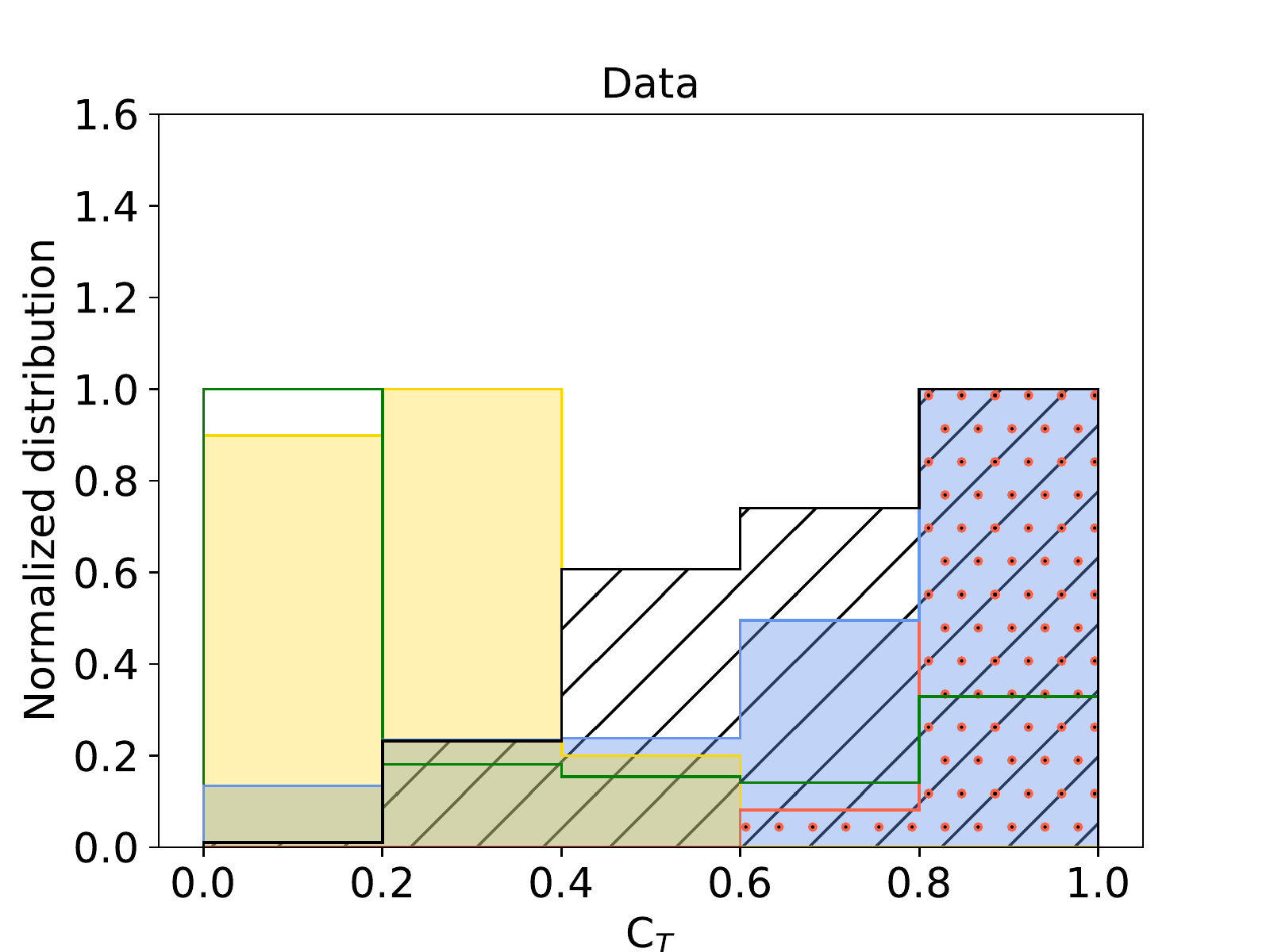}
\par}
\caption{Comparison between the covering factor parameter space. Left panel: combined probability distributions for all the models used in this work. Right panel: combined probability distributions derived for the entire Sy sample using each model.}
\label{covering_factor_parameter_space}
\end{figure*}
\label{covering_factor}

\begin{figure*}
\centering
\renewcommand{\arraystretch}{2.0}
    \begin{tabular}{c c c}
    \textbf{{\hspace{0.9cm} smooth F06 torus models}} & \textbf{{\hspace{0.9cm} clumpy N08 torus models}} & \textbf{{ \hspace{0.9cm} clumpy H10 torus models}}\\
      \includegraphics[width=5.8cm]{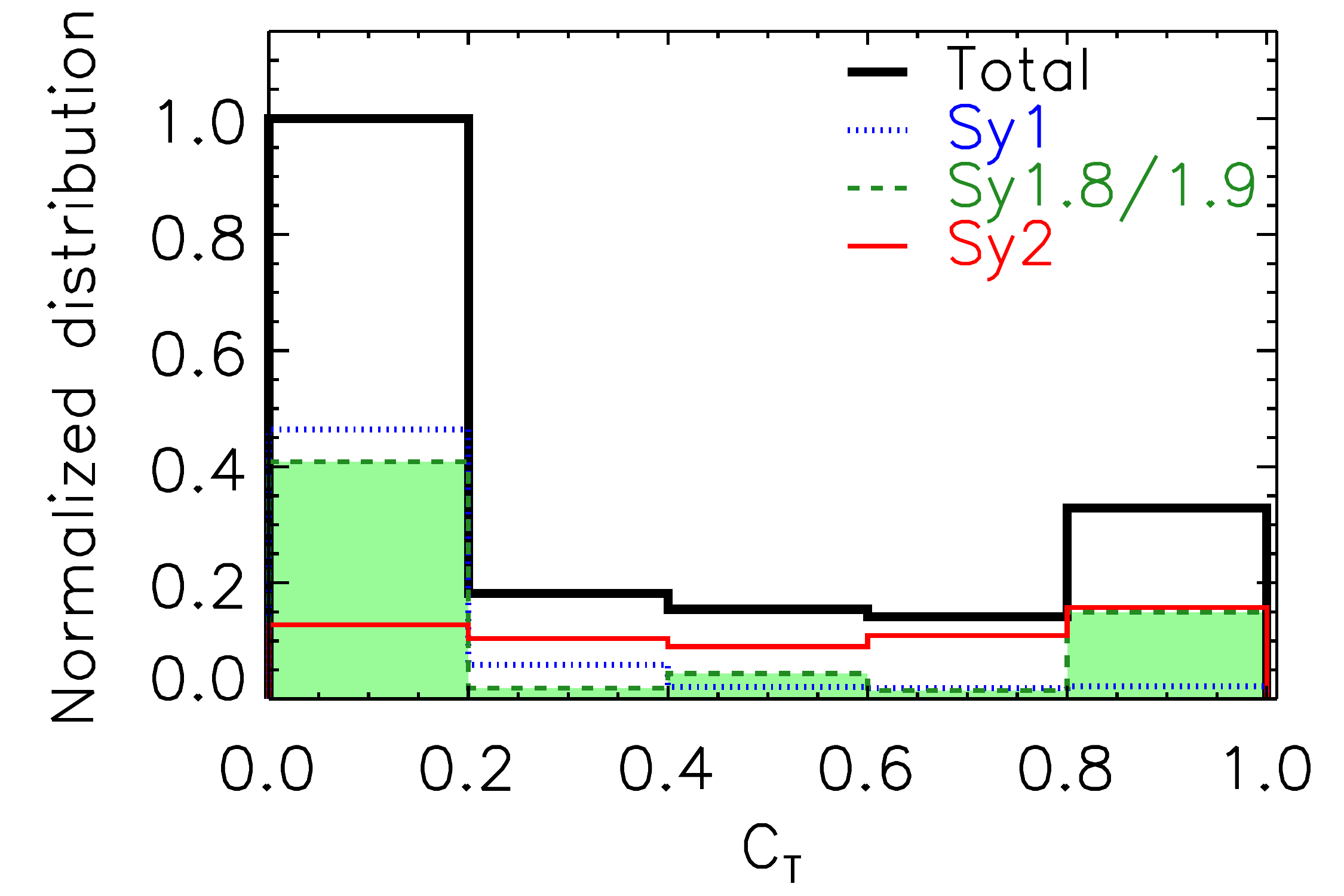}		 & \includegraphics[width=5.8cm]{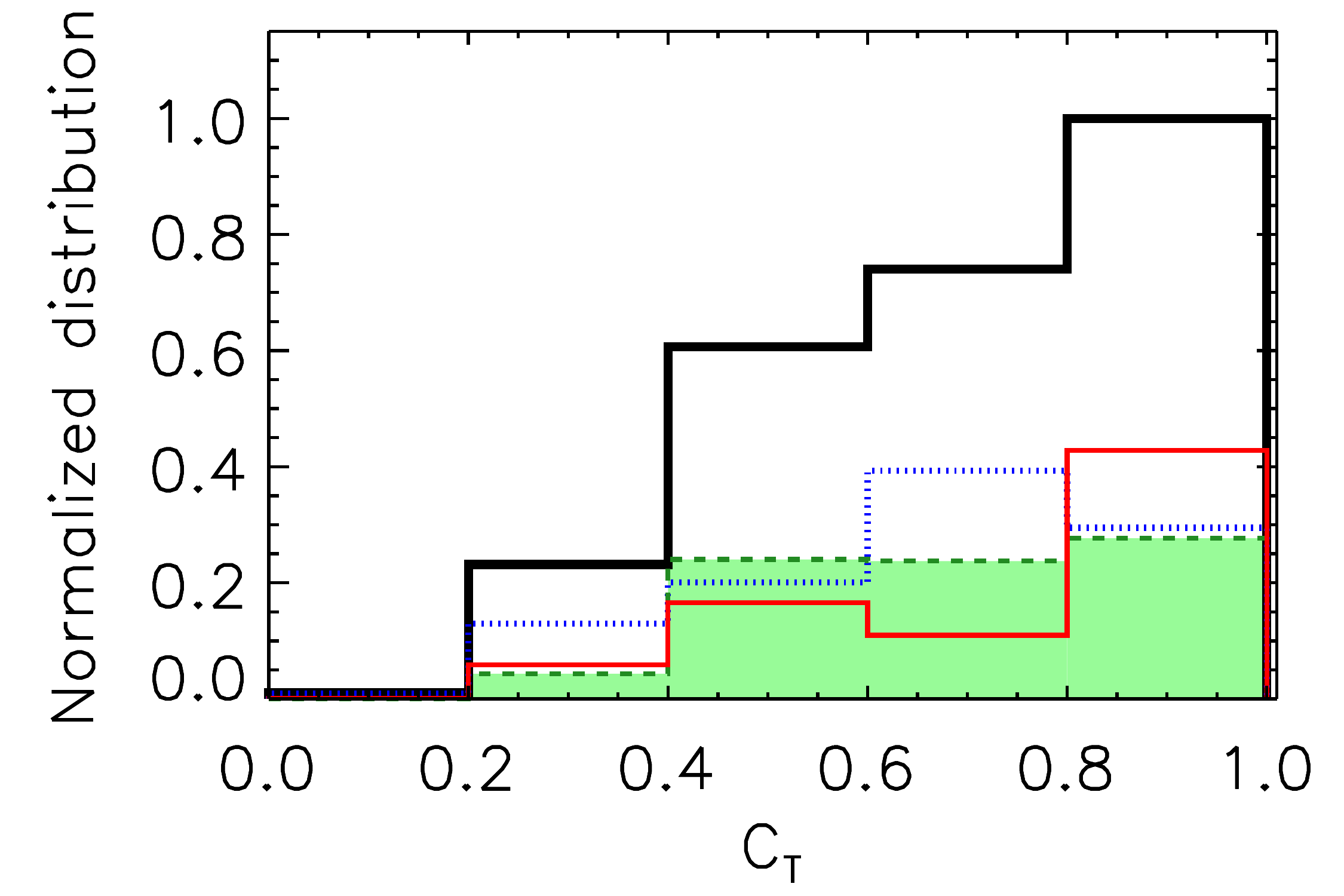} &
    \includegraphics[width=5.8cm]{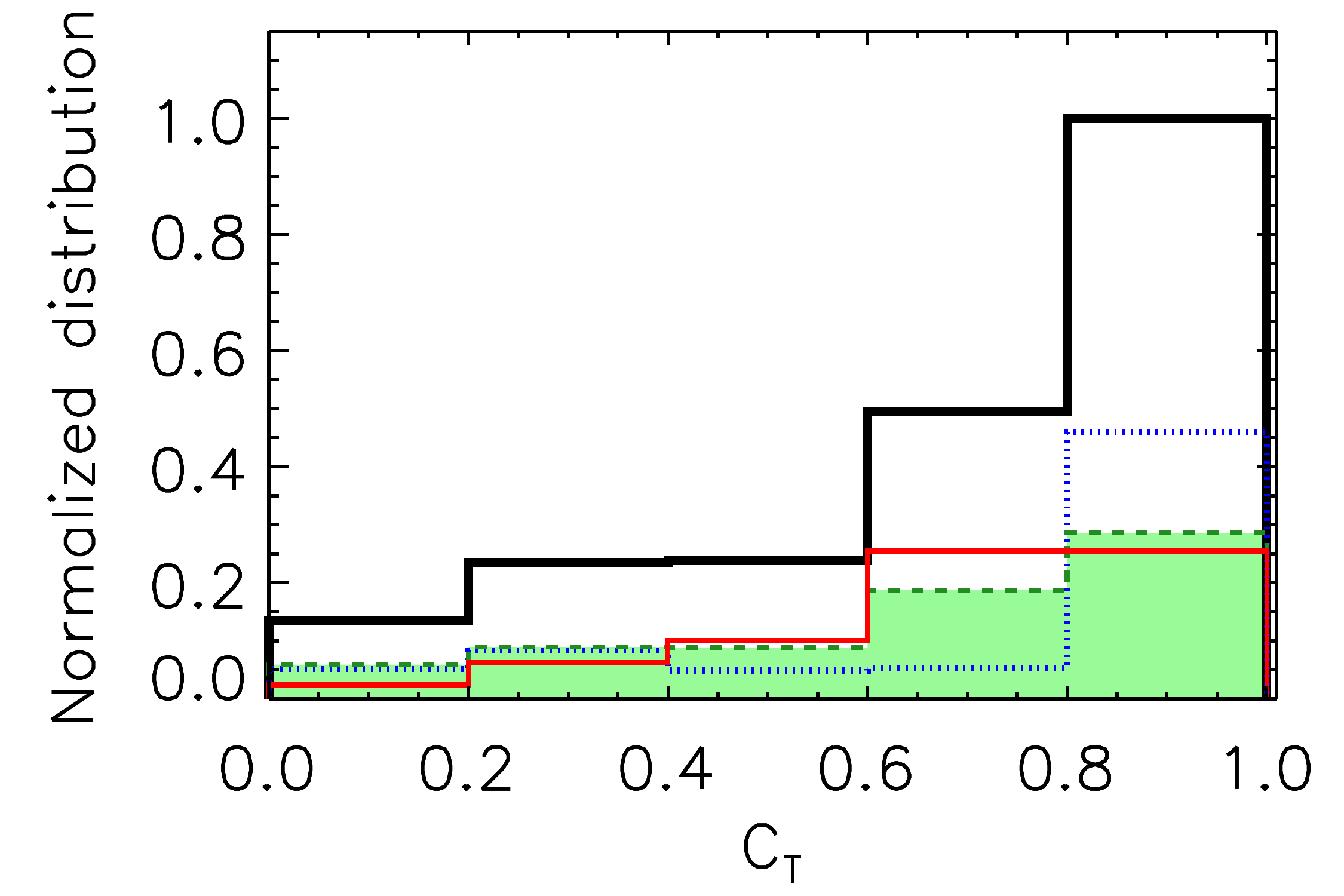}\\	
         \textbf{{\hspace{0.8cm} two-phase S16 torus models}} & \textbf{{\hspace{0.8cm} clumpy disc$+$wind H17 models}}  \\
		\includegraphics[width=5.8cm]{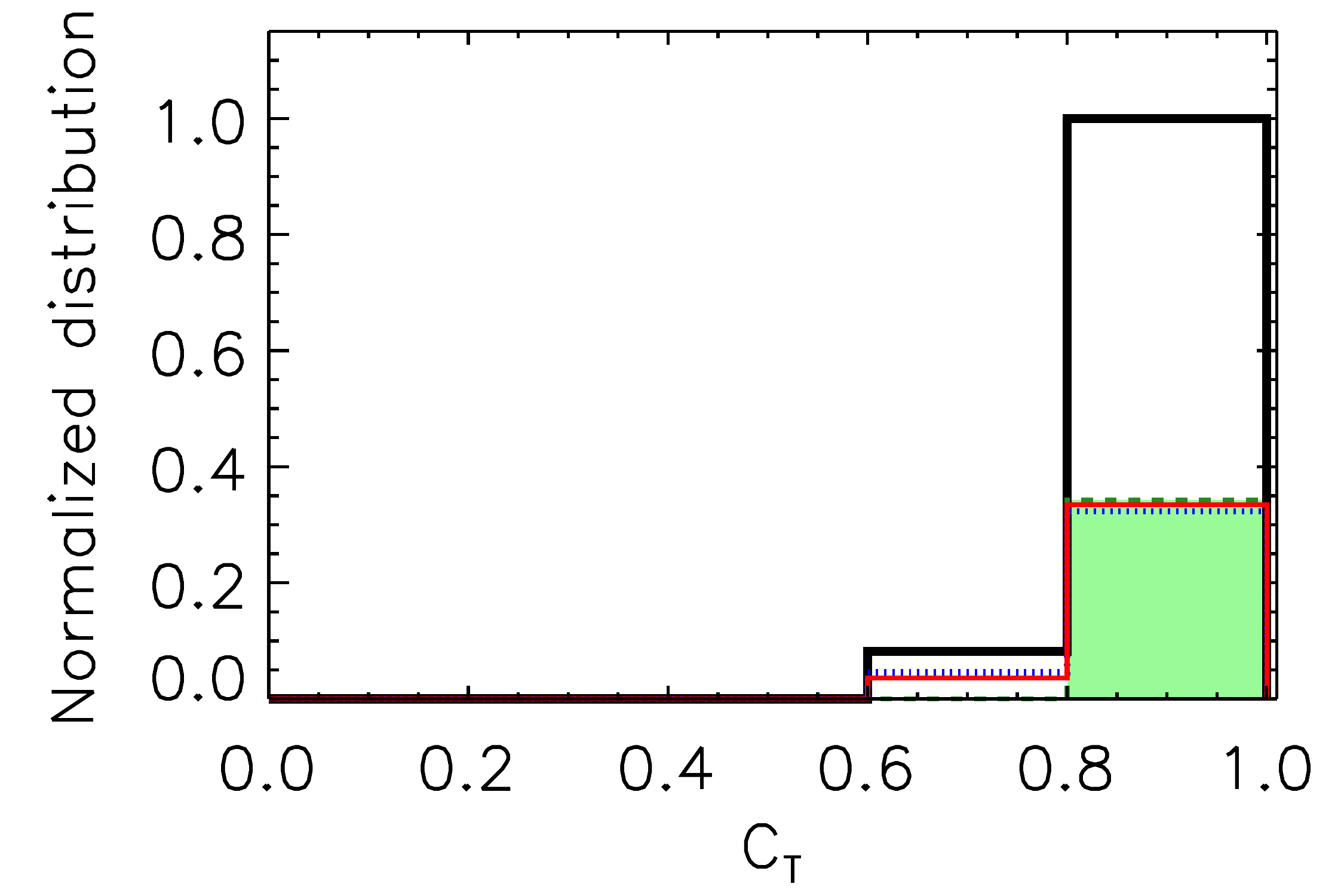} &
      \includegraphics[width=5.8cm]{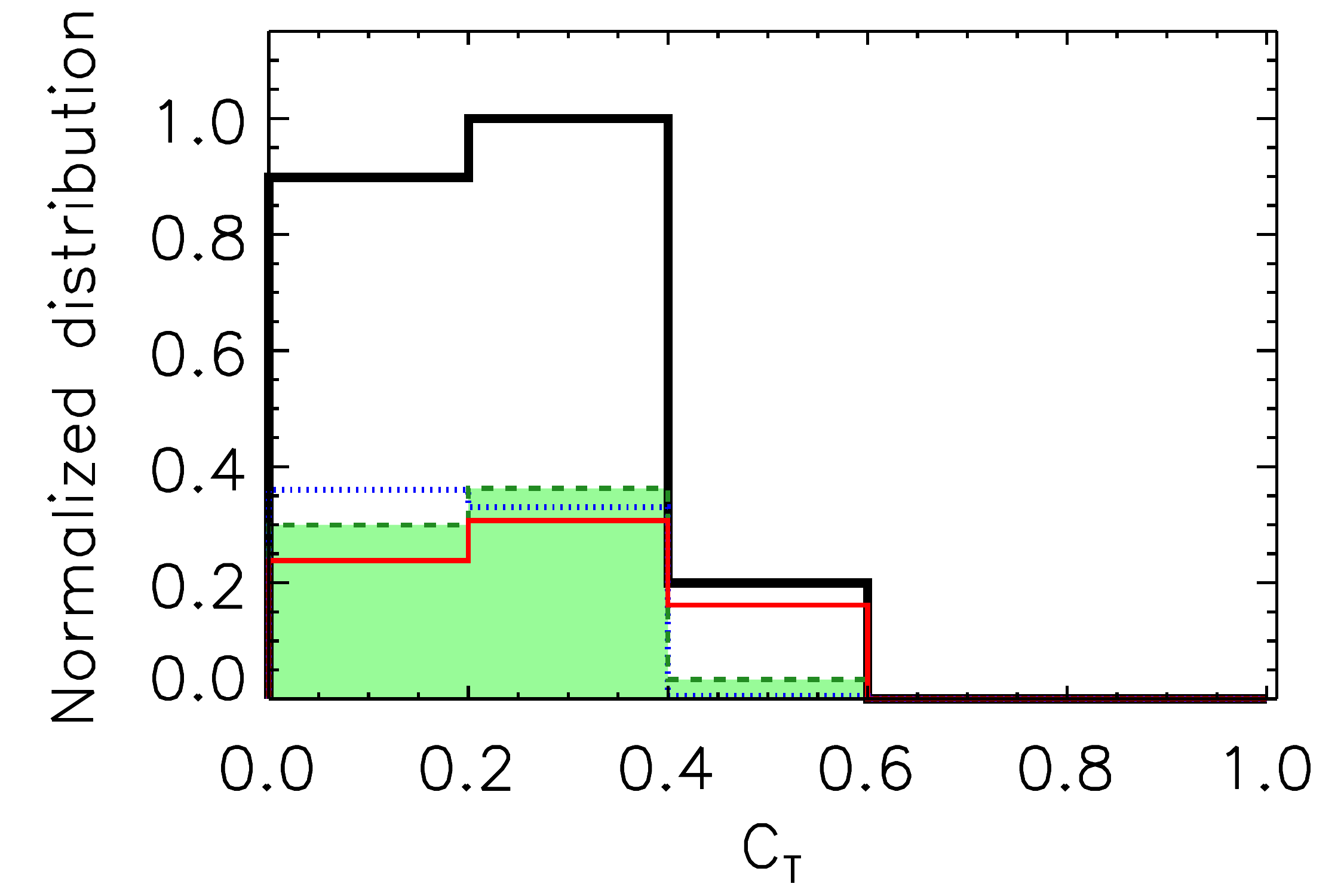}	\\
    \end{tabular}
\caption{Same as Fig. \ref{i_distribution} but for the covering factor.}
\label{covering_factor_distributions}
\end{figure*}

\subsection{Derived torus covering factor, size and mass}
\label{AGN_proper} 

\subsubsection{The Covering Factor of the nuclear obscuring material}
\label{ct_sec} 

The nuclear obscuration is strongly dependent on the covering factor (C$_{\rm T}$) which is defined as the fraction of sky covered by the obscuring material. The covering factor is one of the main elements regulating the intensity of the reprocessed AGN radiation (e.g. RA11, \citealt{Ramos17}). 

For the various models, C$_{\rm T}$ can be calculated as:

\begin{equation}
  \begin{aligned}
C_{\rm T} = 1-\int_{0}^{\pi/2} e^{-\tau_{\nu}\rm(\alpha)}~\cos\alpha \,d\alpha
  \end{aligned}
\end{equation}

\noindent where $\tau_{\nu}\rm(\alpha)$ is the line-of-sight optical depth, which depends on the azimuthal angle ($\alpha$). The line-of-sight optical depth is computed from the distribution of clouds for clumpy torus models and from the equatorial opacity, and from the density distribution for smooth torus models (see GM19B and references therein).

Since the covering factor is defined as the fraction of the sky at the AGN centre covered by obscuring material, it strongly depends on the torus dust distribution and geometry assumed by the model (see Fig.\,\ref{covering_factor_parameter_space}; also GB19B). To further investigate the differences between the various models, we produced the C$_{\rm T}$ combined probability distributions of the models sampling the entire space of parameters of each model. For example, the clumpy$+$wind H17 model consists of a clumpy dusty disc plus a hollow dusty cone which will naturally produce lower values of the covering factor (C$_T<$0.6; see yellow distribution in left panel of Fig.\,\ref{covering_factor_parameter_space}) than a dusty torus with a large range of angular sizes. On the other hand, the two-phase S16 models provide large values of the covering factor (C$_T>$0.6; see red dot distribution in the left panel of Fig.\,\ref{covering_factor_parameter_space}). Therefore, caution must be applied when comparing covering factors between various torus models due to the different ranges of parameter space. 

In order to compare the C$_{\rm T}$ probability distributions of the models with those of the observations, we also derived the C$_{\rm T}$ combined probability distributions (for each model) by concatenating the individual C$_{\rm T}$ probability distributions of the various galaxies. Indeed, we find that the C$_{\rm T}$ combined probability distributions of the models and those derived for the entire sample are similar (Fig.\,\ref{covering_factor_parameter_space}). For instance, small values of the C$_{\rm T}$ are derived for the fitted data when using clumpy$+$wind H17 models whereas clumpy H10, clumpy N08 and two-phase S16 models require larger C$_{\rm T}$. However, the C$_{\rm T}$ distributions derived for clumpy$+$wind H17 models tend to have smaller values than the parent distribution. The same applies for the smooth F06 models, whereas the derived C$_{\rm T}$ distributions for clumpy N08 models favour intermediates C$_{\rm T}$ values compared with those of the models which peak at larger values. Considering the different C$_{\rm T}$ ranges covered by the various models, in the following, we have consistently used the same models when comparing covering factors of the Sy groups.

Fig.\,\ref{covering_factor_distributions} shows that generally Sy1 have smaller median values of the covering factor than Sy2. According to the KLD test, there are statistically significant differences in the covering factor of Sy1 and Sy2 for the smooth F06 torus models and the clumpy H17 disc$+$wind models (see Appendix\,\ref{Combined_distribution}). There is a similar trend, but with less significant (using the KLD test), for the clumpy N08 torus models (see e.g. RA11, AH11, \citealt{Ichikawa15} and GB19). Earlier works using clumpy N08 torus models also showed statistically significant differences between the covering factor of Sy1 and Sy2 galaxies (e.g. RA11, AH11, \citealt{Ichikawa15} and GB19). The lower significance found here might be due to the fact that we are not using priors for the angular width of the torus (based on [O\,III] data), unlike previous works.

Finally, we find that the covering factor remains broadly constant within the errors for the majority of the models throughout the luminosity range: log(L$_{\rm bol}$\,erg~s$^{-1}$)$\sim$41.8--45.9. The same applies when using Eddington ratios ($\lambda_{\rm Edd}$: -3.40 to -0.26) instead of the bolometric luminosity.

\subsubsection{Torus/disc size and mass}
\label{size}

Using the radial extent of the torus/disc (Y=R$_{\rm o}$/R$_{\rm d}$) and the dust sublimation radius (R$_{\rm d}$), we can derive the physical radius of the torus/disc (R$_{\rm o}$). The dust sublimation radius also depends on the dust sublimation temperature and the bolometric luminosity. Note that the clumpy H10 torus models and clumpy$+$disc H17 models fixed the Y parameter to a large value for all the SEDs (Y$=$150 and Y$=$500, respectively). 

\begin{figure*}
\centering
    \begin{tabular}{c c c}
    \textbf{{\hspace{0.9cm} smooth F06 torus models}} & \textbf{{\hspace{0.9cm} clumpy N08 torus models}} & \textbf{{\hspace{0.8cm} two-phase S16 torus models}}\\
      \includegraphics[width=5.8cm]{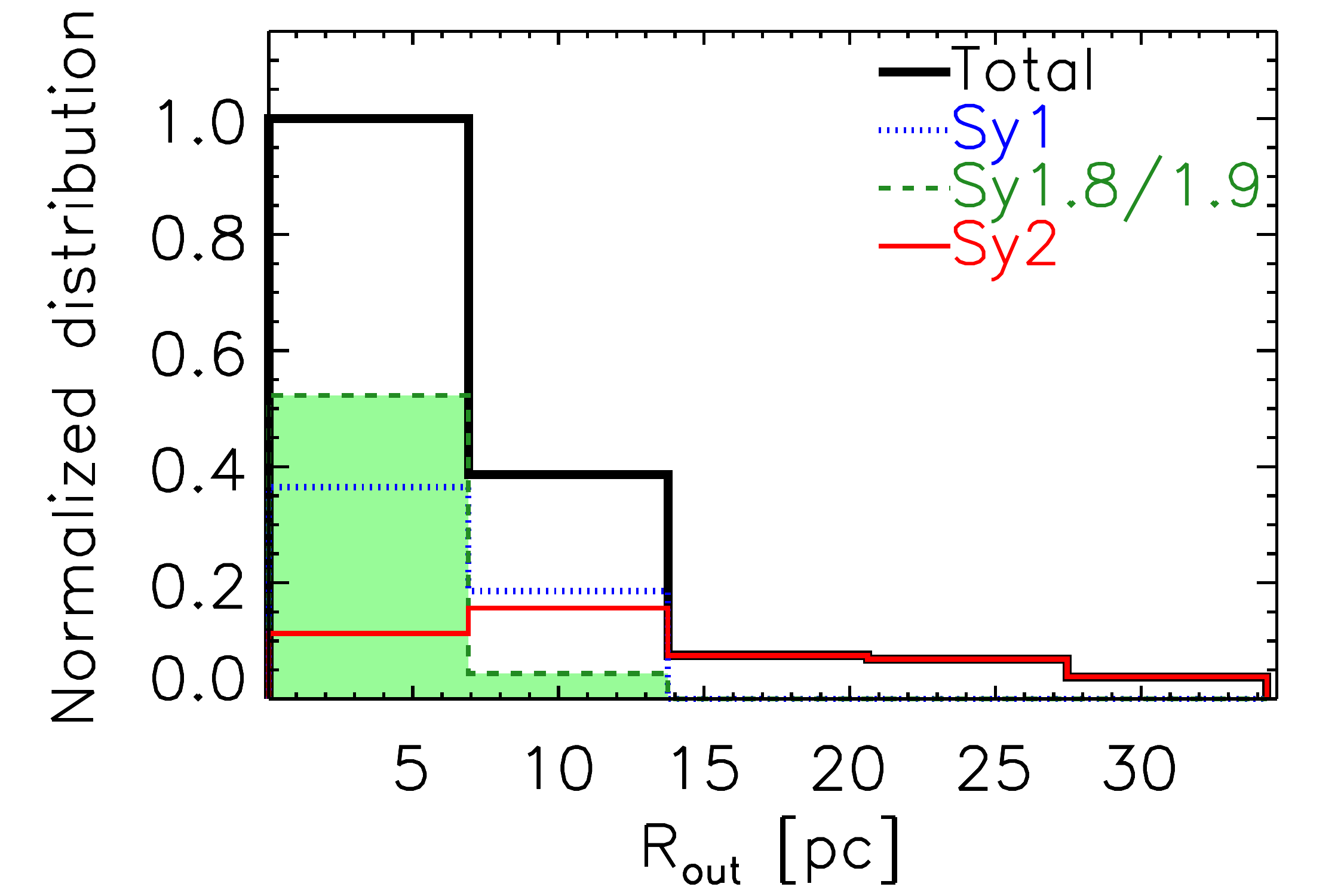}		 & \includegraphics[width=5.8cm]{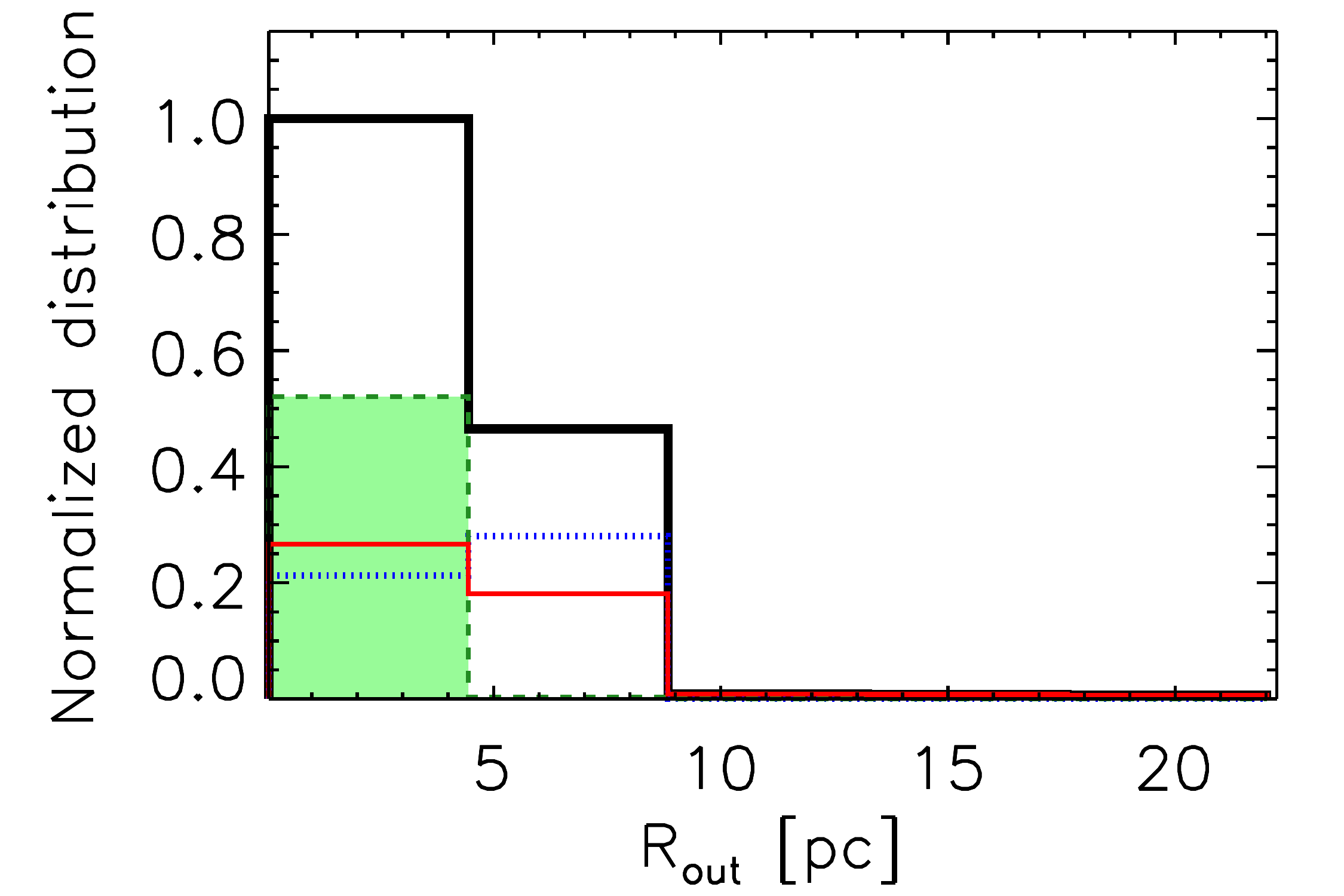} &
       \includegraphics[width=5.8cm]{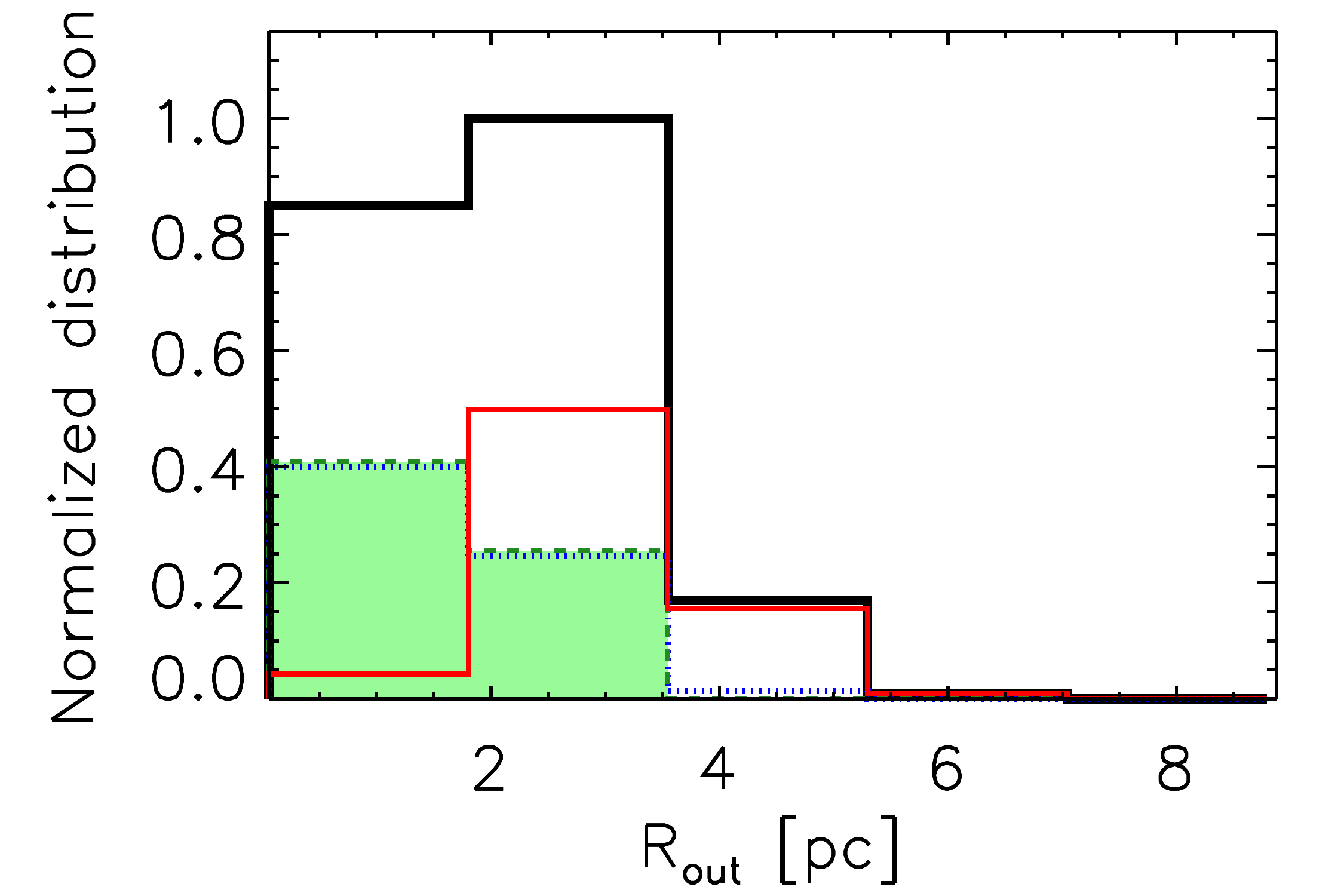} \\    
    \end{tabular}
\caption{Same as Fig. \ref{i_distribution} but for the torus size (R$_{\rm out}$).}
\label{rout_distributions}
\end{figure*}

The radius distributions of Sy2 for the smooth F06 torus models and the two-phase S16 models show a tail towards larger tori in comparison with those of Sy1 (see Fig.\,\ref{rout_distributions}). In particular, in the case of the smooth F06 torus models, we derive median values of the torus size for Sy2 galaxies larger ($\sim$3-5 times) than those of Sy1 and Sy1.8/1.9. Note that using the smooth F06 torus models the radius probability distribution for Sy2 galaxies reaches maximum values of $\sim$30\,pc. However, the clumpy N08 torus models do not show statistically significant differences between Sy1 and Sy2 radii. In general, we find relatively compact (1-15\,pc) torus radii for all the Seyfert galaxies in our sample (see Fig.\,\ref{rout_distributions}). Note that we use the term compact torus for those with sizes below the largest resolution element in the MIR for our sample (i.e. $<$50\,pc).    

\begin{figure*}
\centering
\renewcommand{\arraystretch}{2.0}
    \begin{tabular}{c c c}
    \textbf{{\hspace{0.9cm} smooth F06 torus models}} & \textbf{{\hspace{0.9cm} clumpy N08 torus models}} & \textbf{{ \hspace{0.9cm} clumpy H10 torus models}}\\
      \includegraphics[width=5.8cm]{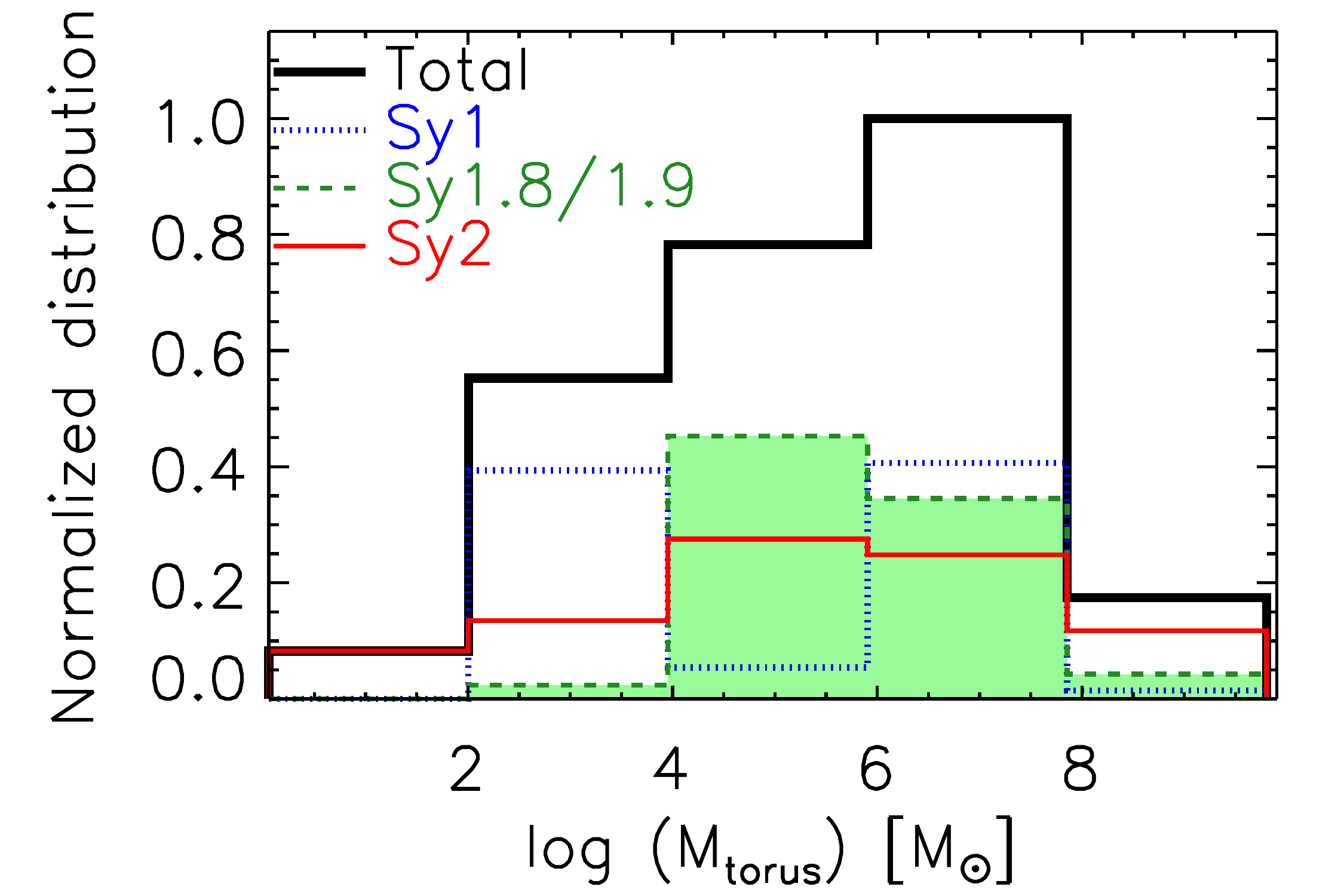}		 & \includegraphics[width=5.8cm]{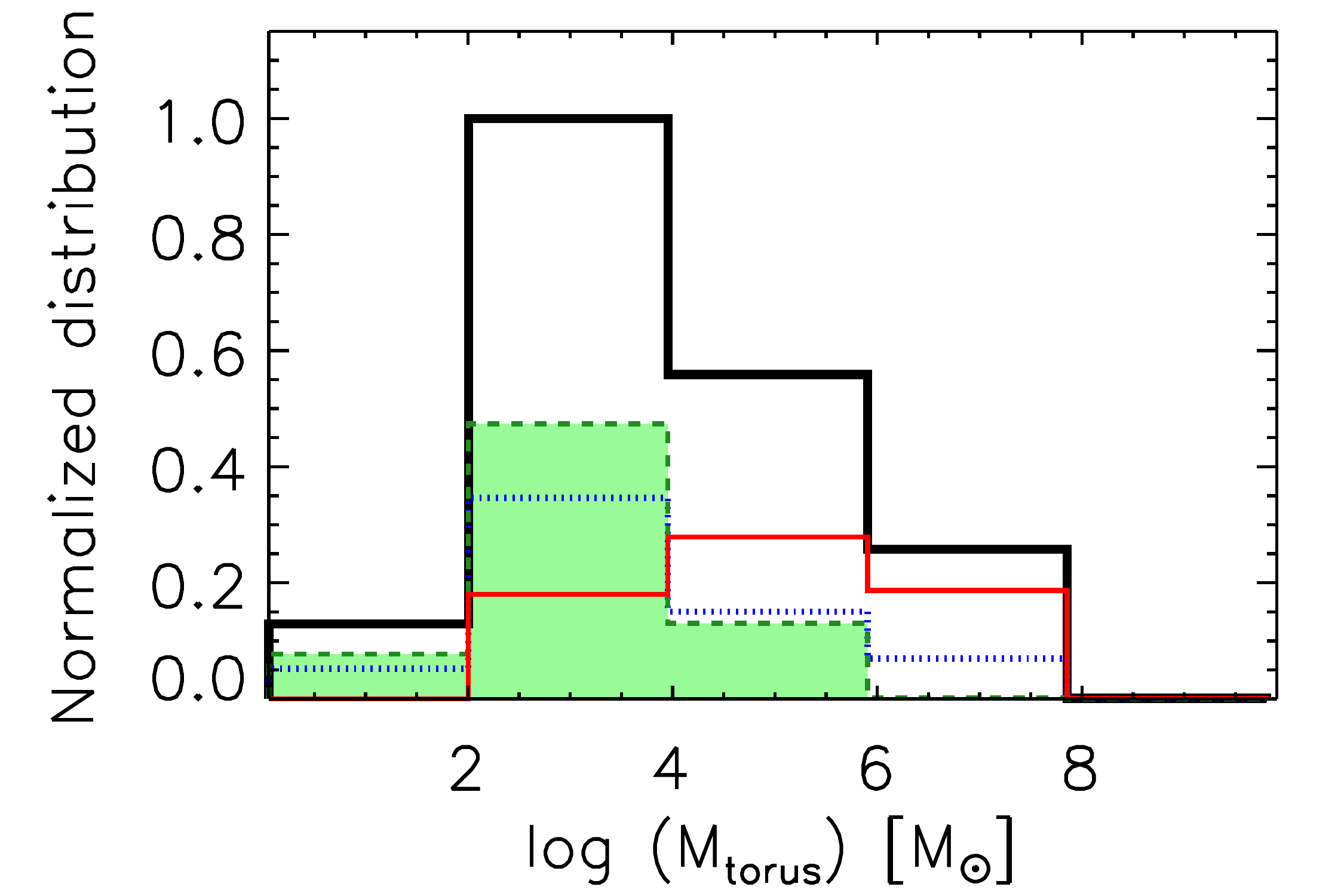} &
    \includegraphics[width=5.8cm]{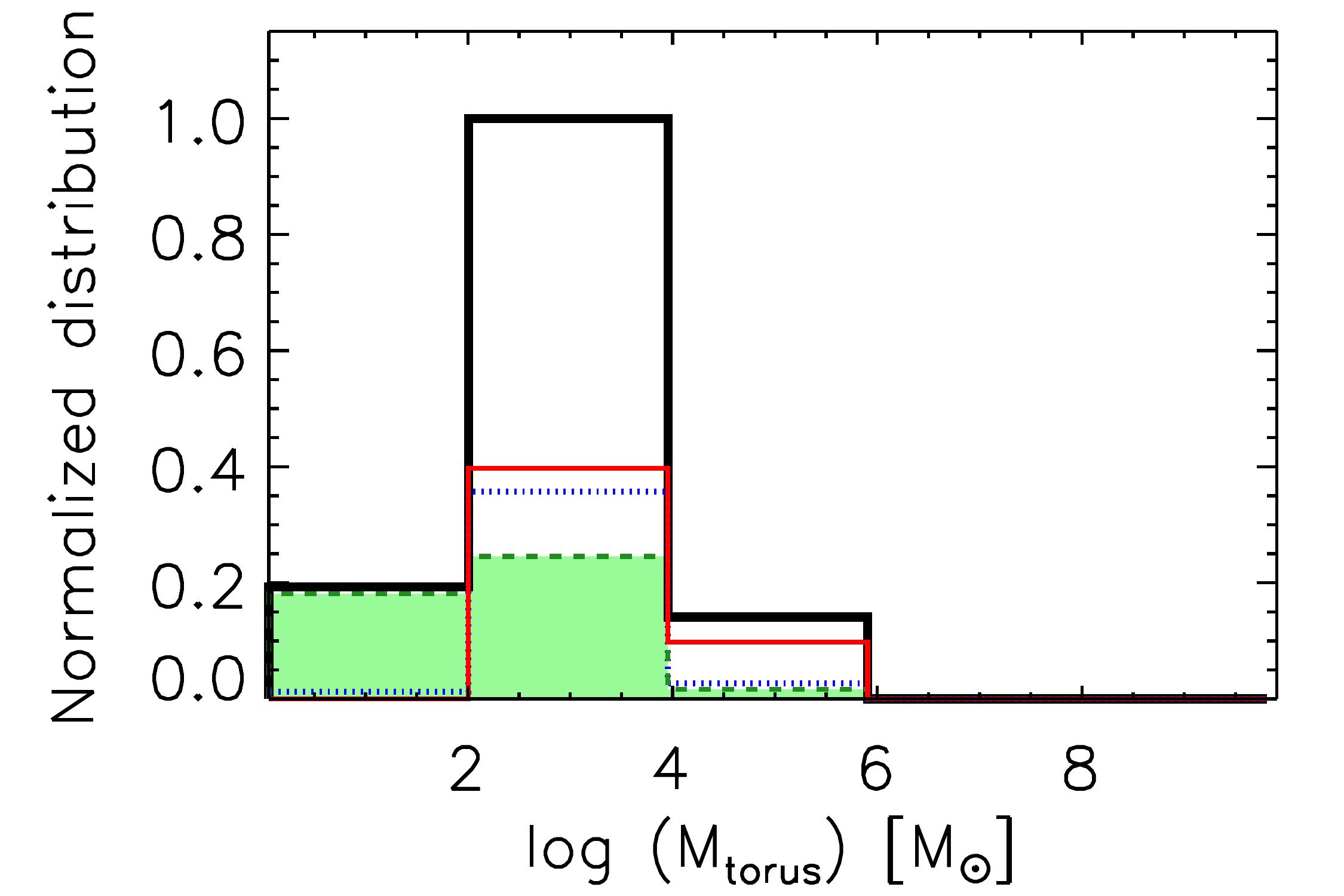}\\
         \textbf{{\hspace{0.8cm} two-phase S16 torus models}} & \textbf{{\hspace{0.8cm} clumpy disc$+$wind H17 models}}  \\
		  \includegraphics[width=5.8cm]{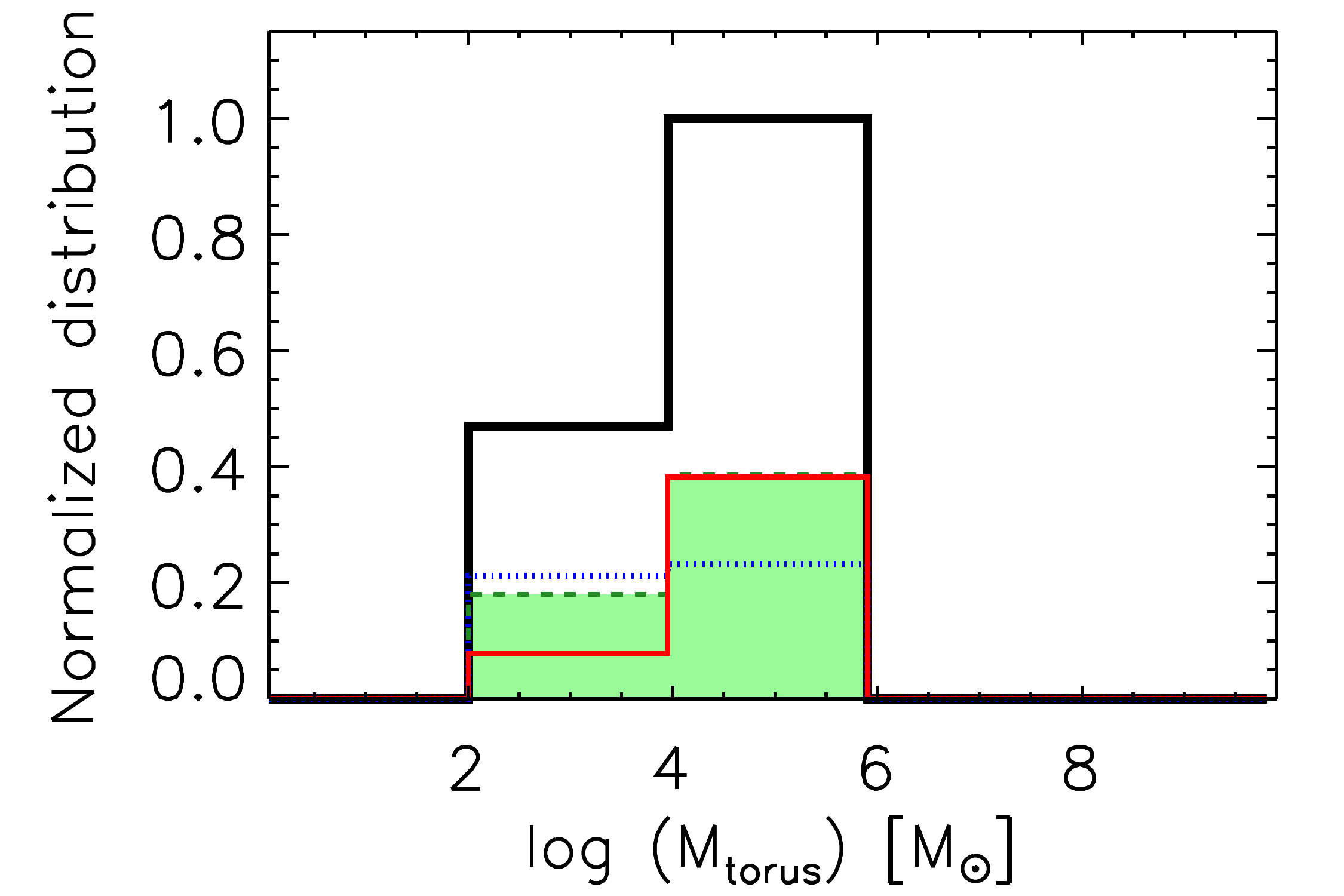}&
      \includegraphics[width=5.8cm]{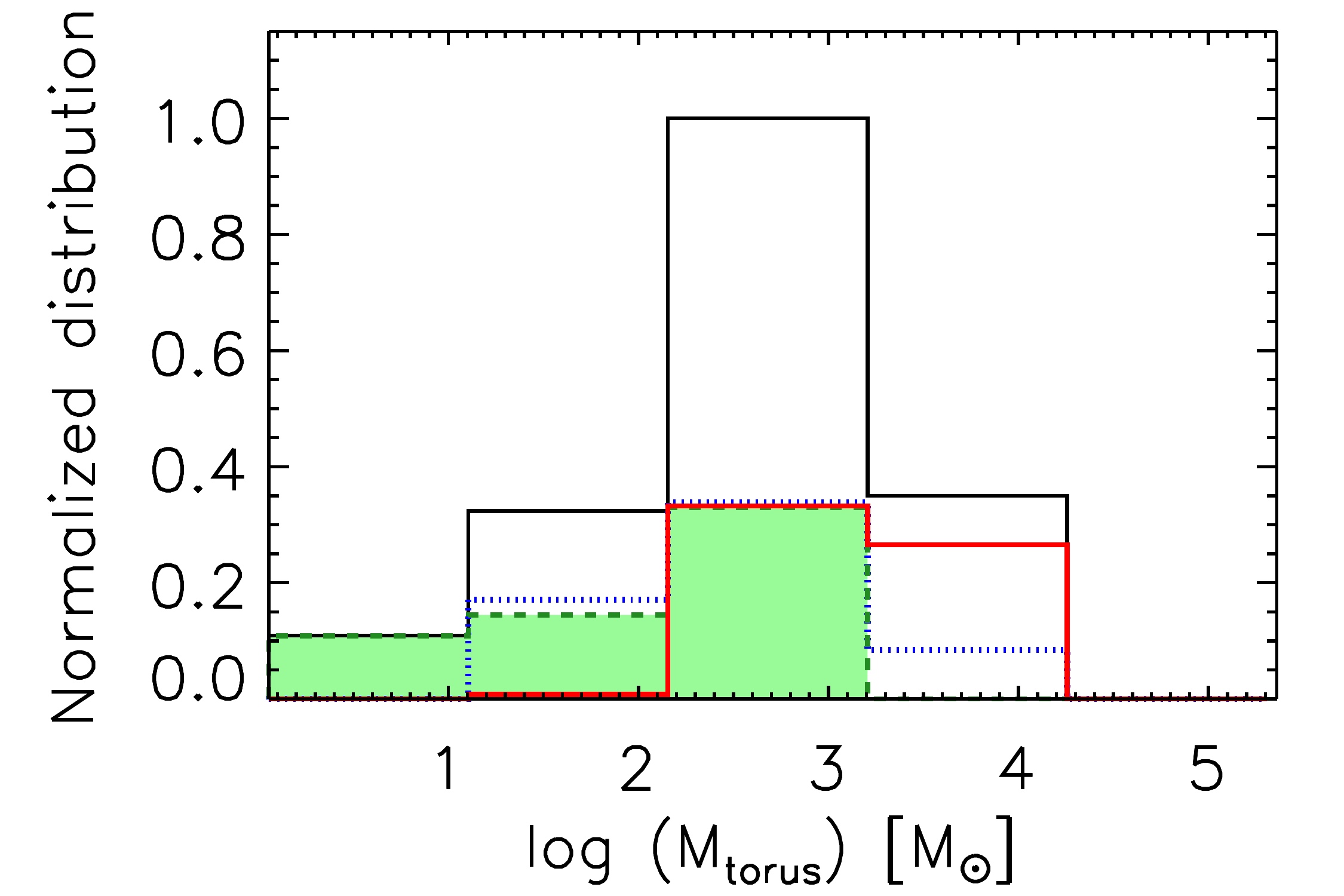}		\\
    \end{tabular}
\caption{Same as Fig. \ref{i_distribution} but for the torus gas mass.}
\label{dustmass_distributions}
\end{figure*}

Using the Galactic dust-to-gas ratio \citep{Bohlin78}, we can also estimate the torus gas mass associated with the fitted nuclear dusty structure by integrating the density distribution function for each model (see Appendix \ref{equations}, see also GM19B and references therein). We computed the torus mass within the fitted dusty structure volume, thus, it may not be representative of the whole torus gas mass distribution which is traced by the cold gas (see e.g. \citealt{Honig19} and Section \ref{derived_high_angular} for further discussion). We find slightly larger values of the torus/disc gas mass for Sy2 than for Sy1 galaxies, but their differences are generally within the errors (see Fig.\,\ref{dustmass_distributions}). The total gas masses of the tori are in the range log(M$_{\rm torus}$)$\sim$2-6\,M$_\sun$ and, the majority of the models used in this work provide similar values of the total gas mass within the errors (median values of log(M$_{\rm torus}$)$\sim$4\,M$_\sun$). The exception are the smooth F06 torus (log(M$_{\rm torus}^{\rm F06}$)=5.6$\pm$1.9\,M$_\sun$) and disc$+$wind H17 models (log(M$_{\rm torus}^{\rm H17}$)=2.6$\pm$0.8\,M$_\sun$), for which we find larger gas masses and smaller masses respectively than for the other torus models.

\begin{figure*}
\centering
\begin{tabular}{c c}
\includegraphics[width=8.4cm, clip, trim=0 0 10 0]{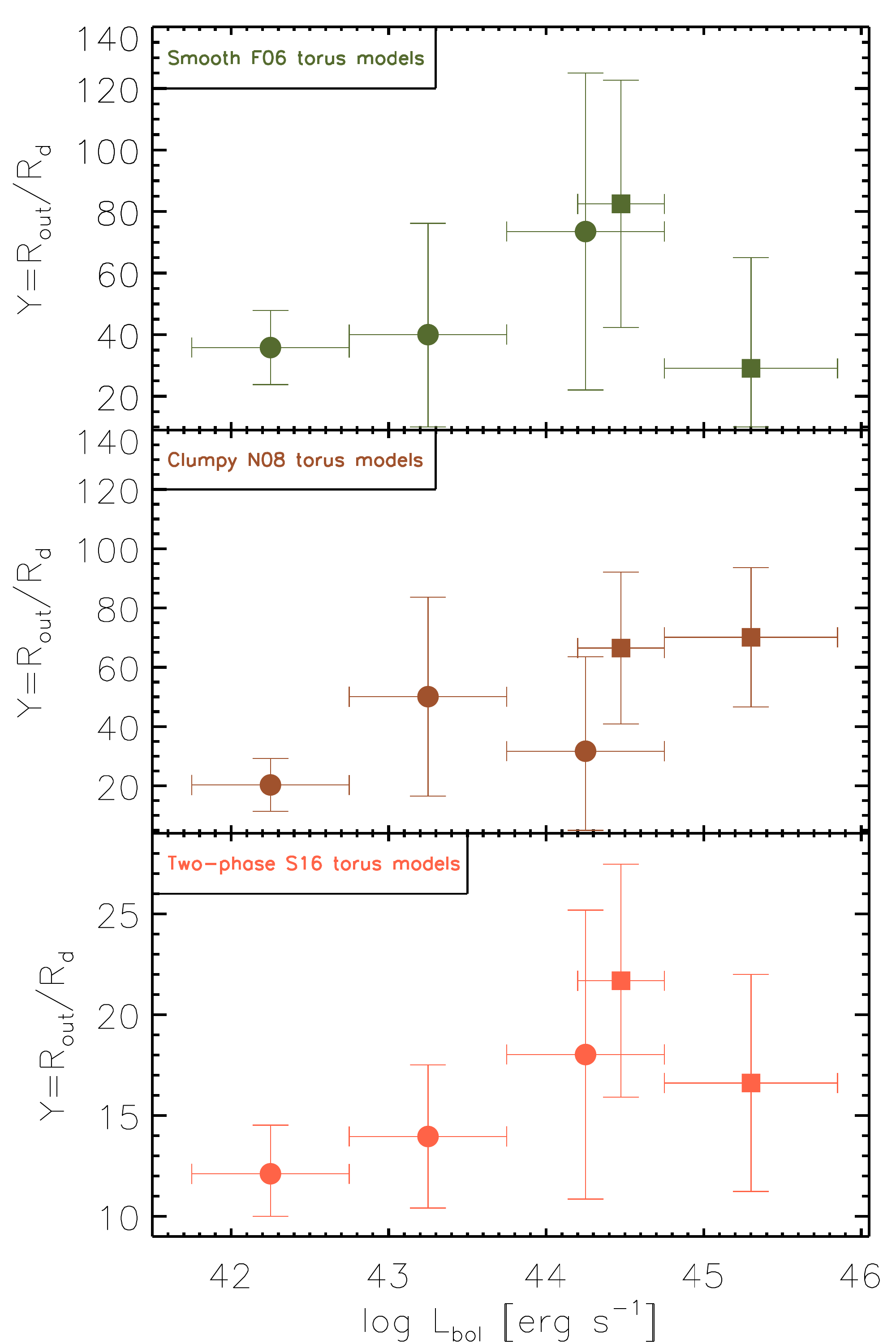}&
\includegraphics[width=8.1cm, clip, trim=20 0 10 0]{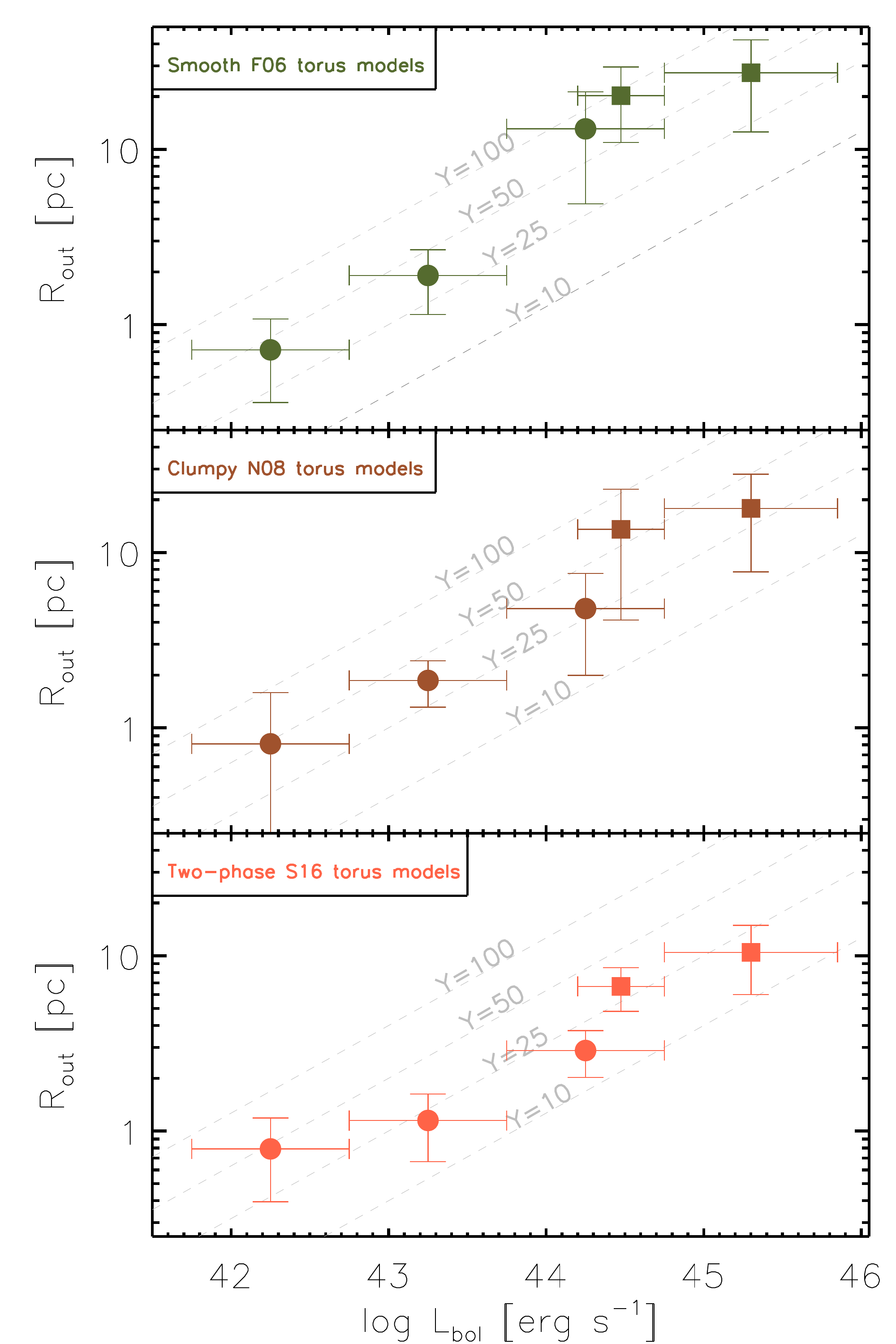} \\ 
\end{tabular}
\caption{Luminosity dependence of the radial extent of the torus/disc (Y=R$_{\rm o}$/R$_{\rm d}$; left panels) and of the torus/disc radius (right panels). From top to bottom panels the result for the various models with free Y parameter used in this work (see text for details). The first 3 bins correspond to the Sy galaxies of BCS$_{40}$ sample. Note that we also include two additional bins of a sample of QSOs with higher bolometric luminosities from \citet{Martinez-Paredes21}.}
\label{rout_lum_distributions}
\end{figure*}

The derived dusty torus/disc sizes ($\sim$1-15\,pc) are similar to those found using MIR imaging and interferometric data (r$<10$~pc; see Section\,\ref{derived_high_angular} for further discussion). However, they are generally smaller than those observed in cold dust by ALMA ($\sim$42\,pc; \citealt{Garcia-Burillo21}), indicating that larger values of the of the radial extent of the torus/disc (Y) than those covered by the models are needed to match the torus sizes measured in ALMA sub-mm observations at Sy-like luminosities. Indeed, we found larger torus sizes ($\sim$1-34\,pc) for the clumpy$+$disc H17 models that use a large value of the radial extent (fixed value of Y=500)\footnote{As expected smaller torus sizes ($\sim$0.3-10.3\,pc) are found when using the clumpy H10 torus model (Y=150)}. Thus, we compare the fitted values of Y with the range covered by the models. The left panel of Fig.\,\ref{rout_lum_distributions} shows that the three models compared here do not favour the largest values of Y. 

Finally, we also investigate the relation between the bolometric luminosity and the torus/disc size dividing our sample into several luminosity bins (see Fig.\,\ref{rout_lum_distributions}). In the first bin we include the three sources with log(L$_{\rm bol}$\,erg~s$^{-1}$)$<$42.75, while the rest of the sample was divided into two bins of equal logarithmic width (1~dex). Note that we also include data from QSOs (i.e. two additional bins from \citealt{Martinez-Paredes21}) to expand the range of luminosities beyond our original Sy sample. These authors used the same methodology as here to fit the high angular resolution NIR-to-MIR SEDs of a sample of type 1 QSOs with log(L$_{\rm bol}$\,erg~s$^{-1}$)$\sim$44.2-45.9. Therefore, these two bins do not include type 2 AGNs. All these models show the same trend throughout the entire luminosity range (log(L$_{\rm bol}$\,erg~s$^{-1}$)$\sim$41.8--45.9): the higher the luminosity the larger the size (see right panel of Fig.\,\ref{rout_lum_distributions}). The same applies when using the BH mass instead of the bolometric luminosity. However, we do not find a clear dependence of Y for higher luminosities (see left panel of Fig.\,\ref{rout_lum_distributions}).

On the other hand, the torus/disc size depends on the Y parameter and dust sublimation radius ($\propto $L$_{\rm bol}^{1/2}$). Therefore, to further investigate the relationship of the torus size with the luminosity, we compare our results with the expected torus sizes at a given bolometric luminosity and Y parameter (dashed grey lines in right panel of Fig.\,\ref{rout_lum_distributions}). Considering the almost constant Y values (within the errors) for each luminosity bin and model in the left panel of Fig.\,\ref{rout_lum_distributions}, the torus size--luminosity correlations might be caused, at least in part, by the sublimation radius dependence with the bolometric luminosity. Futhermore, we find that the derived torus/disc masses also depend on the bolometric luminosities, as expected, given the relation between the torus size and luminosity.  

\section{Discussion}
\label{discussion}

\subsection{The covering factor}

It has been suggested that the bolometric luminosity (e.g. \citealt{Lawrence91,Simpson05}) and Eddington ratio (e.g. \citealt{Buchner17,Ricci17c}) may be the key parameters determining the covering factor. However, according to our results, we do not find a clear dependence of the torus model covering factor with the bolometric luminosity (or the Eddington ratio; see also GB19), although the ranges probed by our sample are relatively reduced (log(L$_{\rm bol}$\,erg~s$^{-1}$)$\sim$41.8--45.9; $\lambda_{\rm Edd}$: -3.40 to -0.26). This lack of dependence was also reported by \citet{Mateos16,Mateos17,Netzer16,Stalevski16,Lani17,Ichikawa18}; GM19B and GB19. Regarding the covering factor, we find that Sy2 galaxies generally have larger values of the covering factor (and angular width of the torus) than Sy1s, for the majority of the models used. This was first reported by RA11 (see also e.g. AH11, \citealt{Ichikawa15} and GB19) but using clumpy N08 torus models only.

Using high-spatial resolution NIR-to-MIR data of an ultra-hard X-ray selected sample of Seyferts, this work confirms that the covering factor of Sy1 and Sy2 galaxies are different. Therefore, our findings indicate that the Seyfert type classification depends not only on the dusty structure inclination but also on the intrinsic differences (e.g. covering factor) of type-1 and type-2 AGN. 

\begin{figure*}
\centering
\par{
\includegraphics[width=8.8cm]{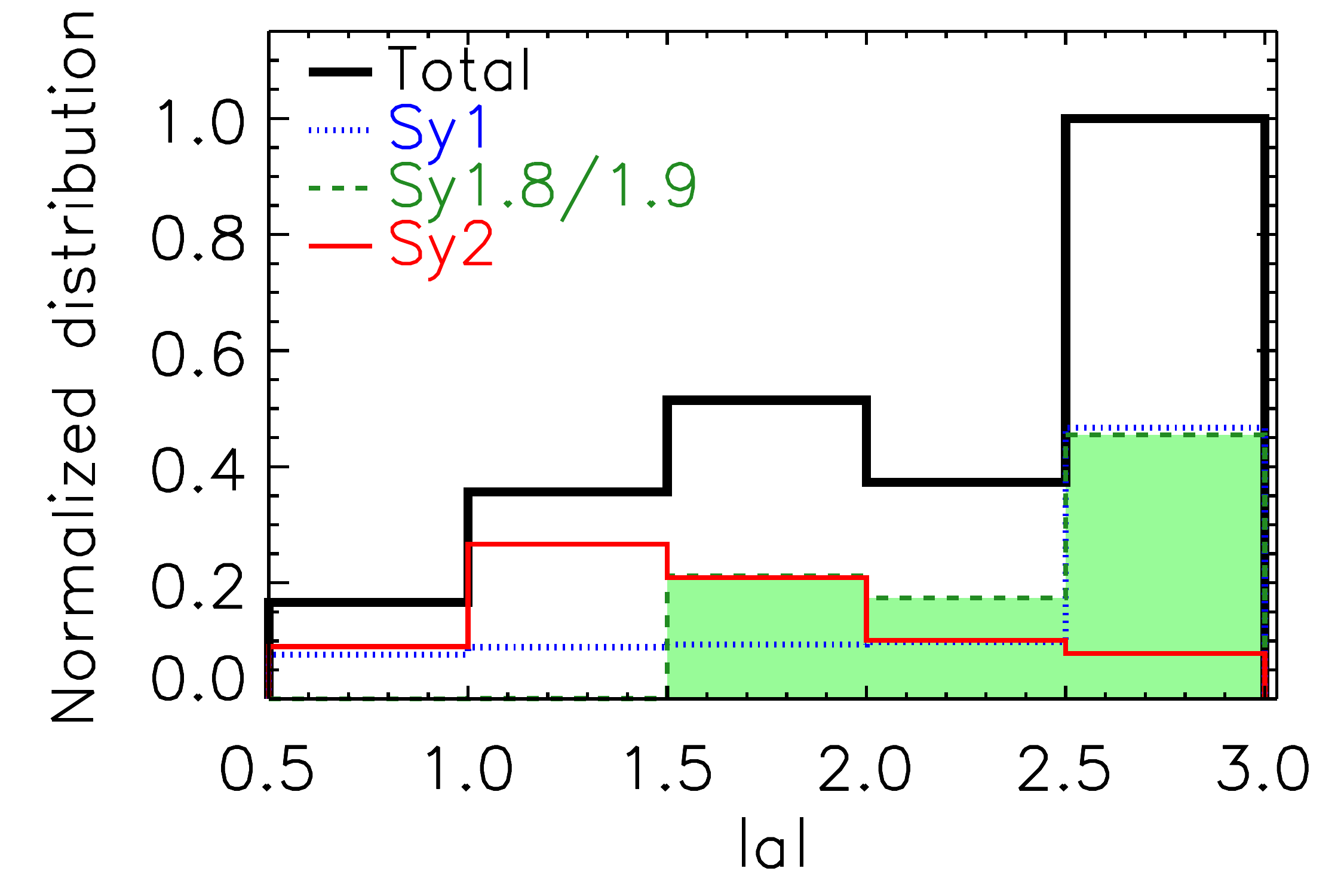}
\includegraphics[width=8.8cm]{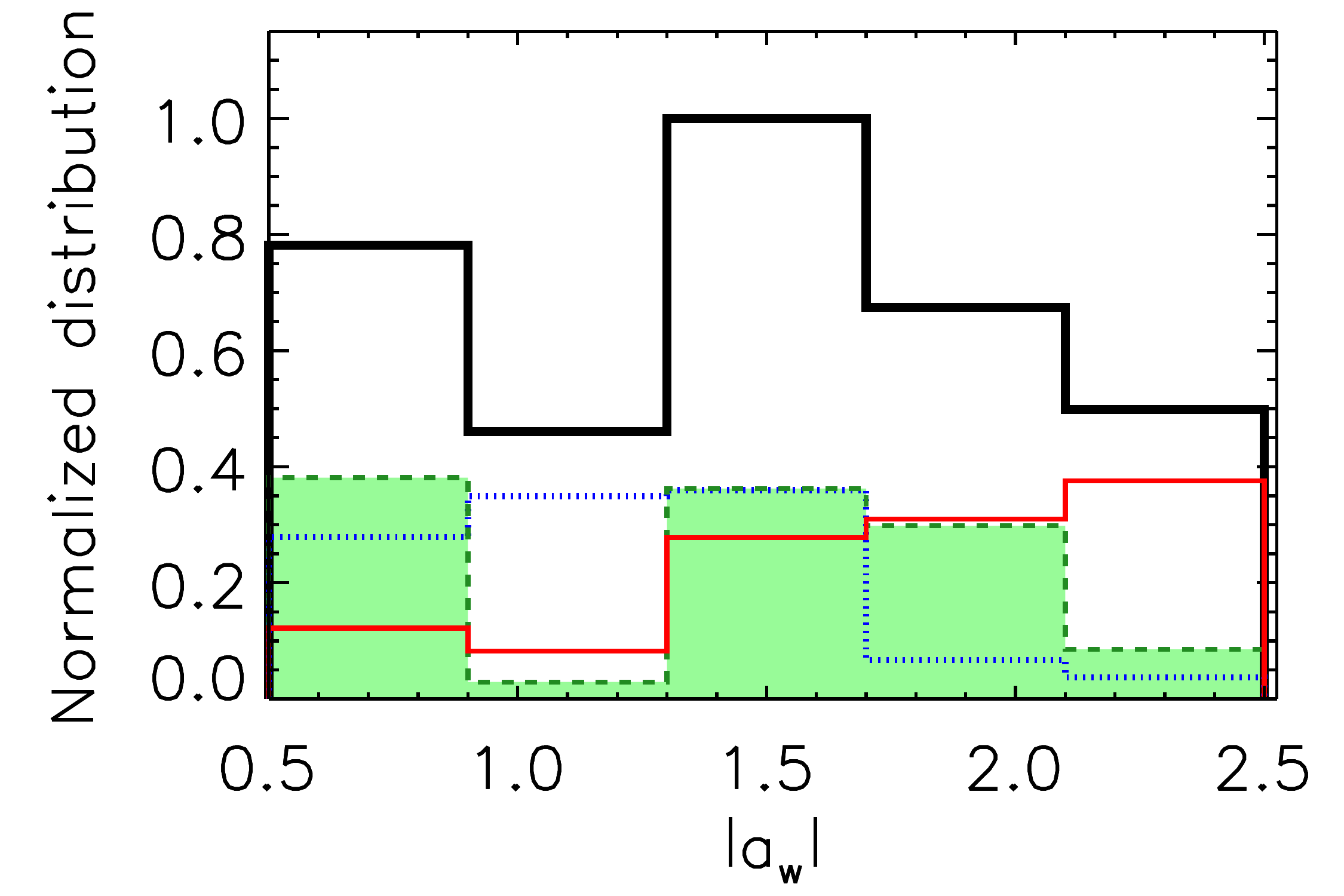}
\par}
\caption{Comparison between the clumpy disc$+$wind H17 model parameters combined probability distributions for the optical classification. Left panel: for the radial dust-cloud distribution power law index (a). Right panel: for the dust-cloud distribution power law along the wind (a$_{\rm w}$). Blue dot-dashed, green dashed, and red solid lines represent the parameter distributions of Sy1, Sy1.8/1.9, and Sy2 galaxies, respectively.}
\label{hoenig17_distribution_a_aw}
\end{figure*}

\subsection{Mid-IR versus sub-mm torus observations}
\label{derived_high_angular}

As shown in Section \ref{size}, we find differences on the outer radius of the torus with the AGN type and luminosity, although they depend on the torus models used. In this section we further explore these parameters by comparing the torus/disc size and mass derived from the fitted nuclear IR SED with those measured from IR and sub-mm data (i.e. VLT/SINFONI, NOEMA, and ALMA). For all the models, we derive relatively compact dusty torus/disc sizes ($\sim$1-15\,pc). This is in agreement with the torus sizes reported in previous works using the clumpy N08 torus models (see e.g. \citealt{Ramos09}; RA11; AH11; \citealt{Lira13,Ichikawa15,Fuller16}; GB19). The derived torus sizes in this work are of the same order of magnitude as those upper-limit sizes derived from MIR observations. For example, using MIR direct imaging, \citet{Packham05} and \citet{Radomski2008} found that the MIR size of the torus is less than $\sim$4~pc (diameter) for Circinus and Centaurus\,A. Furthermore, modelled MIR interferometric data (e.g. \citealt{Jaffe04,Tristram07,Tristram09,Burtscher09,Burtscher13,Raban09,Lopez-Gonzaga16}) also show a relatively compact torus of r$<10$~pc. However, recent works using ALMA sub-mm observations of low-luminosity AGN and Seyfert galaxies measure large molecular discs with physical scales (diameters) ranging from 10 to 130\,pc, with a typical value of 42\,pc (e.g. \citealt{Herrero18,Herrero19,Herrero21,Combes19,Garcia-Burillo21}). The larger sizes measured in the sub-mm compared to those inferred from IR observations are expected since sub-mm sizes correspond to the colder and, thus, more external material within the torus (e.g. \citealt{Lopez-Rodriguez18,Honig19,Herrero21,Nikutta21}).

\begin{figure*}
    \centering
    \includegraphics[width=1.07\columnwidth, clip, trim=0 0 40 20]{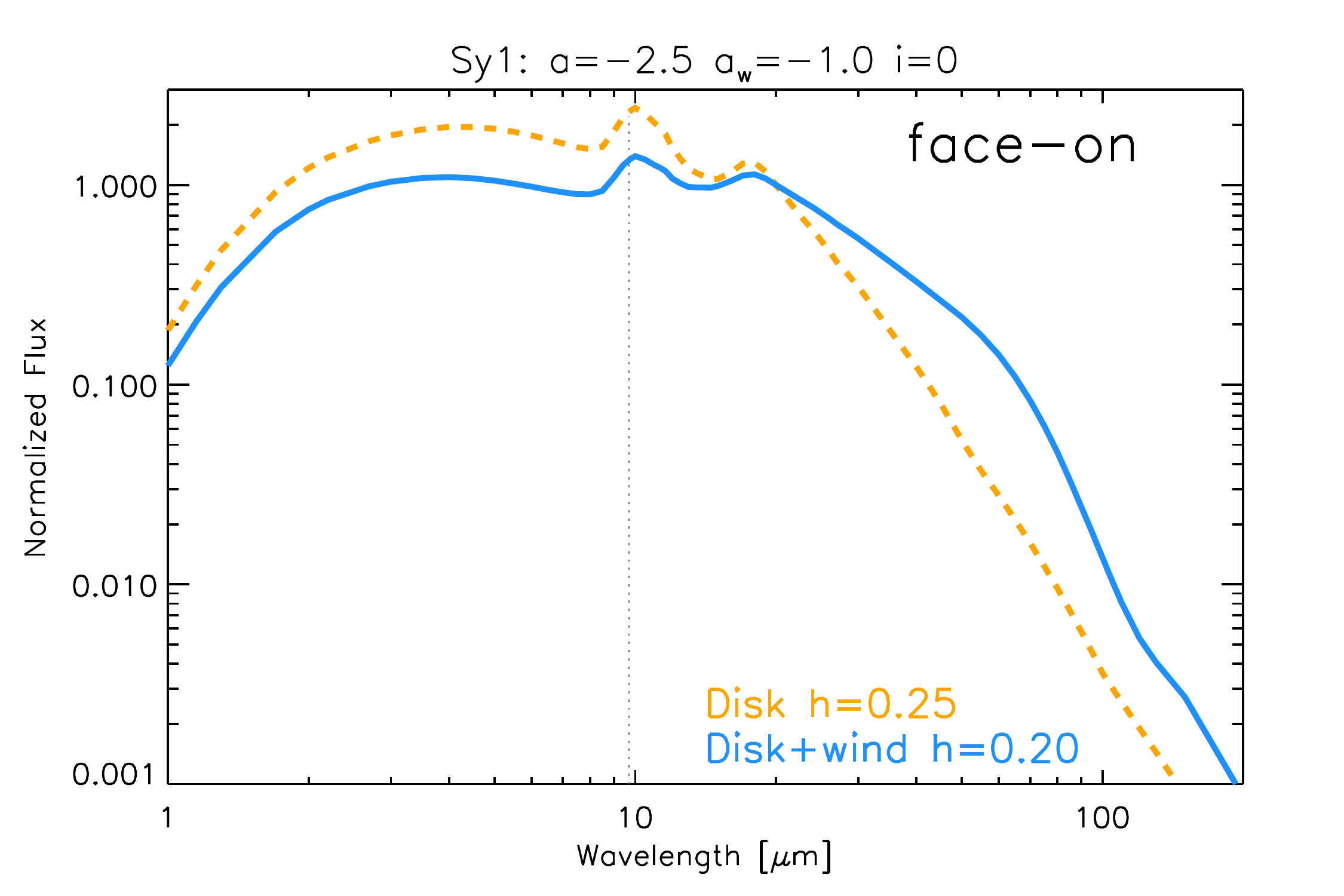}
    \includegraphics[width=0.93\columnwidth, clip, trim=80 0 40 20]{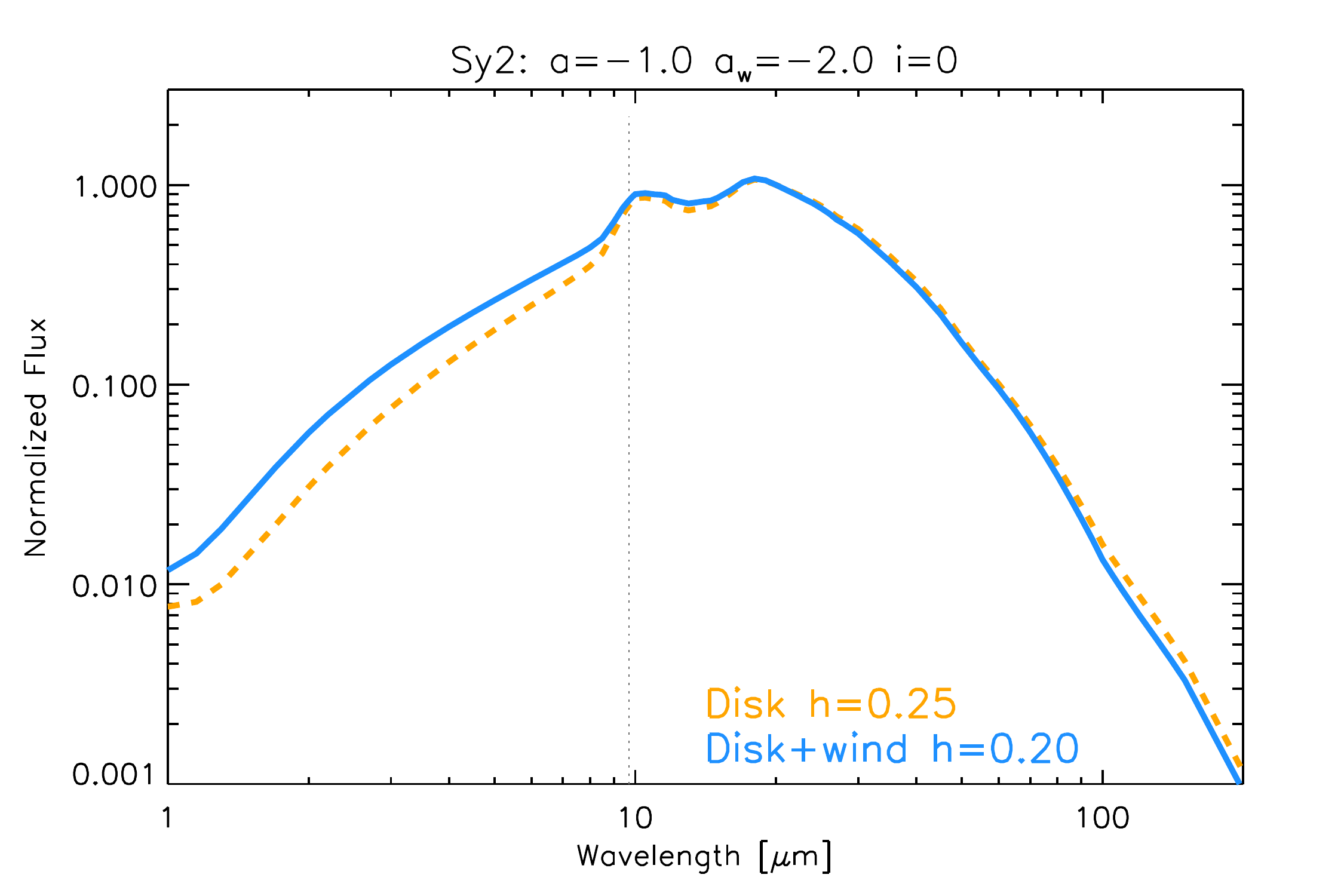}
    \includegraphics[width=1.07\columnwidth, clip, trim=0 0 40 20]{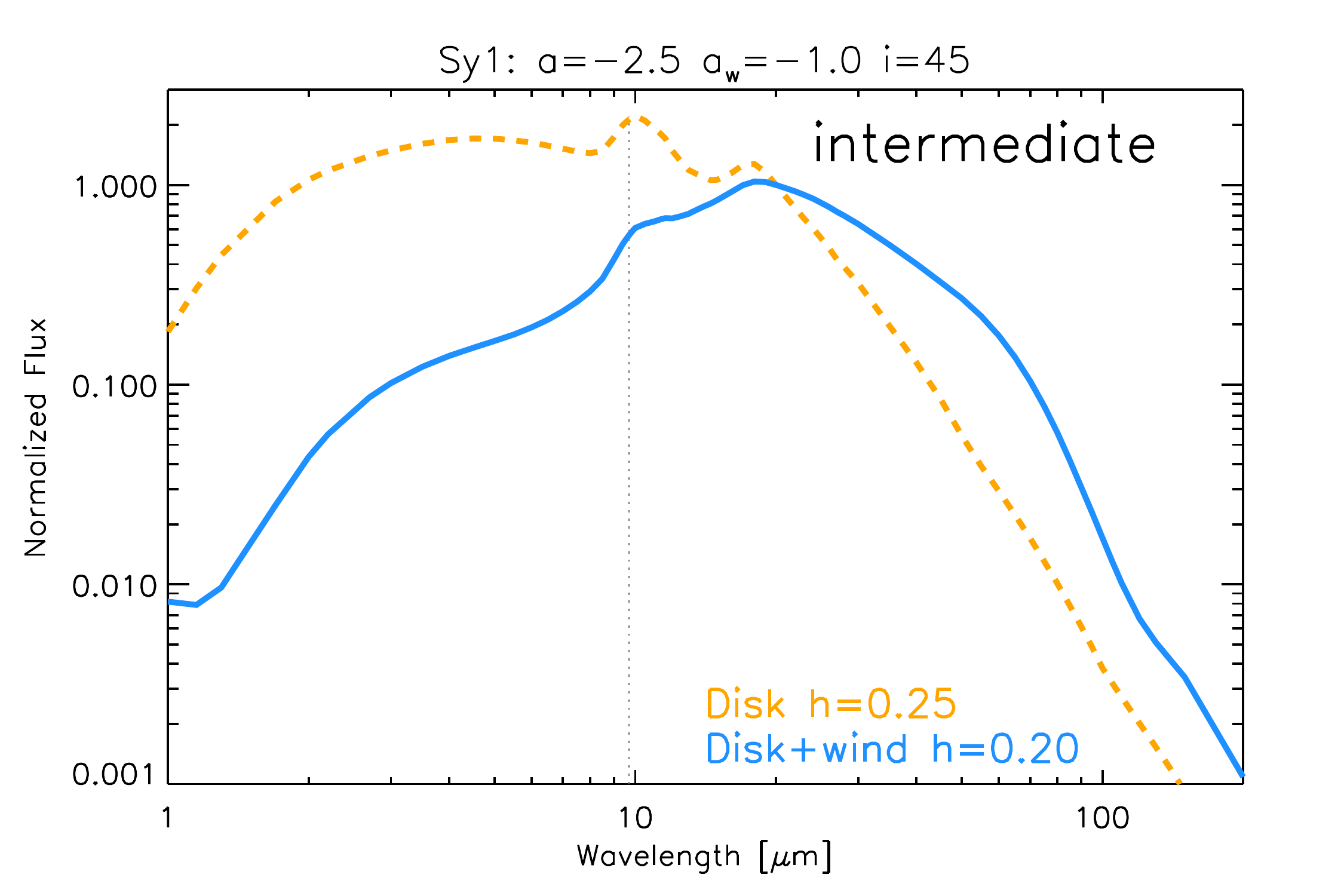}
    \includegraphics[width=0.93\columnwidth, clip, trim=80 0 40 20]{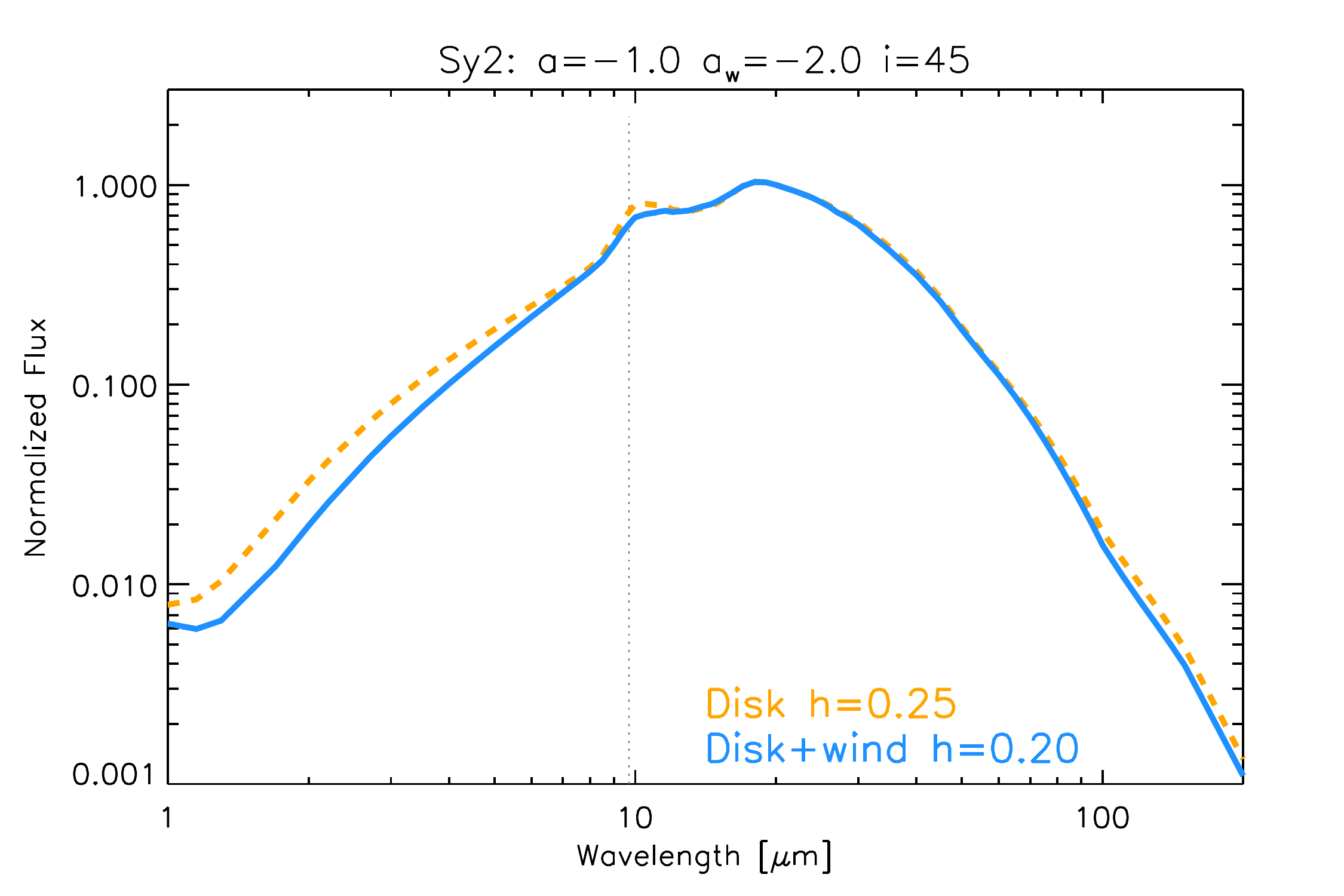}
    \includegraphics[width=1.07\columnwidth, clip, trim=0 0 40 20]{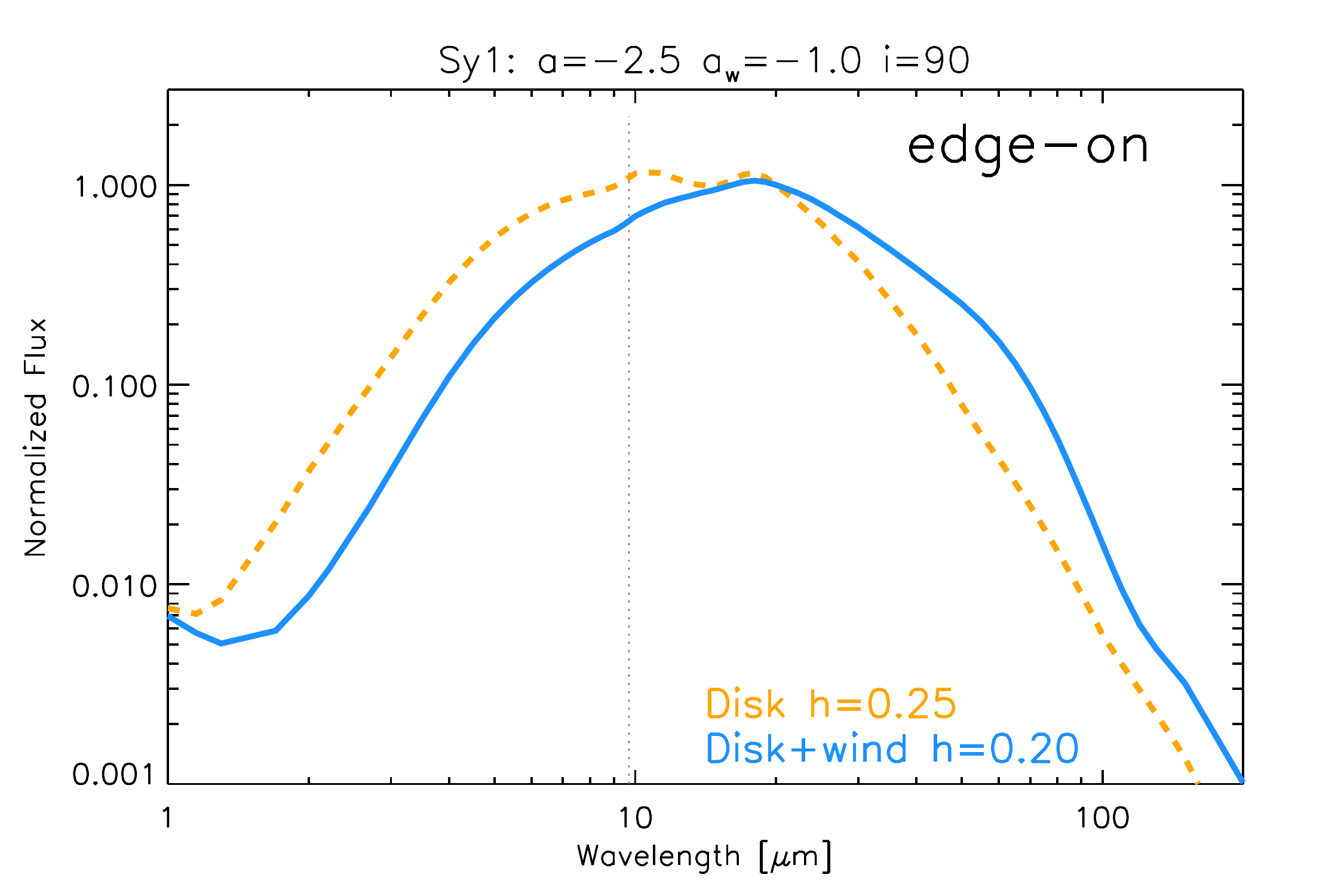}
    \includegraphics[width=0.93\columnwidth, clip, trim=80 0 40 20]{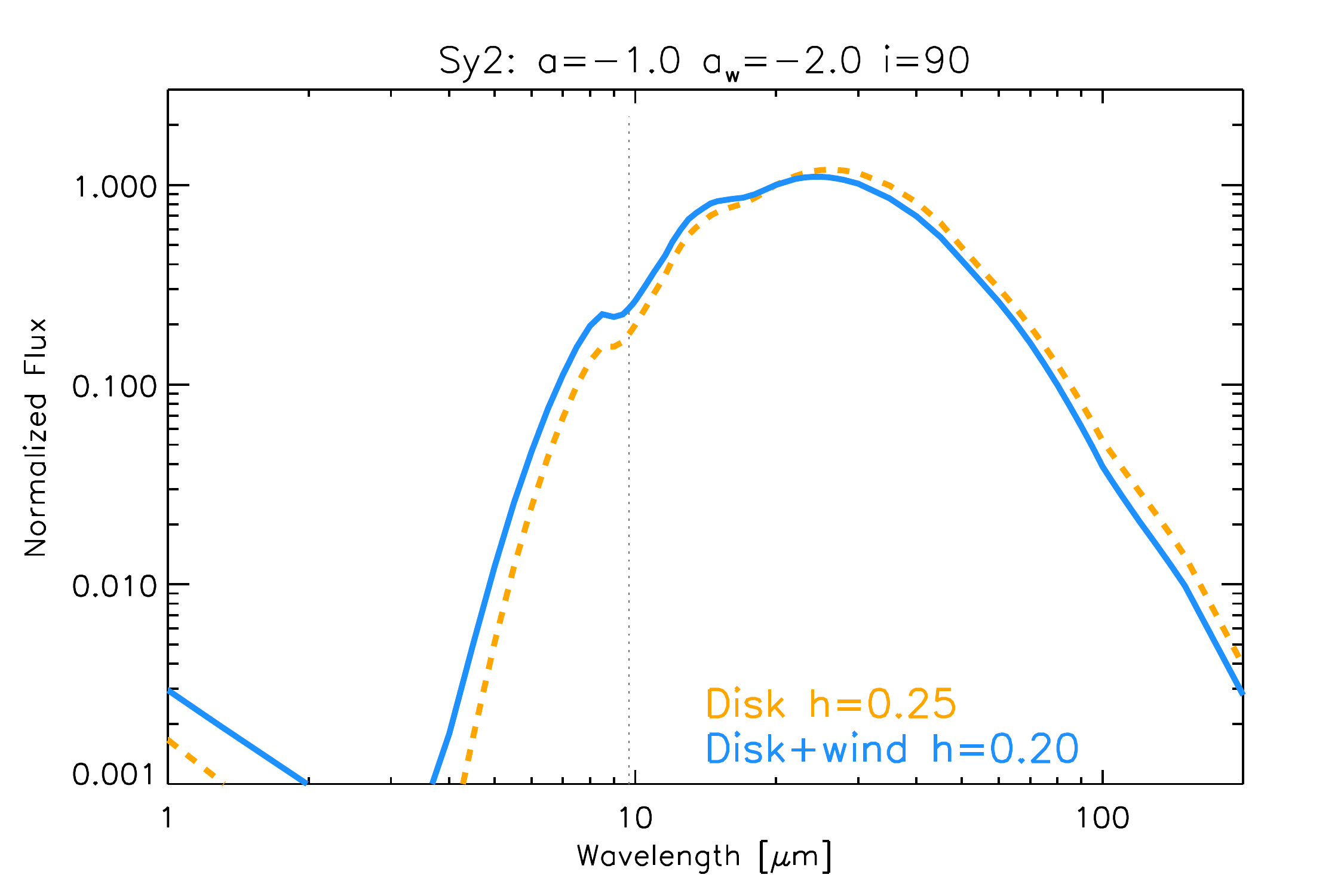}
    \caption{Clumpy disc$+$wind H17 (blue solid lines) and clumpy disc H17 (orange dashed lines) model SEDs (normalized at 20~$\mu$m). Left panel: Sy1 configuration, which consists of a concentrated disc (a=-2.5) and an extended wind (a$_{\rm w}$=-1.0). Right panel: Sy2 configuration, which consists of a relatively extended disc (a=-1.0) and a concentrated wind (a$_{\rm w}$=-2.0). From top to bottom panels: face-on, intermediate (45$^\circ$) and edge-on values of the inclination angle for the clumpy disk and disk$+$wind dusty structure. Note that all the models use N=7, and the clumpy disc$+$wind H17 models use f$_{\rm wd}$=0.6, $\theta$=45$^\circ$ and $\sigma$=10$^\circ$ (see main text for further details on the parameters of the models). }
    \label{selfobscuration}
\end{figure*}

ALMA observations made possible to estimate the molecular gas masses of nearby Seyferts (including a large fraction of the galaxies in this work) and low-luminosity AGN, in the range 10$^5$-10$^7$~M$_\odot$ (see e.g. \citealt{Herrero18,Alonso-Herrero20,Combes19,Garcia-Burillo21}). Using the H$_2$~1-0S(1) emission line at 2.12~$\mu$m of a sample of Sy galaxies, \citet{Hicks09} derived the torus/disc gas masses within the inner 30~pc (radius): M$_{\rm gas}^{\rm H_2}$=0.9--9$\times$10$^6$~M$_\odot$. As expected, the torus masses derived from the fitted nuclear IR SED are relatively smaller ($\sim$10$^2$--10$^6$~M$_\odot$) than those derived from high angular resolution sub-mm data due to the different inferred IR sizes of the torus and those measured in the sub-mm. Therefore, our result is consistent with a temperature-driven stratified disc/torus, where the inner radius is dominated by the hot and warm dust emitting at NIR and MIR wavelengths while the sub-mm observations trace a more extended (and massive) colder component \citep[see also][]{Garcia-Burillo21}.

\subsection{Torus dust composition and geometry}
\label{discussion_dust_compo}
Our findings indicate that torus models better reproduce the NIR-to-MIR emission of AGN with relatively high hydrogen column density (i.e. Sy2s), whereas those of Sy1/Sy1.8/1.9 (with low hydrogen column density) are best fitted by the disc$+$wind H17 model (see Section\,\ref{results}). We also showed that the disc$+$wind dust models improve the spectral fit toward the NIR emission for Sy1/1.8/1.9 galaxies \citep[see also][]{Garcia-Gonzalez17, Gonzalez-Martin19B,Isbell21,Martinez-Paredes21}.

The origin of the NIR bump in the SED remains unclear. Direct emission from the accretion disc of the AGN might be an important contribution of the NIR emission for Sy1 galaxies (e.g. \citealt{caballero16}, GB19, \citealt{Landt19} and references therein), but in this work we have remove it from their SEDs. An alternative explanation for the observed NIR excess in Sy1s is an extra contribution of a hot pure-graphite component (T$_{\rm sub}^{\rm graphites}$>T$_{\rm sub}^{\rm silicates}$) heated by the AGN and located in the inner regions of torus \citep{Mor09}. \citet{Bernete17} found a tight correlation between the hard X-ray fluxes (Nuclear Spectroscopic Telescope Array; NuSTAR) and the NIR emission of a sample of 24 unobscured type 1 AGN, suggesting that the observed NIR bump is produced by AGN-heated hot dust (T>T$_{\rm sub}^{\rm silicates}$).

Clumpy disc$+$wind H17 models predict that the included polar dust (ranging from few pc to tens of pc) mainly contributes to the MIR and sub-mm emission. However, the clumpy disc$+$wind H17 models also include very hot dusty clouds close to the AGN that can reach T$\sim$1900\,K (pure-graphite dust; i.e. T>T$_{\rm sub}^{\rm silicates}$) whose emission peaks at NIR wavelengths. Thus, in Section \ref{results} we tested if the NIR bump can be explained by including graphite grains. To do so, we repeated the fitting process using only the SEDs of the clumpy disc component of these models (clumpy H17D disc models). The fits have slightly smaller residuals in the NIR range of Sy1 galaxies than other torus models (see Fig.\,\ref{Residuals_histogram}). This might be related, at least in part, to the addition of large graphite grains in the dust composition of the disc. 

Therefore, it is important to include large pure-graphite grains that are able to survive at high temperatures, and physically motivated dust sublimation models for reproducing, at least in part, the nuclear IR emission of Sy1s. However, the clumpy flared disc from H17D models still produce larger NIR residuals than those of the models including the polar dust component (i.e. clumpy disc$+$wind H17 models). We also note that torus models (i.e. without including the polar dusty wind component) can produce NIR and MIR model images with emission strongly elongated in the polar directions for certain torus parameters \citep[e.g.][and references therein]{Lopez-Rodriguez18,Nikutta21}. However, \citet{Nikutta21} found that the observed elongations in IR interferometric data of Seyfert galaxies are difficult to reproduce with a single component torus model (see also \citealt{Stalevski17}).

\subsection{A clumpy disc$+$wind versus clumpy disc IR emission}

To further investigate how including the polar dust component modifies the predicted IR emission, we compare the SEDs of the clumpy disc$+$wind H17 models with those of the clumpy disc H17D models. We define two representative set of parameters for Sy1 and Sy2 based on the average values found for each subgroup (see Appendix\,\ref{Combined_distribution}). Using the combined probability distributions of the clumpy disc$+$wind H17 models, we find centrally peaked wind components (i.e. ${a_{\rm w}^{\rm Sy2}}<{a_{\rm w}^{\rm Sy1}}$) for Sy2 and less extended disc components (i.e. ${a_{\rm}^{\rm Sy1}}<{a_{\rm}^{\rm Sy2}}$) for Sy1 galaxies (see Fig.\,\ref{hoenig17_distribution_a_aw} and Appendix\,\ref{Combined_distribution}). Therefore, we select representative SEDs for Sy1 and Sy2 galaxies using different cloud radial distributions for the disc and the wind, but keeping the other parameters to the same values (N=7, h=0.20, f$_{\rm wd}$=0.6, $\theta$=45$^\circ$ and $\sigma$=10$^\circ$; see Table\,\ref{hoenig17_tab_parameters} and corresponding Appendix\,\ref{nuclear_fits} for a description of the model parameters). In particular, we use two configurations of the radial distributions of the clouds: a) Sy1 configuration with a centrally peaked disc (a=-2.5) and an extended wind (a$_{\rm w}$=-1.0); and b) Sy2 configuration with a relatively extended disc (a=-1.0) and a centrally peaked wind (a$_{\rm w}$=-2.0).

\begin{figure*}
\centering
\par{
\includegraphics[width=16cm, clip, trim=0 0 0 150]{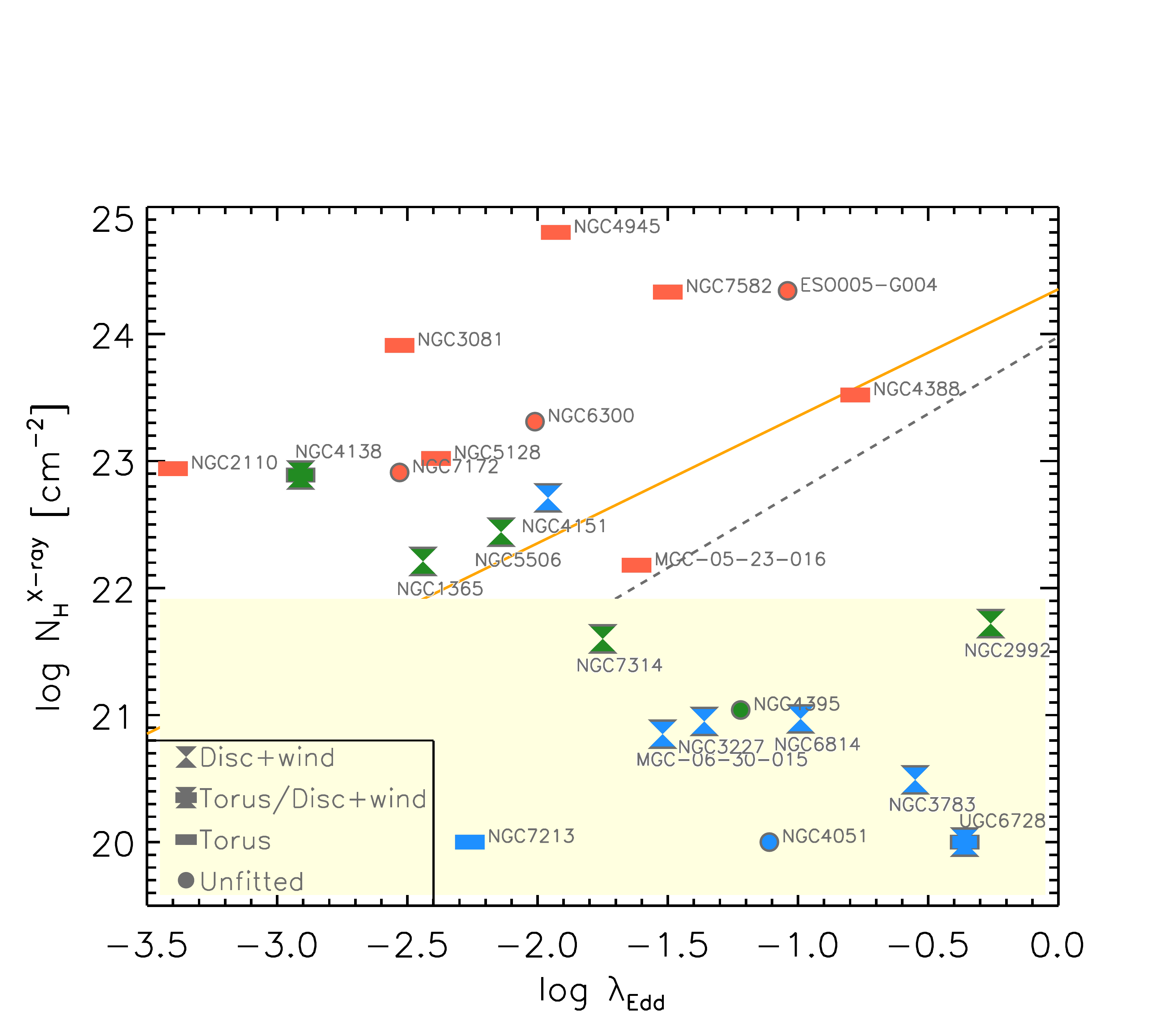}
\par}
\caption{Relation between the X-ray hydrogen column density and the Eddington ratio for the BCS$_{40}$ sample. Blue, green and red symbols represent Sy1, Sy1.8/1.9 and Sy2, respectively. The yellow shaded region corresponds to the region where X-ray column density might not be representative of the molecular gas column density of the torus (i.e. N$_{\rm H}^{\rm X-ray}<$10$^{22}$\,cm$^{-2}$; \citealt{Garcia-Burillo21}). The orange solid line is the limit for IR dusty outflows derived by \citet{Venanzi20} (assuming L$_{\rm AGN}$= 2.2$\times$10$^{43}$ erg s$^{-1}$), and the black dashed line represents the blowout limit predicted by \citet{Fabian08}. Filled bar and hourglass symbols denote galaxies best fitted by torus and disc$+$wind models, respectively. Filled circles represent those galaxies not fitted by any of the models used in this work.}
\label{wind_dependence_nh_edd}
\end{figure*}

Fig.\,\ref{selfobscuration} shows the clumpy disc$+$wind H17 model SEDs (blue solid lines) of the Sy1 and Sy2 configurations for inclinations of 0, 45 and 90$^\circ$ (i.e. face-on, intermediate inclination and edge-on), and the clumpy disc H17D model SEDs (orange dashed lines) using the same parameters\footnote{Note that for clumpy disc H17D SEDs we take the closest value to h=0.20 available (i.e. h=0.25).}. The SEDs for the Sy2 configuration are practically identical regardless of the addition of the polar dust component (see right panels of Fig.\,\ref{selfobscuration}), except at intermediate inclinations (i.e. 45$^\circ$) in the NIR and MIR where the self-obscuration of the inner wall of the dusty structure produced by the cone walls is expected to be relevant (see some of the MIR model images presented by \citealt{Herrero21}). The SEDs of disc$+$wind H17 models are significantly different from those of the clumpy disc H17D models for the Sy1 configuration (see left panels of Fig.\,\ref{selfobscuration}). The far-IR and sub-mm emission of the Sy1 configuration is strongly enhanced by the extra polar dust component. In addition, the torus angular width can play an important role on the self-obscuration of the inner walls. Furthermore, extra self-obscuration takes place by including a dusty wind component. We find that the strong impact of the self-obscuration from the dusty cone takes place in the NIR and MIR range, especially at intermediate inclinations. The polar dust cone walls can produce moderate self-obscuration up to $\sim$10~$\mu$m at all inclinations (see left panels of Fig.\,\ref{selfobscuration}; see also \citealt{Herrero21}). However, the polar dust does not produce strong self-obscuration effects at long wavelengths ($>$20~$\mu$m). Thus, the polar-dust component has a negligible impact in the spectral fit for Sy2 nuclei whereas it produces an enhancement of emission at far-IR and sub-mm wavelengths and an extra self-obscuration at NIR and MIR wavelengths for Sy1 nuclei. This is key to explain the better performance of the disk+wind model for Sy1 nuclei. 

\subsection{Dependence of the best--fitted model with the AGN properties}
\label{bestfits_discussion} 

Recently, \citet{Venanzi20} presented a semi-analytical model to investigate the simulation of radiatively accelerated dusty winds launched by the AGN. In this model the primary mass reservoir for the outflow is the material within the dusty disc. Their simulations show that the wind and its orientation (polar vs. equatorial) depend on the Eddington ratio, AGN luminosity, and nuclear column density. At relatively high column densities (N$_{\rm H}>$10$^{23}$ cm$^{-2}$) the gravity strongly dominates and all the orbits are confined in a compact thick toroidal structure (i.e. the uplift of dusty material is suppressed) for representative values of Sy-like Eddington ratios. At lower values of the column density (N$_{\rm H}<$10$^{23}$ cm$^{-2}$) their model predicted that IR dusty outflows can take place above a certain Eddington ratio. From the observational point of view, \citet{Herrero21} found that 7 of 12 Sy galaxies showed Eddington ratios and nuclear N$_{\rm H}^{\rm ALMA}$ favourable for the launching of the dusty winds, unlike the remaining five galaxies. 

Fig.\,\ref{wind_dependence_nh_edd} shows the line-of-sight hydrogen column density measured at X-rays (N$_{\rm H}^{\rm X-ray}$) versus the Eddington ratios for our sample. The black dashed line represents the blowout limit predicted by \citet{Fabian08}. The orange solid line is the limit for producing IR dusty outflows derived by \citet{Venanzi20} (assuming L$_{\rm AGN}$= 2.2$\times$10$^{43}$ erg s$^{-1}$). Note that we use N$_{\rm H}^{\rm X-ray}$ measurements which are representative of all the material along a pencil-beam line-of-sight to the accretion disc and, thus, it depends on the viewing direction. \citet{Herrero21} show a similar plot but but using N$_{\rm H}$ derived from ALMA observations. \citet{Garcia-Burillo21} compared the hydrogen column densities derived by X-rays and the nuclear integrated values from ALMA for the GATOS\footnote{Galaxy Activity, Torus and Outflow Survey.}  sample and found a good agreement between these N$_{\rm H}$ estimates for obscured Sys (i.e. N$_{\rm H}^{\rm X-ray}>$10$^{22}$\,cm$^{-2}$), whereas the molecular gas column density of the torus probed by ALMA is systematically larger than the N$_{\rm H}^{\rm X-ray}$ for unobscured Sy galaxies (i.e. N$_{\rm H}^{\rm X-ray}<$10$^{22}$\,cm$^{-2}$). In particular, all Sy1 galaxies in \citet{Garcia-Burillo21} have N$_{\rm H}^{\rm ALMA}>$10$^{22}$\,cm$^{-2}$. This could be explained as the X-ray absorption in Sys is related with a smaller pc-scale dust-free gas component compared with the scales probed by ALMA ($\sim$10\,pc; see e.g. \citealt{Garcia-Burillo21}). Therefore, N$_{\rm H}^{\rm X-ray}$ of Sy1 and Sy1.8/1.9 galaxies might be underestimating the molecular gas column density of the torus. To interpret this plot, we add a yellow shaded region highlighting the region where the X-ray column density might be not representative of the torus structure.

We plot in Fig.\,\ref{wind_dependence_nh_edd} galaxies with best fits provided by torus models (i.e. smooth, clumpy, and two-phase torus models) and by disc$+$wind H17 models (see Section\,\ref{results}), using different symbols as shown in the legend. Galaxies with N$_{\rm H}^{\rm X-ray}>$10$^{22}$\,cm$^{-2}$ are located in a region not conducive to launching IR dusty polar outflows, in good agreement with our result that their SEDs are best fitted with torus models. On the other hand, relatively close to the favourable region in the diagram to launching dusty winds, we find a larger number Sys whose SEDs are better fitted with the clumpy disc$+$wind H17 models. Finally, the majority of Sy1 and Sy1.8/1.9 galaxies are located in the the blowout limit region. This might be related, at least in part, to the N$_{\rm H}^{\rm X-ray}$ measurements. Therefore, this dynamical model is able to broadly explain our main result on the dust configurations and it shows the complexity of the AGN torus.

\section{Conclusions}
\label{conclusions}
We presented a detailed comparison of the nuclear dust
emission of an ultra-hard X-ray (14–195~keV) volume-limited (D$_{\rm L}<$40~Mpc) sample of 24 Seyfert galaxies to a set of torus models comprising different dust compositions, distributions and geometries. This sample covers AGN luminosity log(L${_{\textrm{bol}}^{2-10~\textrm{keV}}}$)$\sim$41.75-44.75\,erg~s$^{-1}$) and Eddington ratio ($\lambda_{\rm Edd}$: -3.40 to -0.26) ranges. We include data from QSOs to expand the range of luminosities (log(L${_{\textrm{bol}}^{2-10~\textrm{keV}}}$\,erg~s$^{-1}$)$\sim$44.2-45.9) beyond our original Sy sample. We fitted for the first time the nuclear IR SEDs ($\sim$1-30~$\mu$m) obtained with high angular resolution data with six different torus models to find the model that most closely reproduces the nuclear IR SEDs of type 1 and 2 Seyfert galaxies. Finally, we investigated the relation of the bolometric luminosity, hydrogen column density, and Eddington ratio with different torus parameters. The main results are as follows. 

\begin{enumerate}
   
\item  The various torus models used in this work provide acceptable fits ($\rm{\chi^2_{red}<2}$) to the majority (19/24) of the nuclear IR SEDs. The fraction of best fits provided by
smooth, clumpy and two-phases torus models (i.e. those models that do not include the dusty polar component) and disc$+$wind models is practically the same, 33.3 and 37.5\% (i.e. 8/24 and 9/24 sources), respectively.\\ 

\item  The disc$+$wind models reproduce better the NIR-to-MIR emission of AGN with relatively low X-ray hydrogen column density (median value of log (N$_{\rm H}^{\rm X-ray}$\,cm$^{-2}$)=21.0$\pm$1.0; i.e. Sy1/Sy1.8/1.9), whereas the nuclear IR SED of Sy2 (median value of log (N$_{\rm H}^{\rm X-ray}$\,cm$^{-2}$)=23.5$\pm$0.8) are best fitted by smooth, clumpy and two-phases torus models without including the polar dusty wind component.\\

\item The inclusion of large graphite grains with T$_{\rm sub}\sim$1900\,K, in addition to the self-obscuration produced by the polar component at intermediate inclinations (and/or a thick torus) are crucial to reproduce reproduce the observed nuclear NIR and MIR SED of Sy1/1.8/1.9s.\\

\item In general, we find that the Seyfert galaxies having unfavourable (favourable) conditions, i.e. nuclear hydrogen column density and Eddington ratio, for launching IR dusty polar outflows are best-fitted with smooth, clumpy and two-phase torus (disk$+$wind) models confirming the predictions from simulations.\\

\end{enumerate}
   
Our results indicate that there is a relationship between the choice of model and the hydrogen column density and, thus, the X-ray (unobscured/obscured) and/or optical (Sy1/Sy2) classification. These findings suggest that the torus dusty geometry/grain composition might depend on the amount of nuclear material (N$_{\rm H}$) and AGN properties. This work demonstrates the power of the spectral fitting technique to infer the properties of the inner dusty structure in AGN. In the future, the unprecedented combination of high sensitivity and spatial resolution provided by the James Webb Space Telescope \emph{(JWST)} will be crucial to better understand the nuclear dusty region of AGN using this technique.

\begin{acknowledgements}
IGB and DR acknowledge support from STFC through grant ST/S000488/1. OGM acknowledges support from the UNAM PAPIIT project IN105720. CRA acknowledges financial support from the European Union's Horizon 2020 research and innovation programme under Marie Sk\l odowska-Curie grant agreement No 860744 (BiD4BESt) and from the project ``Feeding and feedback in active galaxies'', with reference PID2019-106027GB-C42, funded by MICINN-AEI/10.13039/501100011033. AAH acknowledges support from grant PGC2018-094671-B-I00 funded by MCIN/AEI/10.13039/501100011033 and by ERDF A way of making Europe. EAD acknowledges support from the Agencia Estatal de Investigaci\'on del Ministerio de Ciencia e Innovaci\'on (AEI-MCINN) under project with reference PID2019-107010GB100. Finally, we thank the anonymous referee for their useful comments.

\end{acknowledgements}

%\Online

\begin{appendix}
\section{Torus model parameters and nuclear IR SED fits}
\label{nuclear_fits}
Using the various torus models, we fit all the nuclear NIR-to-MIR SEDs in our sample (See Section \ref{results}). In Tables \ref{fritz_tab_parameters}-\ref{hoenig17_tab_parameters} we present the different torus model parameters. The values of the model parameters fitted to the individual nuclear IR SEDs are reported in Tables \ref{appendixfitSy1}, \ref{appendixfitSyInt} and \ref{appendixfitSy2}. In addition, the individual fits are shown in Figures (A1-A24).

\begin{table}
\scriptsize
\centering
\caption{Smooth F06 torus model parameters.}
\begin{tabular}{lcc}
\hline
Parameter 									& Symbol 		& Interval 	\\
\hline
Inclination angle of the torus				&	i			&	[0$^\circ$, 90$^\circ$]		\\
Width of clouds angular distribution		&	$\sigma$	&	[20$^\circ$, 60$^\circ$]	\\
Exponent of the logarithmic azimuthal 		&   $\Gamma$	& [0,6]		\\
density distribution						&	&\\
Exponent of the logarithmic radial 		    &   $\beta$	& [-1,0]		\\
profile of the density distribution		&	&\\
Radial extent of the torus 				&	Y			&	[10, 150]		\\
Edge-on optical depth				&	$\tau_{9.7 \mu m}$&	[0.1, 10]		\\
\hline
\end{tabular}					 
\tablefoot{i is measured from the polar axis in this models. Therefore, i$=$90 is face-on and i$=$0 is edge on.}
\label{fritz_tab_parameters}
\end{table}

\begin{table}
\scriptsize
\centering
\caption{Clumpy N08 torus model parameters. }
\begin{tabular}{lcc}
\hline
Parameter 									& Symbol 		& Interval 	\\
\hline
Radial extent of the torus 				&	Y			&	[5, 150]		\\
Width of clouds angular distribution		&	$\sigma$	&	[15$^\circ$, 70$^\circ$]	\\
Number of clouds along an equatorial ray	&	N$_0$		&	[1, 15]		\\
Index of the radial density profile		&	q			&	[0, 3]		\\
Inclination angle of the torus				&	i			&	[0$^\circ$, 90$^\circ$]		\\
Optical depth per single cloud				&	$\tau_{V}$&	[10, 300]		\\
\hline
\end{tabular}					 
\tablefoot{i$=$0 is face-on and i$=$90 is edge on.}
\label{nenkova_tab_parameters}
\end{table}

\begin{table}
\scriptsize
\centering
\caption{Clumpy H10 torus model parameters. }
\begin{tabular}{lcc}
\hline
Parameter 									& Symbol 		& Interval 	\\
\hline
Inclination angle of the torus				&	i			&	[0$^\circ$, 90$^\circ$]		\\
Number of clouds along an equatorial line-of-sight& N$_0$ &  [2.5,10.0]		\\
Half-opening angle of the distribution of clouds		&	$\theta$	&	[5$^\circ$, 60$^\circ$]	\\
Radial dust-cloud distribution power law index 		&   a	& [-2,0]		\\
Opacity of the clouds				&	$\tau_{cl}$&	[30, 80]		\\
\hline
\end{tabular}					 
\tablefoot{i$=$0 is face-on and i$=$90 is edge on. Note that the torus angular width is 90-$\theta_{\rm H10}$.}
\label{hoenig10_tab_parameters}
\end{table}

\begin{table}
\scriptsize
\centering
\caption{Two-phase S16 torus model parameters. }
\begin{tabular}{lcc}
\hline
Parameter 									& Symbol 		& Interval 	\\
\hline
Inclination angle of the torus				&	i			&	[0$^\circ$, 90$^\circ$]		\\
Width of clouds angular distribution		&	$\sigma$	&	[10$^\circ$, 80$^\circ$]	\\
Radial dust density gradient distribution 		&   p	& [0,1.5]		\\
Polar dust density gradient distribution 		&   q	& [0,1.5]		\\
Radial extent of the torus 				&	Y			&	[10, 30]		\\
Average edge-on optical depth				&	$\tau_{9.7\mu m}$&	[3, 11]		\\
\hline
\end{tabular}					 
\tablefoot{i$=$0 is face-on and i$=$90 is edge on.}
\label{stalev_tab_parameters}
\end{table}

\begin{table}
\scriptsize
\centering
\caption{Clumpy disc$+$wind H17 torus model parameters. }
\begin{tabular}{lcc}
\hline
Parameter 									& Symbol 		& Interval 	\\
\hline
Inclination angle of the torus				&	i			&	[0$^\circ$, 90$^\circ$]		\\
Number of clouds along an equatorial line-of-sight& N$_0$ &  [5,10]		\\
Radial dust-cloud distribution power law index 		&   a	& [-3.0,-0.5]		\\
Half-opening angle of the wind		&	$\theta$	&	[30$^\circ$, 45$^\circ$]	\\
Angular width of the hollow wind cone		&	$\sigma$	&	[7$^\circ$, 15$^\circ$]	\\
Dust cloud distribution power law along the wind 		&   a$_w$	& [-2.5,-0.5]		\\
Scale height of the disc		&   h	& [0.1,0.5]		\\
Wind-to-disc ratio of dust clouds		&   f$_{wd}$	& [0.15,0.75]		\\
\hline
\end{tabular}					 
\tablefoot{i$=$0 is face-on and i$=$90 is edge on.}
\label{hoenig17_tab_parameters}
\end{table}

\begin{table*}
\scriptsize
\begin{center}
\caption{Spectral fit results for Type-1 Seyferts.}
\begin{tabular}{llcccccccccccc}
\hline \hline
Obj.	& Mod. &  $\chi^{2}_{red}$ & dof & E(B-V)  &\multicolumn{8}{c}{Parameters}       \\
	    &     &                   &   &   &   &     &     &     &     &      &  &   \\ \hline
				&	F06&		&		&		& $i$ & $\sigma$ & $\Gamma$ & $\beta$ & $Y$ & $\tau_{9.7\mu m}$ &  & \\
				&	N08&		&		&		& $i$ & $N_{0}$  & $\sigma$ & $Y$ & $q$ &$\tau_{v}$ &  & \\
				&	H10&		&		&		& $i$ & $N_{0}$  & $\theta$ & $a$ & $\tau_{cl}$ &  &  &  \\
				&	S16&		&		&		& $i$ & $\sigma$ & $p$ & $q$ & $Y$ & $\tau_{9.7\mu m}$ &  & \\
				&	H17&		&		&		& $i$ & $N_{0}$ & $a$  & $\sigma$ & $\theta$ & $a_{w}$ & $h$ & $f_{w}$ \\
				&	H17D&		&		&		& $i$ & $N_{0}$ & $a$  & $h$\\ \hline \hline
				&		&			&	&		&							&		&	& & & & &	\\
				
MCG-06-30-015          & F06    &   2.07 /   3.25 & 420    / 49     & 0.38$\pm{0.01}$ & $<$28.5 & $<$20.4 & $>$6.0 & -0.75$\pm{0.01}$ & 19.9$\pm{0.3}$ & 1.01$\pm_{0.02}^{0.01}$ &  &  &  \\ 
                       & N08    &   2.56 /   5.29 & 420    / 49     & 0.08$\pm{0.01}$ & $>$89.9 & $>$14.9 & $>$25.2 & $<$5.0 & $>$2.5 & $<$10.0 &  &  &  \\ 
                       & H10    &   1.34 /   2.85 & 421    / 50     & 0.27$\pm{0.01}$ & 51.6$\pm_{0.7}^{0.1}$ & $>$9.9 & $<$53.6 & $<$-1.6 & $>$79.5 &  &  &  &  \\ 
                       & S16    &   1.83 /   4.72 & 420    / 49     & 0* & 79.9$\pm_{6.5}^{0.5}$ & 69.1$\pm_{7.4}^{1.3}$ & $>$1.5 & $>$1.5 & $<$10.0 & 4.04$\pm_{0.02}^{0.17}$ &  &  &  \\ 
                       & H17$\bullet$    &   0.95 /   1.20 & 418    / 47     & 0.26$\pm{0.01}$ & 60.1$\pm_{0.5}^{0.7}$ & 6.9$\pm_{0.3}^{0.1}$ & $<$-3.0 & 10.0$\pm_{0.3}^{0.4}$ & $>$43.0 & $>$-0.5 & 0.28$\pm_{0.01}^{0.03}$ & 0.30$\pm{0.02}$ &  \\ 			
                      & H17D   &   1.12 /   2.16 & 422    / 51     & 0.18$\pm{0.01}$ & $>$66.8 & $>$9.9 & $<$-2.4 & $>$0.3 &  \\ 
NGC3227                & F06    &   1.35 /   2.76 & 232    / 33     & 0.26$\pm_{0.01}^{0.02}$ & $<$33.4 & $<$20.2 & $>$6.0 & -0.75$\pm{0.01}$ & 30.1$\pm_{0.3}^{2.1}$ & $>$4.9 &  &  &  \\ 
                       & N08    &   1.38 /   3.02 & 232    / 33     & 0.66$\pm_{0.02}^{0.01}$ & $<$18.3 & 1.15$\pm_{0.04}^{0.05}$ & $>$47.5 & $<$5.2 & $>$2.4 & $>$296.2 &  &  &  \\ 
                       & H10    &   1.11 /   3.09 & 233    / 34     & 0.42$\pm_{0.01}^{0.02}$ & $>$46.8 & $>$7.5 & $<$54.8 & $>$-0.8 & $>$79.6 &  &  &  &  \\ 
                       & S16    &   1.39 /   5.27 & 232    / 33     & 0* & 35.4$\pm_{1.5}^{4.8}$ & $<$63.3 & 1.0$\pm_{0.05}^{0.03}$ & $>$0.0 & $<$20.1 & $<$4.9 &  &  &  \\ 
                       & H17$\bullet$    &   0.61 /   1.26 & 230    / 31     & 0.22$\pm_{0.04}^{0.03}$ & 70.5$\pm{3.2}$ & $>$9.1 & -1.9$\pm{0.1}$ & $>$13.2 & $>$36.8 & $>$-0.6 & $<$0.2 & 0.3$\pm{0.1}$ &  \\ 
                       & H17D$\bullet$   &   0.77 /   1.64 & 234    / 35     & 0.28$\pm{0.03}$ & $>$49.8 & 6.6$\pm{0.4}$ & -1.68$\pm{0.03}$ & 0.42$\pm{0.02}$ &  \\

NGC3783                & F06    &   1.23 /   2.42 & 418    / 28     & 0.44$\pm{0.01}$ & $<$32.7 & $<$20.4 & $>$0.0 & -0.75$\pm{0.01}$ & $>$33.1 & $>$1.1 &  &  &  \\ 
                       & N08    &   0.89 /   3.95 & 418    / 28     & 0.15$\pm_{0.02}^{0.03}$ & $>$79.6 & 5.0$\pm_{0.3}^{0.6}$ & 36.2$\pm_{5.4}^{11.6}$ & $<$5.3 & $>$2.4 & 46.4$\pm_{2.3}^{1.7}$ &  &  &  \\ 
                       & H10    &   0.83 /   3.20 & 419    / 29     & 0.25$\pm_{0.03}^{0.02}$ & 30.0$\pm_{2.1}^{5.0}$ & $>$8.9 & 52.0$\pm_{2.6}^{0.6}$ & -0.73$\pm_{0.16}^{0.03}$ & $>$79.2 &  &  &  &  \\ 
                       & S16    &   0.86 /   3.91 & 418    / 28     & 0* & $>$68.9 & $>$70.2 & 0.5$\pm_{0.2}^{0.1}$ & $>$1.3 & $<$12.2 & 4.5$\pm_{0.4}^{0.1}$ &  &  &  \\ 
                       & H17$\bullet$    &   0.67 /   0.84 & 416    / 26     & 0.31$\pm_{0.02}^{0.03}$ & $<$28.8 & $<$6.7 & $<$-2.8 & $>$7.8 & $>$37.4 & -1.0$\pm_{0.1}^{0.2}$ & $<$0.1 & $>$0.5 &  \\ 
                       & H17D   &   0.74 /   2.35 & 420    / 30     & 0.21$\pm{0.02}$ & 44.7$\pm_{0.2}^{0.6}$ & 7.0$\pm_{0.5}^{0.1}$ & -1.75$\pm{0.01}$ & 0.42$\pm_{0.01}^{0.04}$ &  \\ 

NGC4051*                & F06    &   1.49 /   2.68 & 231    / 48     & 0* & 26.6$\pm_{1.3}^{0.5}$ & $<$21.1 & $>$5.5 & -0.75$\pm{0.01}$ & $>$26.1 & 4.8$\pm_{0.2}^{0.3}$ &  &  &  \\ 
                       & N08    &   1.90 /   2.48 & 231    / 48     & 0.49$\pm_{0.01}^{0.02}$ & $<$19.8 & $<$1.0 & $>$67.2 & $<$5.1 & $>$2.5 & $>$292.3 &  &  &  \\ 
                       & H10    &   1.38 /   2.38 & 232    / 49     & 0.07$\pm{0.04}$ & 15.0$\pm_{2.3}^{2.9}$ & 7.8$\pm_{0.3}^{0.5}$ & $<$44.4 & $>$-0.5 & $>$78.8 &  &  &  &  \\ 
                       & S16    &   1.53 /   3.23 & 231    / 48     & 0* & 18.5$\pm_{8.6}^{3.0}$ & 70.0$\pm_{2.0}^{3.1}$ & $<$0.6 & $<$0.0 & $<$14.4 & $>$3.7 &  &  &  \\ 
                       & H17    &   1.31 /   2.26 & 229    / 46     & $<$0.04 & 80.5$\pm_{2.5}^{2.9}$ & $>$9.2 & $<$-2.9 & $>$14.1 & $>$43.7 & -1.4$\pm{0.1}$ & $<$0.1 & 0.47$\pm_{0.03}^{0.02}$ &  \\ 
                       & H17D   &   1.28 /   2.06 & 233    / 50     & $<$0.07 & 30.0$\pm_{9.6}^{3.3}$ & 5.4$\pm_{0.2}^{0.6}$ & -1.5$\pm_{0.1}^{1.2}$ & $>$0.7 &  \\ 
NGC4151                & F06    &   1.21 /   2.45 & 162    / 47     & 0.51$\pm_{0.04}^{0.02}$ & $>$67.3 & $<$21.5 & $>$5.8 & -0.75$\pm{0.01}$ & 30.0$\pm_{0.3}^{0.5}$ & $>$1.1 &  &  &  \\ 
                       & N08    &   1.17 /   3.30 & 162    / 47     & 0.66$\pm_{0.03}^{0.04}$ & $<$27.6 & $>$1.0 & $>$15.0 & $<$6.9 & $>$2.3 & $>$284.0 &  &  &  \\ 
                       & H10    &   1.16 /   3.22 & 163    / 48     & 0.43$\pm_{0.02}^{0.04}$ & $>$43.2 & 7.5$\pm_{0.3}^{0.2}$ & $<$57.6 & $>$-0.9 & $>$78.3 &  &  &  &  \\ 
                       & S16    &   1.33 /   4.30 & 162    / 47     & $<$0.08 & 9.9$\pm_{9.2}^{1.2}$ & 78.1$\pm_{1.4}^{0.7}$ & 1.0$\pm{0.1}$ & $<$0.0 & $<$10.5 & $<$5.2 &  &  &  \\ 
                       & H17$\bullet$    &   0.29 /   0.71 & 160    / 45     & 0.14$\pm{0.03}$ & 44.8$\pm_{5.0}^{2.2}$ & $<$5.6 & $<$-3.0 & $>$14.4 & $>$43.1 & -1.56$\pm_{0.02}^{0.03}$ & $<$0.1 & 0.6$\pm_{0.2}^{0.1}$ &  \\ 
                       & H17D   &   1.00 /   2.37 & 164    / 49     & 0.45$\pm_{0.04}^{0.07}$ & $>$60.0 & $<$5.3 & -1.8$\pm_{0.1}^{1.5}$ & $>$0.3 &  \\ 
NGC6814                & F06    &   1.53 /   2.28 & 82     / 37     & 0.16$\pm_{0.02}^{0.01}$ & $>$73.0 & $<$20.8 & $>$6.0 & -0.75$\pm{0.01}$ & 38.9$\pm_{6.6}^{7.7}$ & 4.1$\pm{0.3}$ &  &  &  \\ 
                       & N08    &   1.94 /   3.77 & 82     / 37     & 0.0* & 30.0$\pm_{9.2}^{13.3}$ & $>$14.3 & 25.1$\pm_{1.0}^{1.4}$ & $<$5.1 & $<$0.7 & 30.0$\pm_{4.6}^{2.6}$ &  &  &  \\ 
                       & H10    &   1.65 /   3.07 & 83     / 38     & 0.0* & $<$15.6 & 7.5$\pm_{0.4}^{0.3}$ & 45.6$\pm_{1.2}^{1.1}$ & -0.53$\pm_{0.03}^{0.02}$ & $>$79.2 &  &  &  &  \\ 
                       & S16    &   1.88 /   3.84 & 82     / 37     & 0.0* & 40.9$\pm_{3.8}^{2.5}$ & 50.0$\pm_{2.5}^{2.1}$ & $<$0.0 & $<$0.1 & $<$10.5 & 4.4$\pm_{0.7}^{0.8}$ &  &  &  \\ 
                       & H17$\bullet$    &   0.81 /   1.19 & 80     / 35     & 0.0* & 75.0$\pm_{6.4}^{3.9}$ & $<$5.9 & $<$-2.9 & $>$7.0 & 40.5$\pm_{2.2}^{0.6}$ & -1.1$\pm_{0.1}^{0.2}$ & $<$0.1 & $>$0.6 &  \\ 
                       & H17D   &   1.54 /   2.29 & 84     / 39     & 0.0* & 45.1$\pm_{3.0}^{1.2}$ & 7.0$\pm_{0.9}^{0.1}$ & -1.72$\pm_{0.03}^{1.47}$ & 0.4$\pm_{0.01}^{0.06}$ &  \\ 

NGC7213                & F06    &   6.66 /   9.02 & 418    / 39     & 1.02$\pm_{0.03}^{0.01}$ & $<$0.0 & $<$20.3 & 4.03$\pm_{0.05}^{0.35}$ & $>$-0.0 & $>$149.9 & $<$0.1 &  &  &  \\ 
                       & N08    &   1.05 /   3.09 & 418    / 39     & 0.80$\pm_{0.03}^{0.05}$ & $<$44.2 & 2.9$\pm_{0.8}^{1.2}$ & $<$16.1 & $>$83.0 & 1.07$\pm_{0.06}^{0.03}$ & $>$14.2 &  &  &  \\ 
                       & H10    &   1.23 /   2.85 & 419    / 40     & 0.69$\pm_{0.02}^{0.02}$ & $<$0.0 & $<$2.5 & $>$59.9 & $>$-0.5 & $<$30.1 &  &  &  &  \\ 
                       & S16    &   8.17 /  10.96 & 418    / 39     & 0.05$\pm{0.02}$  & 30.0$\pm_{0.2}^{1.4}$ & $>$79.9 & $<$0.0 & $>$1.5 & $>$30.0 & $<$3.0 &  &  &  \\ 
                       & H17    &   1.65 /   2.18 & 416    / 37     & 0.46$\pm_{0.01}^{0.02}$ & 30.0$\pm_{0.3}^{0.1}$ & $<$5.0 & $>$-1.7 & $>$7.3 & $<$30.0 & -1.0$\pm{0.1}$ & $<$0.1 & 0.45$\pm{0.01}$ &  \\ 
                       & H17D$\bullet$   &   1.00 /   1.61 & 420    / 41     & 0.51$\pm{0.02}$ & $<$0.0 & $<$5.0 & -1.67$\pm_{0.01}^{1.42}$ & 0.25$\pm{0.01}$ &  \\ 

UGC6728                & F06    &   2.72 /   2.54 & 77     / 42     & 0.0* & $>$59.6 & $>$20.0 & $>$0.1 & -0.74$\pm_{0.01}^{0.06}$ & $<$10.2 & 1.52$\pm_{0.03}^{0.02}$ &  &  &  \\ 
                       & N08    &  12.11 /  12.13 & 77     / 42     & 0.0* & $>$89.7 & $>$14.0 & 34.7$\pm_{1.7}^{0.6}$ & $<$5.0 & $>$2.5 & $<$10.0 &  &  &  \\ 
                       & H10$\bullet$    &   1.35 /   1.40 & 78     / 43     & 0.11$\pm{0.01}$ & 36.0$\pm_{3.7}^{6.2}$ & $>$7.4 & $<$7.5 & $<$-1.9 & $>$79.3 &  &  &  &  \\ 
                       & S16    &   5.10 /   5.33 & 77     / 42     & 0.0* & 66.5$\pm_{0.9}^{2.3}$ & 59.6$\pm_{1.8}^{0.7}$ & $>$1.5 & $>$1.5 & $<$10.0 & $>$4.0 &  &  &  \\ 
                       & H17$\bullet$    &   1.05 /   0.87 & 75     / 40     & 0.23$\pm_{0.01}^{0.02}$ & $<$3.2 & $>$7.8 & $<$-3.0 & $>$9.8 & 36.1$\pm_{1.2}^{2.0}$ & -1.6$\pm_{0.1}^{0.2}$ & $<$0.1 & $>$0.6 &  \\
                       & H17D$\bullet$   &   1.19 /   1.17 & 79     / 44     & 0.01$\pm_{0.01}^{0.02}$ & 44.9$\pm_{2.2}^{0.6}$ & $>$9.8 & $<$-2.5 & $>$0.7 &  \\ 

\hline \hline
\end{tabular}
\tablefoot{Best-fit results per object and model. Models are quoted in Col.\,2 as follows. F06: [Fritz06]; N08: [Nenkova08]; H10: [Hoenig10]; S16: [Stalev16]; and H17: [Hoenig17]. The reduced $\rm{\chi^2}$ ($\rm{\chi^2/dof}$) is included in Col.\,3, color excess for the foreground extinction $\rm{E(B-V)}$ is included in Col.\,4, and the final parameters per model are included in Cols.\,5-13. Comparably good fits ($\rm{\chi^2/dof< min(\chi^2/dof)+0.5}$) are marked with filled circles next to the model name. }
\label{appendixfitSy1}
\end{center}
\end{table*}

\begin{table*}
\scriptsize
\begin{center}
\caption{Spectral fit results for Type 1.8/1.9 Seyferts.}
\begin{tabular}{llcccccccccccc}
\hline \hline
Obj.	& Mod. &  $\chi^{2}_{red}$ & dof & E(B-V)  &\multicolumn{8}{c}{Parameters}       \\
	    &     &                   &   &   &   &     &     &     &     &      &  &   \\ \hline
				&	F06&		&		&		& $i$ & $\sigma$ & $\Gamma$ & $\beta$ & $Y$ & $\tau_{9.7\mu m}$ &  & \\
				&	N08&		&		&		& $i$ & $N_{0}$  & $\sigma$ & $Y$ & $q$ &$\tau_{v}$ &  & \\
				&	H10&		&		&		& $i$ & $N_{0}$  & $\theta$ & $a$ & $\tau_{cl}$ &  &  &  \\
				&	S16&		&		&		& $i$ & $\sigma$ & $p$ & $q$ & $Y$ & $\tau_{9.7\mu m}$ &  & \\
				&	H17&		&		&		& $i$ & $N_{0}$ & $a$  & $\sigma$ & $\theta$ & $a_{w}$ & $h$ & $f_{w}$ \\
				&	H17D&		&		&		& $i$ & $N_{0}$ & $a$  & $h$ \\\hline \hline
				&		&			&	&		&							&		&	& & & & &	\\
NGC1365                & F06    &   1.13 /   3.43 & 197    / 38     & $<$0.06 & $<$33.3 & $<$22.4 & $>$5.6 & $>$-0.1 & 20.5$\pm_{0.9}^{1.4}$ & $>$9.0 &  &  &  \\ 
                       & N08    &   1.06 /   3.07 & 197    / 38     & 0.04$\pm_{0.03}^{0.05}$ & 40.1$\pm_{8.5}^{9.2}$ & $>$12.3 & 26.1$\pm_{1.5}^{1.9}$ & 5.9$\pm_{0.5}^{0.7}$ & $<$0.7 & 40.0$\pm_{4.1}^{5.3}$ &  &  &  \\ 
                       & H10    &   1.05 /   2.77 & 198    / 39     & 0.07$\pm_{0.04}^{0.03}$ & $<$21.3 & 7.5$\pm_{1.6}^{0.4}$ & $<$45.6 & $>$-0.2 & $>$77.3 &  &  &  &  \\ 
                       & S16    &   1.12 /   3.58 & 197    / 38     & 0* & 24.6$\pm_{6.9}^{23.0}$ & $>$77.9 & $<$0.0 & $>$1.0 & $>$17.2 & 5.0$\pm_{1.4}^{0.4}$ &  &  &  \\ 
                       & H17$\bullet$    &   0.70 /   0.98 & 195    / 36     & 0* & 52.9$\pm_{2.6}^{7.6}$ & $<$6.8 & $<$-3.0 & $>$12.6 & $>$42.7 & -1.58$\pm_{0.05}^{0.28}$ & $<$0.1 & $>$0.6 &  \\ 
                       & H17D   &   0.92 /   2.15 & 199    / 40     & $<$0.05 & 30.2$\pm_{2.0}^{1.3}$ & $<$5.1 & -1.36$\pm_{0.02}^{1.11}$ & $>$0.7 &  \\ 

NGC2992                & F06    &   2.18 /   3.96 & 233    / 38     & 1.0$\pm_{0.03}^{0.02}$ & $<$69.4 & $<$20.1 & $>$6.0 & $>$-0.0 & $>$51.8 & $>$10.0 &  &  &  \\ 
                       & N08    &   1.25 /   2.36 & 233    / 38     & 0.79$\pm{0.03}$ & $<$62.9 & $>$13.6 & 17.3$\pm_{0.6}^{1.4}$ & 14.7$\pm_{3.8}^{5.0}$ & 1.5$\pm_{0.3}^{0.2}$ & $<$162.5 &  &  &  \\ 
                       & H10    &   1.42 /   2.76 & 234    / 39     & 0.41$\pm{0.03}$ & $>$44.3 & $>$2.8 & $>$5.1 & $>$-0.1 & $>$79.7 &  &  &  &  \\ 
                       & S16    &   1.39 /   4.16 & 233    / 38     & 0* & $>$32.9 & 65.9$\pm_{1.7}^{2.4}$ & $<$0.0 & $<$0.0 & $<$14.7 & $>$10.7 &  &  &  \\ 
                       & H17$\bullet$     &   0.80 /   1.02 & 231    / 36     & 0.25$\pm_{0.04}^{0.06}$ & $>$50.1 & $>$9.1 & $>$-2.2 & 10.0$\pm_{0.3}^{1.1}$ & $>$44.5 & $>$-0.6 & 0.24$\pm{0.02}$ & $>$0.7 &  \\ 
                       & H17D   &   1.65 /   2.76 & 235    / 40     & 0.46$\pm_{0.02}^{0.05}$ & $>$70.5 & $>$5.0 & -1.28$\pm_{0.01}^{1.03}$ & $>$0.3 &  \\ 

NGC4138                & F06    &   4.46 /   5.18 & 82     / 46     & 0* & 16.0$\pm_{1.7}^{0.7}$ & $<$20.3 & 5.8$\pm{0.1}$ & -0.25$\pm_{0.03}^{0.05}$ & $<$10.0 & 6.0$\pm_{0.3}^{0.1}$ &  &  &  \\ 
                       & N08    &   1.82 /   2.36 & 82     / 46     & 0.10$\pm_{0.01}^{0.02}$ & $>$88.8 & 2.99$\pm_{0.29}^{0.04}$ & 24.9$\pm_{6.7}^{11.6}$ & $<$5.1 & $>$2.5 & 59.0$\pm_{1.8}^{1.4}$ &  &  &  \\ 
                       & H10$\bullet$    &   1.24 /   1.62 & 83     / 47     & 0.12$\pm{0.02}$ & $>$43.8 & 7.6$\pm_{2.3}^{1.5}$ & $>$49.5 & -1.16$\pm_{0.09}^{0.04}$ & $>$78.7 &  &  &  &  \\ 
                       & S16    &   4.36 /   5.03 & 82     / 46     & 0* & $>$79.7 & $>$79.5 & $<$0.0 & $>$1.5 & $<$10.1 & $<$3.1 &  &  &  \\ 
                       & H17$\bullet$    &   0.98 /   1.05 & 80     / 44     & 0.16$\pm_{0.01}^{0.02}$ & 45.0$\pm_{4.1}^{2.5}$ & $>$5.8 & $>$-2.7 & $>$13.1 & $>$43.3 & $<$-1.6 & $<$0.1 & 0.31$\pm{0.04}$ &  \\ 
                       & H17D$\bullet$   &   1.05 /   1.31 & 84     / 48     & 0.06$\pm_{0.01}^{0.02}$ & 65.8$\pm_{7.5}^{3.4}$ & 6.9$\pm_{0.8}^{0.4}$ & -2.02$\pm_{0.06}^{0.04}$ & 0.27$\pm_{0.02}^{0.08}$ &  \\ 

NGC4395*                & F06    &   1.72 /   4.95 & 70     / 37     & 0* & $<$0.1 & $<$20.9 & $>$5.9 & $>$-0.0 & 31.3$\pm_{2.4}^{5.3}$ & 5.6$\pm{0.3}$ &  &  &  \\ 
                       & N08    &   1.59 /   4.24 & 70     / 37     & 0.23$\pm{0.08}$ & 63.8$\pm{12.3}$ & $>$11.2 & 20.1$\pm_{2.5}^{3.6}$ & 19.8$\pm_{5.9}^{3.3}$ & $<$0.8 & $<$32.9 &  &  &  \\ 
                       & H10    &   1.61 /   4.31 & 71     / 38     & 0.21$\pm_{0.10}^{0.07}$ & 45.2$\pm_{5.6}^{7.6}$ & 5.0$\pm_{1.2}^{1.0}$ & 42.0$\pm_{8.2}^{4.3}$ & $>$-0.0 & $>$70.0 &  &  &  &  \\ 
                       & S16    &   1.86 /   4.80 & 70     / 37     & 0* & $>$81.5 & $>$73.5 & $<$0.0 & $>$1.4 & $>$28.3 & 3.7$\pm{0.1}$ &  &  &  \\ 
                       & H17    &   0.75 /   3.02 & 68     / 35     & 0.23$\pm_{0.03}^{0.05}$ & 35.9$\pm_{2.2}^{0.6}$ & 7.0$\pm_{0.5}^{0.4}$ & $<$-3.0 & 10.0$\pm_{0.2}^{0.8}$ & $>$44.3 & $>$-0.5 & $<$0.1 & $>$0.7 &  \\ 
                       & H17D   &   1.62 /   4.22 & 72     / 39     & 0.23$\pm_{0.10}^{0.09}$ & 35.5$\pm_{10.0}^{7.70}$ & $<$8.3 & -0.7$\pm_{0.2}^{0.4}$ & $>$0.6 &  \\ 

NGC5506                & F06    &   1.14 /   2.63 & 201    / 24     & 2.31$\pm_{0.03}^{0.05}$ & $<$61.3 & $>$56.4 & $<$5.3 & -0.75$\pm_{0.01}^{0.02}$ & $<$11.3 & 0.86$\pm_{0.06}^{0.07}$ &  &  &  \\ 
                       & N08    &   3.10 /   7.38 & 201    / 24     & 2.05$\pm_{0.03}^{0.04}$ & 59.9$\pm_{4.9}^{7.2}$ & $<$1.0 & $<$15.6 & $<$5.0 & $>$2.5 & $>$296.0 &  &  &  \\ 
                       & H10    &   1.51 /   4.22 & 202    / 25     &2.16$\pm_{0.03}^{0.05}$ & 60.2$\pm_{1.0}^{1.5}$ & 5.0$\pm_{0.2}^{0.4}$ & $>$5.0 & $<$-2.0 & $>$79.2 &  &  &  &  \\ 
                       & S16    &   1.20 /   3.28 & 201    / 24     & 2.60$\pm_{0.04}^{0.03}$ & $>$64.2 & $<$10.1 & $>$1.5 & $<$0.0 & $<$10.1 & $>$11.0 &  &  &  \\ 
                       & H17$\bullet$    &   0.85 /   1.79 & 199    / 22     & 1.90$\pm{0.04}$ & 75.2$\pm_{0.9}^{1.6}$ & $<$5.2 & $<$-3.0 & 10.0$\pm_{0.5}^{0.9}$ & $>$43.1 & -2.06$\pm_{0.06}^{0.05}$ & $<$0.1 & $>$0.7 &  \\ 
                       & H17D   &   1.62 /   2.63 & 203    / 26     & 2.08$\pm_{0.03}^{0.04}$ & 60.3$\pm_{0.7}^{1.1}$ & $>$9.7 & $<$-2.5 & $>$0.3 &  \\ 

NGC7314                & F06    &   2.89 /   6.70 & 73     / 28     & 0.10$\pm_{0.03}^{0.04}$ & $>$63.6 & $<$20.7 & $<$0.0 & $>$-0.0 & $<$10.1 & $>$5.6 &  &  &  \\ 
                       & N08    &   2.29 /   5.29 & 73     / 28     & 0.39$\pm_{0.07}^{0.11}$ & $>$31.3 & $>$12.0 & 38.7$\pm_{5.4}^{11.4}$ & 10.0$\pm_{3.2}^{0.7}$ & $>$0.0 & 40.0$\pm_{6.3}^{8.5}$ &  &  &  \\ 
                       & H10    &   2.19 /   5.09 & 74     / 29     & 0.44$\pm_{0.06}^{0.07}$ & 59.8$_{4.9}^{1.8}$ & 7.3$\pm_{0.4}^{0.3}$ & 45.6$\pm_{1.9}^{1.5}$ & -0.5$\pm{0.1}$ & $>$71.3 &  &  &  &  \\ 
                       & S16    &   2.24 /   5.32 & 73     / 28     & 0.49$\pm_{0.05}^{0.03}$ & $<$0.3 & 78.4$\pm_{0.6}^{0.9}$ & $<$0.0 & $<$0.0 & $<$10.4 & 9.0$\pm_{0.4}^{0.6}$ &  &  &  \\ 
                       & H17$\bullet$    &   1.14 /   1.54 & 71     / 26     & 0.67$\pm_{0.02}^{0.03}$ & 60.2$_{0.8}^{1.2}$ & $>$6.5 & $<$-3.0 & $>$14.6 & $>$44.8 & $>$-0.5 & 0.22$\pm_{0.03}^{0.04}$ & $>$0.7 &  \\ 
                       & H17D   &   2.00 /   4.34 & 75     / 30     & 0.39$\pm_{0.06}^{0.05}$ & $>$79.7 & $>$5.0 & -1.49$\pm_{0.02}^{0.03}$ & $>$0.3 &  \\ 

\hline \hline
\end{tabular}
\tablefoot{Best-fit results per object and model. Models are quoted in Col.\,2 as follows. F06: [Fritz06]; N08: [Nenkova08]; H10: [Hoenig10]; S16: [Stalev16]; and H17: [Hoenig17]. The reduced $\rm{\chi^2}$ ($\rm{\chi^2/dof}$) is included in Col.\,3, color excess for the foreground extinction $\rm{E(B-V)}$ is included in Col.\,4, and the final parameters per model are included in Cols.\,5-13. Comparably good fits ($\rm{\chi^2/dof< min(\chi^2/dof)+0.5}$) are marked with filled circles next to the model name. }
\label{appendixfitSyInt}
\end{center}
\end{table*}

\begin{table*}
\scriptsize
\begin{center}
\caption{Spectral fit results for Type-2 Seyferts.}
\begin{tabular}{llcccccccccccc}
\hline \hline
Obj.	& Mod. &  $\chi^{2}_{red}$ & dof & E(B-V)  &\multicolumn{8}{c}{Parameters}       \\
	    &     &                   &   &   &   &     &     &     &     &      &  &   \\ \hline
				&	F06&		&		&		& $i$ & $\sigma$ & $\Gamma$ & $\beta$ & $Y$ & $\tau_{9.7\mu m}$ &  & \\
				&	N08&		&		&		& $i$ & $N_{0}$  & $\sigma$ & $Y$ & $q$ &$\tau_{v}$ &  & \\
				&	H10&		&		&		& $i$ & $N_{0}$  & $\theta$ & $a$ & $\tau_{cl}$ &  &  &  \\
				&	S16&		&		&		& $i$ & $\sigma$ & $p$ & $q$ & $Y$ & $\tau_{9.7\mu m}$ &  & \\
				&	H17&		&		&		& $i$ & $N_{0}$ & $a$  & $\sigma$ & $\theta$ & $a_{w}$ & $h$ & $f_{w}$ \\
				&	H17D&		&		&		& $i$ & $N_{0}$ & $a$  & $h$\\\hline \hline
				&		&			&	&		&							&		&	& & & & &	\\
ESO005-G004*            & F06    &   2.27 /   2.89 & 84     / 36     & 0.86$\pm_{0.06}^{0.07}$ & 61.4$\pm_{0.5}^{0.4}$ & $<$20.6 & $<$0.0 & $>$-0.0 & $<$10.1 & $>$9.5 &  &  &  \\ 
                       & N08    &   2.24 /   2.74 & 84     / 36     & 0.99$\pm_{0.14}^{0.20}$ & $>$67.4 & $>$13.2 & $>$45.9 & 10.0$_{0.5}^{0.2}$ & $<$1.0 & 22.8$_{1.0}^{7.2}$ &  &  &  \\ 
                       & H10    &   2.07 /   2.62 & 85     / 37     & 0.85$\pm_{0.05}^{0.14}$ & $>$78.0 & $>$9.3 & $<$41.7 & $>$-0.0 & $<$36.5 &  &  &  &  \\ 
                       & S16    &   2.23 /   2.53 & 84     / 36     & 1.12$\pm_{0.07}^{0.05}$ & 70.0$\pm_{2.9}^{1.3}$ & $>$78.4 & $<$0.0 & $<$0.0 & $>$11.5 & $>$10.8 &  &  &  \\ 
                       & H17    &   2.20 /   2.60 & 82     / 34     & 1.42$\pm_{0.06}^{0.07}$ & 85.0$\pm_{1.5}^{2.0}$ & $>$8.5 & $>$-0.6 & $>$13.9 & $>$43.9 & -1.5$\pm_{0.2}^{0.1}$ & $>$0.1 & $>$0.7 &  \\ 
                       & H17D   &   2.17 /   2.64 & 89     / 38     & 0.98$\pm_{0.13}^{0.18}$ & $>$77.9 & $<$5.6 & -0.9$\pm{0.1}$ & 0.4$\pm_{0.1}^{0.2}$ &  \\ 

MCG-05-23-016          & F06    &   0.96 /   1.39 & 421    / 32     & 0.47$\pm{0.02}$ & 34.2$\pm{1.5}$ & $>$27.3 & $<$5.3 & $>$-0.7 & $>$26.0 & 7.4$\pm{0.4}$ &  &  &  \\ 
                       & N08$\bullet$   &   0.66 /   1.61 & 421    / 32     & 0.50$\pm{0.01}$ & $<$12.3 & $>$14.9 & $>$37.0 & $<$5.1 & $<$0.0 & 39.7$\pm_{1.0}^{0.5}$ &  &  &  \\ 
                       & H10$\bullet$   &   0.44 /   0.26 & 422    / 33     & 0.42$\pm{0.03}$ & 19.7$\pm_{4.3}^{2.2}$ & $>$9.3 & 37.3$\pm_{1.7}^{1.1}$ & -0.7$\pm{0.1}$ & 75.1$\pm_{2.7}^{2.6}$ &  &  &  &  \\ 
                       & S16$\bullet$     &   0.40 /   0.97 & 421    / 32     & 0.61$\pm_{0.02}^{0.01}$ & 54.9$\pm_{1.0}^{0.3}$ & 39.9$\pm_{0.9}^{0.7}$ & 1.02$\pm_{0.05}^{0.11}$ & $<$0.0 & $<$12.0 & $<$10.2 &  &  &  \\ 
                       & H17$\bullet$    &   0.60 /   0.42 & 419    / 30     & 0.27$\pm{0.01}$ & 81.1$\pm_{1.9}^{0.8}$ & 7.01$\pm_{0.04}^{0.20}$ & -2.00$\pm_{0.01}^{0.03}$ & $<$7.1 & $>$44.6 & -1.89$\pm_{0.05}^{0.02}$ & $>$0.2 & $>$0.7 &  \\ 
                       & H17D$\bullet$   &   0.36 /   0.31 & 423    / 34     & 0.43$\pm_{0.03}^{0.02}$ & 29.7$\pm_{3.0}^{0.7}$ & 7.7$\pm_{0.2}^{0.7}$ & -1.36$\pm_{0.04}^{0.05}$ & $>$0.7 &  \\ 

NGC2110                & F06    &   1.82 /   4.36 & 447    / 39     & 0.0* & $<$0.0 & $<$20.3 & $<$5.2 & -0.75$\pm{0.01}$ & 30.0$\pm_{0.01}^{0.70}$ & 1.17$\pm_{0.02}^{0.01}$ &  &  &  \\ 
                       & N08$\bullet$    &   0.59 /   1.85 & 447    / 39     & 0.78$\pm_{0.02}^{0.03}$ & 72.1$\pm_{7.3}^{3.5}$ & $>$1.1 & $<$17.7 & $<$5.6 & $>$2.5 & 73.9$\pm_{7.8}^{13.9}$ &  &  &  \\ 
                       & H10$\bullet$    &   0.42 /   1.30 & 448    / 40     & 0.72$\pm{0.03}$ & 48.5$\pm_{6.7}^{8.2}$ & 3.3$\pm_{0.05}^{0.06}$ & $>$57.8 & -1.10$\pm{0.03}$ & $>$78.0 &  &  &  &  \\ 
                       & S16    &   1.24 /   4.67 & 447    / 39     & 0.0* & $>$89.6 & $>$79.7 & 0.8$\pm{0.1}$ & $>$1.4 & 20.0$\pm_{0.1}^{0.5}$ & $<$3.0 &  &  &  \\ 
                       & H17$\bullet$    &   0.37 /   0.90 & 445    / 37     & 0.64$\pm_{0.01}^{0.02}$ & 75.4$\pm_{0.5}^{2.9}$ & $<$5.3 & -2.00$\pm_{0.01}^{0.02}$ & $<$8.0 & $<$35.9 & -2.00$\pm_{0.03}^{-0.06}$ & $<$0.1 & 0.30$\pm_{0.05}^{0.06}$ &  \\ 
                       & H17D$\bullet$   &   0.34 /   0.96 & 449    / 41     & 0.67$\pm_{0.03}^{0.02}$ & 46.2$\pm_{5.9}^{5.1}$ & $<$5.6 & -2.03$\pm_{0.01}^{0.03}$ & 0.32$\pm_{0.02}^{0.03}$ &  \\

NGC3081                & F06$\bullet$    &   0.92 /   1.43 & 186    / 19     & 0.0* & 24.2$\pm{0.04}$ & $<$20.3 & $>$6.0 & $<$-0.5 & $>$27.8 & $>$9.8 &  &  &  \\ 
                       & N08$\bullet$    &   0.98 /   0.73 & 186    / 19     & 0.26$\pm_{0.04}^{0.05}$ & $<$12.0 & $>$10.5 & 44.5$\pm_{3.8}^{4.0}$ & $<$5.2 & 1.1$\pm_{0.4}^{0.5}$ & $>$71.4 &  &  &  \\ 
                       & H10$\bullet$    &   0.92 /   0.66 & 187    / 20     & 0.30$\pm_{0.03}^{0.04}$ & 63.0$\pm_{4.2}^{3.5}$ & $>$9.6 & $>$51.4 & $<$-1.1 & $>$79.2 &  &  &  &  \\ 
                       & S16$\bullet$    &   0.67 /   0.60 & 186    / 19     & 0.0* & 20.2$\pm_{12.7}^{6.9}$ & 75.3$\pm_{3.6}^{1.0}$ & 0.61$\pm_{0.04}^{0.05}$ & $<$0.0 & $<$10.4 & 6.2$\pm_{0.4}^{0.9}$ &  &  &  \\ 
                       & H17$\bullet$    &   0.91 /   0.75 & 184    / 17     & 0.29$\pm_{0.03}^{0.02}$ & 59.8$\pm_{0.6}^{1.9}$ & $>$9.2 & -1.35$\pm_{0.05}^{0.04}$ & $>$12.5 & $<$31.8 & -1.5$\pm{0.1}$ & $<$0.3 & $<$0.2 &  \\ 
                       & H17D$\bullet$   &   0.77 /   0.66 & 189    / 21     & 0.17$\pm_{0.03}^{0.01}$ & 71.5$\pm_{0.07}^{0.03}$ & 7.04$\pm_{0.07}^{0.16}$ & -1.64$\pm_{0.03}^{0.01}$ & 0.25$\pm_{0.01}^{0.02}$ &  \\ 

NGC4388                & F06    &   2.09 /   3.52 & 211    / 31     & 1.25$\pm_{0.07}^{0.08}$ & 49.7$\pm_{1.1}^{0.7}$ & $>$27.4 & $<$0.0 & $>$-0.0 & $>$12.0 & 6.0$\pm_{0.1}^{0.8}$ &  &  &  \\ 
                       & N08    &   2.24 /   4.65 & 211    / 31     & 1.73$\pm_{0.05}^{0.04}$ & $>$0.0 & $>$14.9 & $>$48.4 & 19.9$\pm_{0.5}^{0.4}$ & $>$0.0 & $<$71.7 &  &  &  \\ 
                       & H10$\bullet$    &   1.36 /   1.99 & 212    / 32     & 1.14$\pm_{0.05}^{0.06}$ & 30.2$\pm_{0.8}^{1.5}$ & $>$7.5 & $<$5.1 & $>$-0.4 & $>$79.6 &  &  &  &  \\ 
                       & S16    &   2.01 /   3.94 & 211    / 31     & 1.27$\pm{0.10}$ & 76.2$\pm_{0.9}^{0.7}$ & 24.9$\pm_{0.6}^{0.7}$ & $>$1.5 & $<$0.0 & 18.2$\pm{1.4}$ & $>$10.9 &  &  &  \\ 
                       & H17    &   1.64 /   2.27 & 209    / 29     & 1.24$\pm_{0.07}^{0.09}$ & 60.4$\pm_{0.5}^{0.4}$ & $>$9.9 & -1.50$\pm_{0.02}^{0.01}$ & $<$7.0 & $>$44.3 & -0.97$\pm_{0.02}^{0.01}$ & $>$0.4 & $>$0.7 &  \\ 
                       & H17D   &   1.90 /   2.42 & 213    / 33     & 1.69$\pm_{0.06}^{0.05}$ & 49.8$\pm_{0.9}^{1.0}$ & $>$9.9 & -1.5$\pm{0.01}$ & $>$0.6 &  \\ 

NGC4945                & F06$\bullet$    &   1.15 /   1.32 & 76     / 45     & 1.43$\pm_{0.15}^{0.14}$ & 49.4$\pm_{4.2}^{1.5}$ & 25.1$\pm_{0.9}^{2.8}$ & $<$0.0 & $<$-1.0 & $>$41.6 & 6.0$\pm_{0.1}^{0.2}$ &  &  &  \\ 
                       & N08$\bullet$    &   0.79 /   0.97 & 76     / 45     & 1.02$\pm_{0.14}^{0.16}$ & $<$0.3 & $>$14.3 & 64.9$\pm_{1.5}^{2.5}$ & 20.5$\pm_{1.9}^{3.2}$ & $<$0.1 & 42.2$\pm_{2.8}^{5.6}$ &  &  &  \\ 
                       & H10$\bullet$    &   1.23 /   1.22 & 77     / 46     & 1.64$\pm{0.06}$ & 14.5$\pm_{12.7}^{0.8}$ & $>$9.9 & 13.5$\pm_{3.2}^{3.5}$ & -0.50$\pm_{0.01}^{0.06}$ & $>$79.6 &  &  &  &  \\ 
                       & S16$\bullet$    &   0.96 /   1.07 & 76     / 45     & 1.49$\pm_{0.17}^{0.16}$ & 77.9$\pm_{0.6}^{0.5}$ & 22.9$\pm_{0.7}^{1.8}$ & $>$1.4 & $<$0.0 & $>$27.9 & $>$10.7 &  &  &  \\ 
                       & H17    &   1.90 /   1.98 & 74     / 43     & 1.58$\pm_{0.06}^{0.05}$ & $>$89.8 & 9.4$\pm{0.4}$ & -2.00$\pm{0.01}$ & $>$14.5 & $>$43.8 & -2.00$\pm_{0.04}^{0.02}$ & $>$0.5 & 0.60$\pm{0.01}$ &  \\ 
                       & H17D   &   1.87 /   1.75 & 78     / 47     & 2.06$\pm_{0.08}^{0.01}$ & 36.6$\pm_{1.0}^{1.3}$ & $>$9.9 & -1.50$\pm{0.01}$ & $>$0.7 &  \\ 

NGC5128                & F06$\bullet$    &   1.40 /   1.96 & 184    / 24     & 2.16$\pm{0.09}$ & 31.5$\pm_{3.3}^{8.9}$ & $<$27.0 & $>$5.7 & -0.64$\pm{0.04}$ & $>$11.5 & 5.0$\pm_{0.5}^{0.7}$ &  &  &  \\ 
                       & N08    &   1.74 /   2.32 & 184    / 24     & 2.05$\pm_{0.04}^{0.02}$ & $<$60.4 & $>$14.6 & $>$21.6 & $<$5.0 & $>$2.4 & $<$35.6 &  &  &  \\ 
                       & H10$\bullet$    &   1.00 /   1.62 & 185    / 25     & 1.71$\pm_{0.02}^{0.04}$ & $<$0.5 & $>$9.9 & $<$36.4 & -0.82$\pm_{0.10}^{0.07}$ & $>$79.0 &  &  &  &  \\ 
                       & S16$\bullet$    &   1.06 /   1.88 & 184    / 24     & 2.14$\pm{0.04}$ & $>$23.1 & 40.8$\pm_{0.6}^{0.8}$ & $>$0.0 & $<$0.0 & $<$10.3 & $>$10.9 &  &  &  \\ 
                       & H17$\bullet$    &   1.19 /   1.70 & 182    / 22     & 1.89$\pm{0.03}$ & 85.4$\pm_{2.9}^{1.6}$ & $>$8.2 & $<$-2.9 & $>$12.9 & $>$43.0 & -1.78$\pm{0.05}$ & $<$0.1 & $>$0.7 &  \\ 
                       & H17D$\bullet$   &   1.09 /   1.80 & 186    / 26     & 1.87$\pm_{0.03}^{0.02}$ & $<$0.0 & $>$9.9 & -1.07$\pm{0.02}$ & $>$0.7 &  \\ 

NGC6300*                & F06    &   5.65 /   6.97 & 81     / 47     & 1.69$\pm_{0.05}^{0.08}$ & $>$87.3 & $>$20.0 & $<$0.0 & $<$-1.0 & $>$147.2 & $>$9.9 &  &  &  \\ 
                       & N08    &   2.35 /   3.65 & 81     / 47     & 0.86$\pm_{0.14}^{0.09}$ & $<$0.0 & $>$14.6 & $>$68.3 & 46.8$\pm_{6.3}^{4.8}$ & $<$0.2 & 40.1$\pm_{0.5}^{0.9}$ &  &  &  \\ 
                       & H10    &   4.18 /   4.53 & 82     / 48     & 1.77$\pm_{0.09}^{0.03}$ & 14.9$\pm_{5.1}^{0.2}$ & $>$10.0 & $>$22.0 & -0.50$\pm_{0.01}^{0.02}$ & $>$79.7 &  &  &  &  \\ 
                       & S16    &   6.55 /   7.70 & 81     / 47     & 1.72$\pm_{0.09}^{0.12}$ & 87.2$\pm_{1.3}^{0.9}$ & $<$13.2 & $>$0.0 & $<$0.0 & $<$10.2 & $<$10.2 &  &  &  \\ 
                       & H17    &   3.46 /   4.47 & 79     / 45     & 1.79$\pm_{0.07}^{0.05}$ & $>$90.0 & $<$5.0 & -1.50$\pm_{0.02}^{0.01}$ & $>$14.9 & $>$44.8 & $<$-2.5 & $<$0.1 & 0.45$\pm{0.01}$ &  \\ 
                       & H17D   &   4.80 /   5.10 & 83     / 49     & 2.06$\pm_{0.06}^{0.05}$ & $<$40.0 & $>$9.9 & -1.74$\pm_{0.02}^{0.03}$ & $>$0.7 &  \\ 

NGC7172*                & F06    &   2.51 /   3.50 & 74     / 34     & 2.76$\pm_{0.09}^{0.08}$ & $>$44.7 & $>$58.4 & $<$0.0 & -0.75$\pm{0.01}$ & $<$10.2 & 0.31$\pm_{0.02}^{0.06}$ &  &  &  \\ 
                       & N08    &   5.48 /   8.03 & 74     / 34     & 1.60$\pm_{0.05}^{0.06}$ & $>$53.6 & $>$14.6 & $>$66.0 & $<$5.0 & $>$2.5 & $<$10.1 &  &  &  \\ 
                       & H10    &   3.47 /   5.01 & 74     / 35     & 2.35$\pm{0.06}$ & 59.7$\pm{1.0}$ & 5.0$\pm_{0.5}^{0.6}$ & $>$57.9 & $<$-2.0 & $>$76.6 &  &  &  &  \\ 
                       & S16    &   4.65 /   7.20 & 74     / 34     & 2.58$\pm{0.07}$ & 29.8$\pm_{9.3}^{2.2}$ & 59.9$\pm_{3.9}^{1.2}$ & $>$1.5 & $>$1.4 & $<$10.2 & $<$3.3 &  &  &  \\ 
                       & H17    &   1.86 /   2.56 & 71     / 32     & 2.18$\pm{0.06}$ & $<$0.4 & $<$5.2 & $<$-3.0 & $>$14.3 & $>$44.3 & -2.33$\pm_{0.05}^{0.06}$ & $<$0.1 & $>$0.7 &  \\  
                       & H17D   &   3.48 /   5.15 & 75     / 36     & 1.49$\pm{0.05}$ & $<$0.0 & $>$9.9 & -1.75$\pm_{0.03}^{0.01}$ & $>$0.7 &  \\ 

NGC7582                & F06$\bullet$    &   0.41 /   0.64 & 196    / 45     & 2.42$\pm_{0.20}^{0.19}$ & 23.9$\pm_{0.8}^{6.8}$ & $>$51.1 & $<$0.0 & $<$-1.0 & $>$147.5 & 4.0$\pm_{1.1}^{2.3}$ &  &  &  \\ 
                       & N08    &   1.07 /   1.29 & 196    / 45     & 1.45$\pm_{0.06}^{0.04}$ & $<$0.2 & $>$14.8 & 55.1$\pm_{0.7}^{1.4}$ & $<$5.0 & 0.5$\pm_{0.3}^{0.1}$ & 20.0$\pm_{0.4}^{0.1}$ &  &  &  \\ 
                       & H10$\bullet$    &   1.08 /   1.11 & 197    / 46     & 1.55$\pm_{0.55}^{0.06}$ & $<$0.0 & $>$9.9 & 30.0$\pm_{1.2}^{0.2}$ & $<$-0.9 & $<$69.5 &  &  &  &  \\ 
                       & S16$\bullet$    &   0.56 /   0.76 & 196    / 45     & 2.52$\pm_{0.09}^{0.04}$ & $>$84.2 & $<$10.1 & $>$1.2 & 1.0$\pm_{0.1}^{0.4}$ & $>$29.0 & 9.0$\pm_{0.3}^{0.4}$ &  &  &  \\ 
                       & H17$\bullet$    &   0.61 /   0.95 & 194    / 43     & 1.50$\pm_{0.05}^{0.06}$ & $>$90.0 & $<$5.0 & -2.00$\pm_{0.01}^{0.02}$ & 10.0$\pm_{0.1}^{0.2}$ & $>$44.7 & $<$-2.5 & $<$0.1 & 0.6$\pm{0.6}$ &  \\ 
                       & H17D$\bullet$   &   0.72 /   1.00 & 198    / 47     & 1.76$\pm{0.04}$ & $<$0.0 & $>$9.9 & -1.75$\pm_{0.02}^{0.01}$ & $>$0.7 &  \\ 

\hline \hline
\end{tabular}
\tablefoot{Best-fit results per object and model. Models are quoted in Col.\,2 as follows. F06: [Fritz06]; N08: [Nenkova08]; H10: [Hoenig10]; S16: [Stalev16]; and H17: [Hoenig17]. The reduced $\rm{\chi^2}$ ($\rm{\chi^2/dof}$) is included in Col.\,3, color excess for the foreground extinction $\rm{E(B-V)}$ is included in Col.\,4, and the final parameters per model are included in Cols.\,5-13. Comparably good fits ($\rm{\chi^2/dof< min(\chi^2/dof)+0.5}$) are marked with filled circles next to the model name. }
\label{appendixfitSy2}
\end{center}
\end{table*}

\begin{figure*}
\centering
    \includegraphics[width=0.75\columnwidth]{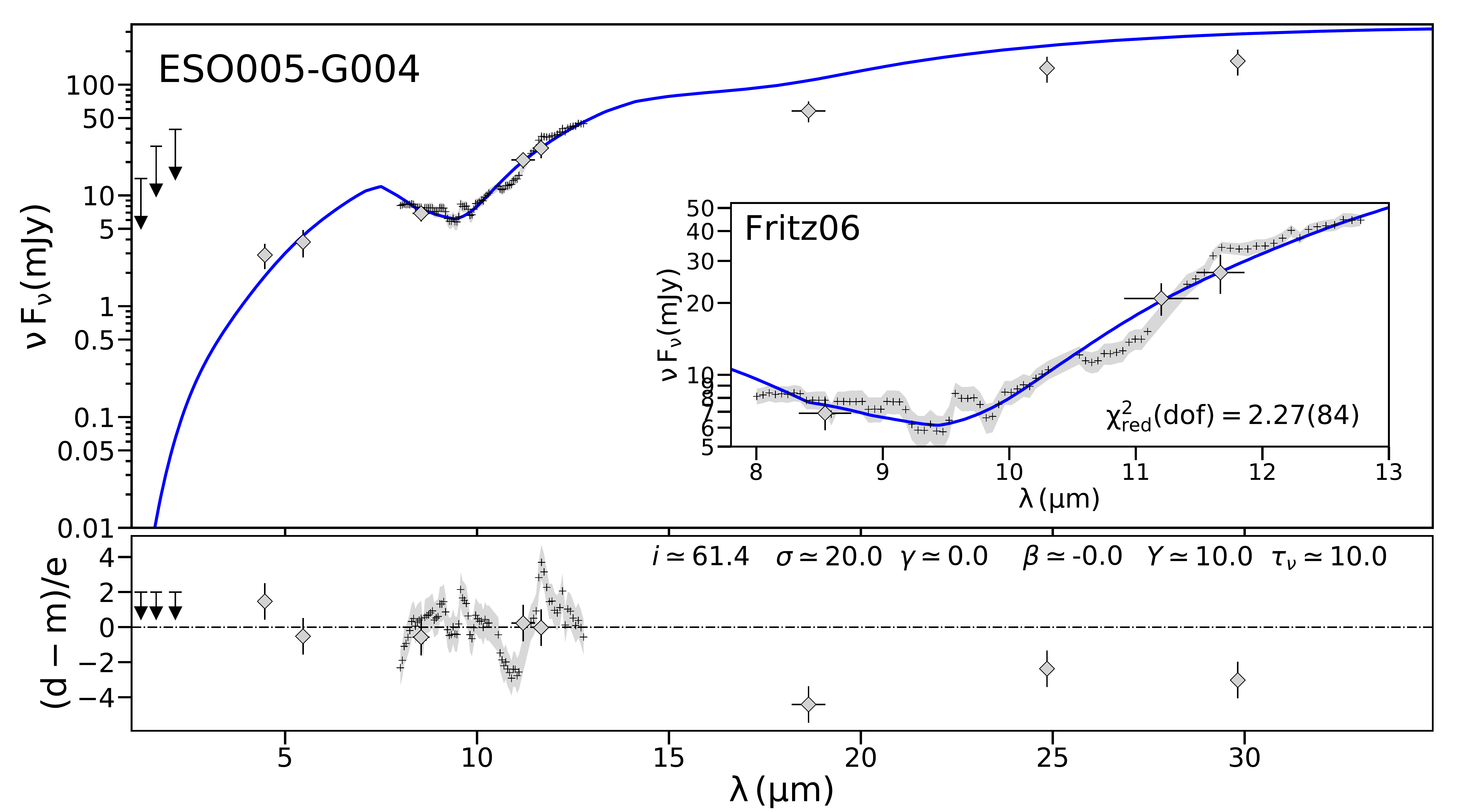}
    \includegraphics[width=0.75\columnwidth]{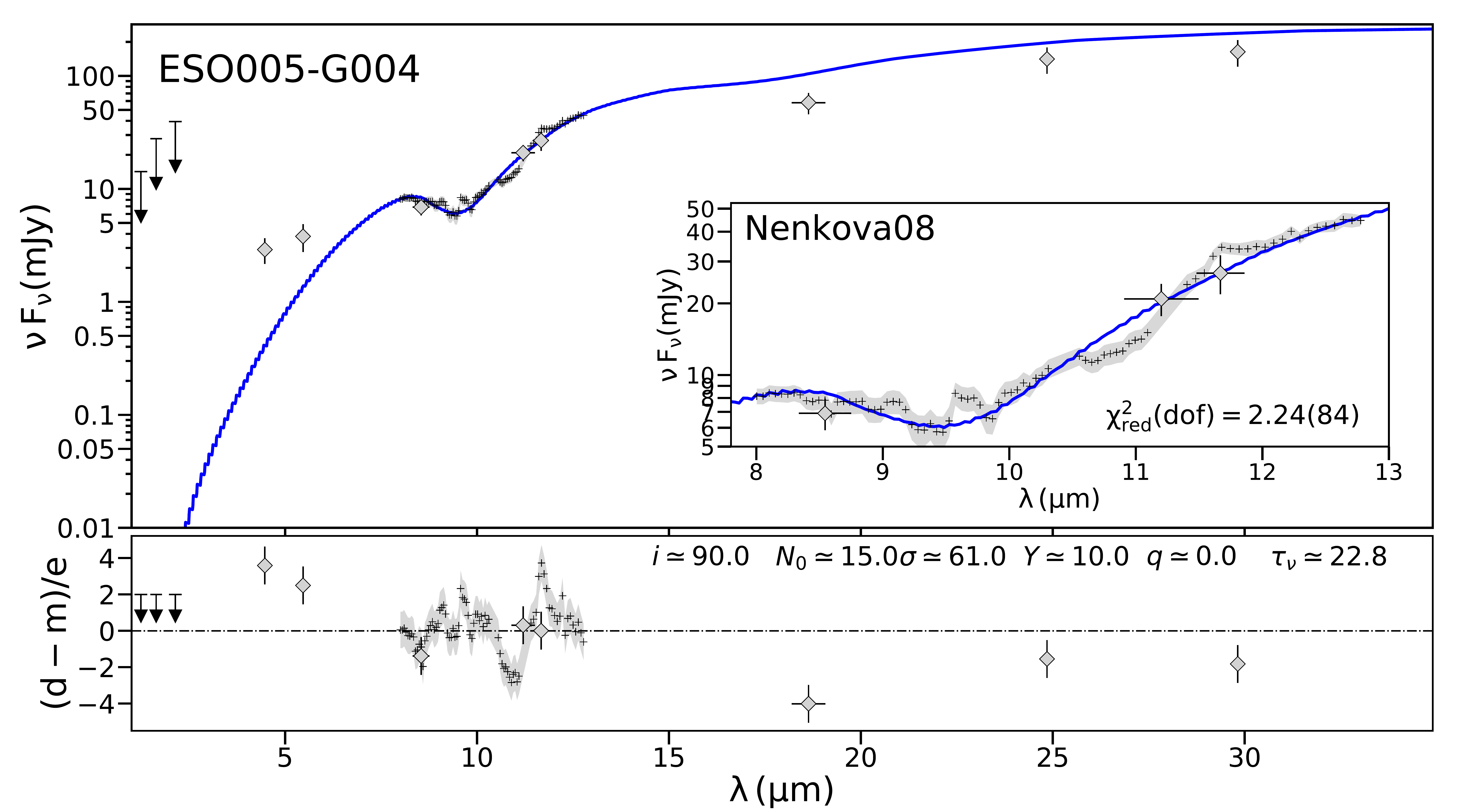}
    \includegraphics[width=0.75\columnwidth]{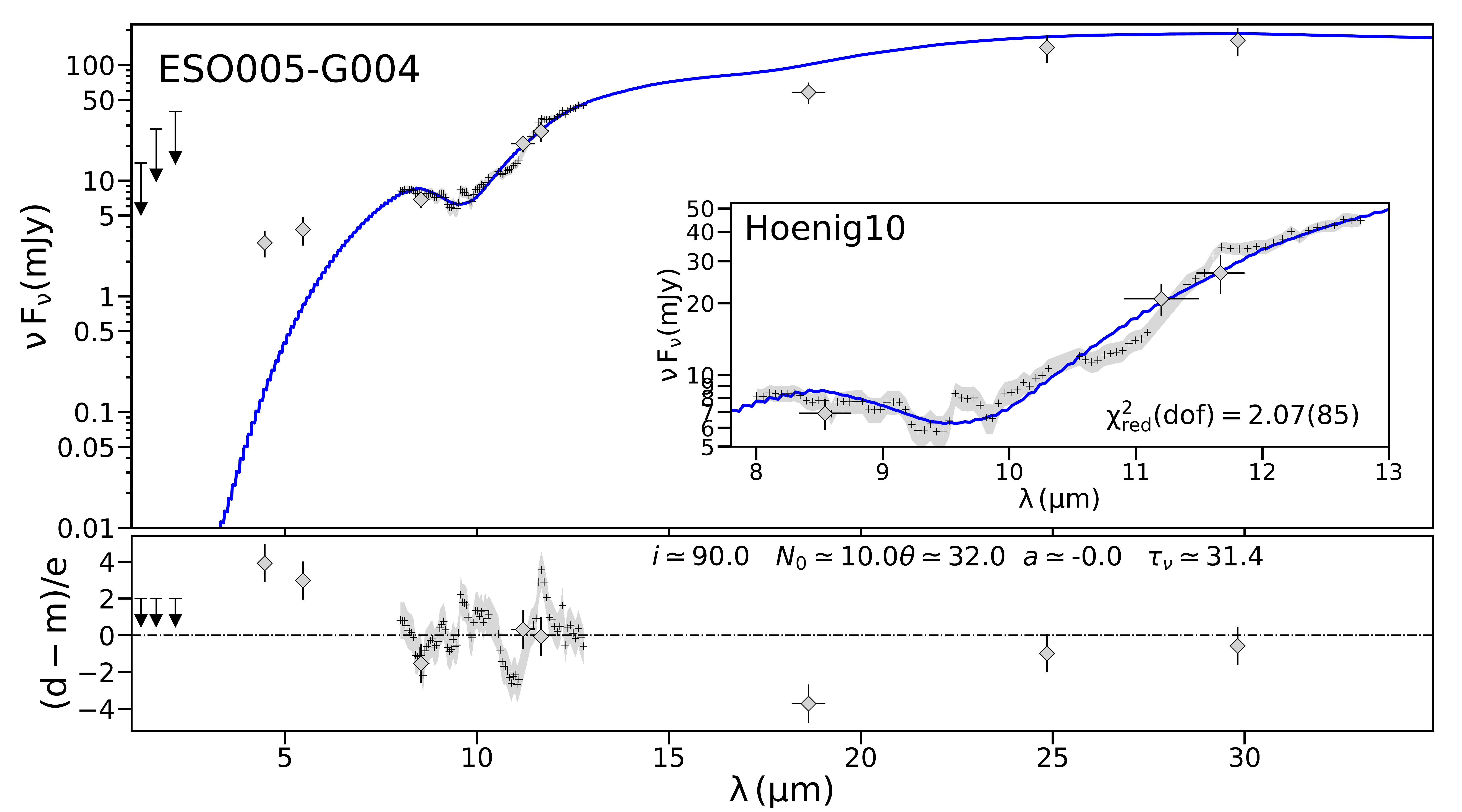}
    \includegraphics[width=0.75\columnwidth]{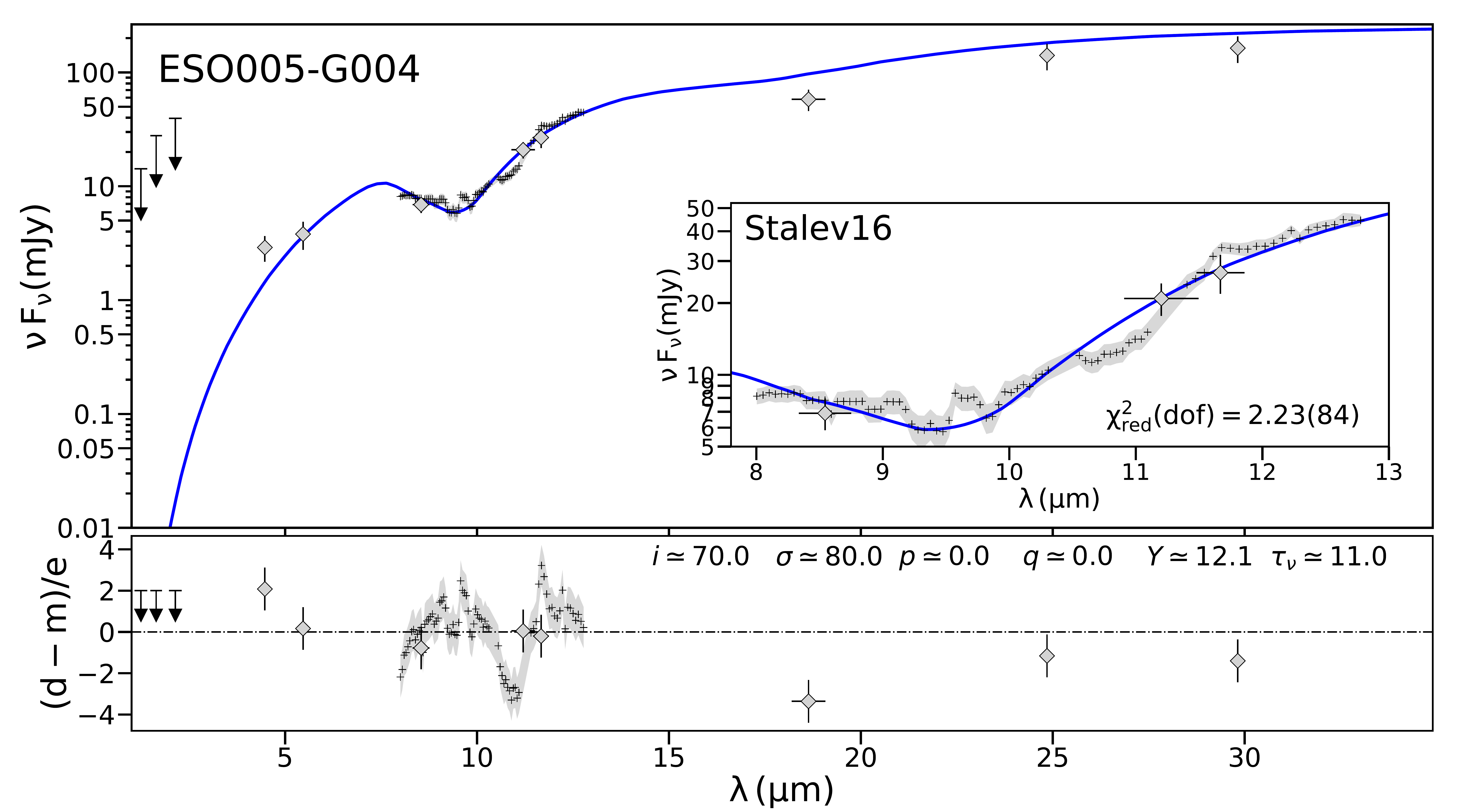}
   \includegraphics[width=0.75\columnwidth]{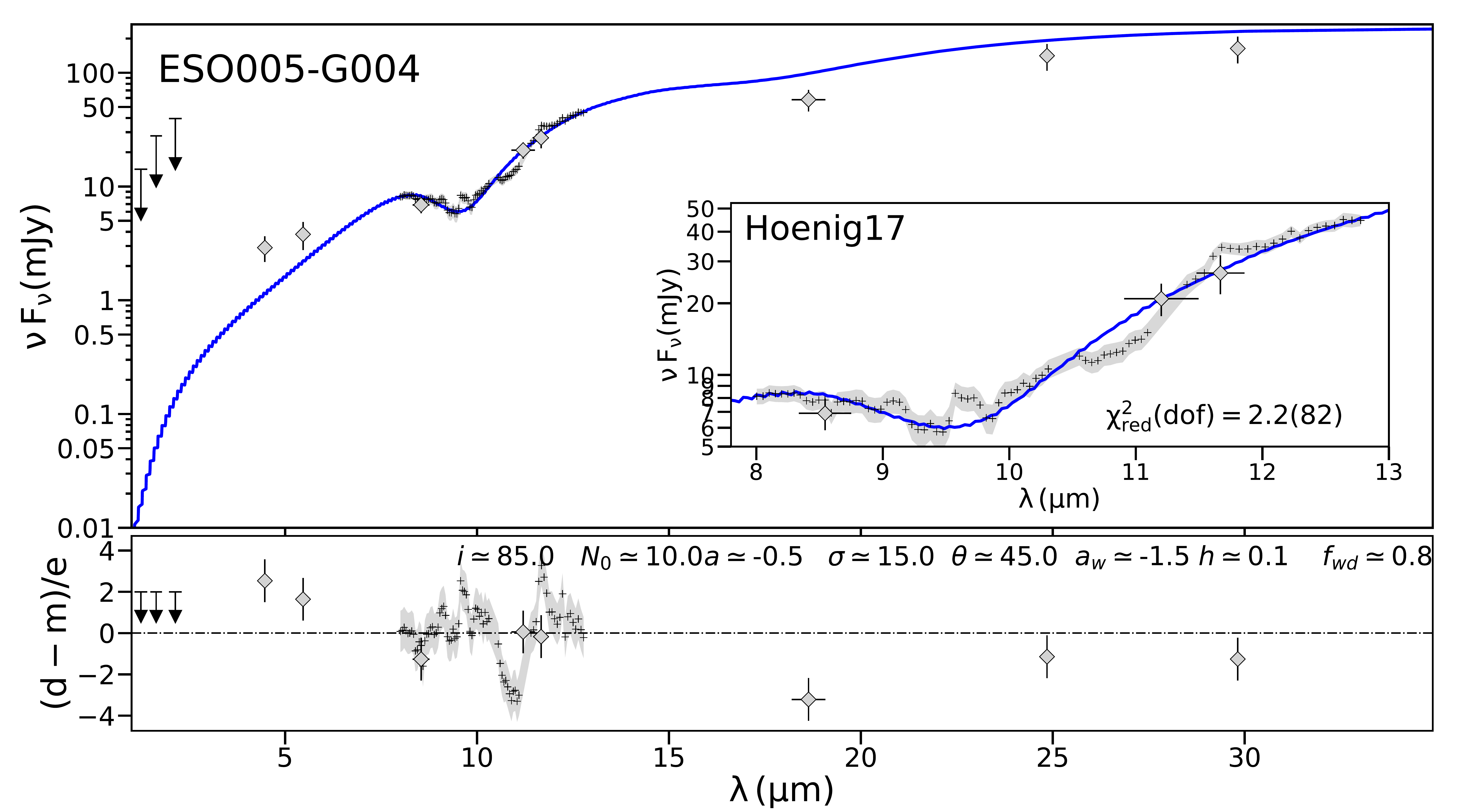}
    \includegraphics[width=0.75\columnwidth]{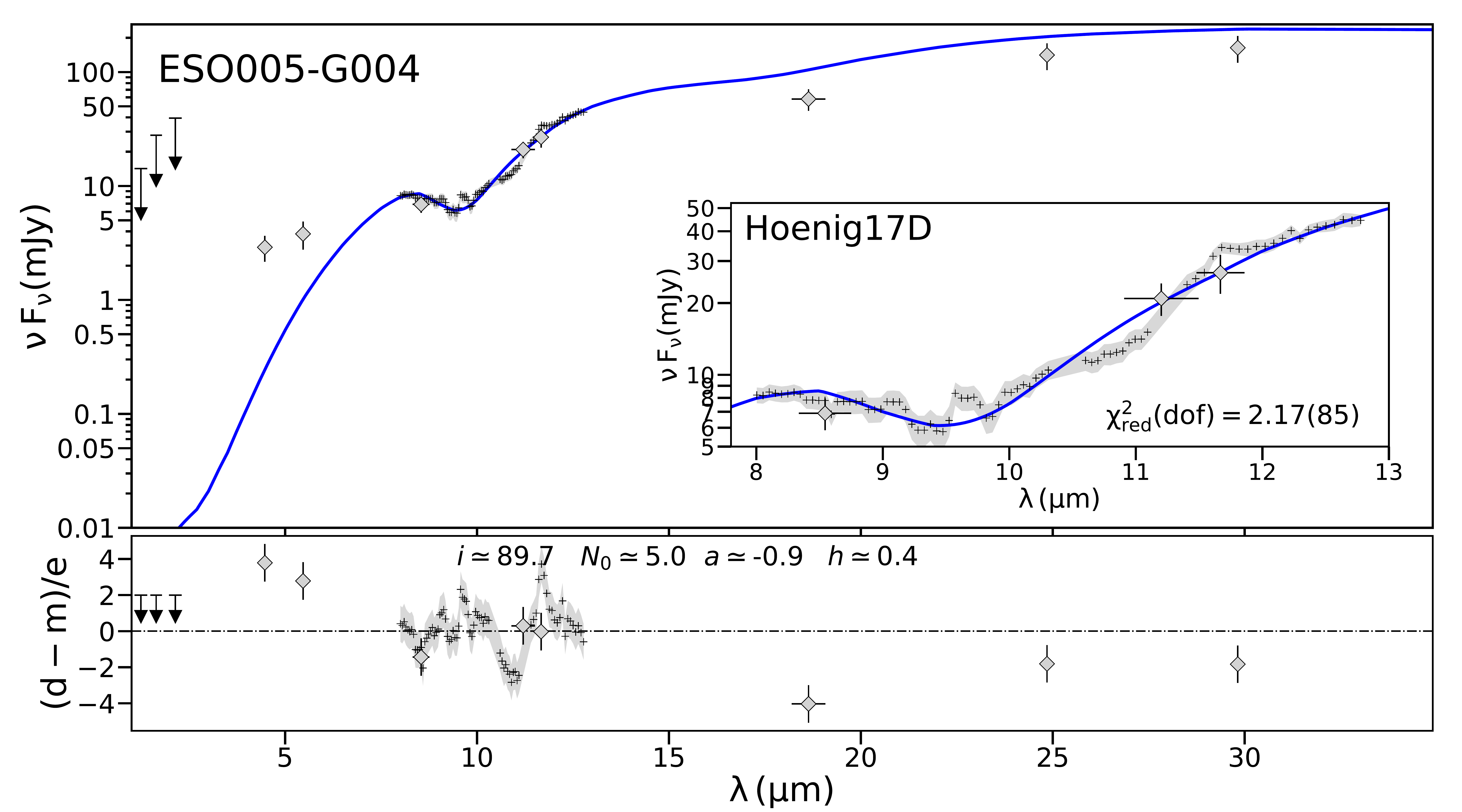}

    \caption{Nuclear IR SED of ESO005-G004. Solid blue line correspond to the best fit per torus model. Grey diamond and black crosses are the photometric data points and N-band spectrum, respectively. The black arrows represent low angular resolution data, which are treated as upper limits..}
    \label{fig:ESO005-G004}
\end{figure*}
\begin{figure*}
    \centering
    \includegraphics[width=0.75\columnwidth]{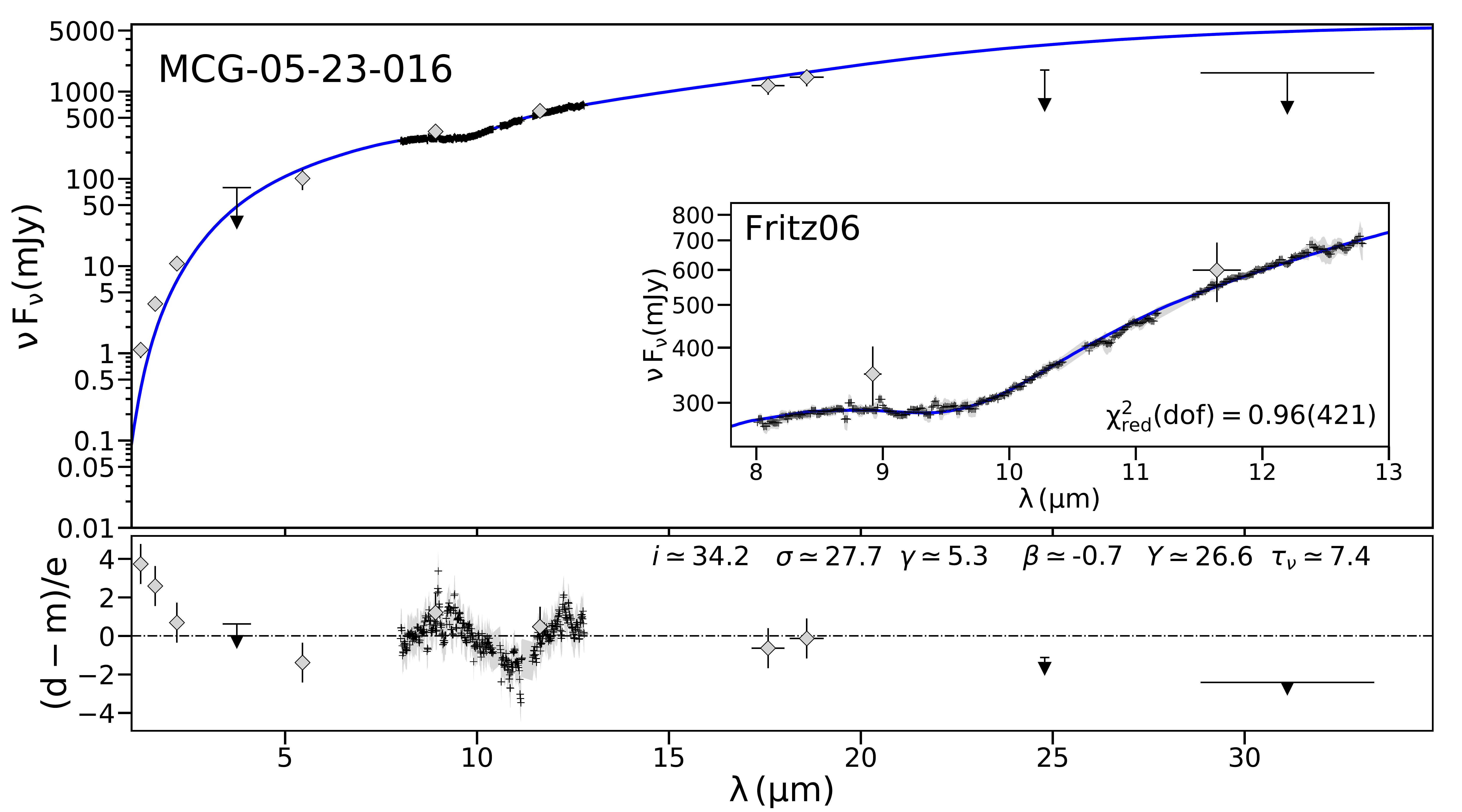}
    \includegraphics[width=0.75\columnwidth]{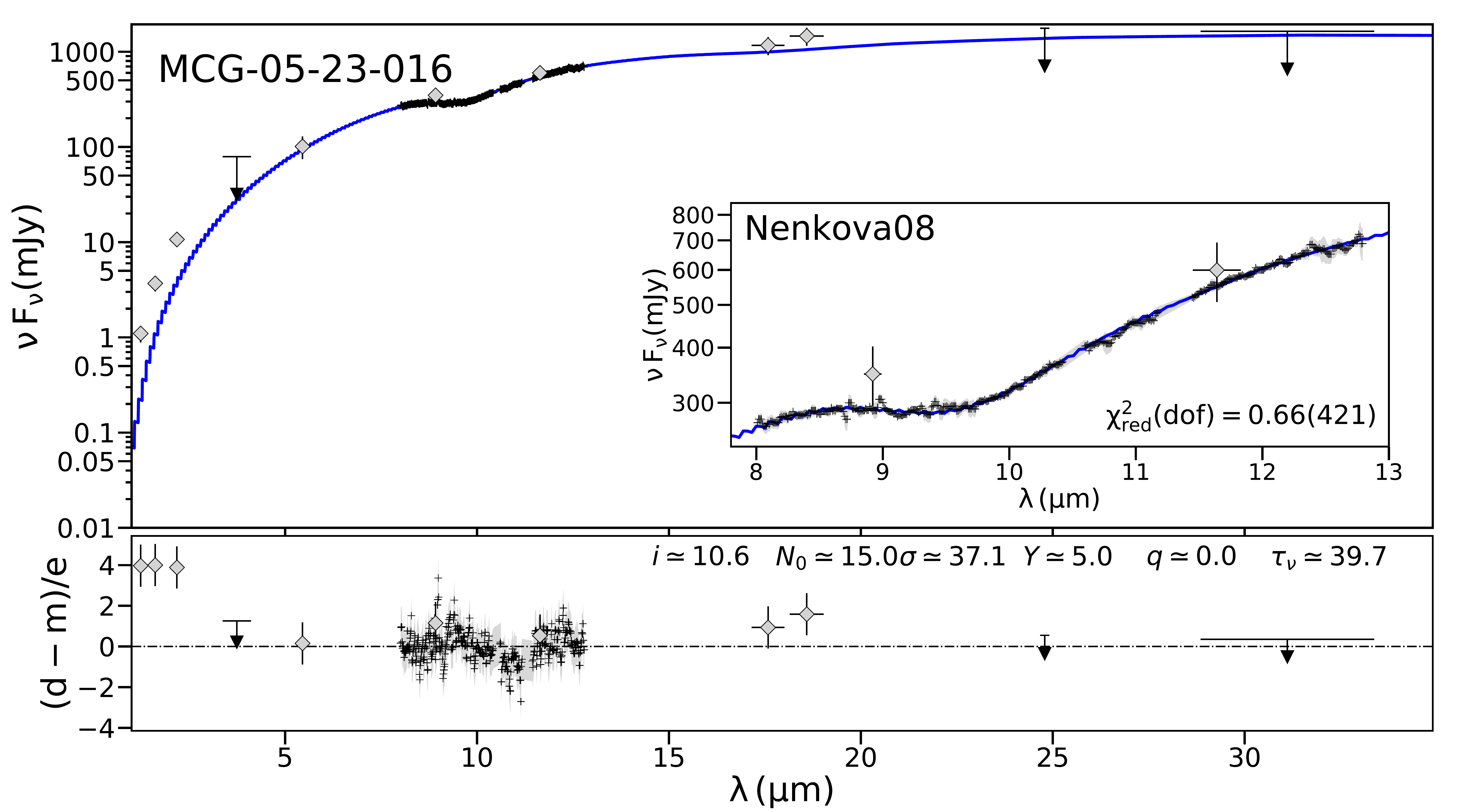}
    \includegraphics[width=0.75\columnwidth]{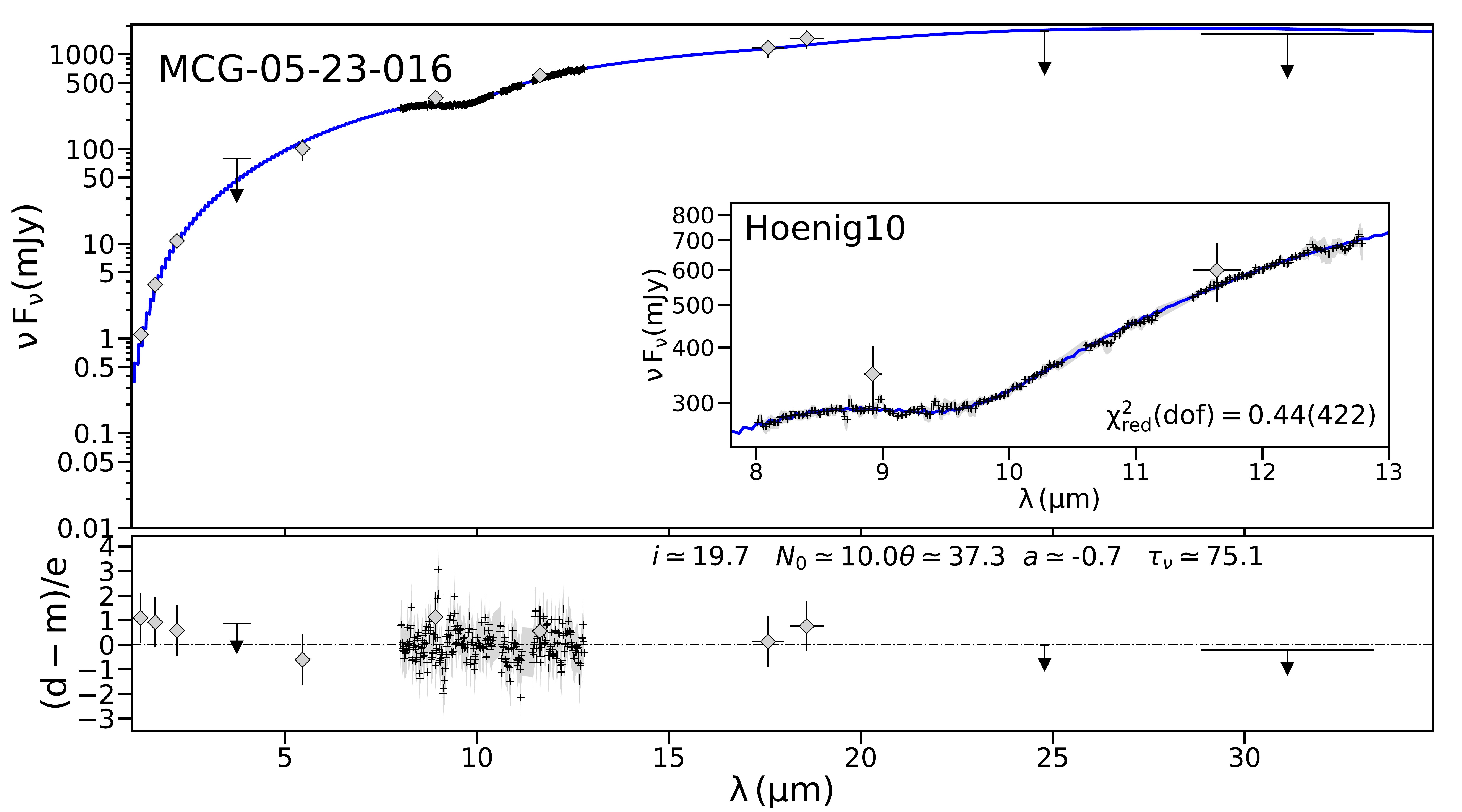}
    \includegraphics[width=0.75\columnwidth]{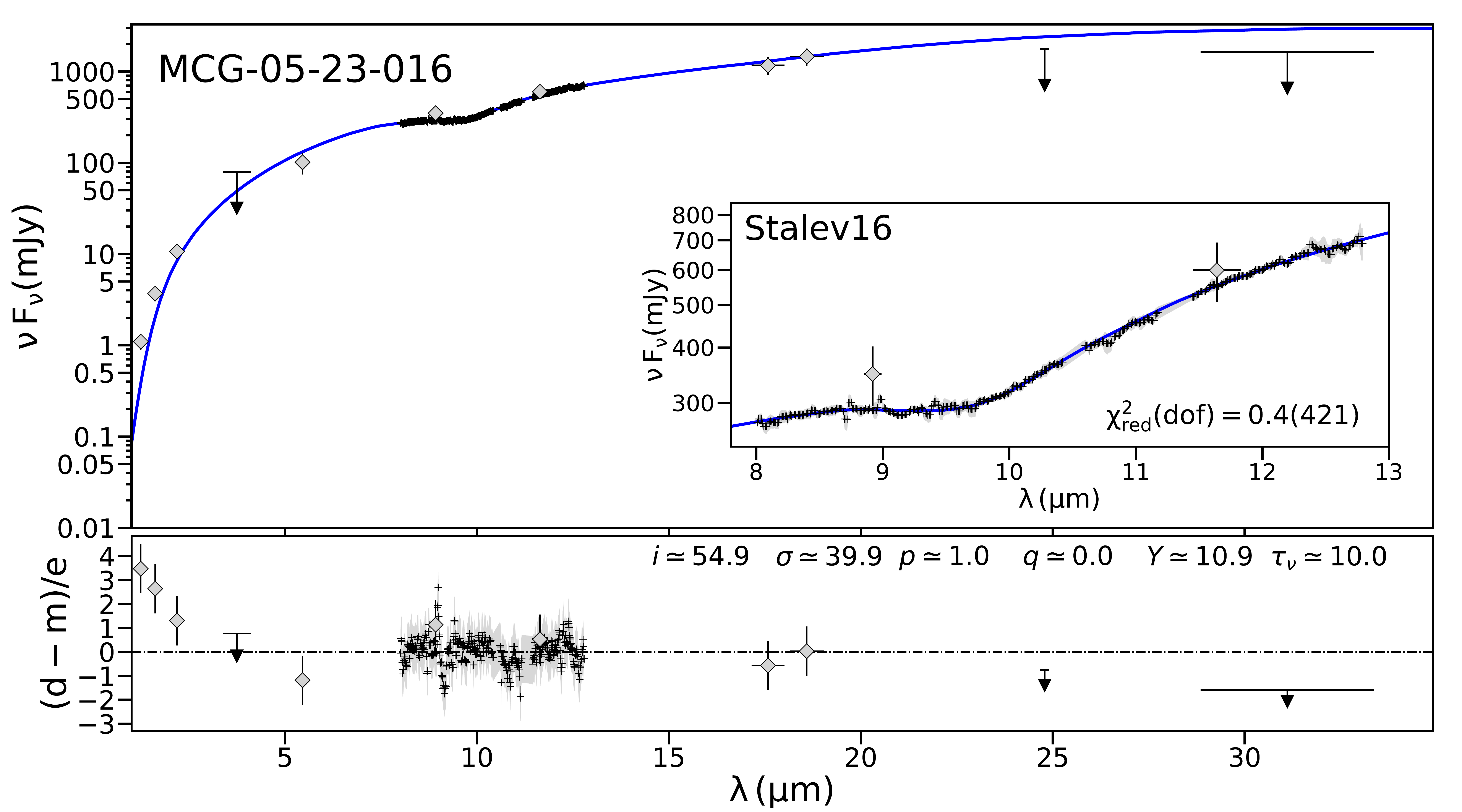}
    \includegraphics[width=0.75\columnwidth]{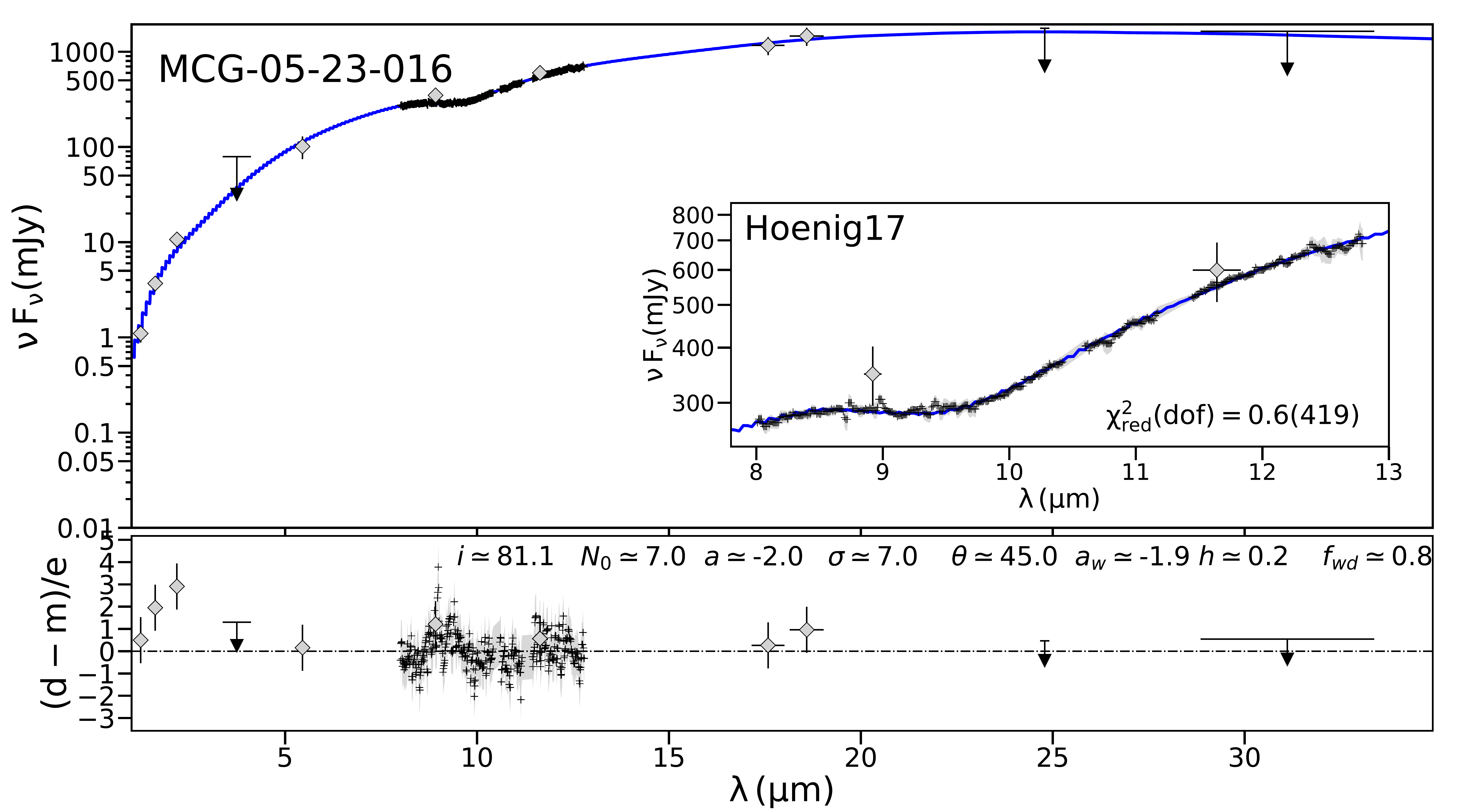}
    \includegraphics[width=0.75\columnwidth]{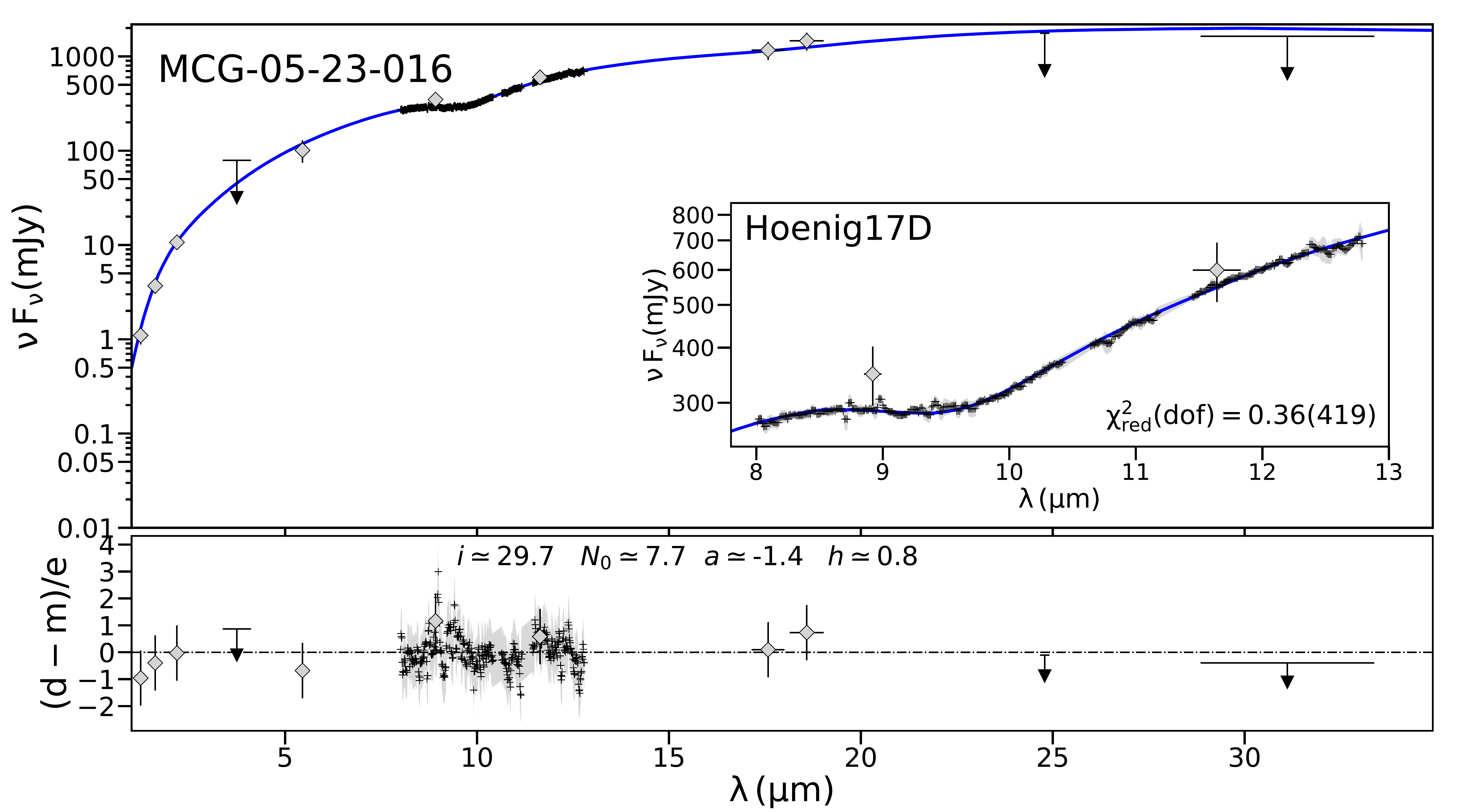}
    \caption{Same as Fig. \ref{fig:ESO005-G004} but for MCG-05-23-016.}
    \label{fig:MCG-05-23-016}
\end{figure*}

\begin{figure*}
    \centering
    \includegraphics[width=0.75\columnwidth]{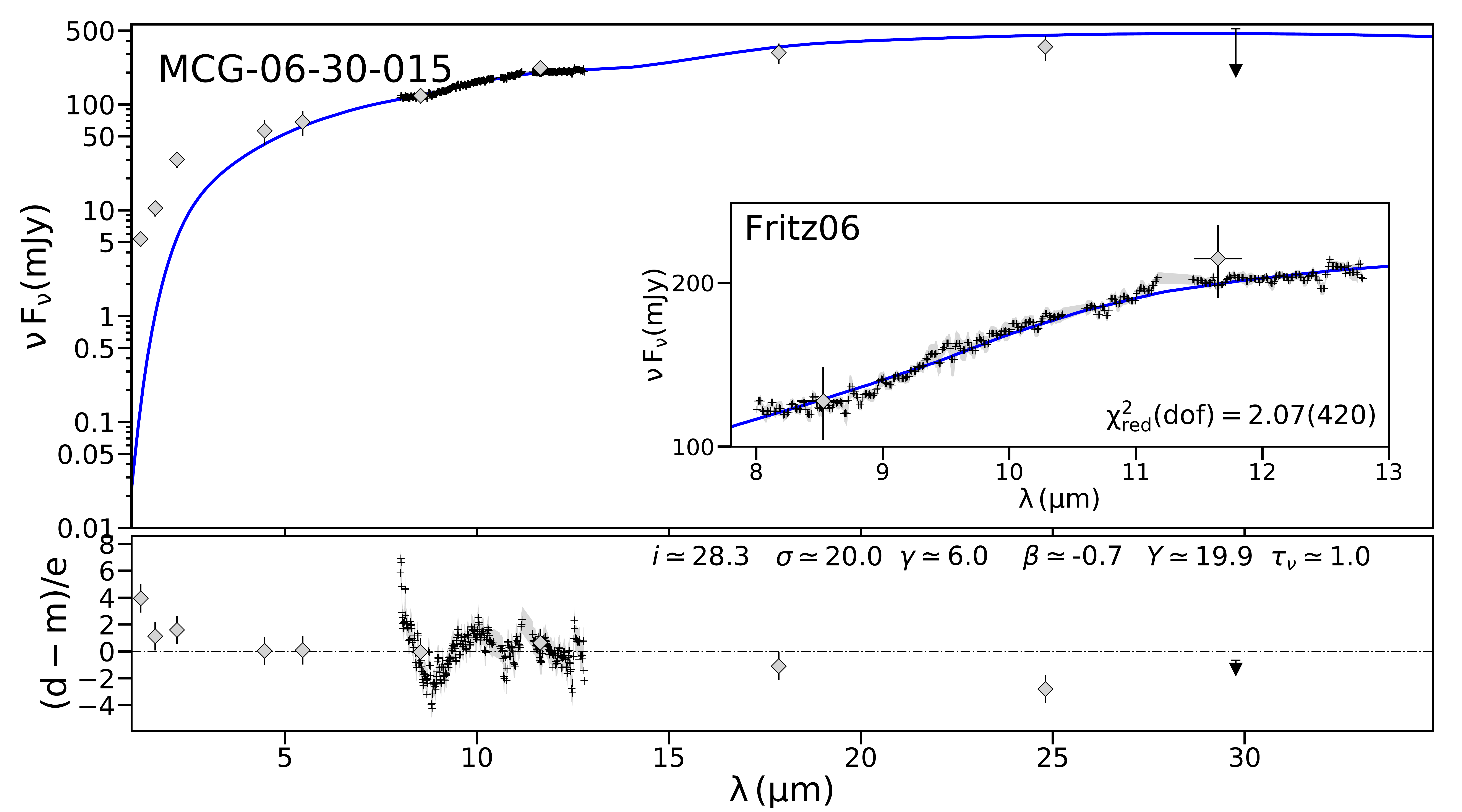}
    \includegraphics[width=0.75\columnwidth]{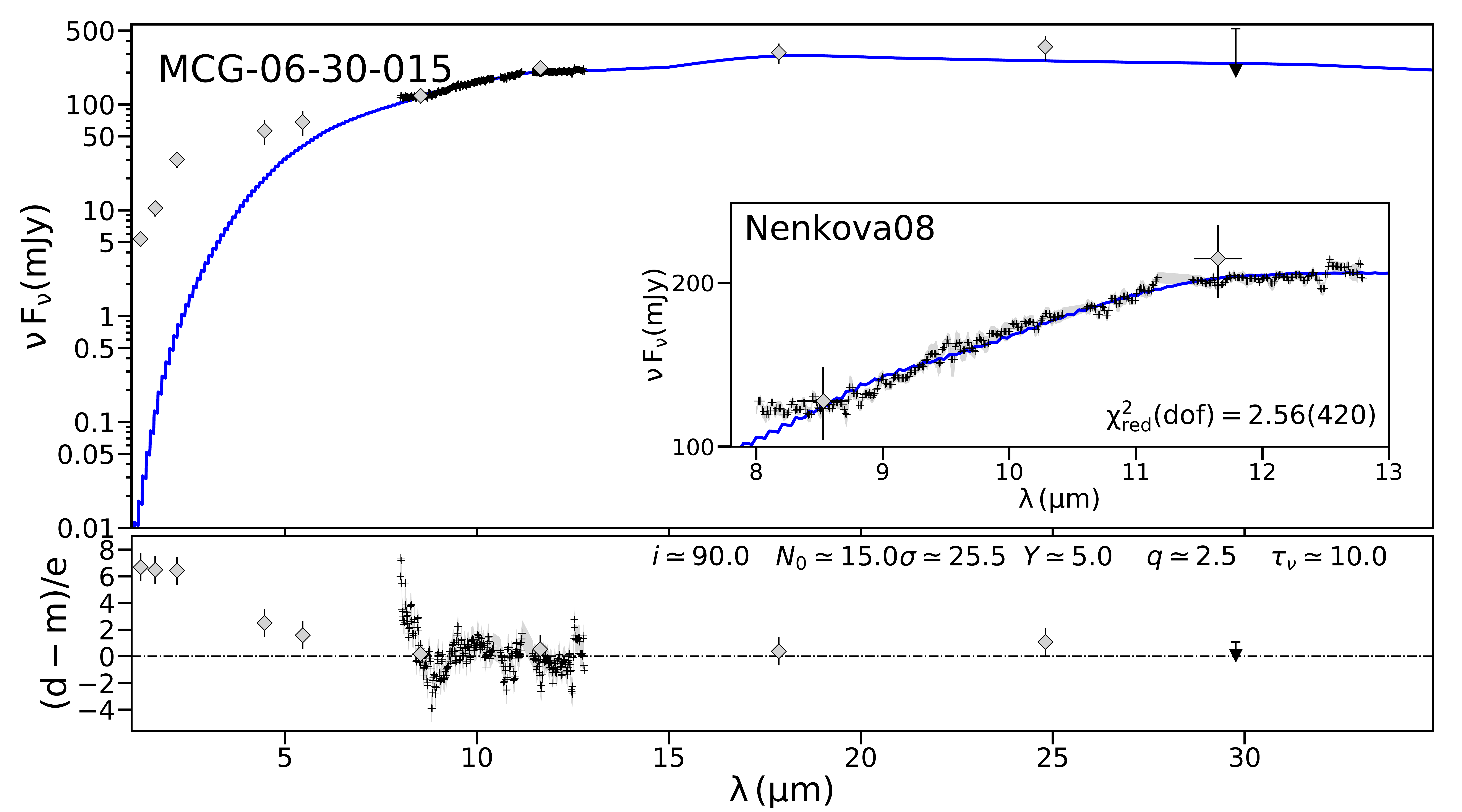}
    \includegraphics[width=0.75\columnwidth]{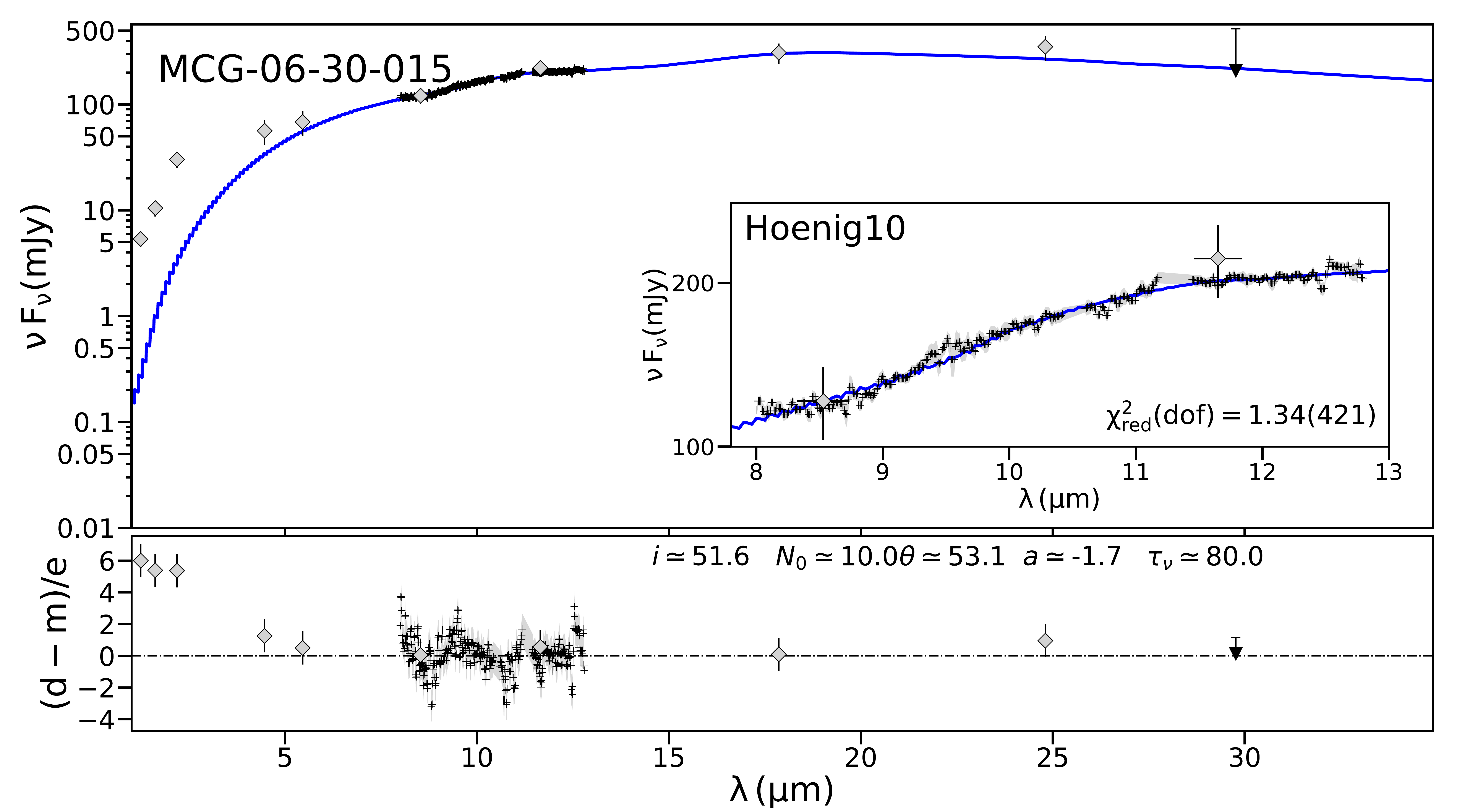}
    \includegraphics[width=0.75\columnwidth]{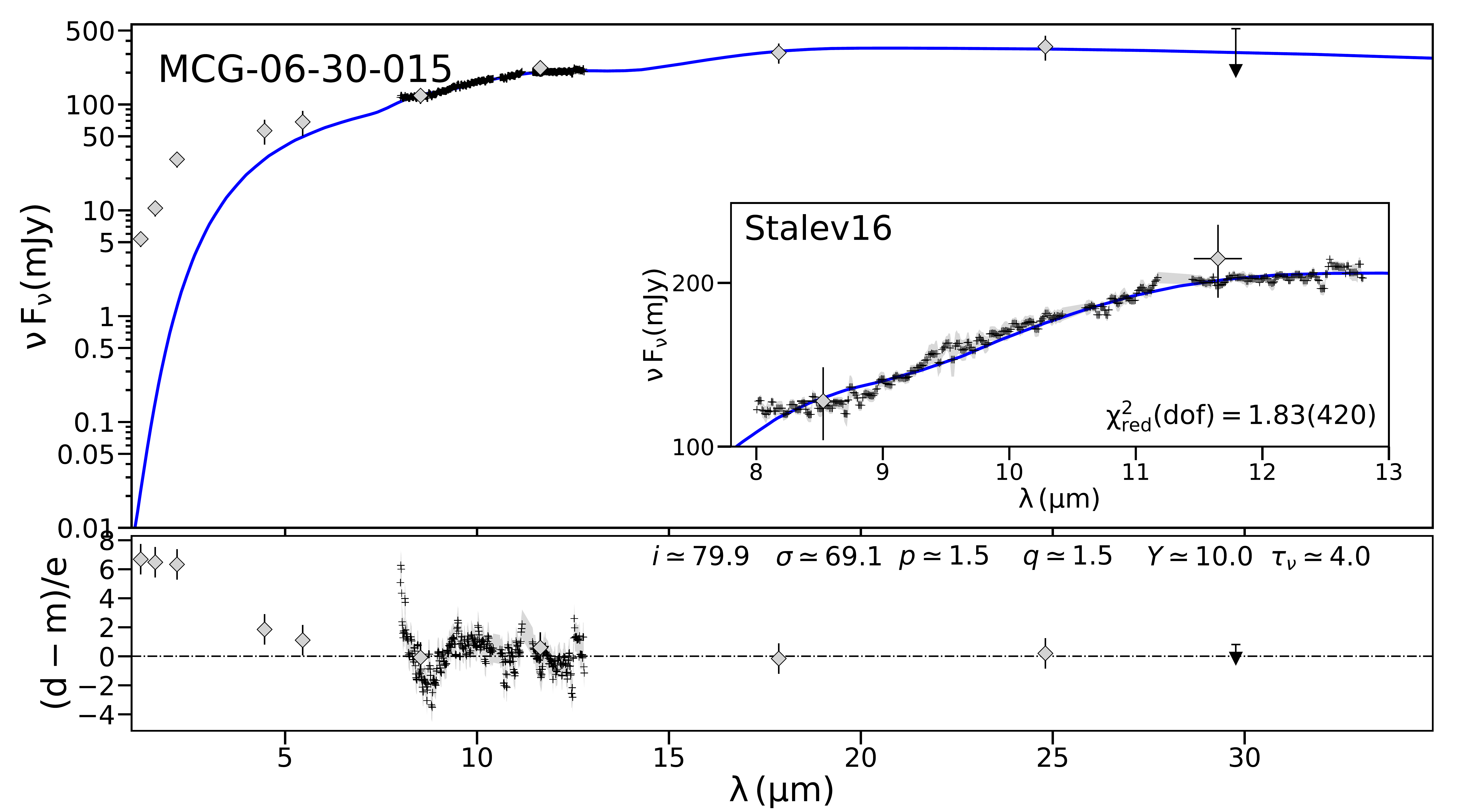}
    \includegraphics[width=0.75\columnwidth]{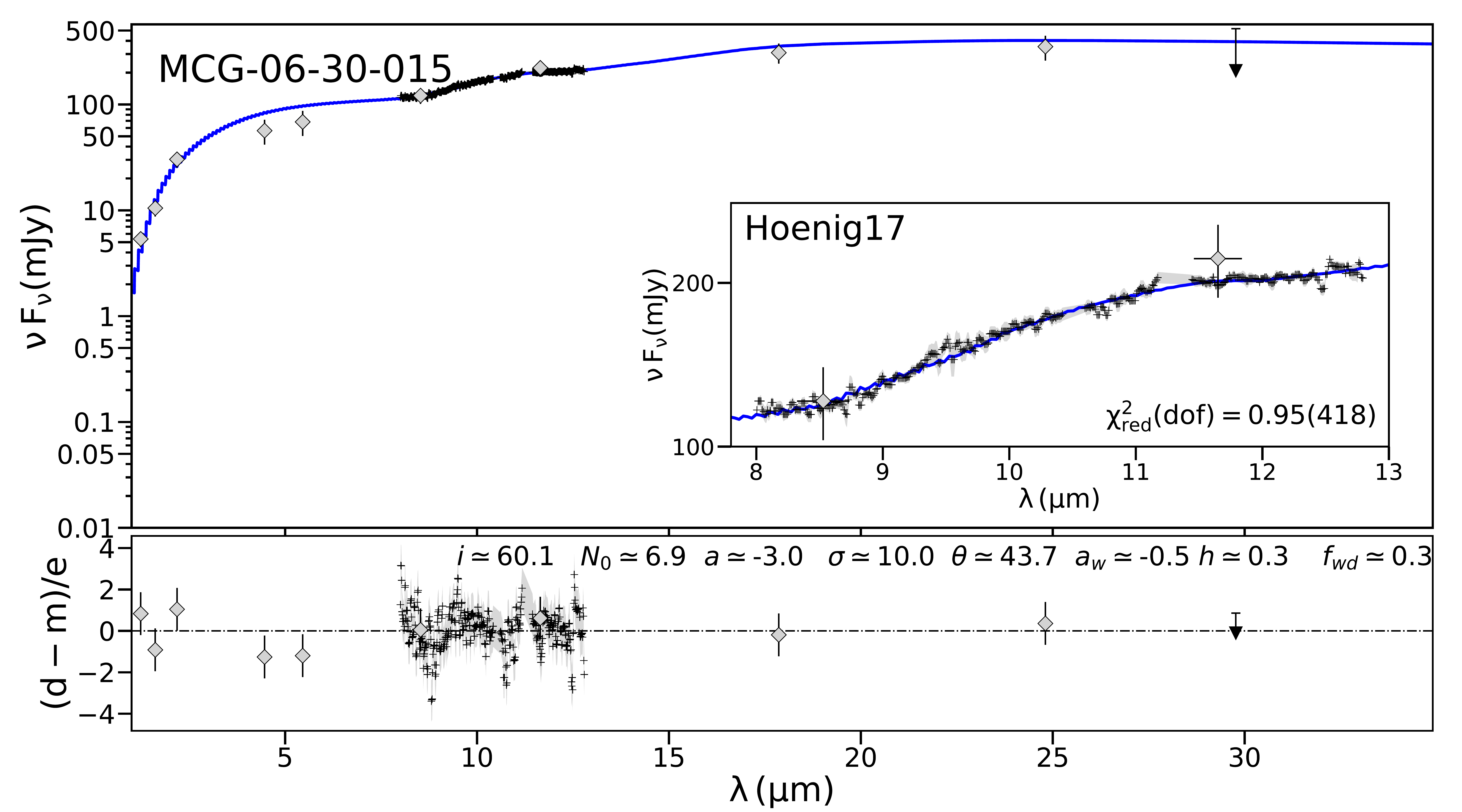}
    \includegraphics[width=0.75\columnwidth]{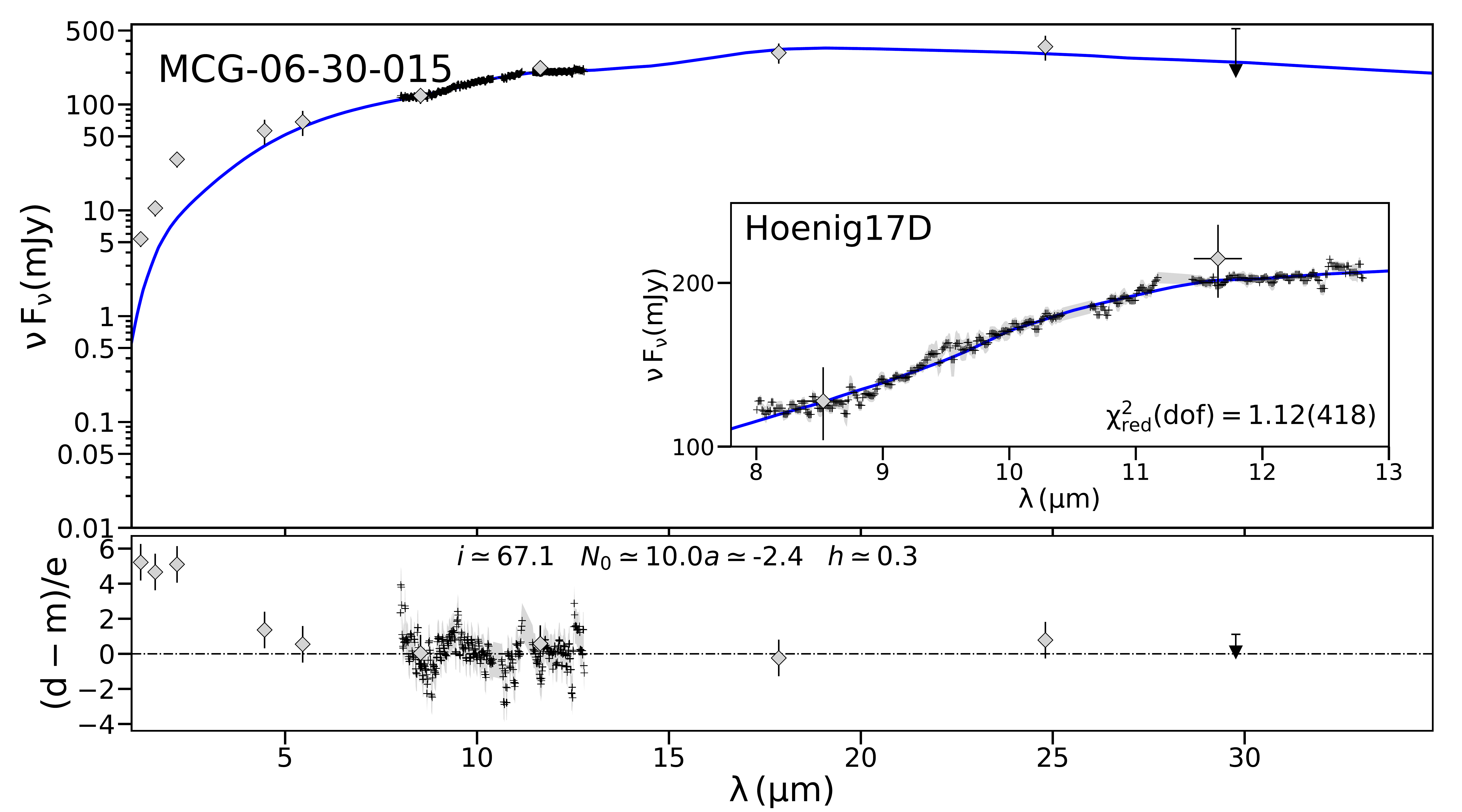}
    \caption{Same as Fig. \ref{fig:ESO005-G004} but for MCG-06-30-015.}
    \label{fig:MCG-06-30-015}
\end{figure*}

\begin{figure*}
    \centering
    \includegraphics[width=0.75\columnwidth]{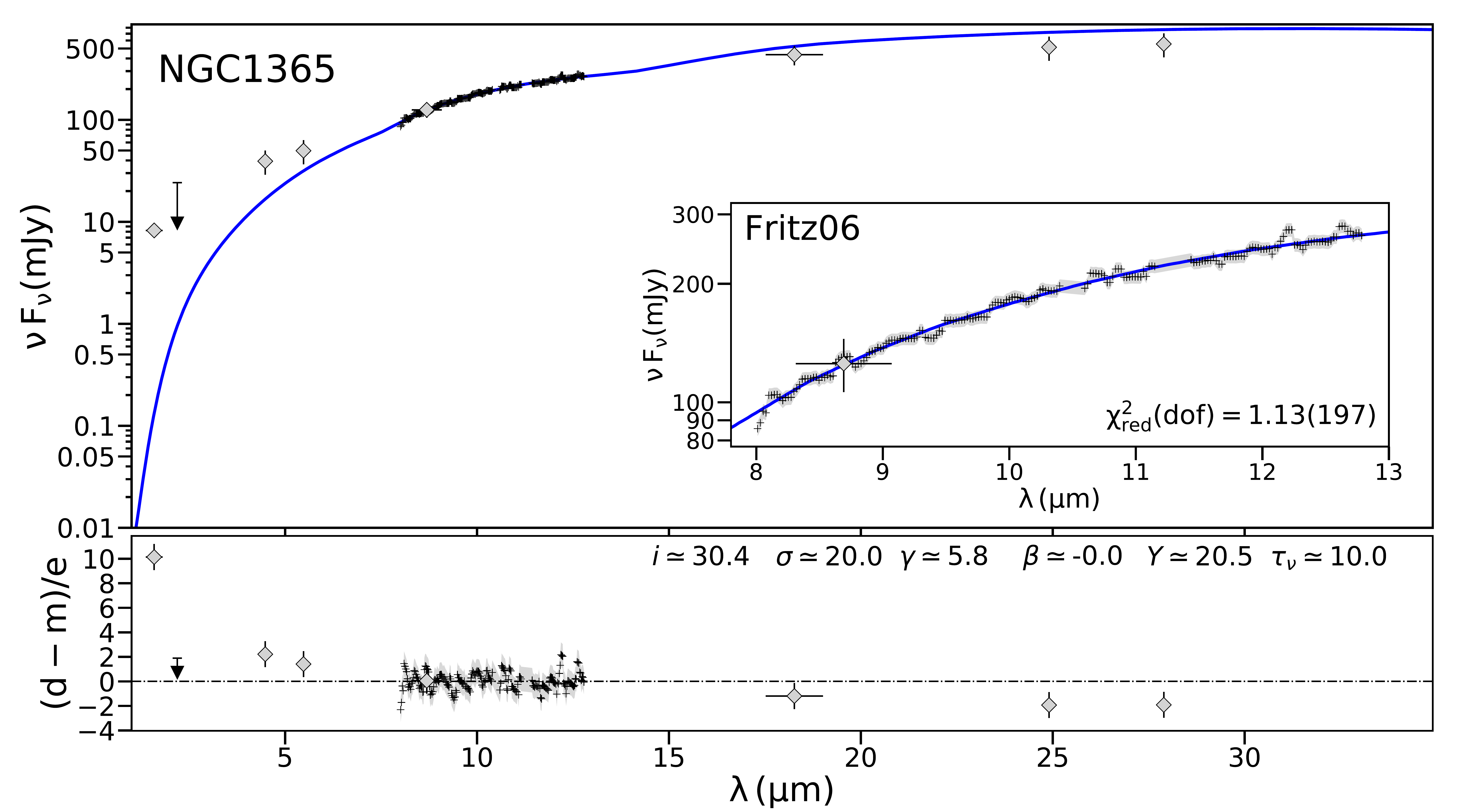}
    \includegraphics[width=0.75\columnwidth]{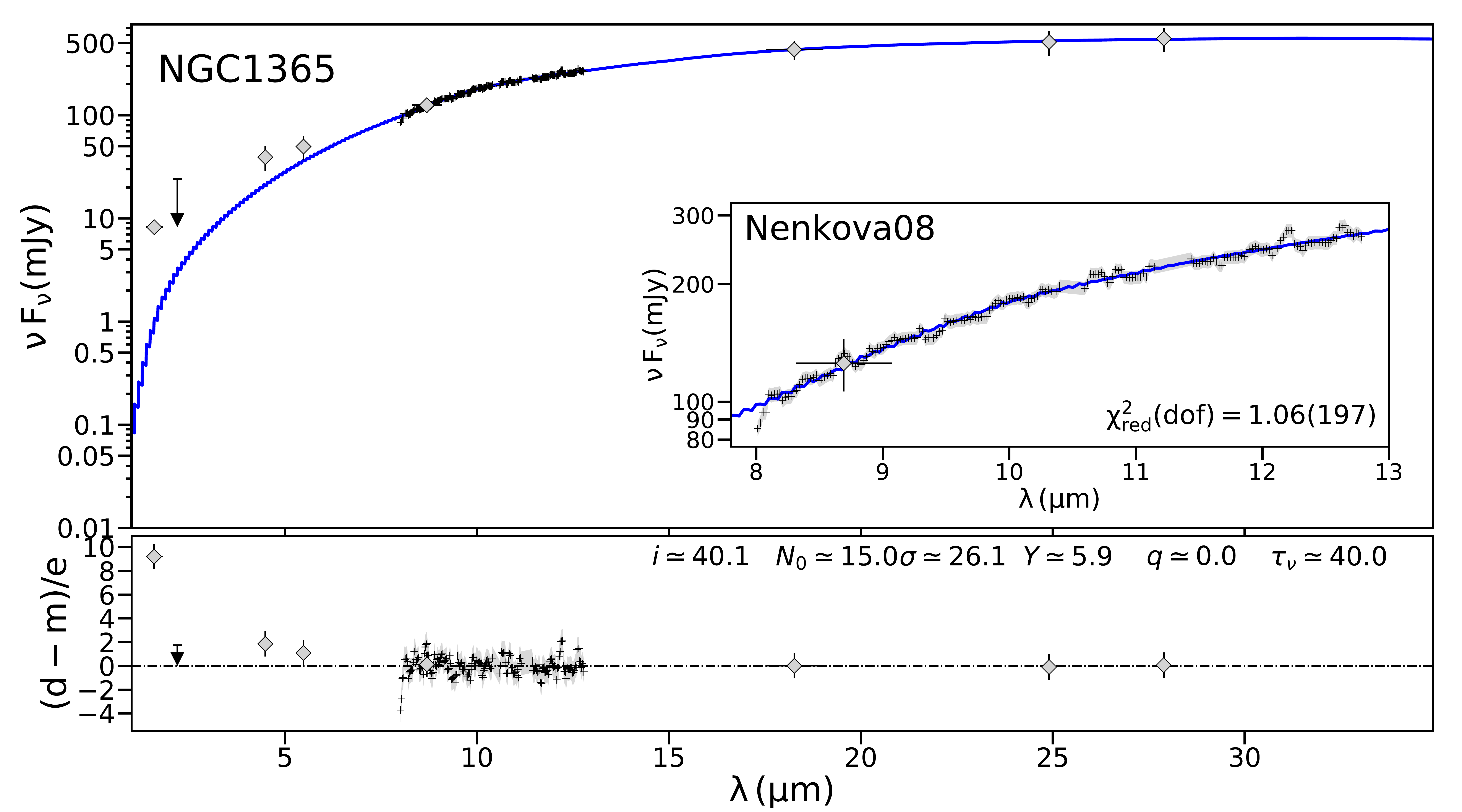}
    \includegraphics[width=0.75\columnwidth]{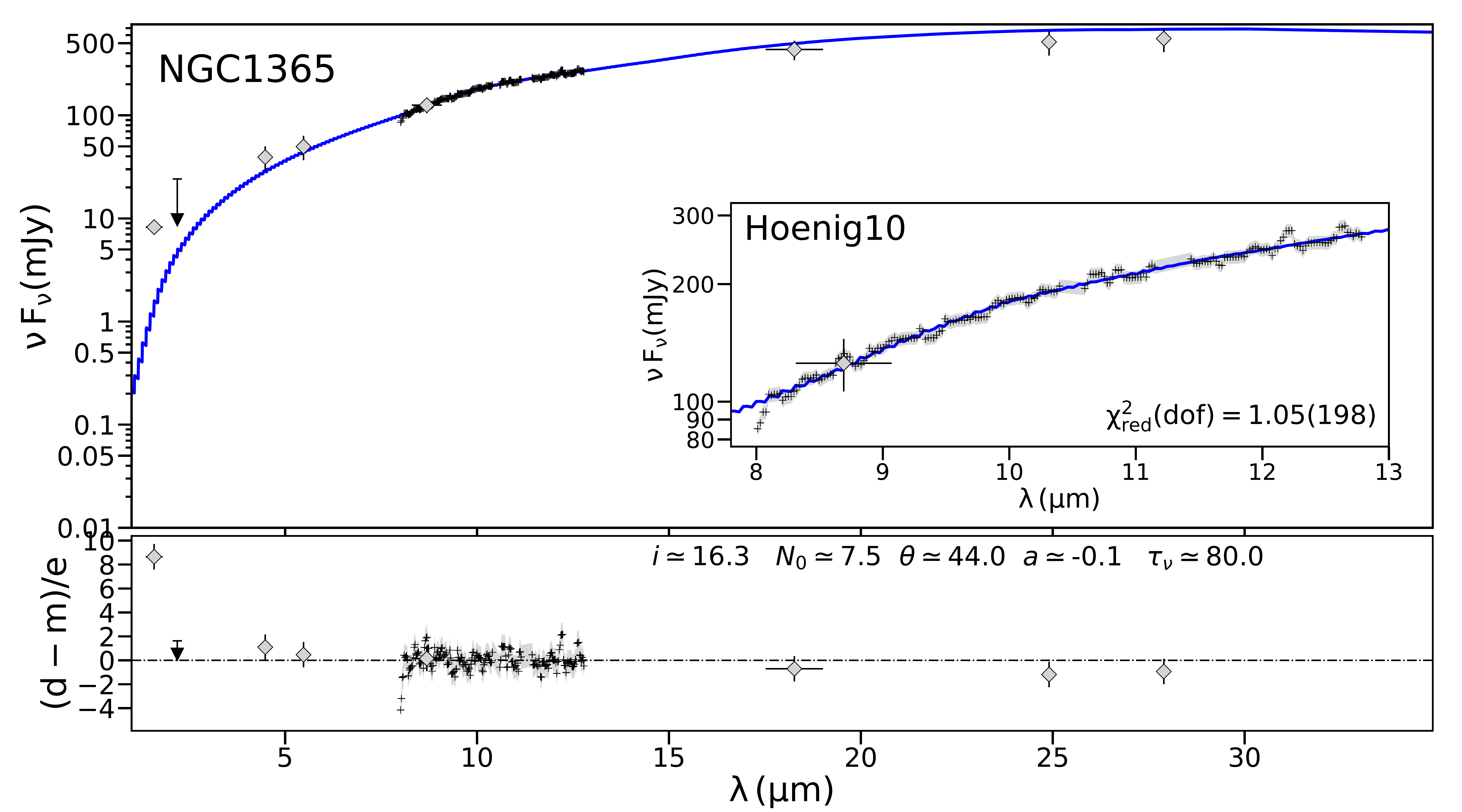}
    \includegraphics[width=0.75\columnwidth]{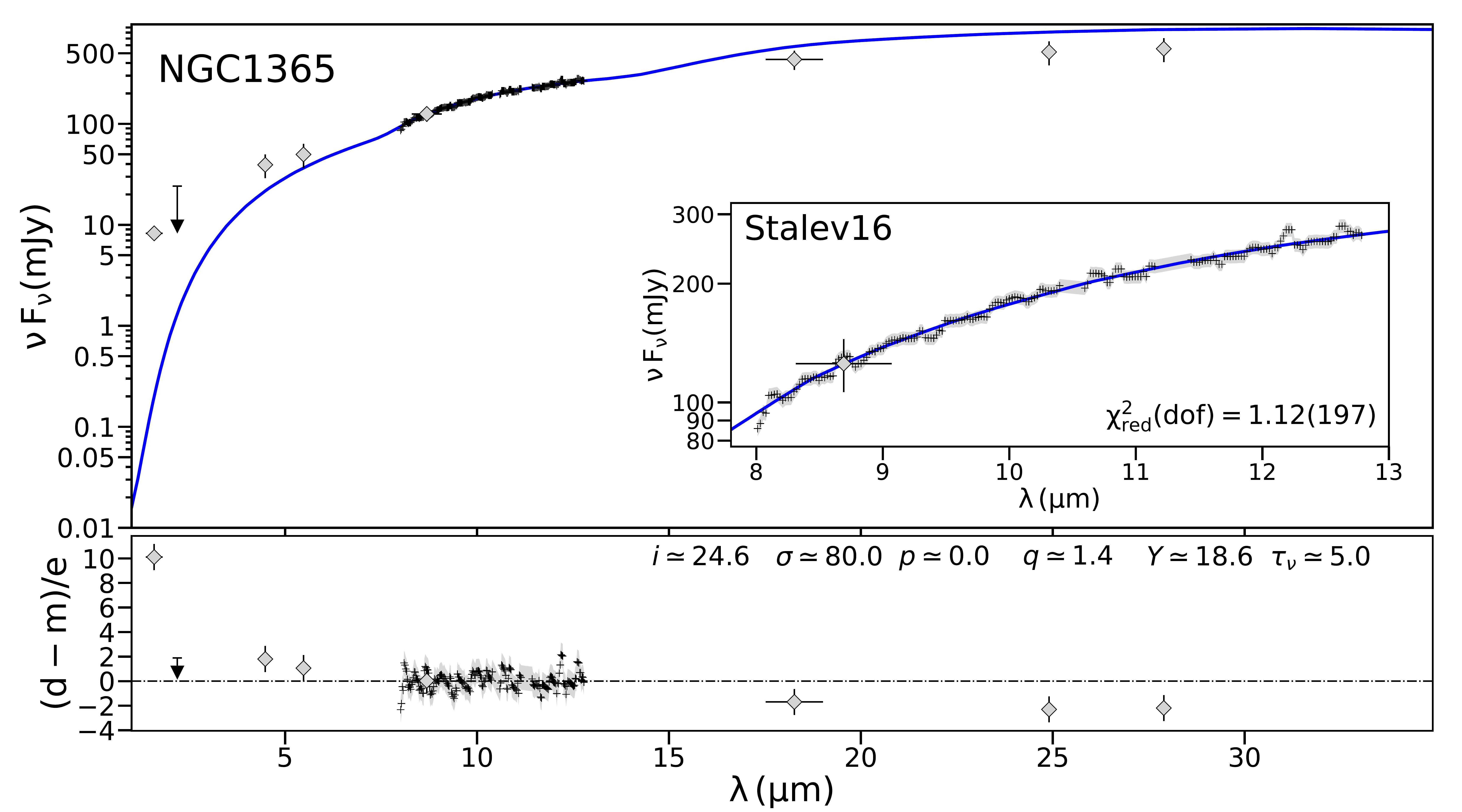}
    \includegraphics[width=0.75\columnwidth]{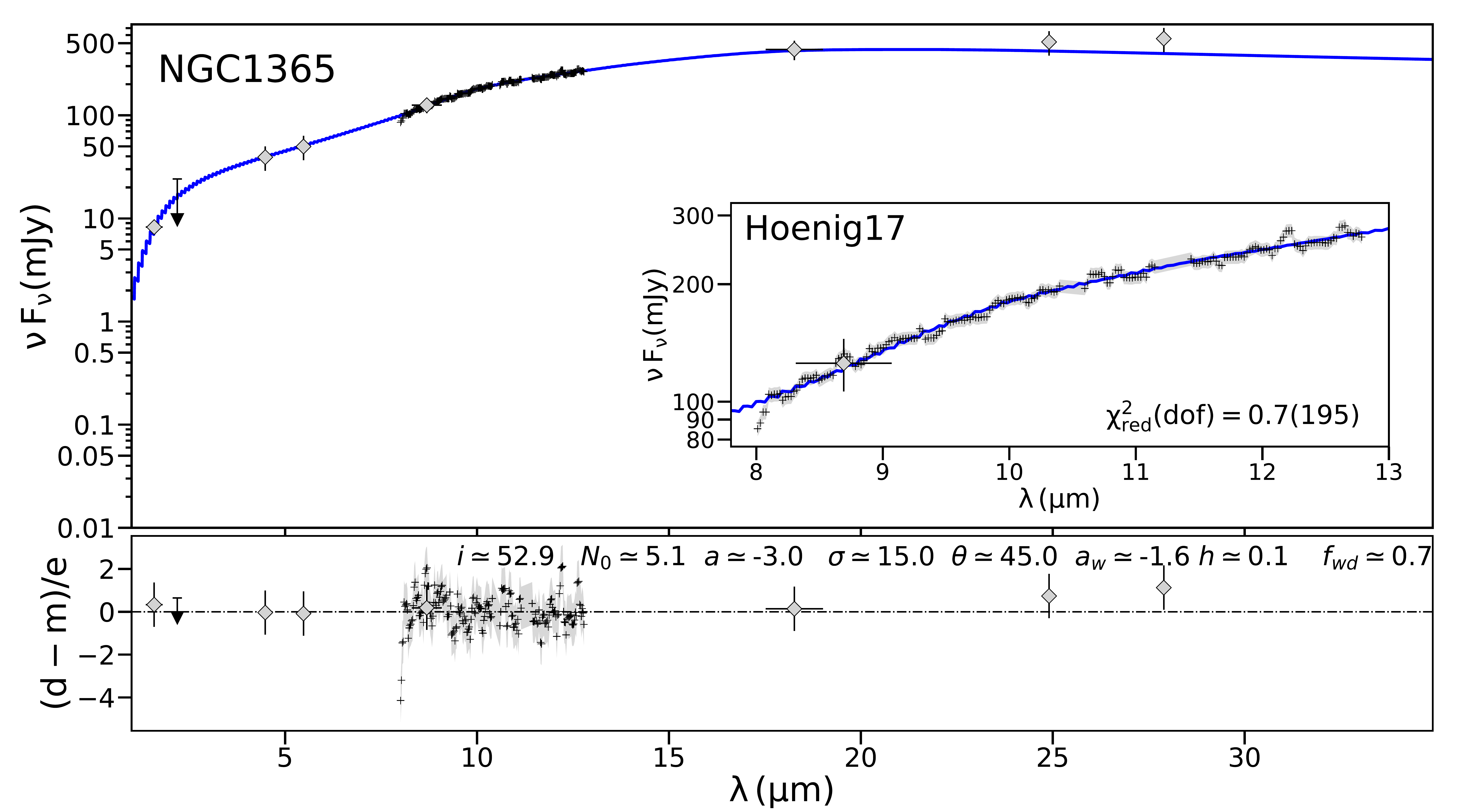}
    \includegraphics[width=0.75\columnwidth]{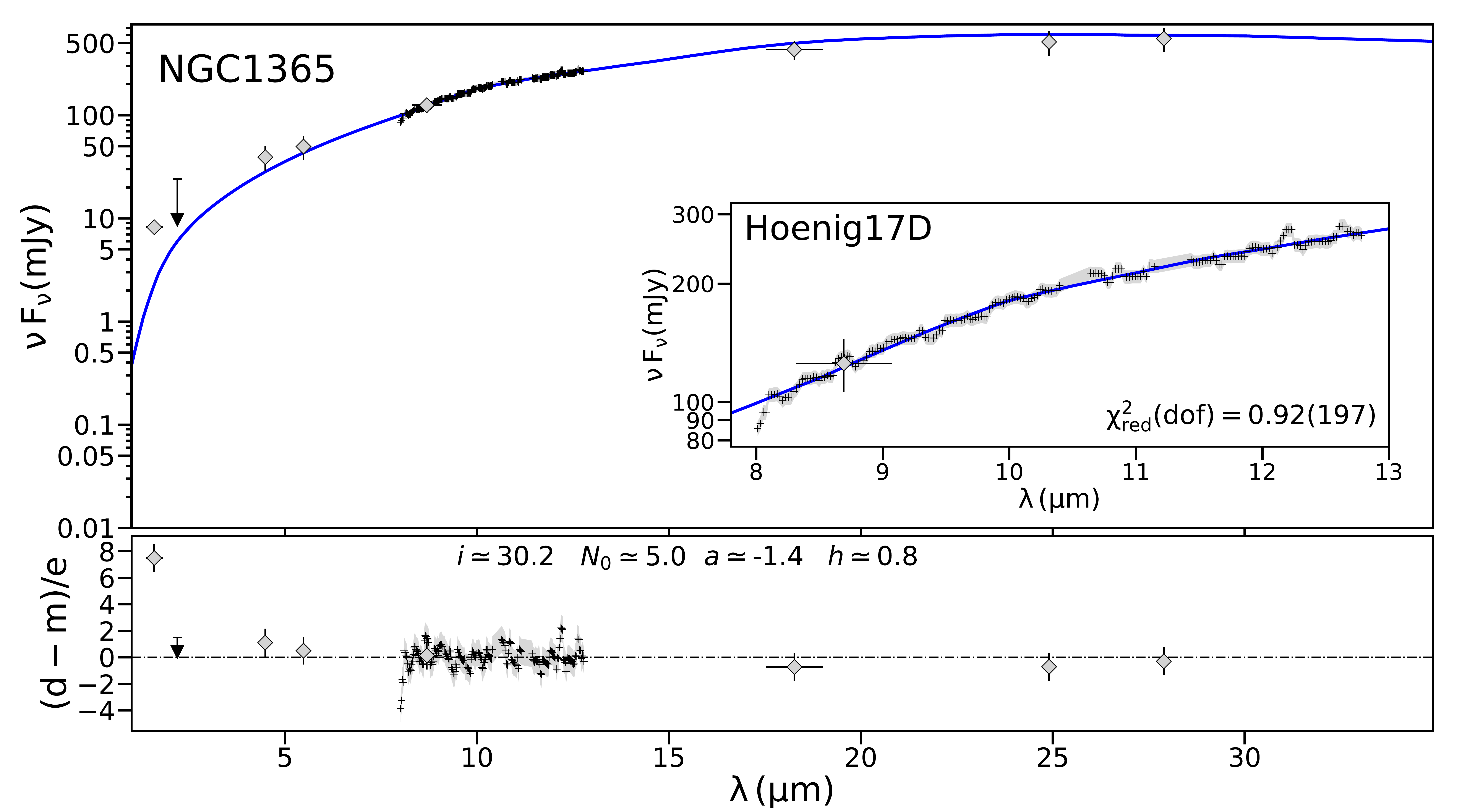}
    \caption{Same as Fig. \ref{fig:ESO005-G004} but for NGC1365.}
    \label{fig:NGC1365}
\end{figure*}
\begin{figure*}
    \centering
    \includegraphics[width=0.75\columnwidth]{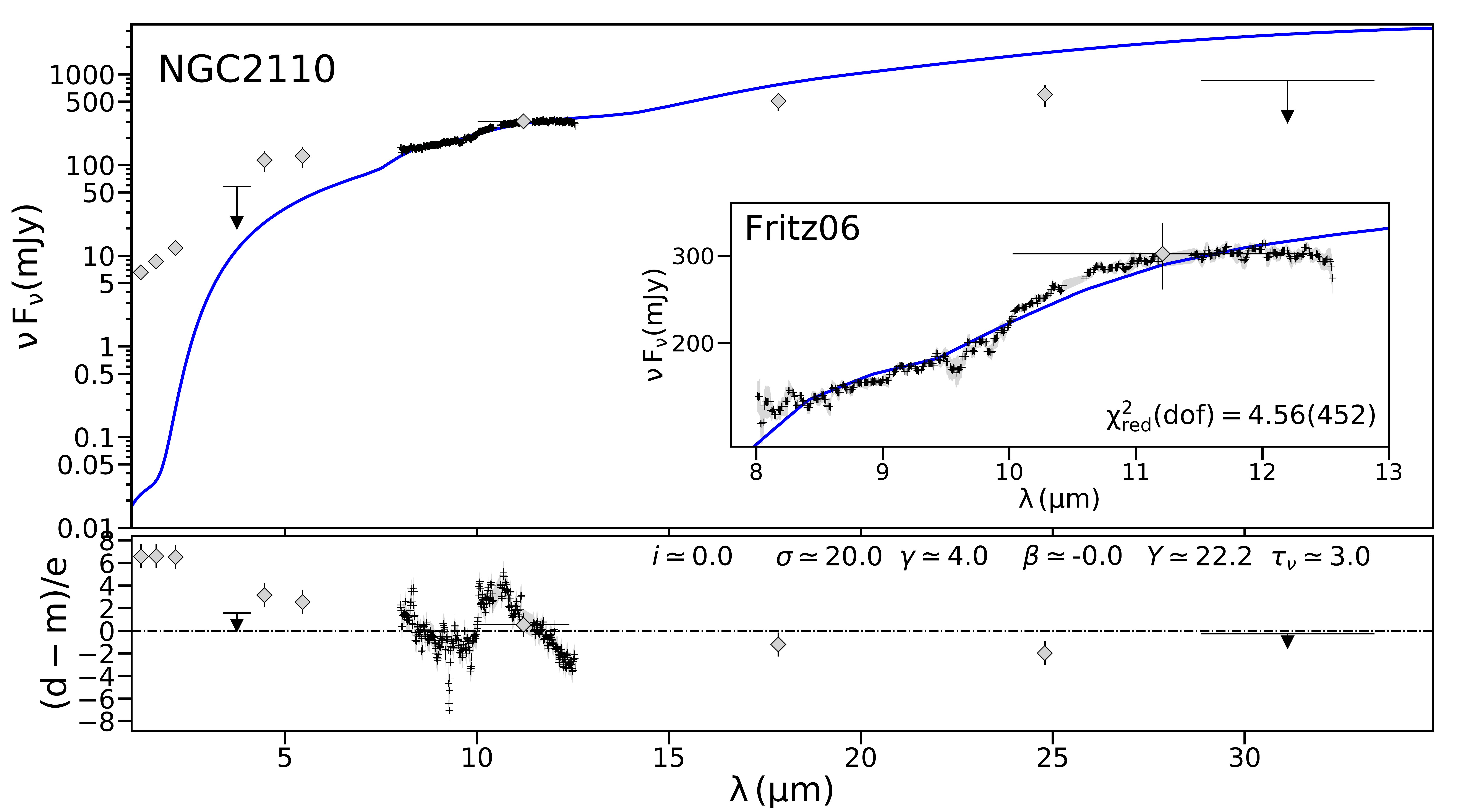}
    \includegraphics[width=0.75\columnwidth]{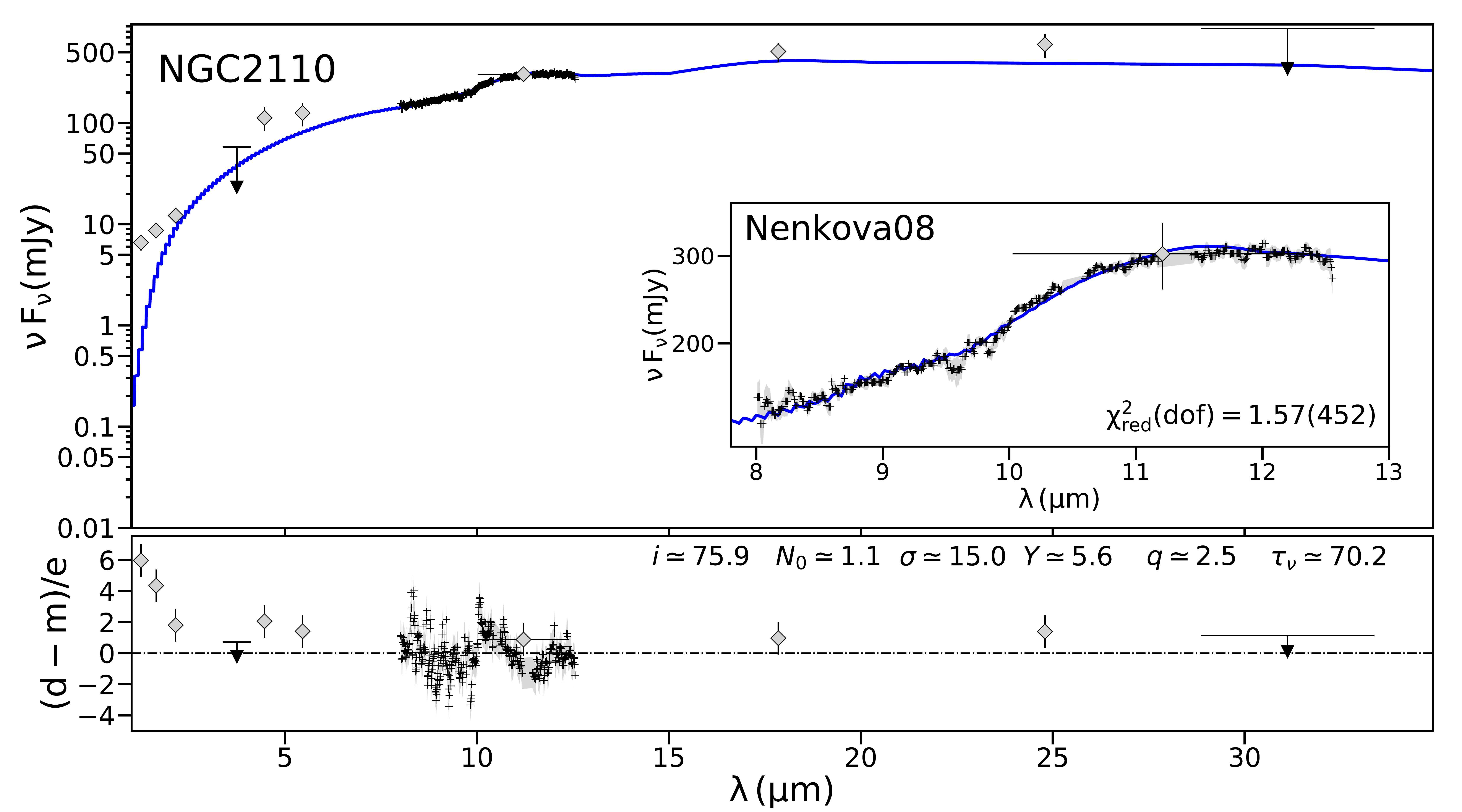}
    \includegraphics[width=0.75\columnwidth]{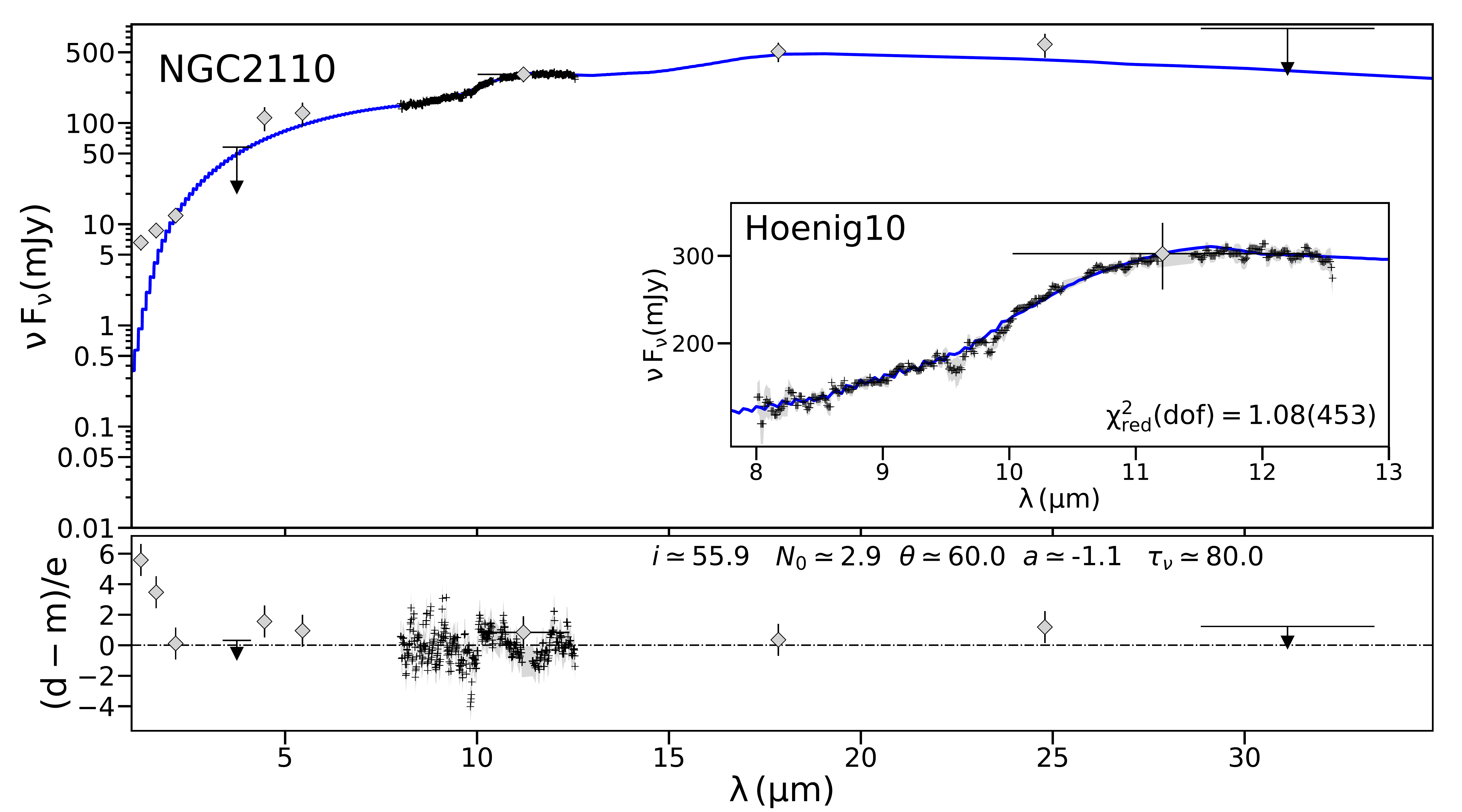}
    \includegraphics[width=0.75\columnwidth]{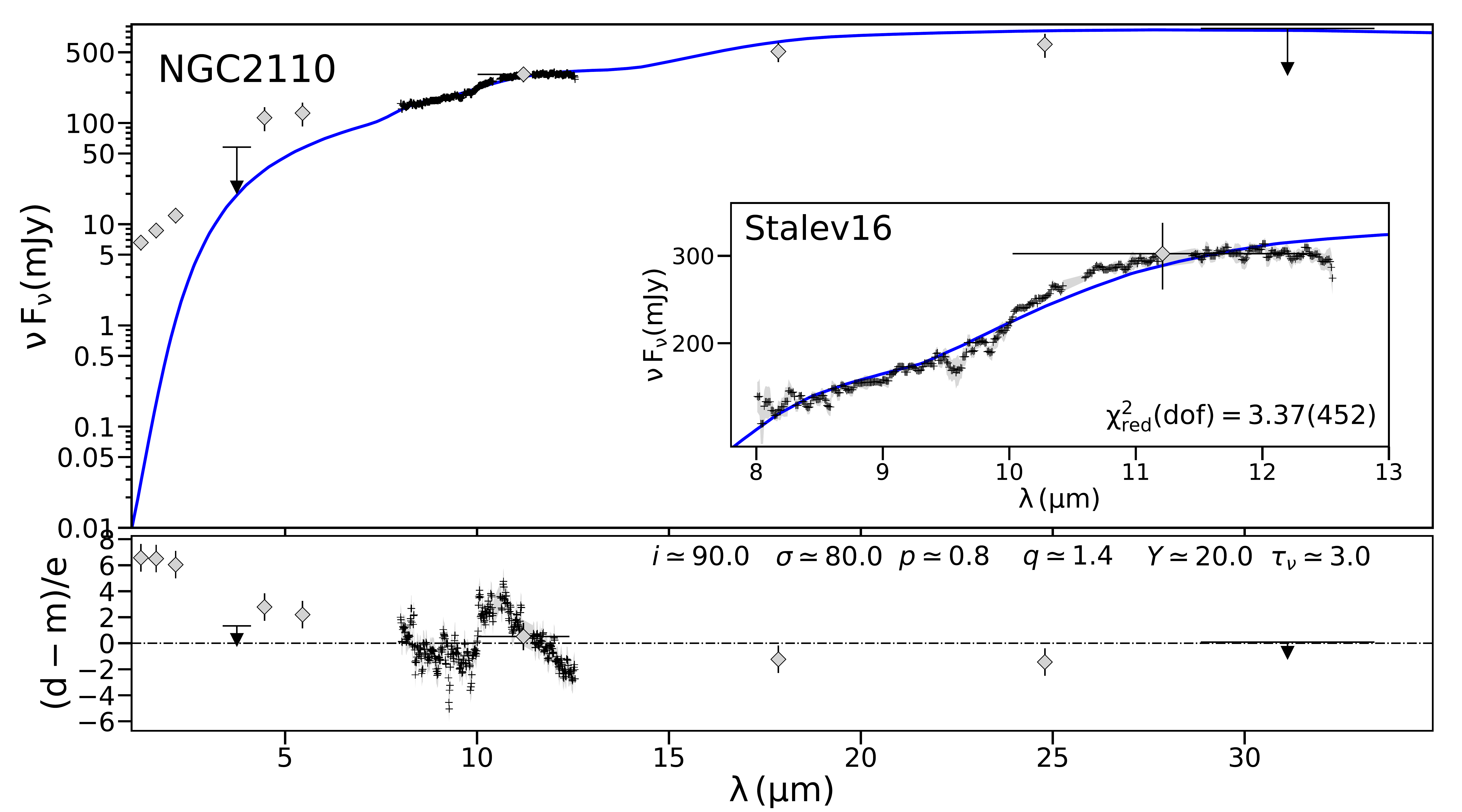}
    \includegraphics[width=0.75\columnwidth]{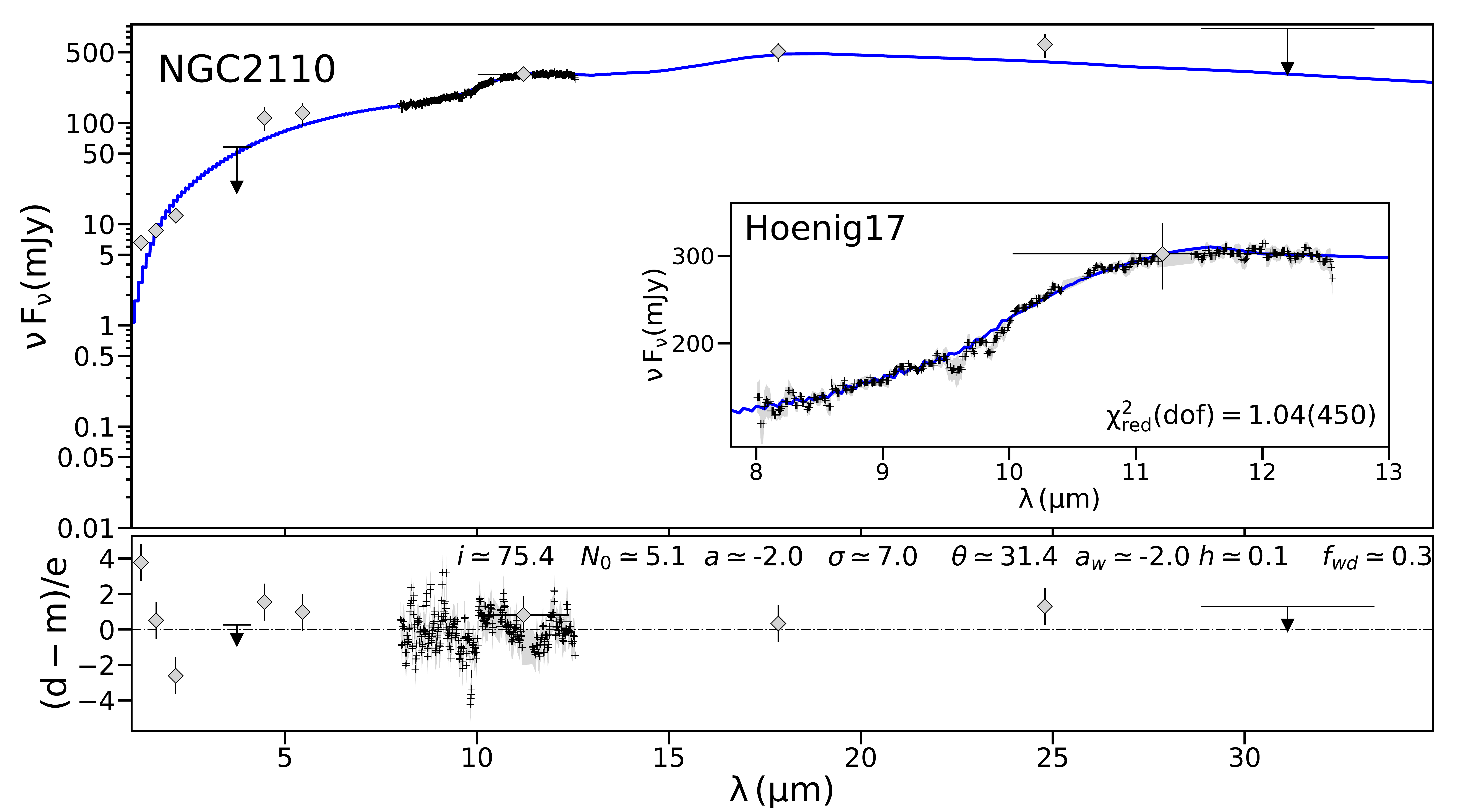}
    \includegraphics[width=0.75\columnwidth]{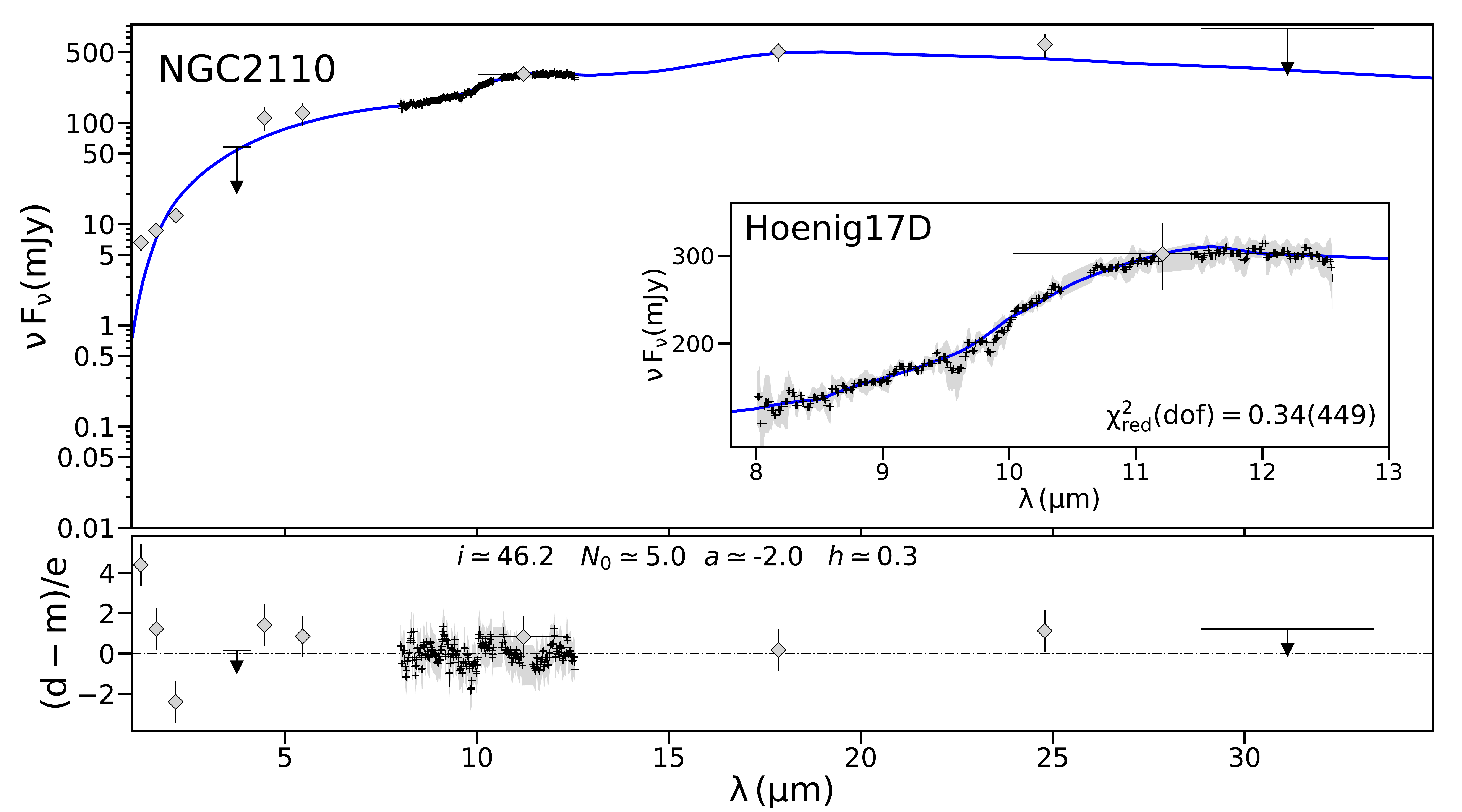}
    \caption{Same as Fig. \ref{fig:ESO005-G004} but for NGC2110.}
    \label{fig:NGC2110}
\end{figure*}
\
\begin{figure*}
    \centering
    \includegraphics[width=0.75\columnwidth]{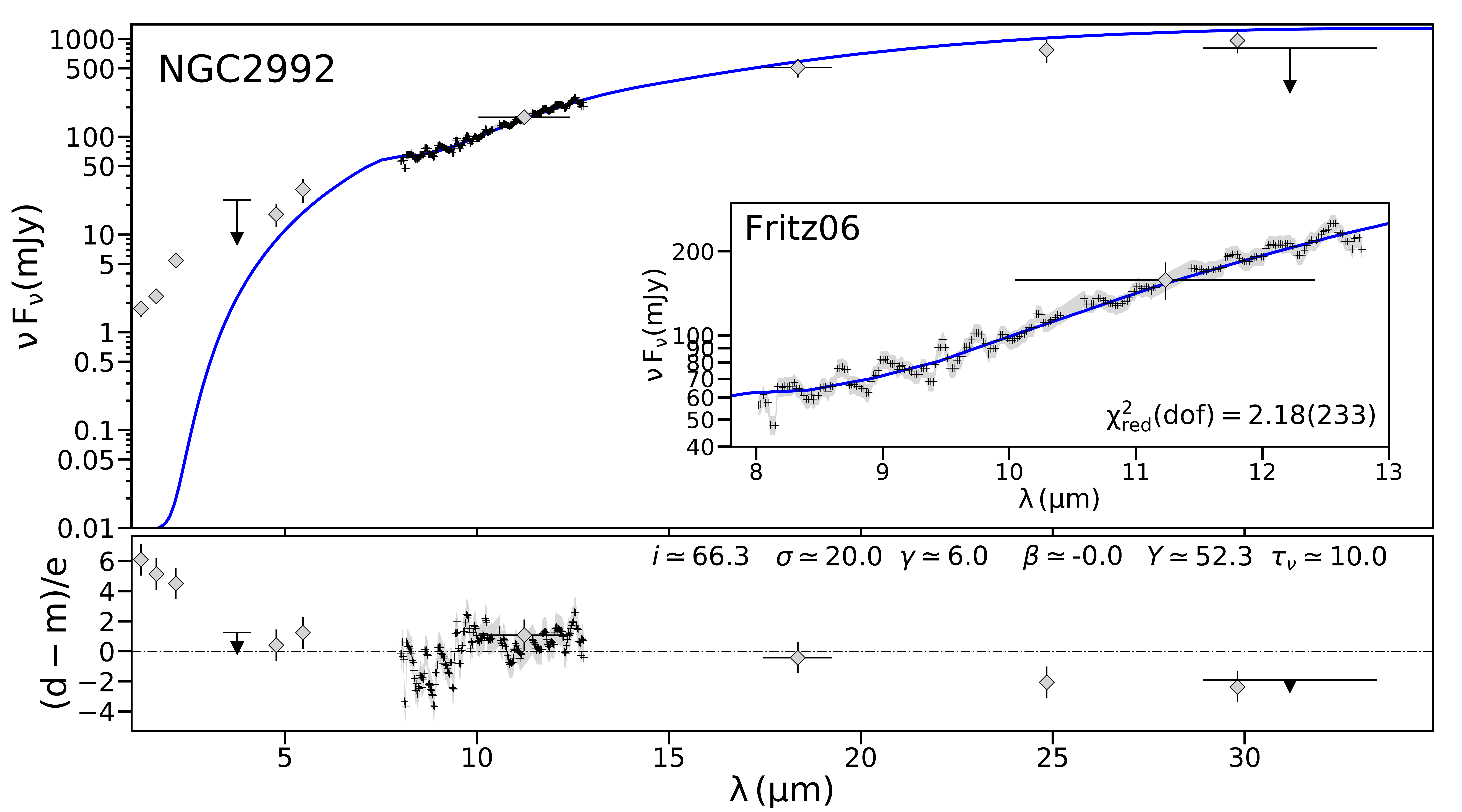}
    \includegraphics[width=0.75\columnwidth]{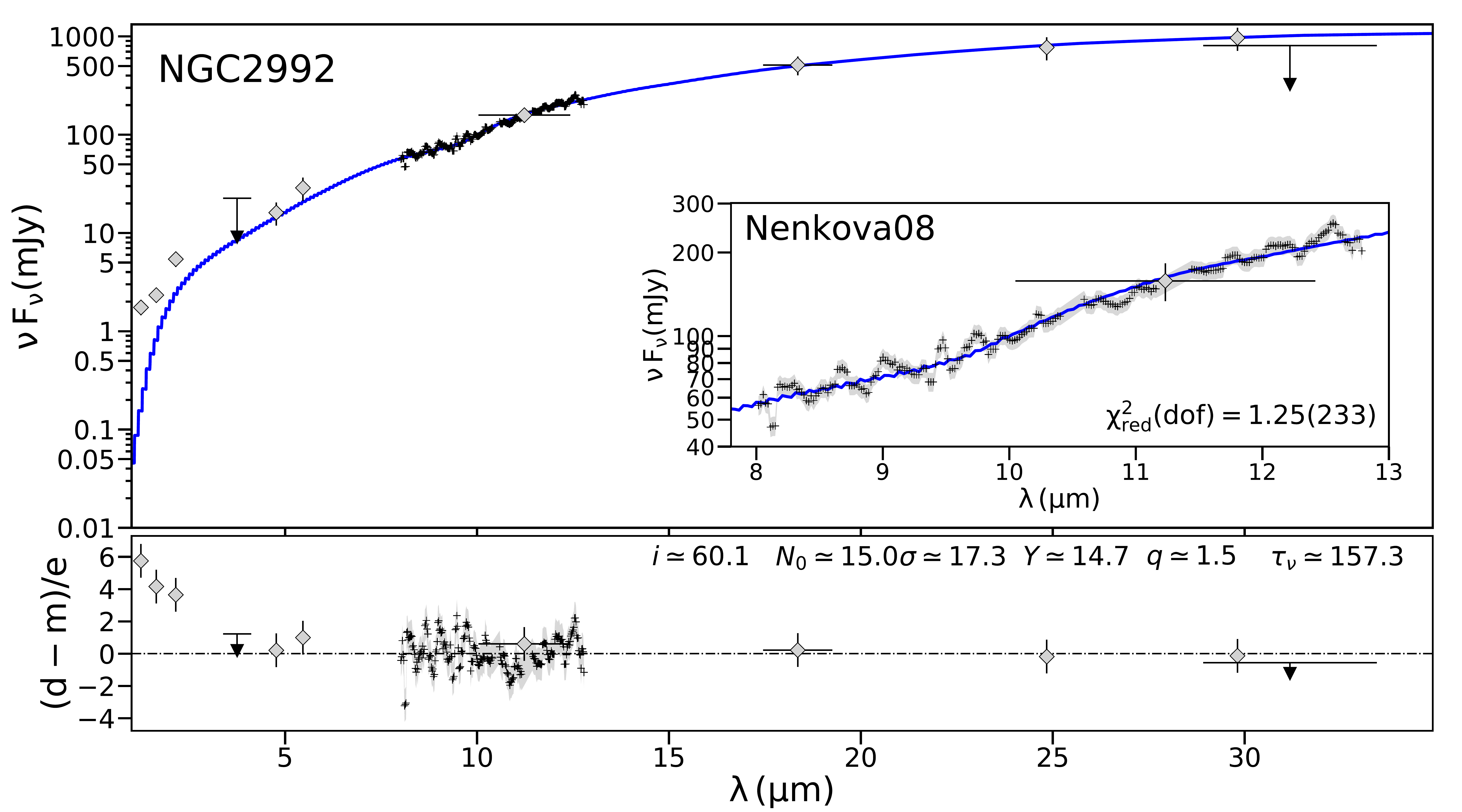}
    \includegraphics[width=0.75\columnwidth]{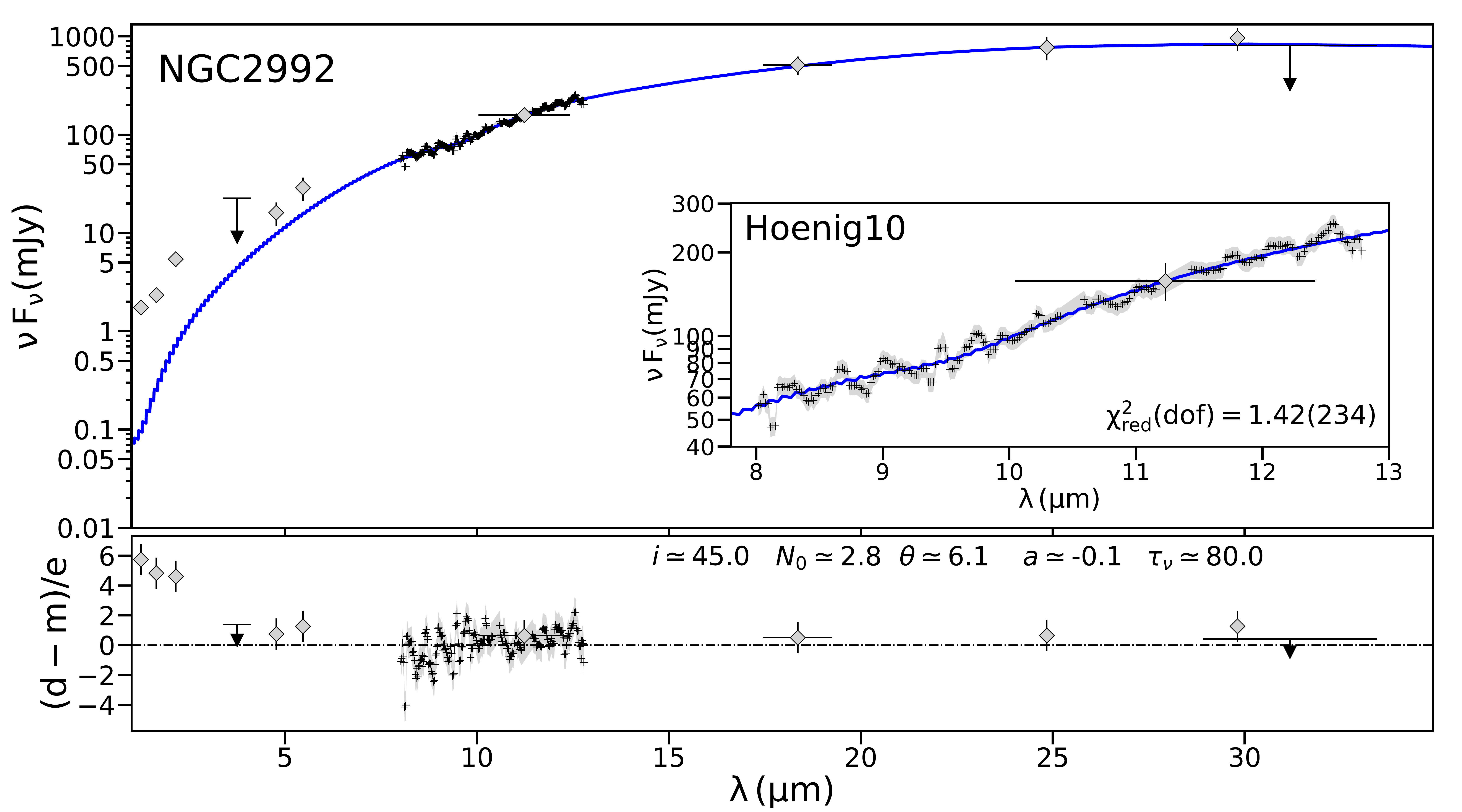}
    \includegraphics[width=0.75\columnwidth]{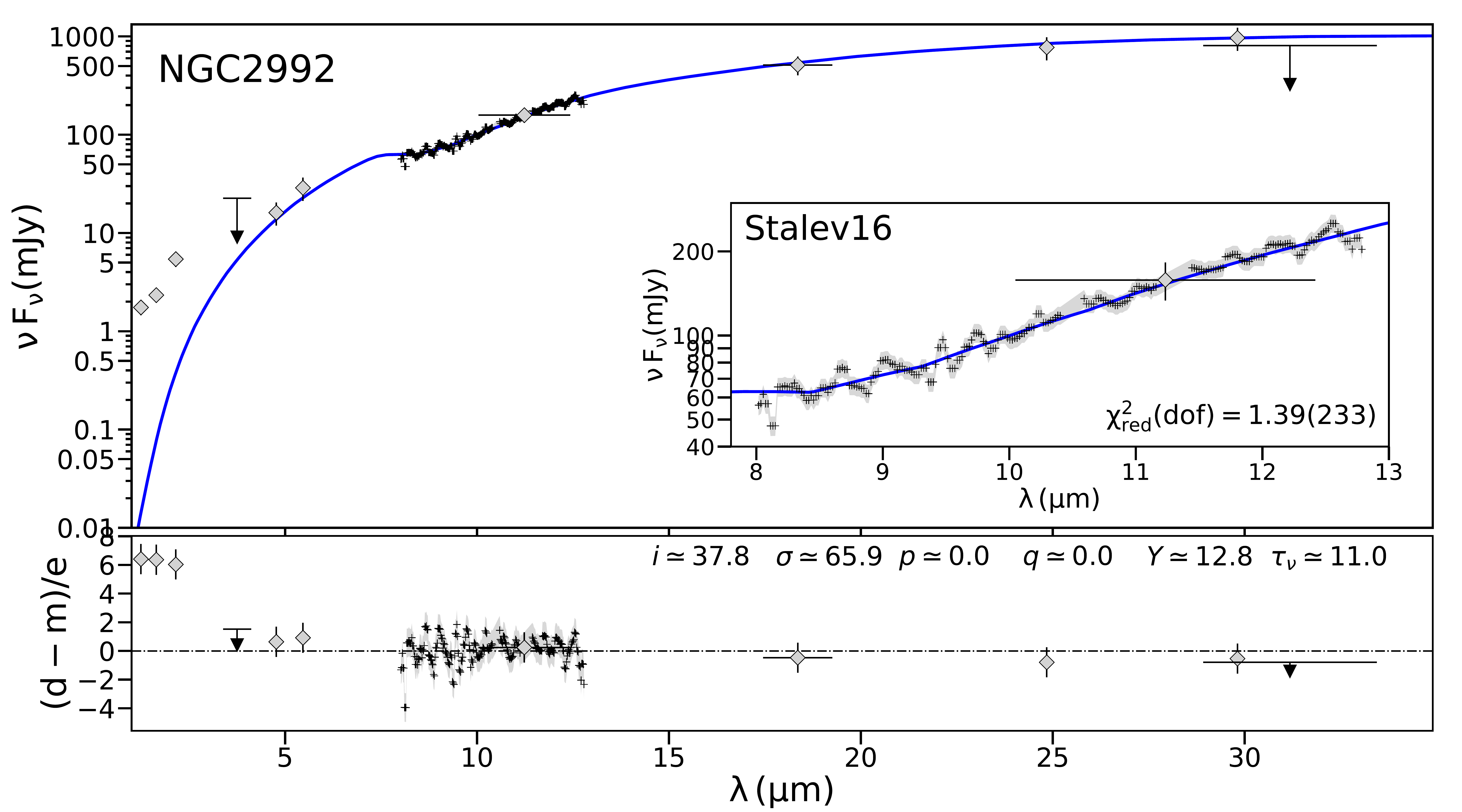}
    \includegraphics[width=0.75\columnwidth]{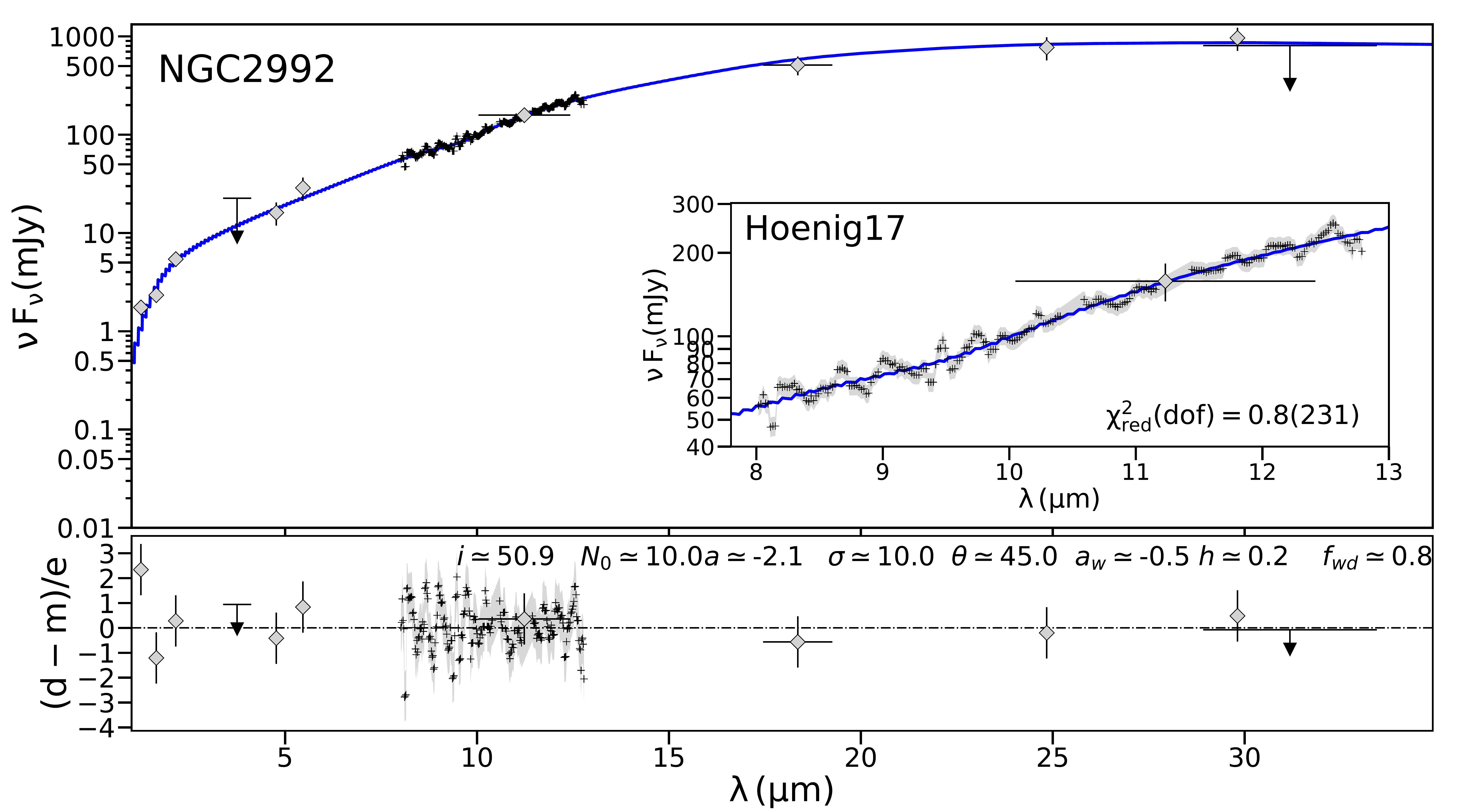}
    \includegraphics[width=0.75\columnwidth]{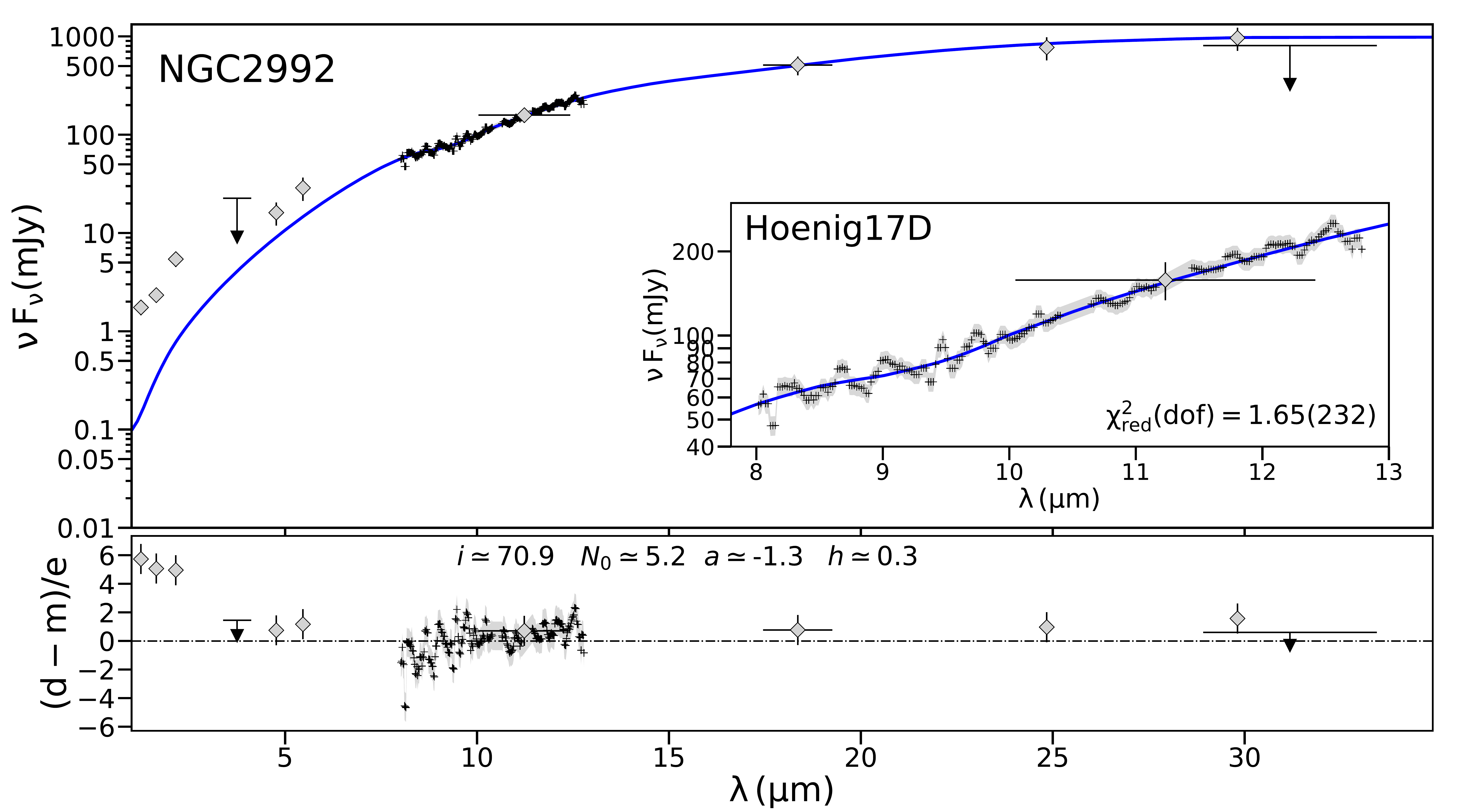}
    \caption{Same as Fig. \ref{fig:ESO005-G004} but for NGC2992.}
    \label{fig:NGC2992}
\end{figure*}

\begin{figure*}
    \centering
    \includegraphics[width=0.75\columnwidth]{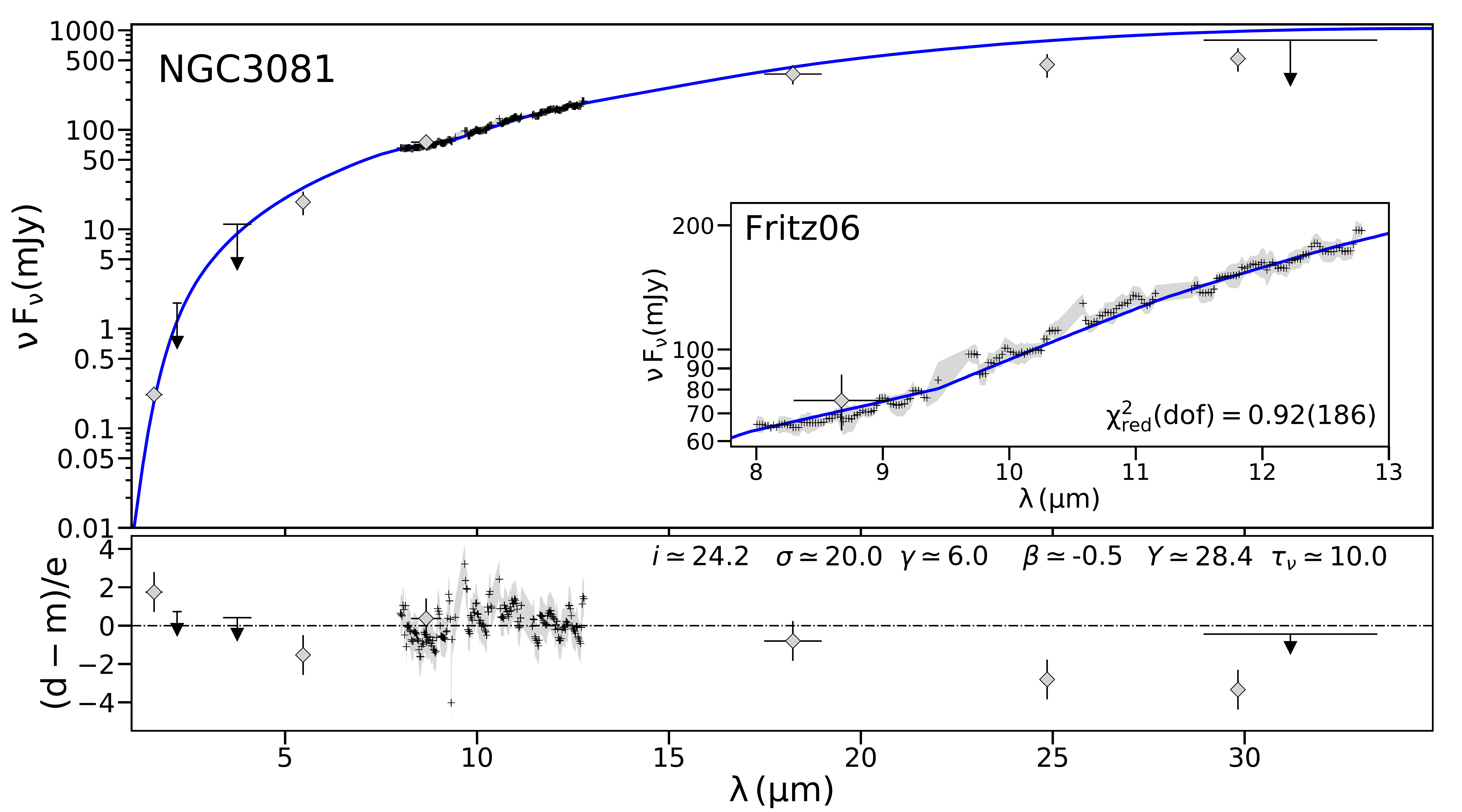}
    \includegraphics[width=0.75\columnwidth]{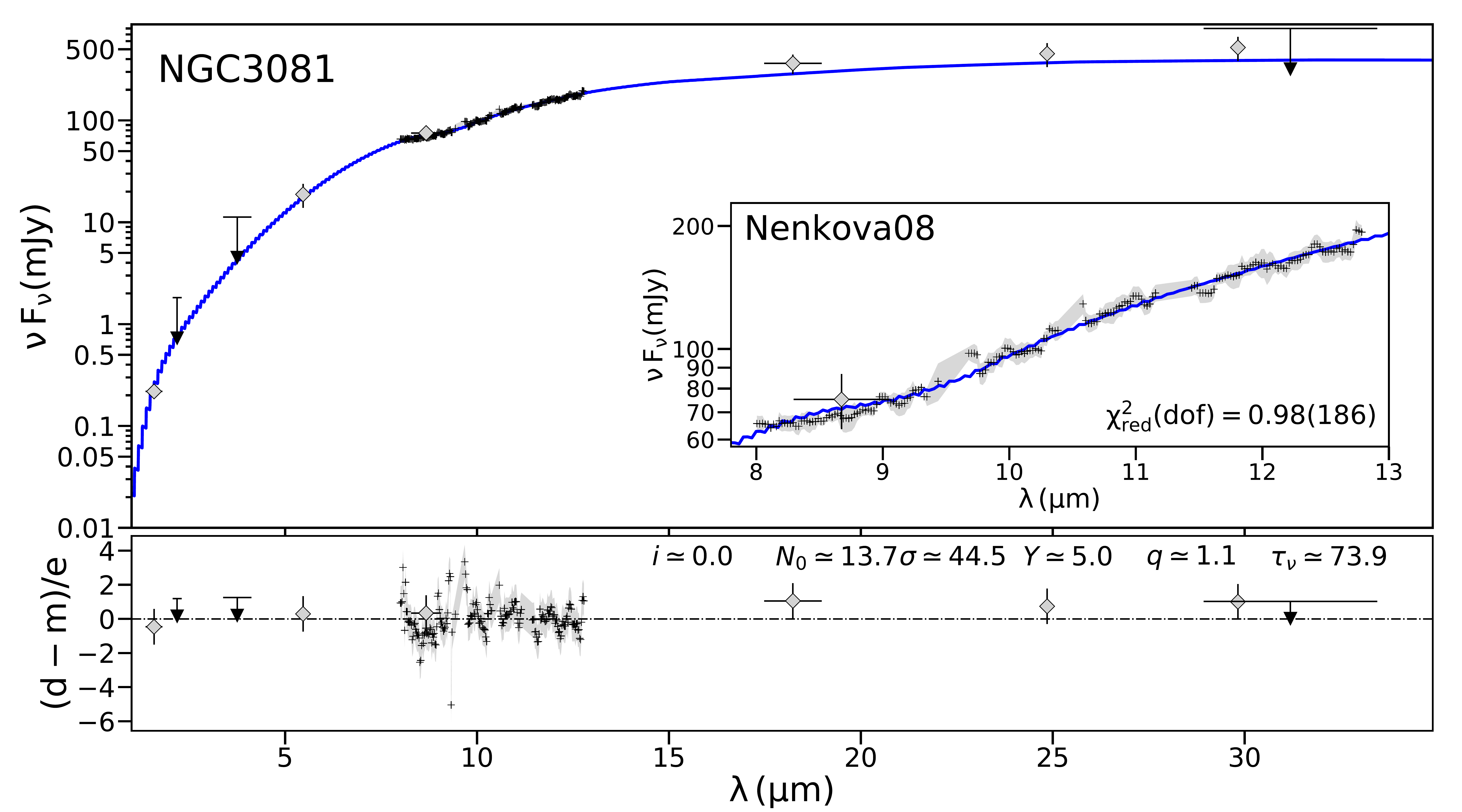}
    \includegraphics[width=0.75\columnwidth]{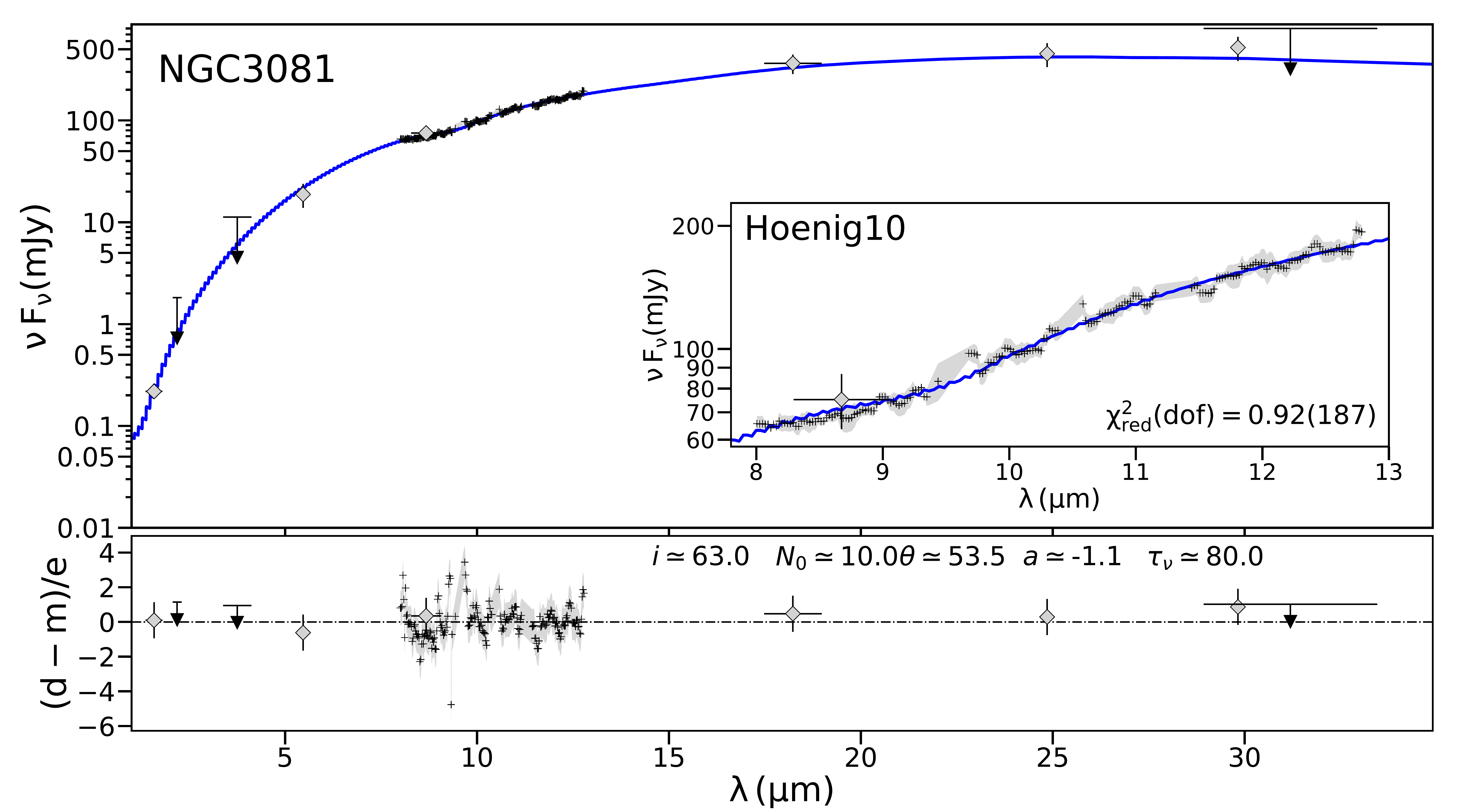}
    \includegraphics[width=0.75\columnwidth]{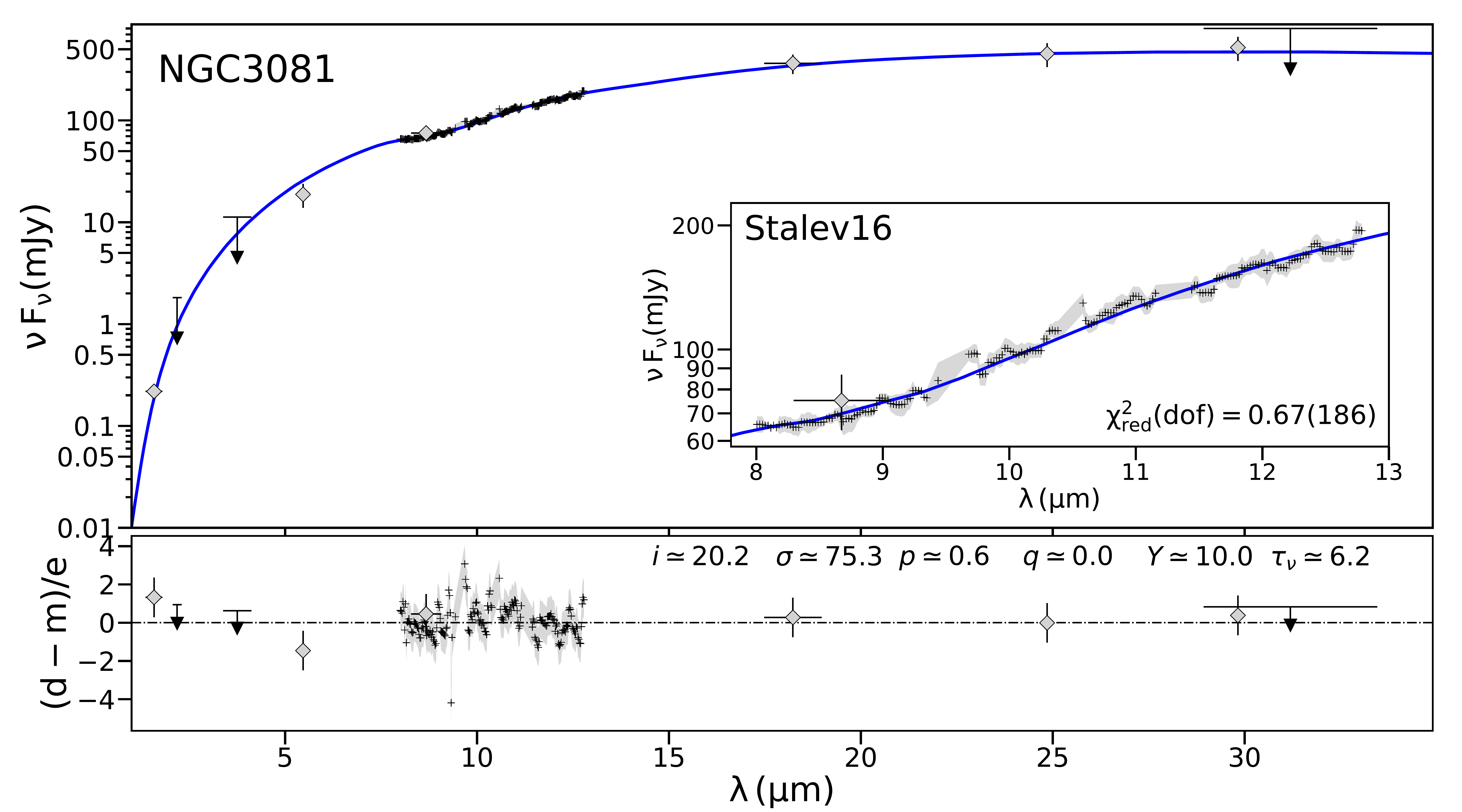}
    \includegraphics[width=0.75\columnwidth]{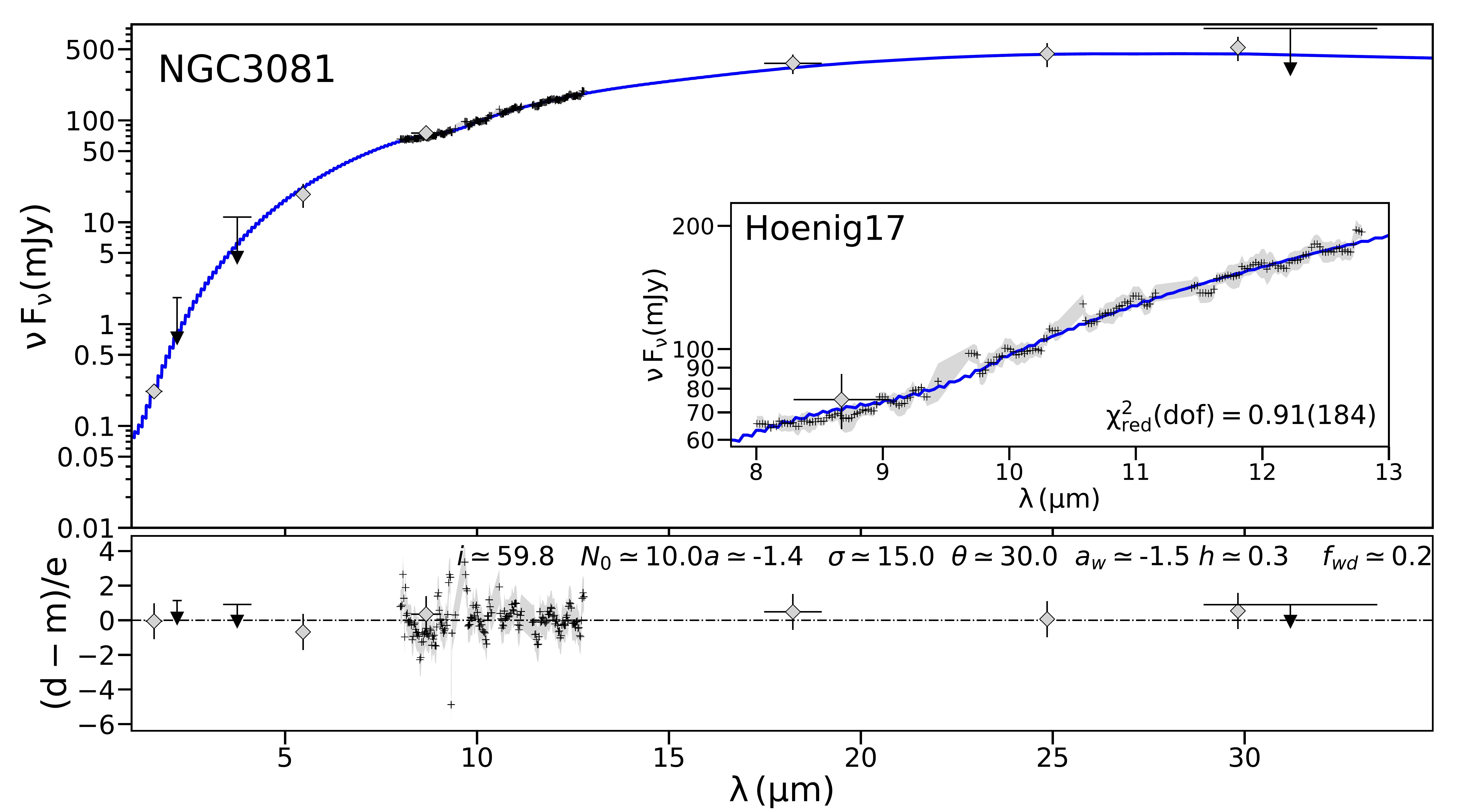}
    \includegraphics[width=0.75\columnwidth]{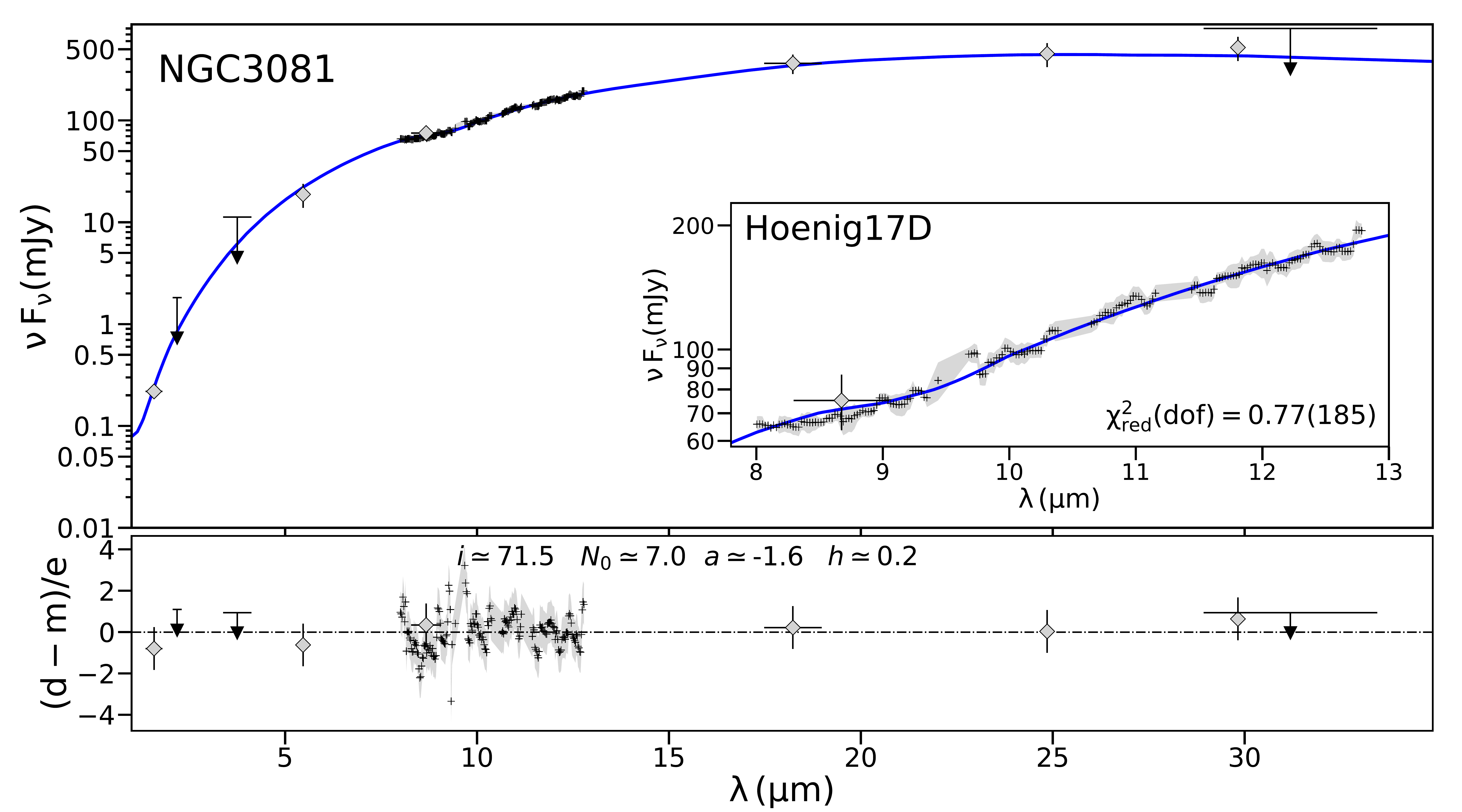}
    \caption{Same as Fig. \ref{fig:ESO005-G004} but for NGC3081.}
    \label{fig:NGC3081}
\end{figure*}

\begin{figure*}
    \centering
    \includegraphics[width=0.75\columnwidth]{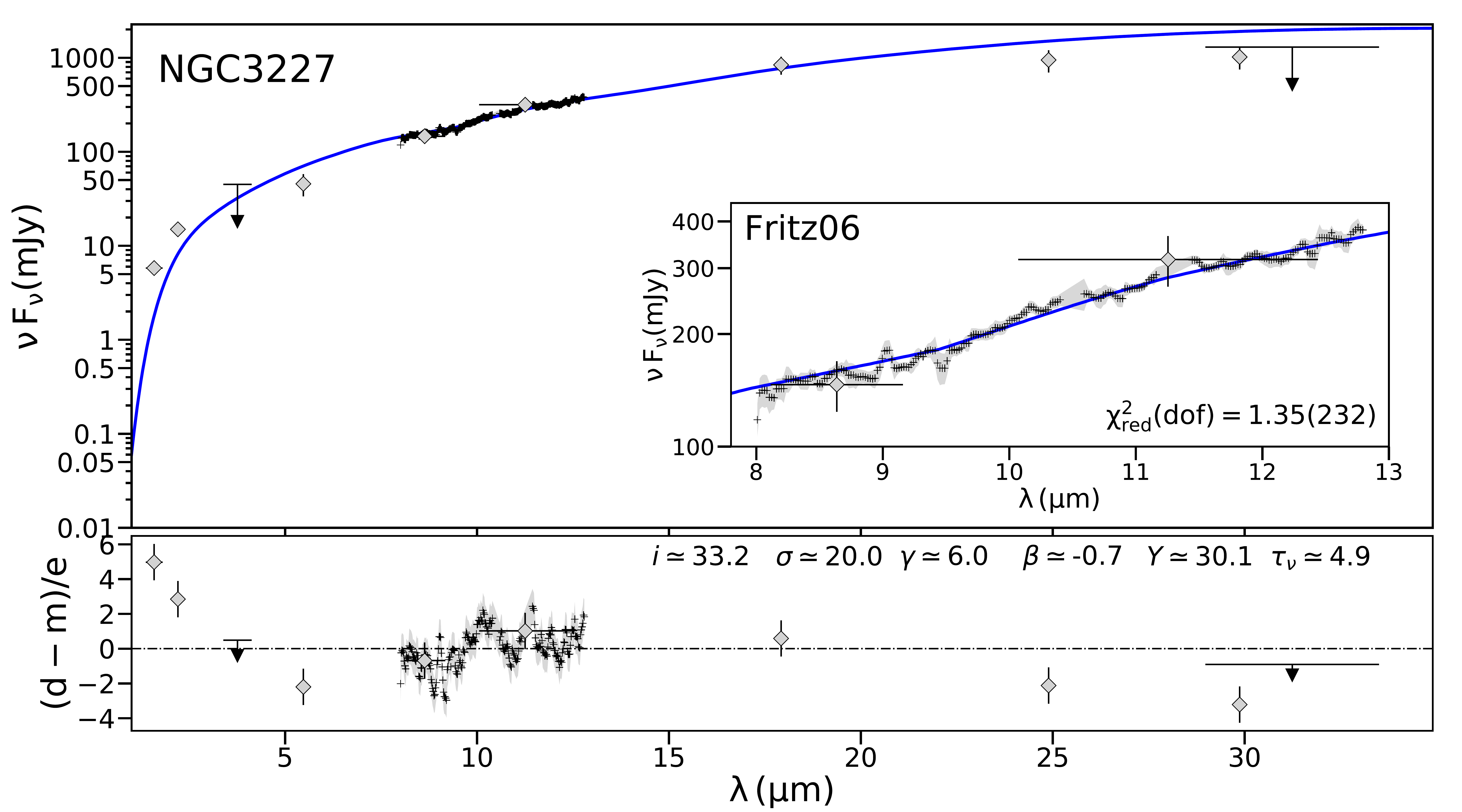}
    \includegraphics[width=0.75\columnwidth]{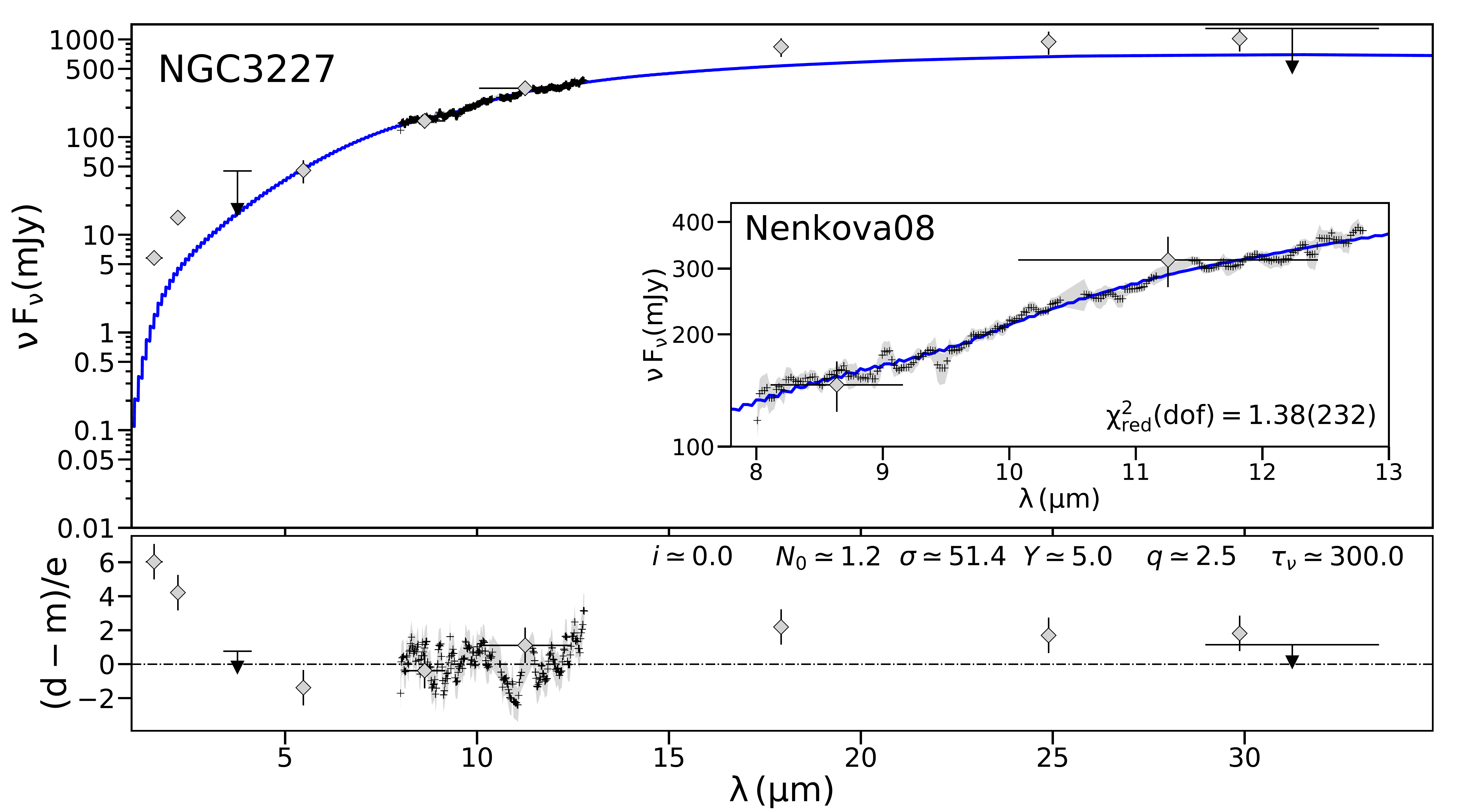}
    \includegraphics[width=0.75\columnwidth]{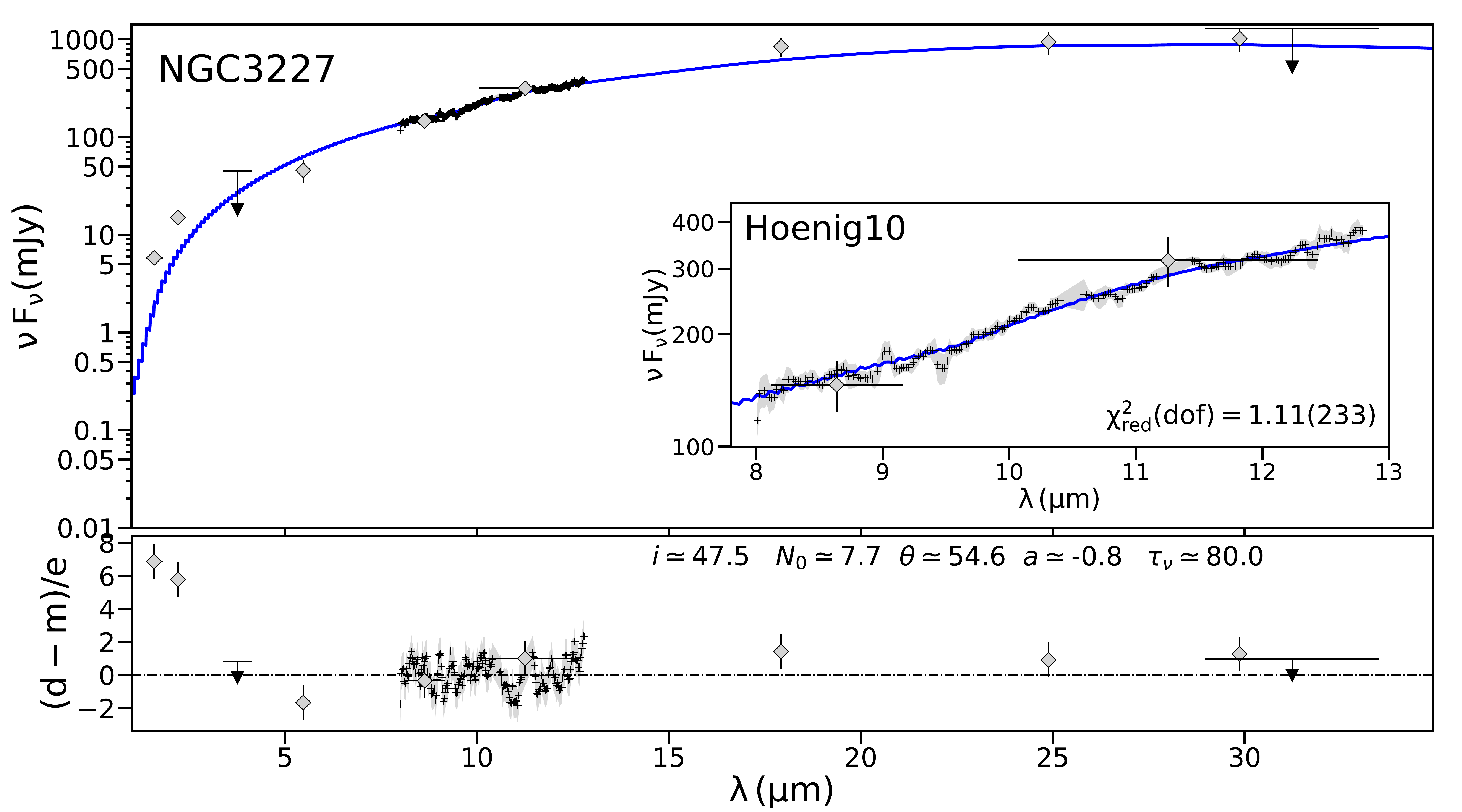}
    \includegraphics[width=0.75\columnwidth]{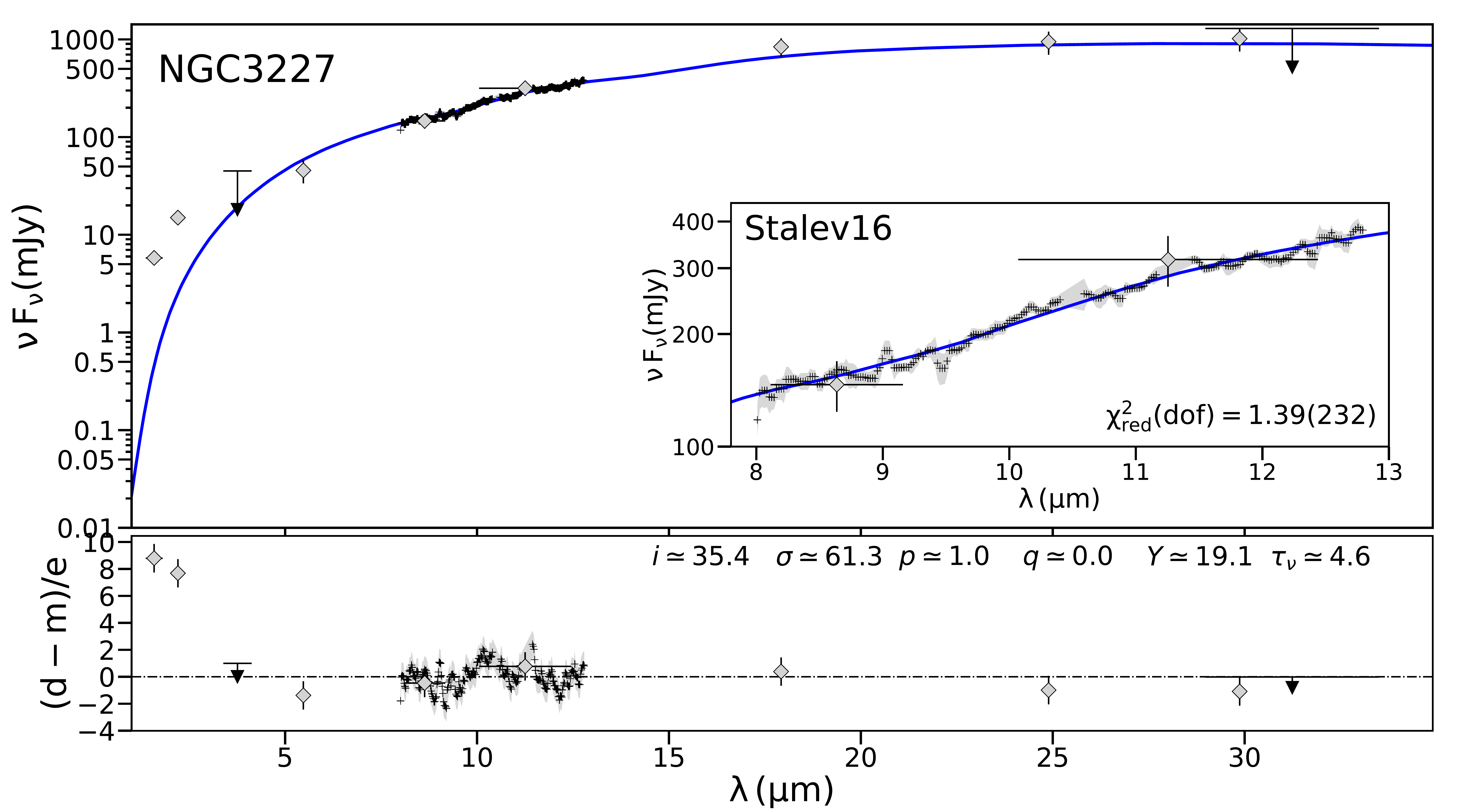}
    \includegraphics[width=0.75\columnwidth]{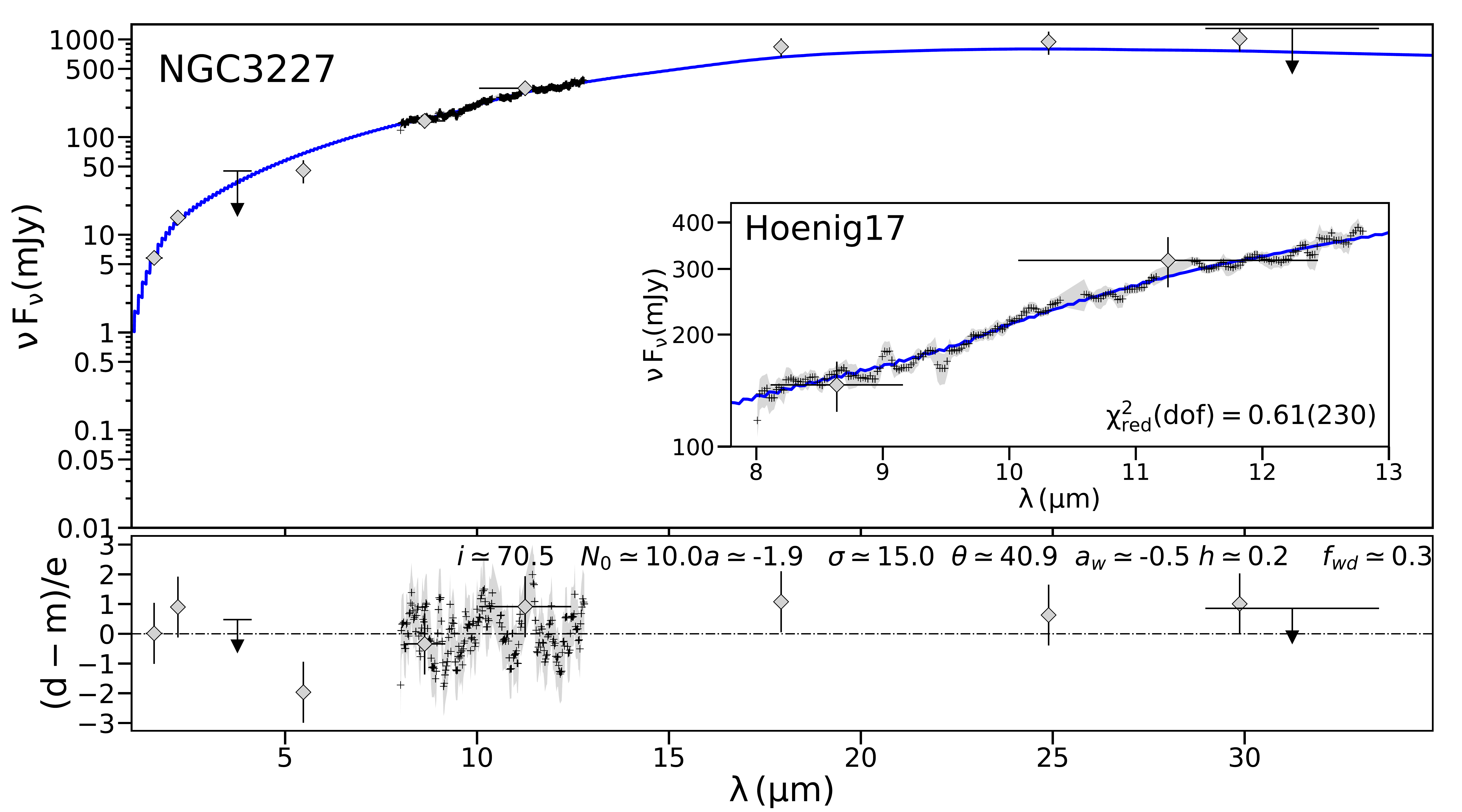}
    \includegraphics[width=0.75\columnwidth]{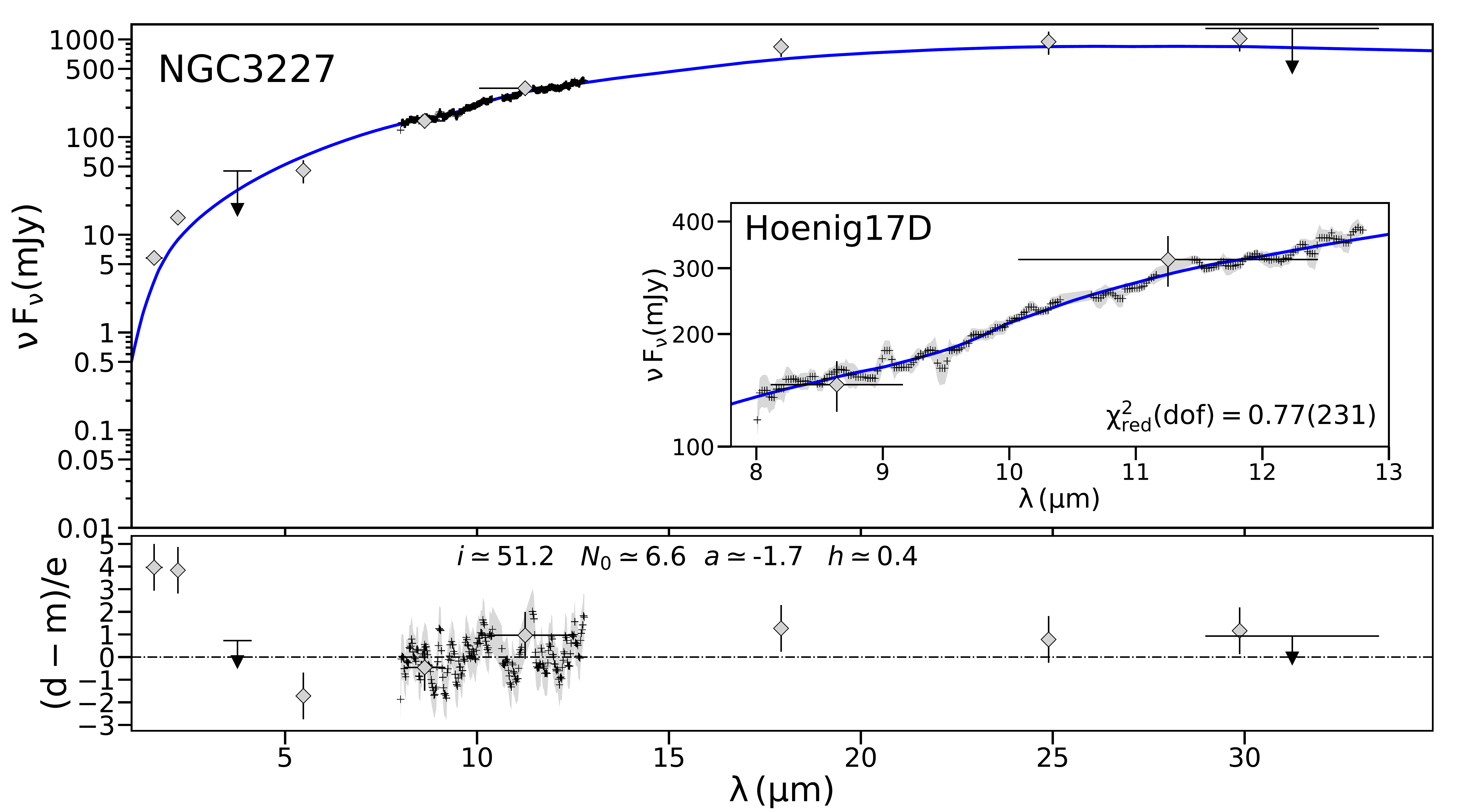}
    \caption{Same as Fig. \ref{fig:ESO005-G004} but for NGC3227.}
    \label{fig:NGC3227}
\end{figure*}

\begin{figure*}
    \centering
    \includegraphics[width=0.75\columnwidth]{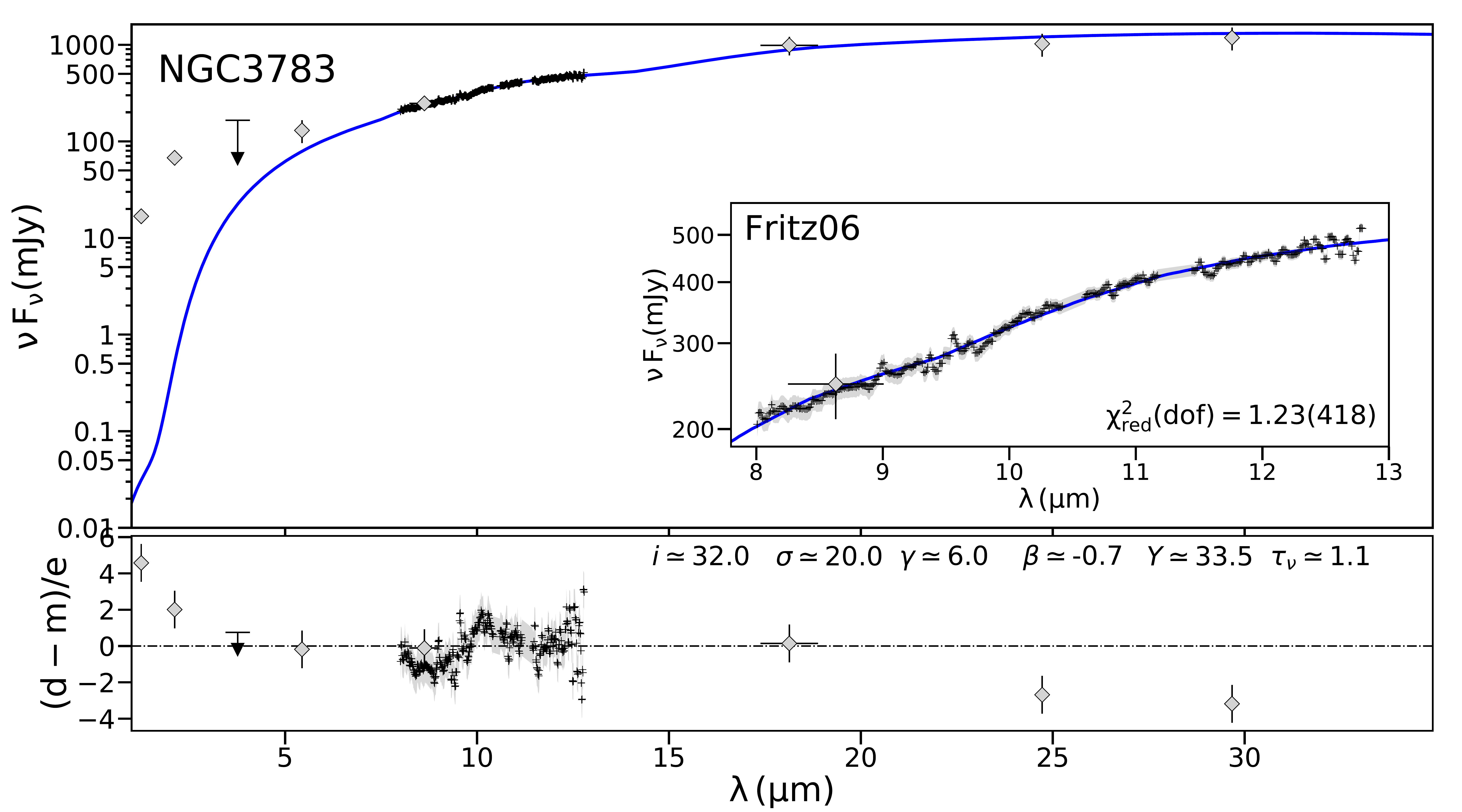}
    \includegraphics[width=0.75\columnwidth]{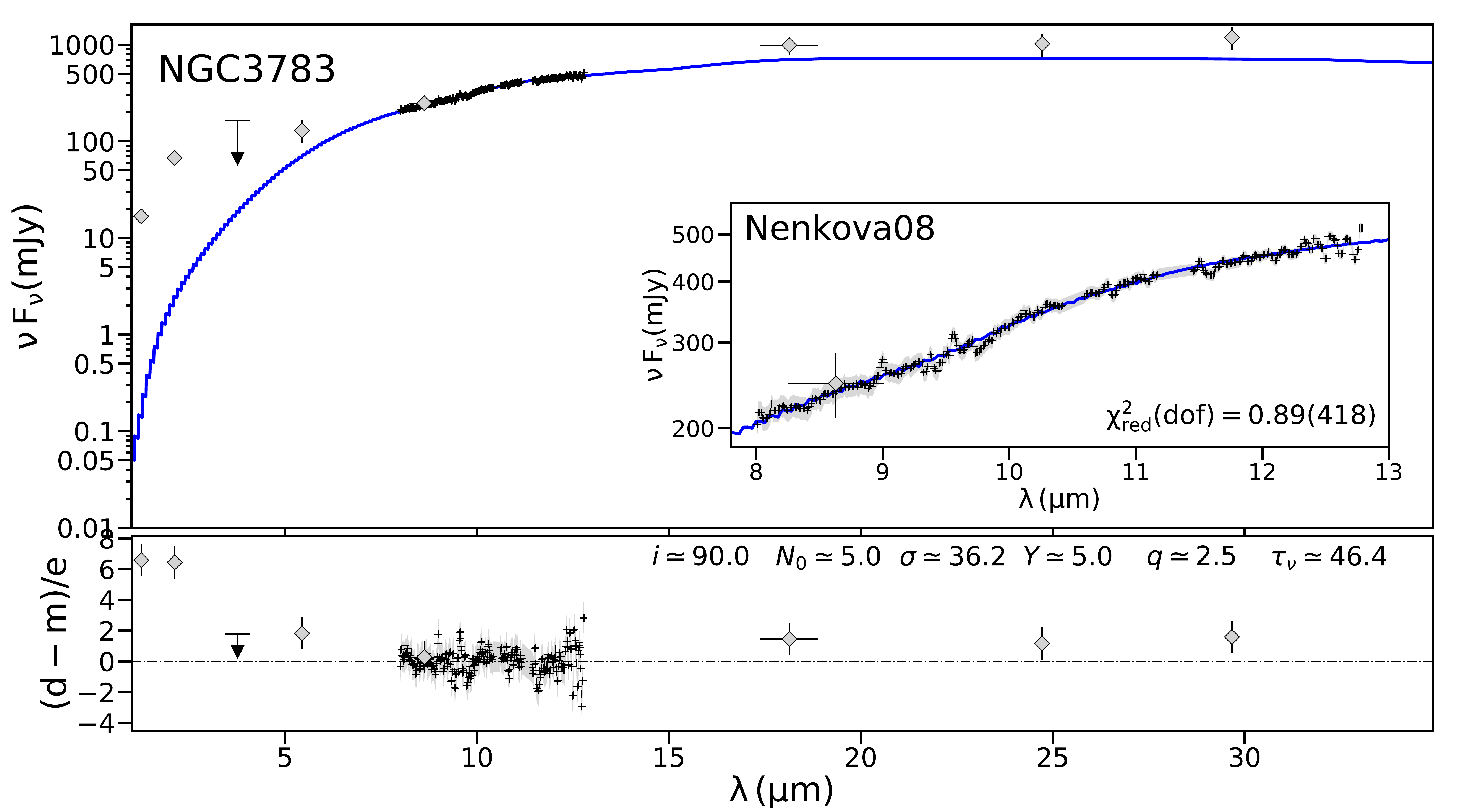}
    \includegraphics[width=0.75\columnwidth]{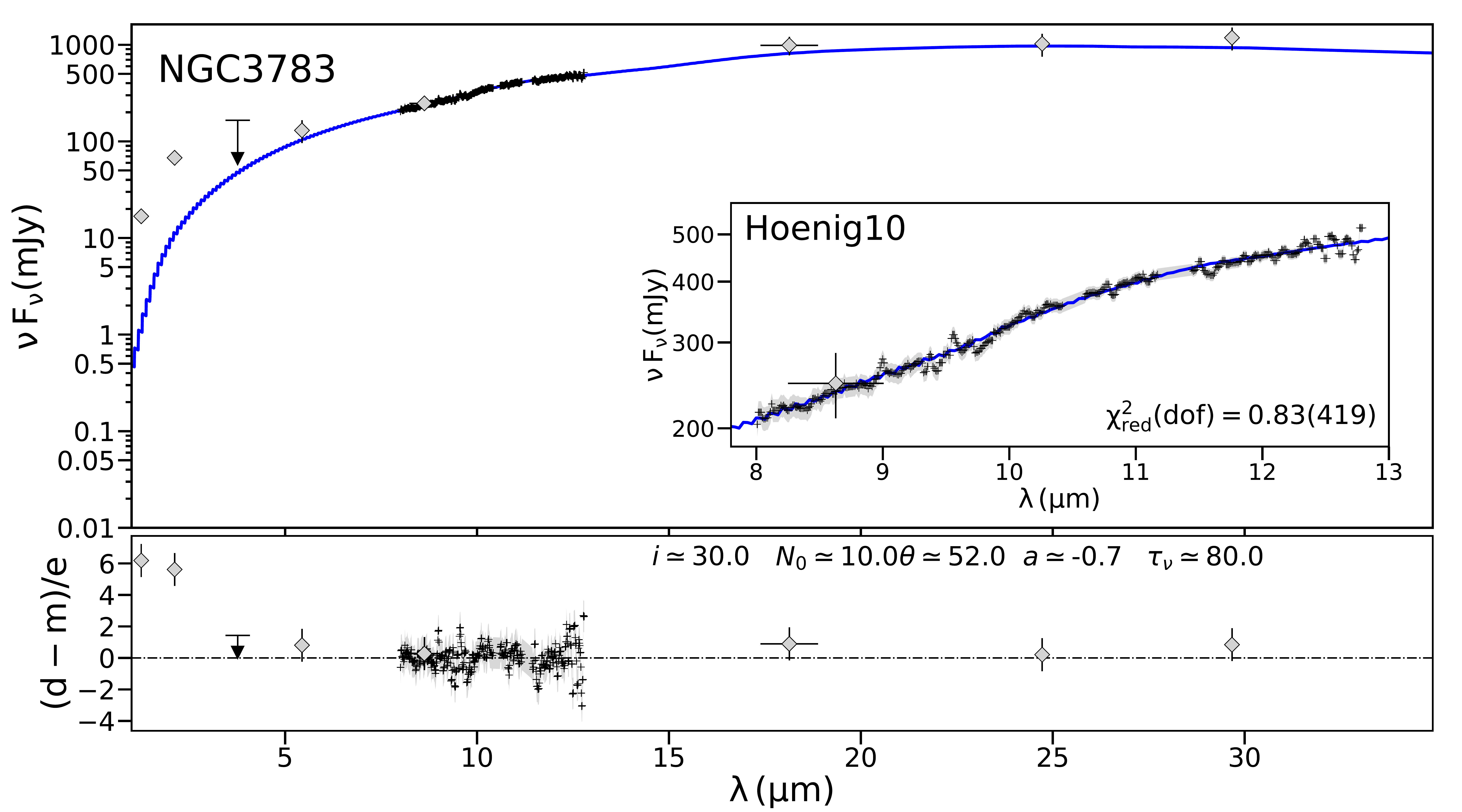}
    \includegraphics[width=0.75\columnwidth]{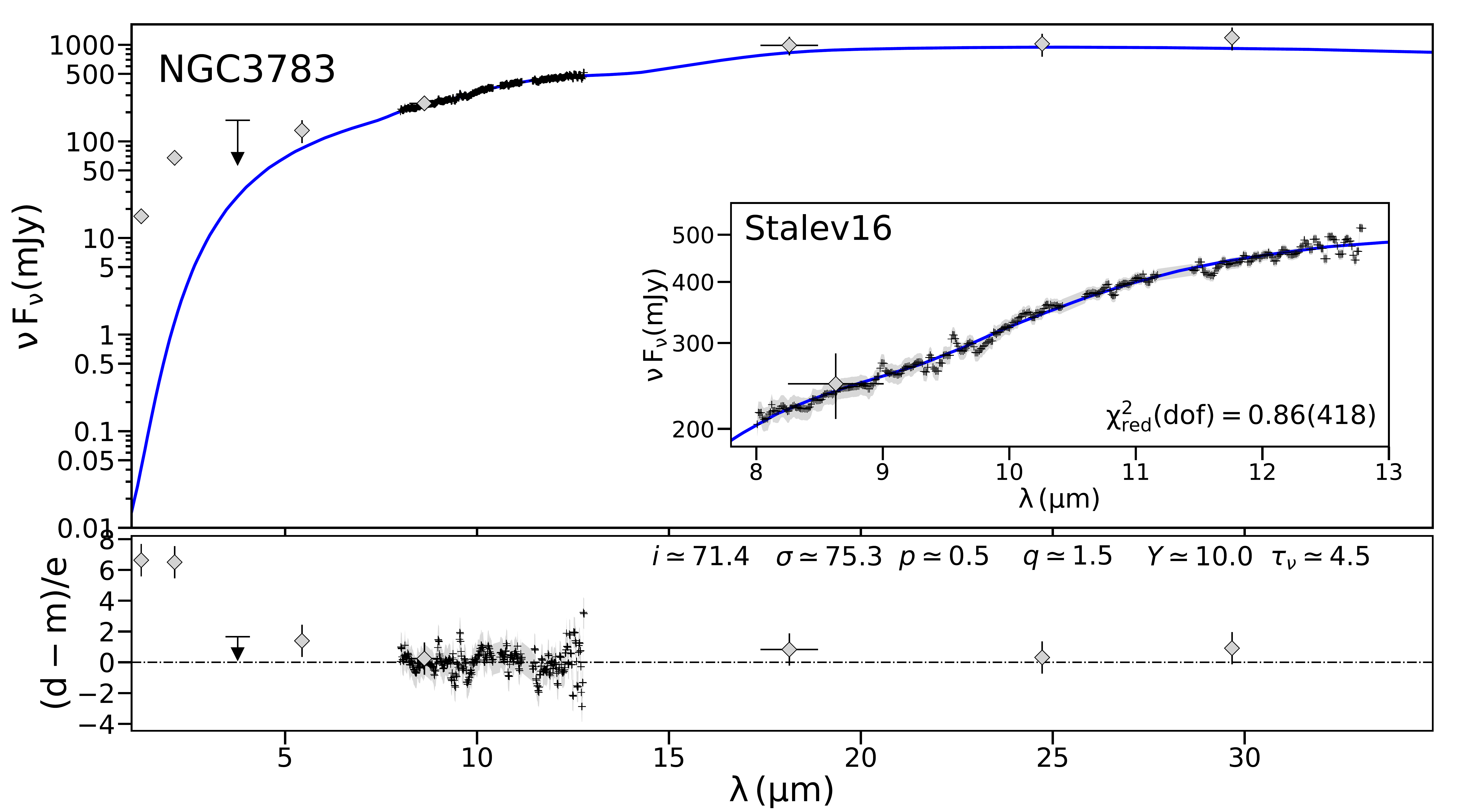}
    \includegraphics[width=0.75\columnwidth]{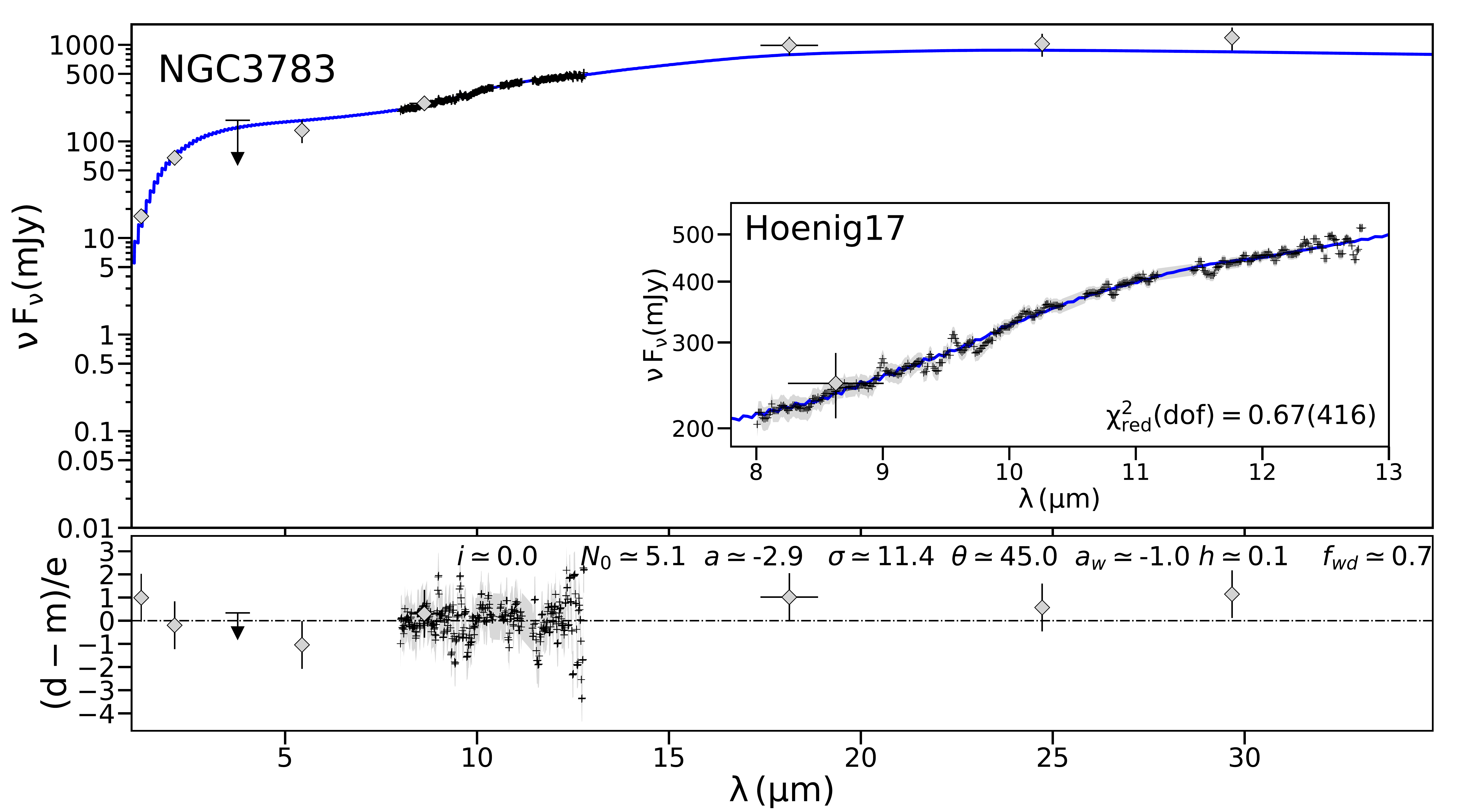}
    \includegraphics[width=0.75\columnwidth]{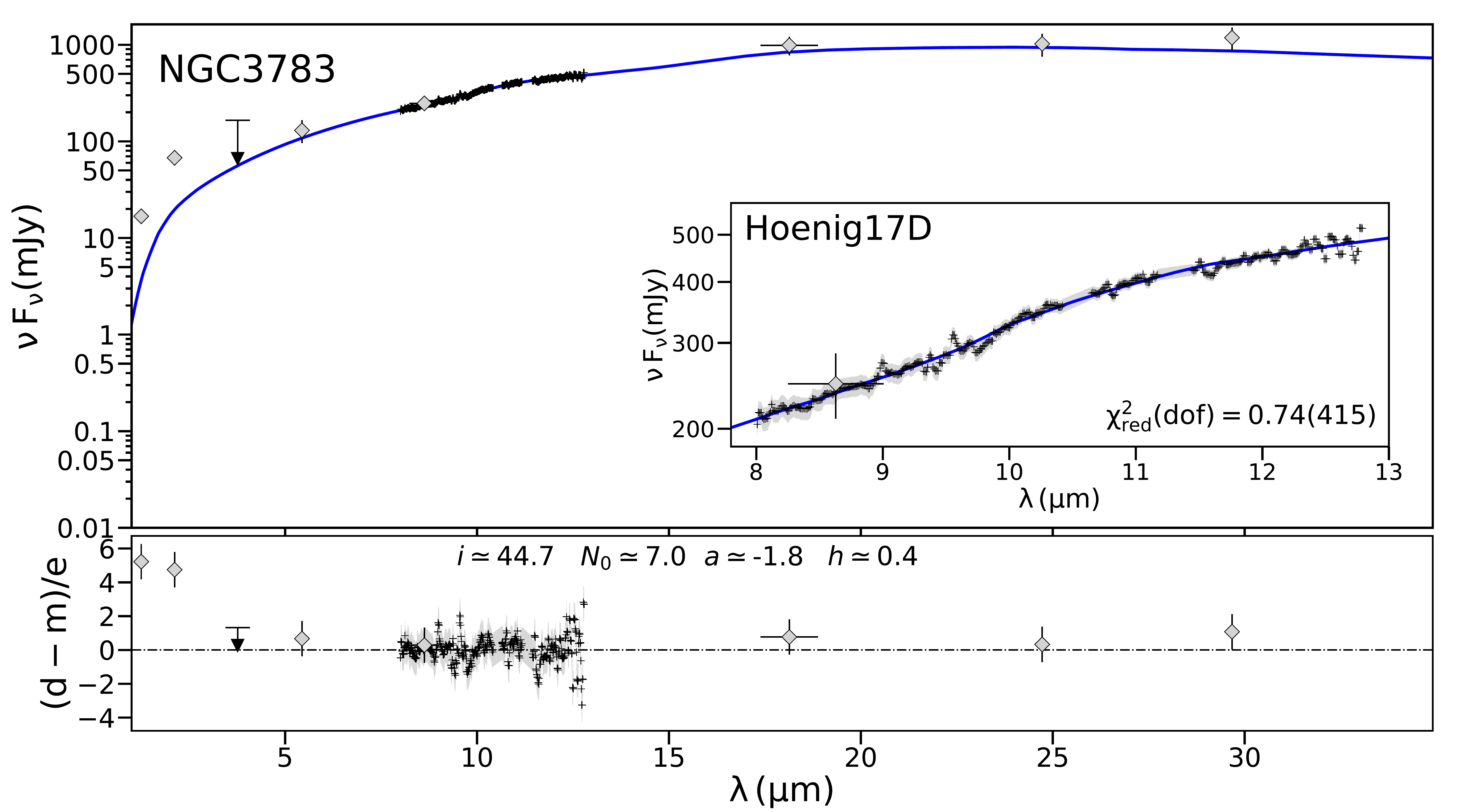}
    \caption{Same as Fig. \ref{fig:ESO005-G004} but for NGC3783.}
    \label{fig:NGC3783}
\end{figure*}

\begin{figure*}
    \centering
    \includegraphics[width=0.75\columnwidth]{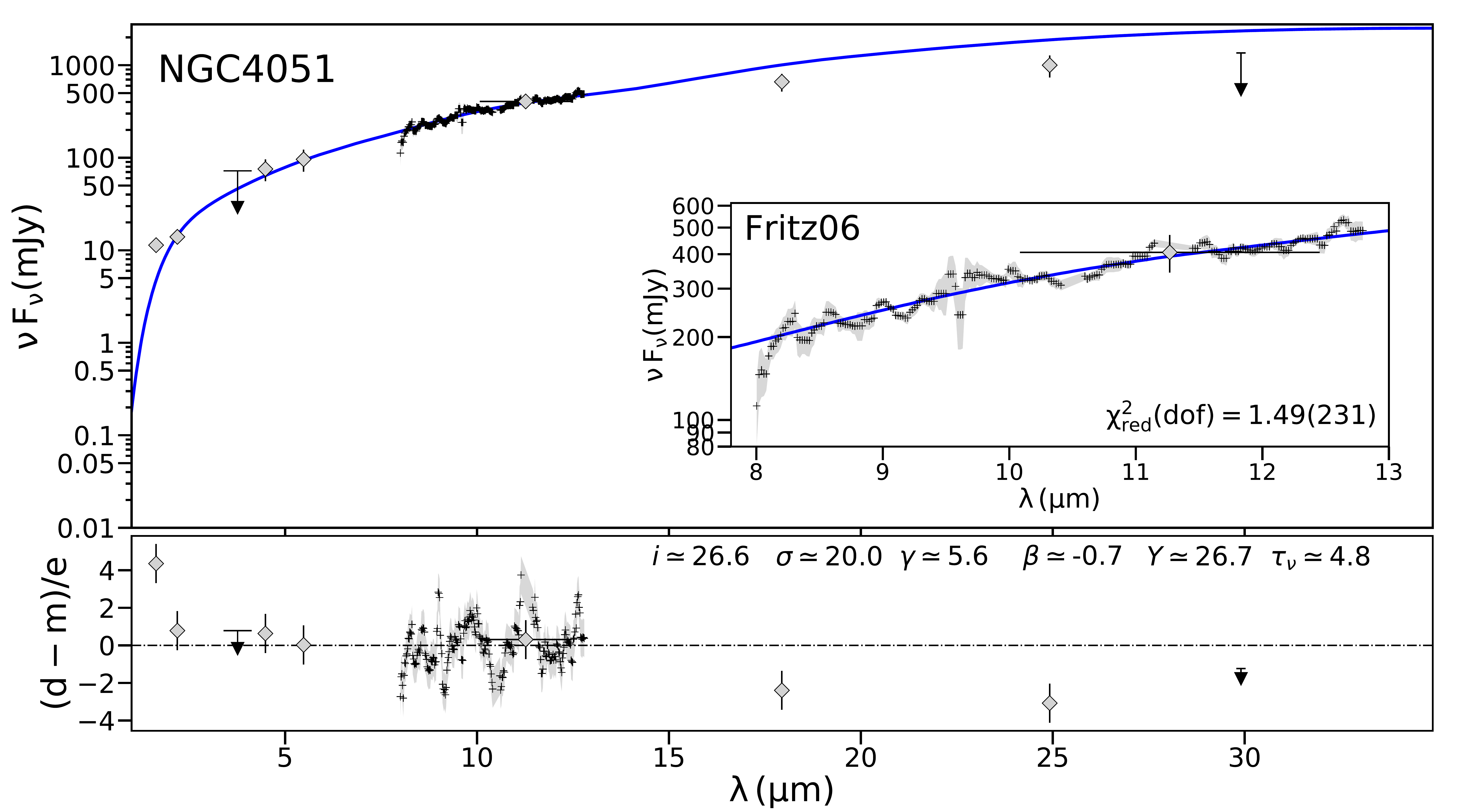}
    \includegraphics[width=0.75\columnwidth]{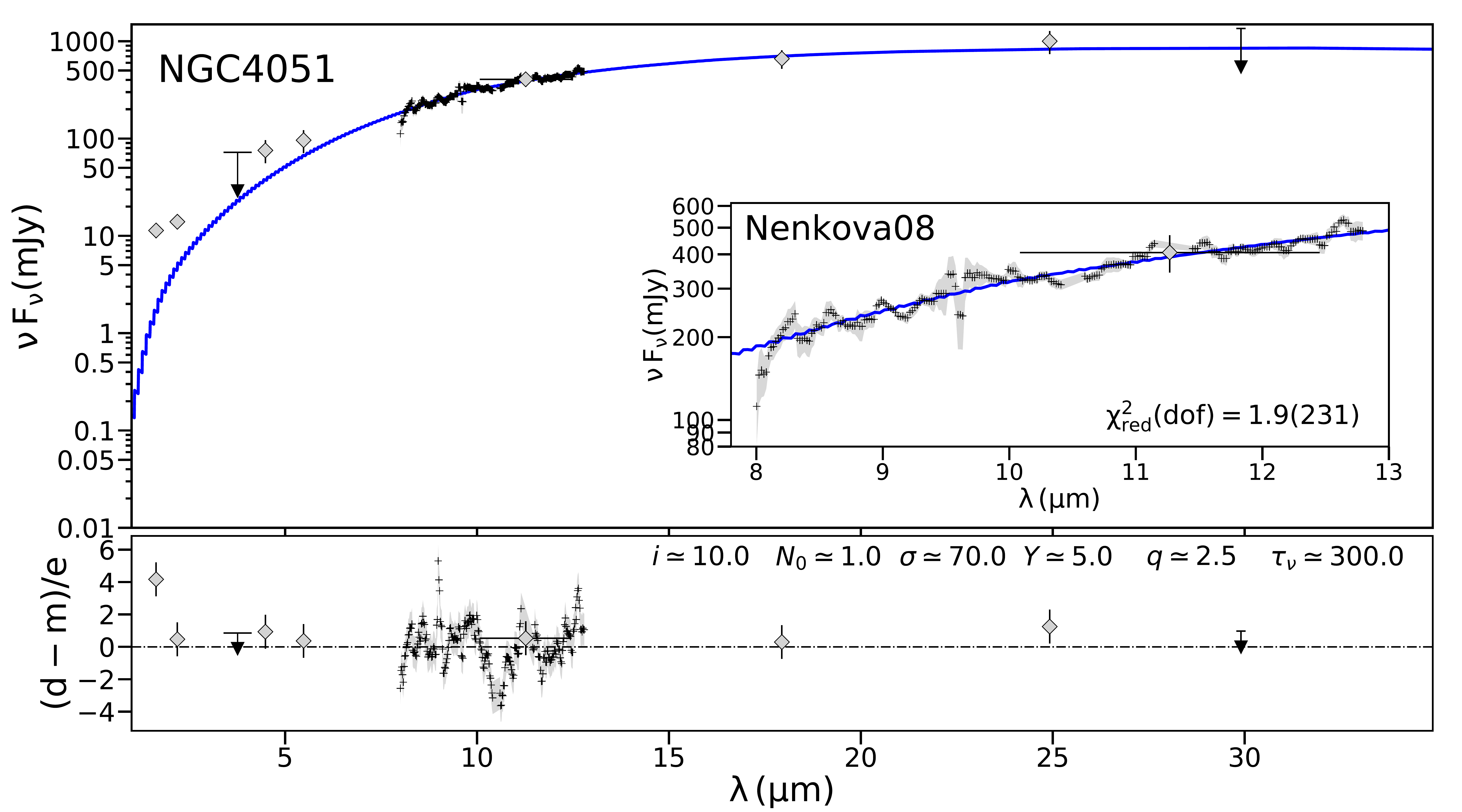}
    \includegraphics[width=0.75\columnwidth]{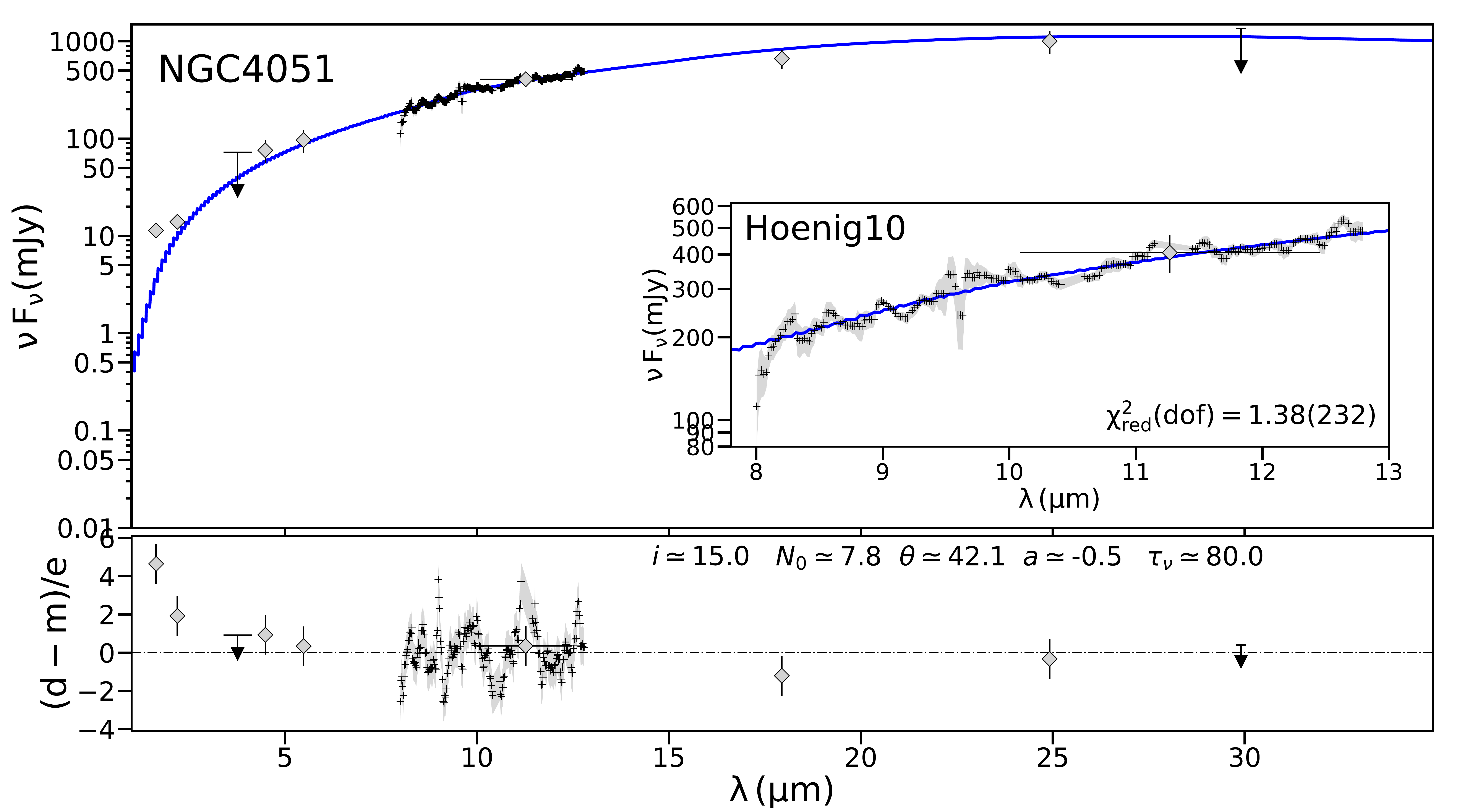}
    \includegraphics[width=0.75\columnwidth]{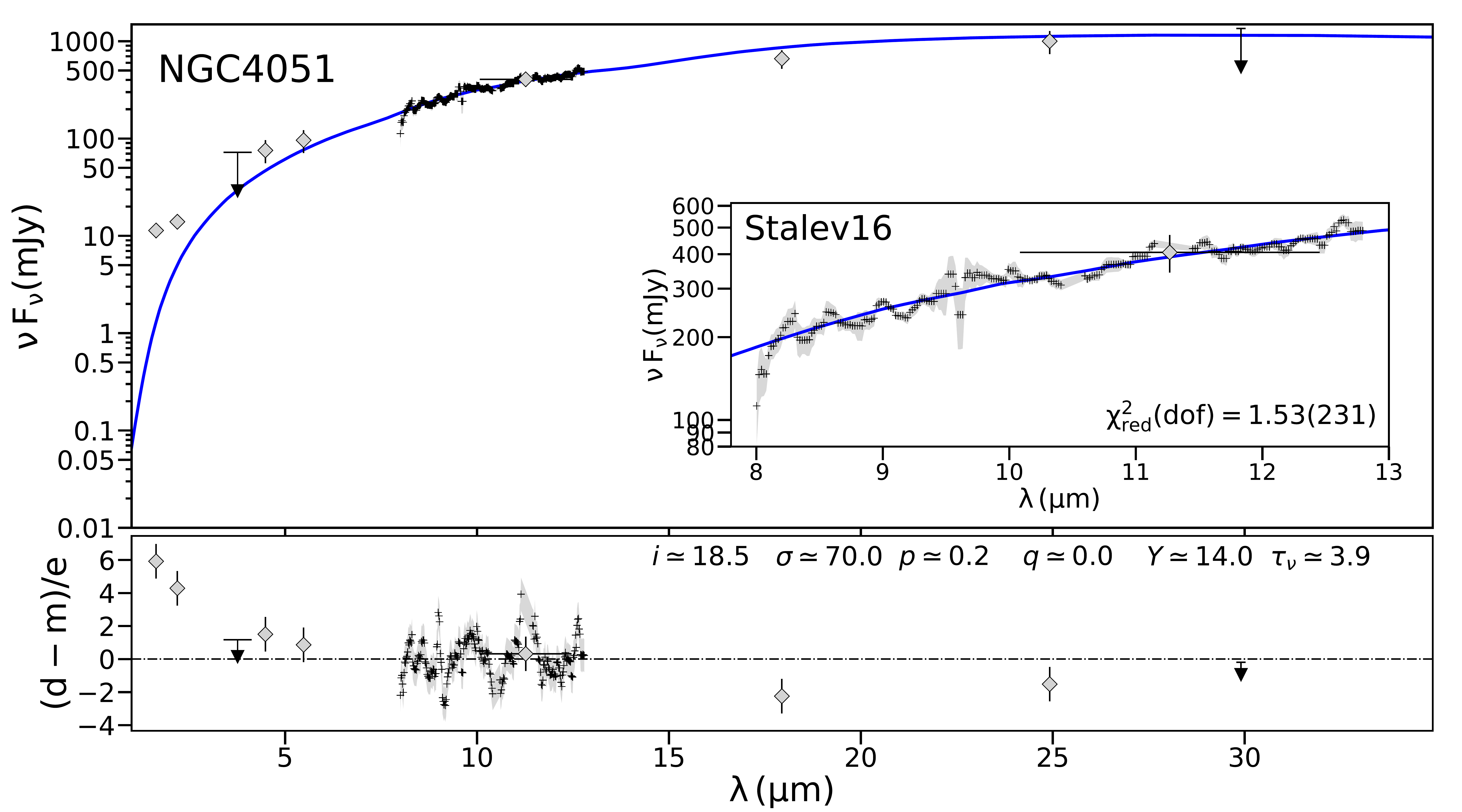}
    \includegraphics[width=0.75\columnwidth]{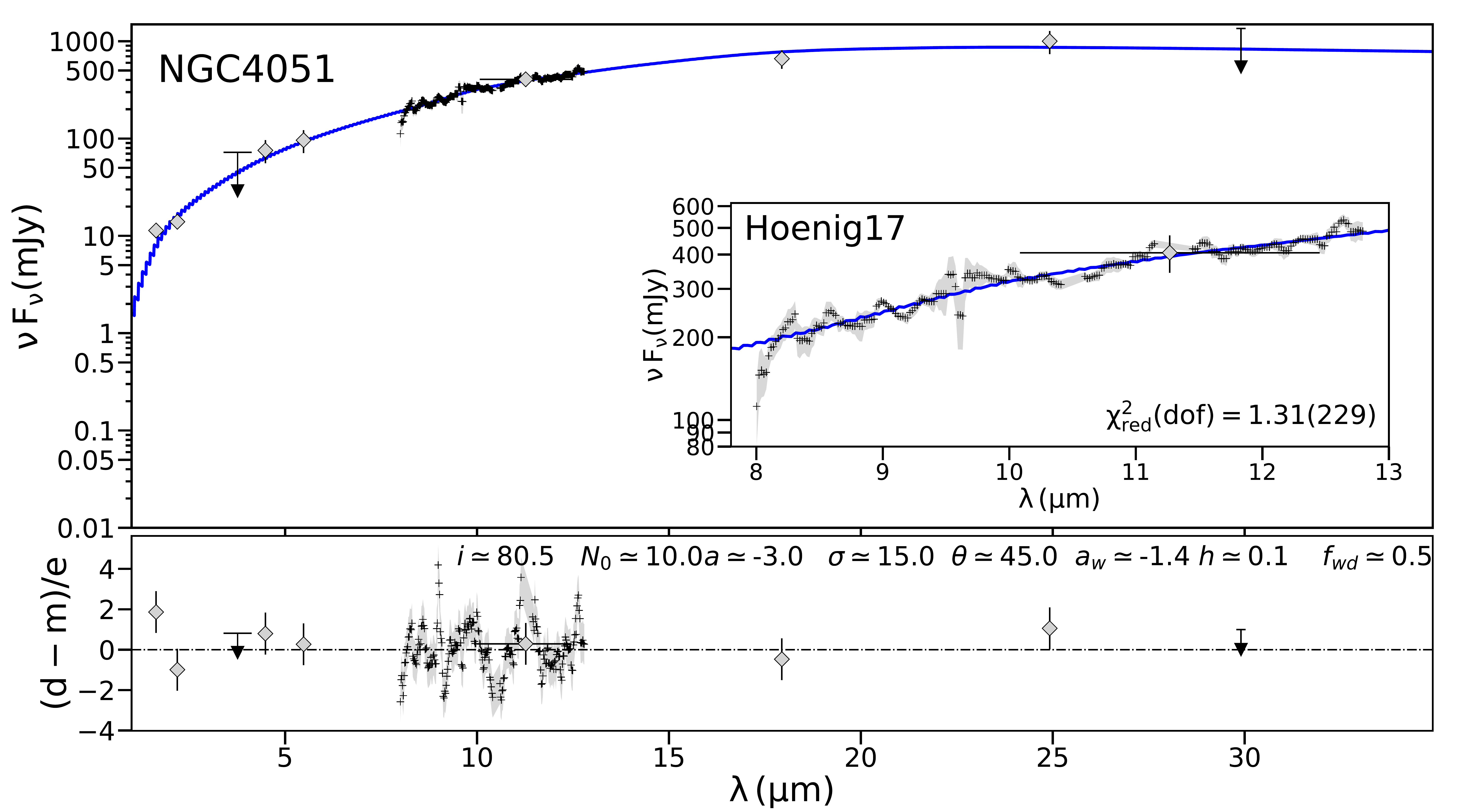}
    \includegraphics[width=0.75\columnwidth]{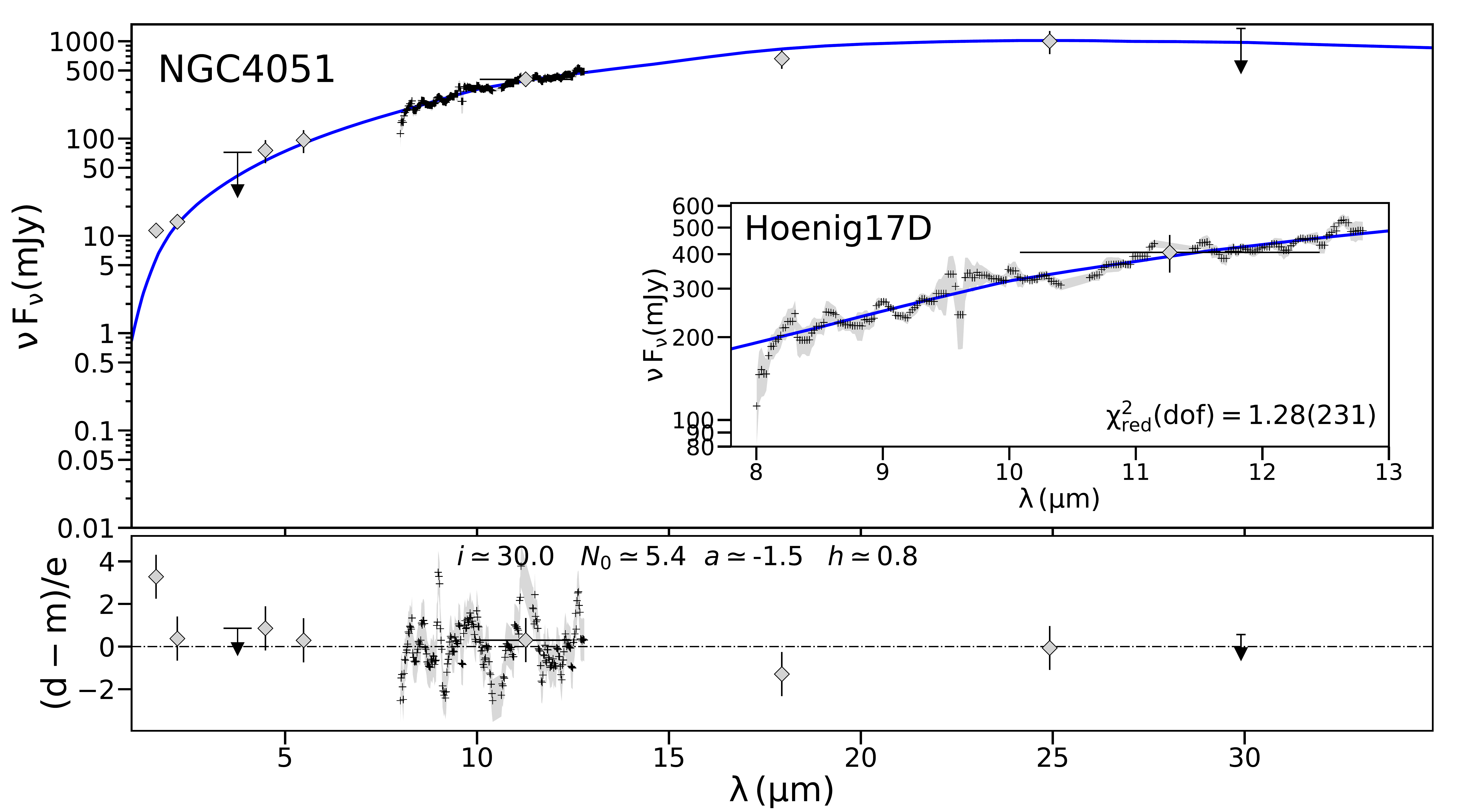}
    \caption{Same as Fig. \ref{fig:ESO005-G004} but for NGC4051.}
    \label{fig:NGC4051}
\end{figure*}

\begin{figure*}
    \centering
    \includegraphics[width=0.75\columnwidth]{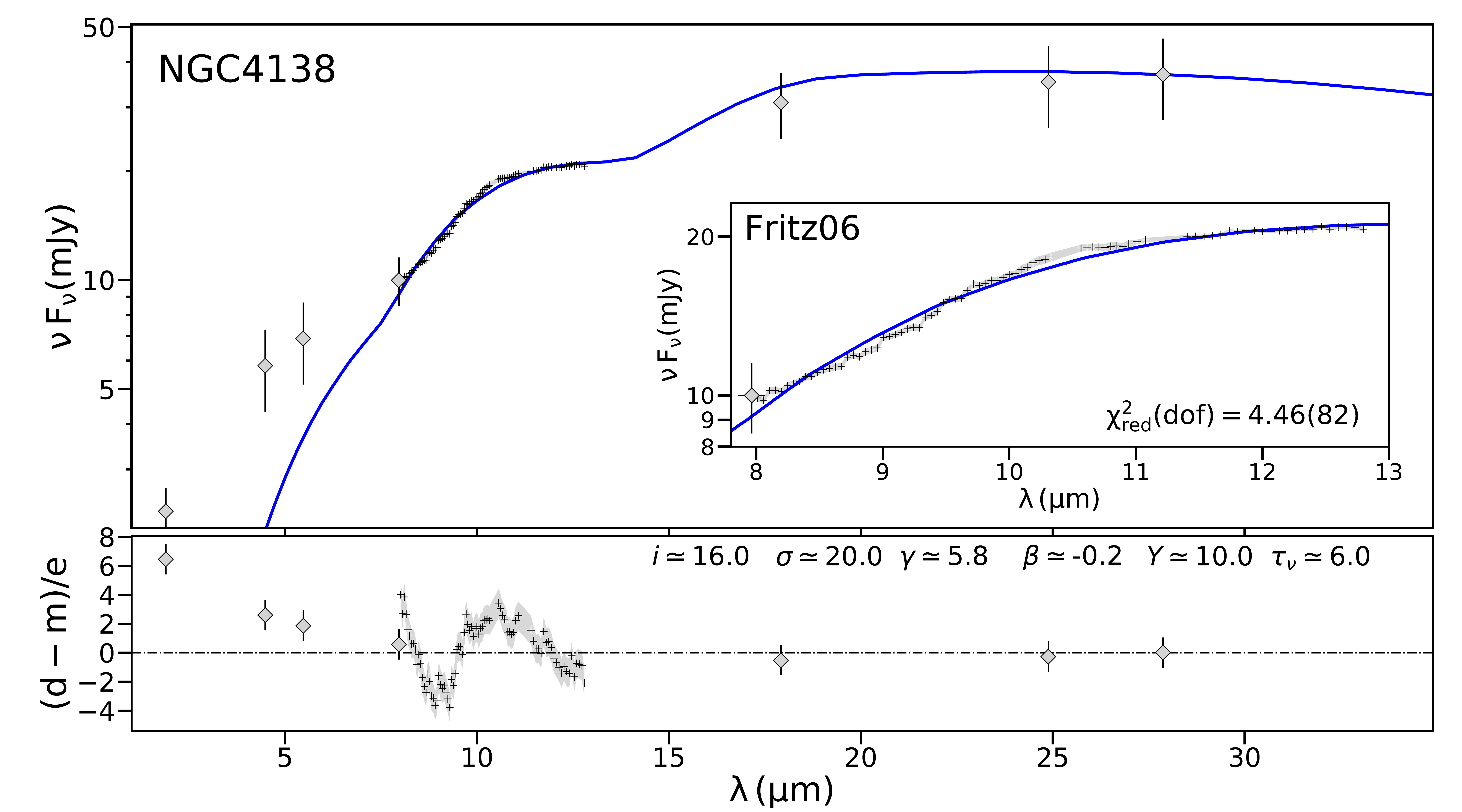}
    \includegraphics[width=0.75\columnwidth]{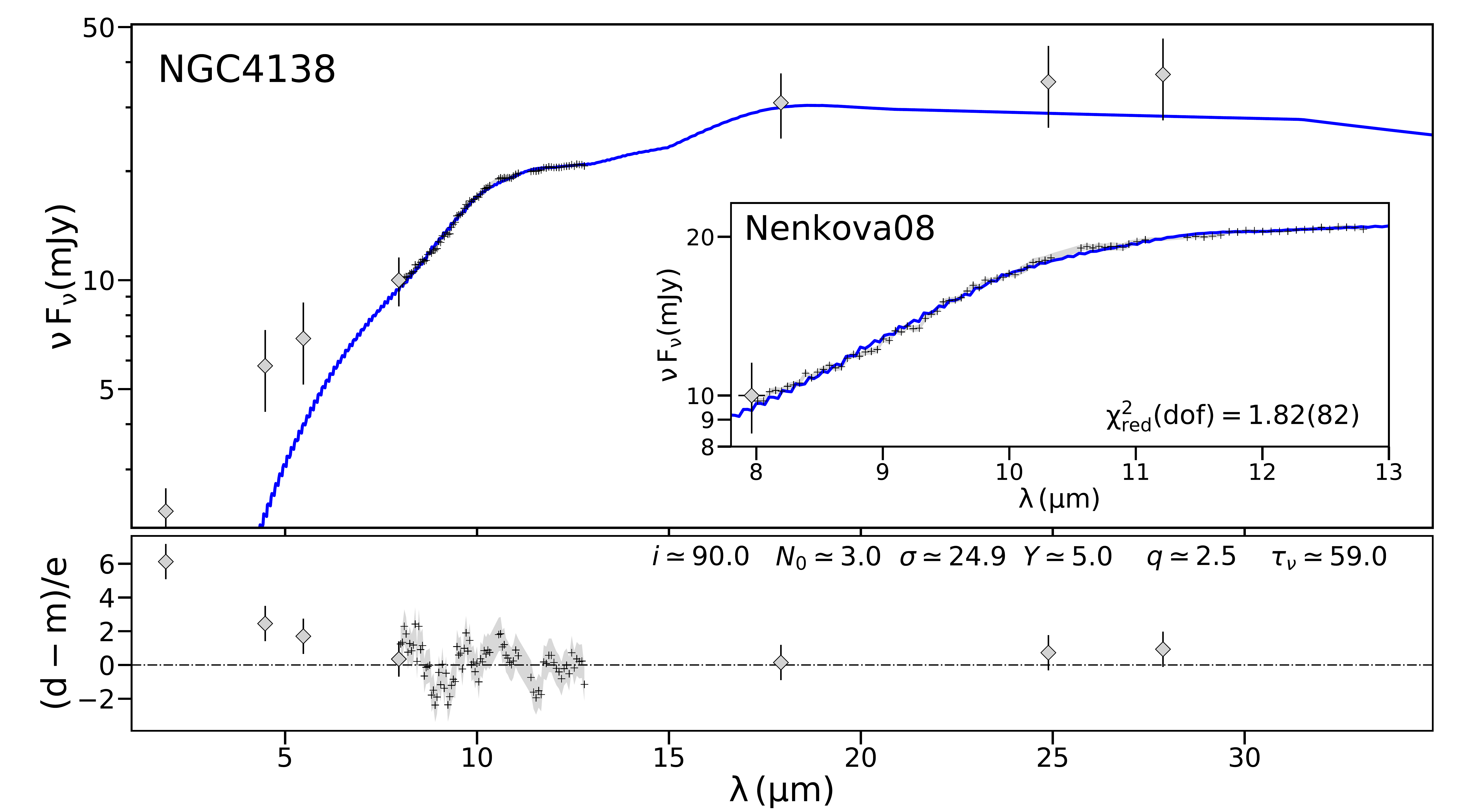}
    \includegraphics[width=0.75\columnwidth]{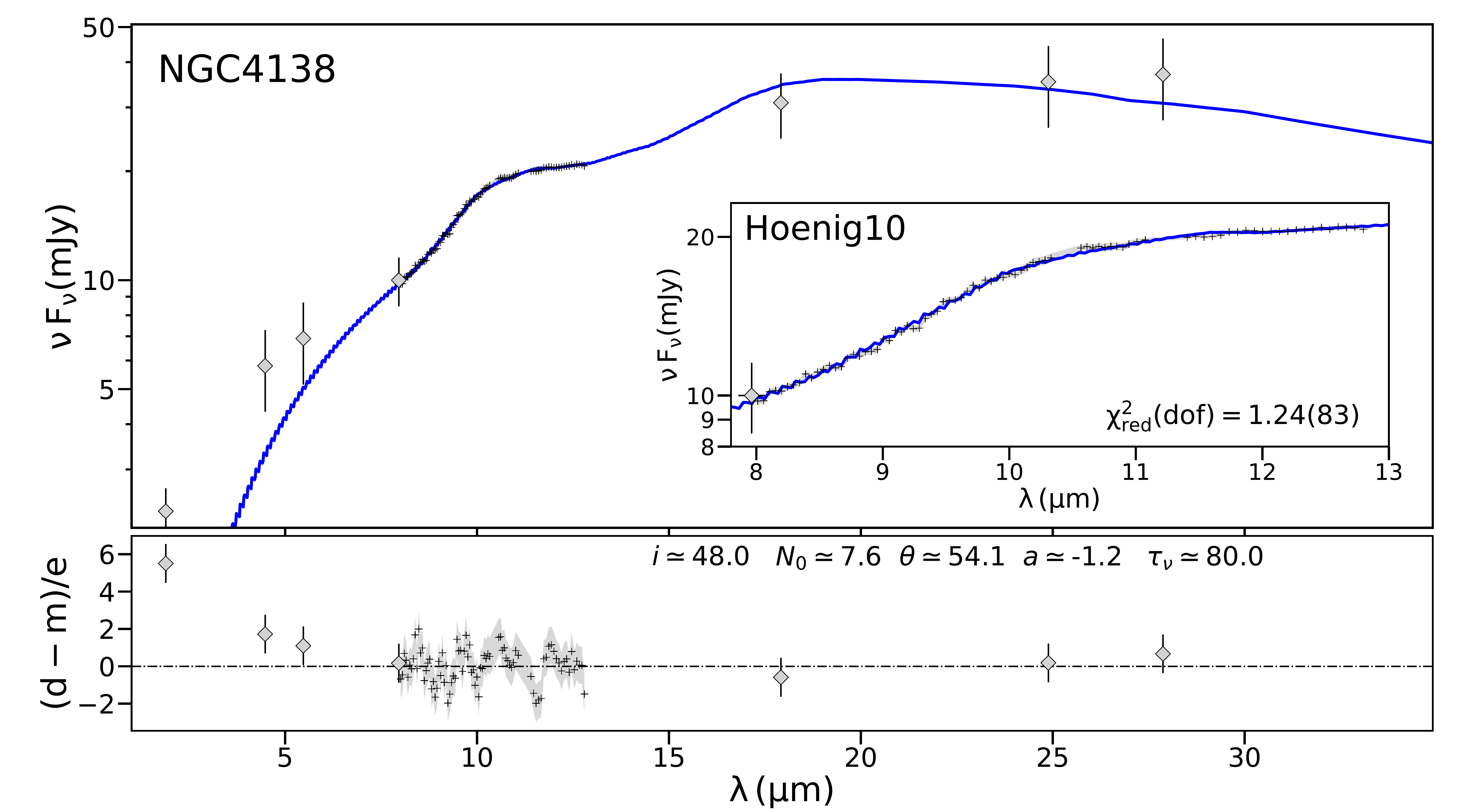}
    \includegraphics[width=0.75\columnwidth]{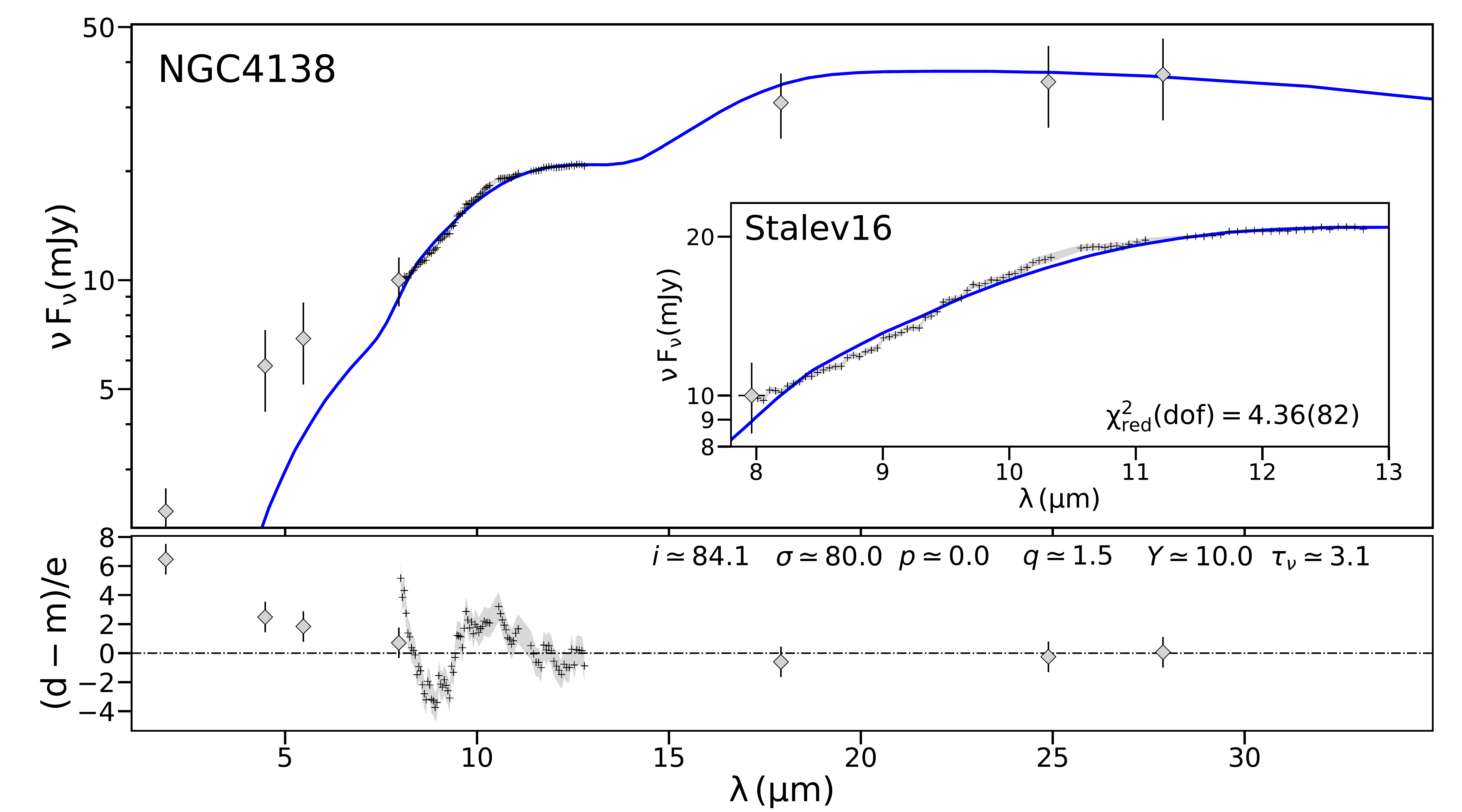}
    \includegraphics[width=0.75\columnwidth]{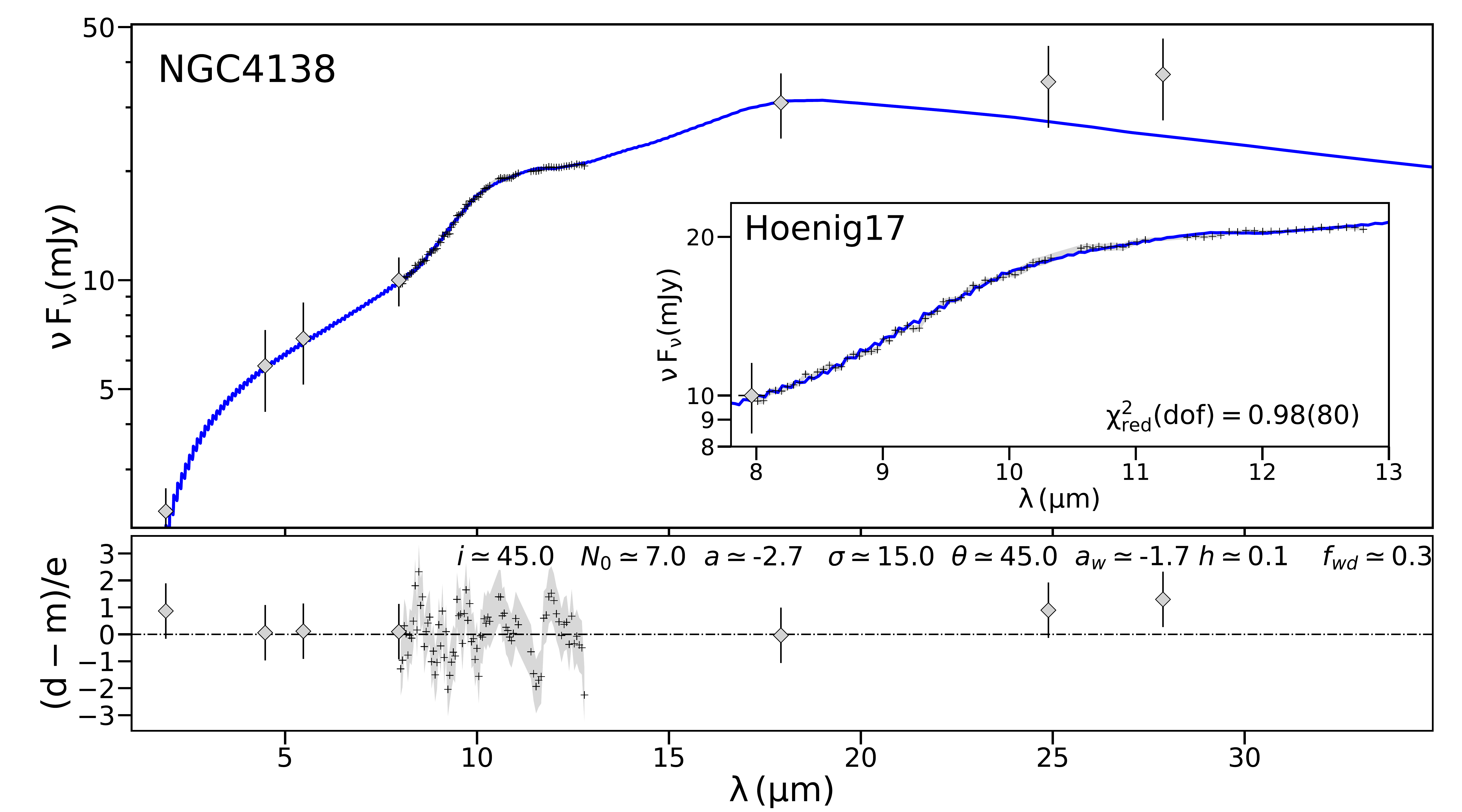}
    \includegraphics[width=0.75\columnwidth]{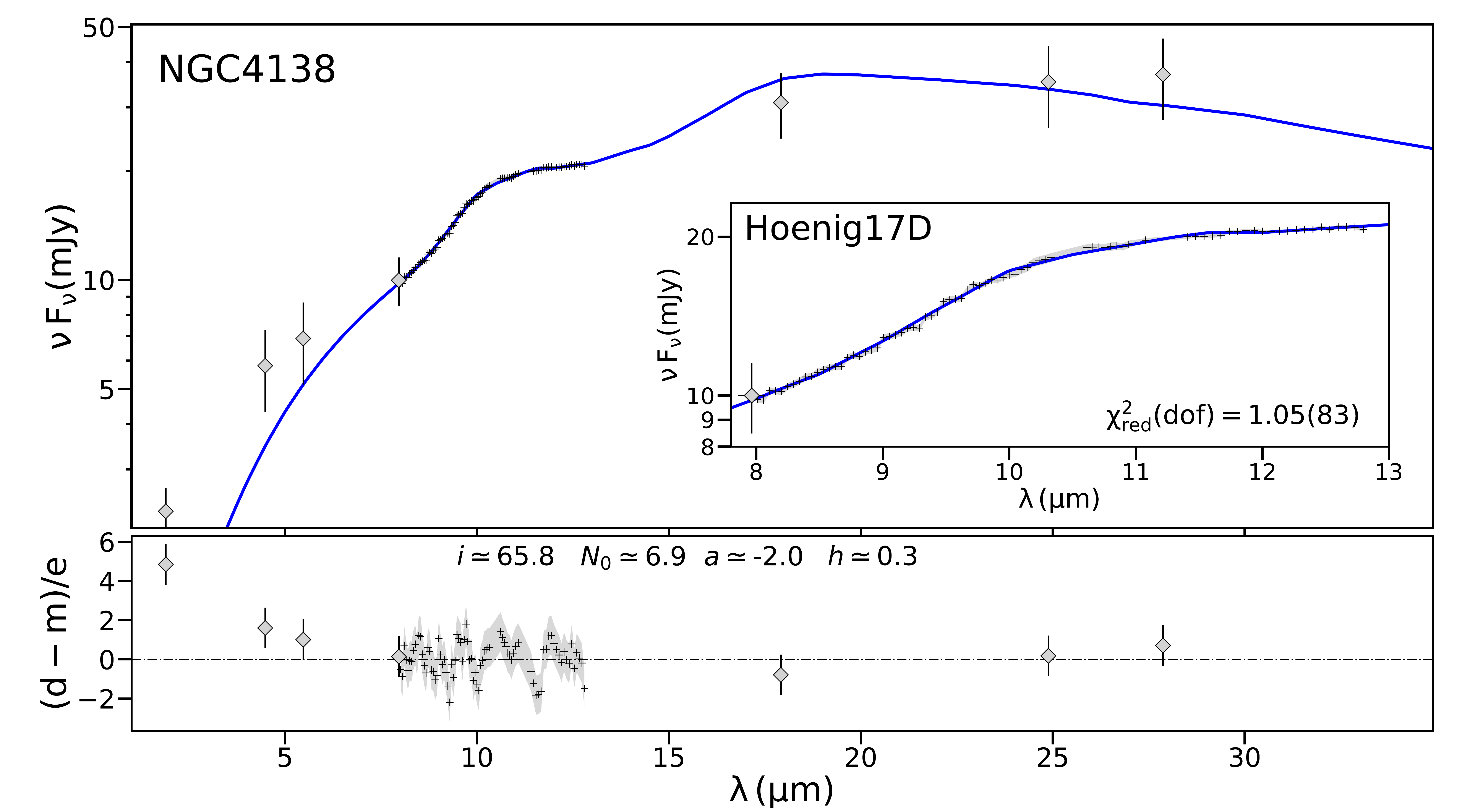}
    \caption{Same as Fig. \ref{fig:ESO005-G004} but for NGC4138.}
    \label{fig:NGC4138}
\end{figure*}
\begin{figure*}
    \centering
    \includegraphics[width=0.75\columnwidth]{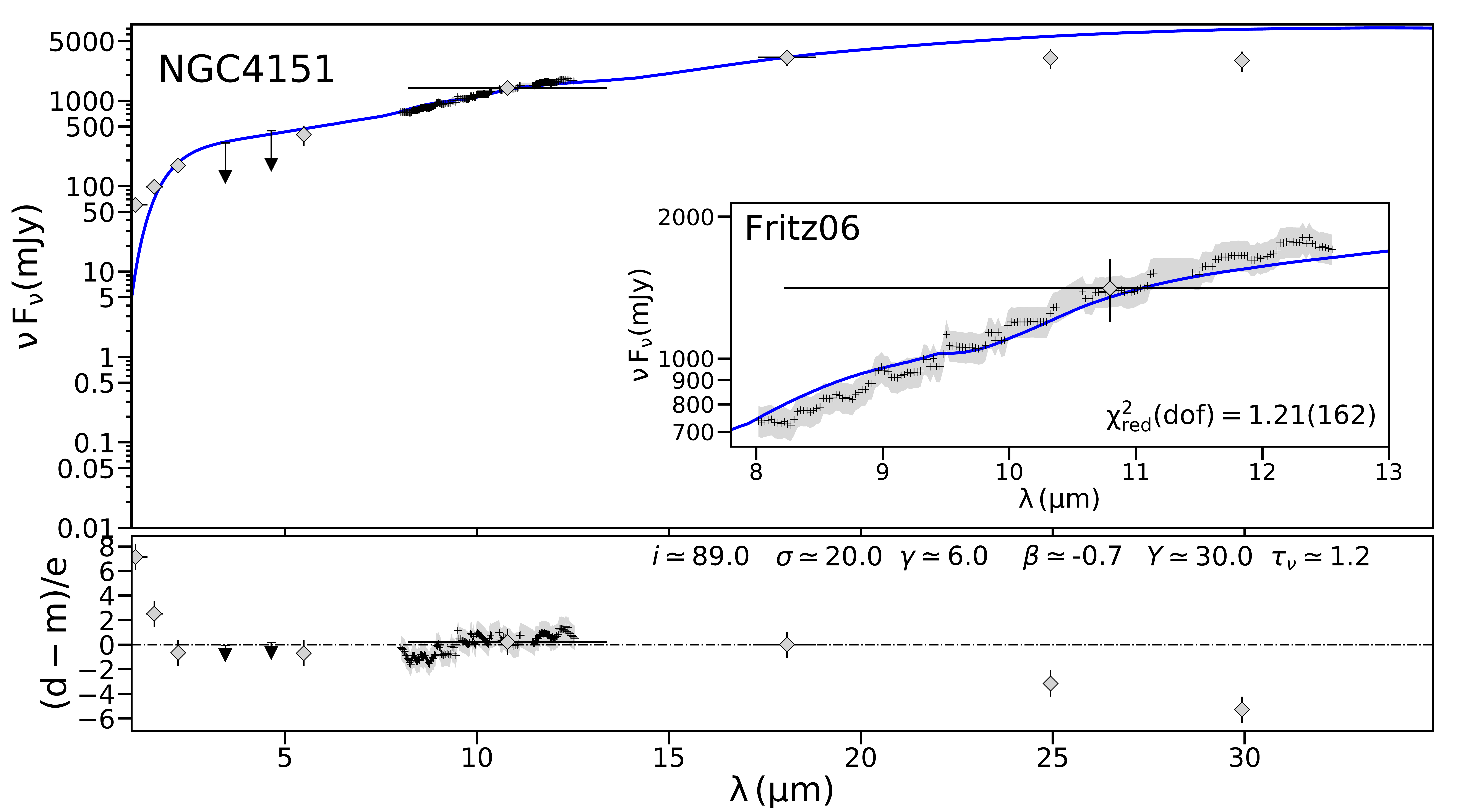}
    \includegraphics[width=0.75\columnwidth]{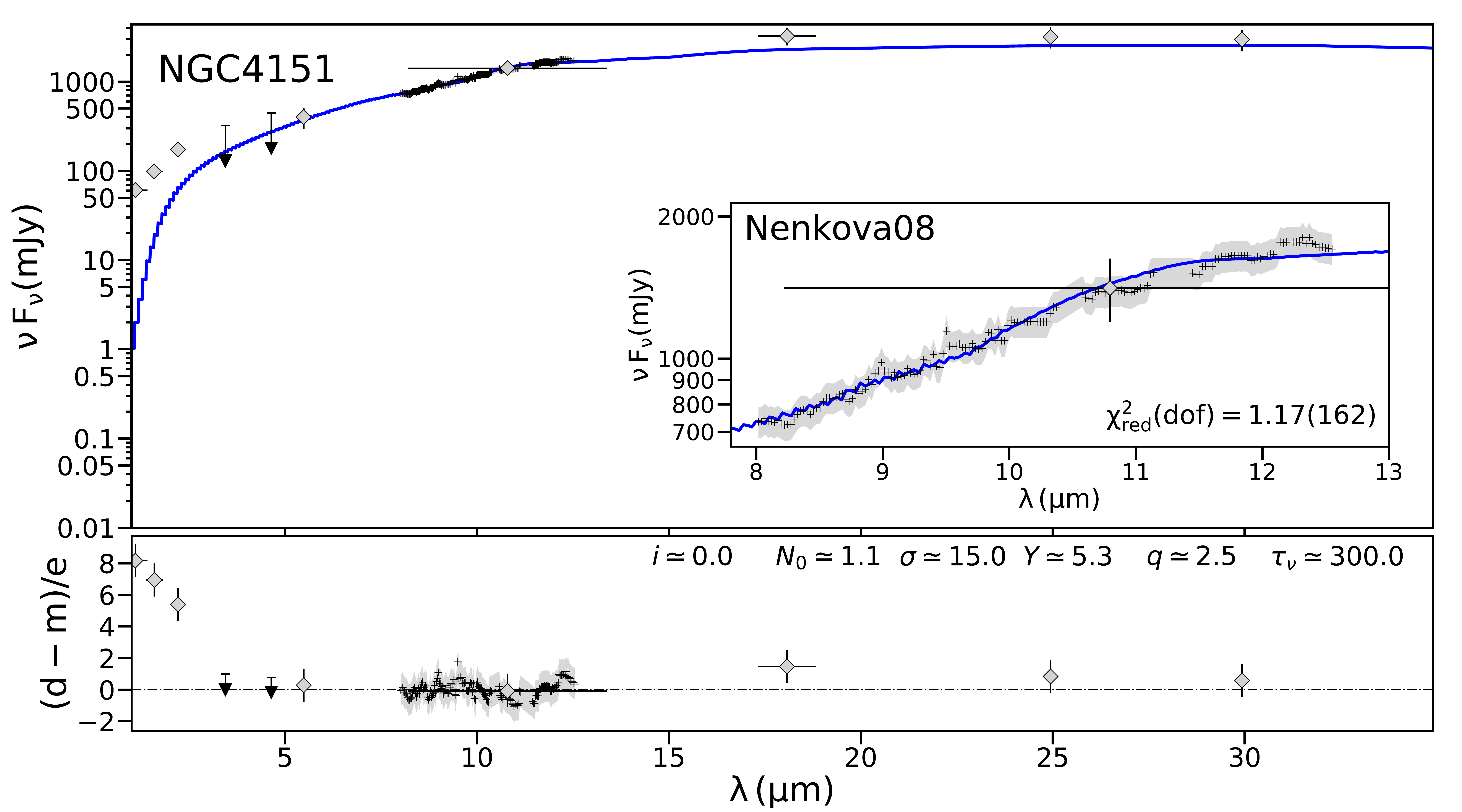}
    \includegraphics[width=0.75\columnwidth]{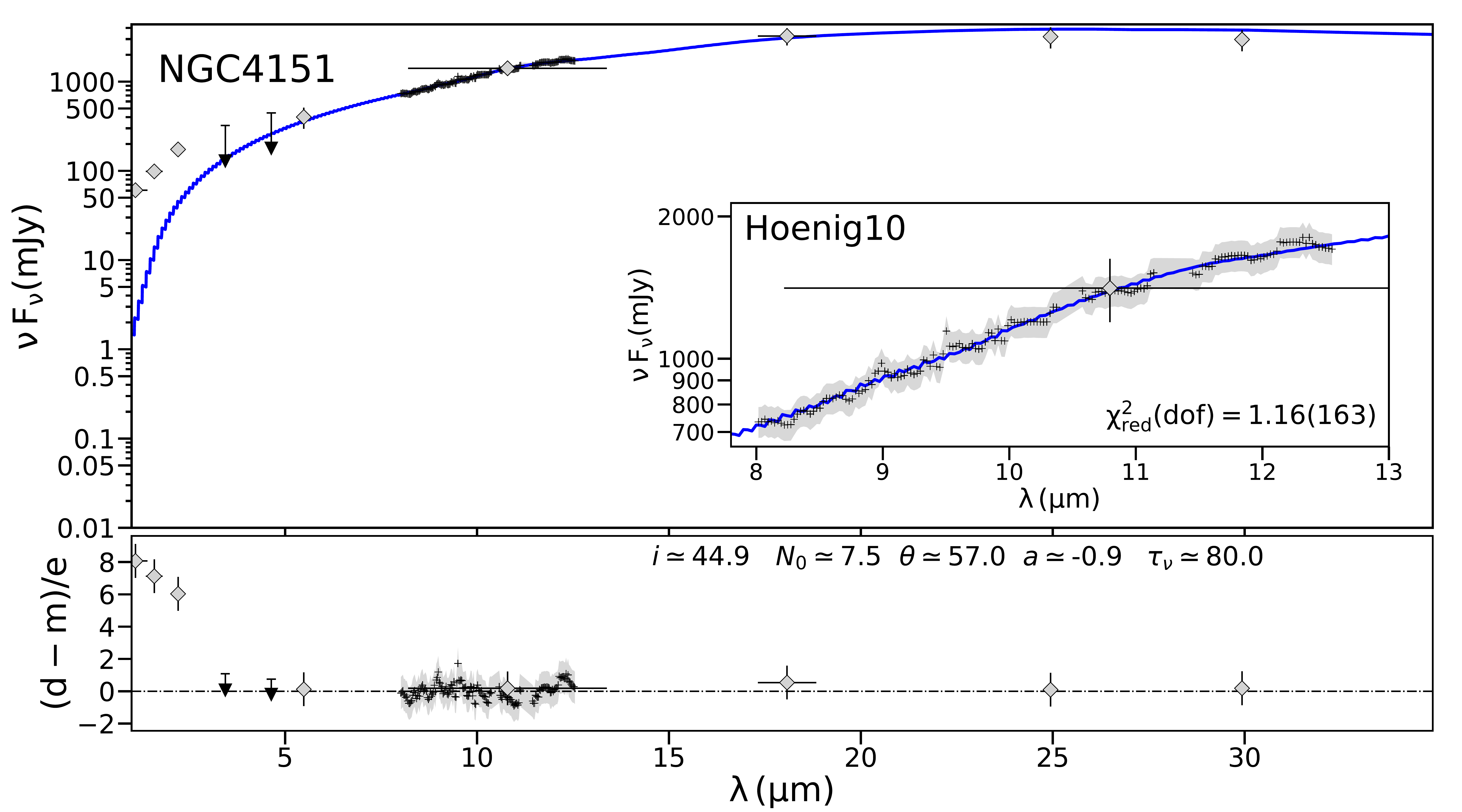}
    \includegraphics[width=0.75\columnwidth]{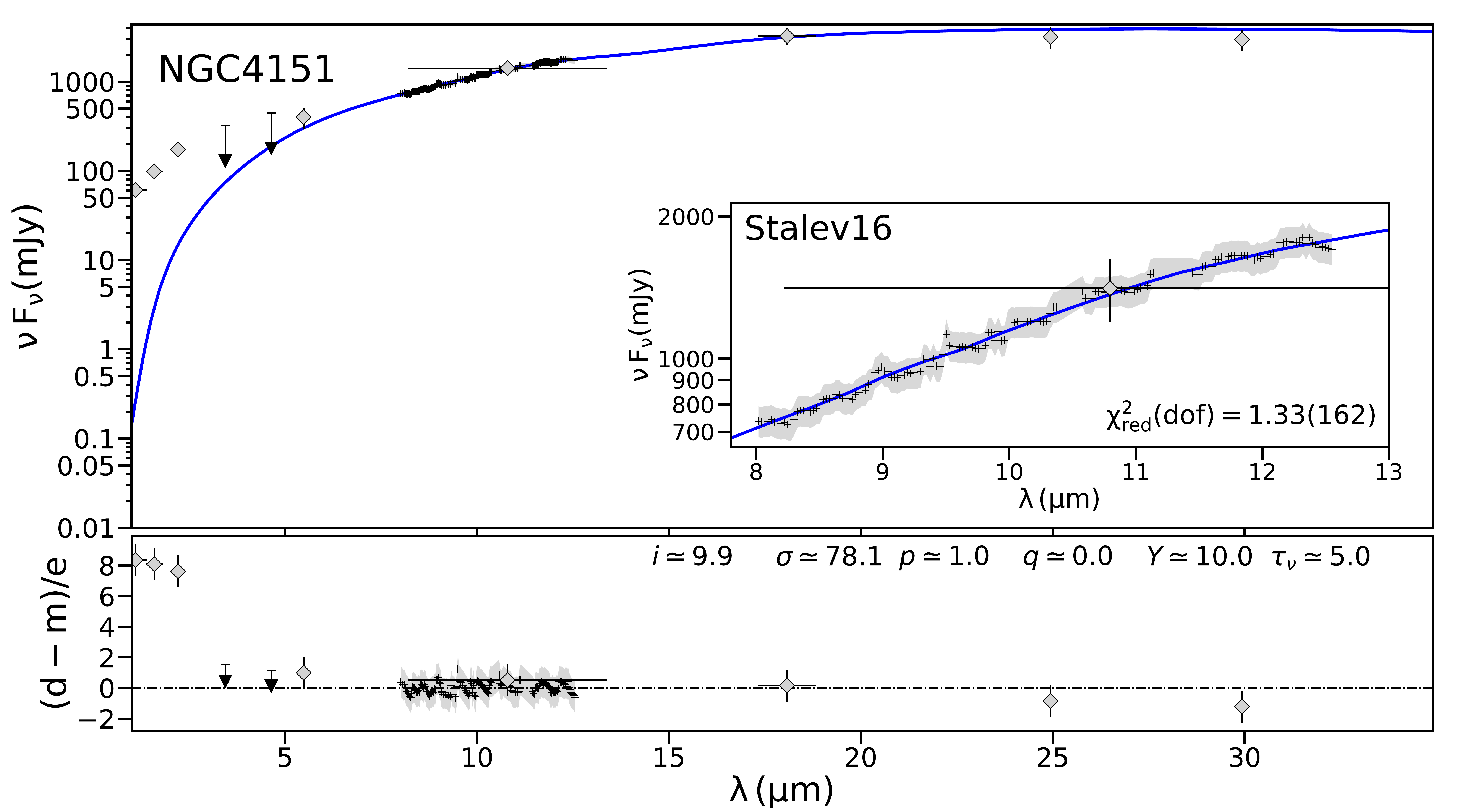}
    \includegraphics[width=0.75\columnwidth]{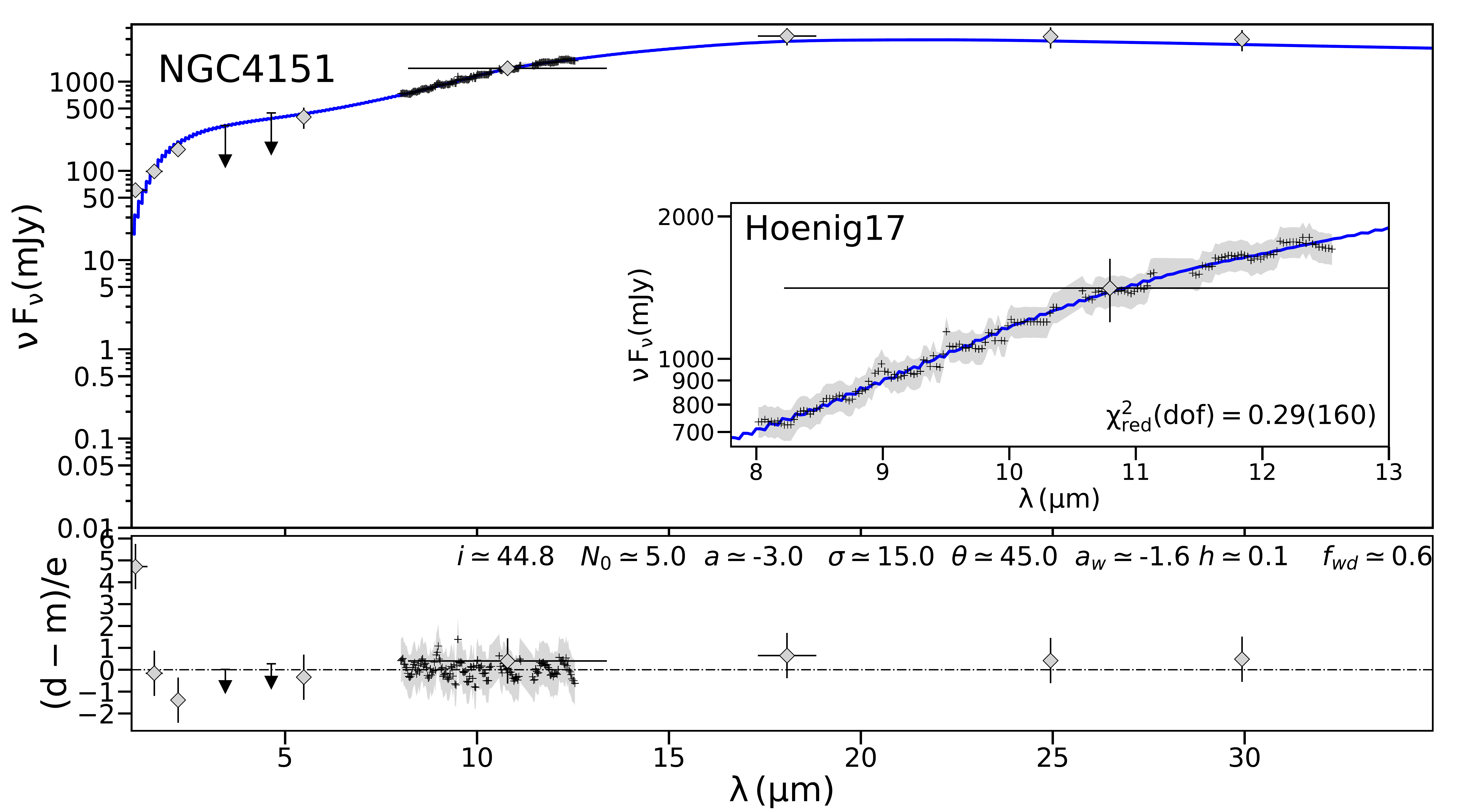}
    \includegraphics[width=0.75\columnwidth]{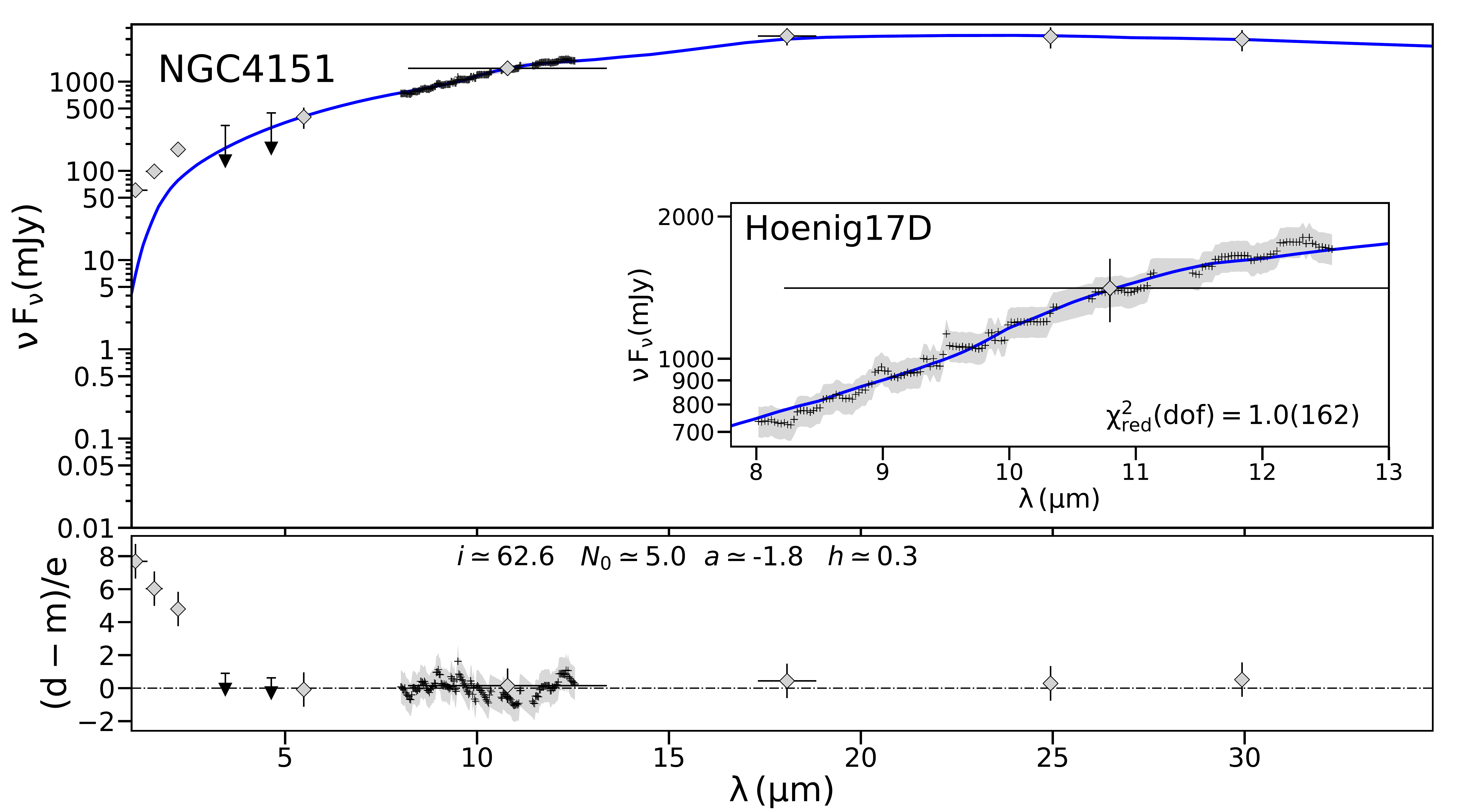}
    \caption{Same as Fig. \ref{fig:ESO005-G004} but for NGC4151.}
    \label{fig:NGC4151}
\end{figure*}
\begin{figure*}
    \centering
    \includegraphics[width=0.75\columnwidth]{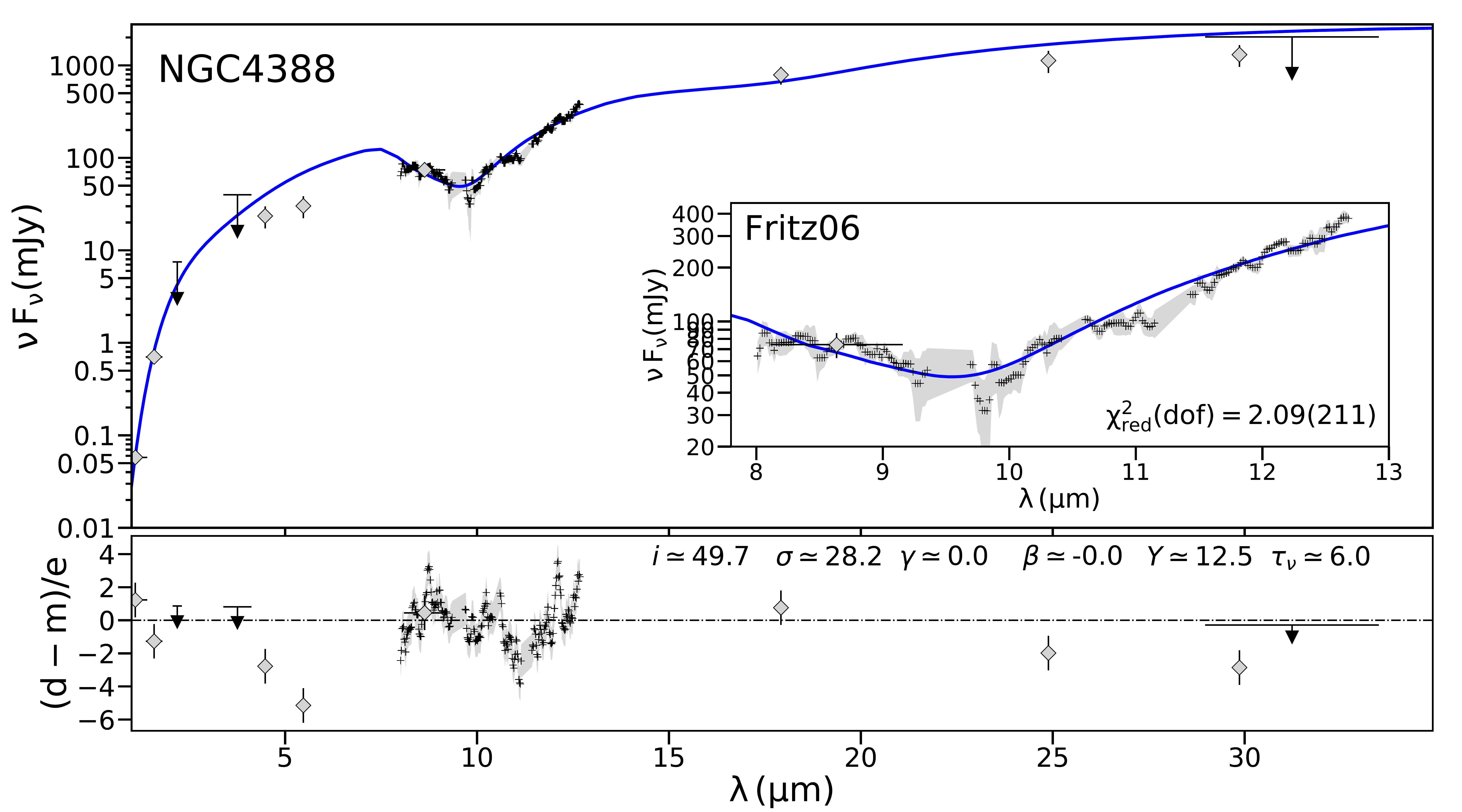}
    \includegraphics[width=0.75\columnwidth]{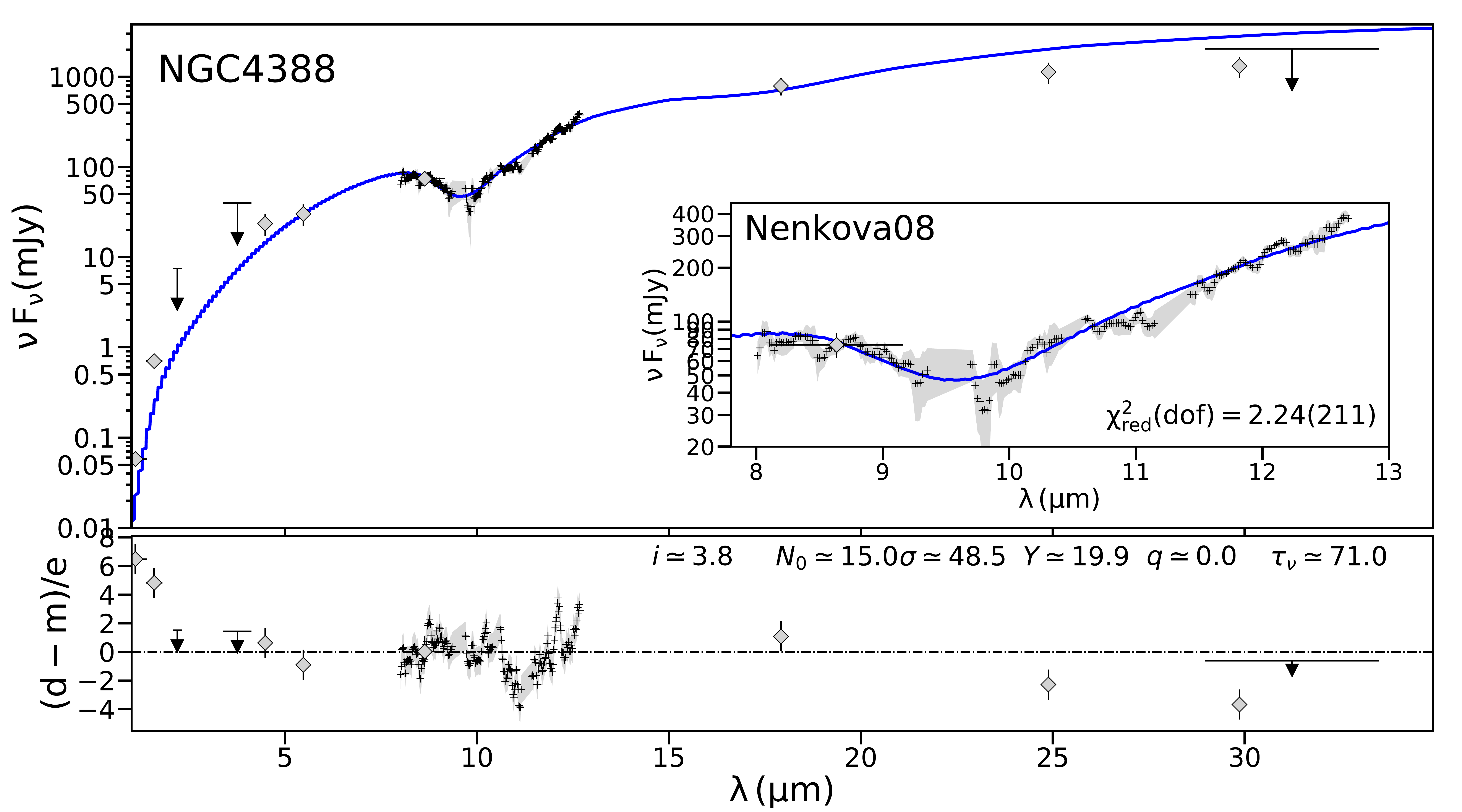}
    \includegraphics[width=0.75\columnwidth]{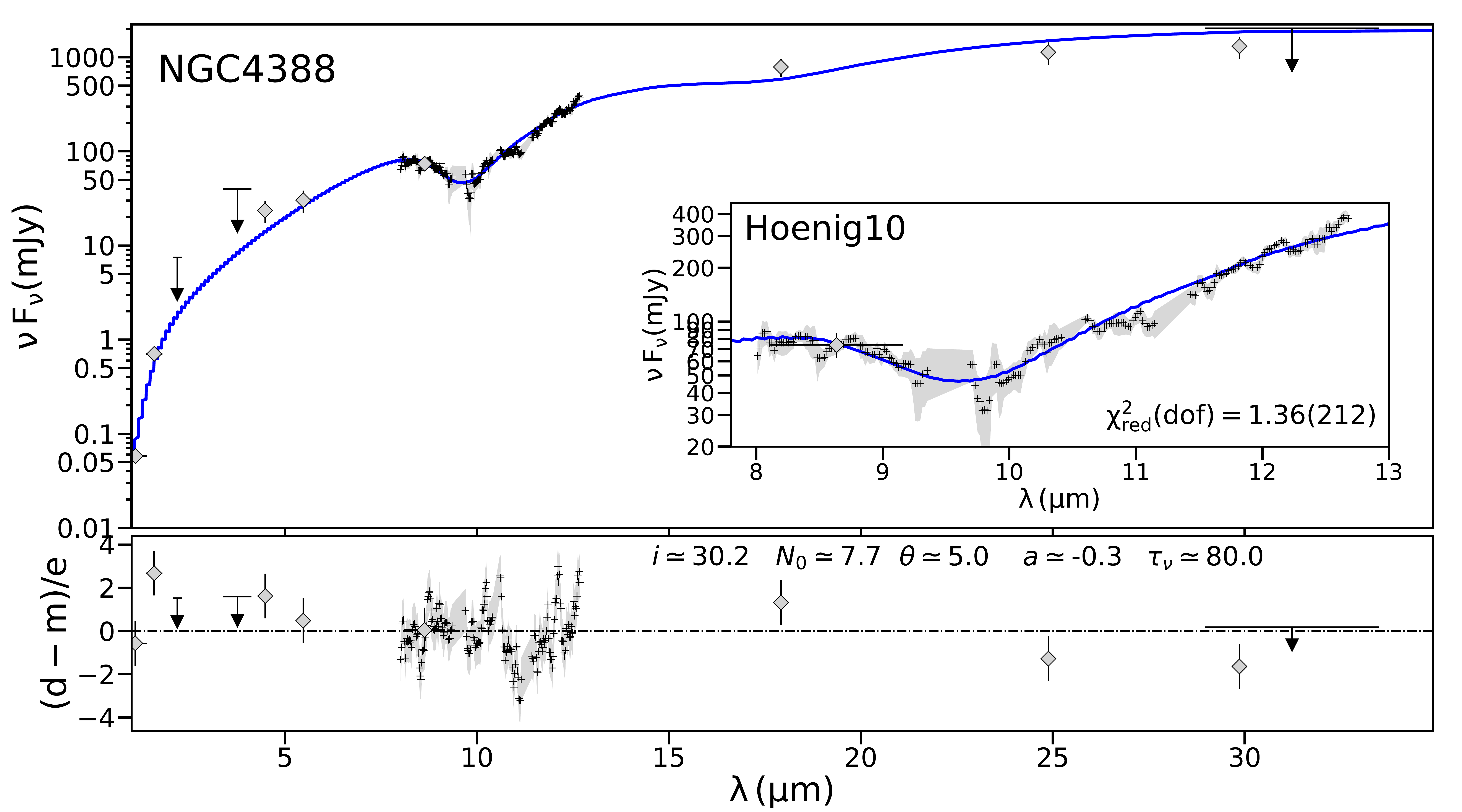}
    \includegraphics[width=0.75\columnwidth]{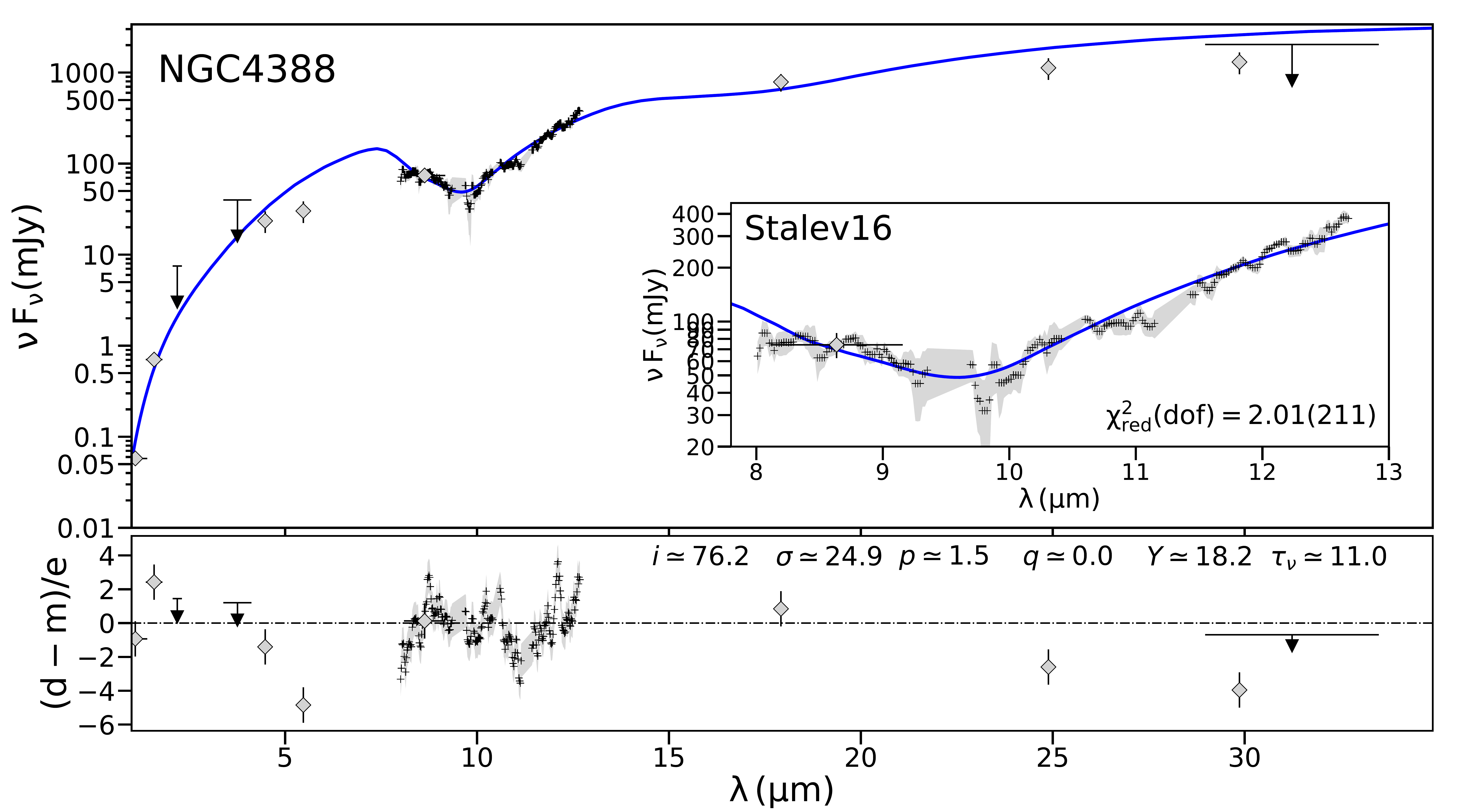}
    \includegraphics[width=0.75\columnwidth]{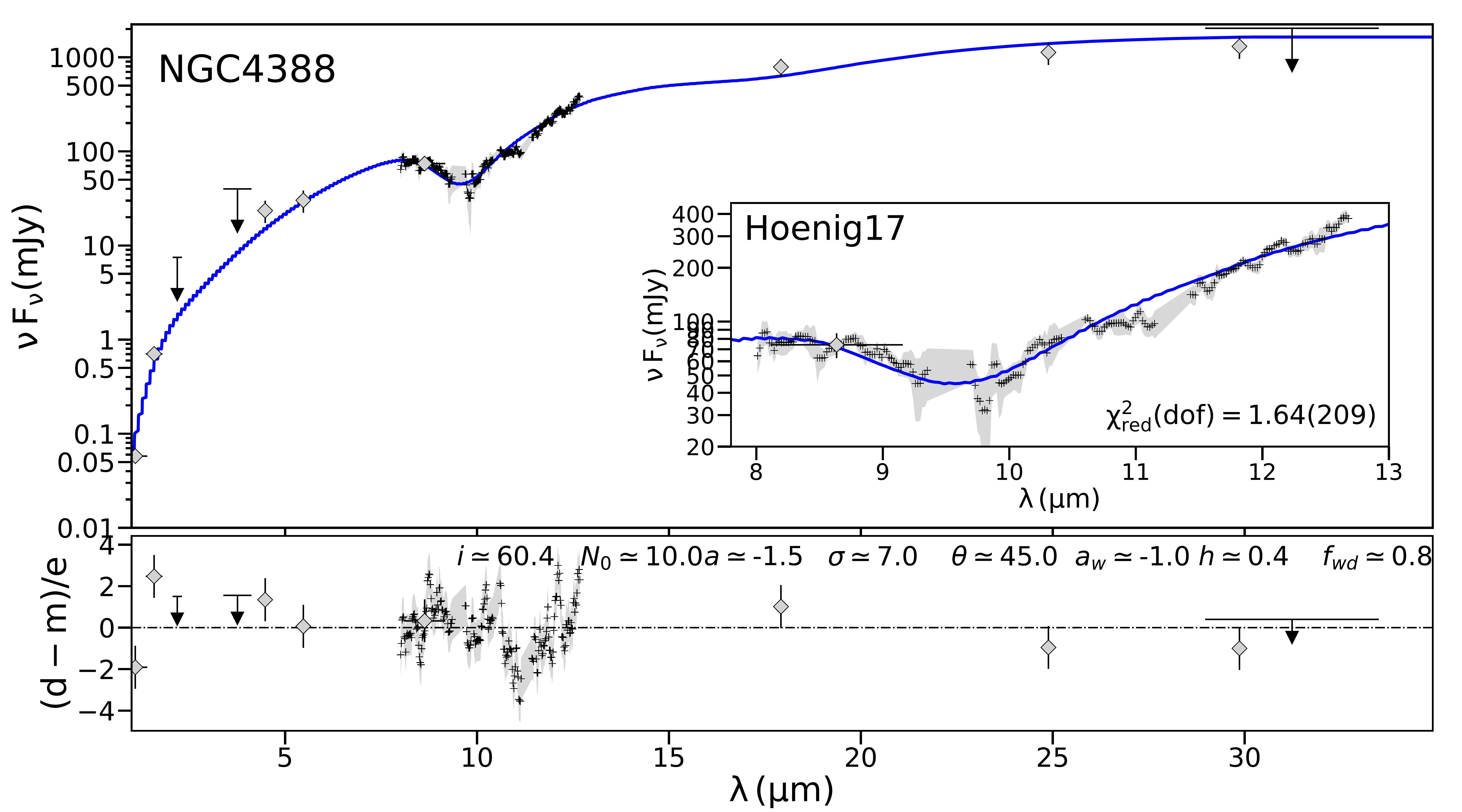}
    \includegraphics[width=0.75\columnwidth]{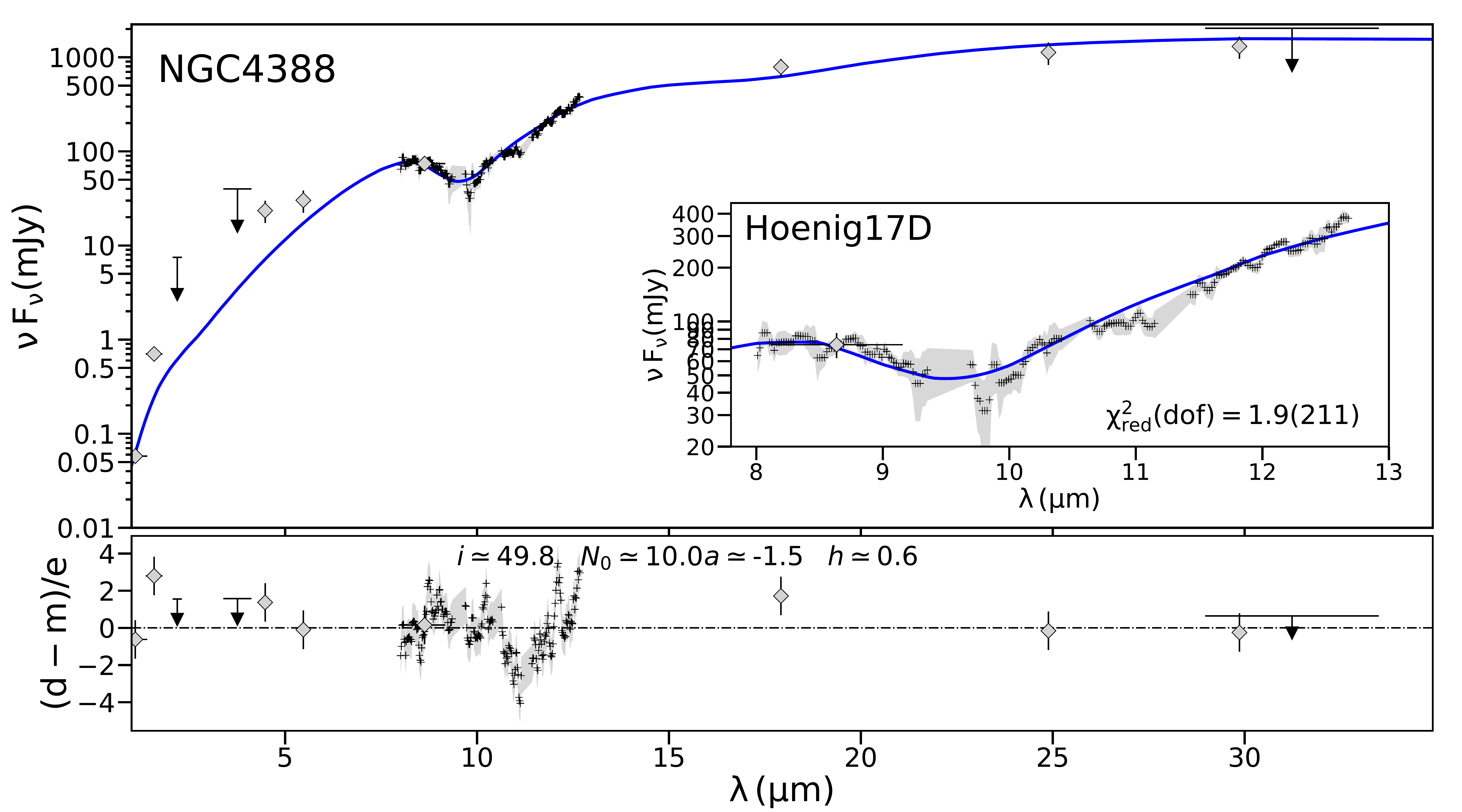}
    \caption{Same as Fig. \ref{fig:ESO005-G004} but for NGC4388.}
    \label{fig:NGC4388}
\end{figure*}

\begin{figure*}
    \centering
    \includegraphics[width=0.75\columnwidth]{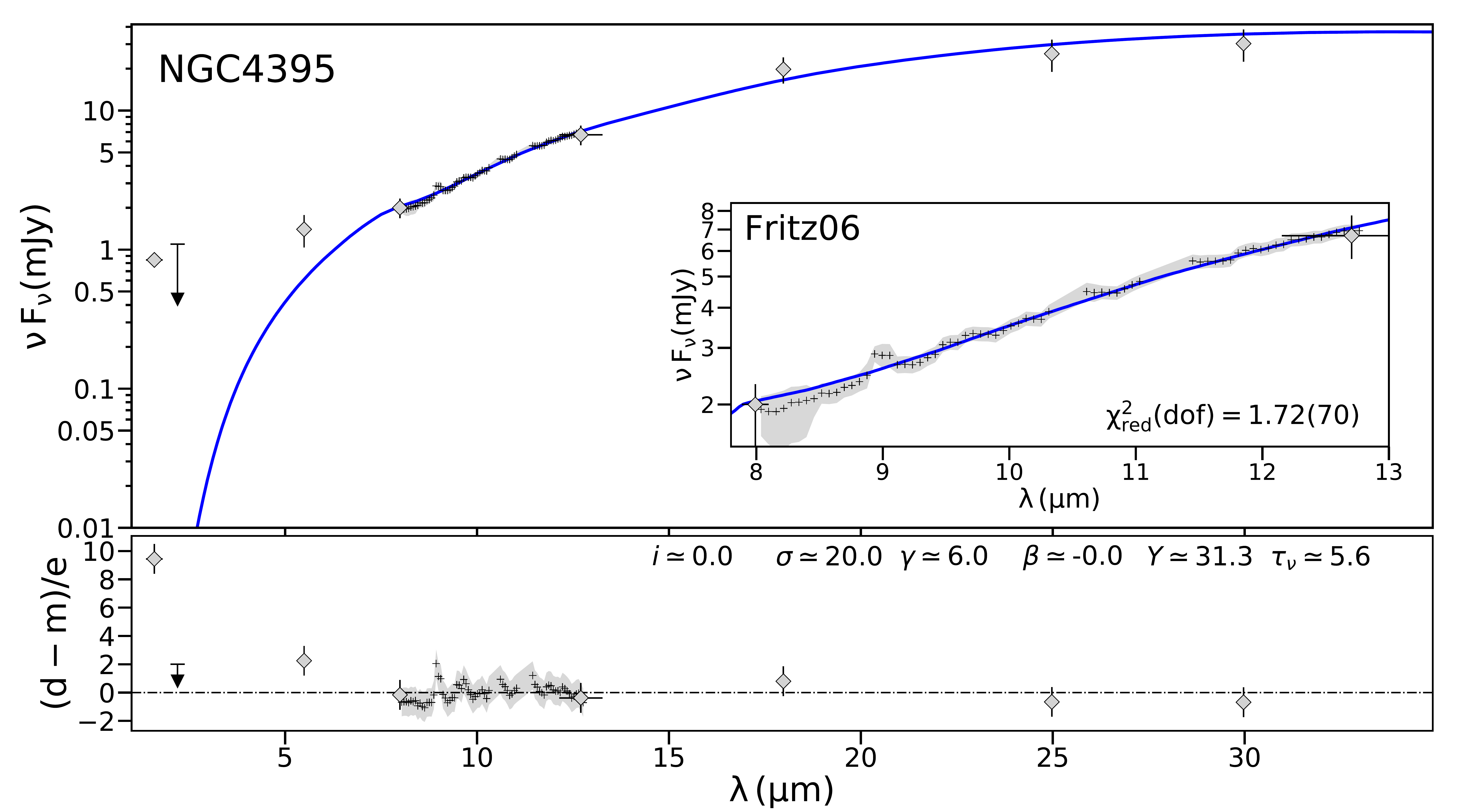}
    \includegraphics[width=0.75\columnwidth]{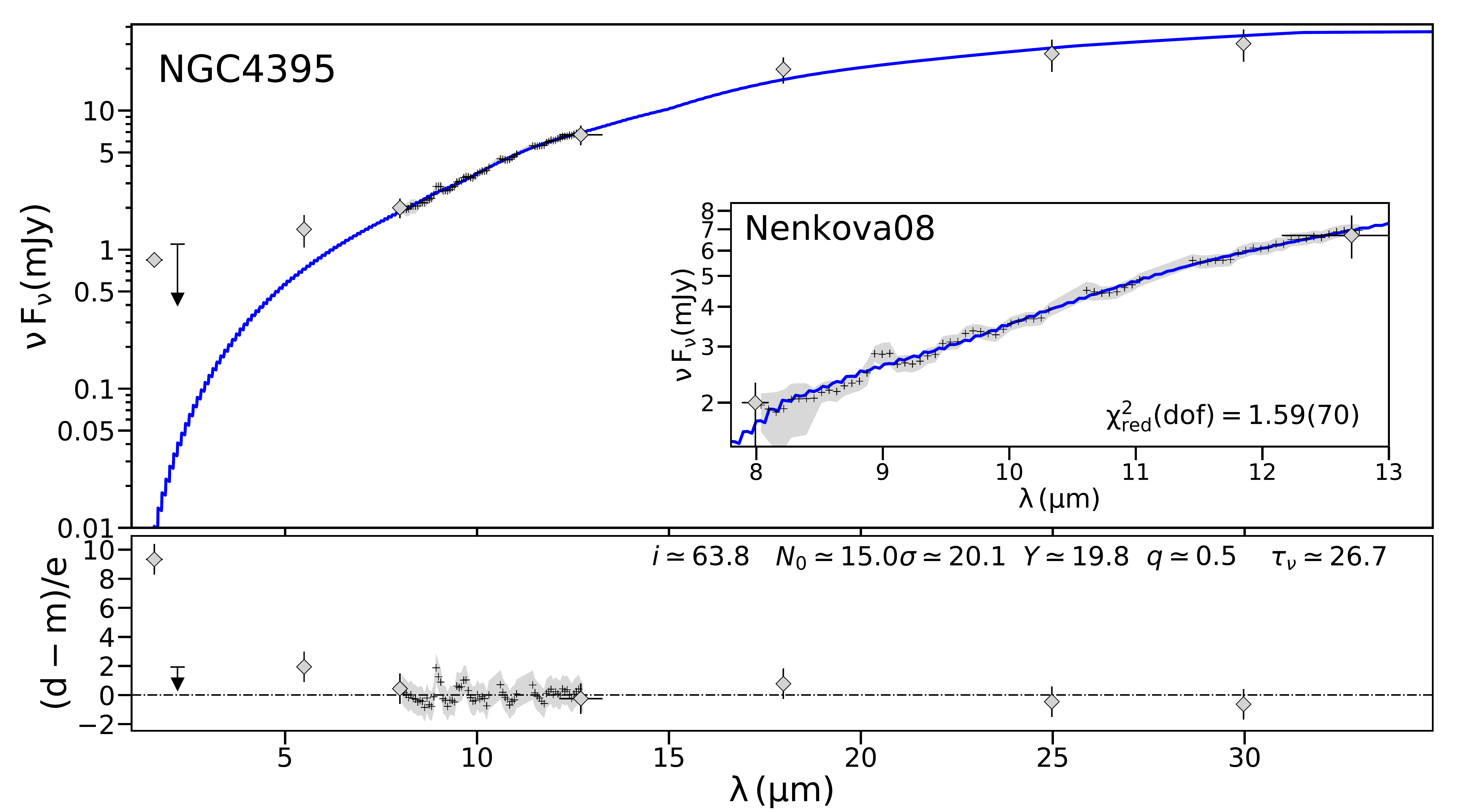}
    \includegraphics[width=0.75\columnwidth]{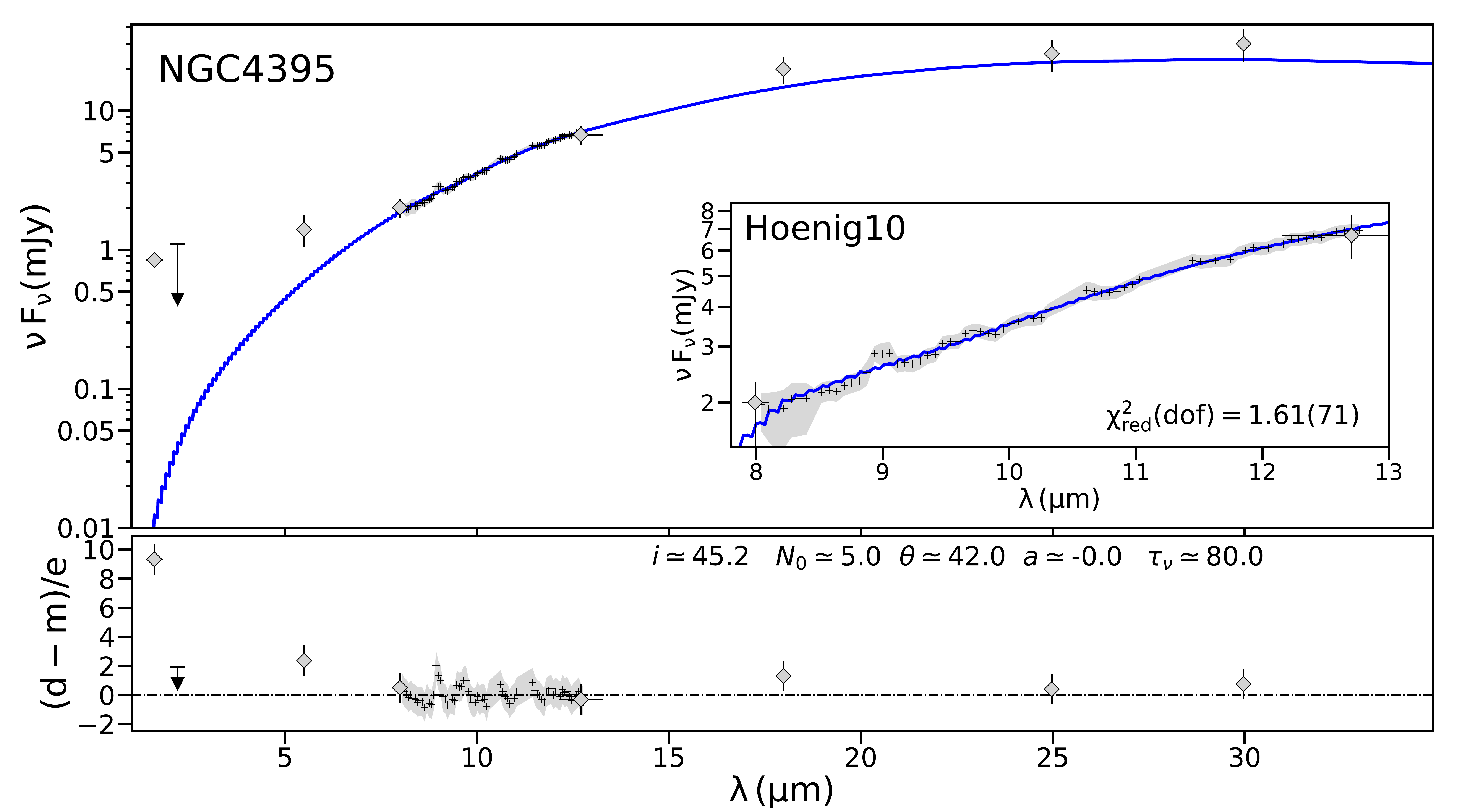}
    \includegraphics[width=0.75\columnwidth]{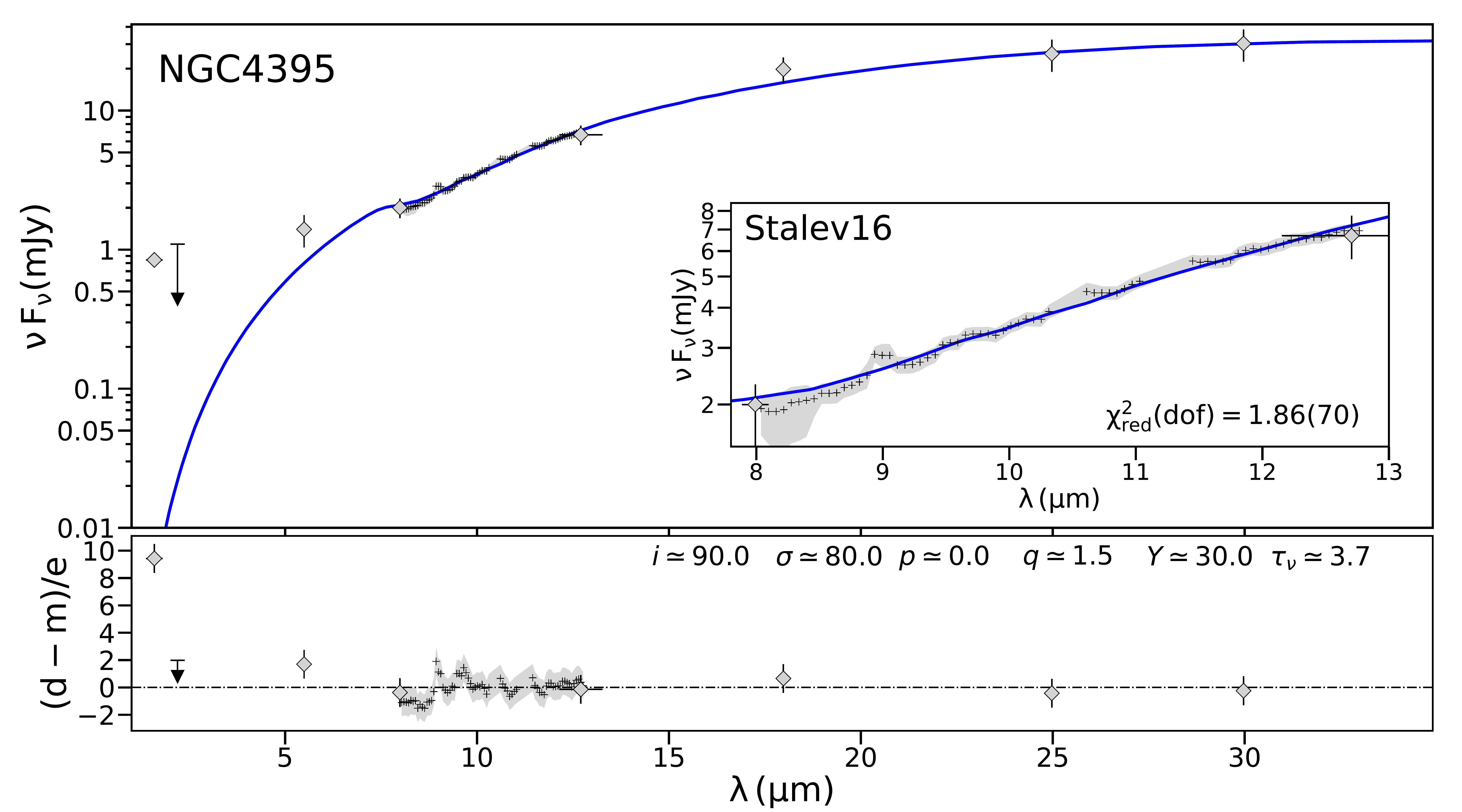}
    \includegraphics[width=0.75\columnwidth]{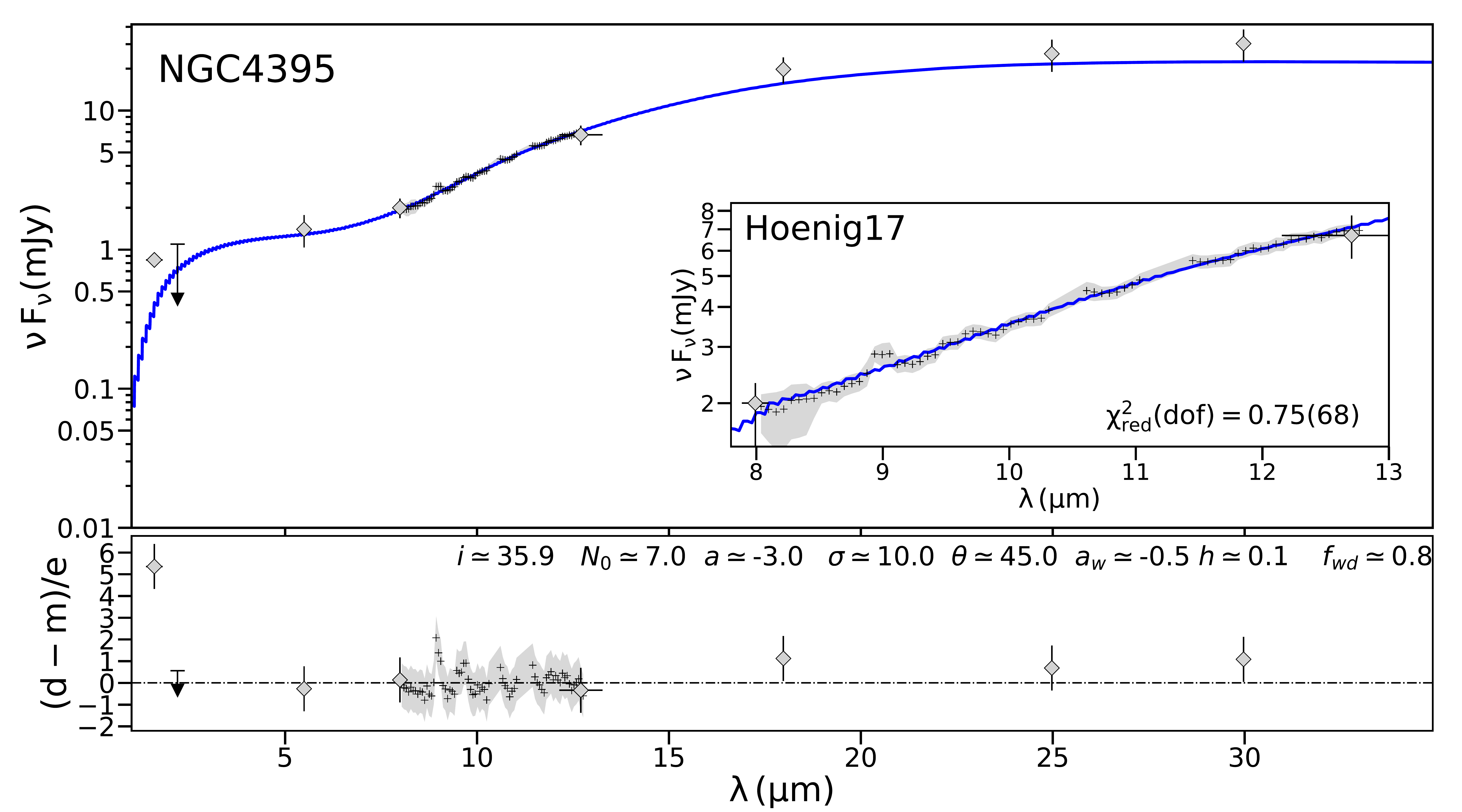}
    \includegraphics[width=0.75\columnwidth]{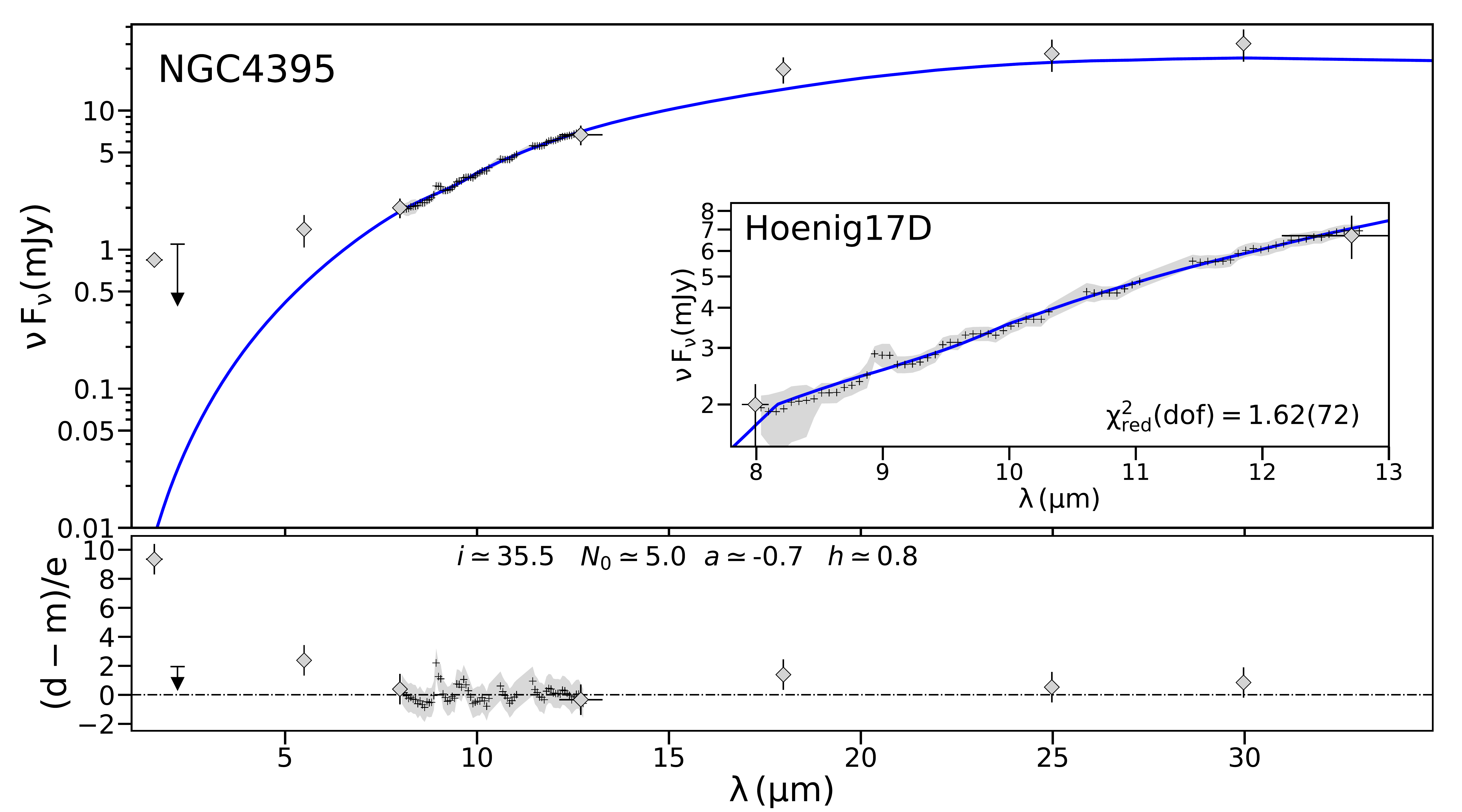}
    \caption{Same as Fig. \ref{fig:ESO005-G004} but for NGC4395.}
    \label{fig:NGC4395}
\end{figure*}

\begin{figure*}
    \centering
    \includegraphics[width=0.75\columnwidth]{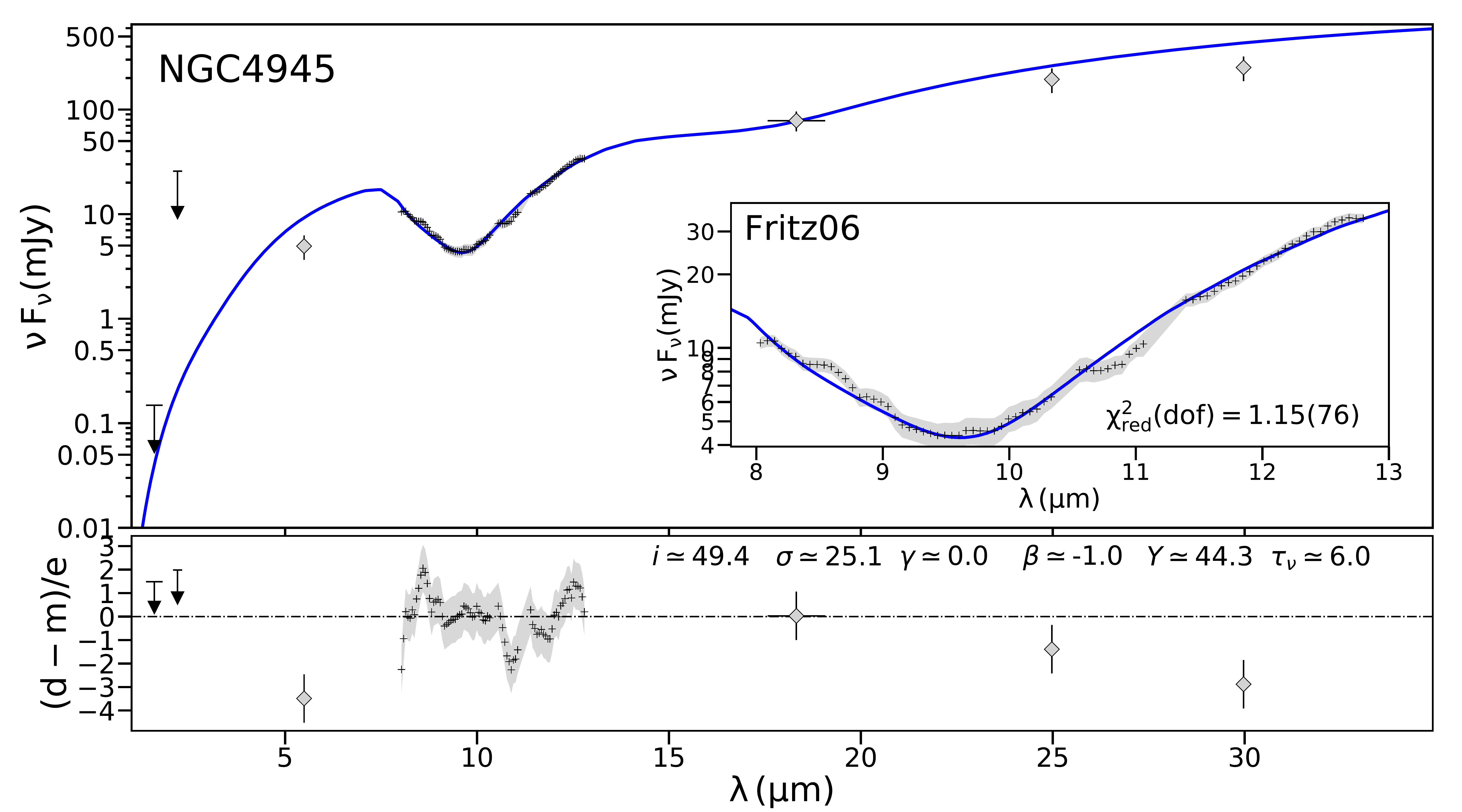}
    \includegraphics[width=0.75\columnwidth]{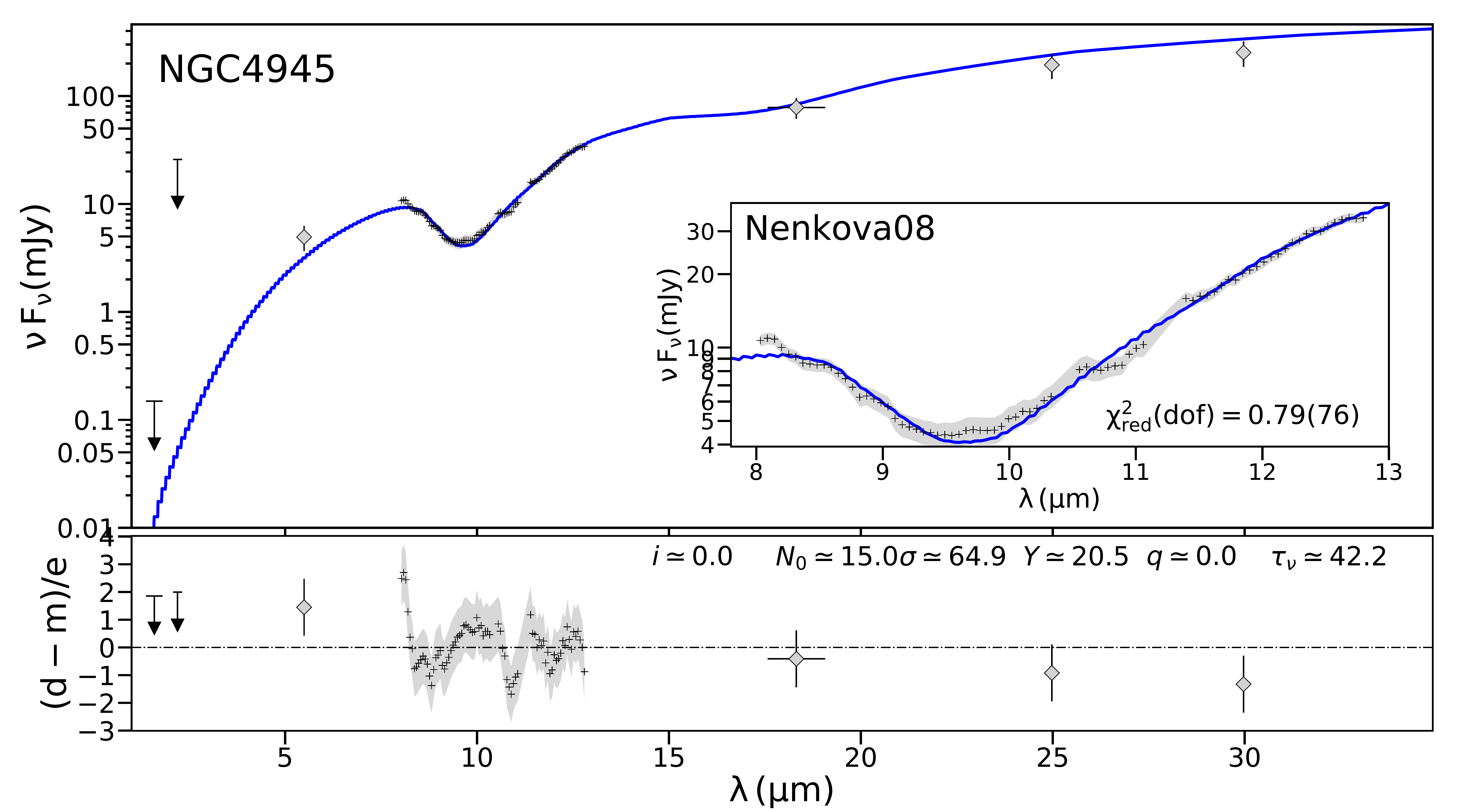}
    \includegraphics[width=0.75\columnwidth]{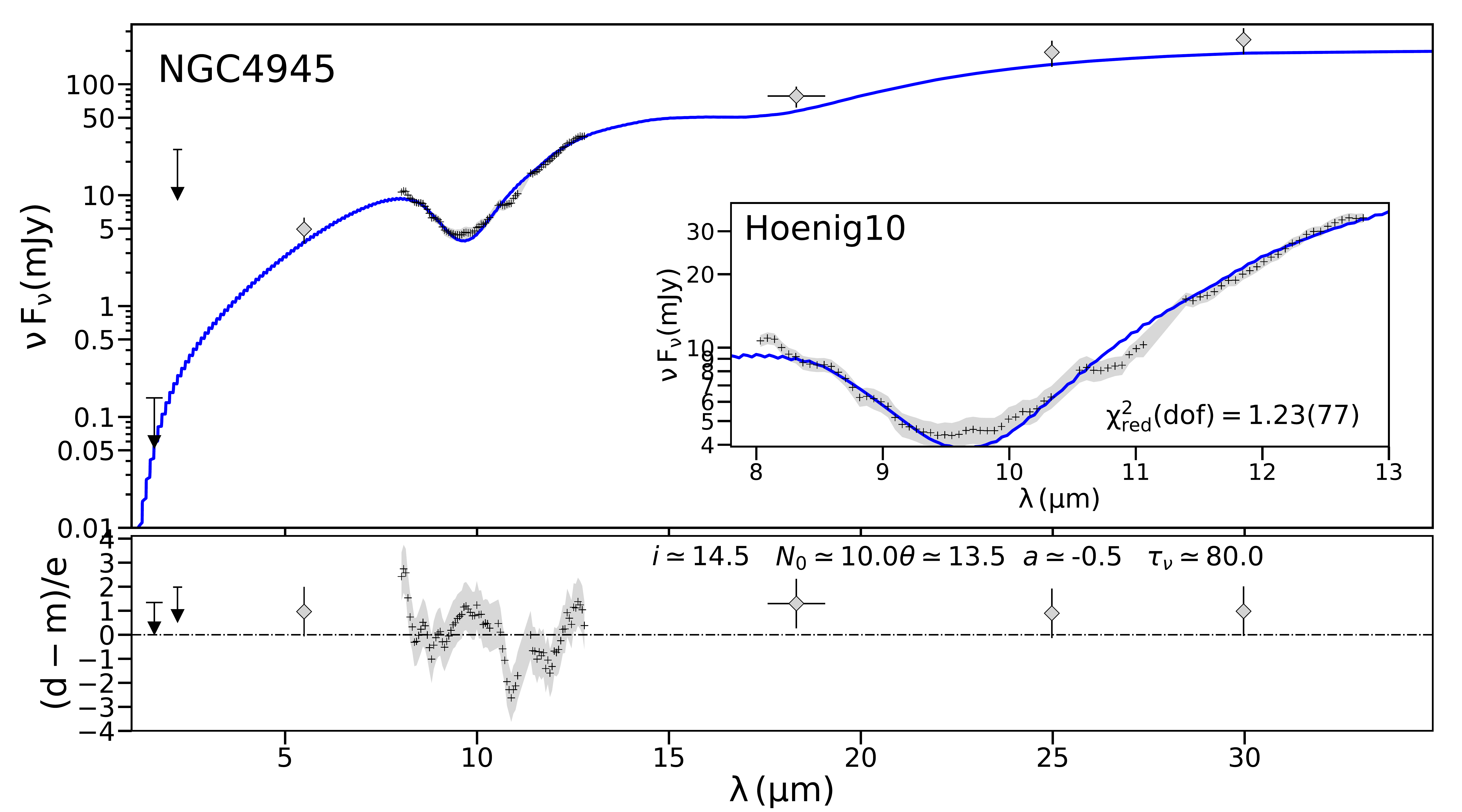}
    \includegraphics[width=0.75\columnwidth]{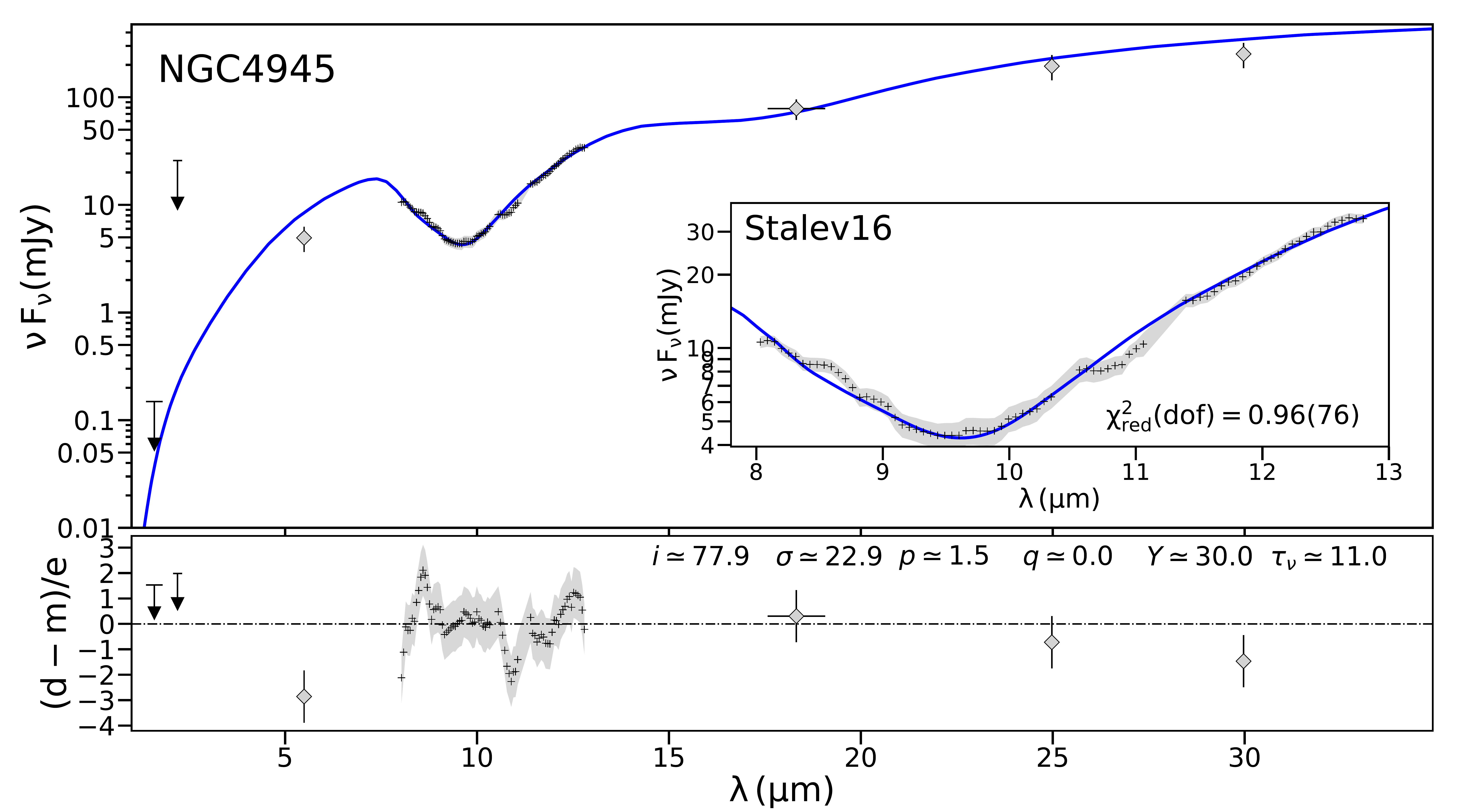}
    \includegraphics[width=0.75\columnwidth]{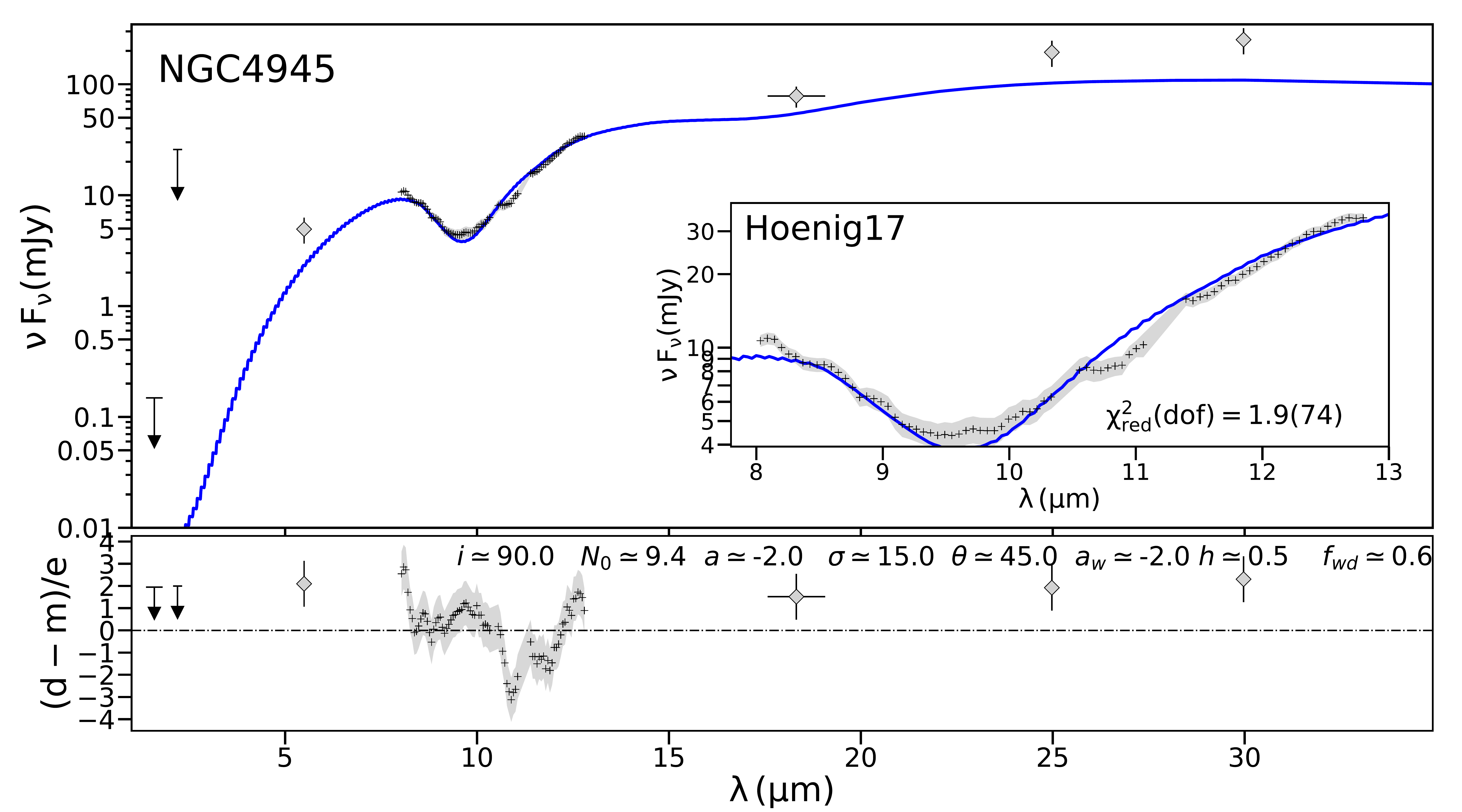}
    \includegraphics[width=0.75\columnwidth]{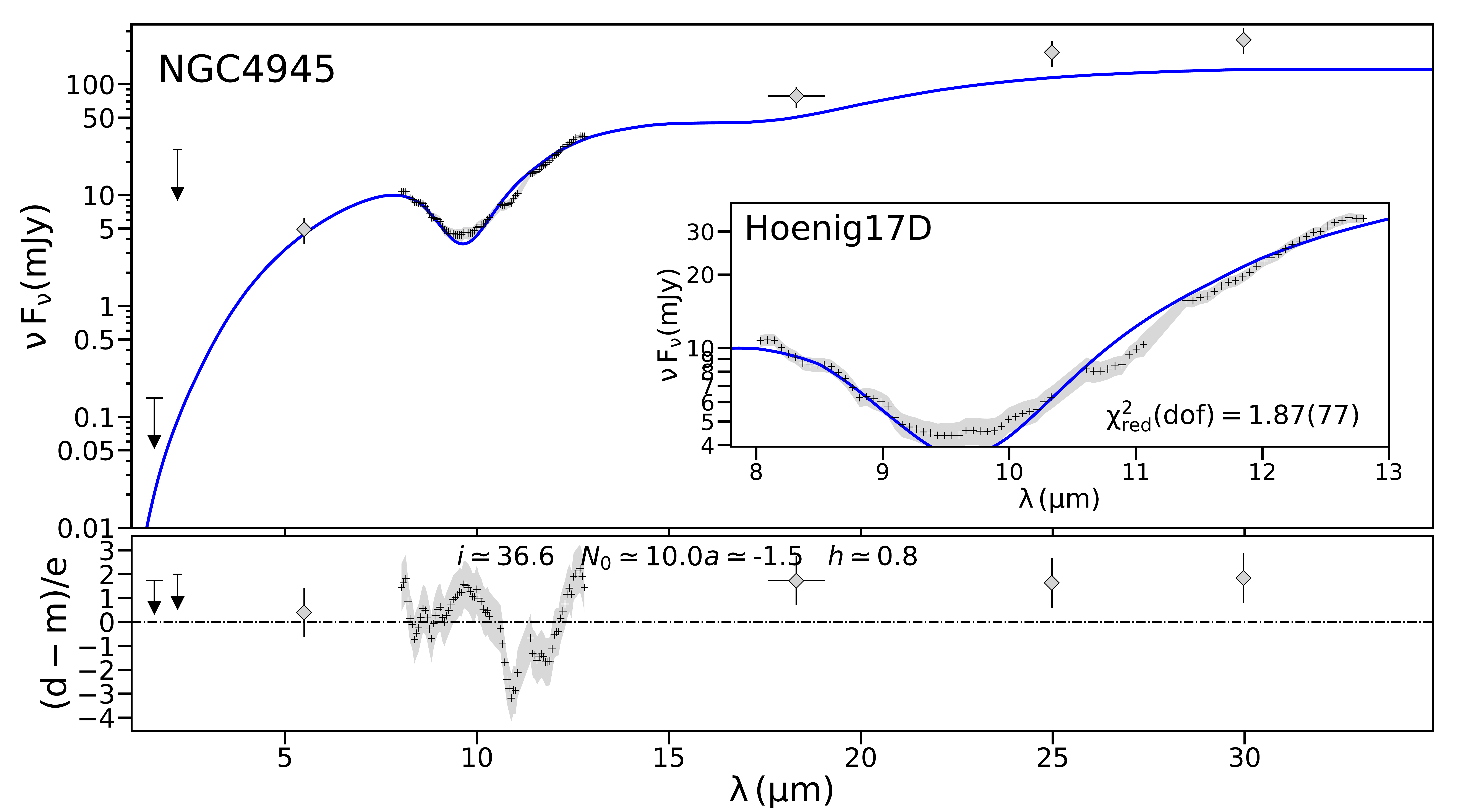}
    \caption{Same as Fig. \ref{fig:ESO005-G004} but for NGC4945.}
    \label{fig:NGC4945}
\end{figure*}

\begin{figure*}
    \centering
    \includegraphics[width=0.75\columnwidth]{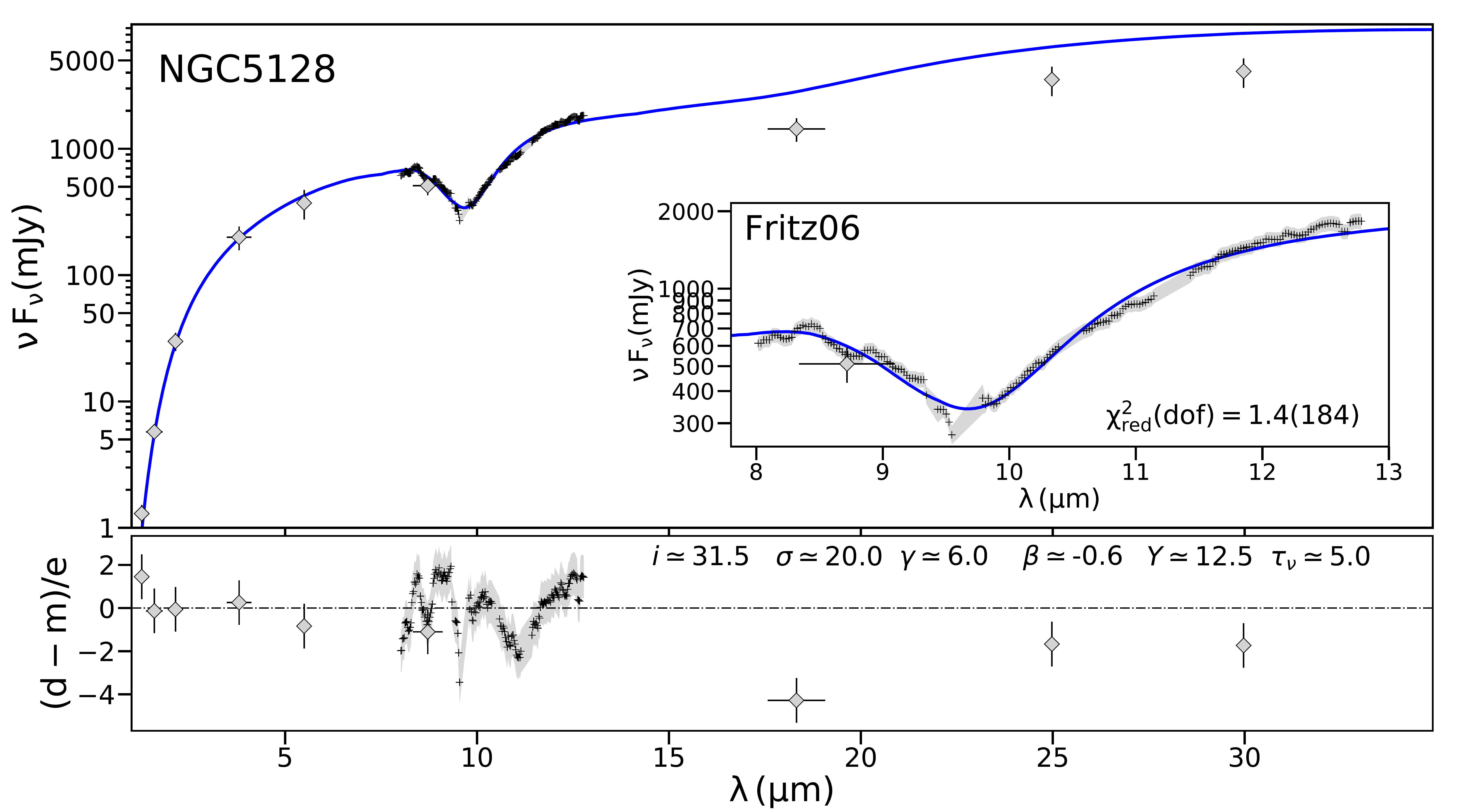}
    \includegraphics[width=0.75\columnwidth]{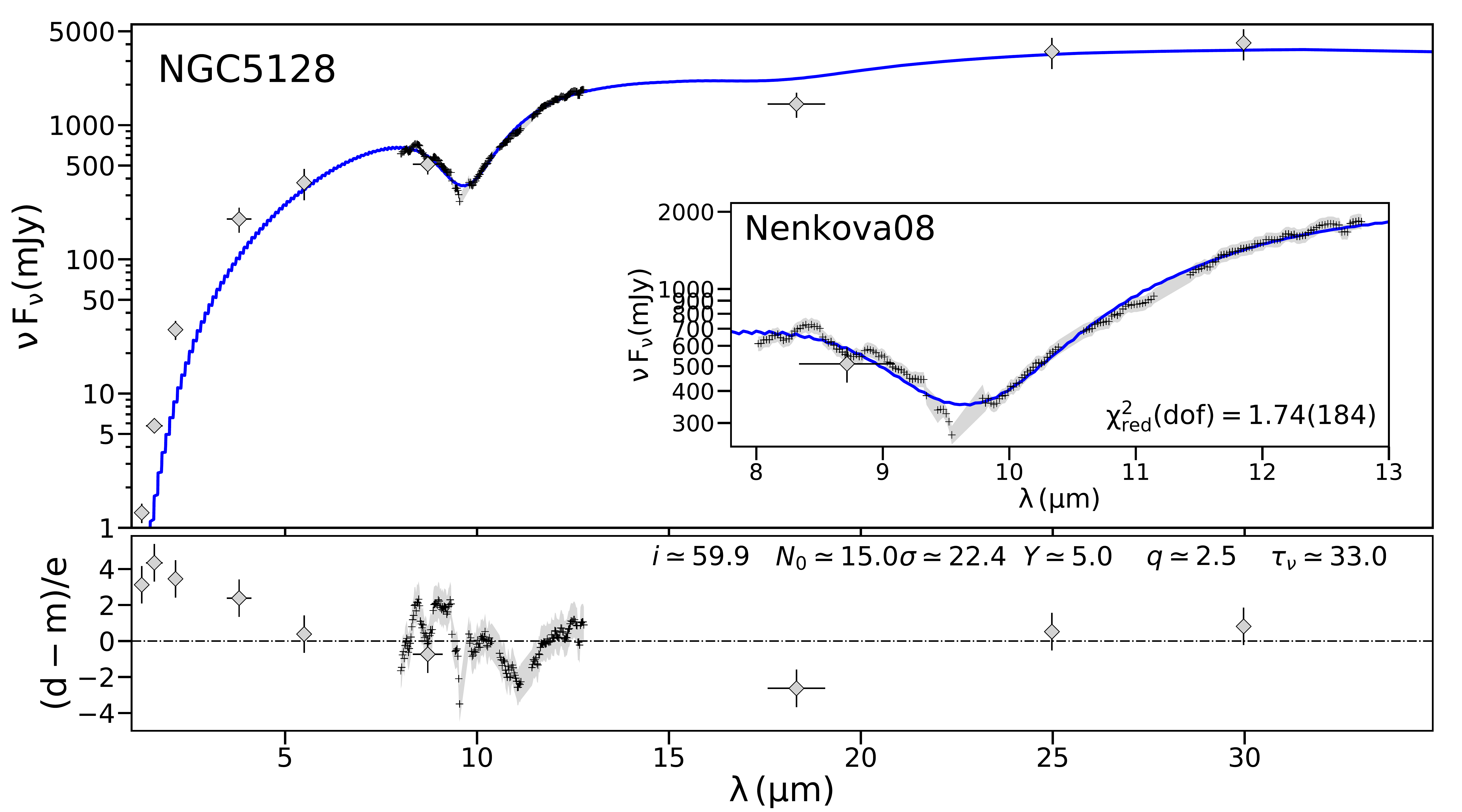}
    \includegraphics[width=0.75\columnwidth]{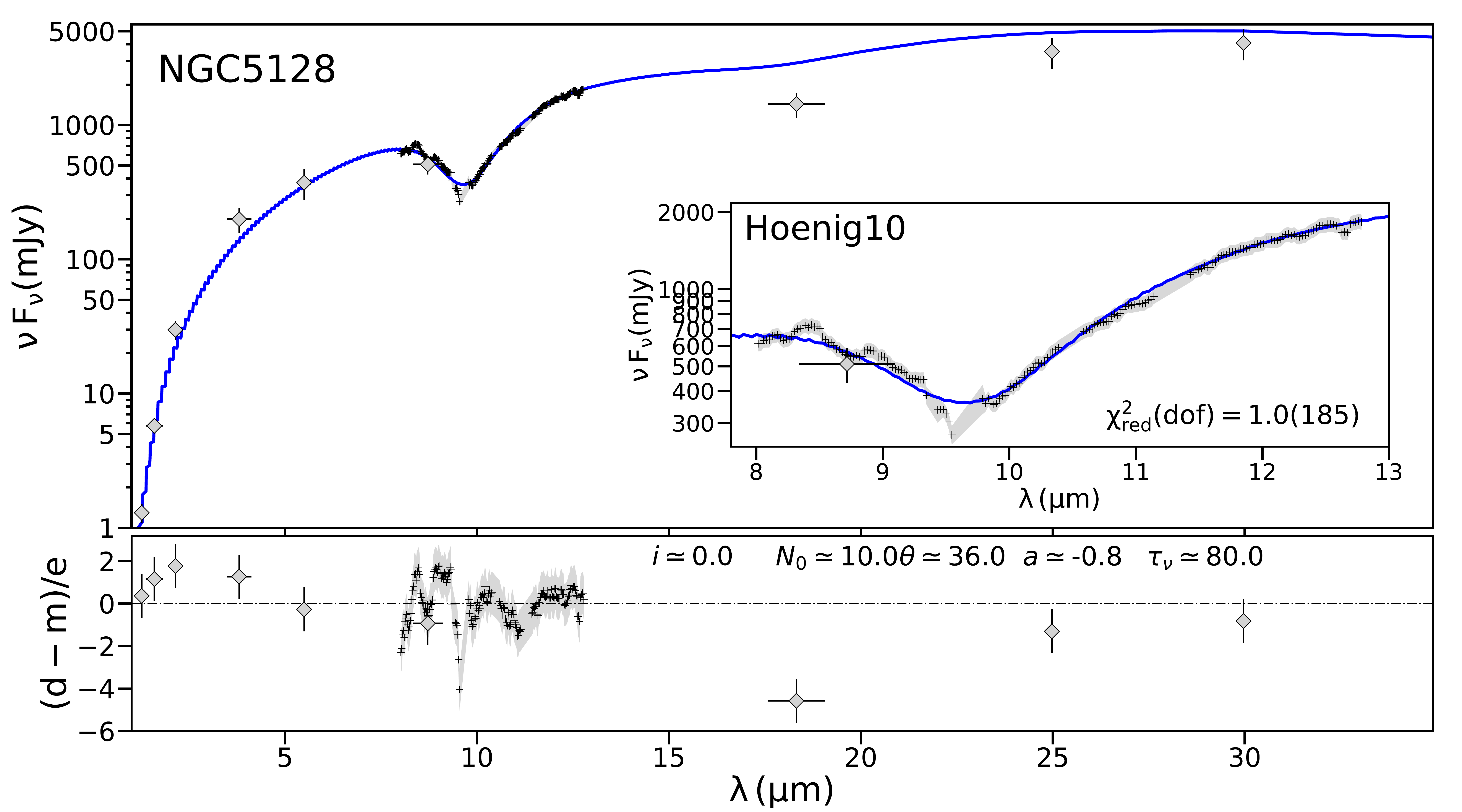}
    \includegraphics[width=0.75\columnwidth]{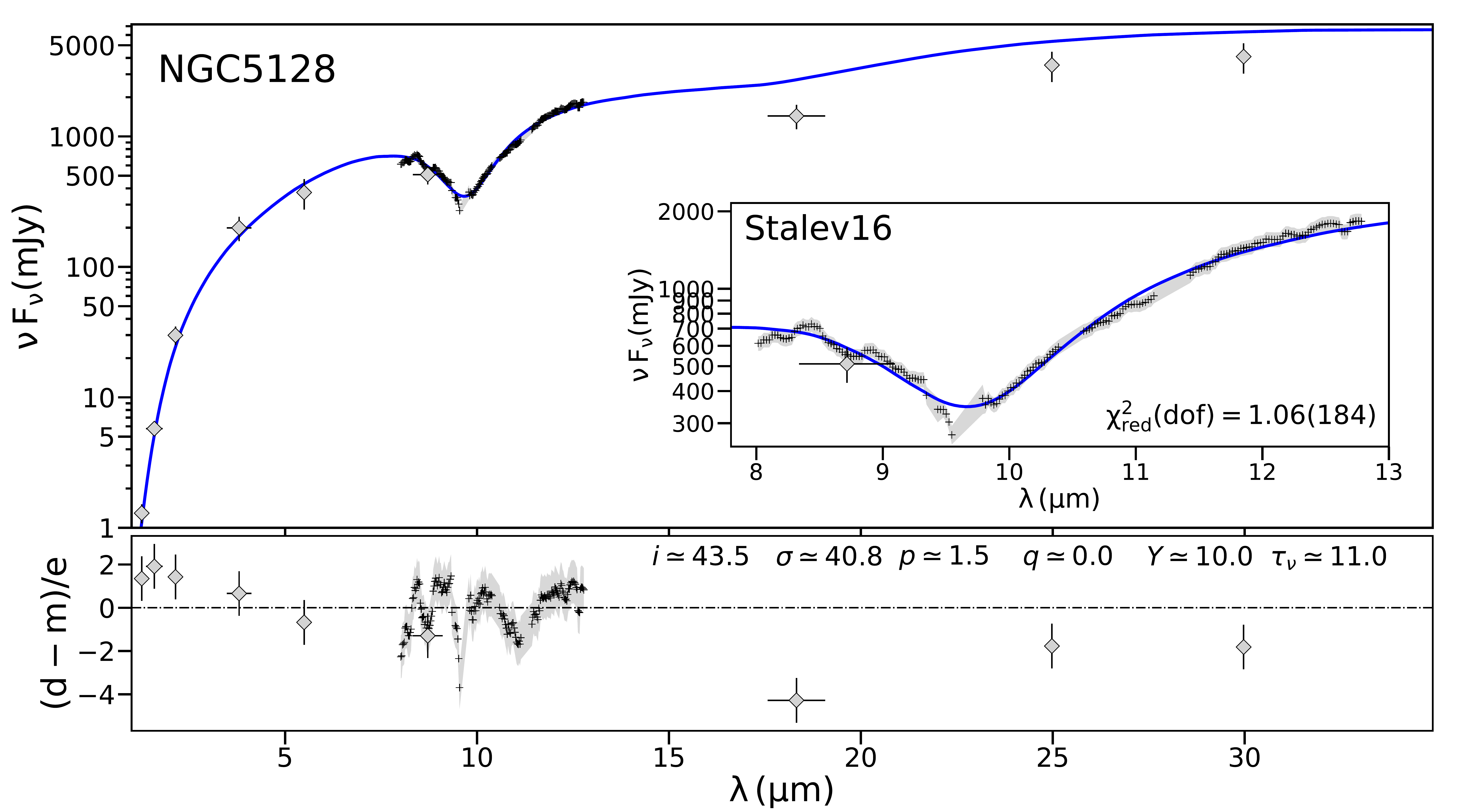}
    \includegraphics[width=0.75\columnwidth]{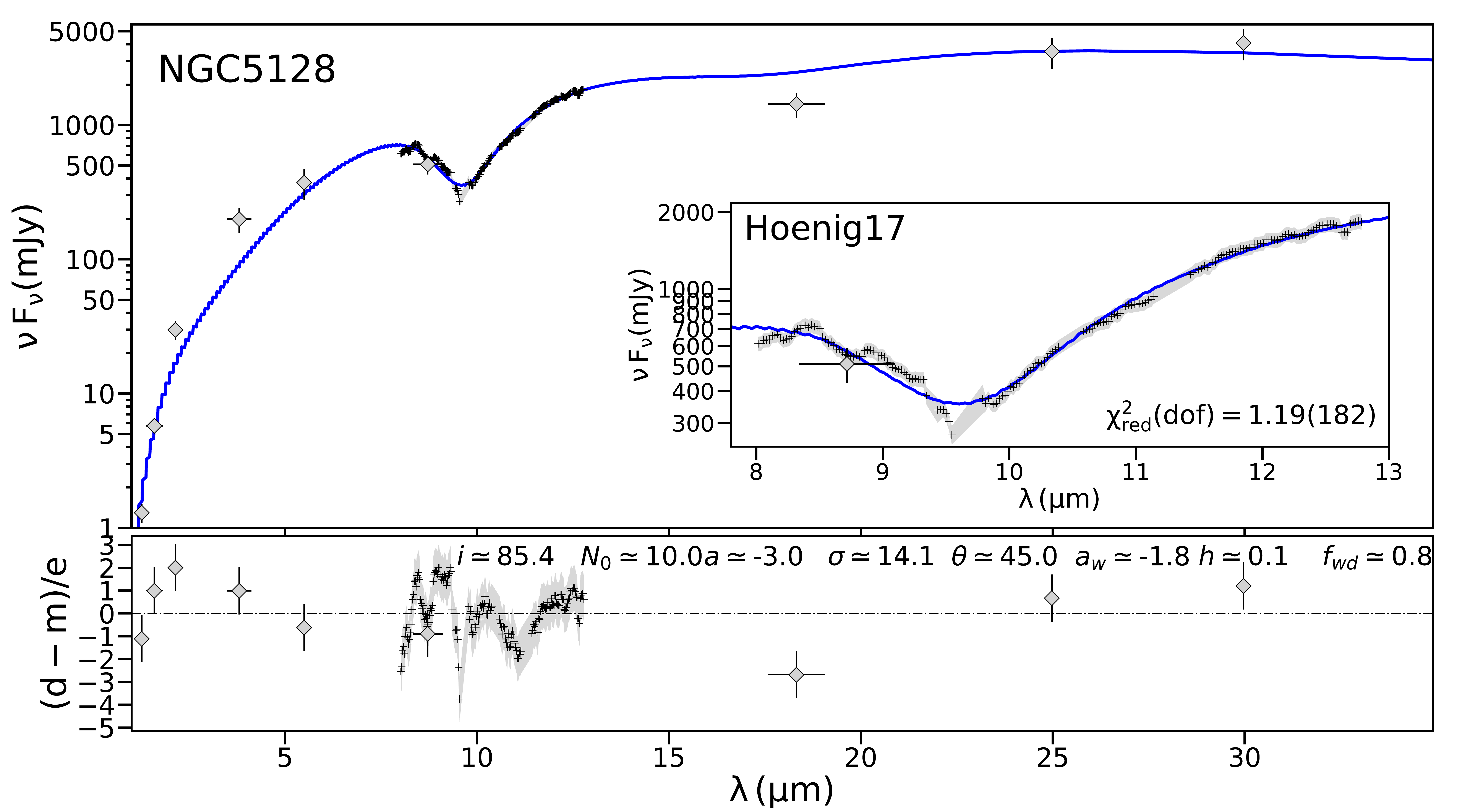}
    \includegraphics[width=0.75\columnwidth]{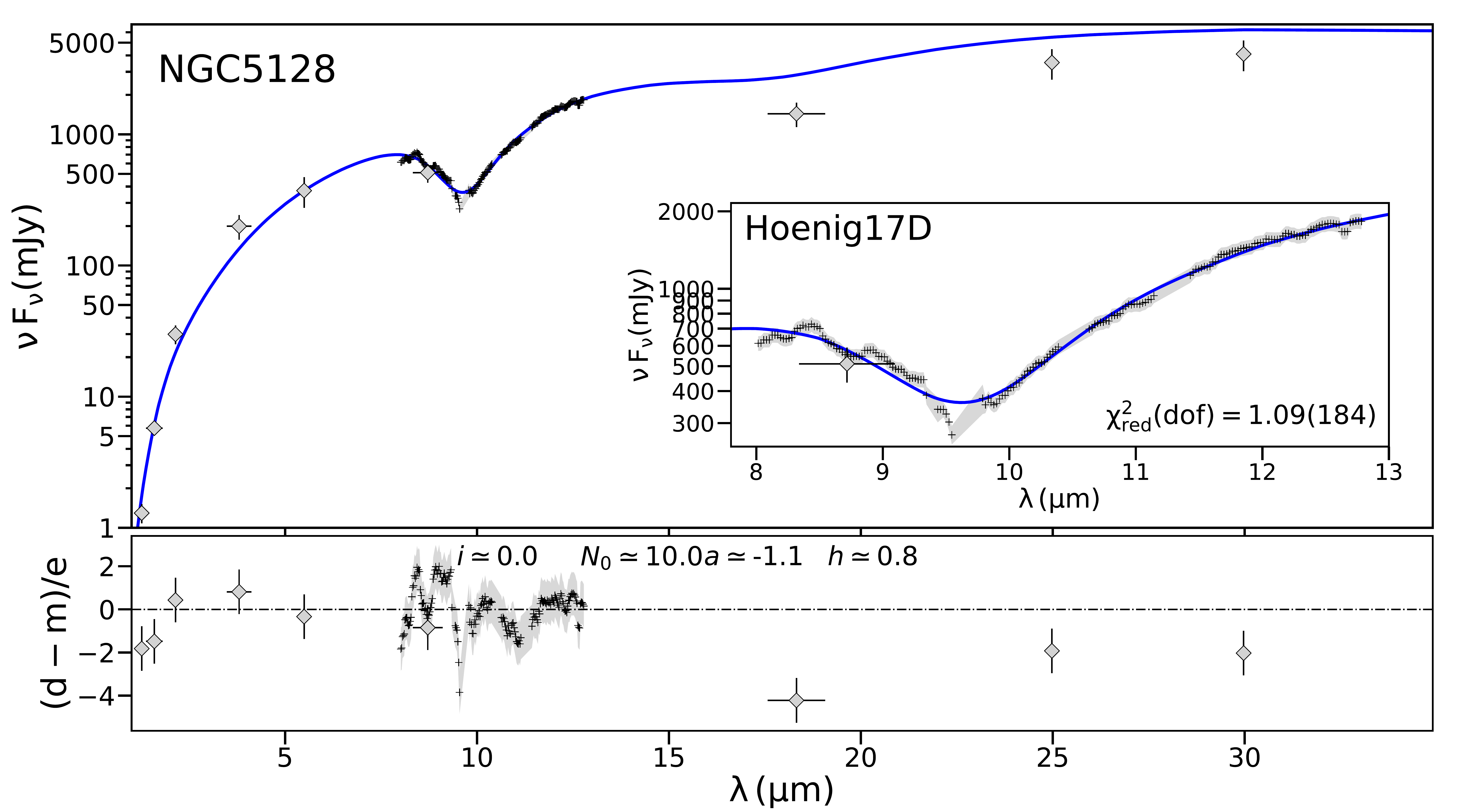}
    \caption{Same as Fig. \ref{fig:ESO005-G004} but for NGC5128.}
    \label{fig:NGC5128}
\end{figure*}

\begin{figure*}
    \centering
    \includegraphics[width=0.75\columnwidth]{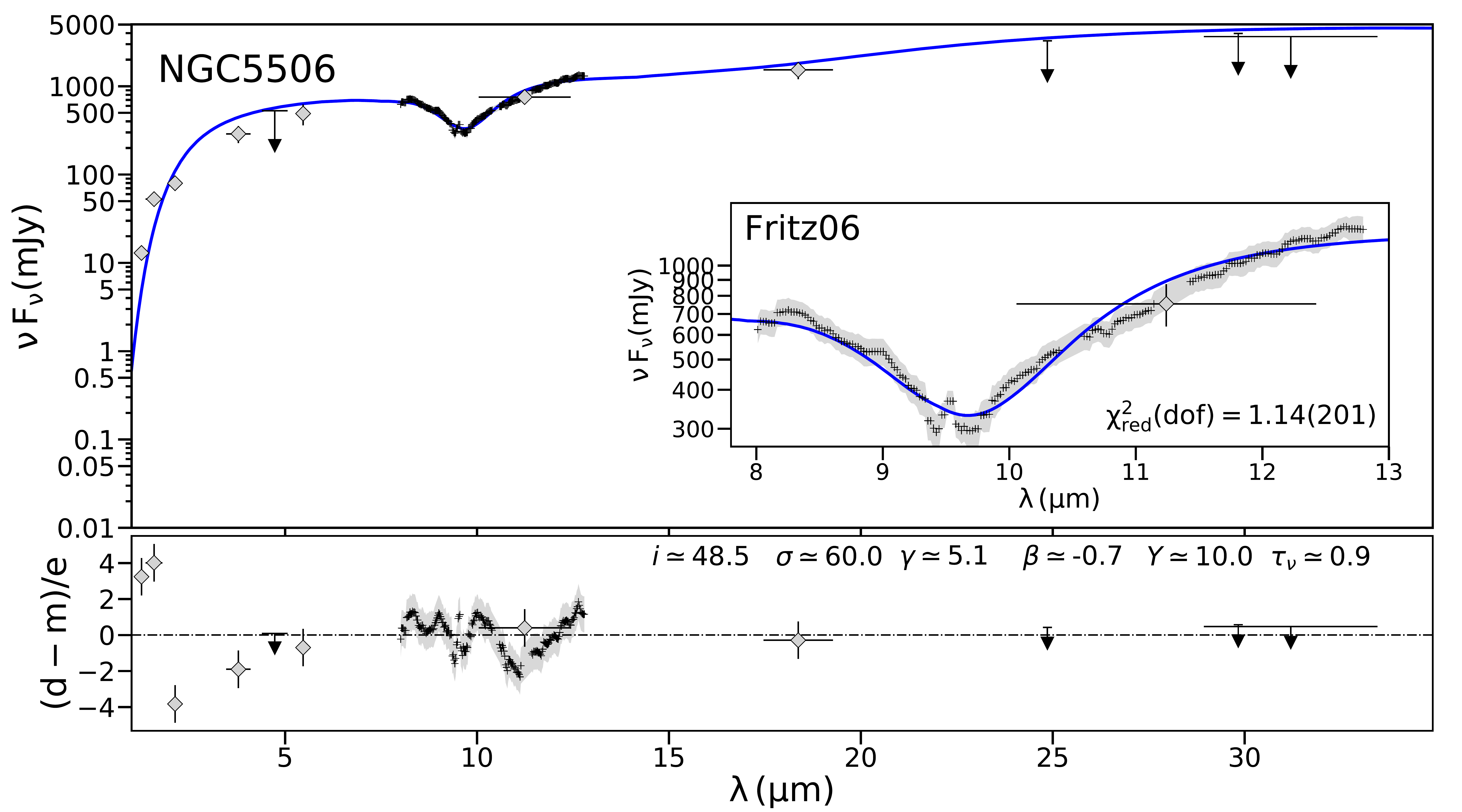}
    \includegraphics[width=0.75\columnwidth]{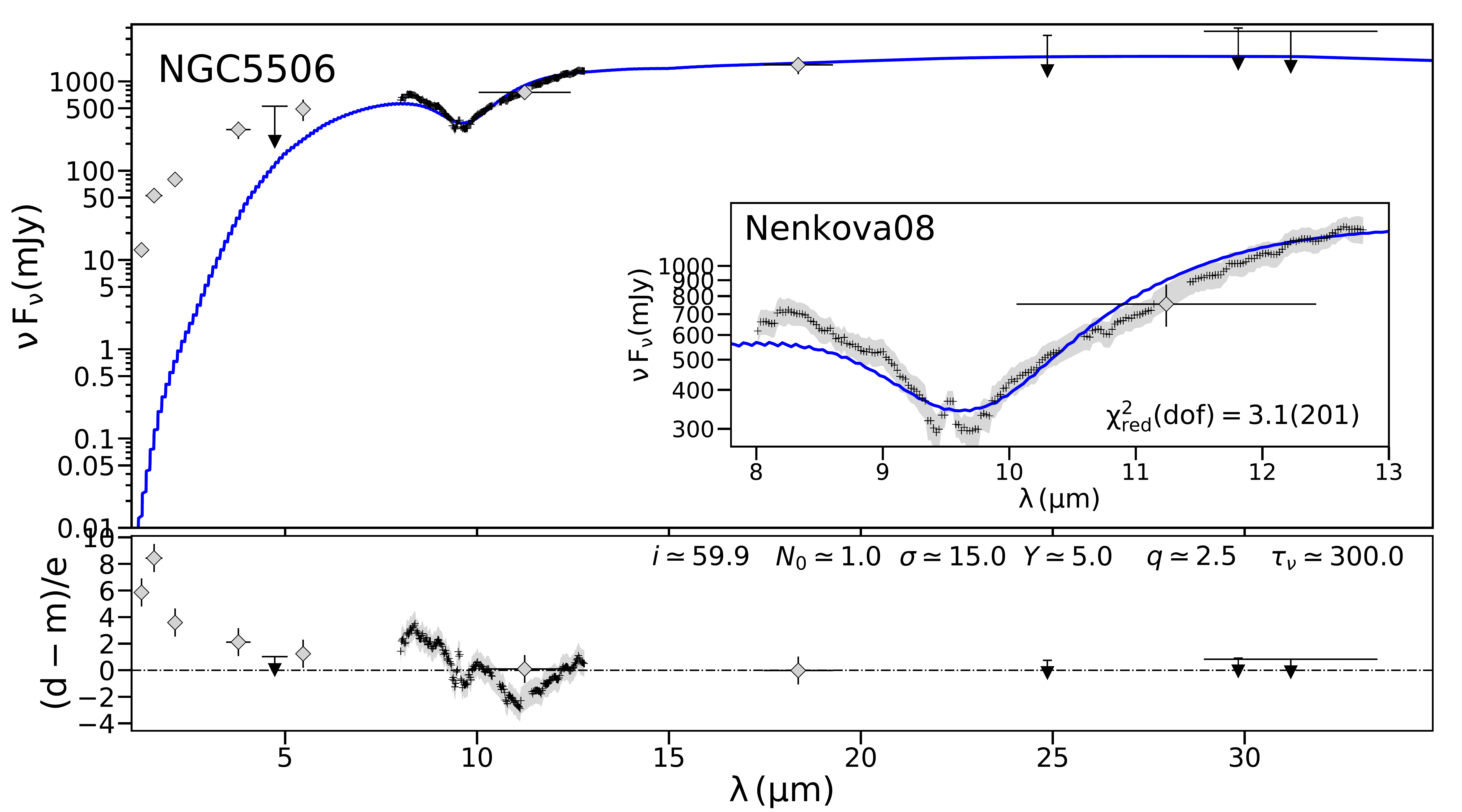}
    \includegraphics[width=0.75\columnwidth]{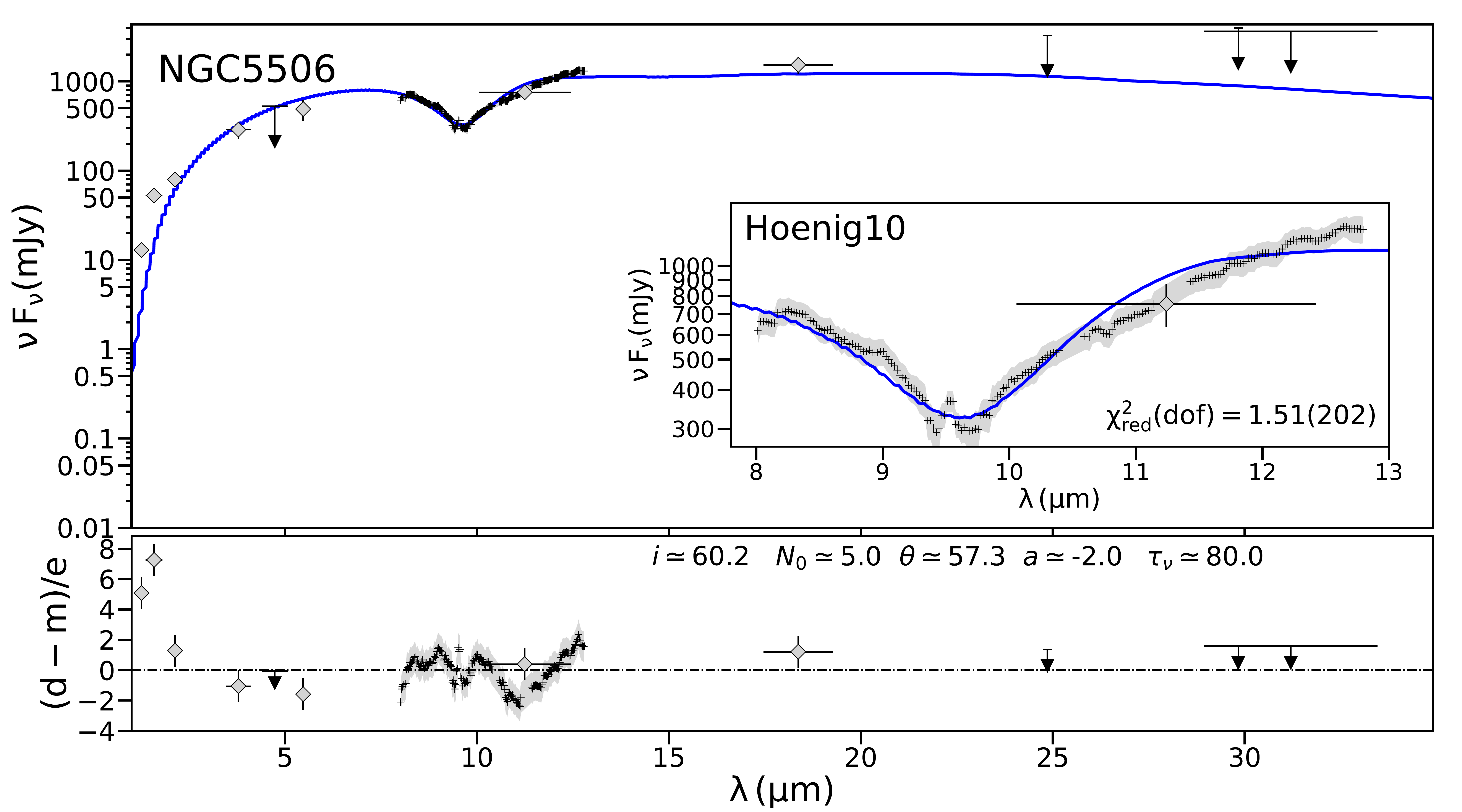}
    \includegraphics[width=0.75\columnwidth]{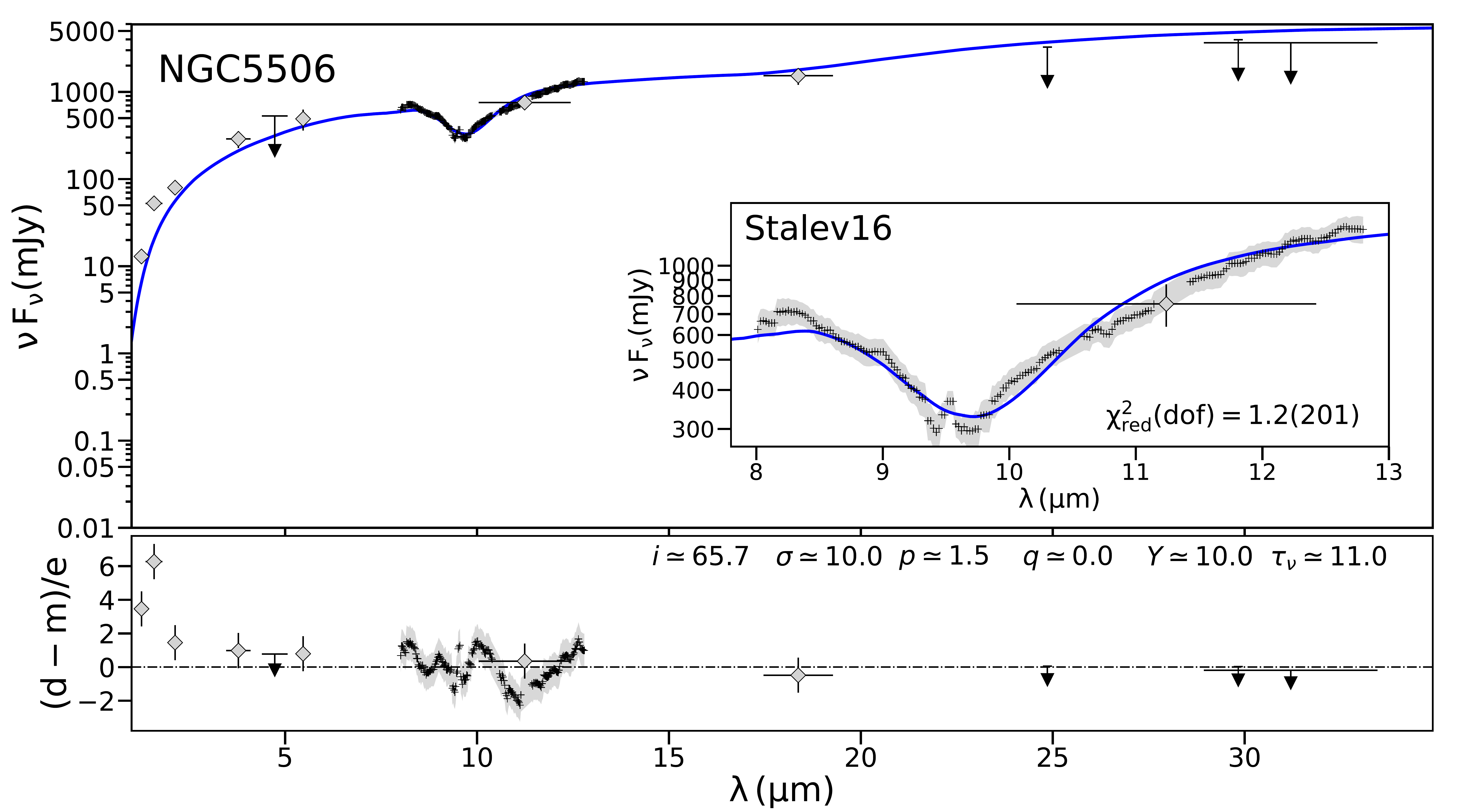}
    \includegraphics[width=0.75\columnwidth]{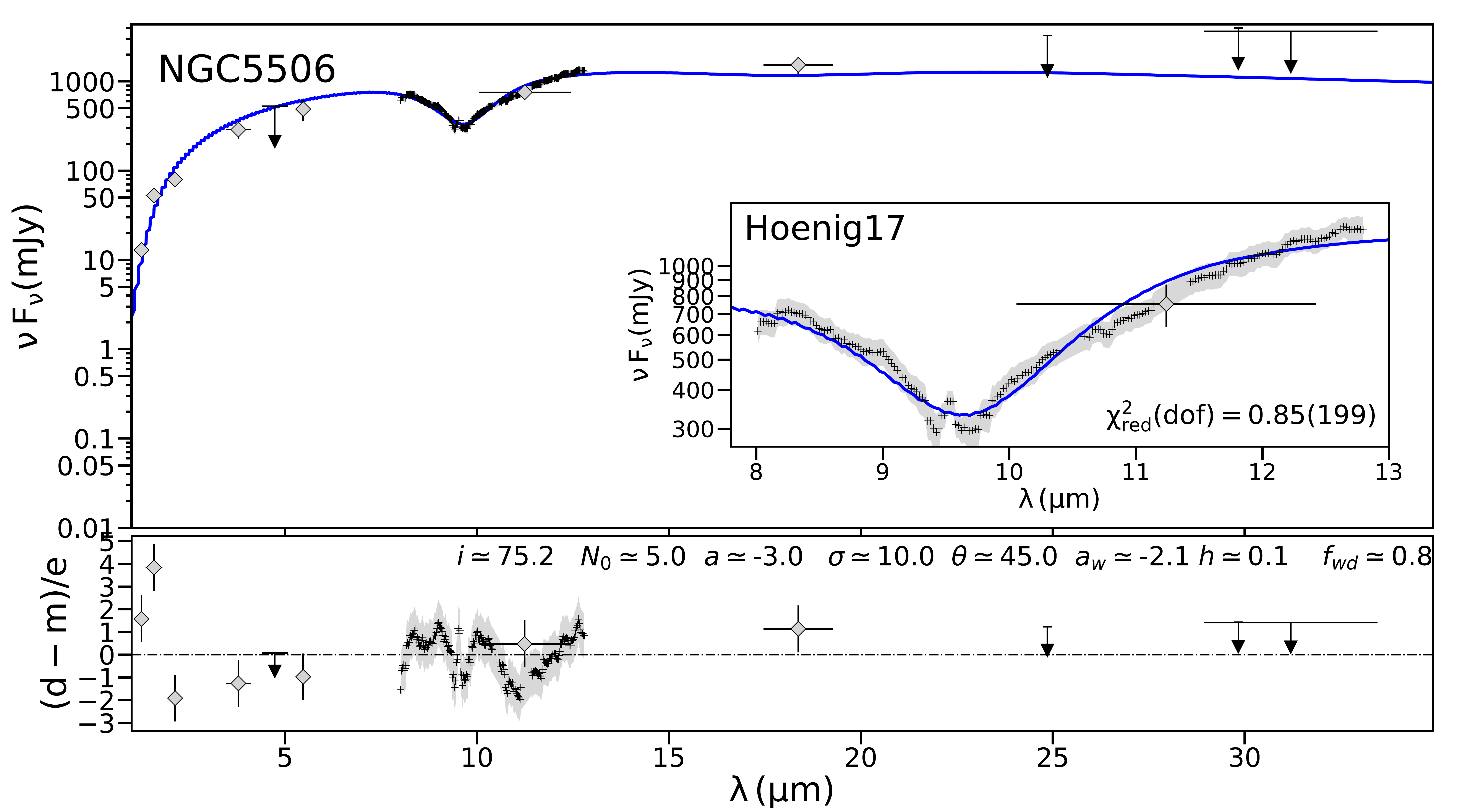}
    \includegraphics[width=0.75\columnwidth]{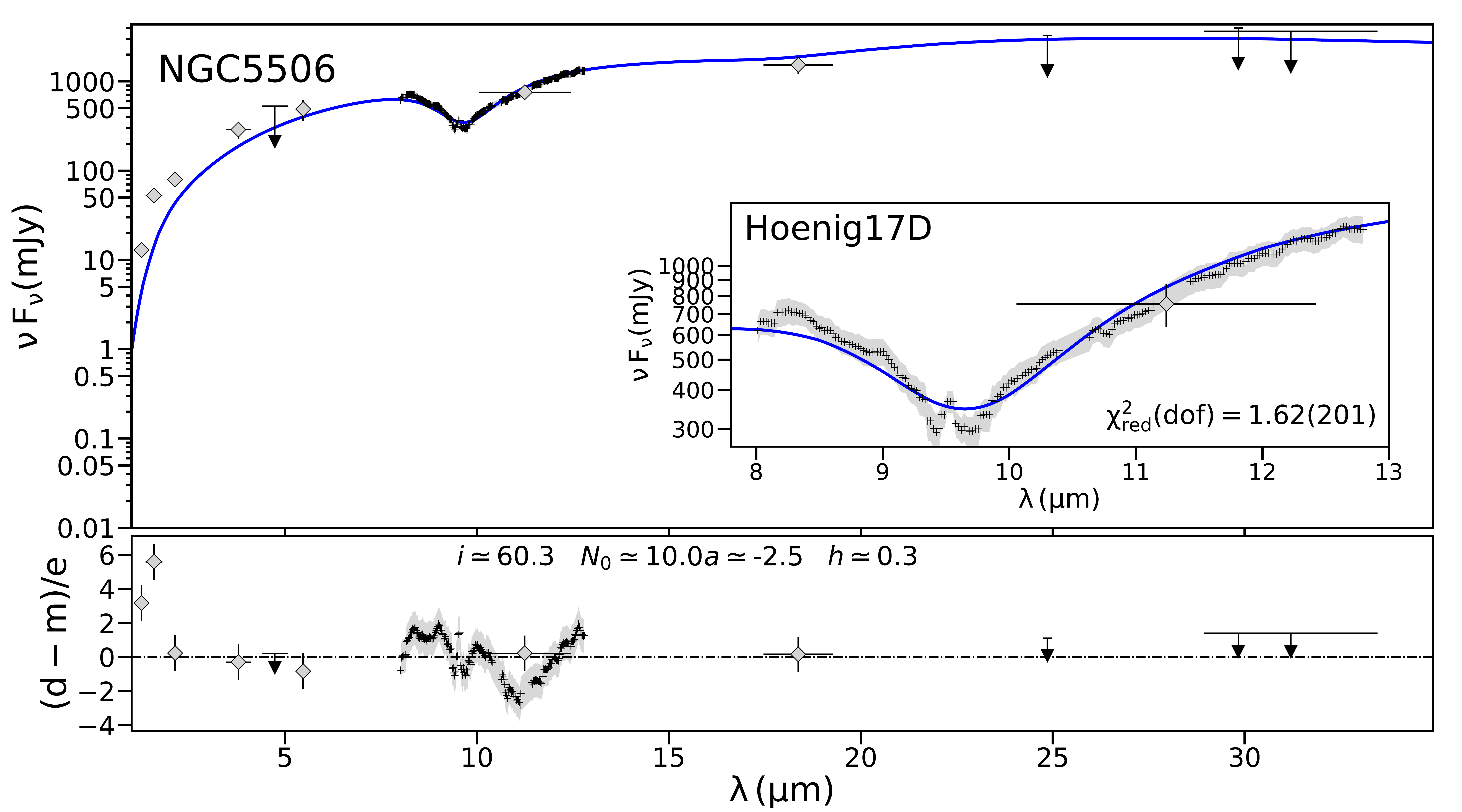}
    \caption{Same as Fig. \ref{fig:ESO005-G004} but for NGC5506.}
    \label{fig:NGC5506}
\end{figure*}

\begin{figure*}
    \centering
    \includegraphics[width=0.75\columnwidth]{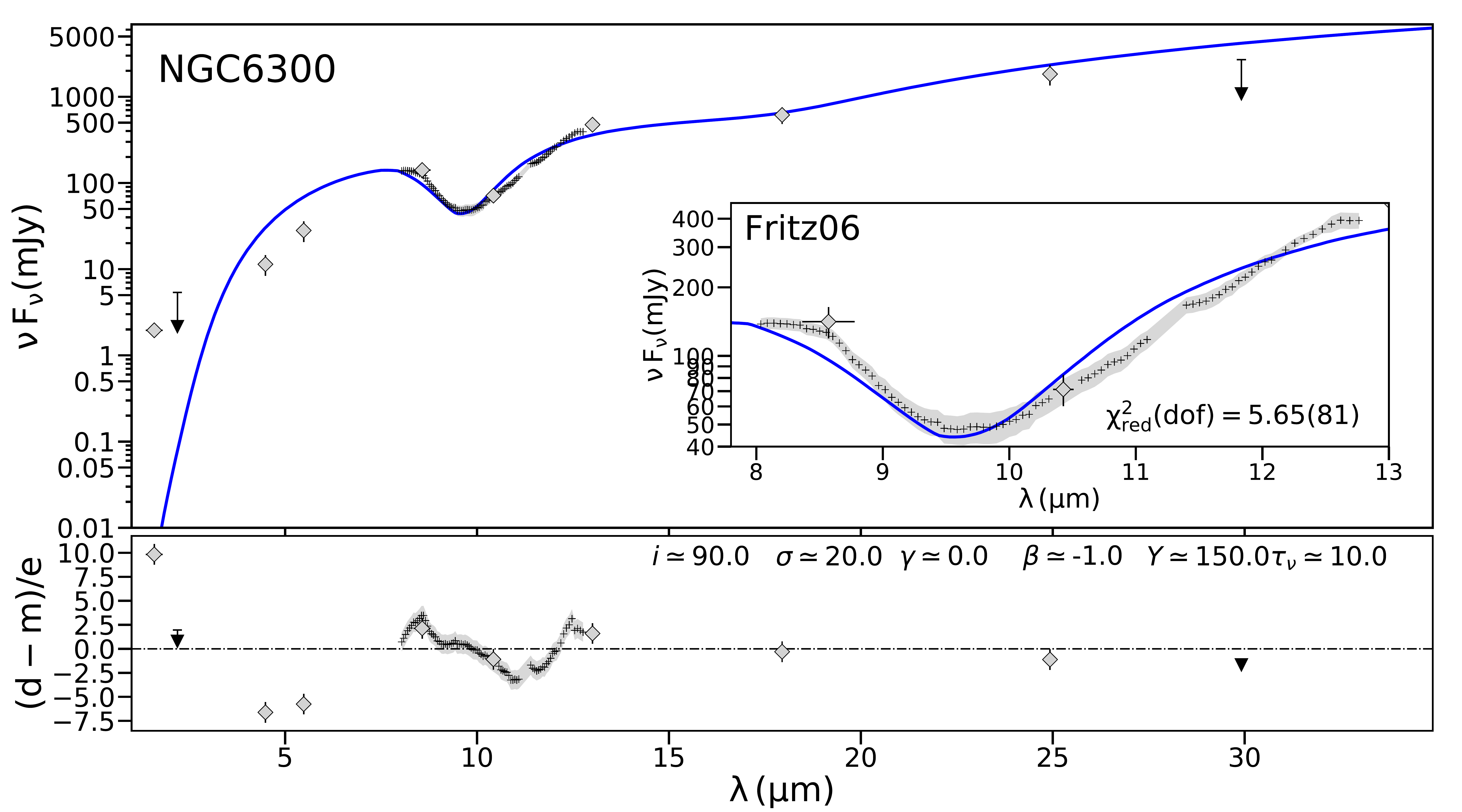}
    \includegraphics[width=0.75\columnwidth]{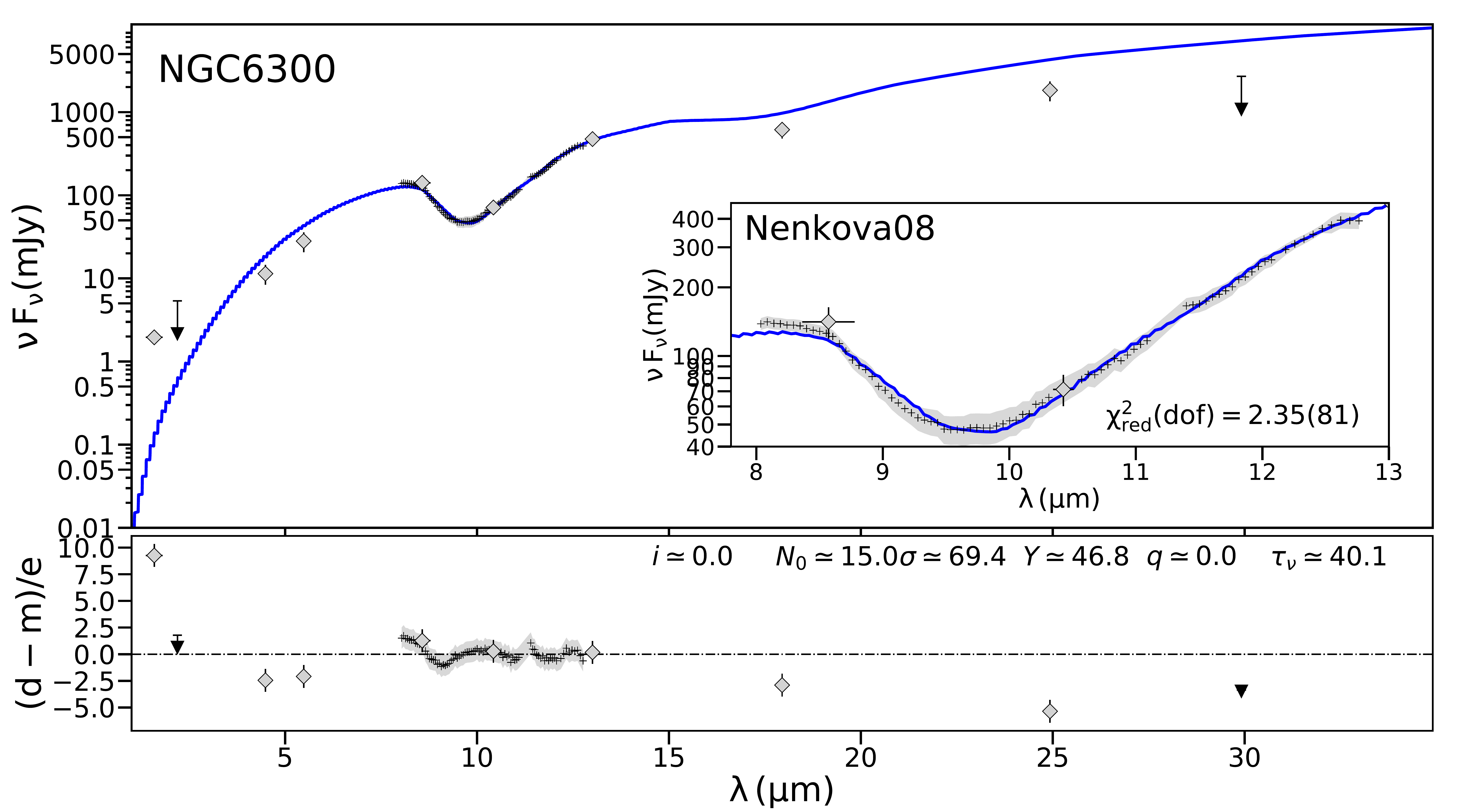}
    \includegraphics[width=0.75\columnwidth]{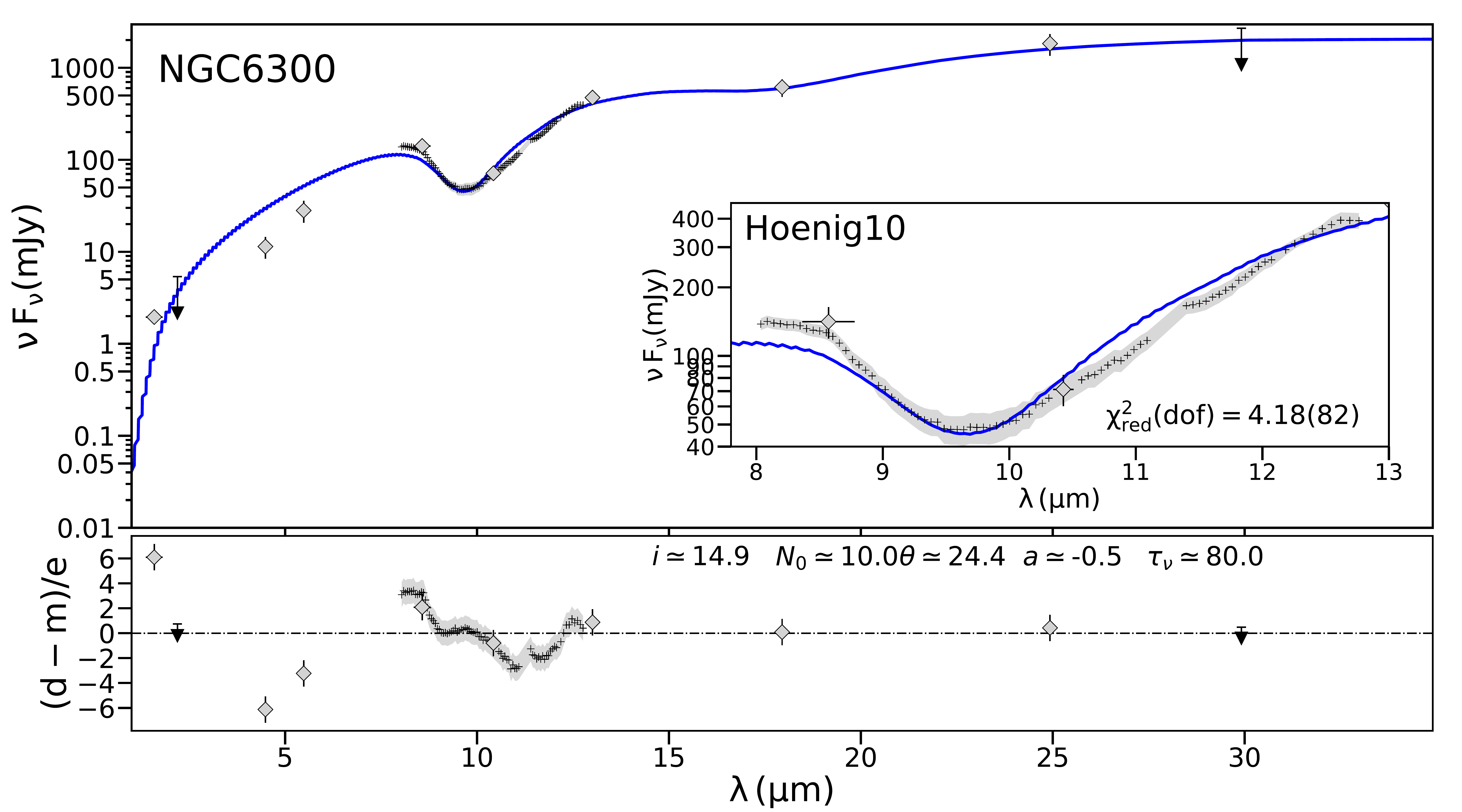}
    \includegraphics[width=0.75\columnwidth]{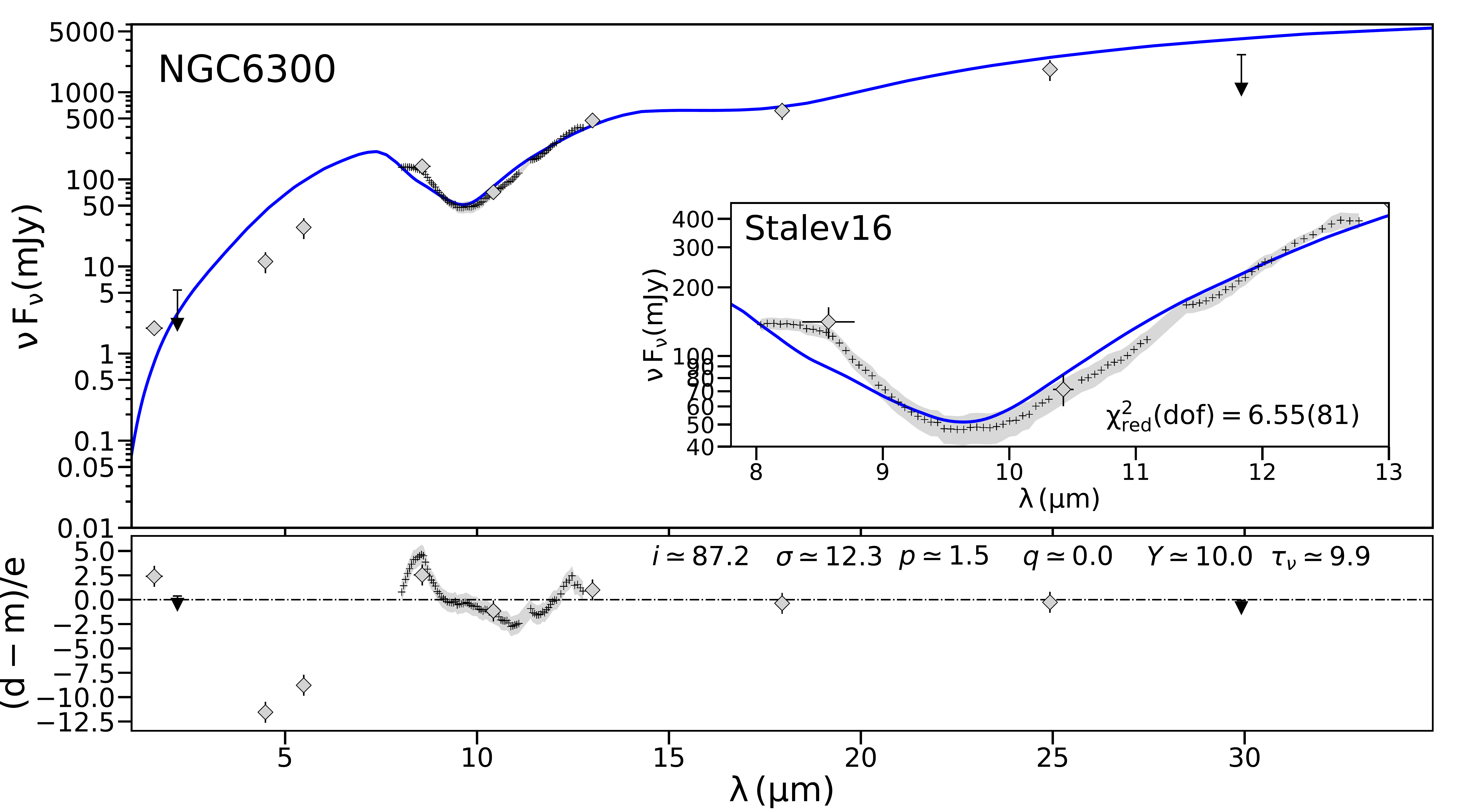}
    \includegraphics[width=0.75\columnwidth]{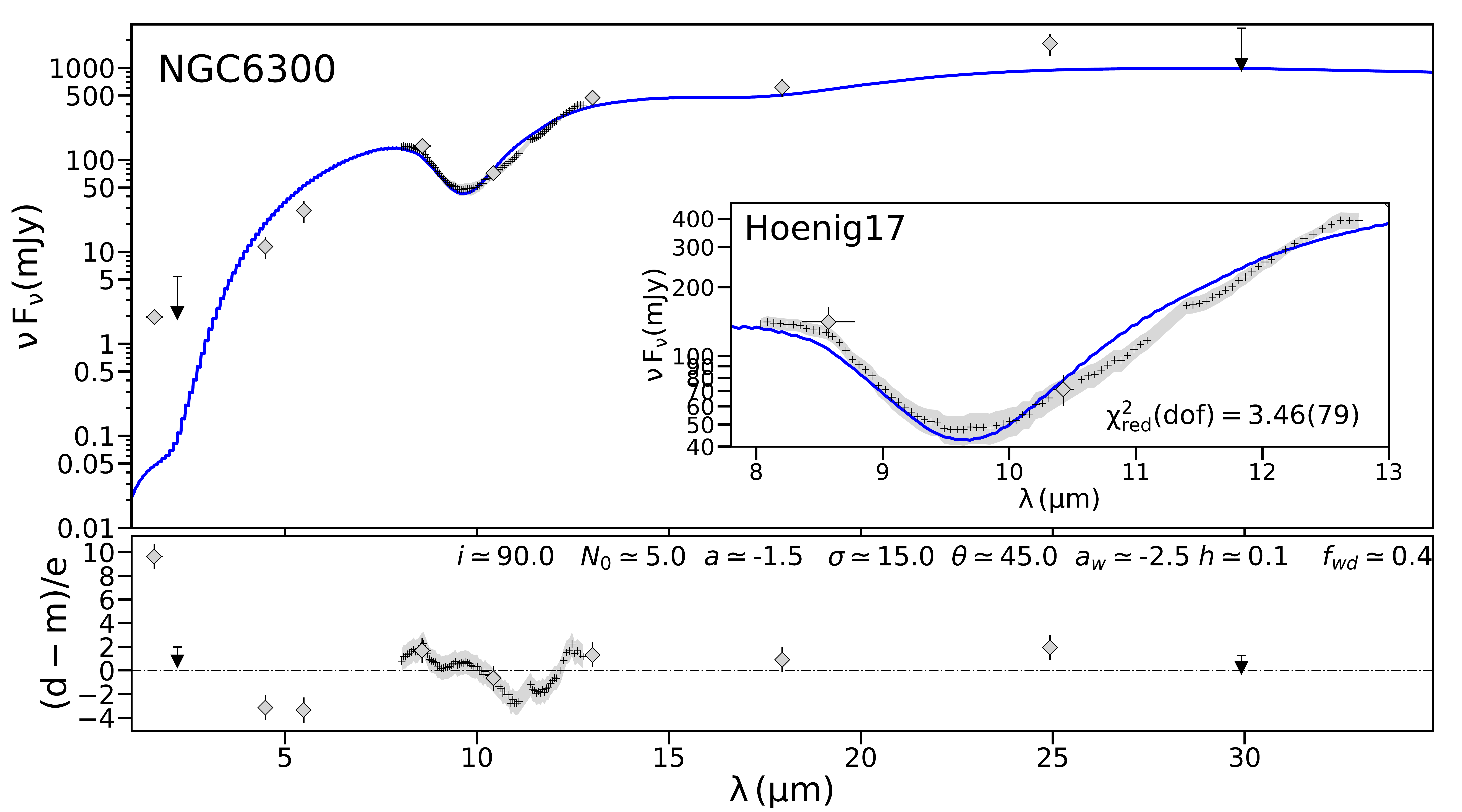}
    \includegraphics[width=0.75\columnwidth]{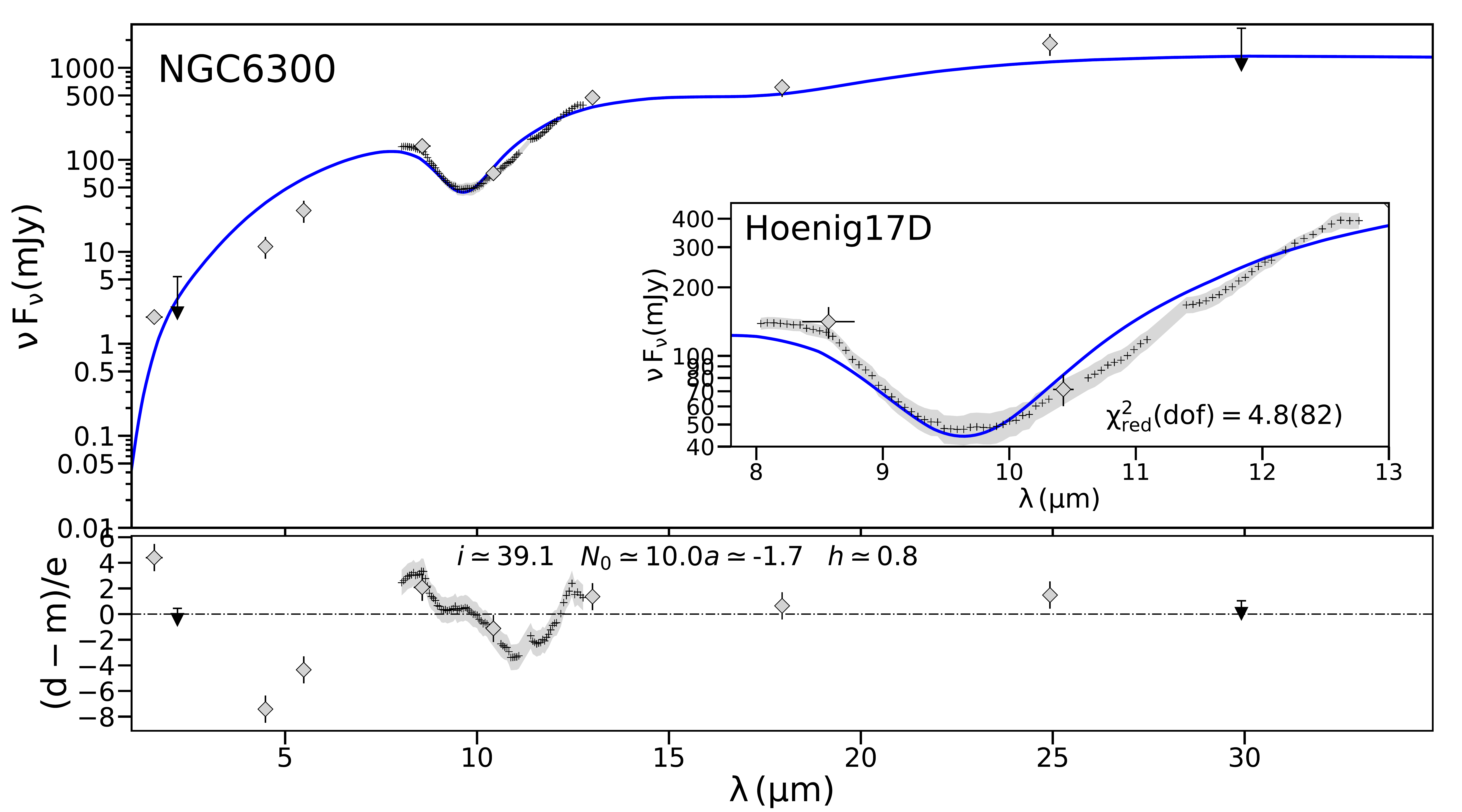}
    \caption{Same as Fig. \ref{fig:ESO005-G004} but for NGC6300.}
    \label{fig:NGC6300}
\end{figure*}

\begin{figure*}
    \centering
    \includegraphics[width=0.75\columnwidth]{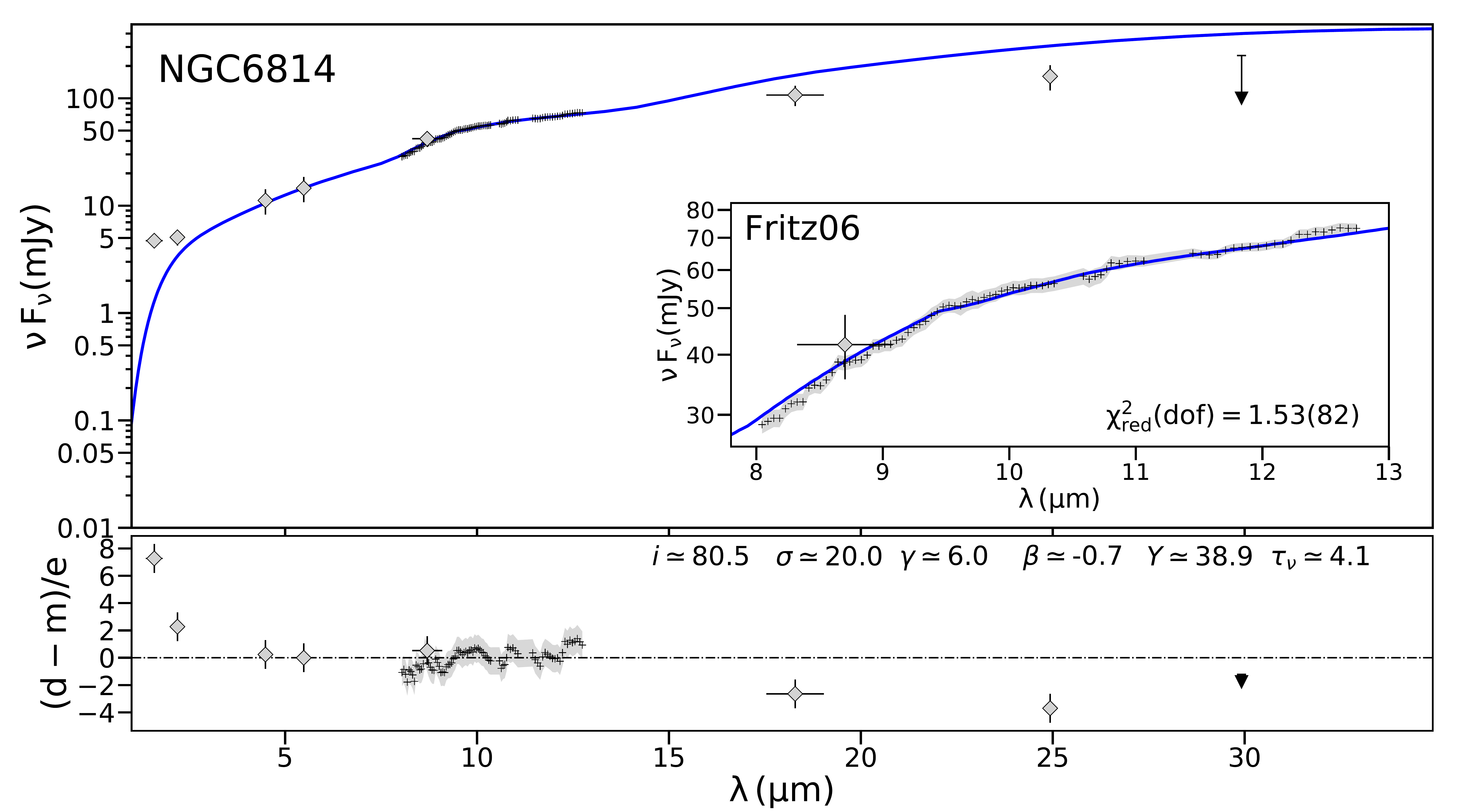}
    \includegraphics[width=0.75\columnwidth]{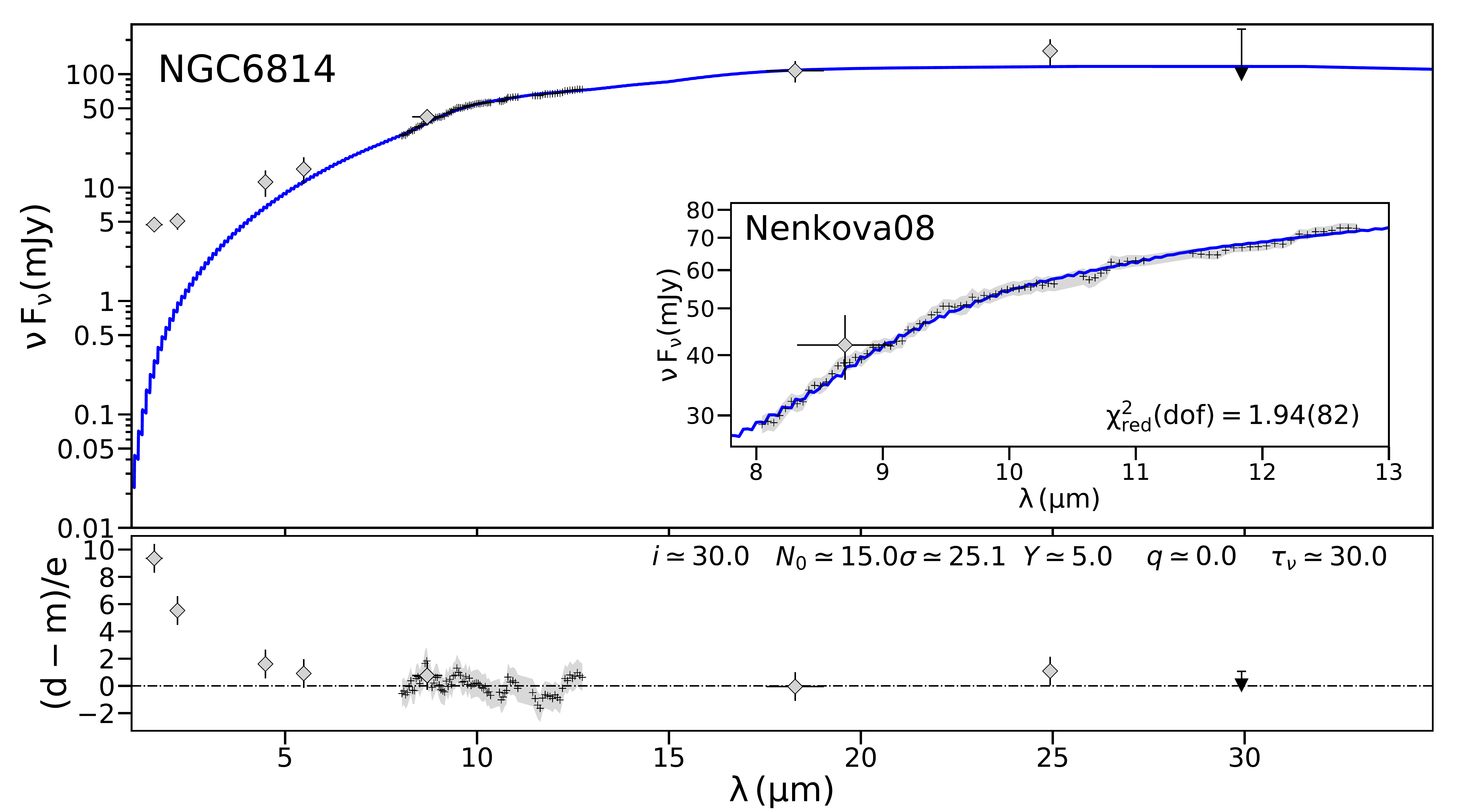}
    \includegraphics[width=0.75\columnwidth]{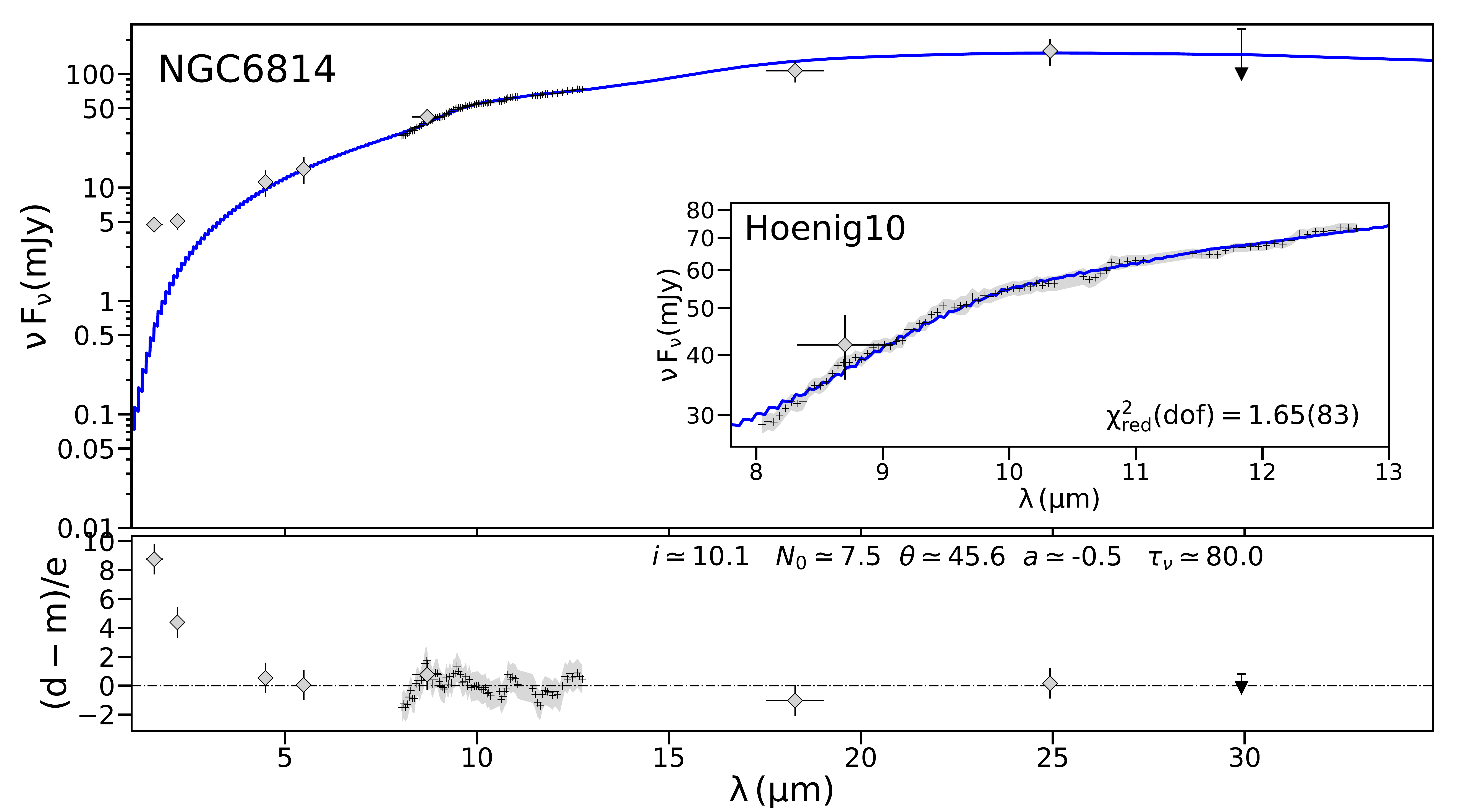}
    \includegraphics[width=0.75\columnwidth]{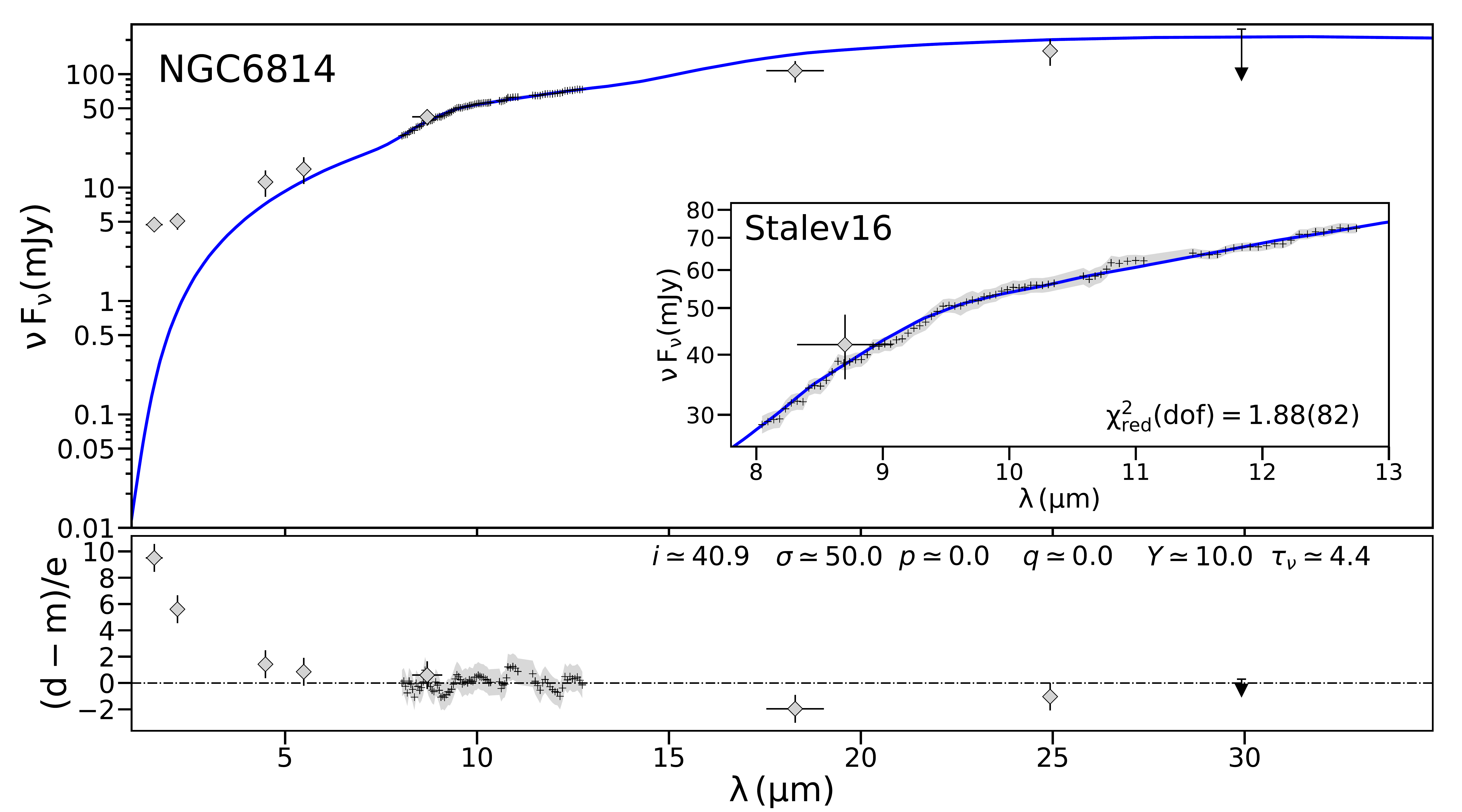}
    \includegraphics[width=0.75\columnwidth]{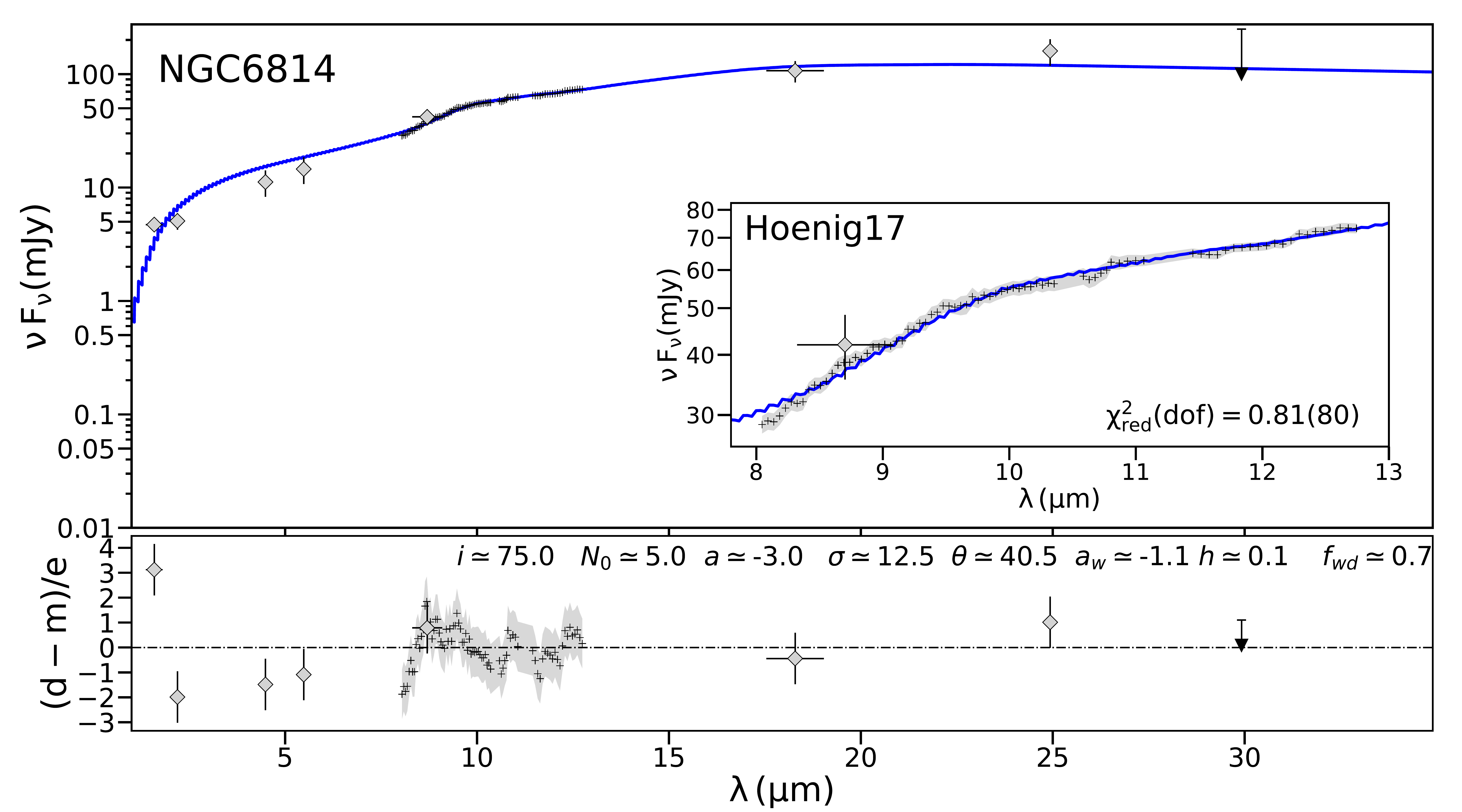}
    \includegraphics[width=0.75\columnwidth]{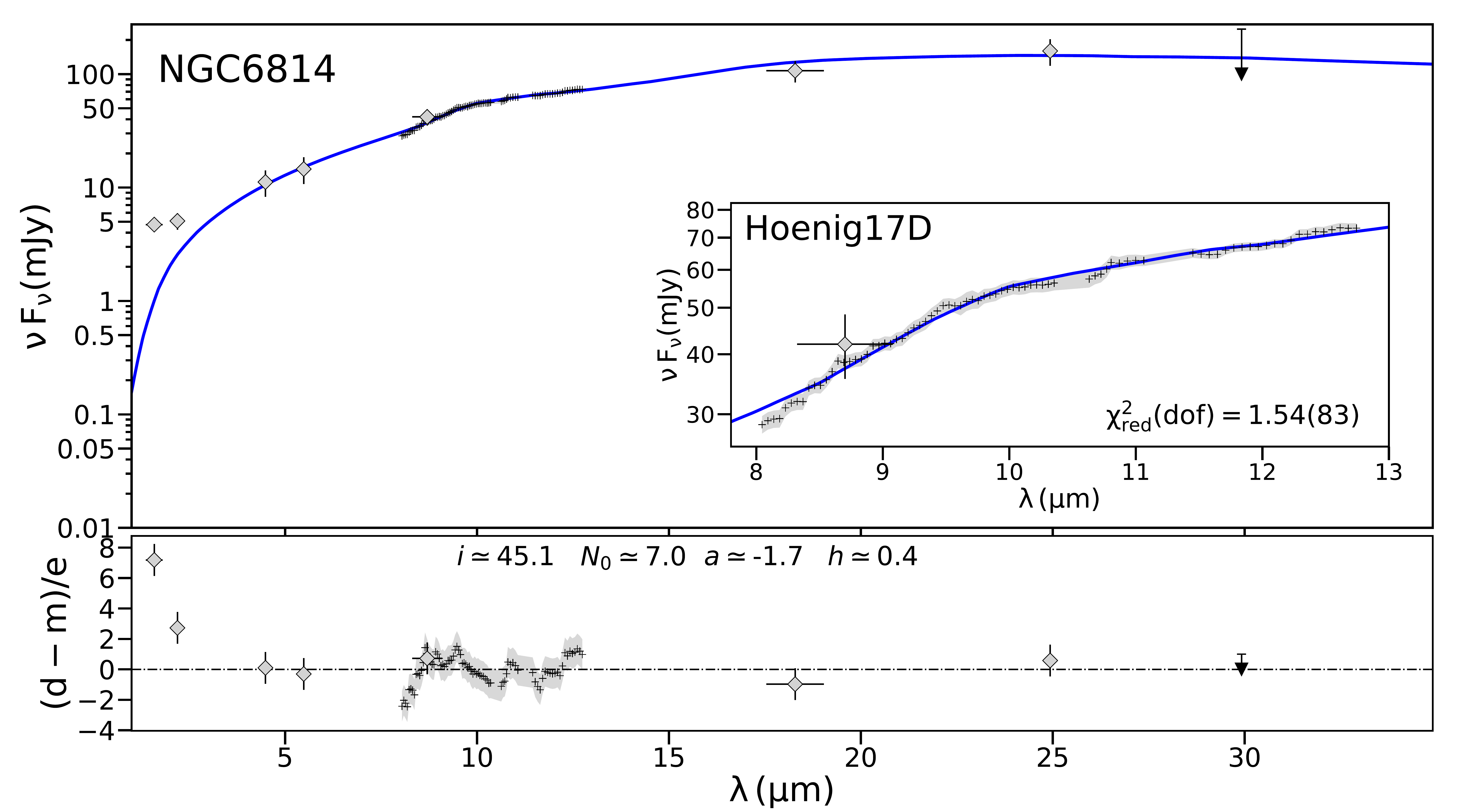}
    \caption{Same as Fig. \ref{fig:ESO005-G004} but for NGC6814.}
    \label{fig:NGC6814}
\end{figure*}

\begin{figure*}
    \centering
    \includegraphics[width=0.75\columnwidth]{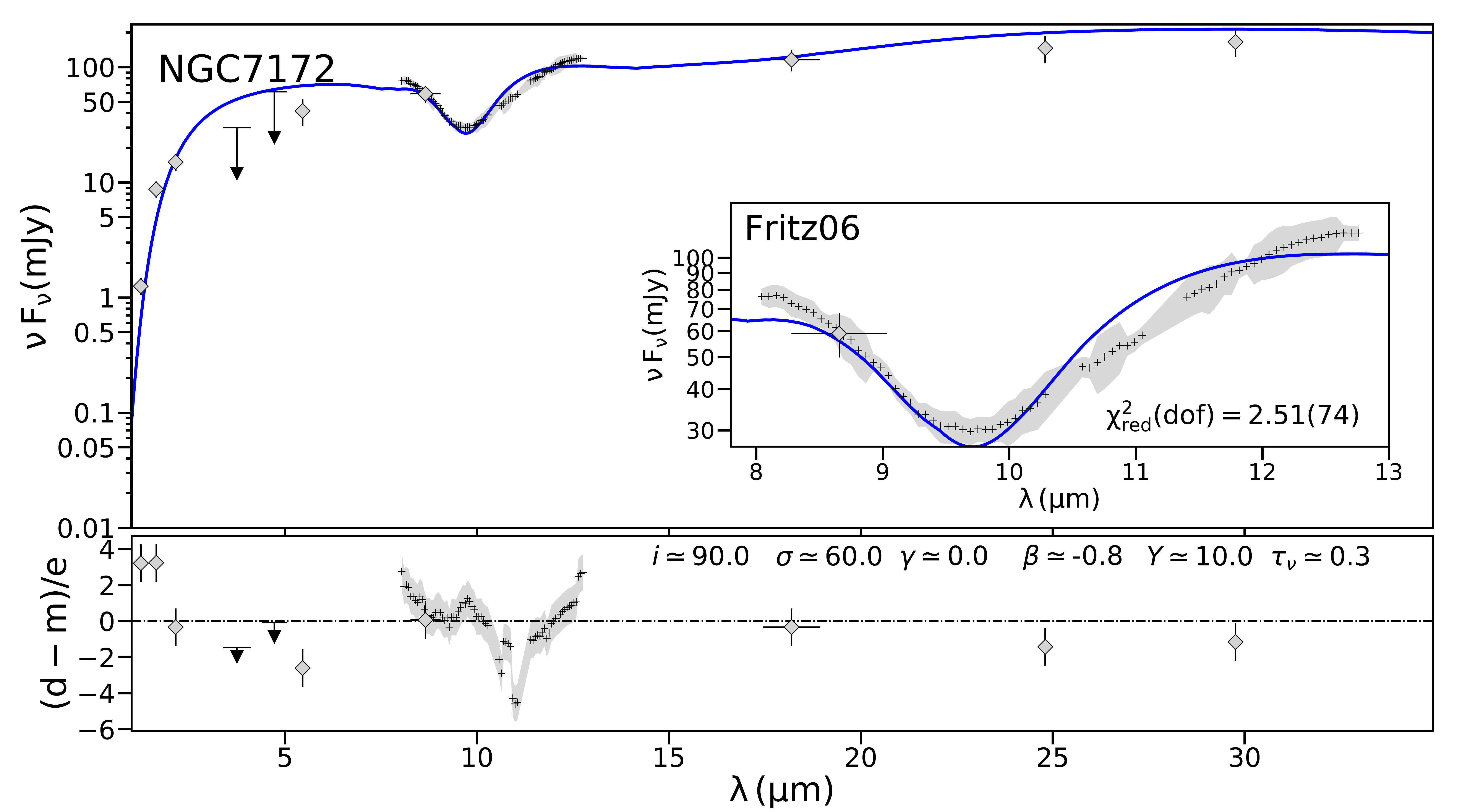}
    \includegraphics[width=0.75\columnwidth]{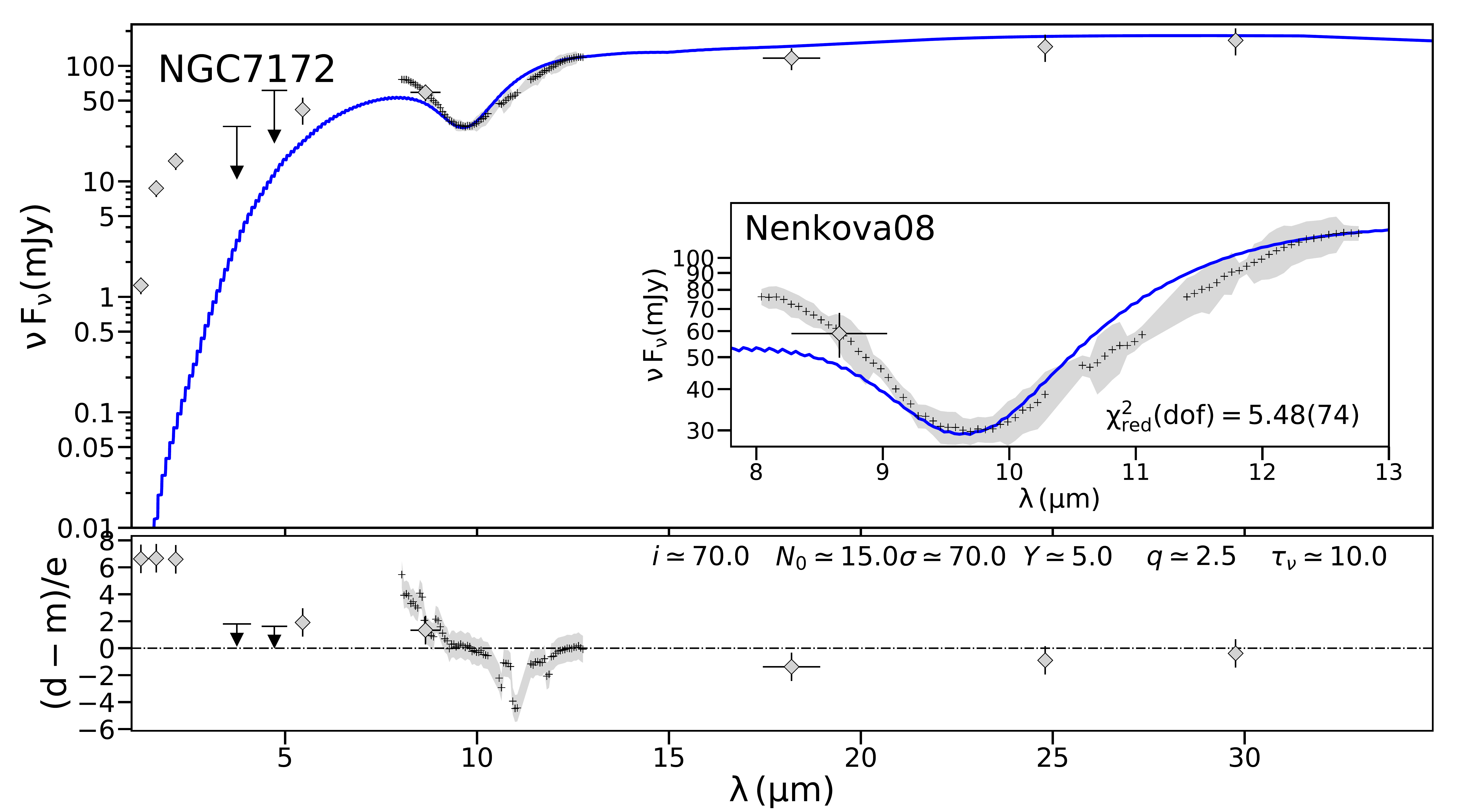}
    \includegraphics[width=0.75\columnwidth]{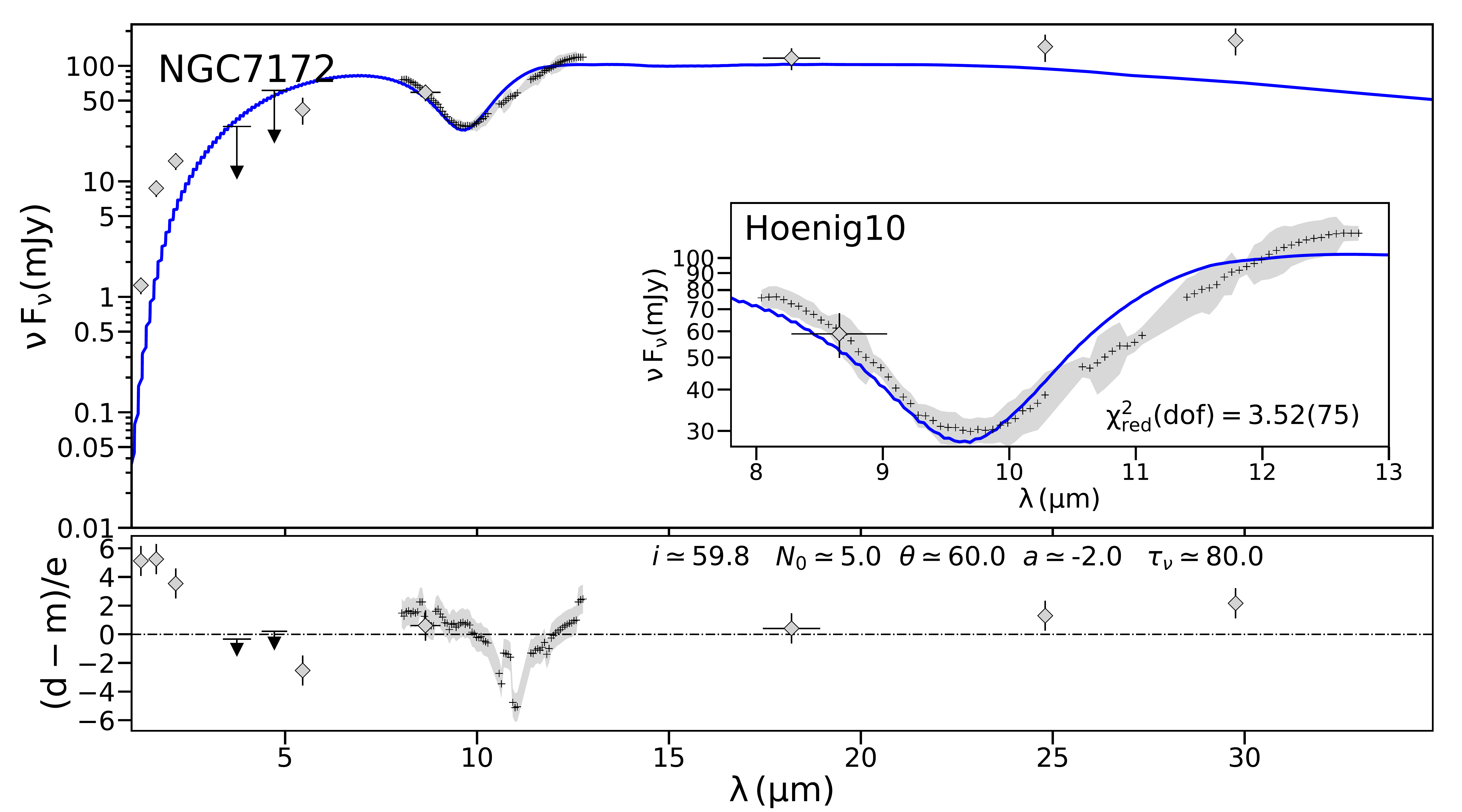}
    \includegraphics[width=0.75\columnwidth]{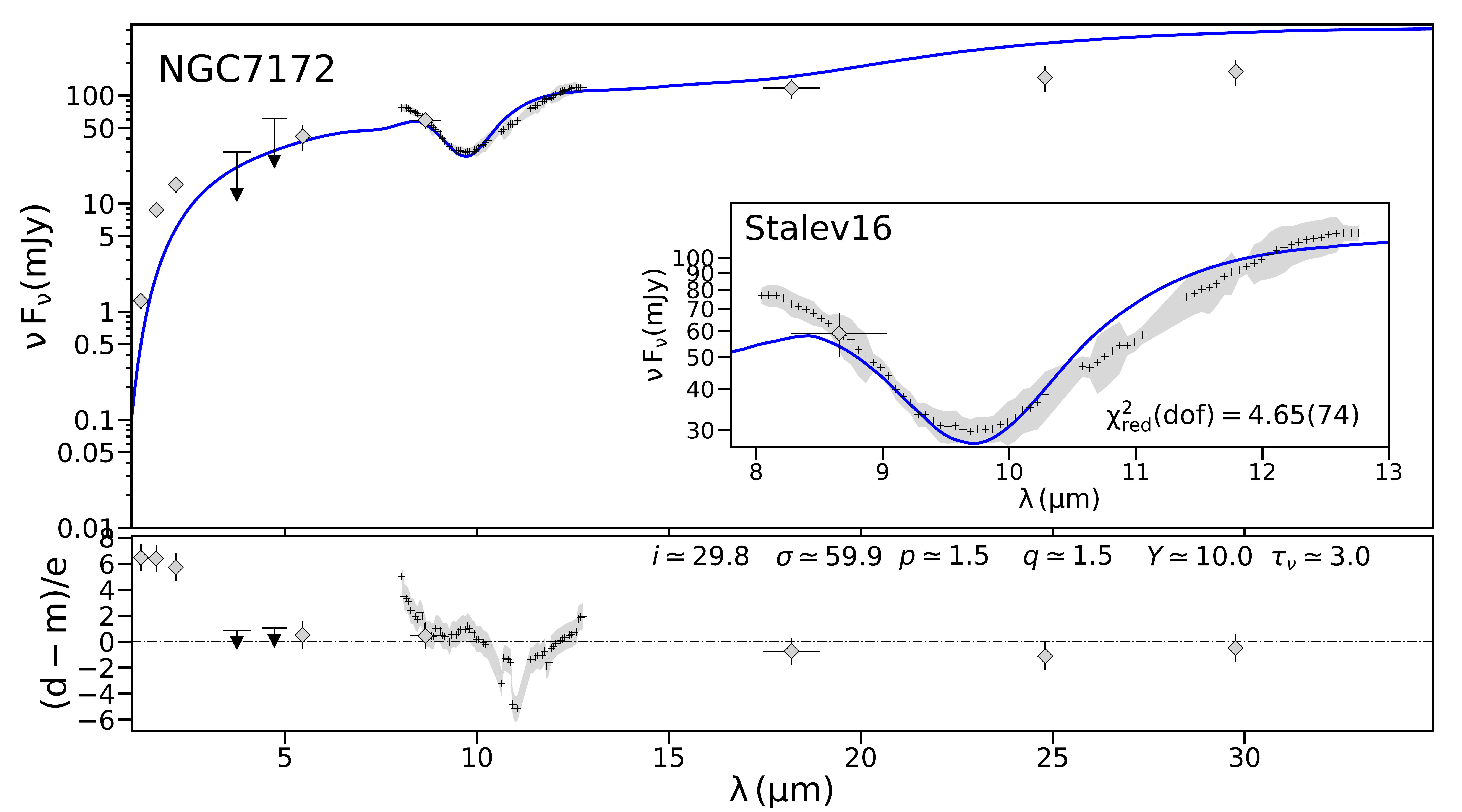}
    \includegraphics[width=0.75\columnwidth]{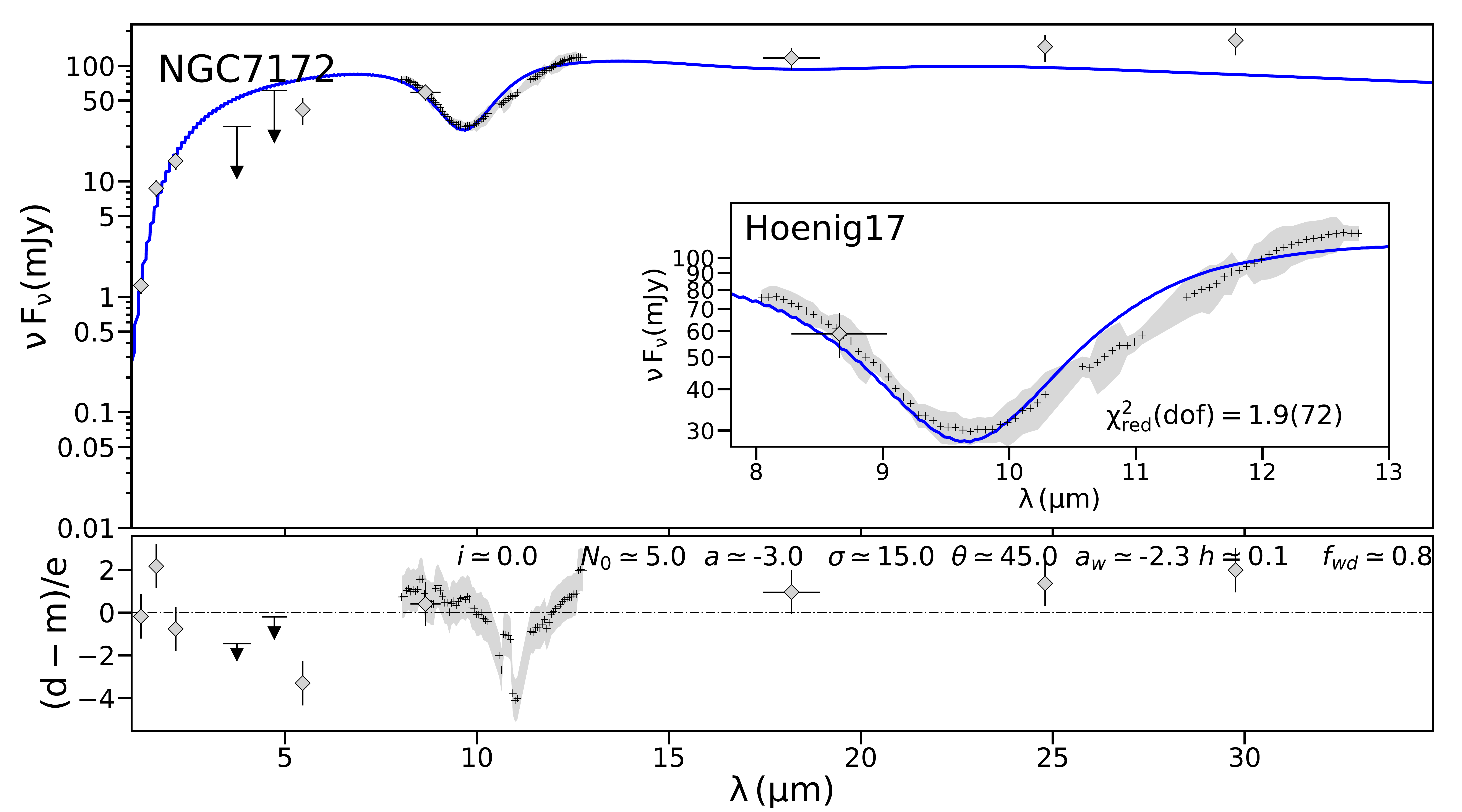}
    \includegraphics[width=0.75\columnwidth]{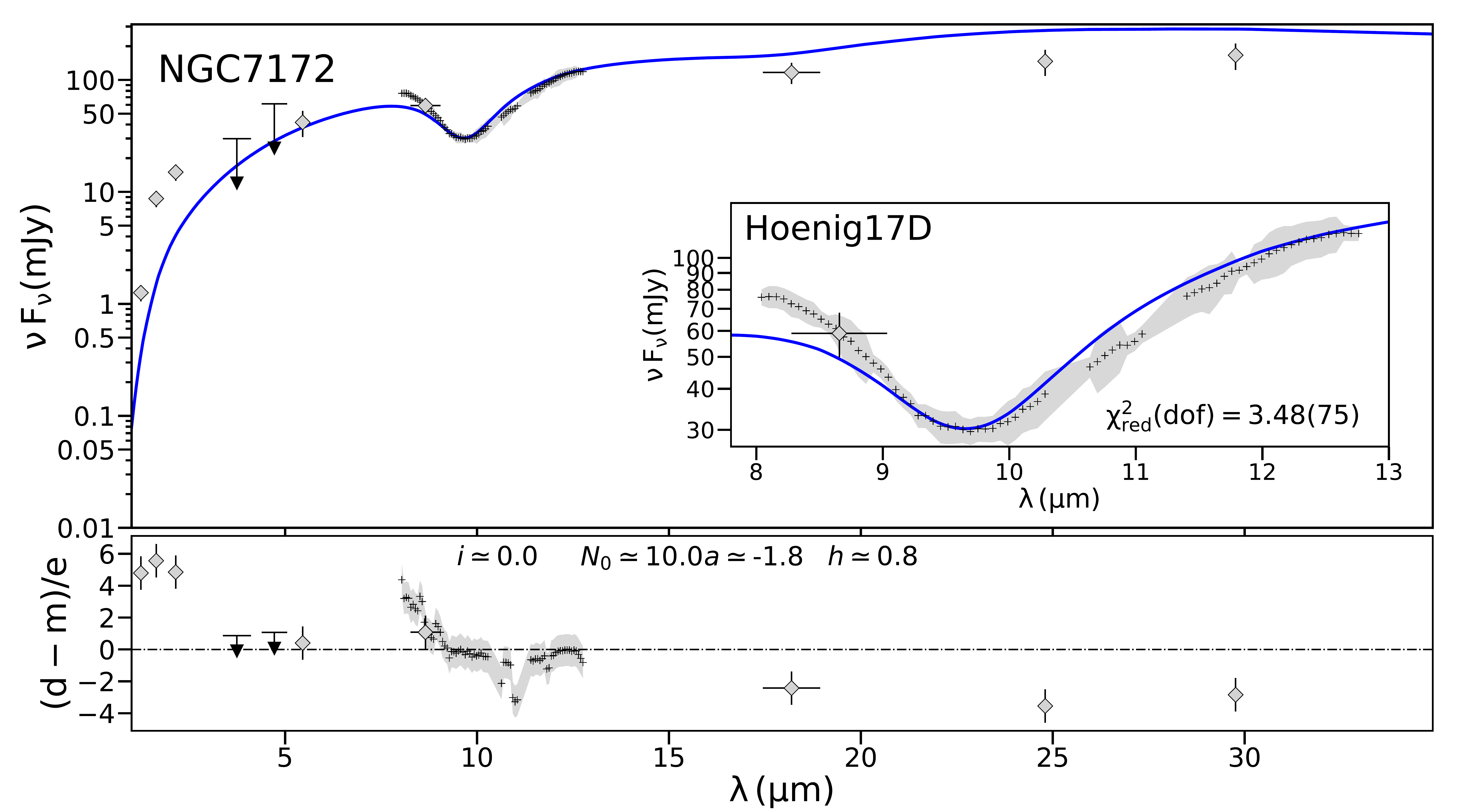}
    \caption{Same as Fig. \ref{fig:ESO005-G004} but for NGC7172.}
    \label{fig:NGC7172}
\end{figure*}

\begin{figure*}
    \centering
    \includegraphics[width=0.75\columnwidth]{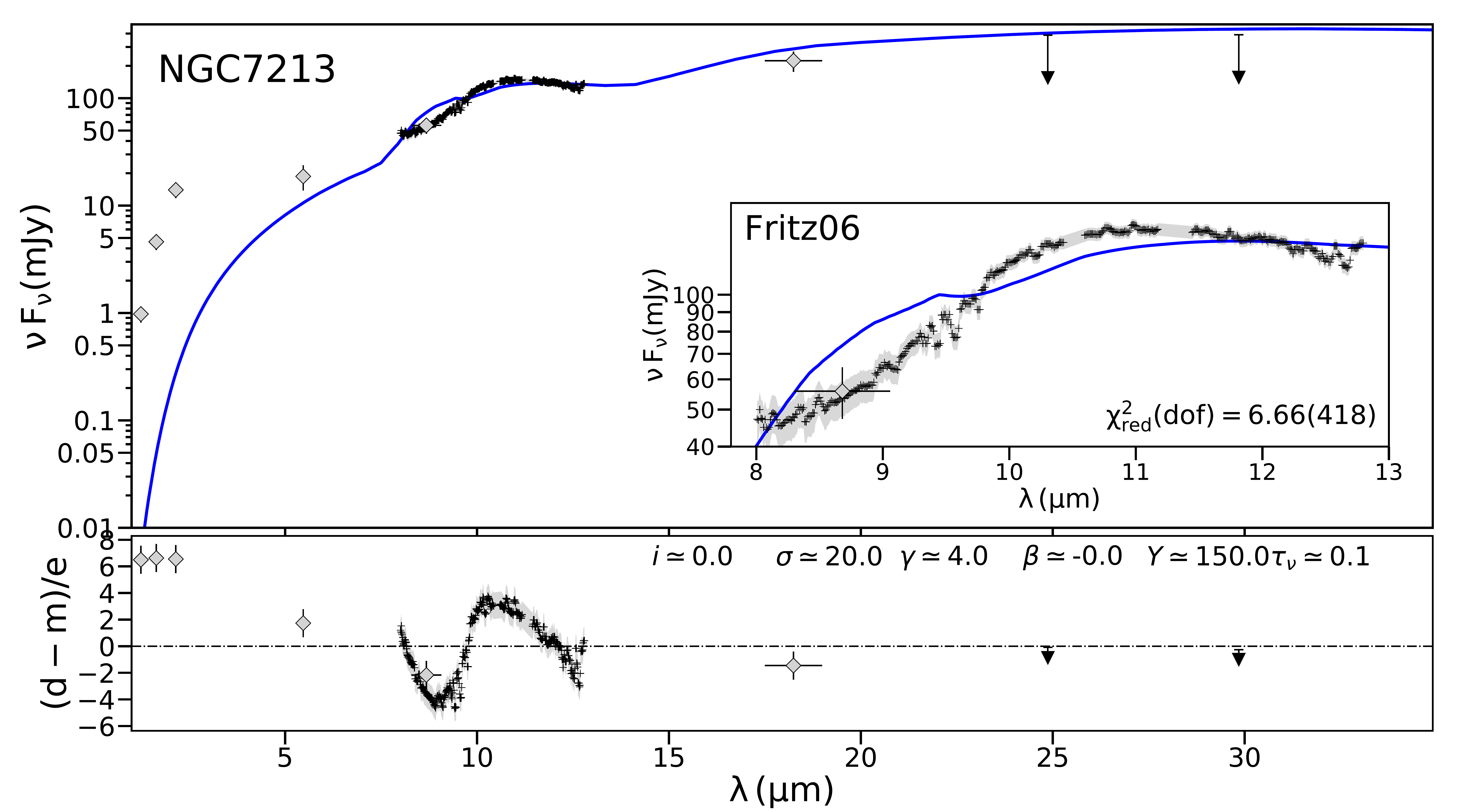}
    \includegraphics[width=0.75\columnwidth]{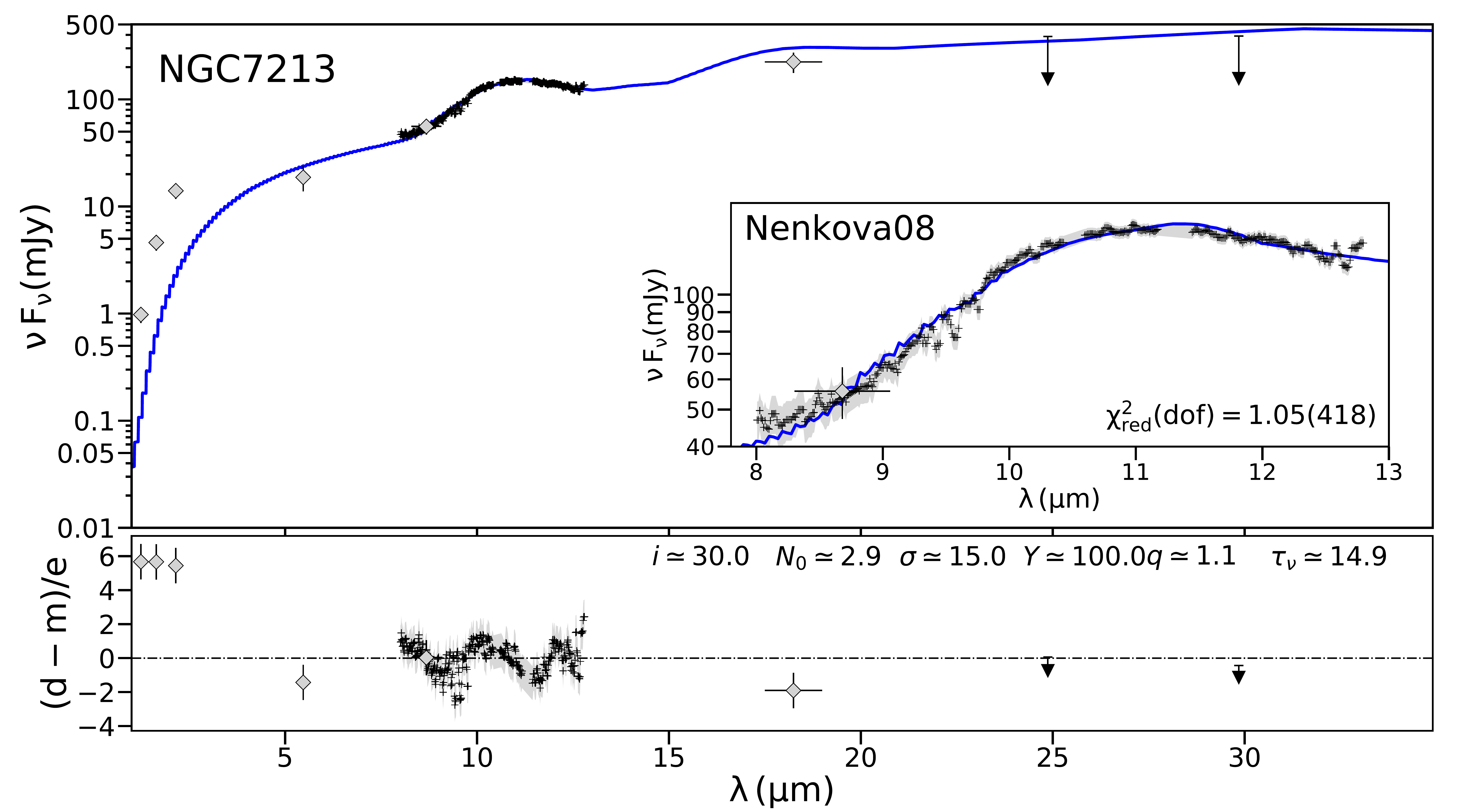}
    \includegraphics[width=0.75\columnwidth]{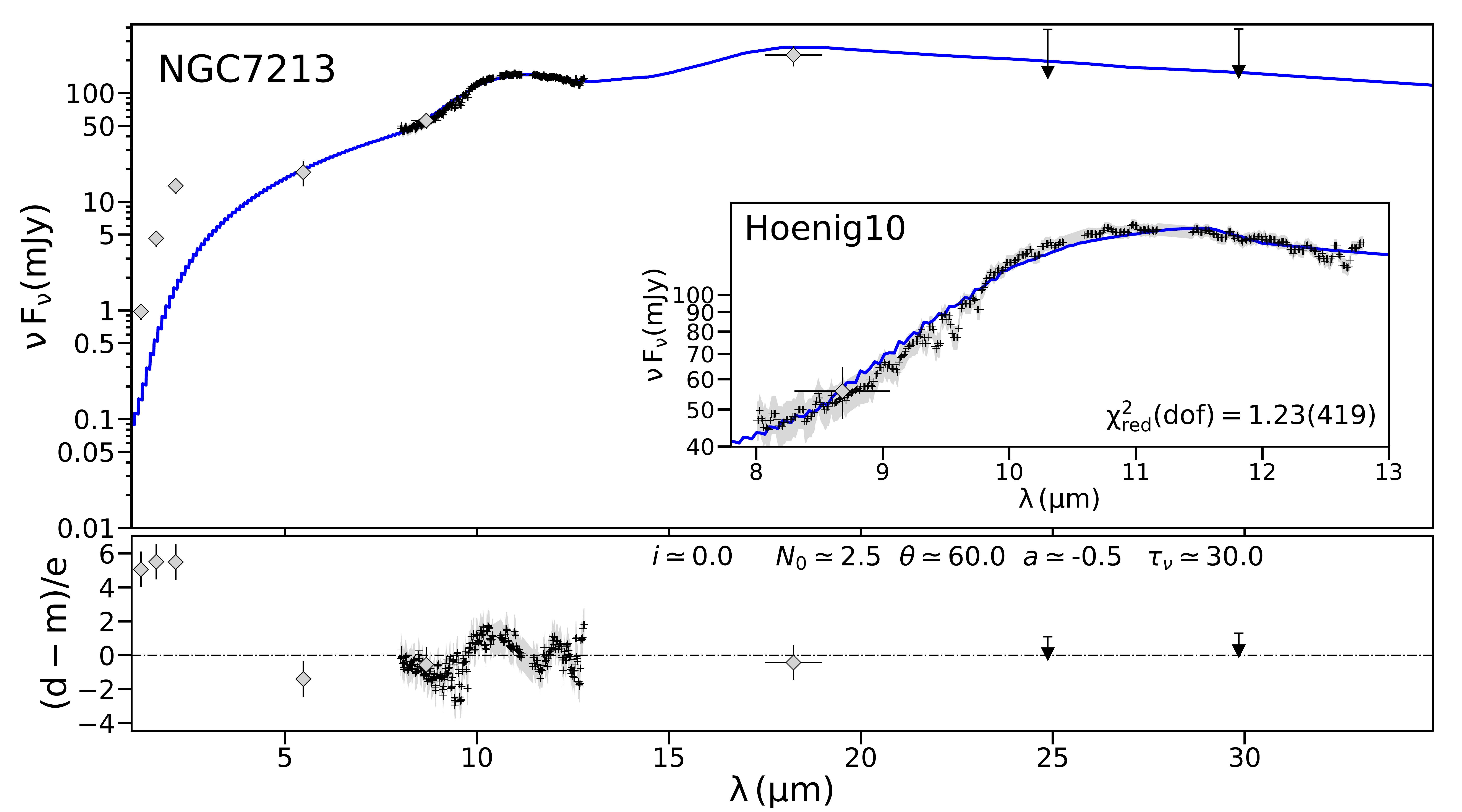}
    \includegraphics[width=0.75\columnwidth]{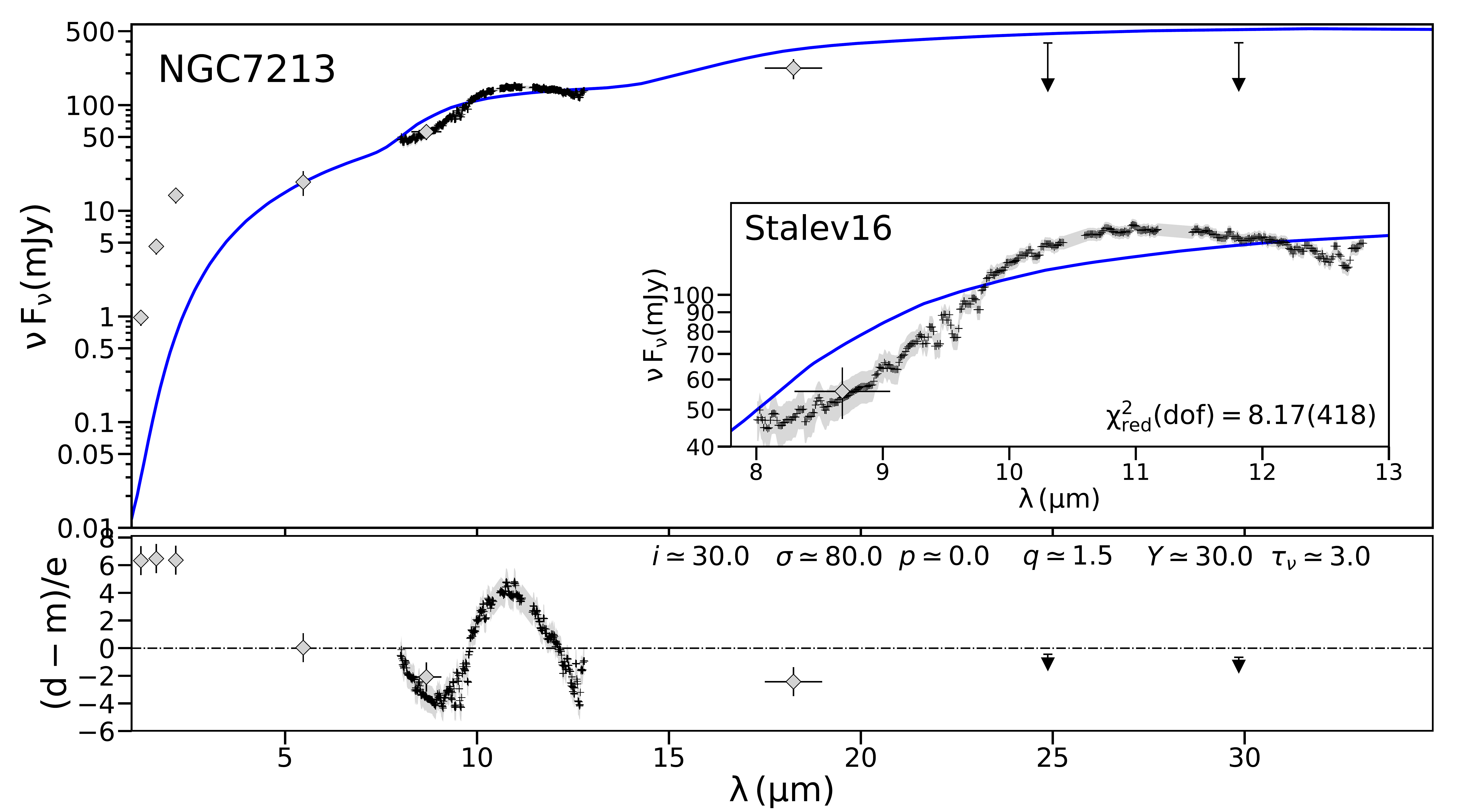}
    \includegraphics[width=0.75\columnwidth]{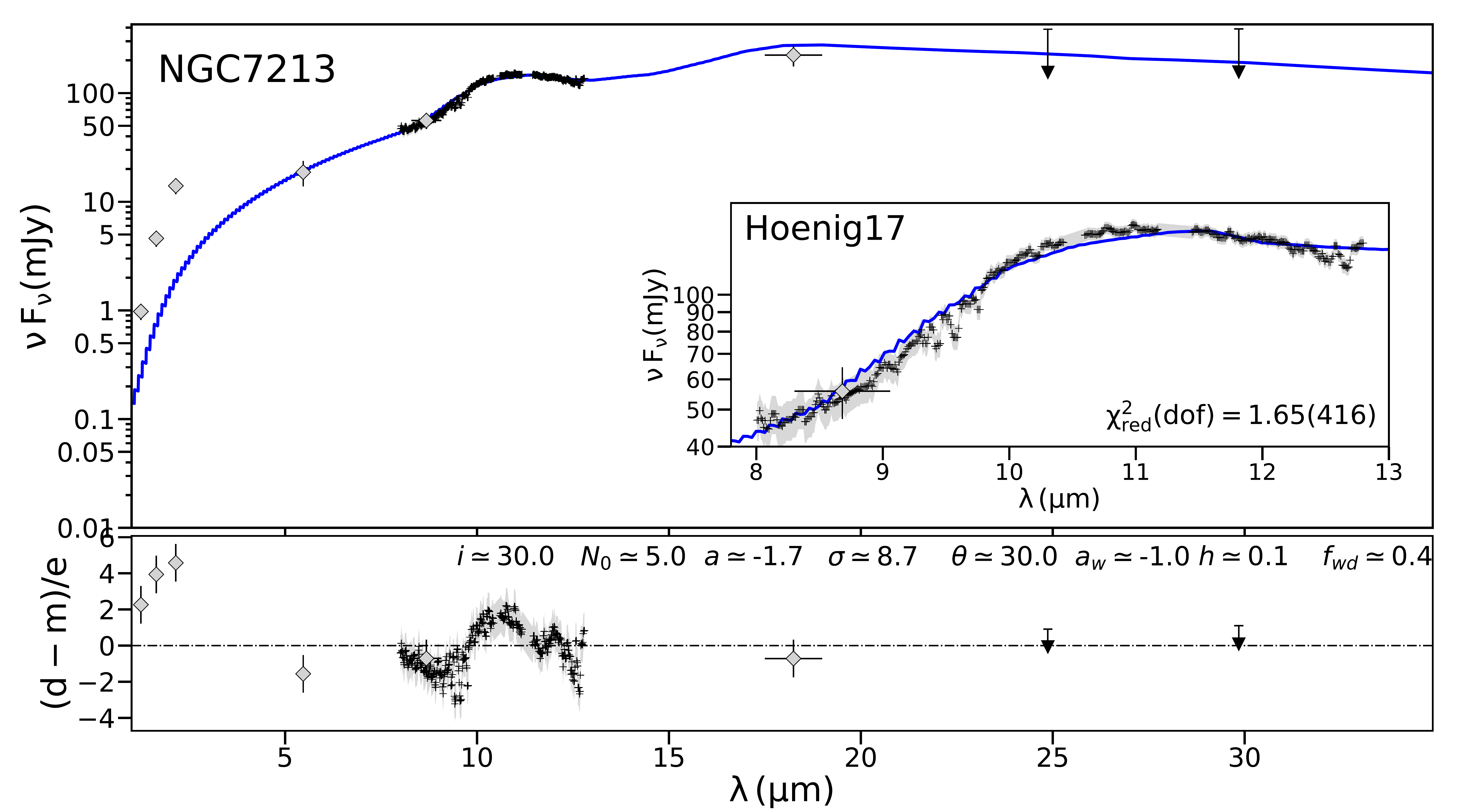}
    \includegraphics[width=0.75\columnwidth]{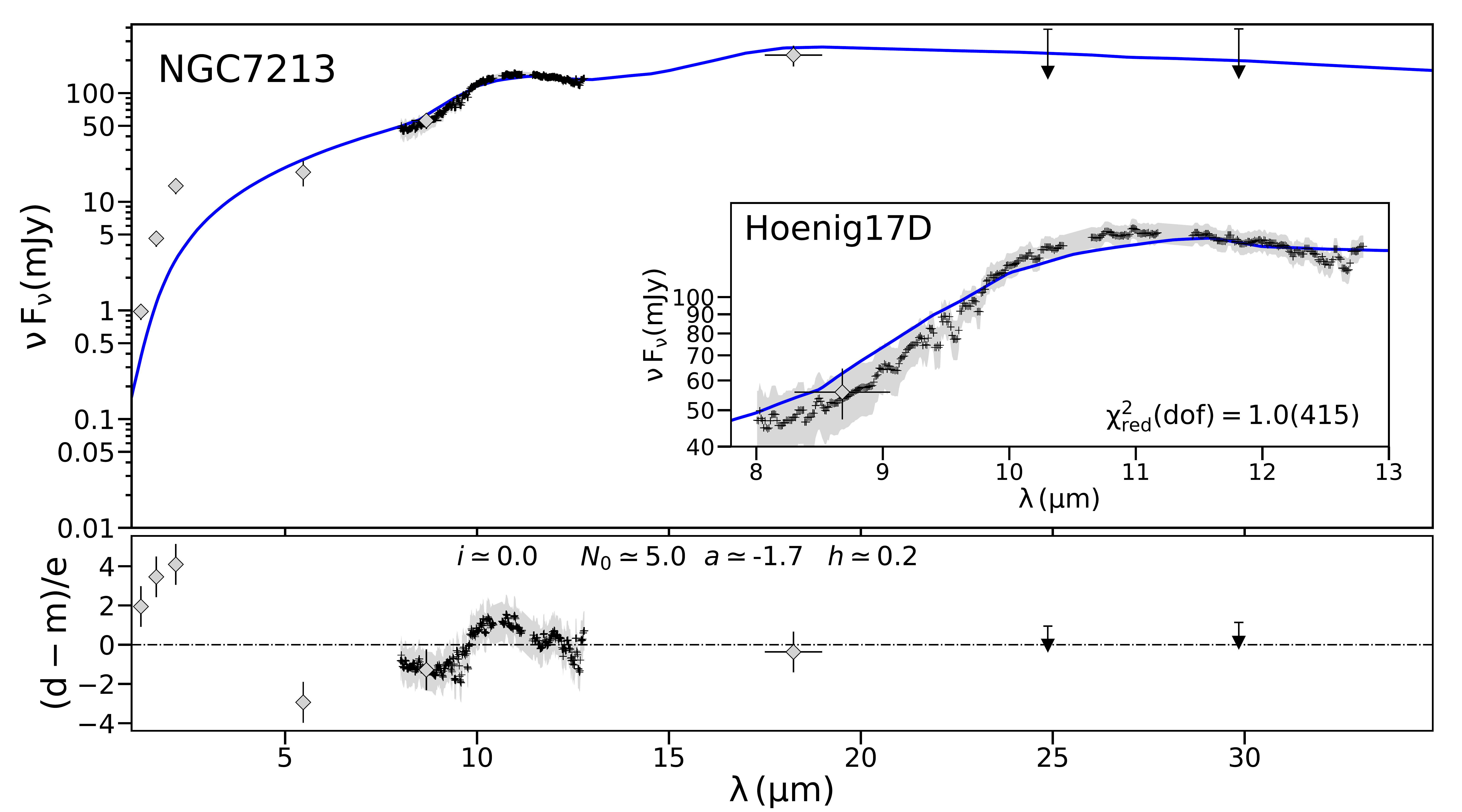}
    \caption{Same as Fig. \ref{fig:ESO005-G004} but for NGC7213.}
    \label{fig:NGC7213}
\end{figure*}

\begin{figure*}
    \centering
    \includegraphics[width=0.75\columnwidth]{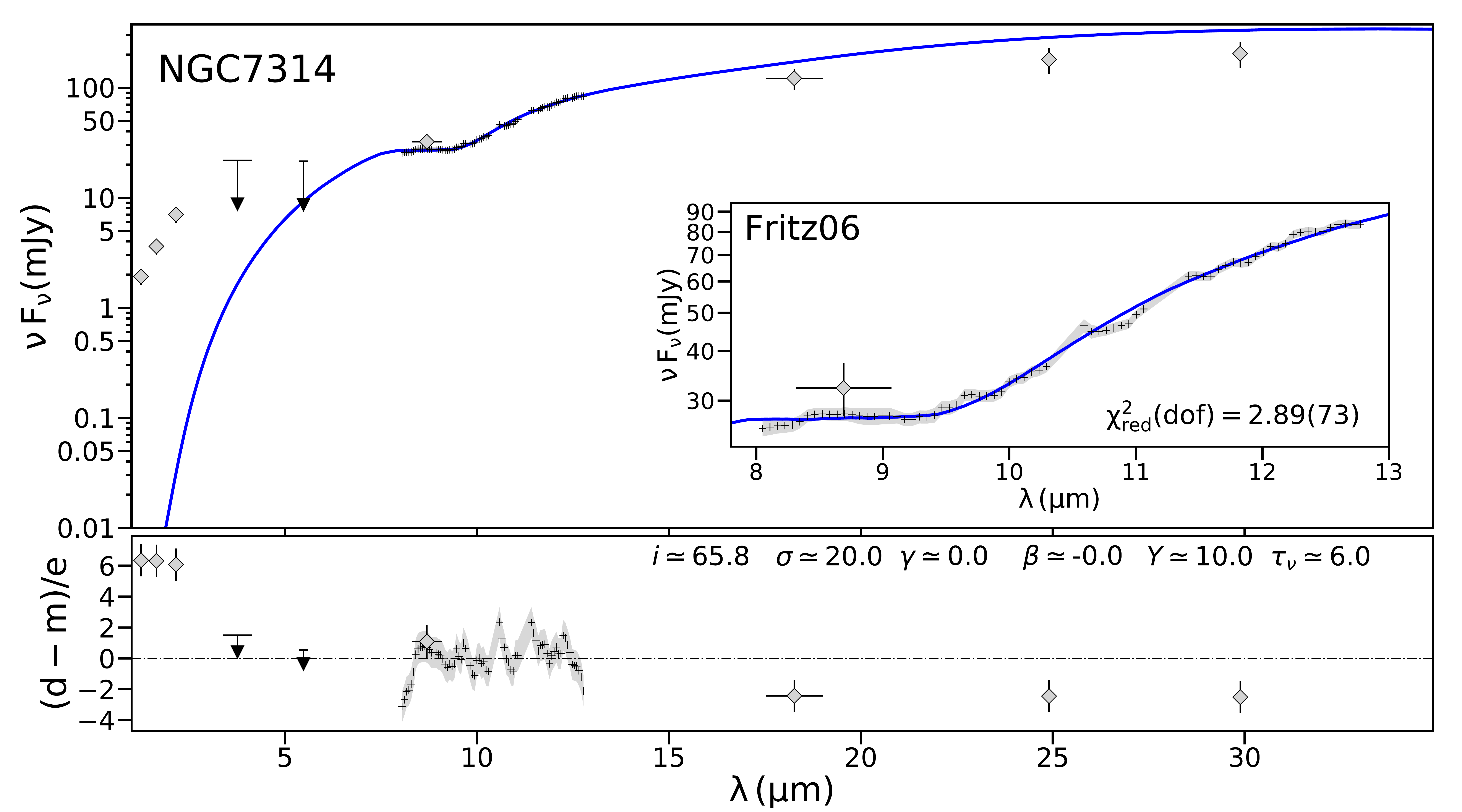}
    \includegraphics[width=0.75\columnwidth]{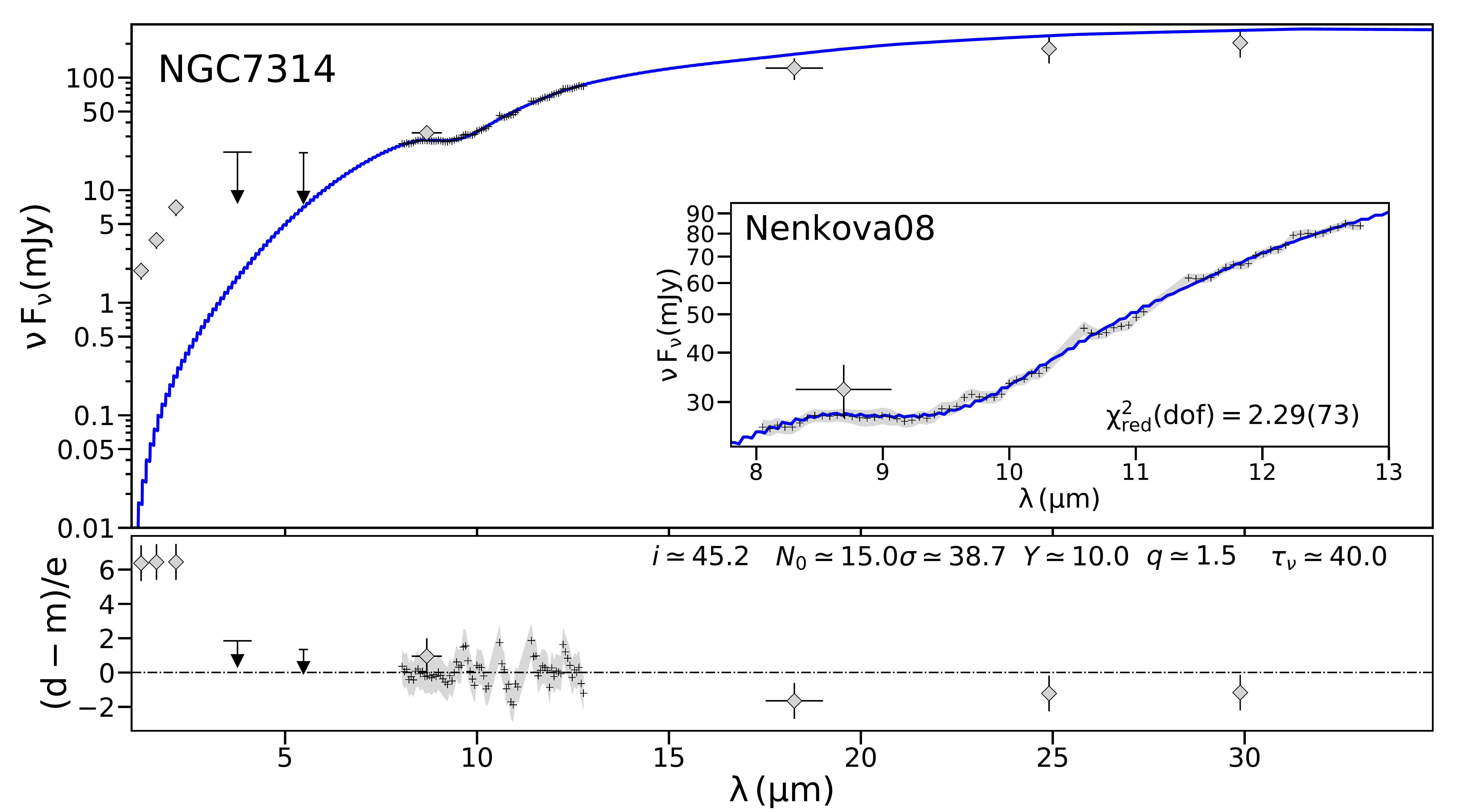}
    \includegraphics[width=0.75\columnwidth]{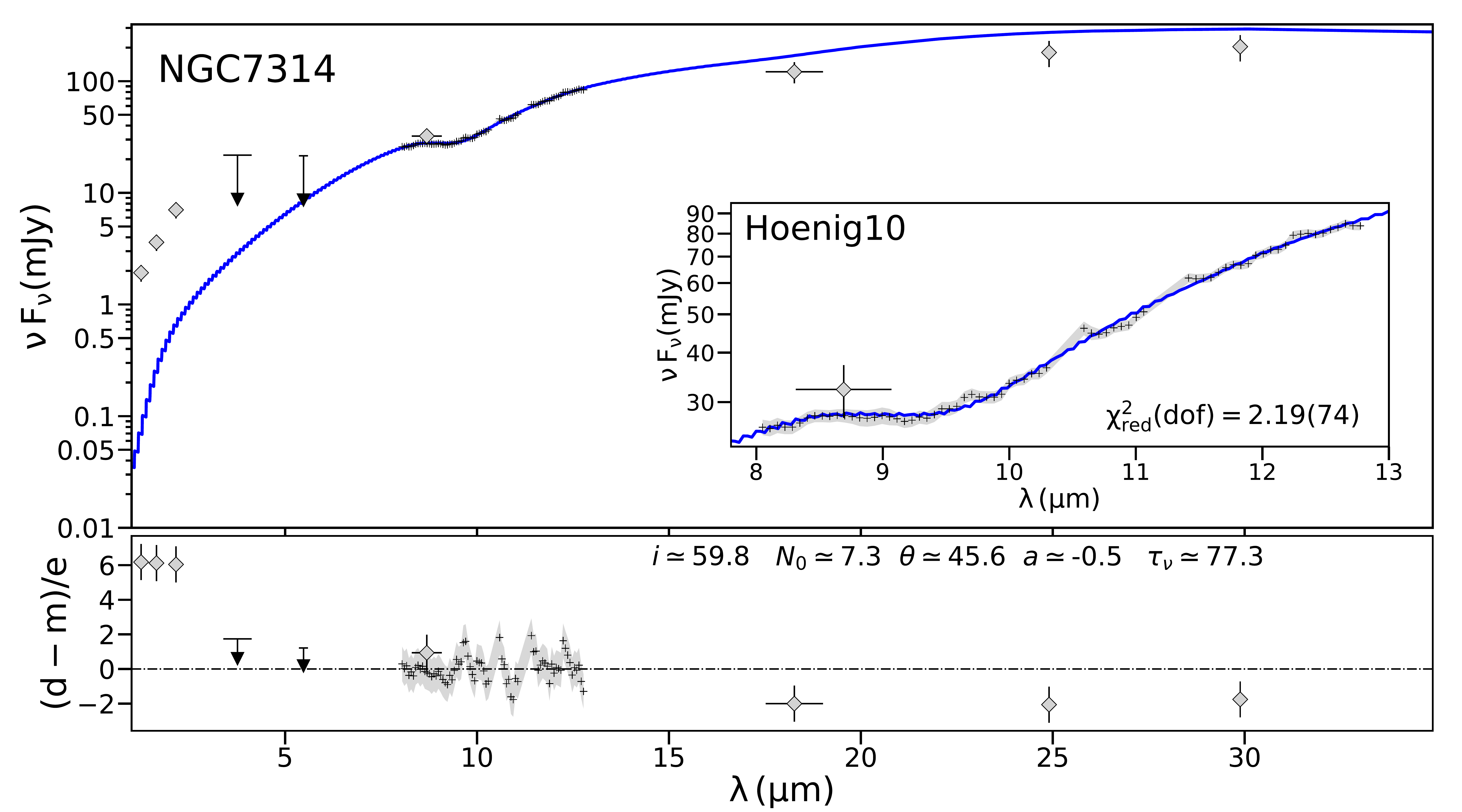}
    \includegraphics[width=0.75\columnwidth]{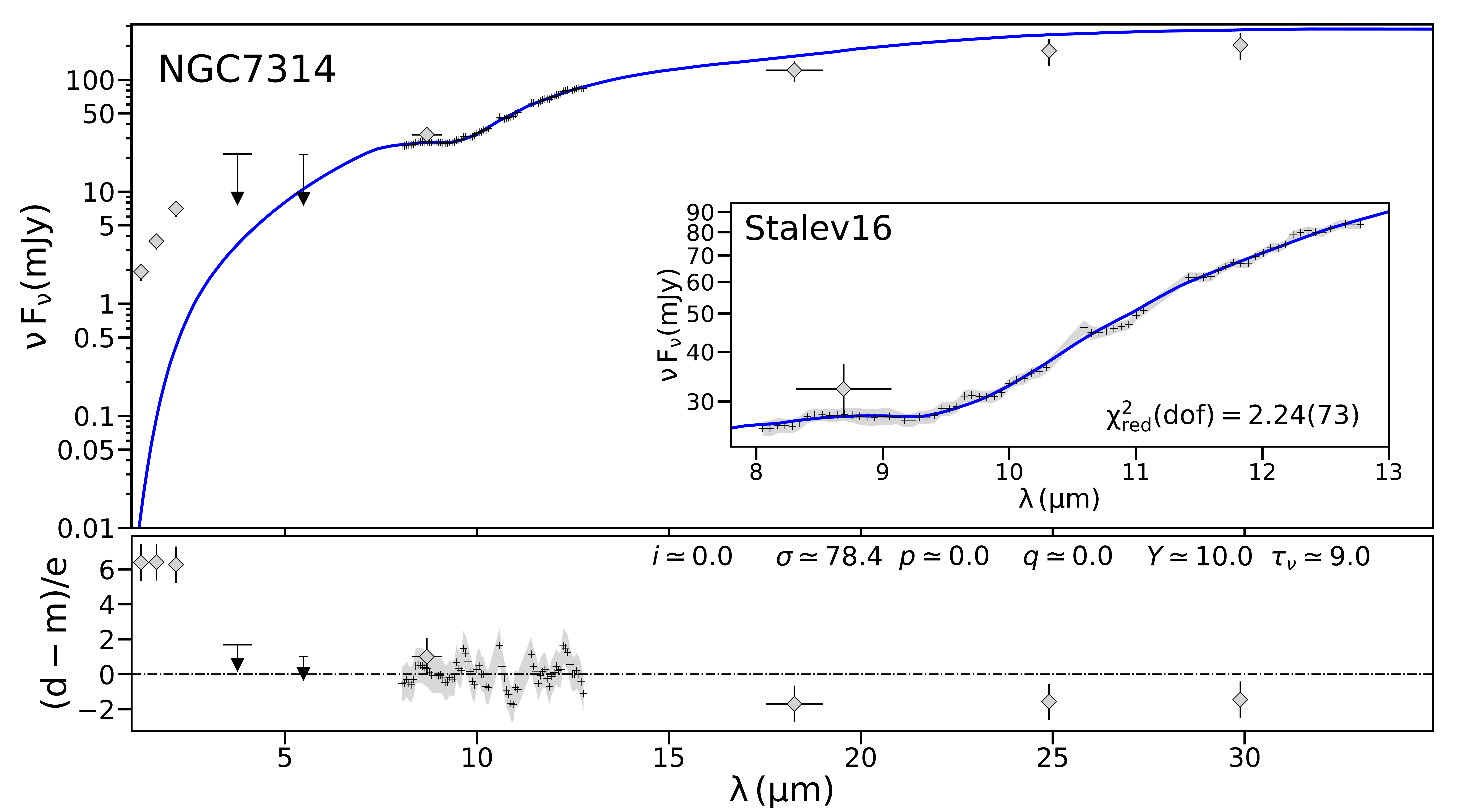}
    \includegraphics[width=0.75\columnwidth]{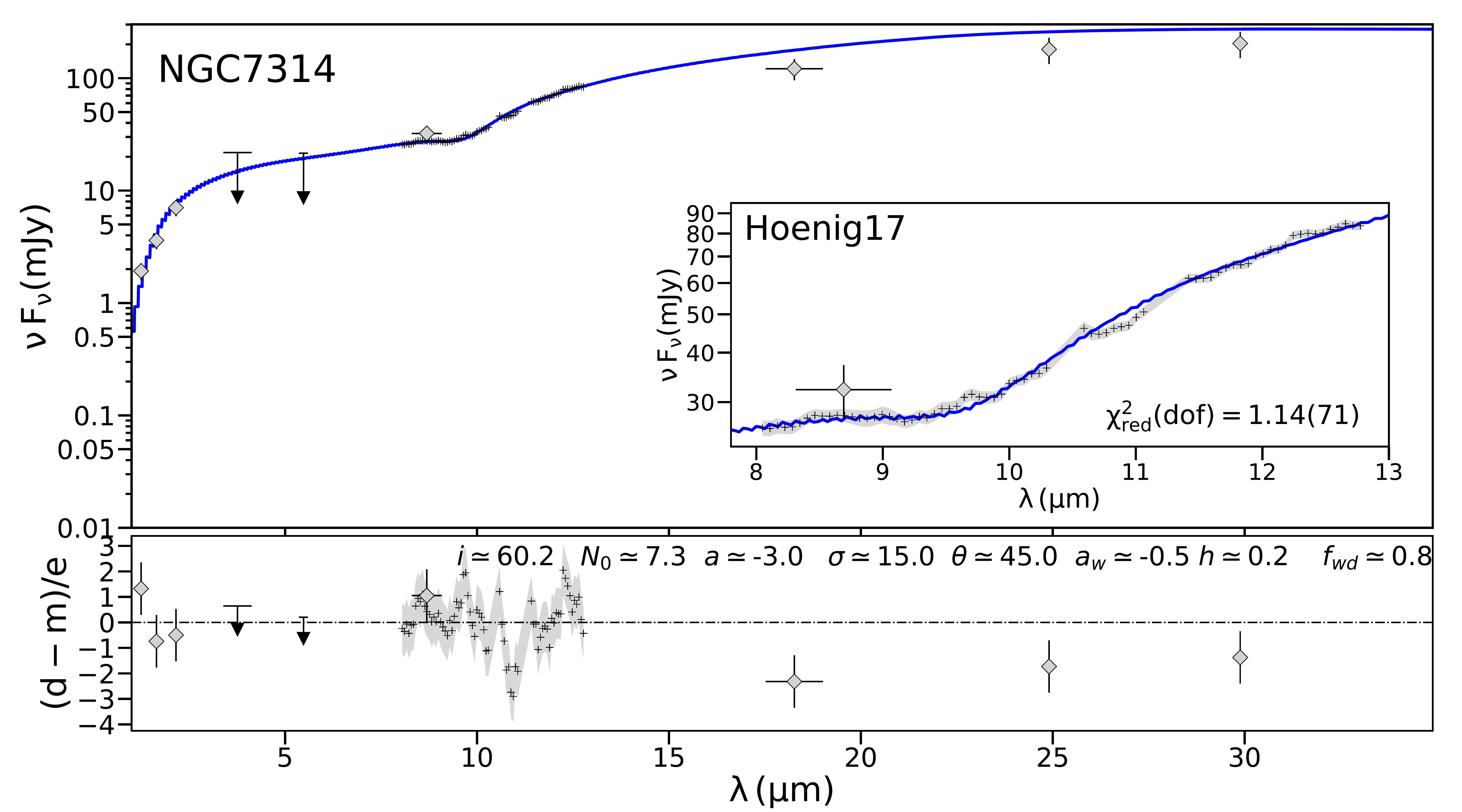}
    \includegraphics[width=0.75\columnwidth]{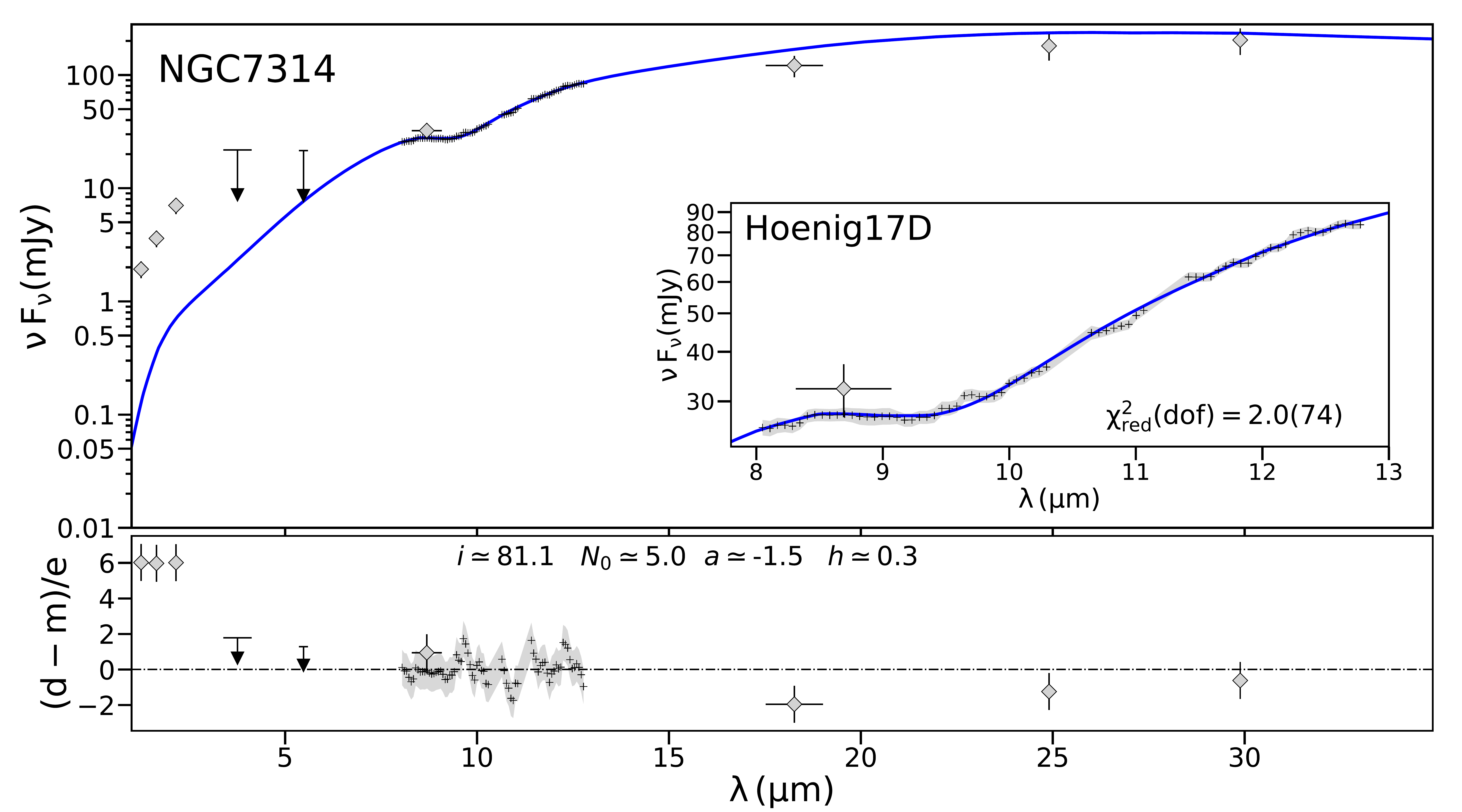}
    \caption{Same as Fig. \ref{fig:ESO005-G004} but for NGC7314.}
    \label{fig:NGC7314}
\end{figure*}

\begin{figure*}
    \centering
    \includegraphics[width=0.75\columnwidth]{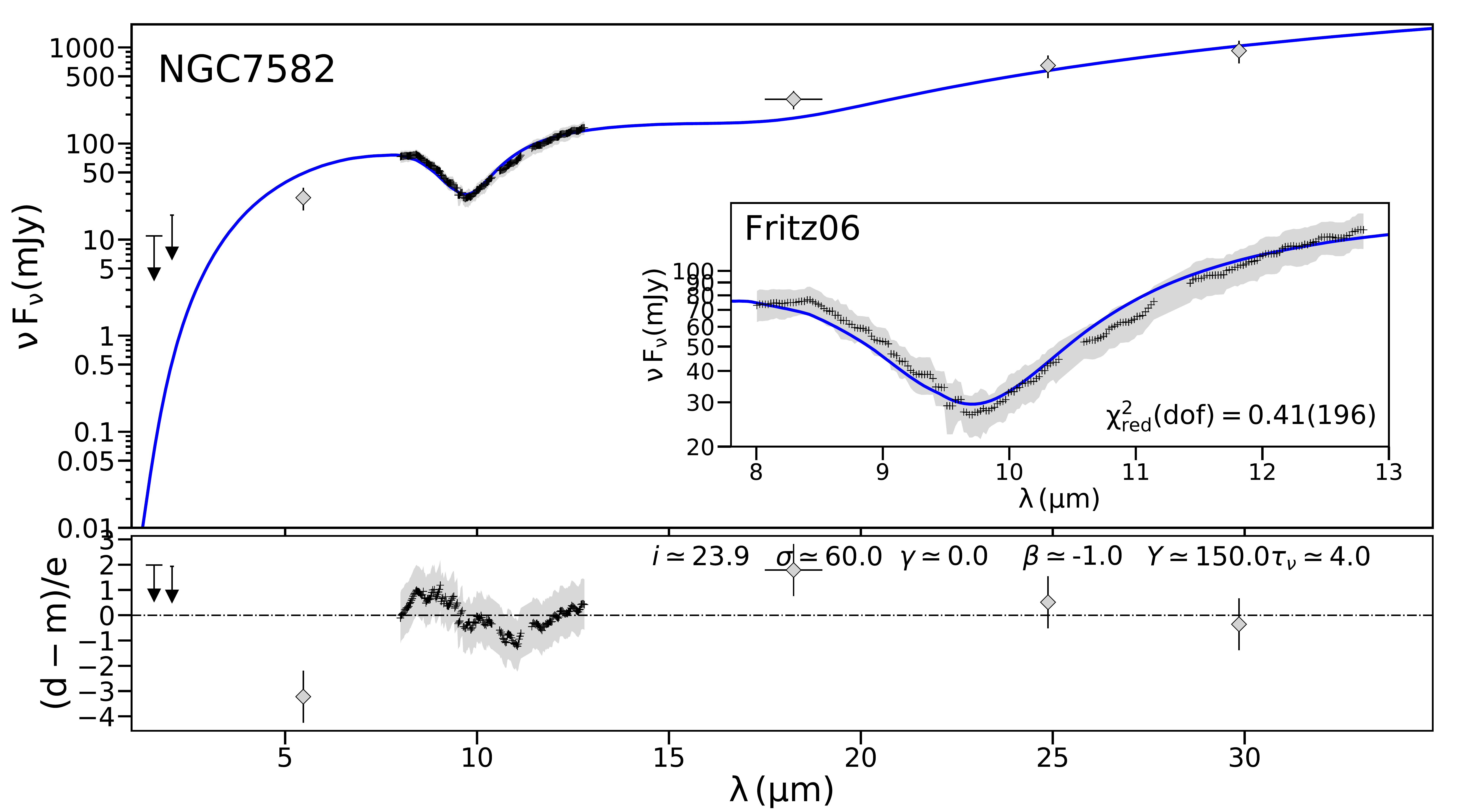}
    \includegraphics[width=0.75\columnwidth]{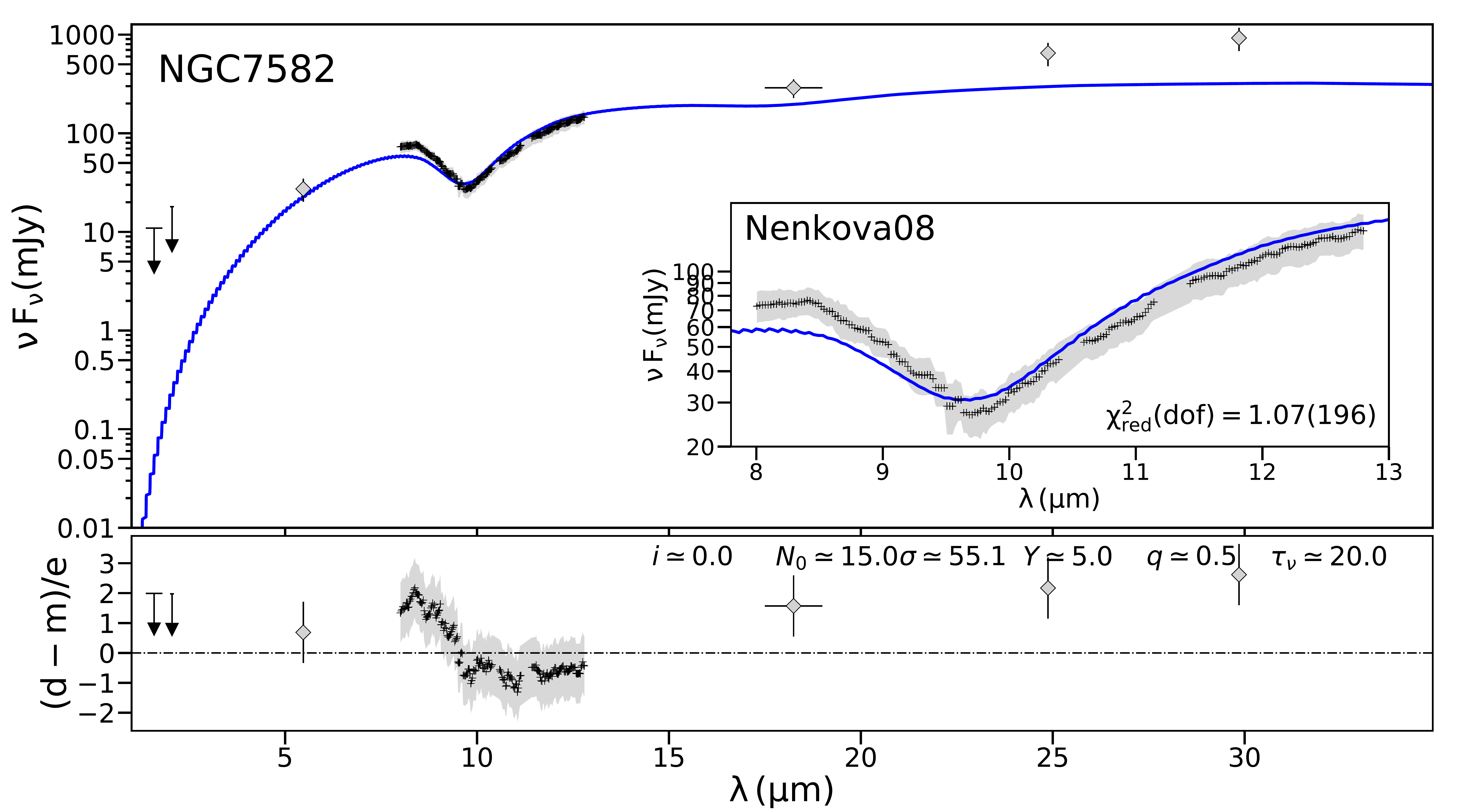}
    \includegraphics[width=0.75\columnwidth]{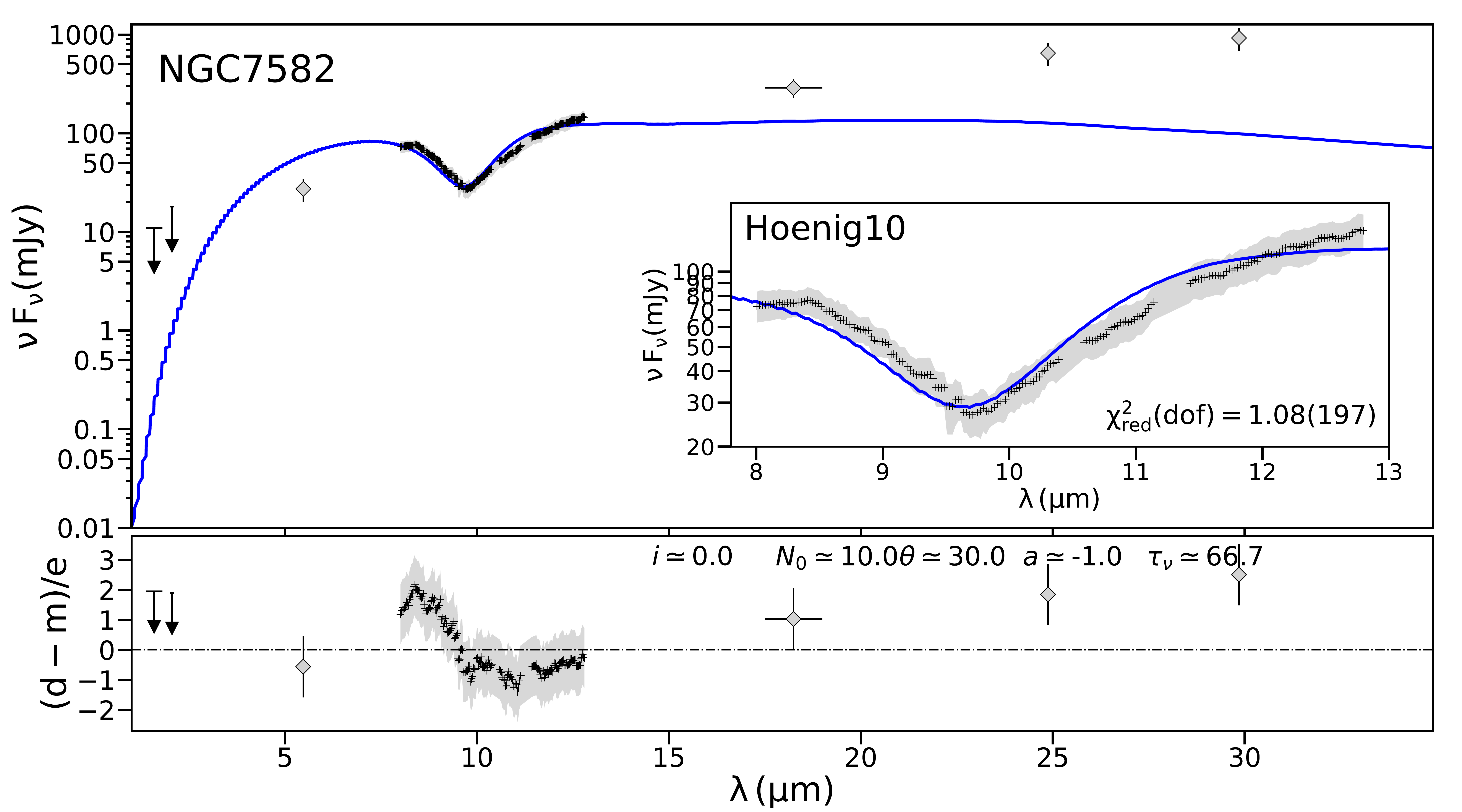}
    \includegraphics[width=0.75\columnwidth]{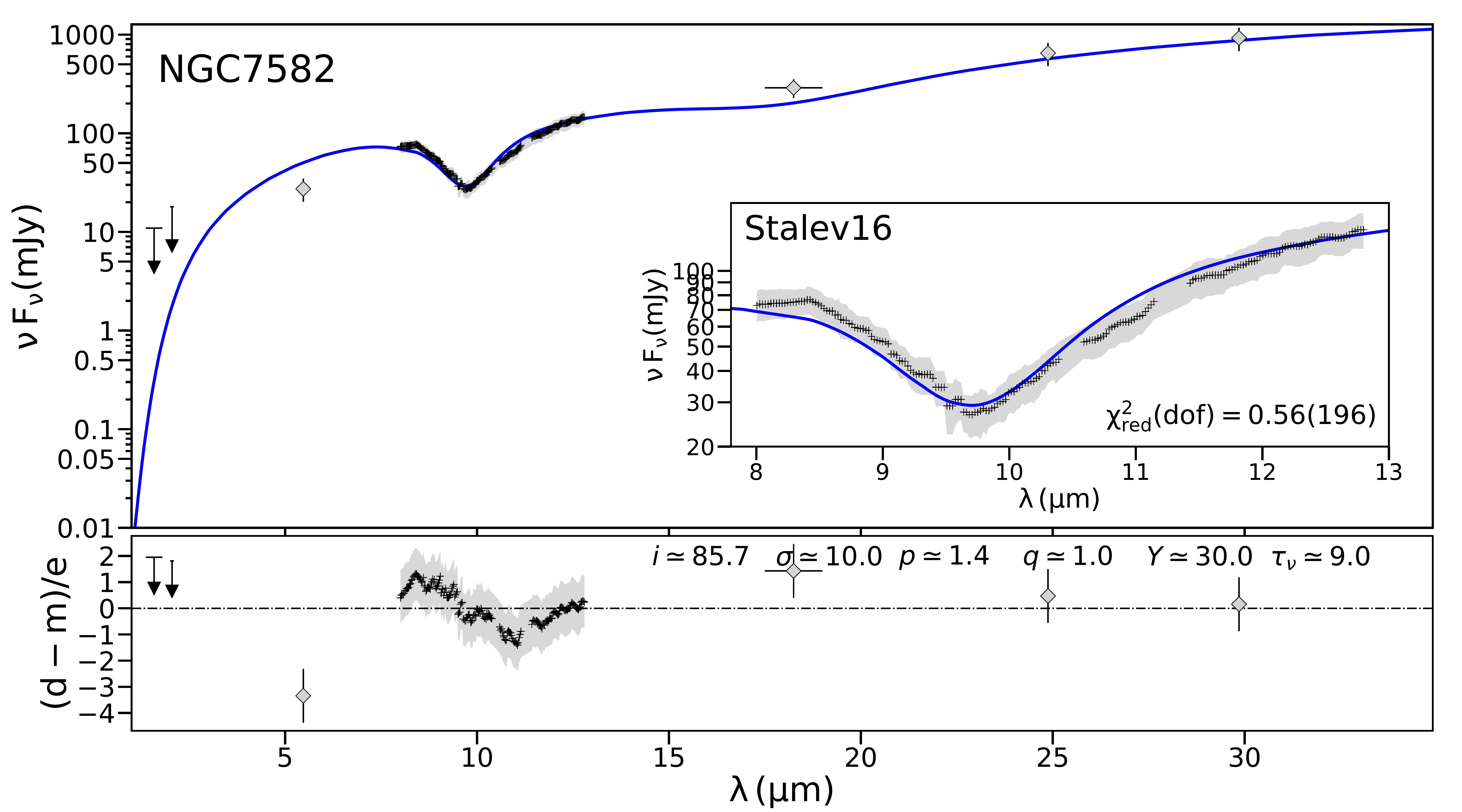}
    \includegraphics[width=0.75\columnwidth]{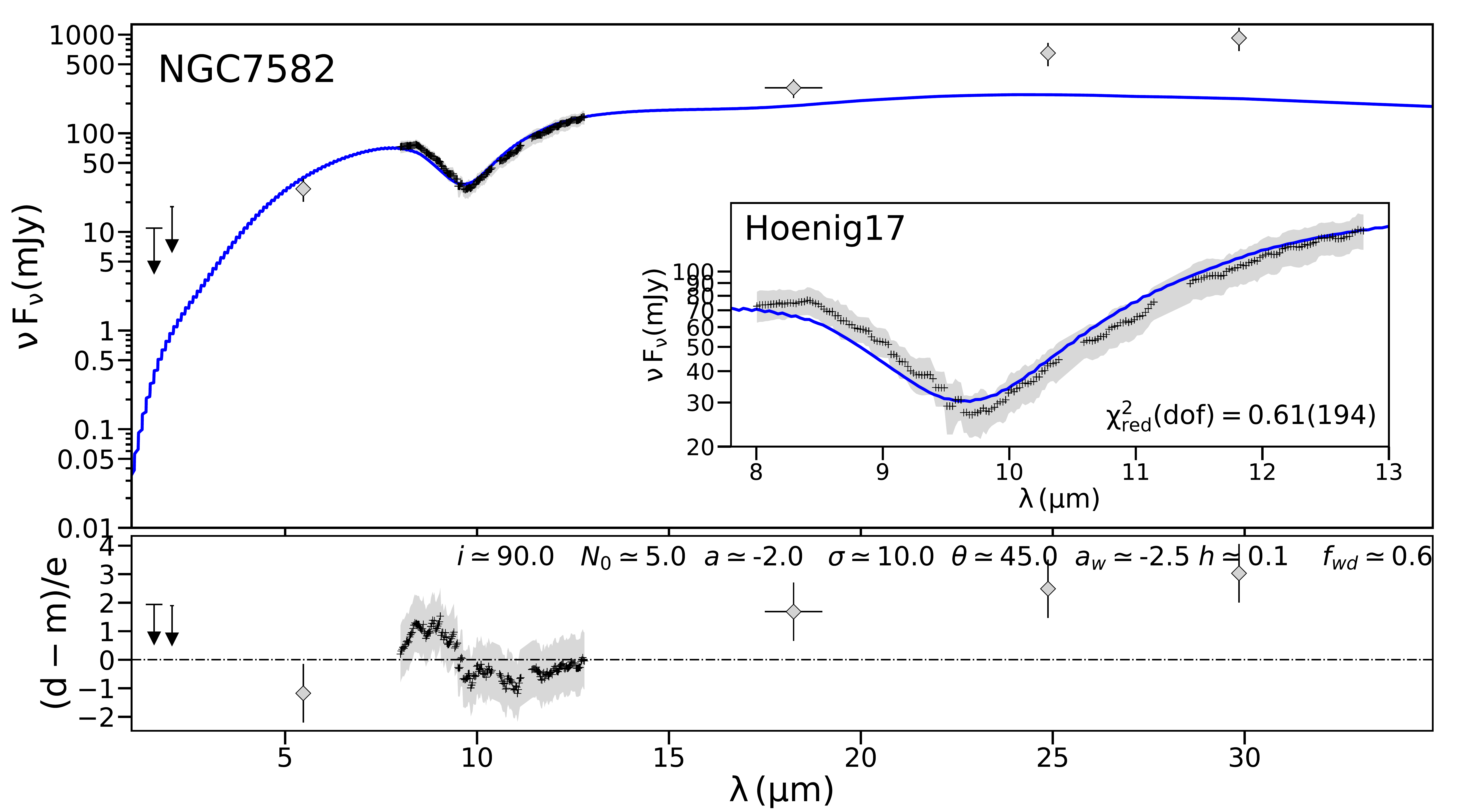}
    \includegraphics[width=0.75\columnwidth]{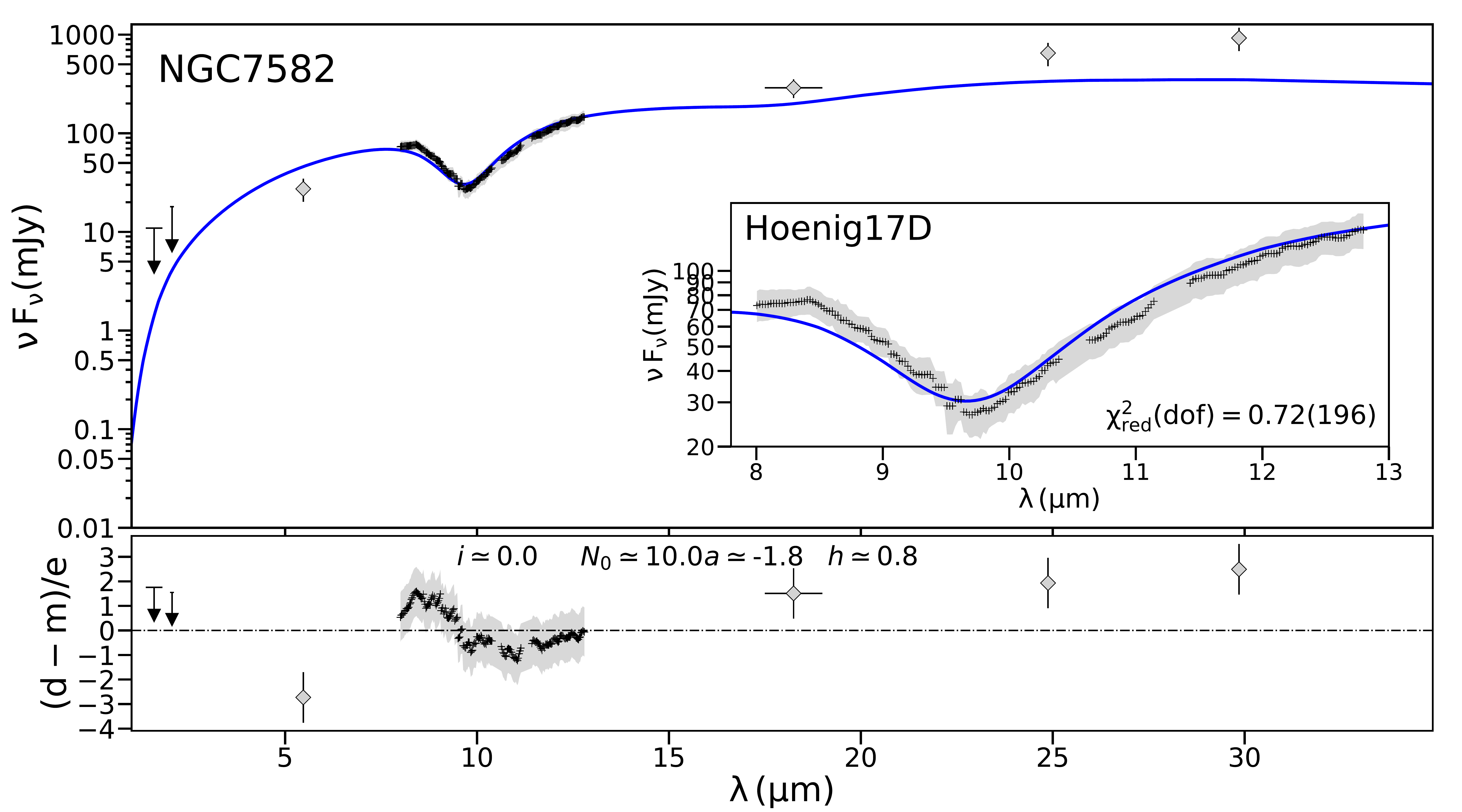}
    \caption{Same as Fig. \ref{fig:ESO005-G004} but for NGC7582.}
    \label{fig:NGC7582}
\end{figure*}

\begin{figure*}
    \centering
    \includegraphics[width=0.75\columnwidth]{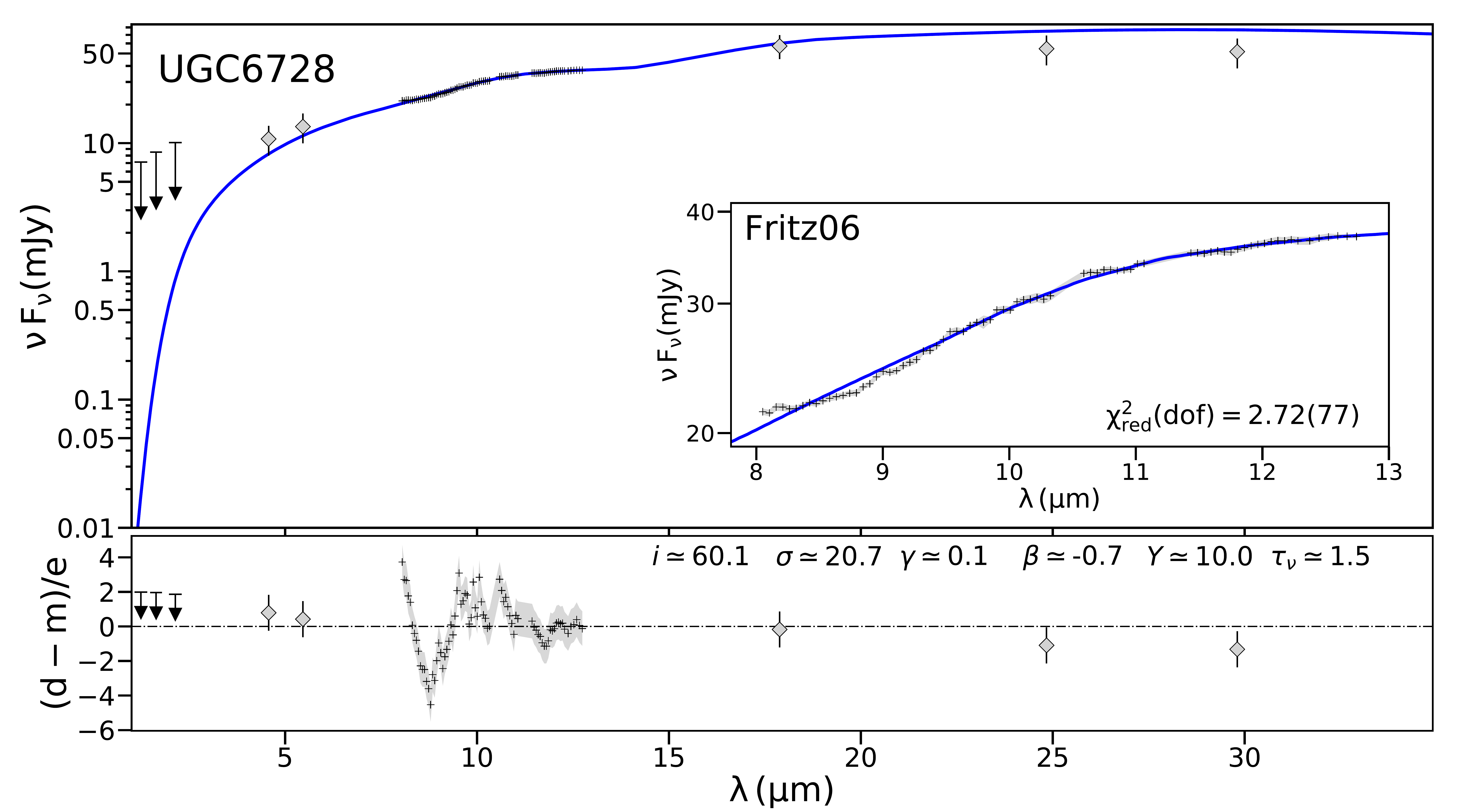}
    \includegraphics[width=0.75\columnwidth]{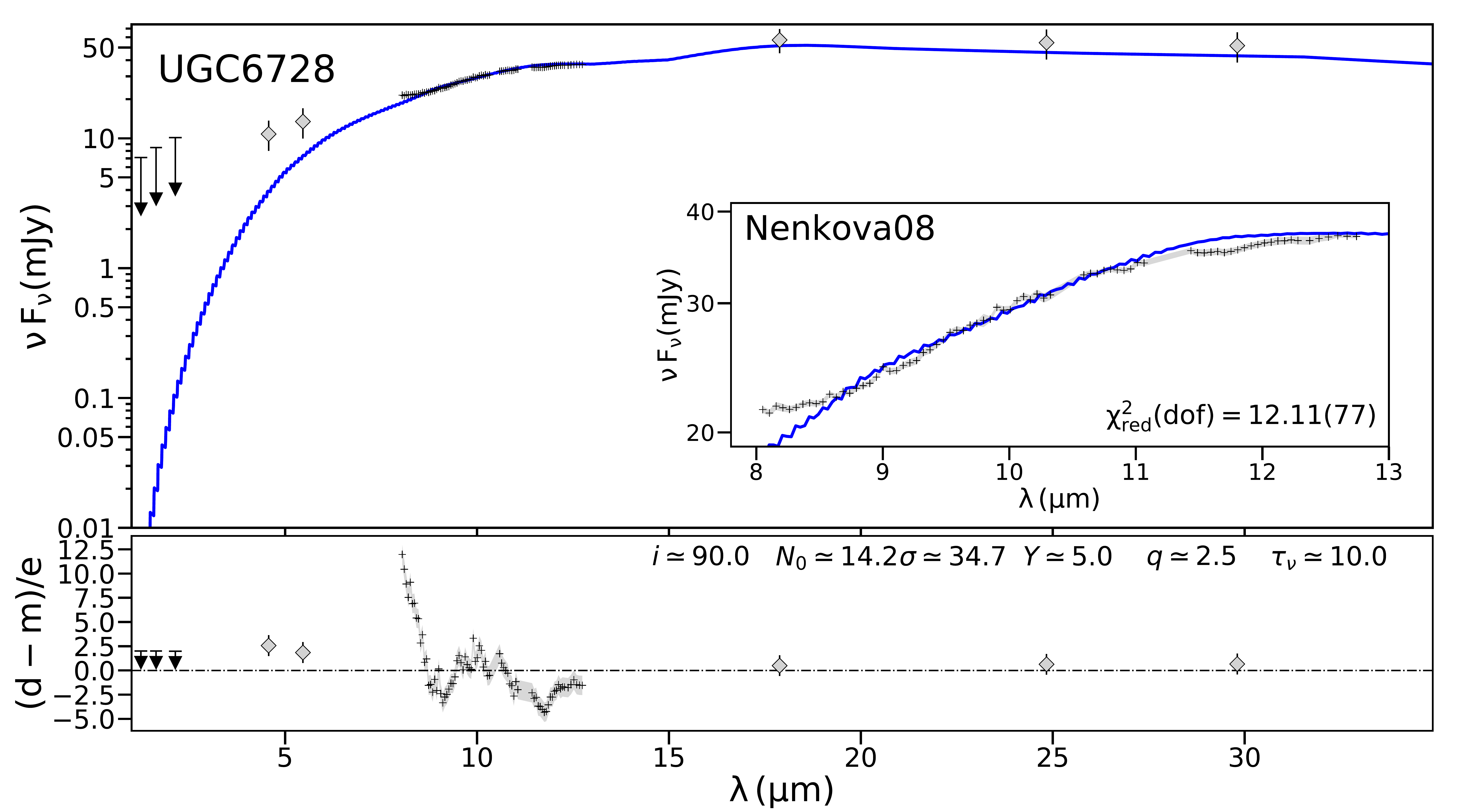}
    \includegraphics[width=0.75\columnwidth]{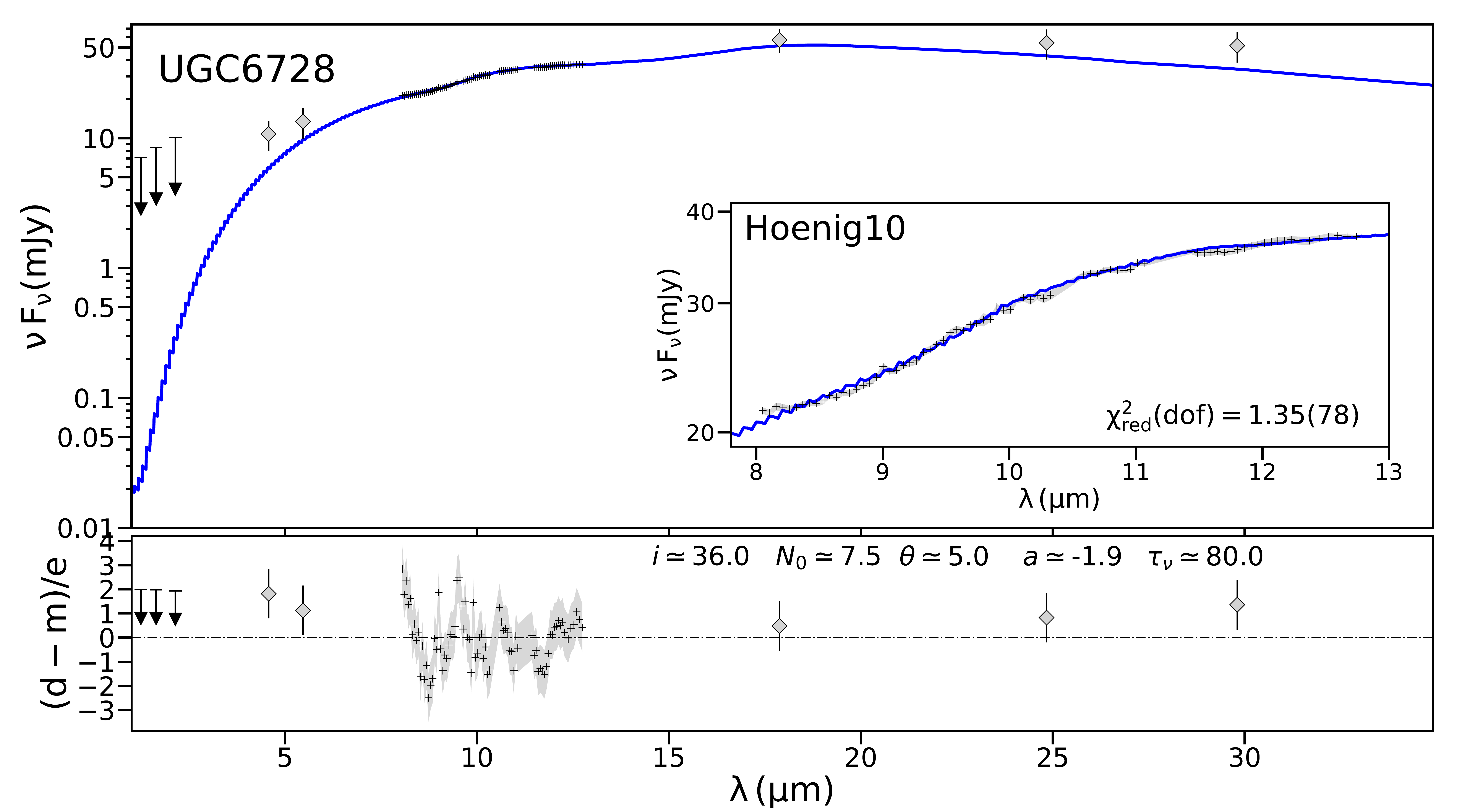}
    \includegraphics[width=0.75\columnwidth]{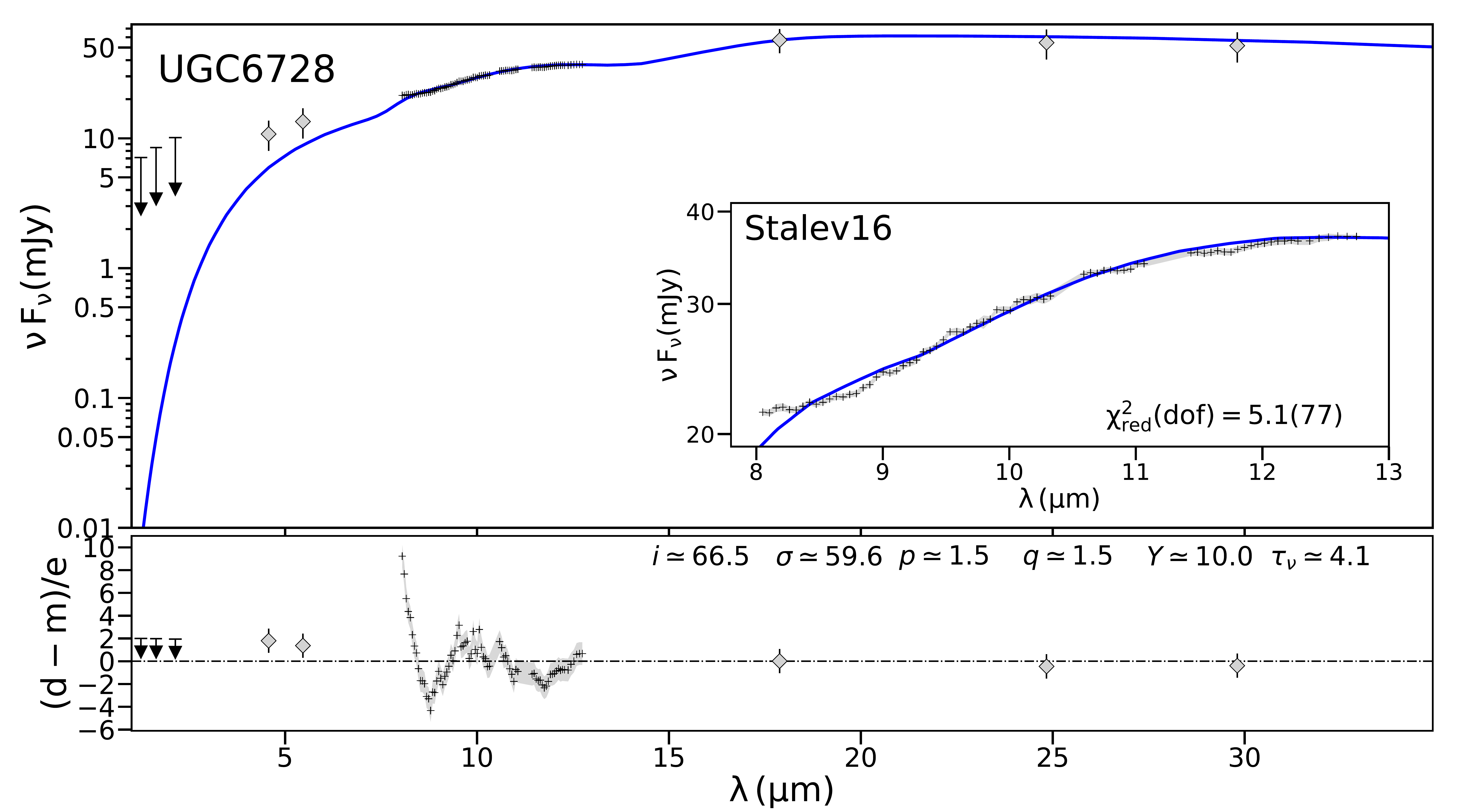}
    \includegraphics[width=0.75\columnwidth]{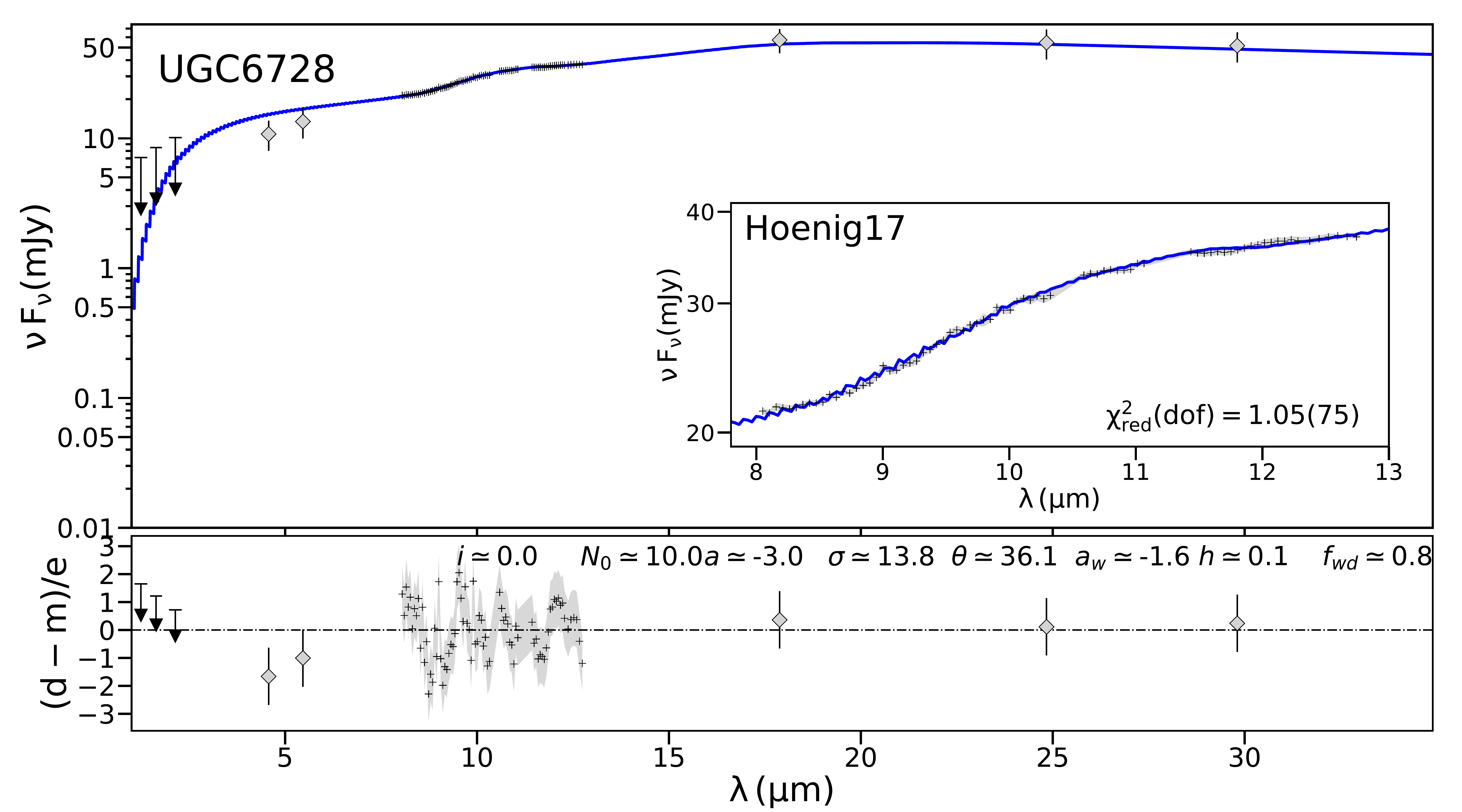}
    \includegraphics[width=0.75\columnwidth]{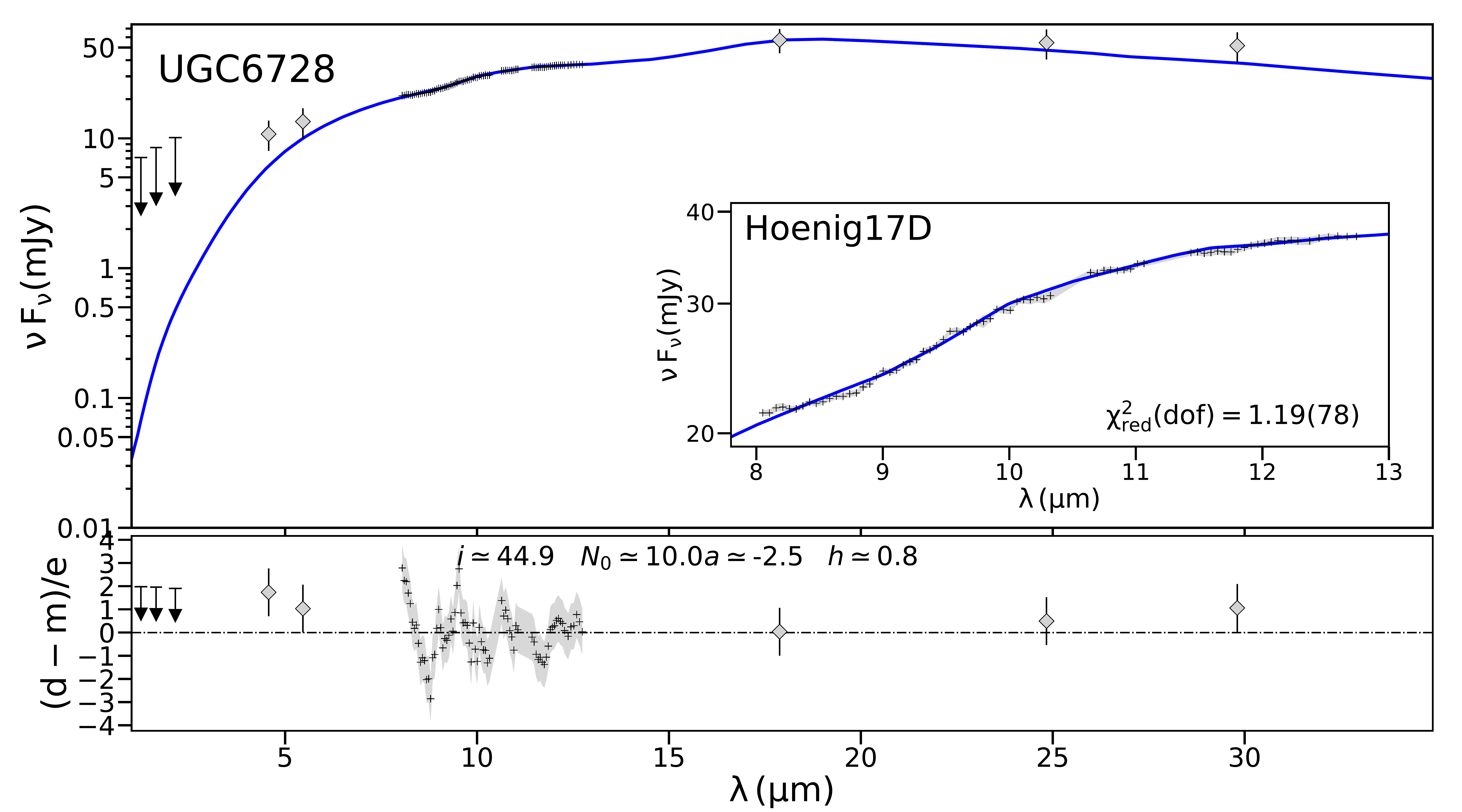}
    \caption{Same as Fig. \ref{fig:ESO005-G004} but for UGC6728.}
    \label{fig:UGC6728}
\end{figure*}

\section{Combined probability distributions}
\label{Combined_distribution}
\begin{figure*}
\centering
\par{
\includegraphics[width=7.5cm]{Figures/fritz06/90minus_i_fritz06-eps-converted-to.pdf}	
\includegraphics[width=7.5cm]{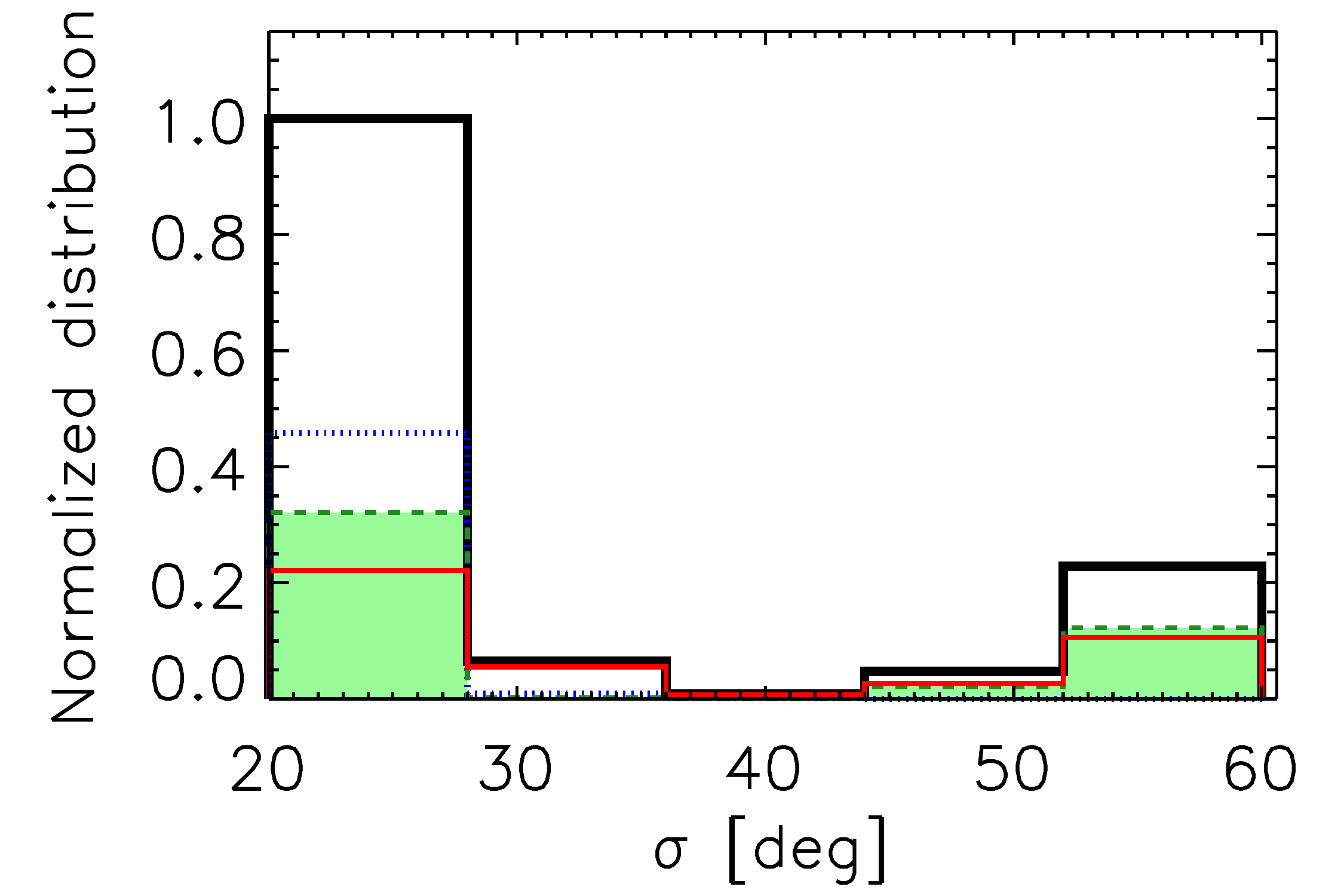}	
\includegraphics[width=7.5cm]{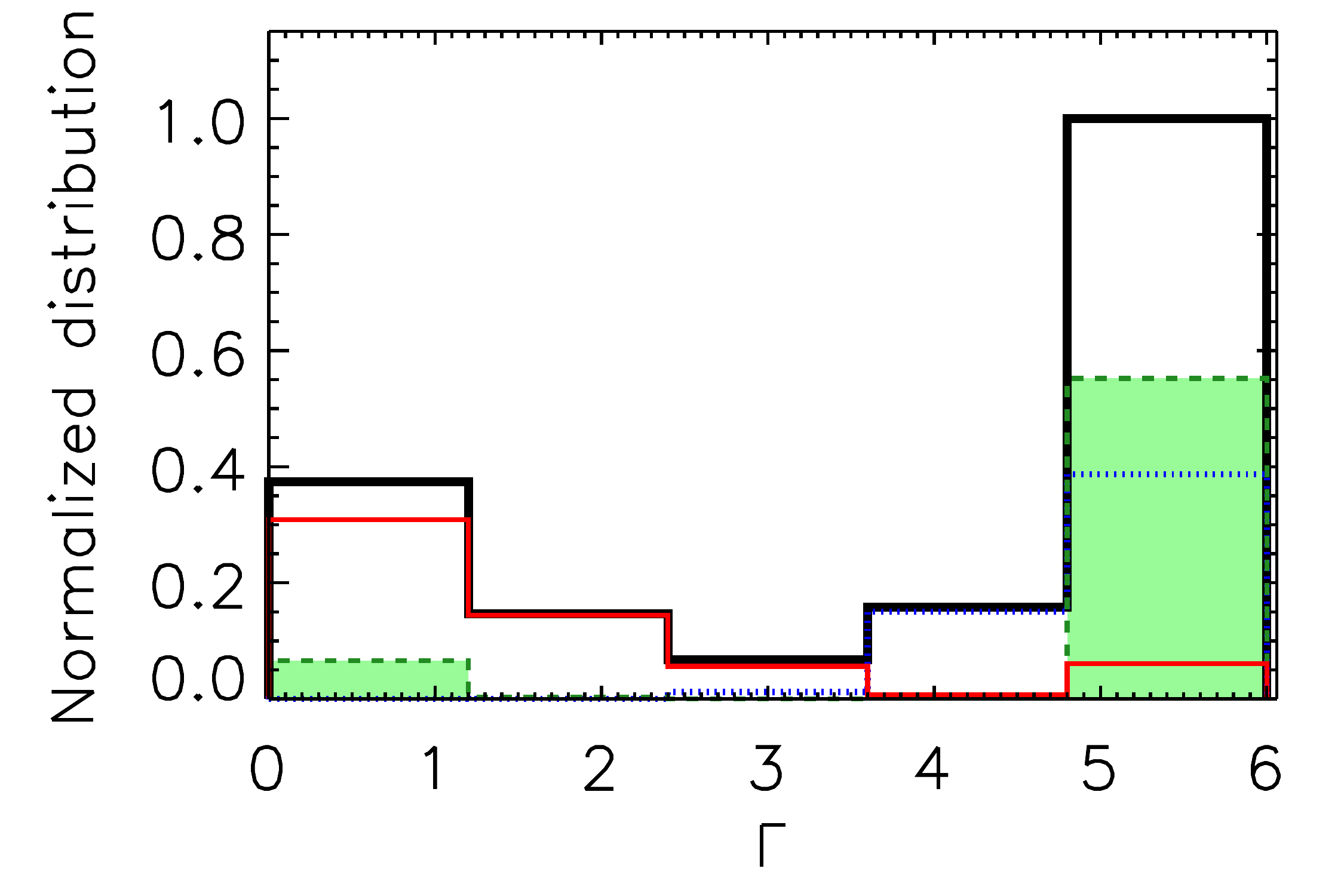}	
\includegraphics[width=7.5cm]{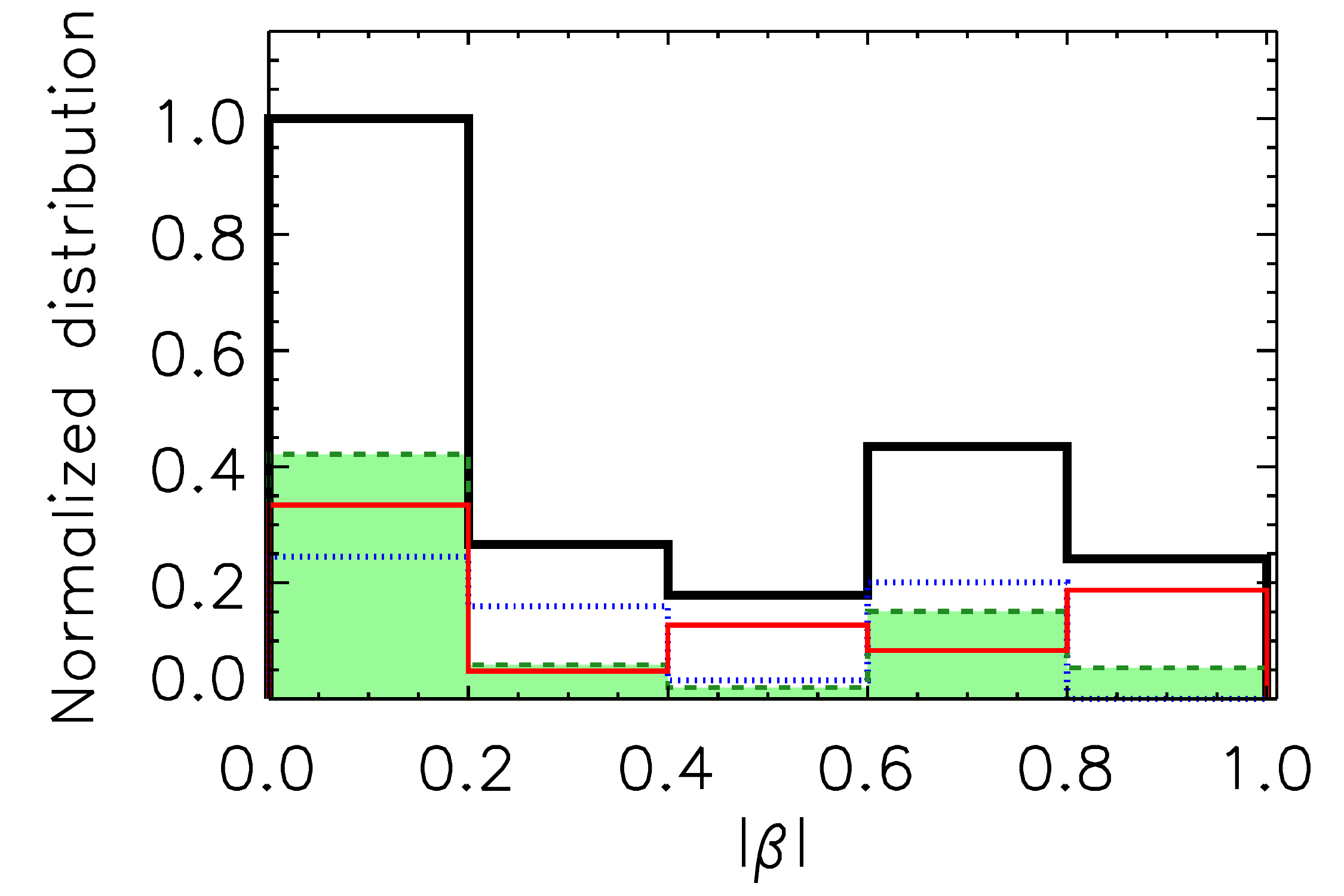}
\includegraphics[width=7.5cm]{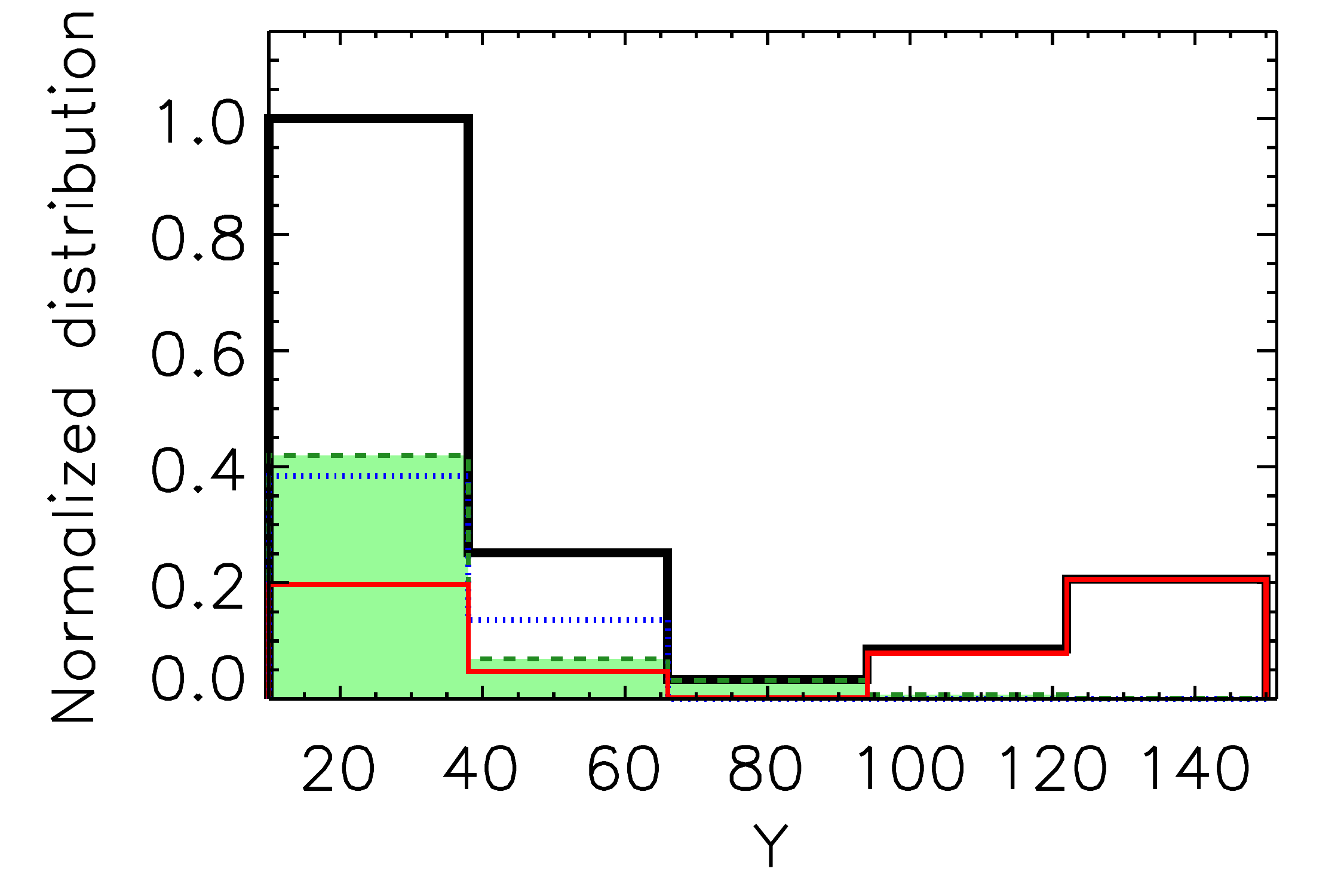}
\includegraphics[width=7.5cm]{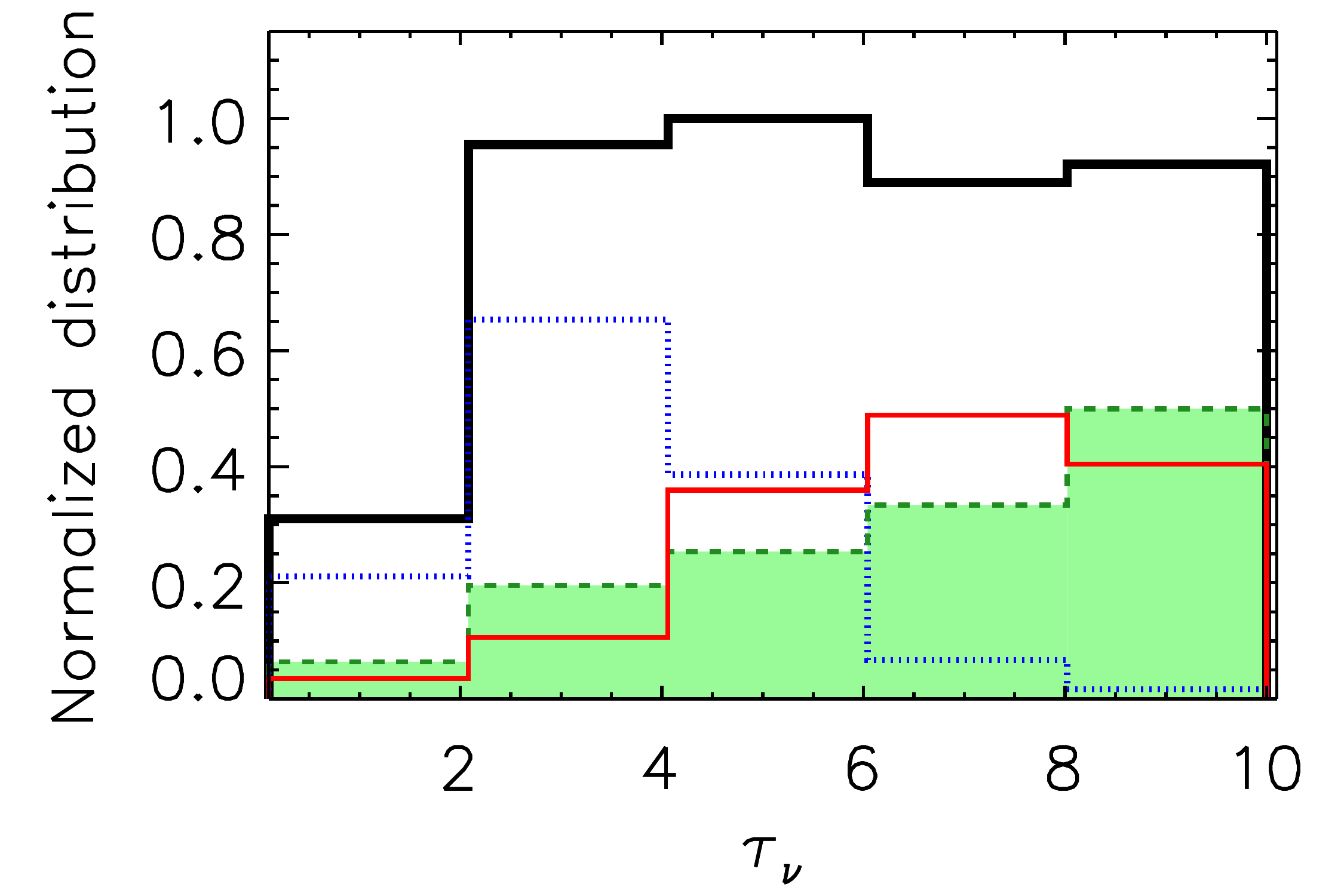}
\par}
\caption{Comparison between the smooth F06 torus model parameter combined probability distributions. Blue dotted, green dashed, red solid and black solid lines represent the parameter distributions of Sy1, Sy1.8/1.9, Sy2 and the entire sample, respectively.}
\label{fritz06_distribution}
\end{figure*}

\begin{figure*}
\centering
\par{
\includegraphics[width=7.5cm]{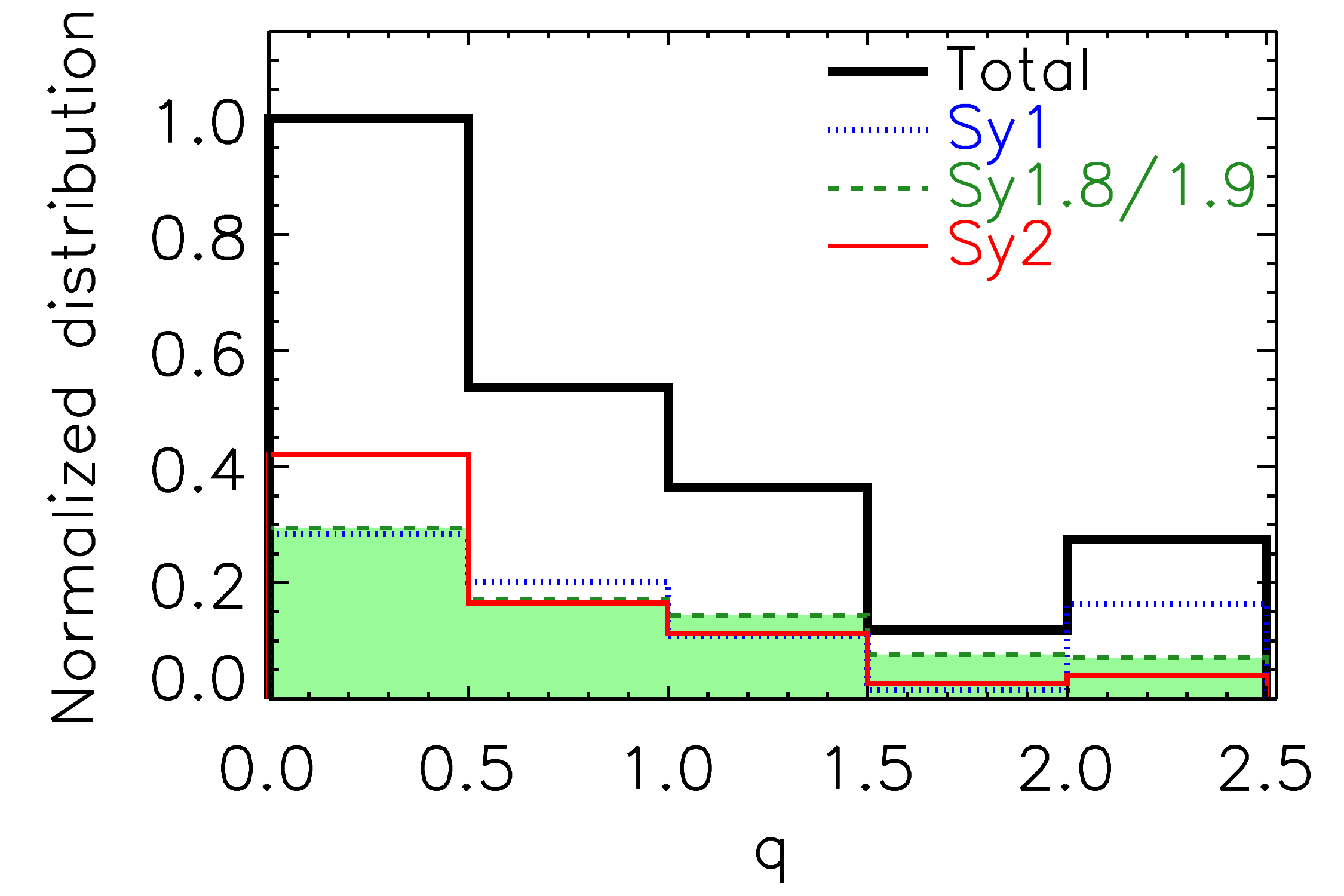}
\includegraphics[width=7.5cm]{Figures/nenkova08/sigma-eps-converted-to.pdf}	
\includegraphics[width=7.5cm]{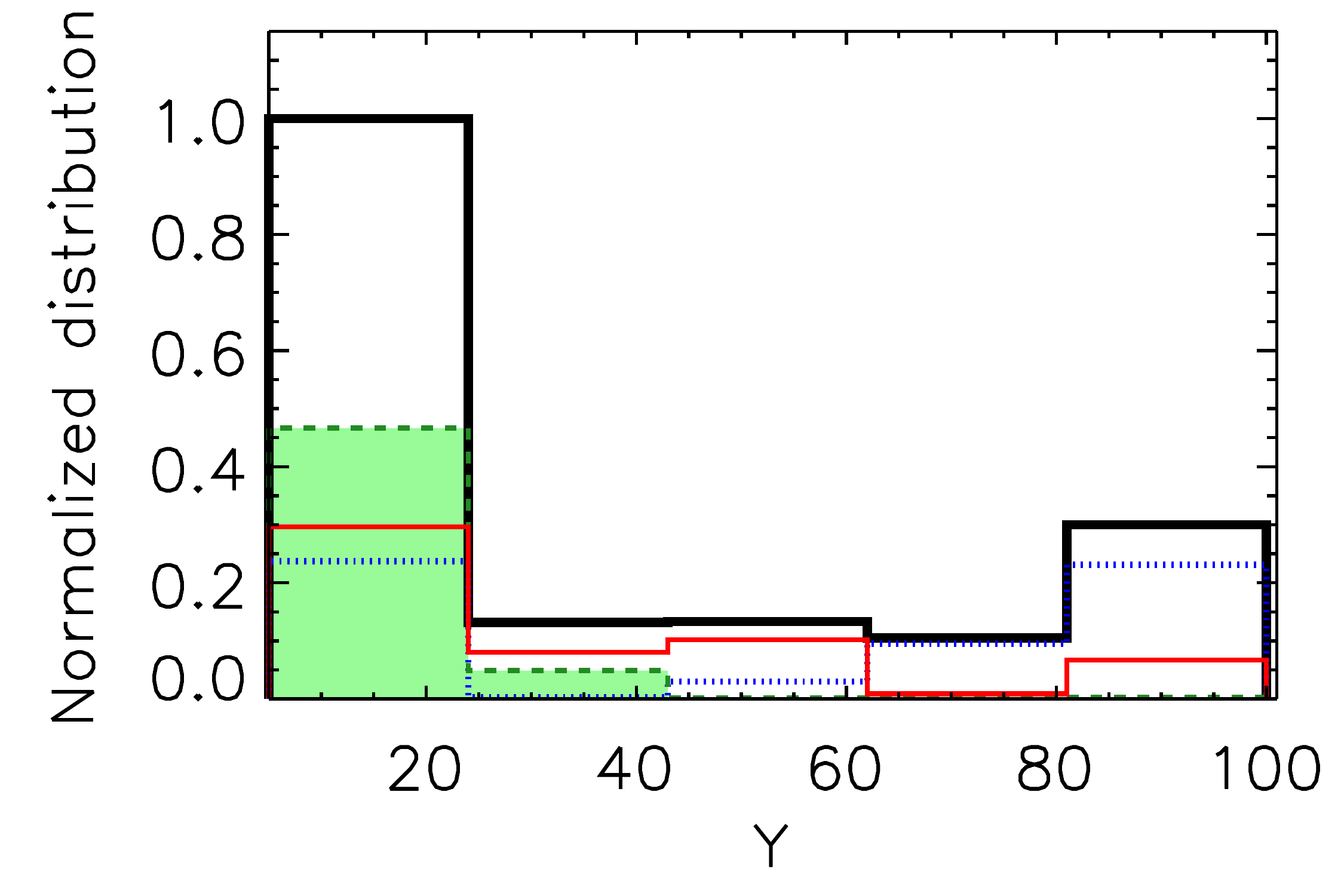}	
\includegraphics[width=7.5cm]{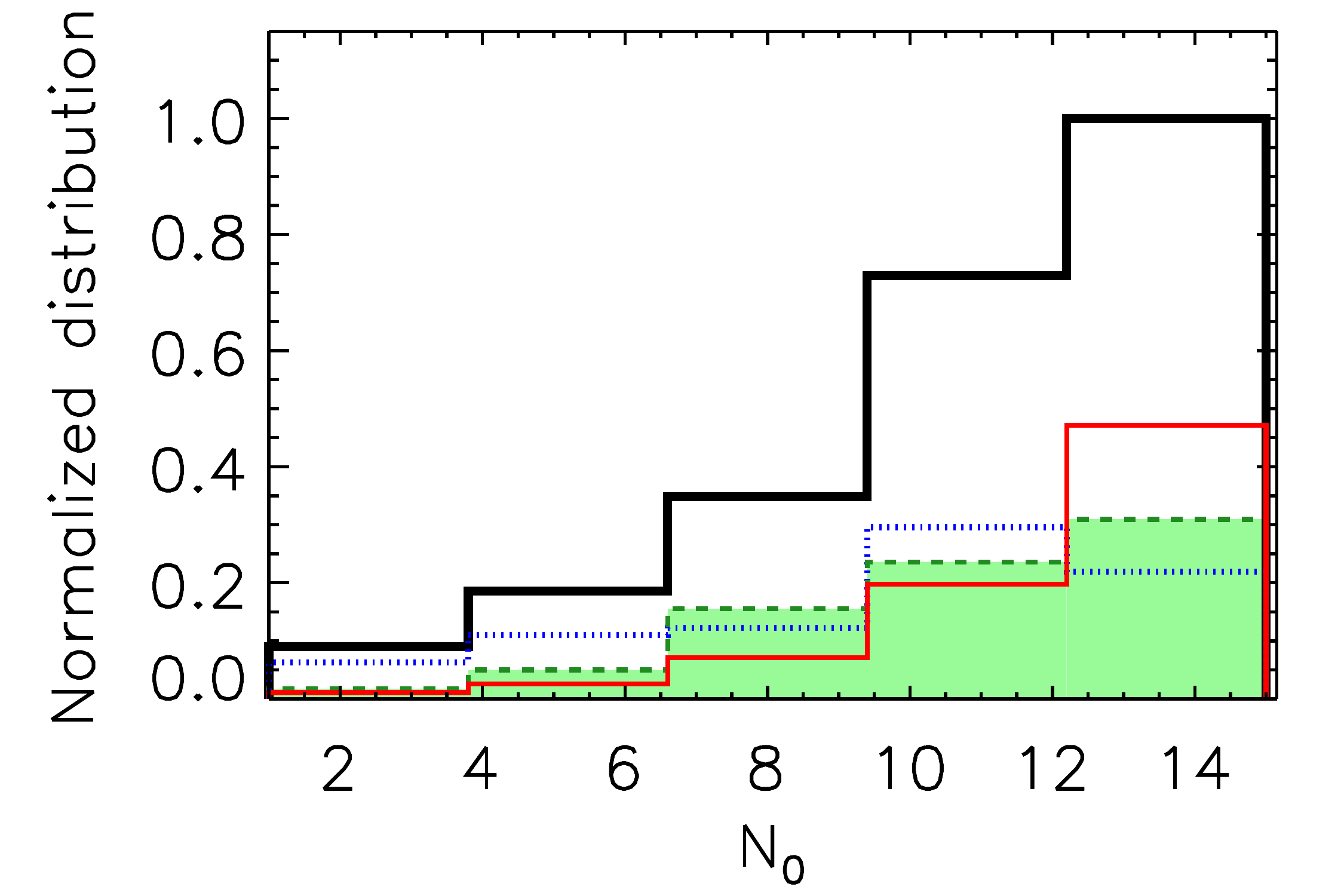}	
\includegraphics[width=7.5cm]{Figures/nenkova08/i-eps-converted-to.pdf}
\includegraphics[width=7.5cm]{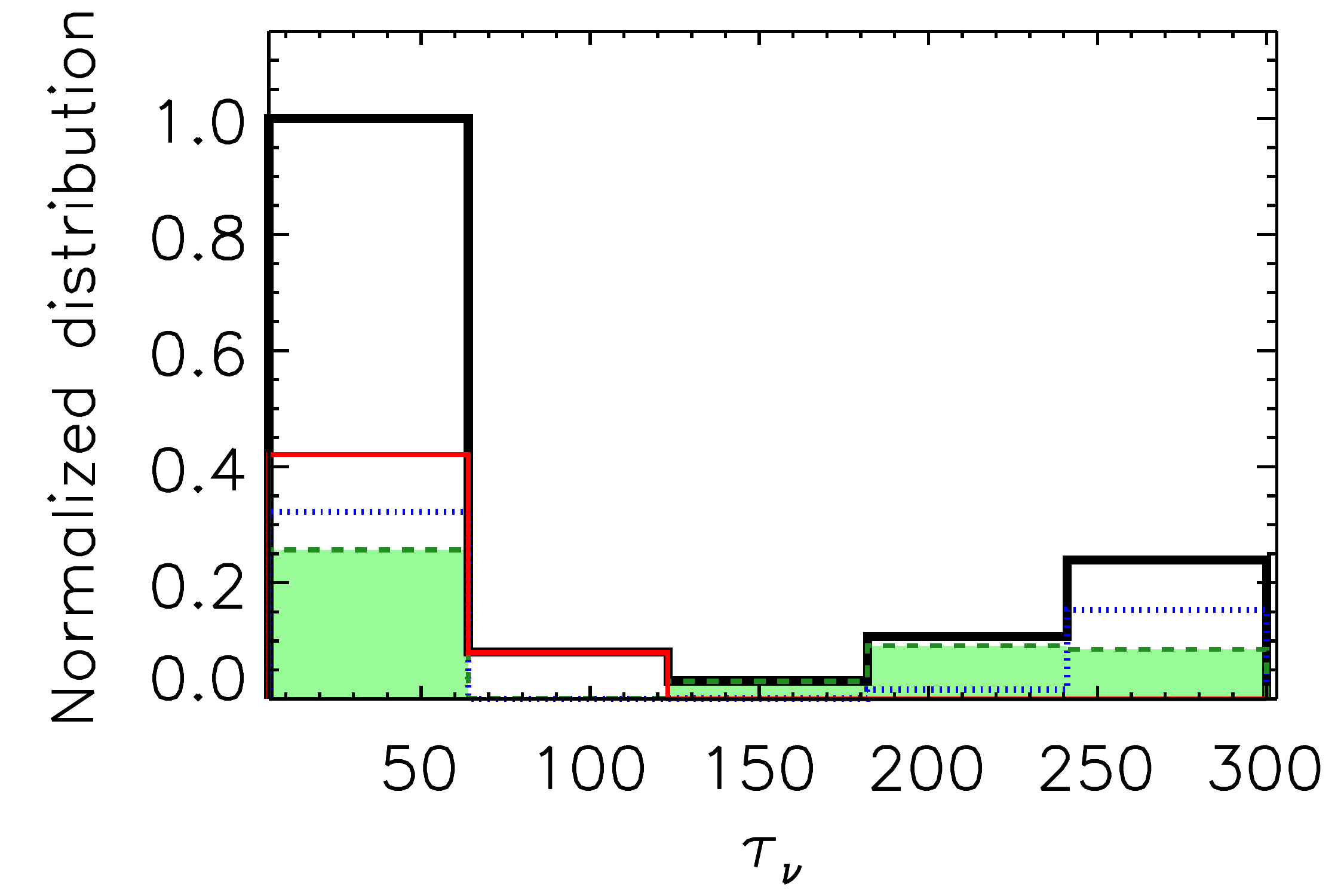}
\par}
\caption{Same as Fig. \ref{fritz06_distribution} but for the clumpy NO8 torus models.}
\label{nenkova08_distribution}
\end{figure*}

\begin{figure*}
\centering
\par{
\includegraphics[width=7.5cm]{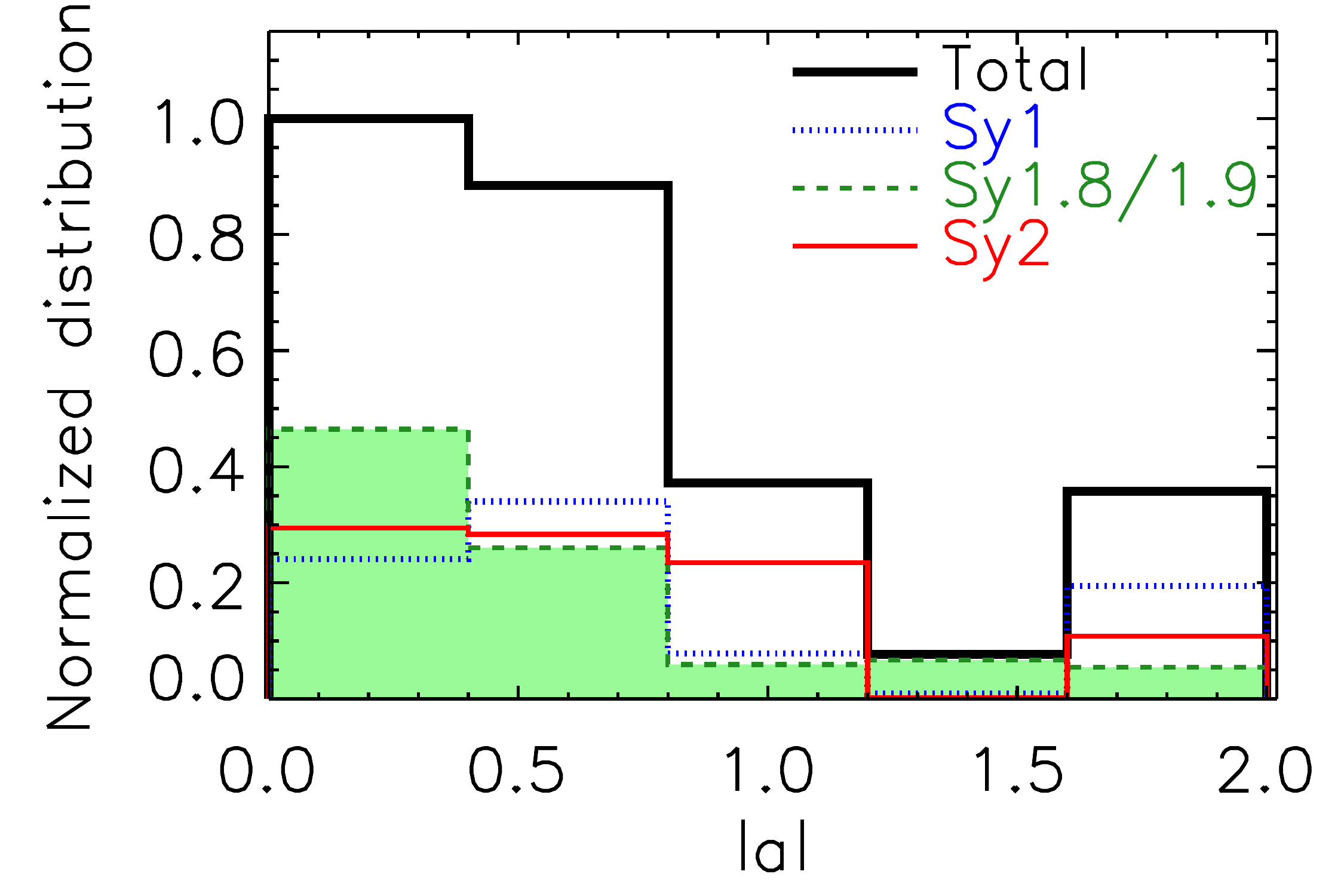}
\includegraphics[width=7.5cm]{Figures/hoenig10/i-eps-converted-to.pdf}	
\includegraphics[width=7.5cm]{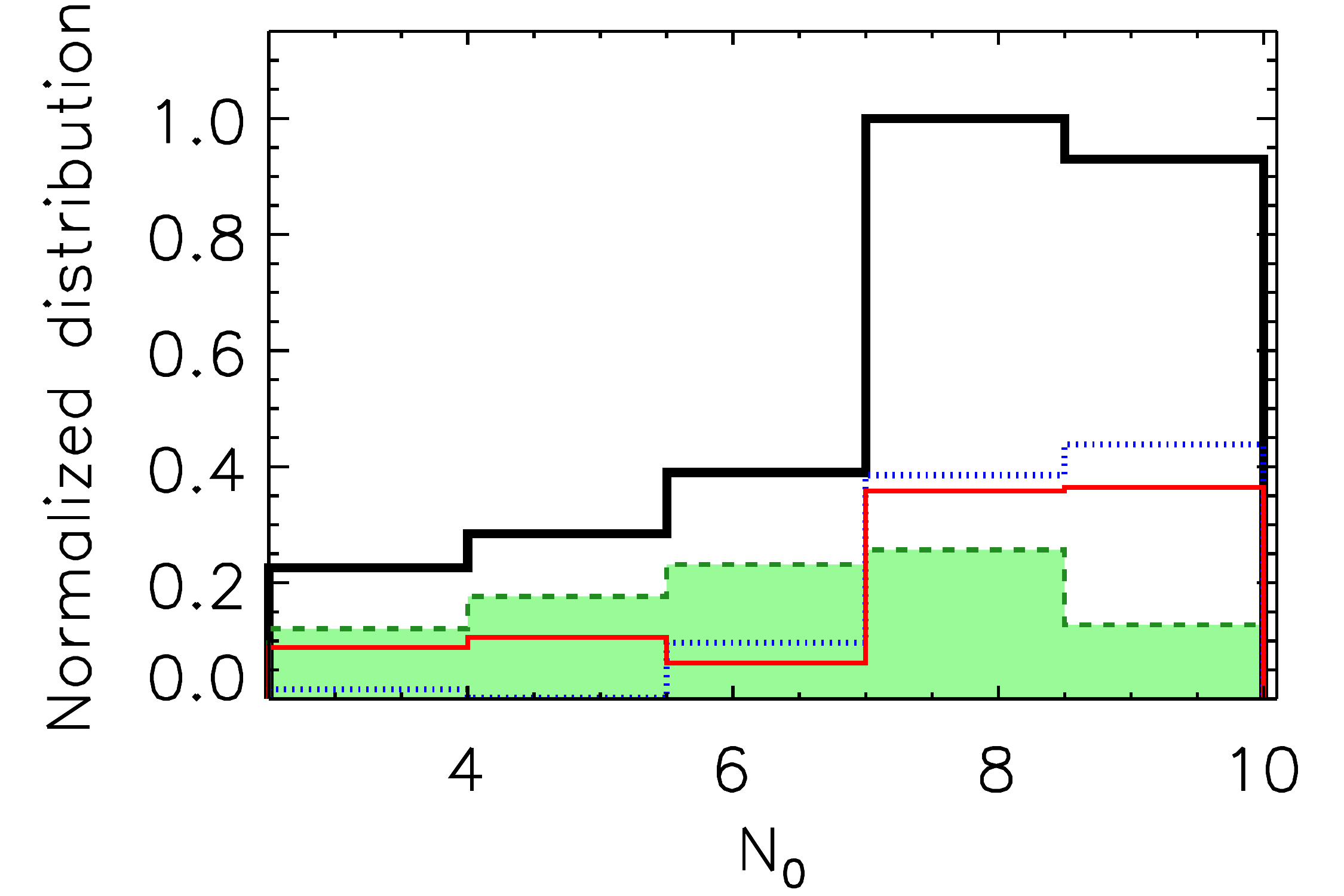}	
\includegraphics[width=7.5cm]{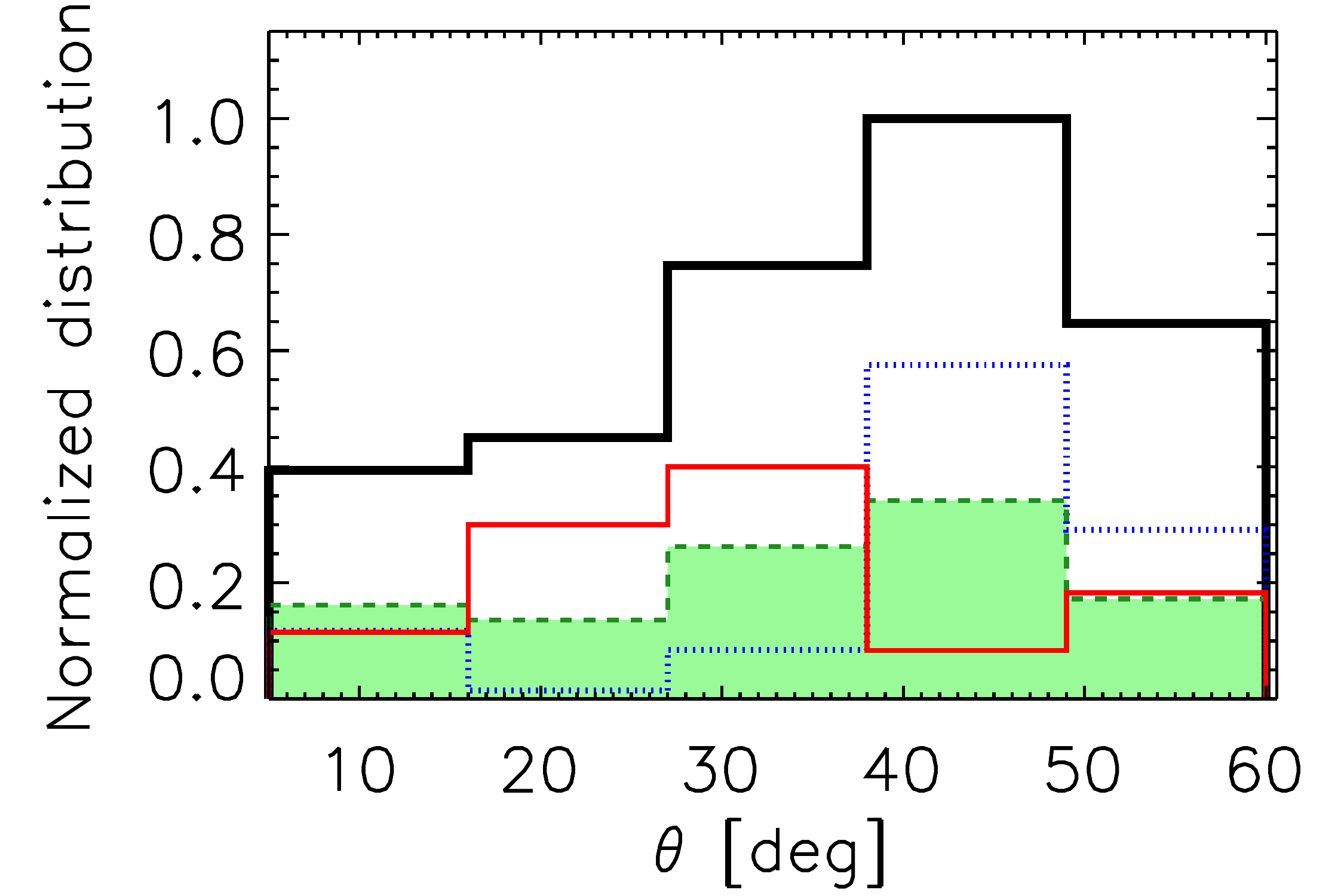}	
\includegraphics[width=7.5cm]{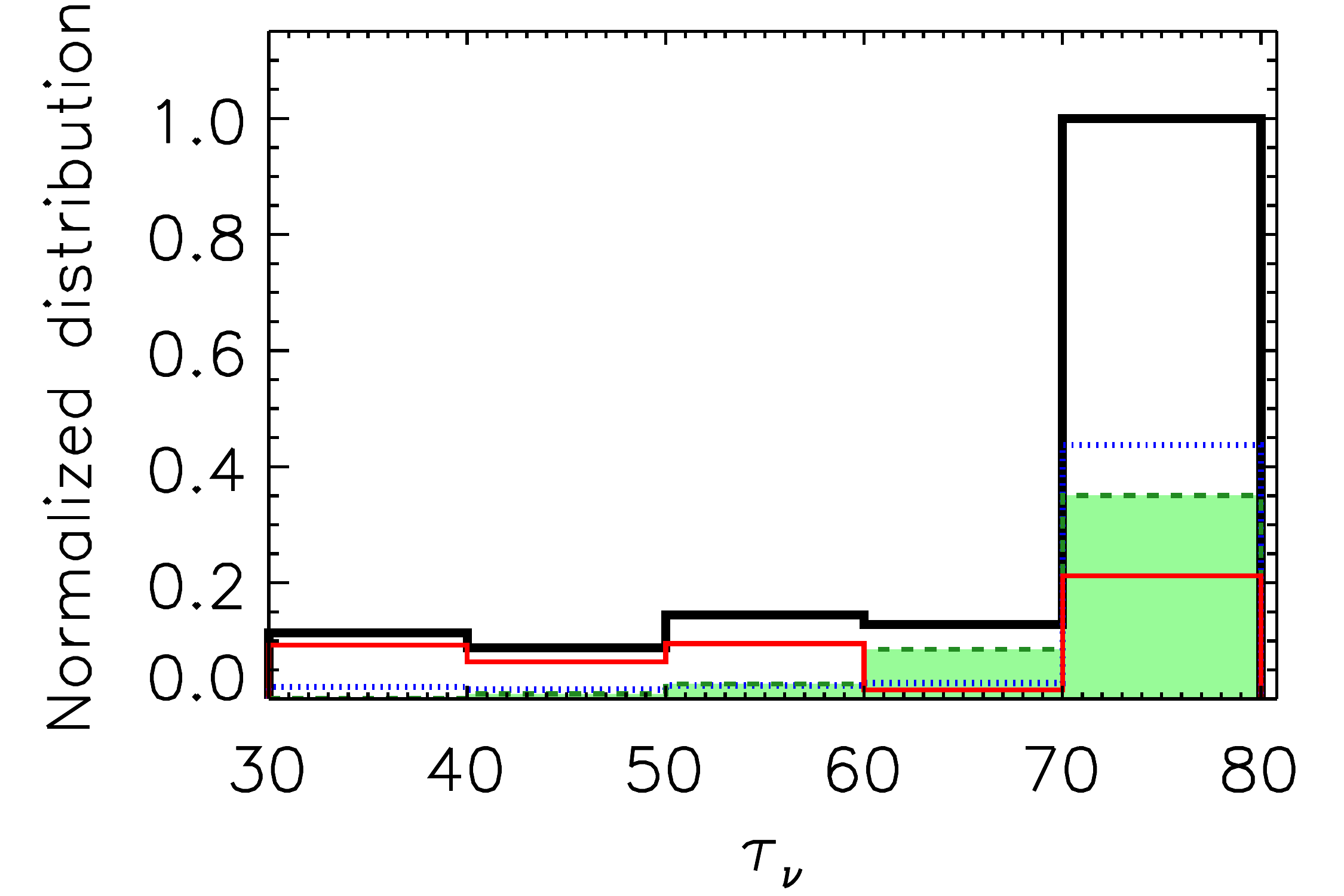}
\par}
\caption{Same as Fig. \ref{fritz06_distribution} but for the clumpy H10 torus models.}
\label{hoenig10_distribution}
\end{figure*}

\begin{figure*}
\centering
\par{
\includegraphics[width=7.5cm]{Figures/hoenig17/a-eps-converted-to.pdf}
\includegraphics[width=7.5cm]{Figures/hoenig17/i-eps-converted-to.pdf}	
\includegraphics[width=7.5cm]{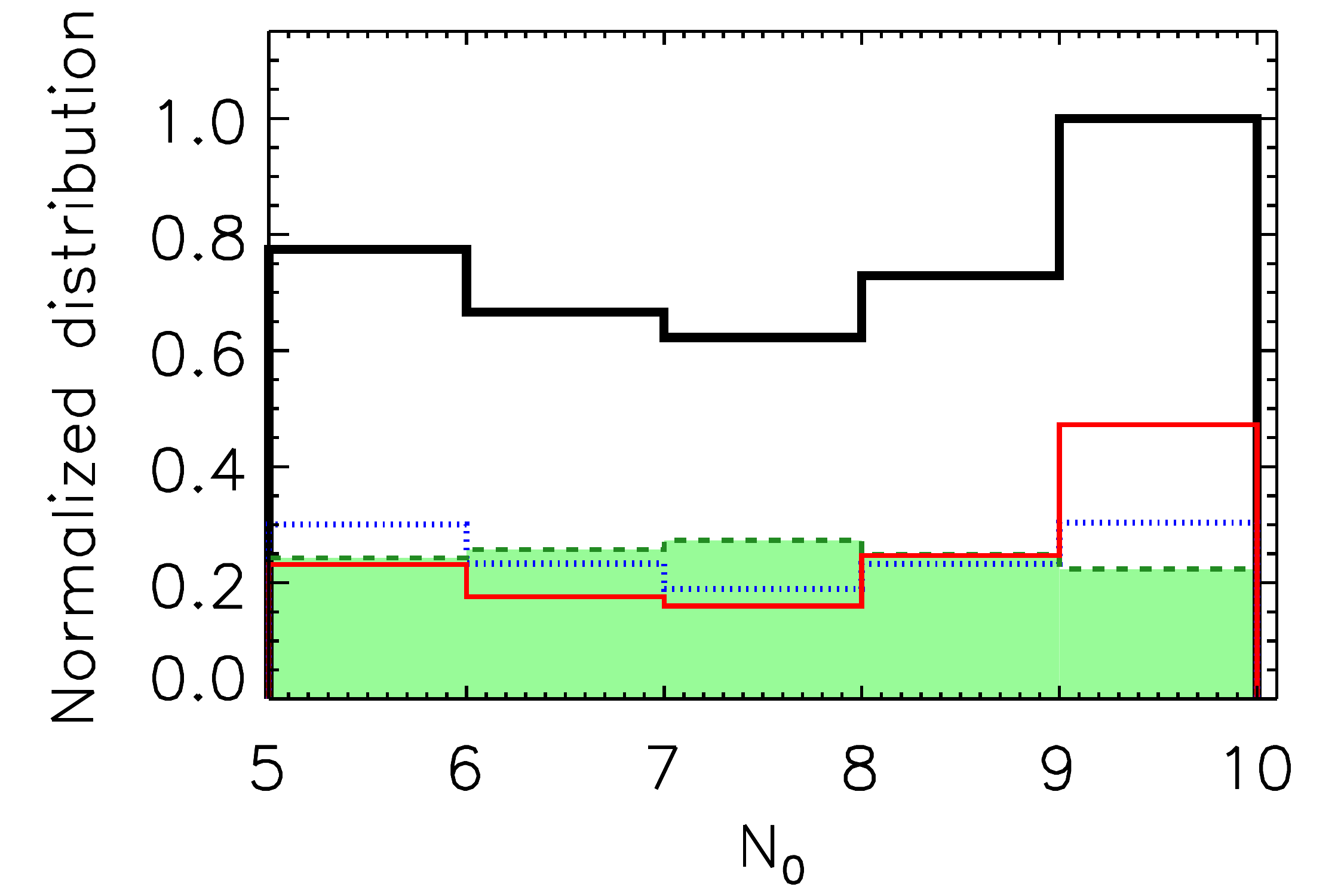}	
\includegraphics[width=7.5cm]{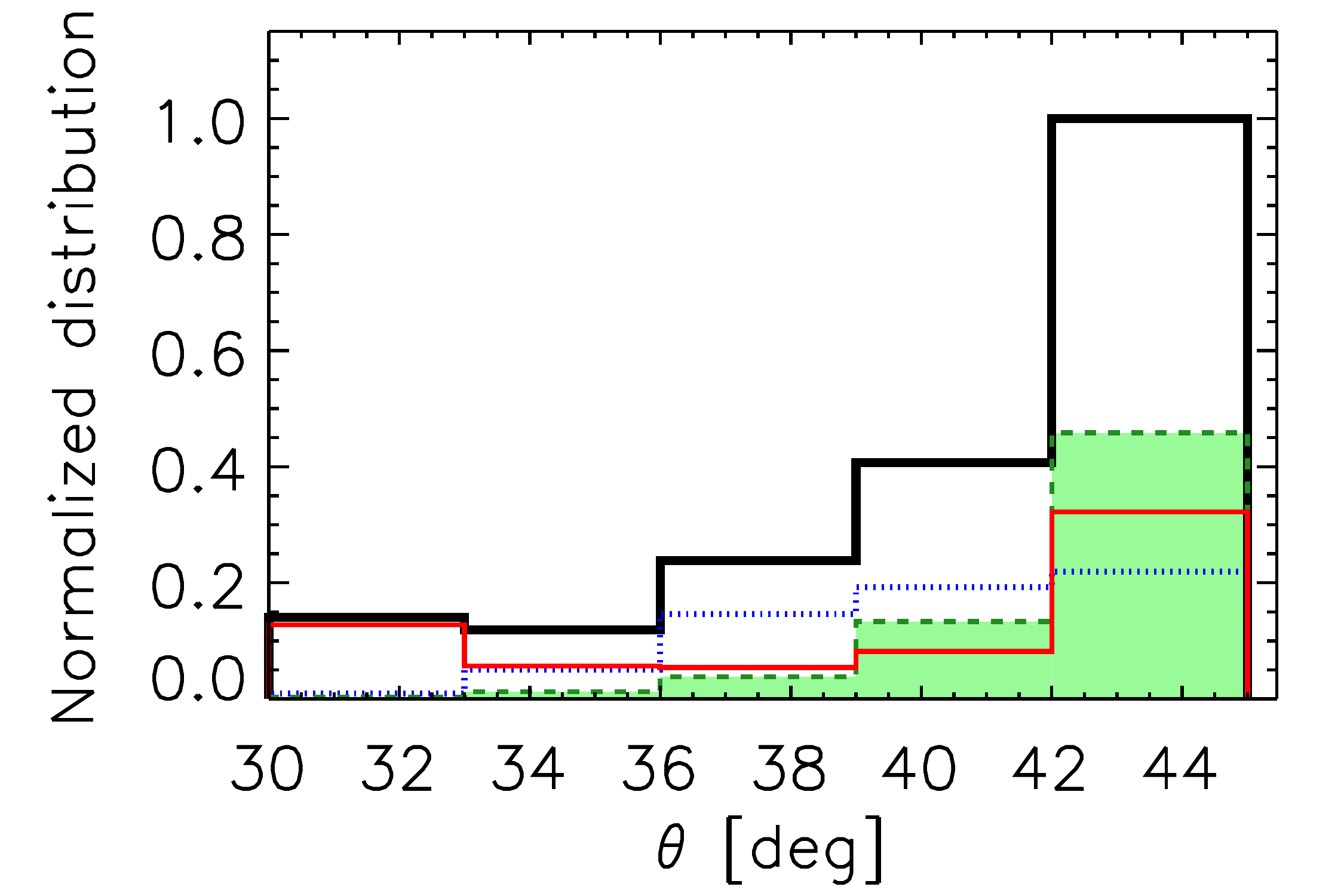}	
\includegraphics[width=7.5cm]{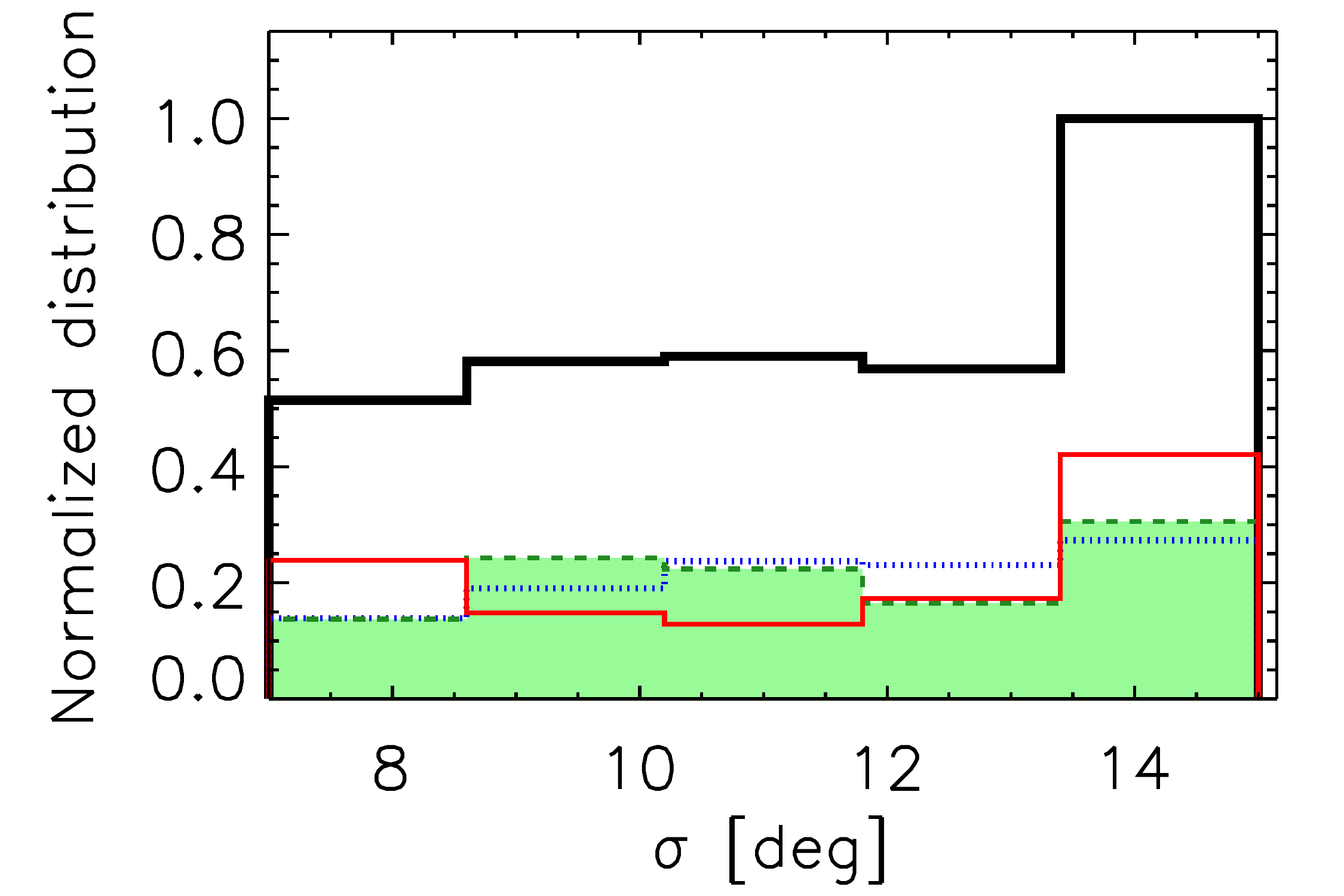}	
\includegraphics[width=7.5cm]{Figures/hoenig17/aw-eps-converted-to.pdf}
\includegraphics[width=7.5cm]{Figures/hoenig17/h-eps-converted-to.pdf}
\includegraphics[width=7.5cm]{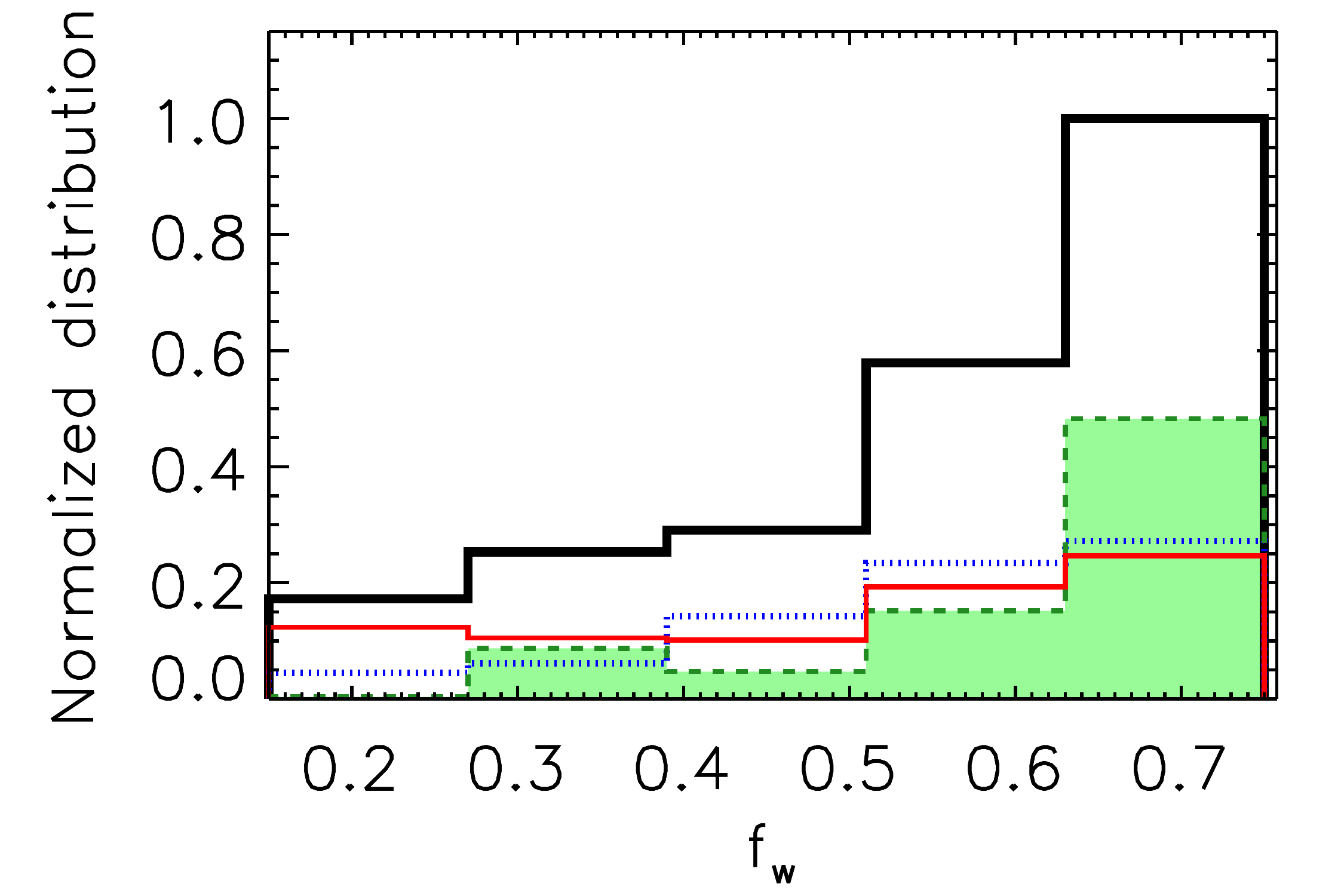}
\par}
\caption{Same as Fig. \ref{fritz06_distribution} but for the clumpy disc$+$wind H17 models.}
\label{hoenig17_distribution}
\end{figure*}

\begin{figure*}
\centering
\par{
\includegraphics[width=7.5cm]{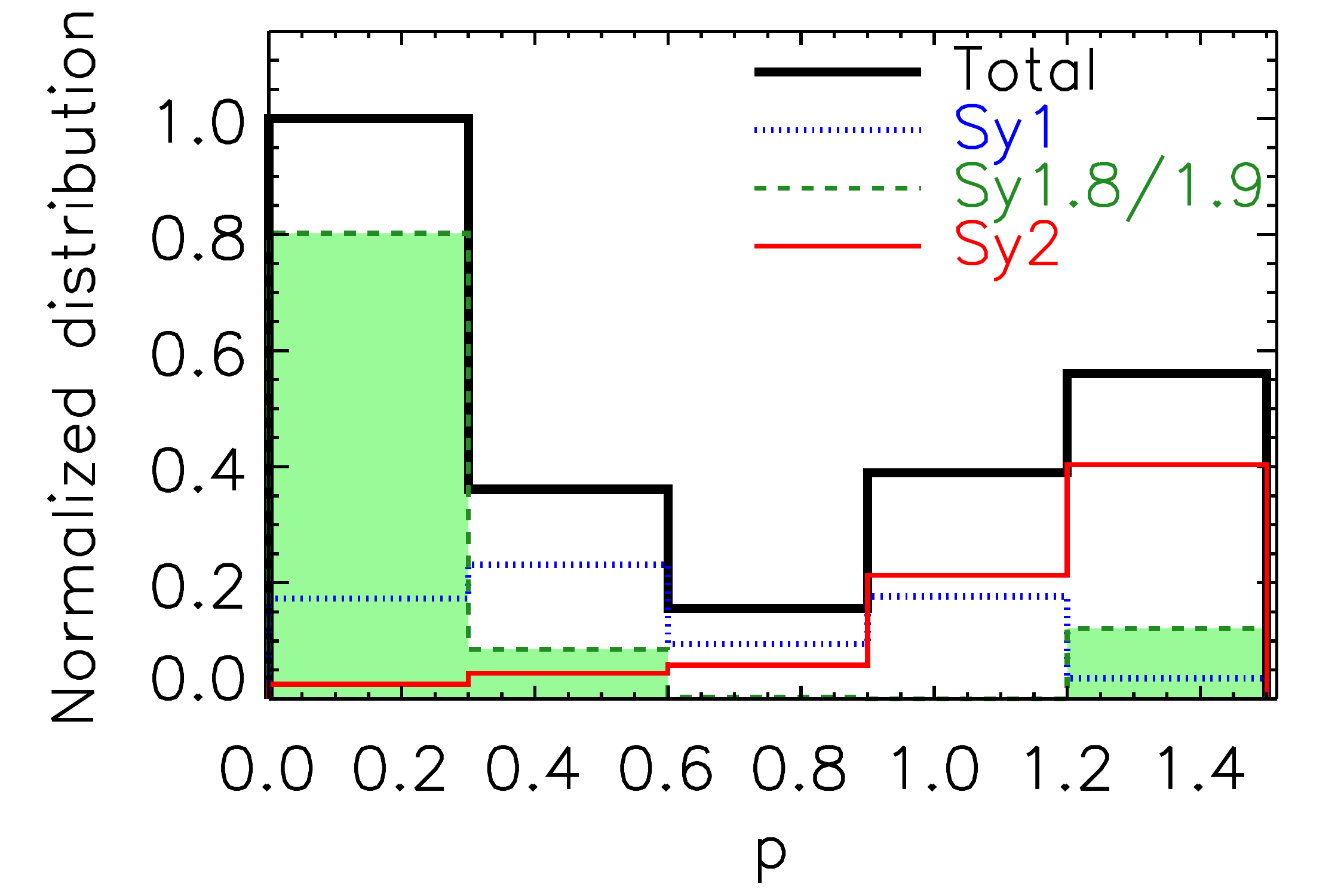}	
\includegraphics[width=7.5cm]{Figures/stalev16/i-eps-converted-to.pdf}	
\includegraphics[width=7.5cm]{Figures/stalev16/sigma-eps-converted-to.pdf}	
\includegraphics[width=7.5cm]{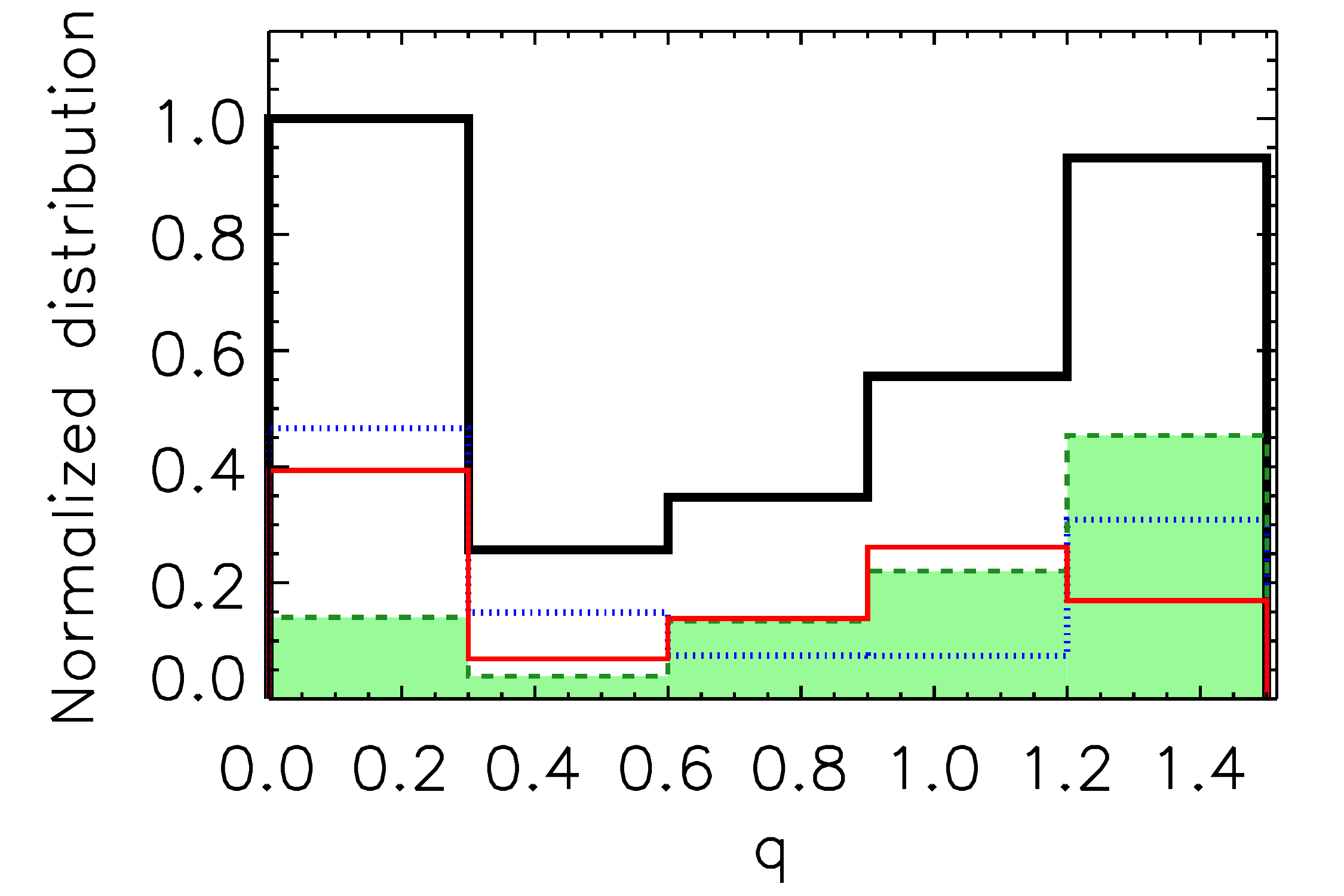}
\includegraphics[width=7.5cm]{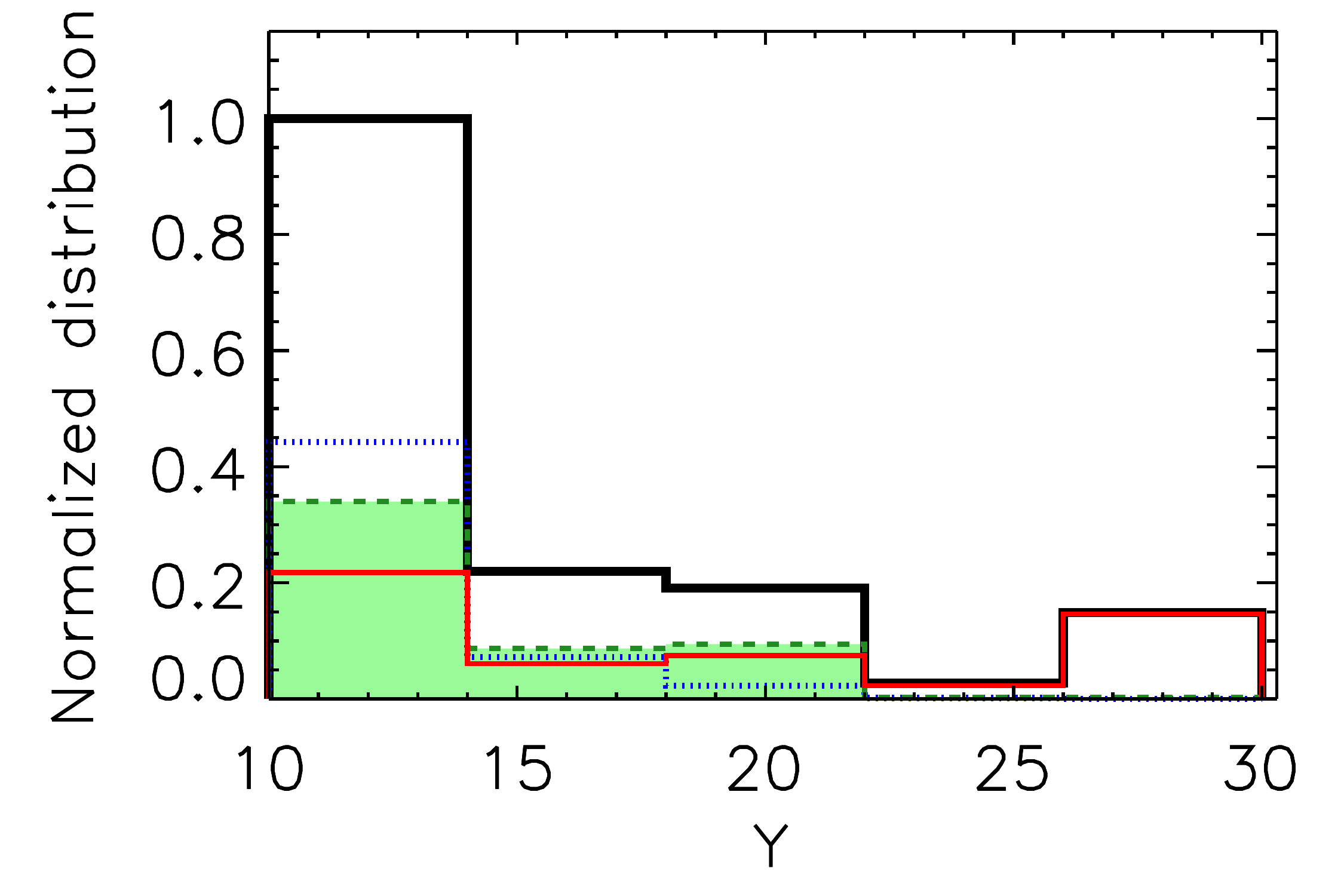}
\includegraphics[width=7.5cm]{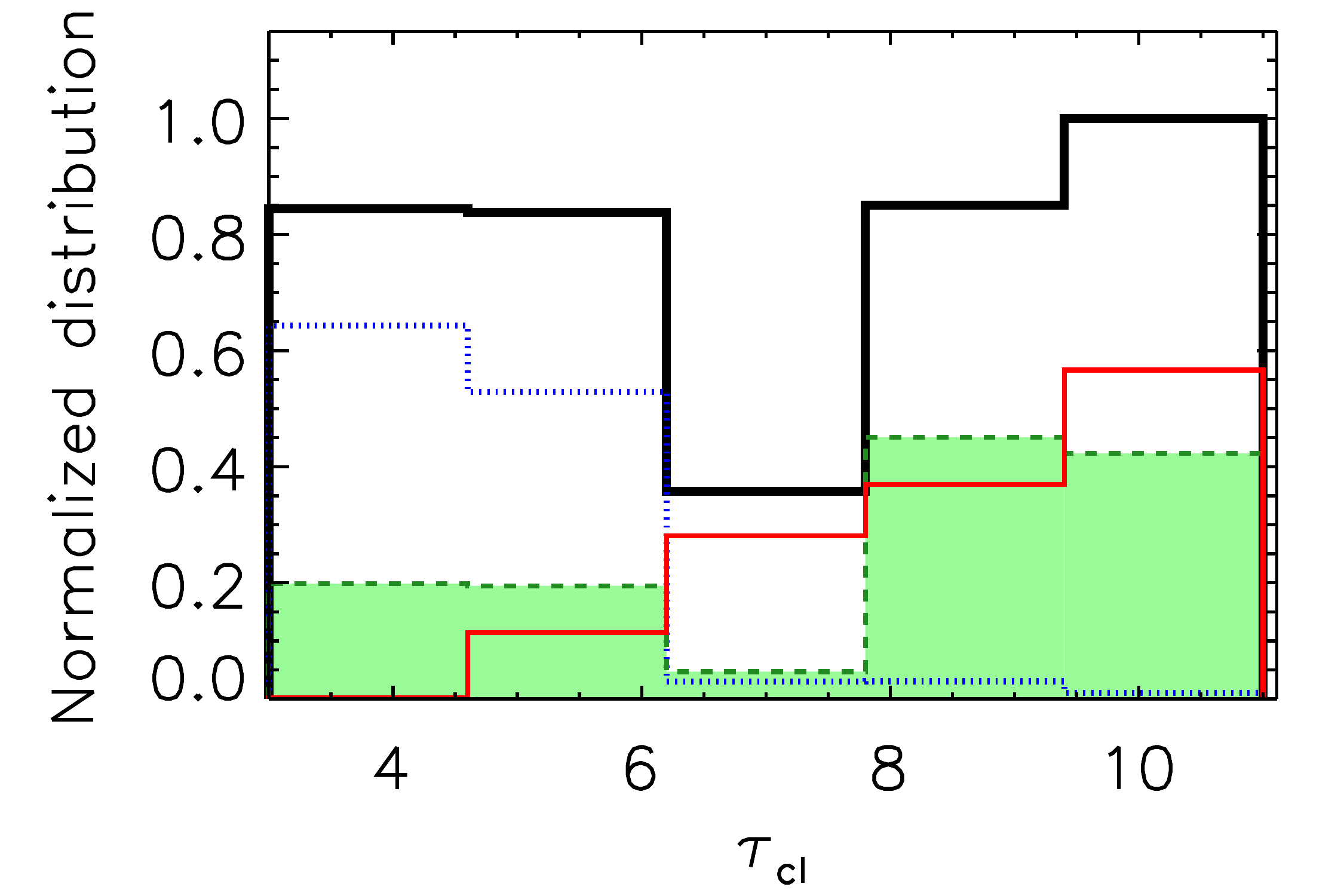}
\par}
\caption{Same as Fig. \ref{fritz06_distribution} but for the two-phases S16 torus models.}
\label{stalev16_distribution}
\end{figure*}

In general, the various torus models provide acceptable fits ($\rm{\chi^2_{red}<2}$) to the majority (19/24) of the nuclear IR SEDs (see Appendix \ref{nuclear_fits}). However, it is difficult to see trend using the individual fit parameters, which can be not well constrained. Therefore, we obtain a global statistical analysis of the torus model parameters of the various Seyfert galaxy types, rather than focusing on the individual fits. Figs. \ref{fritz06_distribution}, \ref{nenkova08_distribution}, \ref{hoenig10_distribution}, \ref{hoenig17_distribution} and \ref{stalev16_distribution} show the combined probability distribution of the different torus model parameters for all the objects in each Sy subgroup. Finally, in Tables \ref{fritz_tab_kdl}, \ref{nenkova_tab_kdl}, \ref{hoenig10_tab_kdl}, \ref{hoenig17_tab_kdl} and \ref{stalev_tab_kdl} we present the KLD results for each torus parameter and Sy subgroup.

\begin{table}
\centering
\caption{KLD results for the smooth F06 torus models.}
\begin{tabular}{lcccccccccccc}
\hline
Subgroups	&i & $\sigma$	& $\Gamma$	& $\beta$	& Y	&$\tau$	&C$_T$\\
	(1) & (2) & (3) & (4) & (5)& (6)& (7)& (8)\\
\hline
Sy1s vs Sy2s		&0.89        	&{\bf{4.70}}	&{\bf{12.04}}	&{\bf{4.15}}	&{\bf{7.23}}	&{\bf{4.52}} &{\bf{1.46}}\\
Sy1s vs Sy1.8/1.9	&{\bf{4.05}}	&{\bf{4.17}}	&{\bf{3.50}}	&{\bf{2.35}}	&{\bf{1.40}}	&{\bf{2.51}}&{\bf{1.65}}\\
Sy2s vs	Sy1.8/1.9	&{\bf{3.79}}	&{\bf{1.81}}	&{\bf{3.60}}	&{\bf{1.01}}	&{\bf{3.17}}	&{\bf{1.01}}&{\bf{1.33}}\\
\hline
\end{tabular}					 
\tablefoot{Comparison of the combined probability distribution of each parameter for the various subgroups. {\bf{In bold we indicate the statistically significant differences.}}}
\label{fritz_tab_kdl}
\end{table}

\begin{table}
\centering
\caption{KLD results for the clumpy N08 torus models.}
\begin{tabular}{lcccccccccccc}
\hline
Subgroups	& $\sigma$	& Y	& N$_0$	& q	&$\tau_{V}$	& i & C$_T$\\
	(1) & (2) & (3) & (4) & (5)& (6)& (7)& (8)\\
\hline
Sy1s vs Sy2s		&{\bf{2.31}}	&{\bf{2.73}}	&0.70			&0.27	&{\bf{4.29}}	&0.92 &0.62\\
Sy1s vs Sy1.8/1.9	&0.20			&{\bf{4.79}}	&0.99	        &0.27	&{\bf{1.17}}	&0.58 &0.59\\
Sy2s vs	Sy1.8/1.9	&{\bf{1.27}}	&{\bf{2.05}}	&0.35			&0.16	&{\bf{5.40}}	&0.87 &0.44\\
\hline
\end{tabular}					 
\tablefoot{Comparison of the combined probability distribution of each parameter for the various subgroups. {\bf{In bold we indicate the statistically significant differences.}}}
\label{nenkova_tab_kdl}
\label{tab_kld}
\end{table}

\begin{table}
\centering
\caption{KLD results for the clumpy H10 torus models.}
\begin{tabular}{lcccccccccccc}
\hline
Subgroups	&i & N$_0$& $\theta$	& a	&$\tau_{V}$	 & C$_T$\\
	(1) & (2) & (3) & (4) & (5)& (6)& (7)\\
\hline
Sy1s vs Sy2s		&{\bf{1.39}}	&{\bf{1.26}}	&{\bf{2.66}}       	&{\bf{1.80}}	&{\bf{1.34}}       &0.67\\
Sy1s vs Sy1.8/1.9	&0.81			&{\bf{2.24}}	&0.69		        &{\bf{2.57}}	&0.31       	   &0.48\\
Sy2s vs	Sy1.8/1.9	&0.53			&0.66	        &0.70				&{\bf{1.71}}	&{\bf{1.56}}       &0.14\\
\hline
\end{tabular}
\tablefoot{Comparison of the combined probability distribution of each parameter for the various subgroups. {\bf{In bold we indicate the statistically significant differences.}}}
\label{hoenig10_tab_kdl}
\end{table}

\begin{table}
\tiny
\centering
\caption{KLD results for the clumpy disc$+$wind H17 models.}
\begin{tabular}{lcccccccccccc}
\hline
Subgroups	&i & N$_0$ & a & $\theta$	& $\sigma$	&a$_w$ &h &f$_w$	 & C$_T$\\
	(1) & (2) & (3) & (4) & (5)& (6)& (7)& (8)& (9)& (10)\\
\hline
Sy1s vs Sy2s		&{\bf{2.15}}	&0.10	&{\bf{2.54}}     &0.95 & 0.19	    &{\bf{1.55}}	&0.68              &0.21  &{\bf{1.26}}\\
Sy1s vs Sy1.8/1.9	&0.92	    	&0.05	&{\bf{2.34}}	 &0.55 & 0.06       &{\bf{1.25}}    &0.82		       &0.44  &0.18\\
Sy2s vs	Sy1.8/1.9	&{\bf{2.93}} 	&0.20   &{\bf{3.92}}	 &0.75 & 0.17		&{\bf{1.69}}	&{\bf{1.72}}       &0.73  &0.98\\
\hline
\end{tabular}					 
\tablefoot{Comparison of the combined probability distribution of each parameter for the various subgroups. {\bf{In bold we indicate the statistically significant differences.}}}
\label{hoenig17_tab_kdl}
\end{table}

\begin{table}
\centering
\caption{KLD results for the two-phase S16 torus models.}
\begin{tabular}{lcccccccccccc}
\hline
Subgroups	&i & $\sigma$	& p	& q	& Y	&$\tau_{cl}$	& C$_T$\\
	(1) & (2) & (3) & (4) & (5)& (6)& (7)& (8)\\
\hline
Sy1s vs Sy2s		&0.47	&{\bf{4.59}}	&{\bf{3.52}}	&0.56	&{\bf{2.87}}	&{\bf{5.27}} &0.56\\
Sy1s vs Sy1.8/1.9	&0.89	&0.96			&{\bf{3.51}}	&0.88	&0.42			&{\bf{2.18}} &{\bf{6.99}}\\
Sy2s vs	Sy1.8/1.9	&0.46	&{\bf{5.78}}	&{\bf{6.34}}	&0.45	&{\bf{1.05}}	&{\bf{1.23}} &{\bf{4.51}}\\
\hline
\end{tabular}					 
\tablefoot{Comparison of the combined probability distribution of each parameter for the various subgroups. {\bf{In bold we indicate the statistically significant differences.}}}
\label{stalev_tab_kdl}
\end{table}

\clearpage

\section{Derived torus/disc sizes and masses}
\label{equations}
For each torus model, we can derive the physical radius of the dusty torus (R$_o$) by using the radial extent of the torus (Y=R$_o$/R$_d$), the bolometric luminosity and the dust sublimation radius (R$_d$) definition (see e.g. RA11, AH11, \citealt{Ichikawa15}, \citealt{Esparza-Arredondo19}, GB19, GM19B, \citealt{Martinez-Paredes21}). Finally, the torus gas mass for each model can be calculated as follows:

\subsection{Smooth F06 torus models}

The torus gas mass associated with the fitted nuclear dusty structure can be calculated as:

\begin{equation}
  \begin{aligned}
{\frac{M_{torus}}{ M_\odot}} = {\frac{4 \pi~m_H~(1.9 \times 10^{21}) ~1.086~(\tau_{9.7}/0.042)~R_d^2}{1.989 \times 10^{30}}}\\\int_{0}^{\pi/2} e^{-\gamma\left|\cos\alpha\right|}\cos\alpha \,d\alpha\\
\int_{1}^{Y_0} r^{(2-q)} \,dr
  \end{aligned}
\end{equation}
\noindent where the 1.9$\times10^{21}$ value is the Galactic dust-to-gas ratio from \citet{Bohlin78}.

\subsection{Clumpy N08 torus models}

The torus gas mass associated with the fitted nuclear dusty structure can be calculated as:

\begin{equation}
 \begin{aligned}
{\frac{M_{torus}}{ M_\odot}} = {\frac{4 \pi~m_H~N_{H}^{equatorial}~R_d^2}{1.989 \times 10^{30}}}\int_{0}^{\pi/2} e^{-\beta^2/\sigma^2}\cos\beta \,d\beta\\
\int_{1}^{Y_0} r^{(2-q)} \,dr
  \end{aligned}
\end{equation}

\noindent where $N_{H}^{equatorial}$ is:

\begin{equation}
N_{H}^{equatorial} = (1.9 \times 10^{21}) ~1.086 ~N_0 ~\tau_{V}~[cm^{-2}]
\end{equation}

\noindent where the 1.9$\times10^{21}$ value is the Galactic dust-to-gas ratio from \citet{Bohlin78}.

\subsection{Clumpy H10 torus models}

The torus gas mass associated with the fitted nuclear dusty structure can be calculated as (assuming b$=$1; see H10):

\begin{equation*}
  \begin{aligned}
{\frac{M_{torus}}{ M_\odot}} = {\frac{\pi^{1/2}~m_H}{R_{cl}^2 1.989 \times 10^{30}}}~\left(N_{H}^{equatorial} Erf (\theta)\right)
  \end{aligned}
\end{equation*}

\noindent where $N_{H}^{equatorial}$ is:

\begin{equation}
N_{H}^{equatorial} = (1.9 \times 10^{21}) ~1.086 ~N_0 ~\tau_{V}~[cm^{-2}]
\end{equation}

\noindent where the 1.9$\times10^{21}$ value is the Galactic dust-to-gas ratio from \citet{Bohlin78}.

\subsection{Clumpy disc$+$wind H17 models}

The disc$+$wind gas mass associated with the fitted nuclear dusty structure can be calculated as (assuming b$=$1; see H10):

\begin{equation*}
  \begin{aligned}
{\frac{M_{torus}}{ M_\odot}} = {\frac{\pi^{1/2}~m_H}{R_{cl}^2 1.989 \times 10^{30}}}~\\
\left(N_{H_w}^{equatorial} (Erf (\theta+\sigma_w)-Erf (\theta)) + N_{H_d}^{equatorial} Erf (\sigma_d)\right)
  \end{aligned}
\end{equation*}

\noindent where $N_{H}^{equatorial}$ is:

\begin{equation}
N_{H}^{equatorial} = (1.9 \times 10^{21}) ~1.086 ~N_0 ~\tau_{V}~[cm^{-2}]
\end{equation}

\noindent where the 1.9$\times10^{21}$ value is the Galactic dust-to-gas ratio from \citet{Bohlin78}.

\subsection{Two-phase S16 torus models}

The torus gas mass associated with the fitted nuclear dusty structure can be calculated as:

\begin{equation}
  \begin{aligned}
{\frac{M_{torus}}{ M_\odot}} = {\frac{4 \pi~m_H~(1.9 \times 10^{21}) ~1.086~(\tau_{9.7}/0.042)~R_d^2}{1.989 \times 10^{30}}}\\\int_{0}^{\pi/2} e^{-\gamma\left|\cos\alpha\right|}\cos\alpha \,d\alpha\\
\int_{1}^{Y_0} r^{(2-q)} \,dr
  \end{aligned}
\end{equation}
\noindent where the 1.9$\times10^{21}$ value is the Galactic dust-to-gas ratio from \citet{Bohlin78}.

\end{appendix}


\begin{thebibliography}{}

\bibitem[\protect\citeauthoryear{Almeyda et al.}{2017}]{Almeyda17} Almeyda T., Robinson A., Richmond M., Vazquez B., Nikutta R., 2017, ApJ, 843, 3. doi:10.3847/1538-4357/aa7687

\bibitem[\protect\citeauthoryear{Alonso-Herrero et al.}{2021}]{Herrero21} Alonso-Herrero A., Garc{\'\i}a-Burillo S., H{\"o}nig S.~F., Garc{\'\i}a-Bernete I., Ramos Almeida C., Gonz{\'a}lez-Mart{\'\i}n O., L{\'o}pez-Rodr{\'\i}guez E., et al., 2021, A\&A, 652, A99. doi:10.1051/0004-6361/202141219



\bibitem[\protect\citeauthoryear{Alonso-Herrero et al.}{2020}]{Alonso-Herrero20} Alonso-Herrero A., Pereira-Santaella M., Rigopoulou D., Garc{\'\i}a-Bernete I., Garc{\'\i}a-Burillo S., Dom{\'\i}nguez-Fern{\'a}ndez A.~J., Combes F., et al., 2020, A\&A, 639, A43. doi:10.1051/0004-6361/202037642

\bibitem[\protect\citeauthoryear{Alonso-Herrero et al.}{2019}]{Herrero19} Alonso-Herrero A., Garc{\'\i}a-Burillo S., Pereira-Santaella M., Davies R.~I., Combes F., Vestergaard M., Raimundo S.~I., et al., 2019, A\&A, 628, A65. doi:10.1051/0004-6361/201935431

\bibitem[\protect\citeauthoryear{Alonso-Herrero et al.}{2018}]{Herrero18} Alonso-Herrero A., et al., 2018, ApJ, 859, 144 


\bibitem[\protect\citeauthoryear{Alonso-Herrero et al.}{2011}]{Herrero11} Alonso-Herrero A., et al., 2011, ApJ, 736, 82 425, 311 (AH11)

\bibitem[\protect\citeauthoryear{Alonso-Herrero et al.}{2003}]{herrero03} Alonso-Herrero A., Quillen A.~C., Rieke G.~H., Ivanov V.~D., Efstathiou A., 2003, AJ, 126, 81. doi:10.1086/375545
\bibitem[\protect\citeauthoryear{Alonso-Herrero et al.}{1998}]{herrero98} Alonso-Herrero A., Simpson C., Ward M.~J., Wilson A.~S., 1998, ApJ, 495, 196. doi:10.1086/305269


\bibitem[\protect\citeauthoryear{Antonucci}{1993}]{Antonucci93} Antonucci R., 1993, ARA\&A, 31, 473

\bibitem[Arnaud(1996)]{Arnaud96} Arnaud, K.~A.\ 1996, Astronomical Data Analysis Software and Systems V, 101, 17

\bibitem[\protect\citeauthoryear{Asmus et al.}{2014}]{Asmus14} Asmus D., H{\"o}nig S.~F., Gandhi P., Smette A., Duschl W.~J., 2014, MNRAS, 439, 1648. doi:10.1093/mnras/stu041

\bibitem[\protect\citeauthoryear{Asmus et al.}{2019}]{Asmus19} Asmus D., 2019, MNRAS, 489, 2177. doi:10.1093/mnras/stz2289


\bibitem[\protect\citeauthoryear{Binggeli, Sandage, \& Tammann}{1985}]{Binggeli85} Binggeli B., Sandage A., Tammann G.~A., 1985, AJ, 90, 1681 

\bibitem[\protect\citeauthoryear{Bock et al.}{2000}]{Bock00} Bock J.~J., Neugebauer G., Matthews K., Soifer B.~T., Becklin E.~E., Ressler M., Marsh K., et al., 2000, AJ, 120, 2904. doi:10.1086/316871

\bibitem[\protect\citeauthoryear{Bohlin, Savage, \& Drake}{1978}]{Bohlin78} Bohlin R.~C., Savage B.~D., Drake J.~F., 1978, ApJ, 224, 132. doi:10.1086/156357

\bibitem[\protect\citeauthoryear{Buchner \& Bauer}{2017}]{Buchner17} Buchner J., Bauer F.~E., 2017, MNRAS, 465, 434 
 
\bibitem[\protect\citeauthoryear{Burtscher et al.}{2013}]{Burtscher13} Burtscher L., et al., 2013, A\&A, 558, A149 

\bibitem[\protect\citeauthoryear{Burtscher et al.}{2009}]{Burtscher09} Burtscher L., Jaffe W., Raban D., Meisenheimer K., Tristram K.~R.~W., R{\"o}ttgering H., 2009, ApJ, 705, L53 

\bibitem[\protect\citeauthoryear{Combes et al.}{2019}]{Combes19} Combes F., Garc{\'\i}a-Burillo S., Audibert A., Hunt L., Eckart A., Aalto S., Casasola V., et al., 2019, A\&A, 623, A79. doi:10.1051/0004-6361/201834560

\bibitem[\protect\citeauthoryear{Dorodnitsyn \& Kallman}{2017}]{Dorodnitsyn17} Dorodnitsyn A., Kallman T., 2017, ApJ, 842, 43. doi:10.3847/1538-4357/aa7264

\bibitem[\protect\citeauthoryear{Edelson \& Malkan}{1986}]{Edelson86} Edelson R.~A., Malkan M.~A., 1986, ApJ, 308, 59. doi:10.1086/164479

\bibitem[\protect\citeauthoryear{Efstathiou 
\& Rowan-Robinson}{1995}]{Efstathiou95} Efstathiou A., Rowan-Robinson M., 1995, MNRAS, 273, 649 

\bibitem[\protect\citeauthoryear{Emmanoulopoulos et al.}{2016}]{Emmanoulopoulos16} Emmanoulopoulos D., Papadakis I.~E., Epitropakis A., Pech{\'a}{\v{c}}ek T., Dov{\v{c}}iak M., McHardy I.~M., 2016, MNRAS, 461, 1642. doi:10.1093/mnras/stw1359

\bibitem[\protect\citeauthoryear{Esparza-Arredondo et al.}{2019}]{Esparza-Arredondo19} Esparza-Arredondo D., Gonz{\'a}lez-Mart{\'\i}n O., Dultzin D., Ramos Almeida C., Fritz J., Masegosa J., Pasetto A., et al., 2019, ApJ, 886, 125. doi:10.3847/1538-4357/ab4ced

\bibitem[Esparza-Arredondo et al.(2021)]{Esparza-Arredondo21} Esparza-Arredondo, D., Gonzalez-Mart{\'\i}n, O., Dultzin, D., et al.\ 2021, \aap, 651, A91. doi:10.1051/0004-6361/202040043

\bibitem[\protect\citeauthoryear{Fabian, Vasudevan, \& Gandhi}{2008}]{Fabian08} Fabian A.~C., Vasudevan R.~V., Gandhi P., 2008, MNRAS, 385, L43. doi:10.1111/j.1745-3933.2008.00430.x

\bibitem[Fritz et al.(2006)]{Fritz06} Fritz, J., Franceschini, A., \& Hatziminaoglou, E.\ 2006, \mnras, 366, 767

\bibitem[\protect\citeauthoryear{Fuller et al.}{2016}]{Fuller16} Fuller L., et al., 2016, MNRAS, 462, 2618 

\bibitem[\protect\citeauthoryear{Gallimore et al.}{2016}]{Gallimore16} Gallimore J.~F., Elitzur M., Maiolino R., Marconi A., O'Dea C.~P., Lutz D., Baum S.~A., et al., 2016, ApJL, 829, L7. doi:10.3847/2041-8205/829/1/L7

\bibitem[\protect\citeauthoryear{G{\'a}mez Rosas et al.}{2022}]{Rosas22} G{\'a}mez Rosas V., Isbell J.~W., Jaffe W., Petrov R.~G., Leftley J.~H., Hofmann K.-H., Millour F., et al., 2022, Natur, 602, 403. doi:10.1038/s41586-021-04311-7

\bibitem[\protect\citeauthoryear{Garc{\'\i}a-Bernete et al.}{2019}]{Bernete19} Garc{\'\i}a-Bernete I., et al., 2019, MNRAS, 486, 4917 (GB19)

\bibitem[\protect\citeauthoryear{Garc{\'{\i}}a-Bernete et al.}{2017}]{Bernete17} Garc{\'{\i}}a-Bernete I., Ramos Almeida C., Landt H., Ward M.~J., Balokovi{\'c} M., Acosta-Pulido J.~A., 2017, MNRAS, 469, 110 

\bibitem[\protect\citeauthoryear{Garc{\'{\i}}a-Bernete et al.}{2016}]{Bernete16} Garc{\'{\i}}a-Bernete I., et al., 2016, MNRAS, 463, 3531

\bibitem[\protect\citeauthoryear{Garc{\'{\i}}a-Bernete et al.}{2015}]{Bernete2015} Garc{\'{\i}}a-Bernete I., et al., 2015, MNRAS, 449, 1309 

\bibitem[\protect\citeauthoryear{Garc{\'\i}a-Burillo et al.}{2021}]{Garcia-Burillo21} Garc{\'\i}a-Burillo S., Alonso-Herrero A., Ramos Almeida C., Gonz{\'a}lez-Mart{\'\i}n O., Combes F., Usero A., H{\"o}nig S., et al., 2021, A\&A, 652, A98. doi:10.1051/0004-6361/202141075


\bibitem[\protect\citeauthoryear{Garc{\'\i}a-Burillo et al.}{2016}]{Garcia-Burillo16} Garc{\'{\i}}a-Burillo S., et al., 2016, ApJ, 823, L12 

\bibitem[\protect\citeauthoryear{Garc{\'\i}a-Gonz{\'a}lez et al.}{2017}]{Garcia-Gonzalez17} Garc{\'\i}a-Gonz{\'a}lez J., Alonso-Herrero A., H{\"o}nig S.~F., Hern{\'a}n-Caballero A., Ramos Almeida C., Levenson N.~A., Roche P.~F., et al., 2017, MNRAS, 470, 2578. doi:10.1093/mnras/stx1361

\bibitem[Gonz{\'a}lez-Mart{\'\i}n et al.(2019A)]{Gonzalez-Martin19A} Gonz{\'a}lez-Mart{\'\i}n, O., Masegosa, J., Garc{\'\i}a-Bernete, I., et al.\ 2019A, \apj, 884, 10 (GM19A)


\bibitem[Gonz{\'a}lez-Mart{\'\i}n et al.(2019B)]{Gonzalez-Martin19B} Gonz{\'a}lez-Mart{\'\i}n, O., Masegosa, J., Garc{\'\i}a-Bernete, I., et al.\ 2019B, \apj, 884, 11 (GM19B)

\bibitem[\protect\citeauthoryear{Gonz{\'a}lez-Mart{\'\i}n et al.}{2013}]{gonzalez-martin13} Gonz{\'a}lez-Mart{\'\i}n O., Rodr{\'\i}guez-Espinosa J.~M., D{\'\i}az-Santos T., Packham C., Alonso-Herrero A., Esquej P., Ramos Almeida C., et al., 2013, A\&A, 553, A35. doi:10.1051/0004-6361/201220382

\bibitem[\protect\citeauthoryear{Hern{\'a}n-Caballero et al.}{2016}]{caballero16} Hern{\'a}n-Caballero A., Hatziminaoglou E., Alonso-Herrero A., Mateos S., 2016, MNRAS, 463, 2064. doi:10.1093/mnras/stw2107

\bibitem[\protect\citeauthoryear{Hickox et al.}{2014}]{Hickox14} Hickox R.~C., Mullaney J.~R., Alexander D.~M., Chen C.-T.~J., Civano F.~M., Goulding A.~D., Hainline K.~N., 2014, ApJ, 782, 9. doi:10.1088/0004-637X/782/1/9

\bibitem[\protect\citeauthoryear{Hicks et al.}{2009}]{Hicks09} Hicks E.~K.~S., Davies R.~I., Malkan M.~A., Genzel R., Tacconi L.~J., M{\"u}ller S{\'a}nchez F., Sternberg A., 2009, ApJ, 696, 448 

\bibitem[\protect\citeauthoryear{H{\"o}nig}{2019}]{Honig19} H{\"o}nig S.~F., 2019, ApJ, 884, 171. doi:10.3847/1538-4357/ab4591

\bibitem[H{\"o}nig \& Kishimoto(2017)]{Hoenig17} H{\"o}nig, S.~F., \& Kishimoto, M.\ 2017, \apjl, 838, L20

\bibitem[\protect\citeauthoryear{H{\"o}nig et al.}{2013}]{Hoenig13} H{\"o}nig S.~F., Kishimoto M., Tristram K.~R.~W., Prieto M.~A., Gandhi P., Asmus D., Antonucci R., et al., 2013, ApJ, 771, 87. doi:10.1088/0004-637X/771/2/87

\bibitem[H{\"o}nig \& Kishimoto(2010)]{Hoenig10B} H{\"o}nig, S.~F., \& Kishimoto, M.\ 2010, \aap, 523, A27

\bibitem[H{\"o}nig et al.(2010)]{Hoenig10A} H{\"o}nig, S.~F., Kishimoto, M., Gandhi, P., et al.\ 2010, \aap, 515, A23

\bibitem[H{\"o}nig et al.(2006)]{Hoenig06} H{\"o}nig, S.~F., Beckert, T., Ohnaka, K., \& Weigelt, G.\ 2006, \aap, 452, 459 

\bibitem[\protect\citeauthoryear{Hopkins \& Quataert}{2010}]{HopkinsQuataert10} Hopkins P.~F., Quataert E., 2010, MNRAS, 407, 1529. doi:10.1111/j.1365-2966.2010.17064.x

\bibitem[\protect\citeauthoryear{Hopkins et al.}{2005}]{Hopkins05} Hopkins P.~F., Hernquist L., Cox T.~J., Di Matteo T., Martini P., Robertson B., Springel V., 2005, ApJ, 630, 705. doi:10.1086/432438

\bibitem[\protect\citeauthoryear{Ichikawa et al.}{2018}]{Ichikawa18} Ichikawa K., et al., 2018, arXiv, arXiv:1811.02568 

\bibitem[\protect\citeauthoryear{Ichikawa et al.}{2015}]{Ichikawa15} Ichikawa K., et al., 2015, ApJ, 803, 57 

\bibitem[\protect\citeauthoryear{Ichikawa et al.}{2012}]{Ichikawa2012} Ichikawa K., Ueda Y., Terashima Y., Oyabu 
S., Gandhi P., Matsuta K., Nakagawa T., 2012, ApJ, 754, 45 

\bibitem[\protect\citeauthoryear{Imanishi et al.}{2018}]{Imanishi18} Imanishi M., Nakanishi K., Izumi T., Wada K., 2018, ApJL, 853, L25. doi:10.3847/2041-8213/aaa8df

\bibitem[\protect\citeauthoryear{Impellizzeri et al.}{2019}]{Impellizzeri19} Impellizzeri C.~M.~V., Gallimore J.~F., Baum S.~A., Elitzur M., Davies R., Lutz D., Maiolino R., et al., 2019, ApJL, 884, L28. doi:10.3847/2041-8213/ab3c64

\bibitem[\protect\citeauthoryear{Isbell et al.}{2022}]{Isbell22} Isbell J.~W., Meisenheimer K., Pott J.-U., Stalevski M., Tristram K.~R.~W., Sanchez-Bermudez J., Hofmann K.-H., et al., 2022, arXiv, arXiv:2205.01575

\bibitem[\protect\citeauthoryear{Isbell et al.}{2021}]{Isbell21} Isbell J.~W., Burtcher L., Asmus D., Pott J.-U., Couzy P., Stalevski M., G{\'a}mez Rosas V., et al., 2021, arXiv, arXiv:2101.07006

\bibitem[\protect\citeauthoryear{Jaffe et al.}{2004}]{Jaffe04} Jaffe W., et al., 2004, Natur, 429, 47 

\bibitem[\protect\citeauthoryear{Koss et al.}{2017}]{Koss17} Koss M., Trakhtenbrot B., Ricci C., Lamperti I., Oh K., Berney S., Schawinski K., et al., 2017, ApJ, 850, 74. doi:10.3847/1538-4357/aa8ec9

\bibitem[\protect\citeauthoryear{Krolik \& Begelman}{1988}]{Krolik88} Krolik J.~H., Begelman M.~C., 1988, ApJ, 329, 702. doi:10.1086/166414

\bibitem[\protect\citeauthoryear{Kudoh, Wada, \& Norman}{2020}]{Kudoh20} Kudoh Y., Wada K., Norman C., 2020, ApJ, 904, 9. doi:10.3847/1538-4357/abba39

\bibitem[\protect\citeauthoryear{Kullback 	\& Leibler}{1951}]{Kullback51} Kullback, S., \& Leibler, A., 1951, Ann. Math. Stat, 22, 79

\bibitem[\protect\citeauthoryear{Landt et al.}{2019}]{Landt19} Landt H., Ward M.~J., Kynoch D., Packham C., Ferland G.~J., Lawrence A., Pott J.-U., et al., 2019, MNRAS, 489, 1572. doi:10.1093/mnras/stz2212

\bibitem[\protect\citeauthoryear{Lani, Netzer, \& Lutz}{2017}]{Lani17} Lani C., Netzer H., Lutz D., 2017, MNRAS, 471, 59 

\bibitem[\protect\citeauthoryear{Lawrence}{1991}]{Lawrence91} Lawrence A., 1991, MNRAS, 252, 586. doi:10.1093/mnras/252.4.586

\bibitem[\protect\citeauthoryear{Leftley et al.}{2018}]{Leftley18} Leftley J.~H., Tristram K.~R.~W., H{\"o}nig S.~F., Kishimoto M., Asmus D., Gandhi P., 2018, ApJ, 862, 17 

\bibitem[\protect\citeauthoryear{Lira et al.}{2013}]{Lira13} Lira P., Videla L., Wu Y., Alonso-Herrero A., Alexander D.~M., Ward M., 2013, ApJ, 764, 159 

\bibitem[\protect\citeauthoryear{L{\'o}pez-Gonzaga et al.}{2016}]{Lopez-Gonzaga16} L{\'o}pez-Gonzaga N., Burtscher L., Tristram K.~R.~W., Meisenheimer K., Schartmann M., 2016, A\&A, 591, A47 

\bibitem[\protect\citeauthoryear{Lopez-Rodriguez et al.}{2018}]{Lopez-Rodriguez18} Lopez-Rodriguez E., Fuller L., Alonso-Herrero A., Efstathiou A., Ichikawa K., Levenson N.~A., Packham C., et al., 2018, ApJ, 859, 99. doi:10.3847/1538-4357/aabd7b

\bibitem[\protect\citeauthoryear{Lu, Kumar, \& Evans}{2016}]{Lu16} Lu W., Kumar P., Evans N.~J., 2016, MNRAS, 458, 575. doi:10.1093/mnras/stw307


\bibitem[\protect\citeauthoryear{Mart{\'\i}nez-Paredes et al.}{2021}]{Martinez-Paredes21} Mart{\'\i}nez-Paredes M., Gonz{\'a}lez-Mart{\'\i}n O., HyeongHan K., Geier S., Garc{\'\i}a-Bernete I., Ramos Almeida C., Alonso-Herrero A., et al., 2021, ApJ, 922, 157. doi:10.3847/1538-4357/ac1d55

\bibitem[\protect\citeauthoryear{Mart{\'\i}nez-Paredes et al.}{2020}]{Martinez-Paredes20} Mart{\'\i}nez-Paredes M., Gonz{\'a}lez-Mart{\'\i}n O., Esparza-Arredondo D., Kim M., Alonso-Herrero A., Krongold Y., Hoang T., et al., 2020, ApJ, 890, 152. doi:10.3847/1538-4357/ab6732

\bibitem[\protect\citeauthoryear{Mateos et al.}{2017}]{Mateos17} Mateos S., et al., 2017, ApJ, 841, L18 

\bibitem[\protect\citeauthoryear{Mateos et al.}{2016}]{Mateos16} Mateos S., et al., 2016, ApJ, 819, 166 

\bibitem[\protect\citeauthoryear{Mathis, Rumpl, \& Nordsieck}{1977}]{Mathis77} Mathis J.~S., Rumpl W., Nordsieck K.~H., 1977, ApJ, 217, 425. doi:10.1086/155591

\bibitem[\protect\citeauthoryear{Mor, Netzer, \& Elitzur}{2009}]{Mor09} Mor R., Netzer H., Elitzur M., 2009, ApJ, 705, 298 

\bibitem[\protect\citeauthoryear{Mould et al.}{2000}]{Mould00} 
Mould J.~R., et al., 2000, ApJ, 529, 786 

\bibitem[Nenkova et al.(2008a)]{Nenkova08A} Nenkova, M., Sirocky, M.~M., Ivezi{\'c}, {\v Z}., \& Elitzur, M.\ 2008, \apj, 685, 147-159

\bibitem[Nenkova et al.(2008b)]{Nenkova08B} Nenkova, M., Sirocky, M.~M., Nikutta, R., Ivezi{\'c}, {\v Z}., \& Elitzur, M.\ 2008, \apj, 685, 160-180 

\bibitem[\protect\citeauthoryear{Netzer et al.}{2016}]{Netzer16} Netzer H., Lani C., Nordon R., Trakhtenbrot B., Lira P., Shemmer O., 2016, ApJ, 819, 123

\bibitem[\protect\citeauthoryear{Neugebauer et al.}{1979}]{Neugebauer79} Neugebauer G., Oke J.~B., Becklin E.~E., Matthews K., 1979, ApJ, 230, 79. doi:10.1086/157063

\bibitem[\protect\citeauthoryear{Nikutta et al.}{2021}]{Nikutta21} Nikutta R., Lopez-Rodriguez E., Ichikawa K., Levenson N.~A., Packham C.~C., 2021, IAUS, 356, 44. doi:10.1017/S1743921320002550

\bibitem[\protect\citeauthoryear{Packham et 
al.}{2005}]{Packham05} Packham C., Radomski J.~T., Roche P.~F., 
Aitken D.~K., Perlman E., Alonso-Herrero A., Colina L., Telesco C.~M., 
2005, ApJ, 618, L17 

\bibitem[Pei(1992)]{Pei92} Pei, Y.~C.\ 1992, \apj, 395, 130

\bibitem[\protect\citeauthoryear{Perna, Lazzati, \& Fiore}{2003}]{Perna03} Perna R., Lazzati D., Fiore F., 2003, ApJ, 585, 775. doi:10.1086/346109

\bibitem[\protect\citeauthoryear{Pier 
\& Krolik}{1992}]{Pier92} Pier E.~A., Krolik J.~H., 1992, ApJ, 401, 99 

\bibitem[\protect\citeauthoryear{Raban et al.}{2009}]{Raban09} Raban D., Jaffe W., R{\"o}ttgering H., Meisenheimer K., Tristram K.~R.~W., 2009, MNRAS, 394, 1325 

\bibitem[\protect\citeauthoryear{Radomski et al.}{2008}]{Radomski2008} Radomski J.~T., et al., 2008, ApJ, 681, 141 

\bibitem[\protect\citeauthoryear{Radomski et al.}{2003}]{Radomski03} Radomski J.~T., Pi{\~n}a R.~K., Packham C., Telesco C.~M., De Buizer J.~M., Fisher R.~S., Robinson A., 2003, ApJ, 587, 117. doi:10.1086/367612

\bibitem[\protect\citeauthoryear{Ramos Almeida \& Ricci}{2017}]{Ramos17} Ramos Almeida C., Ricci C., 2017, NatAs, 1, 679

\bibitem[\protect\citeauthoryear{Ramos Almeida et 
al.}{2014}]{Ramos14} Ramos Almeida C., et al., 2014, MNRAS, 
445, 1130 

\bibitem[\protect\citeauthoryear{Ramos Almeida et 
al.}{2011b}]{Ramos11b} Ramos Almeida C., et al., 2011b, ApJ, 731, 
92 (RA11)

\bibitem[\protect\citeauthoryear{Ramos Almeida et 
al.}{2009}]{Ramos09} Ramos Almeida C., et al., 2009, ApJ, 702, 
1127

\bibitem[\protect\citeauthoryear{Ricci et al.}{2017c}]{Ricci17c} Ricci C., et al., 2017c, Natur, 549, 488

\bibitem[\protect\citeauthoryear{Ricci et al.}{2017}]{Ricci17} Ricci C., et al., 2017, ApJS, 233, 17 

\bibitem[\protect\citeauthoryear{Ricci et al.}{2015}]{Ricci15} Ricci C., Ueda Y., Koss M.~J., Trakhtenbrot B., Bauer F.~E., Gandhi P., 2015, ApJ, 815, L13 

\bibitem[\protect\citeauthoryear{Schartmann et al.}{2014}]{Schartmann14} Schartmann M., Wada K., Prieto M.~A., Burkert A., Tristram K.~R.~W., 2014, MNRAS, 445, 3878. doi:10.1093/mnras/stu2020

\bibitem[\protect\citeauthoryear{Schartmann et al.}{2008}]{Schartmann08} Schartmann M., Meisenheimer K., Camenzind M., Wolf S., Tristram K.~R.~W., Henning T., 2008, A\&A, 482, 67

\bibitem[\protect\citeauthoryear{Siebenmorgen, Heymann, 
\& Efstathiou}{2015}]{Siebenmorgen2015} Siebenmorgen R., Heymann F., Efstathiou A., 2015, A\&A, 583, A120 

\bibitem[\protect\citeauthoryear{Simpson}{2005}]{Simpson05} Simpson C., 2005, MNRAS, 360, 565. doi:10.1111/j.1365-2966.2005.09043.x

\bibitem[\protect\citeauthoryear{Stalevski, Asmus, \& Tristram}{2017}]{Stalevski17} Stalevski M., Asmus D., Tristram K.~R.~W., 2017, MNRAS, 472, 3854. doi:10.1093/mnras/stx2227


\bibitem[\protect\citeauthoryear{Stalevski et al.}{2016}]{Stalevski16} Stalevski M., Ricci C., Ueda Y., Lira P., Fritz J., Baes M., 2016, MNRAS, 458, 2288 

\bibitem[\protect\citeauthoryear{Stalevski et 
al.}{2012}]{Stalevski2012} Stalevski M., Fritz J., Baes M., Nakos T., 
Popovi{\'c} L.~{\v C}., 2012, MNRAS, 420, 2756

\bibitem[\protect\citeauthoryear{Takasao, Shuto, \& Wada}{2022}]{Takasao22} Takasao S., Shuto Y., Wada K., 2022, ApJ, 926, 50. doi:10.3847/1538-4357/ac38a8

\bibitem[\protect\citeauthoryear{Tristram et al.}{2014}]{Tristram14} Tristram K.~R.~W., Burtscher L., Jaffe W., Meisenheimer K., H{\"o}nig S.~F., Kishimoto M., Schartmann M., et al., 2014, A\&A, 563, A82. doi:10.1051/0004-6361/201322698
 
\bibitem[\protect\citeauthoryear{Tristram et al.}{2009}]{Tristram09} Tristram, K. R. W. et al. 2009, A\&A, 502, 67

\bibitem[\protect\citeauthoryear{Tristram et al.}{2007}]{Tristram07} Tristram K.~R.~W., et al., 2007, A\&A, 474, 837 


\bibitem[\protect\citeauthoryear{Tueller et 
al.}{2008}]{Tueller2008} Tueller J., Mushotzky R.~F., Barthelmy S., 
Cannizzo J.~K., Gehrels N., Markwardt C.~B., Skinner G.~K., Winter L.~M.,
2008, ApJ, 681, 113 

\bibitem[\protect\citeauthoryear{Ueda et al.}{2015}]{Ueda15} 
Ueda Y., et al., 2015, ApJ, 815, 1 

\bibitem[\protect\citeauthoryear{Vasudevan et al.}{2010}]{Vasudevan10} Vasudevan R.~V., Fabian A.~C., Gandhi P., Winter L.~M., Mushotzky R.~F., 2010, MNRAS, 402, 1081. doi:10.1111/j.1365-2966.2009.15936.x

\bibitem[\protect\citeauthoryear{Vasudevan \& Fabian}{2009}]{Vasudevan09} Vasudevan R.~V., Fabian A.~C., 2009, MNRAS, 392, 1124. doi:10.1111/j.1365-2966.2008.14108.x

\bibitem[\protect\citeauthoryear{Venanzi, H{\"o}nig, \& Williamson}{2020}]{Venanzi20} Venanzi M., H{\"o}nig S., Williamson D., 2020, ApJ, 900, 174. doi:10.3847/1538-4357/aba89f

\bibitem[\protect\citeauthoryear{Wada}{2012}]{Wada12} Wada K., 2012, ApJ, 758, 66 

\bibitem[\protect\citeauthoryear{Wada \& Norman}{2002}]{Wada02} Wada K., Norman C.~A., 2002, ApJ, 566, L21 

\bibitem[\protect\citeauthoryear{Waxman \& Draine}{2000}]{Waxman00} Waxman E., Draine B.~T., 2000, ApJ, 537, 796. doi:10.1086/309053

\bibitem[\protect\citeauthoryear{Weaver et al.}{2010}]{Weaver2010} 
Weaver K.~A., et al., 2010, ApJ, 716, 1151 

\bibitem[\protect\citeauthoryear{Winter et al.}{2010}]{Winter2010} 
Winter L.~M., Lewis K.~T., Koss M., Veilleux S., Keeney B., Mushotzky 
R.~F., 2010, ApJ, 710, 503 

\bibitem[\protect\citeauthoryear{Winter et al.}{2009}]{Winter2009} 
Winter L.~M., Mushotzky R.~F., Reynolds C.~S., Tueller J., 2009, ApJ, 690, 
1322 
 
\end{thebibliography}
\end{document}